%% file: paper.tex
\def\layout{scipost}

\def\lutp{LU-TP 22-16}
\def\mcnet{MCNET-22-04}

\def\sp{scipost}
\ifx\layout\sp
\documentclass[submission, PhysCodeb]{SciPost}
\hypersetup{
    colorlinks,
    linkcolor={red!50!black},
    citecolor={blue!50!black},
    urlcolor={blue!80!black}
}

\usepackage[bitstream-charter]{mathdesign}
\DeclareSymbolFont{usualmathcal}{OMS}{cmsy}{m}{n}
\DeclareSymbolFontAlphabet{\mathcal}{usualmathcal}
\definecolor{NavyBlue}{HTML}{000080}

\makeatletter
\def\@maketitle{%
  \newpage
  \null
  \hspace*{\fill}\lutp\\
  \hspace*{\fill}\mcnet
  \vskip 2em%
  \begin{center}%
  \let \footnote \thanks
  {\Large \textbf{\@title} \par}%
    \vskip 1.5em%
    {\large
      \lineskip .5em%
      \begin{tabular}[t]{c}%
        \@author
      \end{tabular}\par}%
    {\large \@date}%
  \end{center}%
  \par
  \vskip 1.5em}
\makeatother

\renewenvironment{abstract}
{\section*{Abstract}
\bfseries}
{}
\else
\documentclass[a4paper]{article}
\usepackage[english]{babel}
\usepackage[utf8x]{inputenc}
\usepackage[T1]{fontenc}
\usepackage[comma,square,numbers,sort&compress]{natbib}
\usepackage[dvipsnames]{xcolor}

\makeatletter
\def\@maketitle{%
  \newpage
  \null
  \hspace*{\fill}\lutp\\
  \hspace*{\fill}\mcnet
  \vskip 2em%
  \begin{center}%
  \let \footnote \thanks
  {\LARGE {\@title} \par}%
    \vskip 1.5em%
    {\large
      \lineskip .5em%
      \begin{tabular}[t]{c}%
        \@author
      \end{tabular}\par}%
  \end{center}%
  \par
  \vskip 1.5em}
\makeatother

\usepackage[a4paper,top=3cm,bottom=2cm,left=3cm,right=3cm,marginparwidth=1.75cm]{geometry}
\usepackage{amsmath}
\usepackage{amssymb}
\usepackage[colorlinks=true, allcolors=blue]{hyperref}

\fi

\usepackage{xspace}
\usepackage{adjustbox}
\usepackage{relsize}
\usepackage{booktabs}
\usepackage{multirow}
\usepackage{enumitem}
\usepackage{makeidx}
\makeindex

\usepackage{upgreek}
\usepackage{graphicx}
\usepackage{import}
\usepackage{authblk}
\usepackage{tcolorbox}
\usepackage{float}
\usepackage{tikz}
\usepackage{multicol}
\usepackage[printonlyused,withpage]{acronym}
\usepackage{cleveref}

\makeatletter
\AtBeginDocument{%
  \renewcommand*{\AC@hyperlink}[2]{%
    \begingroup
      \hypersetup{hidelinks}%
      \hyperlink{#1}{#2}%
    \endgroup
  }%
}
\makeatother

\newtcolorbox{codebox}[0]{%
    colbacktitle=NavyBlue!10,
    fontupper=\ttfamily,
    colback=gray!10,
    colframe=gray!10,
    sharp corners }

\usepackage{tikz}
\usetikzlibrary{tikzmark}
\usetikzlibrary{calc} 
\usetikzlibrary{decorations.pathmorphing}

\usepackage{tikz-feynman}
\tikzfeynmanset{compat=1.0.0}
\tikzstyle{nobox} = [rectangle, minimum width=0cm, minimum height=0.0cm, text centered, draw=black!00, fill=black!00, inner sep=1pt, inner ysep=1pt]
\tikzstyle{description} = [rectangle, minimum width=1cm, minimum height=0.5cm, draw=black, fill=green!05, inner sep=2pt, inner ysep=2pt, inner xsep=2pt]

\include{definitions}

\begin{document}
\title{A comprehensive guide to the physics and usage of \pythia}

\author[1]{Christian~Bierlich}
\author[1]{Smita~Chakraborty}
\author[2]{Nishita~Desai}
\author[1]{Leif~Gellersen}
\author[3,4]{Ilkka~Helenius}
\author[9]{Philip~Ilten}
\author[1]{Leif~L\"onnblad}
\author[5]{Stephen~Mrenna}
\author[1]{Stefan~Prestel}
\author[6,7]{Christian~T.~Preuss}
\author[1]{Torbj\"orn~Sj\"ostrand}
\author[6]{Peter~Skands}
\author[1,3]{Marius Utheim}
\author[8]{Rob~Verheyen}

\affil[1]{Dept. of Astronomy and Theoretical Physics,
  Lund University, S{\" o}lvegatan 14A, S-223 62 Lund, Sweden}
\affil[2]{Tata Institute of Fundamental Research, Homi Bhabha Road, Mumbai 400005, India}
\affil[3]{University of Jyvaskyla, Department of Physics, P.O. Box 35, FI-40014 University of Jyvaskyla, Finland}
\affil[4]{Helsinki Institute of Physics, P.O. Box 64, FI-00014 University of Helsinki, Finland}
\affil[5]{Fermilab, Batavia, Illinois, USA}
\affil[6]{School of Physics and Astronomy, Monash University,
  Wellington Rd, Clayton VIC-3800, Australia}
\affil[7]{Institute for Theoretical Physics, ETH, CH-8093 Z\"urich, Switzerland}
\affil[8]{Dept. of Physics and Astronomy, UCL, Gower St, Bloomsbury, London WC1E 6BT, United Kingdom}
\affil[9]{Dept. of Physics, University of Cincinnati, Cincinnati, OH 45221, USA}

\maketitle
\begin{abstract}
This manual describes the \pythia event generator, the most recent
 version of an evolving physics tool used to answer fundamental questions
 in particle physics. The program is most often used to generate high-energy-physics
collision ``events'', \ie sets of particles produced in association with the collision
of two incoming high-energy particles, but has several uses beyond that. The guiding
philosophy is to produce and re-produce properties of experimentally obtained collisions
as accurately as possible. The program includes a wide ranges of reactions within and
beyond the Standard Model, and extending to heavy ion physics. Emphasis is put
on phenomena where strong interactions play a major role.

The manual contains both pedagogical and practical components. All included physics models
are described in enough detail to allow the user to obtain a cursory overview of used assumptions
and approximations, enabling an informed evaluation of the program output. A number of the most
central algorithms are described in enough detail that the main results of the program
can be reproduced independently, allowing further development of existing models or the addition of new ones.

Finally, a chapter dedicated fully to the user is included towards the end, providing pedagogical examples
of standard use cases, and a detailed description of a number of external interfaces.
The program code, the online manual, and the latest version of this print manual can be
found on the \pyt web page:\\
\begin{center}
	\url{https://www.pythia.org/}
\end{center}
\end{abstract}
\newpage
\vspace{10pt}
\noindent\rule{\textwidth}{1pt}
\ifx\layout\sp
\tableofcontents\thispagestyle{fancy}
\else
\tableofcontents
\fi
\vspace{10pt}
\noindent\rule{\textwidth}{1pt}
\newpage
\part{Introduction}
\label{part:intro}
\input{introduction/introduction}
\input{introduction/program-structure}

\newpage
\part{Physics content}
\label{part:physics}
\input{physics/introduction}
\input{physics/hard-proc}
\input{physics/parton-showers}

\input{physics/match-merge}
\input{physics/soft-proc}
\input{physics/hadronization}
\input{physics/particle-decays}
\newpage
\part{Using \pythia}
\label{part:use}
\input{using-pythia/introduction}
\input{using-pythia/stand-alone}
\input{using-pythia/interface-external}

\part{Summary and Outlook}
\label{part:summary}
\input{summary/summary}
\newpage
\input{summary/acknowledgements}
\newpage

\part*{Appendices}
\addcontentsline{toc}{part}{Appendices}
\begin{appendix}
\input{physics/hard-proc-table}
\end{appendix}
\clearpage

\begingroup
\def\section*#1{}
\part*{References}
\ifx\layout\sp
\else
\addcontentsline{toc}{part}{References}
\fi
\bibliography{corrected-bib,aux-bib}
\endgroup

\clearpage

\part*{}
\ifx\layout\sp
\else
\addcontentsline{toc}{part}{Index}
\fi
\printindex

\newpage
\ifx\layout\sp
\addcontentsline{toc}{section}{List of acronyms}
\else
\addcontentsline{toc}{part}{List of acronyms}
\fi
\include{acronym-list}

\end{document}

%% file: definitions.tex
\def\mrm{\mathrm}
\def\emrm#1{\ensuremath{\mrm{#1}}}
\def\ebmrm#1{\ensuremath{\overline{\mrm{#1}}}}

\def\gstrong{\ensuremath{g_\mrm{s}}}
\def\alphas{\alpha_\mrm{s}}
\def\alphaem{\alpha_\mrm{em}}

\def\gp#1{\emrm{#1}\xspace}
\def\gpbar#1{\ebmrm{#1}\xspace}
\def\particlepair#1{\ensuremath{\mrm{\gp{#1}\gp{#1}}}\xspace}
\def\particleantipair#1{\ensuremath{\mrm{\gp{#1}\gpbar{#1}}}\xspace}

\def\q{\gp{q}}
\def\qbar{\gpbar{q}}
\def\Q{\gp{Q}}
\def\Qbar{\gpbar{Q}}
\def\g{\gp{g}}
\def\u{\gp{u}}
\def\ubar{\gpbar{u}}
\def\d{\gp{d}}
\def\dbar{\gpbar{d}}
\def\c{\gp{c}}
\def\cbar{\gpbar{c}}
\def\s{\gp{s}}
\def\sbar{\gpbar{s}}
\def\t{\gp{t}}
\def\tbar{\gpbar{t}}
\def\b{\gp{b}}
\def\bbar{\gpbar{b}}
\def\j{\gp{j}}

\def\f{\gp{f}}
\def\fbar{\gpbar{f}}

\def\quarkpair#1{\particleantipair{#1}}
\def\qqbar{\quarkpair{q}}

\def\ssbar{\quarkpair{s}}
\def\ccbar{\quarkpair{c}}
\def\bbbar{\quarkpair{b}}

\def\QQbar{\quarkpair{Q}}

\def\electron{\gp{e}}
\def\ele{\electron}
\def\eplus{\ensuremath{\mrm{e}^+}\xspace}
\def\eminus{\ensuremath{\mrm{e}^-}\xspace}
\def\epm{\ensuremath{\mrm{e}^\pm}\xspace}
\def\epem{\ensuremath{\mrm{e}^+\mrm{e}^-}}

\def\muon{\gp{\mu}}
\def\muplus{\ensuremath{\mrm{\mu}^+}\xspace}
\def\muminus{\ensuremath{\mrm{\mu}^-}\xspace}

\def\tauon{\gp{\tau}}
\def\tauplus{\ensuremath{\mrm{\tau}^+}\xspace}
\def\tauminus{\ensuremath{\mrm{\tau}^-}\xspace}

\def\neu{\gp{\nu}}

\def\pp{\particlepair{p}}
\def\ppbar{\particleantipair{p}}
\def\ep{\ensuremath{\mrm{\electron}\gp{p}}\xspace}
\def\pA{\ensuremath{\mrm{\gp{p}\gp{A}}}\xspace}
\def\AA{\particlepair{A}}
\def\pPb{\ensuremath{\mrm{\gp{p}\gp{Pb}}}\xspace}
\def\PbPb{\particlepair{Pb}}

\def\A{\ensuremath{\gp{A}}\xspace}
\def\G{\ensuremath{\gp{G}}\xspace}
\def\H{\ensuremath{\gp{H}}\xspace}

\def\Hp{\ensuremath{\gp{H}^+}\xspace}
\def\W{\ensuremath{\gp{W}}\xspace}
\def\Wp{\ensuremath{\gp{W}^+}\xspace}
\def\Wm{\ensuremath{\gp{W}^-}\xspace}
\def\Wpm{\ensuremath{\gp{W}^\pm}\xspace}
\def\Z{\ensuremath{\gp{Z}}\xspace}

\def\gam{\ensuremath{\gp{\gamma}\xspace}}
\def\V{\ensuremath{\gp{V}}\xspace}

\def\pion{\gp{\pi}}
\def\piz{\ensuremath{\pion^0}\xspace}
\def\pip{\ensuremath{\pion^+}\xspace}
\def\pim{\ensuremath{\pion^-}\xspace}

\def\rhomeson{\gp{\rho}}
\def\rhoz{\ensuremath{\rhomeson^0}\xspace}

\def\kaon{\gp{K}}
\def\kaonbar{\gpbar{K}}
\def\kz{\ensuremath{\kaon^0}\xspace}
\def\kzbar{\ensuremath{\kaonbar^0}\xspace}
\def\kp{\ensuremath{\kaon^+}\xspace}
\def\km{\ensuremath{\kaon^-}\xspace}
\def\kshort{\ensuremath{\kaon^0_\mrm{S}}\xspace}
\def\klong{\ensuremath{\kaon^0_\mrm{L}}\xspace}

\def\phiz{\ensuremath{\phi^0}\xspace}
\def\Jpsi{\ensuremath{\mathrm{J}/\psi}\xspace}

\def\p{\ensuremath{\gp{p}}}
\def\pbar{\ensuremath{\gpbar{p}}}
\def\n{\ensuremath{\gp{n}}}
\def\nbar{\ensuremath{\gpbar{n}}}
\def\D{\gp{D}}
\def\Dbar{\gpbar{D}}
\def\B{\gp{B}}
\def\Bbar{\gpbar{B}}

\def\Pom{\ensuremath{\mathbb{P}}\xspace}
\def\Reg{\ensuremath{\mathbb{R}}\xspace}

\def\keV{{\ensuremath{\mathrm{ke\kern-.05emV}}}\xspace}
\def\MeV{{\ensuremath{\mathrm{Me\kern-.05emV}}}\xspace}
\def\GeV{{\ensuremath{\mathrm{Ge\kern-.07emV}}}\xspace}
\def\TeV{{\ensuremath{\mathrm{T\kern-.05em e\kern-.07emV}}}\xspace}
\def\kT{{\ensuremath {k_{\perp}}}\xspace}
\newcommand{\pT}[1][]{{\ensuremath {p_{\perp{#1}}}}\xspace}
\newcommand{\pTs}[1][]{{\ensuremath p_{\perp{#1}}^2}\xspace}
\def\pTo{{\ensuremath p_{\perp 0}}\xspace}
\def\pTos{{\ensuremath p_{\perp 0}^2}\xspace}
\def\pTmin{{\ensuremath p_{\perp\mathrm{min}}}\xspace}

\def\pTmax{{\ensuremath p_{\perp\mathrm{max}}}\xspace}

\def\ECM{{\ensuremath E_{\mathrm{CM}}}\xspace}
\def\ECMs{{\ensuremath E_{\mathrm{CM}}^2}\xspace}

\def\fm{{\ensuremath\mathrm{fm}}\xspace}

\def\ME{\ensuremath{\mathcal{M}}\xspace}
\def\antfun{\ensuremath{A}}
\def\antfun#1{\ensuremath{A_{#1}}}
\def\antfunbar#1{\ensuremath{\bar{A}_{#1}}}
\def\sctantfun#1{\ensuremath{A^{\mrm{sct}}_{#1}}}
\def\sctantfunbar#1{\ensuremath{\bar{A}^{\mrm{sct}}_{#1}}}
\def\subantfun#1{\ensuremath{A^{\mrm{gl}}_{#1}}}

\def\sctantfungxsplit{\sctantfun{\q/\g \mrm{X}}}

\def\subantfunqqemit{\subantfun{\g/\q\qbar}}
\def\subantfunqgemit{\subantfun{\g/\q \g}}
\def\subantfunggemit{\subantfun{\g/\g\g}}
\def\subantfungxsplit{\subantfun{\q/\g \mrm{X}}}

\def\dglap#1#2{\ensuremath{P_{#1 \to #2}}}

\def\prob{\ensuremath{\mathcal{P}}}
\def\fpdf{\ensuremath{f}}
\def\RPDF{\ensuremath{R_\mrm{PDF}}}
\def\FPS{\ensuremath{F_\Phi}}
\def\SC{\ensuremath{\mathcal{S}}}
\def\Pmec{\ensuremath{P_\mrm{MEC}}}

\def\colfac{\ensuremath{\mathcal{C}}}
\def\sinv#1{\ensuremath{s_{#1}}}
\def\yinv#1{\ensuremath{y_{#1}}}
\def\sij{\ensuremath{s_{ij}}}

\def\musq#1{\ensuremath{\mu^2_{#1}}}
\def\msq#1{\ensuremath{m^2_{#1}}}

\def\Nc{\ensuremath{N_{\mrm{c}}}}
\def\CA{\ensuremath{C_{\mrm{A}}}}
\def\CF{\ensuremath{C_{\mrm{F}}}}
\def\TR{\ensuremath{T_{\mrm{R}}}}
\def\thetasct{\ensuremath{\Theta^{\mrm{sct}}}}
\def\qsqres#1{\ensuremath{Q^2_{\mrm{res},#1}}}
\def\kallen#1#2#3{\ensuremath{\lambda(#1,#2,#3)}}
\def\QMS{\ensuremath{Q_\mrm{MS}}}
\def\rivet{\textsc{Rivet}\xspace}
\def\hepmc{\textsc{HepMC}\xspace}
\def\pythia{\textsc{Pythia}~8.3\xspace}
\def\pyt{\textsc{Pythia}\xspace}
\def\ariadne{\textsc{Ariadne}\xspace}
\def\dipsy{\textsc{Dipsy}\xspace}
\def\vincia{\textsc{Vincia}\xspace}
\def\dire{\textsc{Dire}\xspace}
\def\professor{\textsc{Professor}\xspace}
\def\autotunes{\textsc{Autotunes}\xspace}
\def\apprentice{\textsc{Apprentice}\xspace}
\def\lha{\textsc{LHA}\xspace}
\def\lhef{\textsc{LHEF}\xspace}
\def\slha{\textsc{SLHA}\xspace}
\def\lhahdffive{\textsc{LHAHDF5}\xspace}
\def\lhapdf{\textsc{LHAPDF}\xspace}
\def\powheg{\textsc{Powheg}\xspace}
\def\powhegbox{\textsc{Powheg Box}\xspace}
\def\powhegboxres{\textsc{Powheg Box Res}\xspace}
\def\mcatnlo{\textsc{MC@NLO}\xspace}
\def\mg5amc{\textsc{MadGraph5\_aMc@NLO}\xspace}
\def\amcatnlo{\textsc{aMc@NLO}\xspace}
\def\madgraphamc{\textsc{MadGraph5\_aMc@NLO}\xspace}
\def\madgraph{\textsc{MadGraph}\xspace}
\def\geneva{\textsc{Geneva}\xspace}
\def\sherpa{\textsc{Sherpa}\xspace}
\def\jetset{\textsc{Jetset}\xspace}
\def\lepto{\textsc{Lepto}\xspace}
\def\jewel{\textsc{Jewel}\xspace}
\def\jetscape{\textsc{JetScape}\xspace}
\def\madonia{\textsc{MadOnia}\xspace}
\def\docker{\textsc{Docker}\xspace}
\def\evtgen{\textsc{EvtGen}\xspace}
\def\helaconia{\textsc{HelacOnia}\xspace}
\def\fastjet{\textsc{FastJet}\xspace}
\def\hdffive{\textsc{HDF5}\xspace}
\def\mpich{\textsc{MPICH}\xspace}
\def\openmp{\textsc{OpenMP}\xspace}
\def\monash{\textsc{Monash 2013}\xspace}
\def\fritiof{\textsc{Fritiof}\xspace}
\def\angantyr{\pythia/\textsc{Angantyr}\xspace}
\def\Angantyr{\textsc{Angantyr}\xspace}
\def\yoda{\textsc{Yoda}\xspace}
\def\root{\textsc{Root}\xspace}
\def\cpp{\hbox{C\raisebox{0.2ex}{\smaller \kern-.1em+\kern-.2em+}}\xspace}
\def\python{\textsc{Python}\xspace}
\def\fortran{\textsc{Fortran}\xspace}
\def\binder{\textsc{Binder}\xspace}
\def\pybind{\textsc{PyBind11}\xspace}

\def\ckkwl{\textsc{Ckkw-l}\xspace}
\def\umeps{\textsc{Umeps}\xspace}
\def\nlthree{\textsc{Nl}3\xspace}
\def\unlops{\textsc{Unlops}\xspace}
\def\msbar{\ensuremath{\overline{\textrm{MS}}}\xspace}

\def\llangle{\left\langle}
\def\rrangle{\right\rangle}
\def\deriv{\ensuremath{\mathrm{d}}}
\def\avg#1{\ensuremath{\left\langle #1 \right\rangle}}
\def\abs#1{\ensuremath{\left| #1 \right|}}

\def\uone{\ensuremath{\mathbf{U(1)}}\xspace}
\def\su#1{\ensuremath{\mathbf{SU(#1)}}\xspace}

\newcommand{\Eg}{\mbox{\itshape E.g.}\xspace}
\newcommand{\eg}{\mbox{\itshape e.g.}\xspace}
\newcommand{\ie}{\mbox{\itshape i.e.}\xspace}
\newcommand{\etal}{\mbox{\itshape et al.}\xspace}
\newcommand{\etc}{\mbox{\itshape etc.}\xspace}
\newcommand{\cf}{\mbox{\itshape cf.}~}

\newcommand{\vs}{\mbox{\itshape vs.}\xspace}

\tikzstyle{block} = [rectangle, minimum width=1.0cm, minimum height=0.75cm, thin, draw=black]
\tikzset{blackarrow/.style={-stealth, semithick, draw=black}}
\tikzset{redarrow/.style={-stealth, semithick, draw=red}}
\tikzset{greenarrow/.style={-stealth, semithick, draw=green}}

\def\pgref#1{p.~\pageref{#1}}
\newcommand{\citeref}[2][]{ref.~\cite[#1]{#2}}
\def\citerefs#1{refs.~\cite{#1}}
\def\oneref#1{ref.~\cite{#1}}
\def\citeone#1{ref.~\cite{#1}}

\def\appref#1{appendix~\ref{#1}}
\def\htmlmanual{\href{https://www.pythia.org/latest-manual/}{\texttt{online manual}}\xspace}

\def\setting#1{\texttt{#1}\index{Settings (expicitly mentioned)!#1@\texttt{#1}}}
\def\settingval#1#2{\texttt{#1 = #2}\index{Settings (expicitly mentioned)!#1@\texttt{#1}}}

\makeatletter
\newenvironment{nonfloatfig}{\par\noindent\minipage{\textwidth}
\def\@captype{table}%
\centering\medskip}{\medskip\endminipage}
\makeatother

%% file: introduction/introduction.tex
This manual is organized into three major parts.
This first part contains introductory material about event generators in general
and the basic technical details of event generation.
The second part presents a more detailed description of the physics
implemented inside of \pyt. The physics is divided according to how it
appears in the program flow itself, though the lines drawn can be
fuzzy: the hard process (including external calculations); parton
showering; multiparton interactions; beam remnants; and hadronization.
There are also dedicated sections on the \dire and \vincia parton showers, as
well as the treatment of heavy-ion collisions.
Some of the details have not been thoroughly documented before, while
others have appeared in prior publications.
The third part is about how the user interacts with \pyt. In many
applications,
\pyt is part of a code stack or work flow, with other programs
calling into \pyt or \textit{vice versa}. This part describes both basic 
standalone usage and documents typical interfaces
in detail.

\section{Preliminaries}
\pythia~\cite{Sjostrand:2014zea} \index{Pythia@\pyt} is a 
scientific code library that is widely
used for the generation of events in
high-energy collisions between particles, where effects of the
strong nuclear force, governed by \ac{QCD}, are of high importance. 
It is written mainly in \cpp
and interweaves a comprehensive set of detailed physics models for
the evolution from a few-body hard-scattering process to a complex
multi-particle final state. Parts of the physics have been rigorously
derived from theory,  
while other parts are based on phenomenological models, with parameters 
to be determined from data.
Currently, the largest user community 
comes from the \ac{LHC} experimental collaborations,
but the program is also used for a multitude of other phenomenological or
experimental studies in astro-, nuclear, and particle physics.
Main tasks performed by the program
include 
investigations of experimental consequences of theoretical hypotheses,
interpretation of experimental data --- including estimation of
systematic uncertainties and unfolding --- 
development of search  strategies, and detector design and performance
studies. 
It also plays an important role as a versatile vessel for exploring new
theoretical ideas and new algorithmic approaches, ranging from minor 
user modifications to full-fledged developments 
of novel physics models. 

\subsection{What is an ``event generator''~?}\label{sec:components}
\index{Event generator}

In particle physics, the outcome of a collision between two incoming
particles, or of the isolated decay of a particle, is called
an ``event''. At the most basic level, an event
therefore consists of a number of outgoing
particles such as might be recorded in a snapshot taken by an
idealized detector, with conservation laws implying that the total summed
energies and momenta of the
final-state particles should match those of the initial state, as
should any discrete quantum numbers that are conserved by the 
physics process(es) in question. 

Due to the randomness of quantum
processes, the number of outgoing particles and their properties vary
from event to event. The probability distributions for these
properties can be inferred by 
studying an ensemble of events in data.
Conversely, given a set of theoretically calculated (or modelled)
probability distributions, it is possible to produce ensembles of
simulated events to compare to data. 

A numerical algorithm that can produce (or ``generate'') random sequences
of such simulated events, one after the other, is called
an  ``event generator''. 
The simulations can be based on known or hypothetical laws of nature. 
This allows for the exploration and comparison of competing
paradigms, and studies of the
sensitivity of proposed physical observables to the differences. 
Only rarely do the algorithms represent exact solutions however, so
a common issue is to consider whether ans\"atze and
approximations made, and the level of detail offered by a given
modelling, are adequate for the problem at hand. The detailed physics
descriptions contained in the main parts of this report are intended
to assist with this task.

Returning to the structure of a high-energy physics event, in its
crudest form, it is a list of the sub-atomic particles produced in a
collision along with a measure of the probability for that event to
occur. In \pyt, the list is referred to as the ``event record'', and
it includes the
four-momentum, production point, and many other properties of each
particle, \cf\cref{sec:eventAnalysis} for details. It typically
also includes quite a bit 
of history information showing intermediate stages of the event modelling.
The measure of the relative probability of a given event within a
sample is given by the weight of that event relative to the sum of
weights for the sample. For the typical case of unweighted
events, this is just the inverse of the total number of events in the
sample; cases that give rise to weighted events are summarized
in \cref{sec:using-weights}. The total 
cross section for the sample is also computed, allowing for 
the conversion of relative probabilities into cross sections.

Note that, although the starting point is often a relatively simple
cross section computed in fixed-order perturbation
theory, the total probability distribution for simulated events, fully
differentially in all relevant phase-space variables and quantum
numbers of the produced set of final-state particles, can typically
not be expressed analytically. Instead, it is evaluated directly,
using numerical methods, with \ac{MCMC} algorithms
based on pseudo-random number generators as the main ingredient. The
mathematical basis of the main ones used in \pyt is covered
in \cref{sec:mc}. 

The aim of the event generator is ambitious: to predict all of the
observable properties of a high-energy collision or decay process.
The full properties 
of an event, however, cannot currently be calculated from first
principles alone. Many different, complex phenomena, which are likely
related, are described by a proliferation of models that each focus on a
limited dynamical range.  As a result, the predictions of an event
generator like \pythia depend upon ${\cal O}(100)$ parameters.  The
values of these parameters are inferred from 
comparisons to data.  A collection of such parameter values is
referred to as a {\it tune}.

Event-generator predictions are useful, because they serve as a proxy
for what an event would look like before interacting with any
measurement devices.  As such, it can be used to investigate the
consequences of new and old phenomena, and study the loss,
mismeasurement, and misidentification of particles in experiments.
Thus, it is an important tool for interpreting collider data.  Event
generators are realized as computer codes. In modern times, most of
the larger projects are developed in the \cpp programming language.

\subsection{The structure of a simulated event} 
\index{Event structure}
\begin{figure}[tp]
\centering
\includegraphics[width=0.73\textwidth]{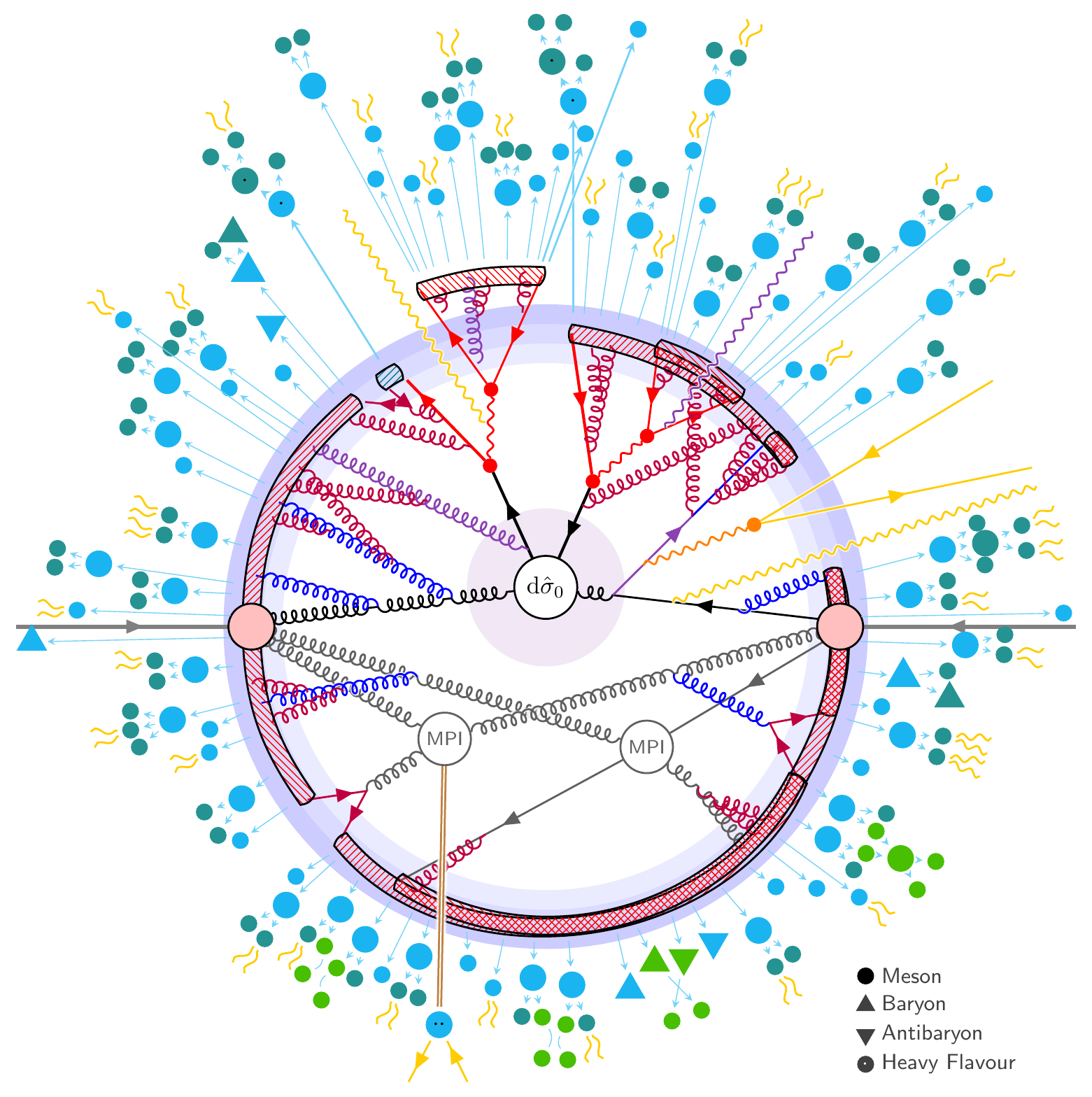}%
\includegraphics[width=0.27\textwidth]{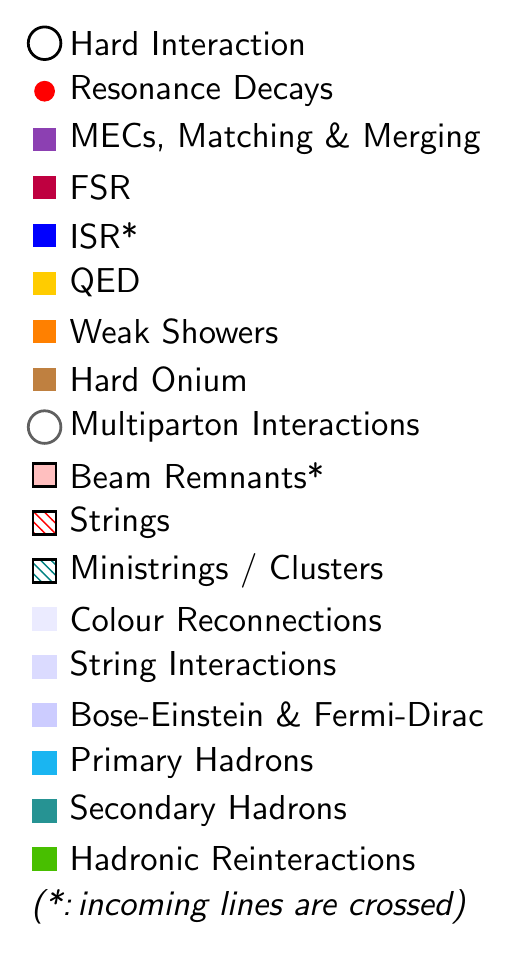}
\caption{Schematic of the structure of a $\p\p\to \t\tbar$ event, 
as modelled by \pyt.
To keep the layout relatively clean, a few minor
simplifications have been made: 1) shower branchings and
final-state hadrons are slightly less numerous than in real \pyt
events, 2) recoil effects are not depicted accurately, 
3) weak decays of light-flavour hadrons are not included
(thus, \eg a $K^0_S$ meson would be depicted as stable in
this figure), and 4) 
incoming momenta are depicted as crossed ($p\to -p$). The latter means
that the beam remnants and the pre- and 
post-branching incoming lines for ISR branchings should be interpreted
with ``reversed'' momentum, directed outwards towards the periphery of
the figure; this avoids beam remnants and outgoing ISR emissions
having to criss-cross the central part of the diagram. 
\label{fig:eventSchematic}}
\end{figure}

The main goal of \pyt is to simulate particle production in
high-energy collisions over the full range of energy scales accessible
to experiments, in as much detail as possible. 
However, hadron collisions and hadroproduction in particular are
exceedingly complex, and no comprehensive theory exists 
currently that can predict event properties over this full range. For practical
purposes, the wide range of phenomena are factored into a number of
components. 
A natural division for these components is a time-ordering or,
equivalently,
an energy or transverse momentum ordering, where the
best understood physics is calculated at the shortest time scales and
largest energies, and
the least understood physics is modelled at the longest time scales
and lowest energies.
This division is well motivated and often underpinned by factorization
theorems, but it is not entirely unambiguous and sometimes is
open to corrections. 
 
The ordering in time is not completely intuitive, at least
not in a directional sense from past to future. We should rather speak
of time windows centred on a hard collision that then expand forwards
and backwards in time, introducing successive phenomena,
until we are left with a pair of incoming protons from accelerator
beams, for example, and a number of outgoing particles. In momentum
space, we normally speak of the ``hardness'' scale that characterizes
each (sub)process, and often use a measure of transverse momentum
$p_\perp$ to quantify this. 

For simplicity, we will here concentrate on the sufficiently complex
case of hadron-hadron collisions, with an explicit
schematic of a fully simulated  $\p\p\to\t\tbar$ event given
in \cref{fig:eventSchematic}. The radial coordinate illustrates hardness scales, starting with the hardest subprocess near
the centre (labelled $\mathrm{d}\hat{\sigma}_0$), and ending with
stable final-state particles and the incoming beam particles at 
the periphery. 

In our hardness- or time-ordered picture, the components of a
high-energy collision are: 
\begin{enumerate}
\item A hard scattering of two partons, one from each incoming
  hadron, into a few outgoing particles.  The initial
  partons are selected using parton distribution functions for the
  incoming hadrons, and the kinematics of the outgoing particles are
  based on matrix elements calculated in perturbation theory.
  Such calculations introduce a factorization
  scale and a renormalization scale. Partons with momenta below these
  scales are not included in the hard scattering, but will be introduced 
  by other stages of the event generation. In the current usage
  of \pyt, it is common 
  to import the results of parton-level calculations from external packages,
  though a number of simple processes are calculated internally.
  Hard-scattering predictions depend on a few, universal input
  parameters that are determined from data, such as the value of the strong
  coupling at the \Z boson mass and parton distribution functions.
\item The hard process may produce a set of short-lived resonances,
  such as \Z or \Wpm gauge bosons or top quarks, whose decay to
  normal particles has to be considered
  in close association with the hard process itself.
\item Fixed-order radiative corrections may be incorporated via
  (combinations of) matrix-element corrections, matching, and/or
 merging strategies, \cf\cref{section:matchmerge}.
 In \cref{fig:eventSchematic}, the violet
 shaded region surrounding the hard process represents the range of
 scales covered by a (generic) matrix-element merging strategy active
 above some given $p_{\perp\mathrm{min}}$ scale.  
\item \ac{ISR} of additional particles (partons, photons, and others)
  starting from the scattering initiators using numerical
  resummation of soft and collinear gluon emission.   This (together
  with its final-state equivalent below) is commonly referred to as the parton shower.
\item \ac{FSR} of additional particles from the
  hard scattering itself and also from any resonance decays.
\item \index{Pileup}In competition with ISR and FSR, further
scattering processes 
between additional partons from the
incoming beams may take place, in a 
  phenomenon known as \ac{MPI}. This is not to
  be confused with ``pileup'', which generally refers to several
  distinct hadron-hadron collisions recorded in the same detector
  snapshot. 
\item At some stage after the MPIs and perhaps before resonance
decays, strings begin to form, as the non-perturbative limit of colour
dipoles. These dipoles, however, are typically defined by
\emph{colour connections} that are assigned in the
$N_c\to\infty$ limit, and are not 
 unique for $N_c=3$. As discussed further in \cref{sec:colRec}, the
 associated colour-space ambiguities can be modelled via \ac{CR}. 
 It is also possible that long-range dynamical
 interactions could physically alter the colour flow and/or change the
 configuration of the expanding strings before they fragment.
 Depending on the characteristic timescales
 involved (often not specified explicitly in simple CR models), such
 effects may also be referred to as colour reconnections, but could
 also come under the rubric of string  interactions. 
\item The strong interaction now results in the confinement of QCD
  partons into colour-singlet subsystems known as strings or, in
  small-mass limiting cases, clusters.  What is currently left of the
  incoming hadron constituents 
  are combined into beam remnants. In \cref{fig:eventSchematic}, the
  transition between the partonic and hadronic stages  
of the event generation is highlighted by
 the concentric annuli shaded blue.
\item The strings fragment into hadrons based on the Lund string
  model. Optionally, effects of overlapping strings may be taken into
  account, \eg by collecting them into so-called ``ropes'' and/or allowing
  interactions between them. 
\item Identical particles that are close in phase space may exhibit
  Bose-Einstein enhancements (for integer-spin particles) or Fermi-Dirac
  suppressions (for half-integer-spin particles). 
\item Unstable hadrons produced in the fragmentation process decay
  into other particles until only stable particles remain (with
  some user flexibility to define what is stable).
\item In densely populated regions of phase space, the produced
particles may rescatter, reannihilate, and/or recombine with one
another.  
\end{enumerate}
The introduction of heavy-ion beams introduces an additional layer of
complexity wrapped around this picture.  Lepton-lepton collisions are much simpler, since they do not involve many of the complications arising from hadron beams.
 
\subsection{To what types of problems can \pyt be applied~?}\index{Pythia@\pyt}

\pyt can be applied to a large set of phenomenological problems in particle
physics, and to related problems in astro-particle, nuclear, and
neutrino physics. Historically, the core of \pyt is the Lund string model of
hadronization. This model is most appropriate when the invariant masses
of the hadronizing systems are above 10 GeV or so. For lower-mass
systems, the model is less firmly reliable. Low-mass systems may 
still occur in \pyt, typically then as subsystems within a larger event, \eg,
produced by heavy-flavour decays, colour reconnections, and/or hadronic
rescattering. For the very lowest-mass systems, which produce just one
or two hadrons, a simple cluster-style model, called ministrings, is
implemented, otherwise the normal string fragmentation is applied.
In addition to string hadronization, \pyt of course also incorporates 
state-of-the-art models for a wide range of other particle-physics
phenomena. Here, we provide a non-exclusive list of various applications of
the \pyt machinery.

We emphasize that the majority of these models are based on dedicated
original work done by authors, students, and sometimes external
contributors, representing a significant and sustained intellectual
effort. When quoting results obtained with \pyt, we therefore ask that
users make an effort to cite, alongside this manual, 
such original works as would be deemed directly relevant to the study
at hand, \ie without whose implementation in \pyt the 
study could not have been done. Appropriate references can be
found throughout the manual. 

\begin{itemize}
\item Lepton-lepton, lepton-hadron, and hadron-hadron collisions with
configurable beam properties, such as beam energies and crossing
angles, to simulate one or many Standard-Model processes encoded
in \pyt.  This is the standard application of \pyt, but not the only
one. 
\item The same as above, except using parton-level configurations for
the hard process input from an external source. 
\item Ordinary particle decays, where the particles are produced by 
another physics program.  This includes the limiting case of a
particle gun (\ie a single particle with user-defined momentum). 
\item \ac{BSM} particle decays, including decay chains.
\item Resonance decays including the effects of final-state parton
showering and hadronization.
\item Hadronization of (colour-singlet) partonic configurations, as
  may arise from ordinary or exotic particle decays.
\item Generation of \ac{LHE} formatted files from the internal hard
processes for other physics studies. 
\item Ion-ion collisions for ion geometries well described with a
Woods-Saxon potential (non-deformed, $A > 16$) for
$\sqrt{s_\mathrm{NN}} > 10$ GeV. 
\item Astro-particle phenomena like dark-matter annihilation into
Standard-Model particles. 
\item User-inspired modifications of standard \pyt modules as allowed
by the \texttt{UserHooks} methods and those for semi-internal
processes and/or semi-internal resonances. 
\end{itemize}
As always, \textit{caveat emptor}.

\subsection{Historical evolution of the \pyt program}\index{Pythia@\pyt}
\index{Jetset@\jetset}

To bring some of the main development lines into context, we here provide a
brief summary of the historical evolution of 
the \pyt program and its ancestor, \jetset.
Detailed descriptions of the various
physics components will be found in subsequent sections, including
relevant references; a more elaborate review of the historical
evolution of \pyt can be found in \citeone{Sjostrand:2019zhc}.

In the late seventies the Lund group began to study strong interactions,
and notably the hadronization subsequent to a collision process.
A linear confinement potential was assumed to be realized by a string
stretched out between a pulled-apart colour--anticolour pair, as a simple
one-dimensional representation of a three-dimensional flux tube or
vortex line. In order to allow detailed studies, two PhD students were
entrusted to code up this model, and also include effects such as
particle decays. This program was given the name \jetset.\index{Jetset@\jetset} 
The model and code were gradually extended to encompass more physics, 
in particular with reference to $\eplus\eminus$ physics. The key addition
was a model for $\eplus\eminus \to \q\qbar\g$, wherein the colour field
was assumed to stretch as one string piece from the $\q$ end to the $\g$
and then as a second piece on from the $\g$ to the $\qbar$ end,
with no direct connection between the $\q$ and $\qbar$. This model
received experimental support at PETRA in 1980~\cite{JADE:1981ofk}, 
thereby starting the success story of the Lund event generators. The idea of 
subdividing the full colour topology into a set of colour--anticolour dipoles 
rapidly prompted extensions also to other collision processes, notably
to $\pp$ ones, with the \pyt generator built on top of \jetset. Later,
it also came to develop into the dipole picture of parton showers,
and to foreshadow related techniques for higher-order matrix-element
calculations.

In part, the continued evolution was driven by interactions with
the experimental communities and their priorities.  An early
involvement in SSC studies led to an extension of the scope of \pyt 
from QCD physics to encompass a wide selection of \ac{SM}
processes, notably those related to Higgs-boson signatures. At the same time, 
QCD processes needed to be modelled better, which led to the
development of new concepts, 
such as backwards evolution to handle initial-state
radiation, and multiparton interactions and colour reconnection to
describe underlying events and minimum-bias physics. When LHC physics studies began in 1990, these capabilities helped \pyt play a prominent role in
benchmarking the evolving design of the LHC detectors, and additionally
many Beyond-the-Standard-Model scenarios were included to cater to the
demands of the community.

The \ac{LEP} became the first operating collider where \jetset had been used from
the early days of detector design, and the program came to play a key role
in most physics analyses carried out there.
QCD phenomena were a primary focus of experimental studies, and this  
led to an emphasis on issues such as parton-shower algorithms and
matrix-element corrections to them. The \ariadne dipole shower \cite{Lonnblad:1992tz} 
offered a successful alternative to the more traditional internal \jetset one.
With LEP~2, the emphasis shifted from QCD towards electroweak processes such as
$\Wp\Wm$ pair production, which had already been incorporated into \pyt.    This led, naturally, to the integration of the \jetset capabilities into \pyt, with \pyt maintaining the project name and legacy.

Also at HERA, the Lund-based programs came to play a prominent role
from the onset, with codes such as \lepto \cite{Ingelman:1996mq}, \ariadne 
and \textsc{LDC} \cite{Kharraziha:1997dn} built on top of \jetset. 
Photon physics was introduced into \pyt to handle $\gamma\gp{p}$ at HERA 
and $\gamma\gamma$ at LEP~2.

A further area of study is heavy-ion collisions, where early on the
\fritiof~\cite{Nilsson-Almqvist:1986ast} program came to be widely used. 
Some of these ideas have been revived, updated, and implemented in 
the \angantyr model. It is worth noting, also, that many heavy-ion collision models, used notably at the \ac{RHIC},
have been based on \pyt.

The separation above, by collider, gives one way of describing the
evolution of the code(s). Underlying it is a belief in universality,
that many aspects of particle collisions are the same, independent
of the beam type. Therefore, physics developments made in one 
context can also be applied to others. This is why one single code
has found such widespread use.

The early codes were all written in \fortran~77. With the CERN decision
to replace that language by \cpp for LHC applications, \pyt underwent a
similar transformation in 2004 -- 2008. A new organizational structure
was put in place for the new \pyt~8, in an attempt to clean up blemishes
incurred during the years of rapid expansion, but deep down most of the
physics algorithms survived in a new shape.

One area where the evolution has overtaken \pyt is that of matrix elements.
Before it was possible for most users to perform matrix-element calculations on computers,
such expressions were published in articles and hard-coded from these.   Now, with the  physics demand for higher final-state multiplicities and higher-order perturbative accuracy, that is no longer feasible. For all but the simplest processes,
we therefore rely on separate, external matrix-element codes to provide the hard
interactions themselves, \eg via the Les Houches interfaces, to which
we then can add parton showers, underlying events, and hadronization.
Also parton distribution functions are obtained externally, even if a
few of the more commonly used ones are distributed with the code. 

The program has continued to expand also after the transition to \cpp.
Some developments are done from the onset within the \pyt code,
such as the machinery for matching and merging between matrix
elements and parton showers, or the \angantyr framework for heavy-ion
collisions, or the space--time picture of hadronization and hadronic
rescattering. Other have come by the integration of externally developed
packages, such as the \vincia and \dire alternatives to the existing
simpler parton showers already in place. 

In total, the \jetset/\pyt manuals have more than 35\,000 citations by now,
attesting to its widespread use. That use also includes possible future
projects such as ILC, FCC, CLIC and EIC. 
The counting of code citations does not include the numerous
articles describing the development and application of the physics content
in the programs.  This is harder to count, with many borderline cases, but the
order of magnitude is comparable with the one for the code itself.

%% file: introduction/program-structure.tex
\section{Program structure and basic algorithms}

The \pythia general-purpose Monte-Carlo event generator's structure reflects the different physics descriptions and models needed to generate fully exclusive final states as they can be detected at collider experiments. The first part of this section gives a brief overview of the program structure, while the latter parts describe basics of \ac{MC} techniques and process generation employed by \pythia.

\subsection{Program structure and overview}

Internally, \pythia is structurally divided into three main parts: process level, parton level, and hadron level. This reflects the components of an event as introduced in \cref{sec:components}.

The process level represents the hard-scattering process, including the production of short-lived resonances. The hard process is typically described perturbatively, with a limited number of particles, typically at high-energy scales.

The parton level includes initial- and final-state radiation, where various shower models are available. Multiparton interactions are also included at this stage, along with the treatment of beam remnants and the possibility of the colour-reconnection phenomenon. At the end of the parton-level evolution, the event represents a realistic partonic structure, including jets and the description of the underlying event.

The hadron level then takes care of QCD confinement of partons into colour-singlet systems. In \pythia, the hadronization is described by QCD strings fragmenting into hadrons. Furthermore, other aspects like the decay of unstable hadrons and hadron rescattering are dealt with at the hadron level. The physics models of hadronization are typically non-perturbative, and thus require modelling and the tuning of parameters. The output of the hadron level is then a realistic event as it can be observed in a detector.

On top of this general structure, a significant number of shared objects and cross talk is passed between these levels: PDFs are relevant in both the process level and ISR, the matching and merging machinery works on the interface between parton showers and process level, and the \texttt{Info} object is used throughout all levels to store and access central information. Under certain circumstances, like the analysis of heavy-ion collisions using the \angantyr model, multiple parton-level objects can be used for separate subcollisions, which are then combined for hadronization.

\begin{figure}
  \includegraphics[width=\textwidth]{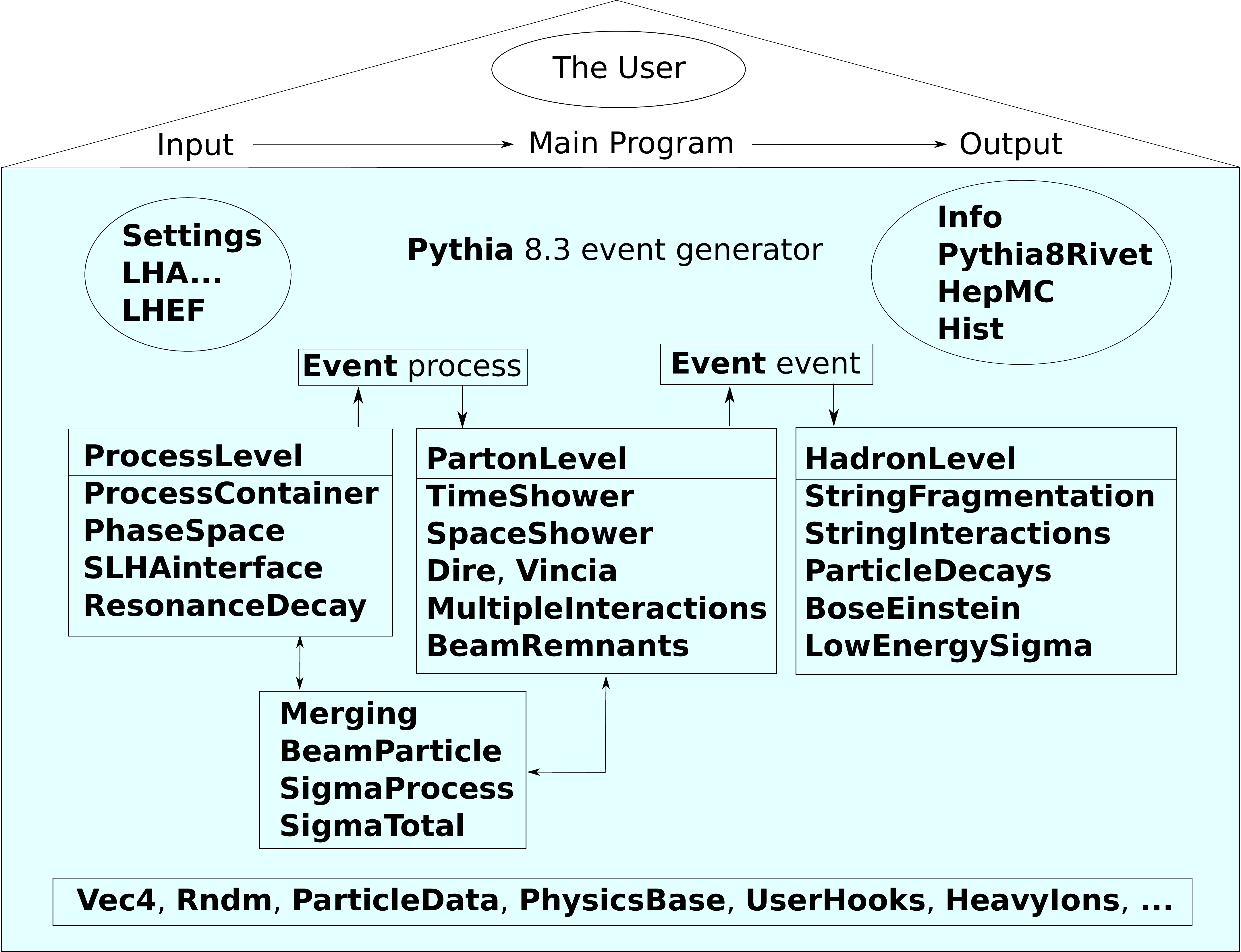}
  \caption{Simplified picture of the \pythia structure, showing some of the important classes in bold. The main program itself creates one or more \texttt{Pythia} objects, and provides input in terms of \texttt{Settings} and potentially-perturbative event input. The main physics components are grouped into \texttt{ProcessLevel}, \texttt{PartonLevel}, and \texttt{HadronLevel}, with additional structure to complement and interconnect them.}
  \label{fig:programStructure}
\end{figure}

From the user's perspective, \pythia is a \cpp library. The actual executable is implemented by the user, based on the requirements regarding input, output, features, and analysis, and many examples come with the \pythia package. For detailed information on how to install and use \pythia, both standalone and with external interfaces, see \cref{part:use}. \Cref{fig:programStructure} gives a rough overview of the \pythia program structure.

\input{introduction/mc-techniques}
\input{introduction/process-generation}

%% file: introduction/mc-techniques.tex
\subsection{Monte-Carlo techniques\label{sec:mc}}

Real events observed in particle colliders are stochastic.
To emulate this, event generators sample from probability
distributions using pseudo-random numbers. Naively,
a pseudo-random number (between 0 and 1) is compared to
a cumulative distribution function to determine an effect, \eg
the angle of a particle in a decay, the type of particle produced in
hadronization, \textit{etc.}
Since real cases are rarely 
this simple, we use this section to describe some of the technical
details of how pseudo-random numbers are used within the program.

\subsubsection{Random-number generation}
\label{sec:rng}
\index{Random numbers}
At the core of all Monte-Carlo methods lies the access to a random
number generator. Truly random numbers require special equipment
and are difficult to obtain at the required pace, so in practice
pseudo-random numbers are used, where deterministic computer algorithms
are used to emulate a random behaviour.
\index{Random numbers!Seed}This also allows a user of the code
to reproduce a given event sample, simply by setting the same
random-number seed. 
Nevertheless, the numbers must appear to be random, \eg evenly distributed between 0 and 1,
have no detectable correlations, and have a long period before they
start to repeat. Many pseudo-random number generators once thought to exhibit no internal 
correlations, have later been revealed to have flaws, so care is needed.

A review of several current generators is found in
\citeref{James:2019nyc}. Common for them is that they can be viewed 
as having an $N$-dimensional state vector $x$, living in a $N$-dimensional
hypercube with periodic boundary conditions such that each number is
in the range between 0 and 1. A new state is obtained by a matrix
multiplication $x_{i+1} = A \times x_i$, where $A$ is a $N \times N$
matrix of integers. There is some sophisticated theory involved in the 
choice of $A$, involving concepts such as Kolmogorov--Anosov mixing
and the Lyapunov exponent. Some of the key results are that $A$ should have
determinant unity, with complex eigenvalues away from the unit circle,
and additionally that multiplication with it should require a minimal
amount of operations so as to keep speed up. 

\index{RANMAR}\index{Marsaglia-Zaman algorithm}\index{Random numbers!RANMAR}The RANMAR default in \pyt is based on the
Marsaglia--Zaman algorithm 
\cite{Marsaglia:1990ig}, but implemented in double precision with $N = 97$.
There remains some tiny correlations 97 numbers apart, which could be
fixed by multiplication by $A$ several times between each set of 97
random numbers actually used \cite{Luscher:1993dy}, but this is not a
necessity for event generators, where typically one is in a completely
different part of the code 97 random numbers later. The RANMAR algorithm
can be initialized to run one of more than 900\,000\,000 different
sequences, each with a period of more than $10^{43}$. By default, the
same sequence is always run, which is useful for checks and debug purposes.

\index{MIXMAX}\index{Random numbers!MIXMAX}The MIXMAX alternative \cite{Savvidy:2014ana} is also
provided as an 
option, and additionally there is an interface allowing the user to link
in an external algorithm of choice.

\subsubsection{Some standard techniques}

There are two main kinds of random-number usage in \pyt, one without
a memory of a previous evolution in ``time'' and one with. The latter
is part of the veto algorithm described in the next subsection. Here
we introduce the former.

\index{Random numbers!Sampling}
The simplest situation is that we know a function $f(x)$ that is
non-negative in the allowed $x$ range
$x_{\mathrm{min}} \leq x \leq x_{\mathrm{max}}$.
We want to select an $x$ at random so that the probability in a small
interval $\d x$ around a given $x$ is proportional to $f(x) \, \d x$.

If it is possible to find a primitive function $F(x)$ with a known
inverse $F^{-1}(x)$, an $x$ can be found as follows:
\begin{align}
& \displaystyle{ \int_{x_{\mathrm{min}}}^{x} f(x) \, \d x =
R \int_{x_{\mathrm{min}}}^{x_{\mathrm{max}}} f(x) \, \d x } & \nonumber \\[1mm]
\Longrightarrow & x = F^{-1}(F(x_{\mathrm{min}}) + 
 R \, (F(x_{\mathrm{max}}) - F(x_{\mathrm{min}}))) ~, &
\end{align}
where $R$ is a random number evenly distributed between 0 and 1.
The statement of the first line is that a fraction $R$ of the
total area under $f(x)$ should be to the left of $x$.
However, seldom are functions of interest so nice that the
method above works. It is therefore necessary to use more complicated
schemes.

\index{Hit-or-miss method}
\index{Random numbers!Hit-or-miss method}
If the maximum of $f(x)$ is known, $f(x) \leq f_{\mathrm{max}}$ in the $x$ range
considered, a hit-or-miss method will yield the correct answer. In this method
$x$ and $y$ are chosen according to
\begin{align}
  x & = x_{\mathrm{min}} + R_1 \, (x_{\mathrm{max}} - x_{\mathrm{min}})
          ~,\nonumber \\
  y & = R_2 \, f_{\mathrm{max}} ~.
\end{align}  
This is repeated until a $y < f(x)$ is selected. The accepted $x$
value is then distributed uniformly in the area below $f(x)$.
Equivalently, the selected $x$ can be accepted with probability
$f(x) / f_{\mathrm{max}}$, without the explicit construction of a $y$.
The efficiency of this method, \ie the average probability that
an $x$ will be retained, is $(\int \, f(x) \, \d x) /
(f_{\mathrm{max}} \, (x_{\mathrm{max}} - x_{\mathrm{min}}))$.
The method is acceptable if this number is not too low, \eg if
$f(x)$ does not fluctuate too wildly or is too sharply peaked.

The algorithm is independent of the absolute normalization of $f(x)$;
in a sense, the procedure automatically rescales the function to have
unit integral. As a by-product of the $x$ selection it is also possible
to do a Monte-Carlo integration of $f(x)$:
\index{Monte Carlo integration}
\begin{equation}
  \int_{x_{\mathrm{min}}}^{x_{\mathrm{max}}} \, f(x) \, \d x \approx
  (x_{\mathrm{max}} - x_{\mathrm{min}}) \, \frac{1}{n_{\mathrm{try}}} \,
  \sum_{i= 1}^{n_{\mathrm{try}}} f(x_i) ~,
 \label{eq:random-integrate}
\end{equation}  
where $x_i$ runs over all $x$ values tried, whether accepted or not.
The error decreases like $1 / \sqrt{n_{\mathrm{try}}}$, also if $x$
represents more than one dimension. More conventional integration
methods converge faster than this in one dimension, but slower in
higher dimensions.

Often $f(x)$ does have narrow spikes, and it may not even be
possible to define an $f_{\mathrm{max}}$. Then one may use a variable
transformation to flatten out the function. A related method is
importance sampling, which works if one can find a function
$g(x)$, with $f(x) \leq g(x)$ over the $x$ range of interest.
Here $g(x)$ is picked to be a ``simple'' function, such that the
primitive function $G(x)$ and its inverse $G^{-1}(x)$ are known.
Then the methods above can be combined:
\index{Random numbers!Sampling}
\index{Importance sampling}
\begin{align}
  x & = G^{-1}(G(x_{\mathrm{min}}) +  R_1 \, (G(x_{\mathrm{max}})
          - G(x_{\mathrm{min}}))) ~, \nonumber \\
  y & = R_2 \, g(x) ~.
 \label{eq:random-importance}
\end{align}  
This is repeated until a $y < f(x)$ is selected. Note that the first 
step selects $(x, y)$ uniformly in the area below $g(x)$, whereas the 
second half is to accept those that also are below $f(x)$.
Using an acceptance probability $f(x) / g(x)$ again removes the need
to introduce an intermediate $y$.

If $f(x)$ has several spikes, it may not be possible to find a
$g(x)$ that both covers all of them and has an invertible primitive
function.  However, assume that we can find a function
$g(x) = \sum_i g_i(x)$, such that $f(x) \leq g(x)$ over the $x$ range
considered, and such that the functions $g_i(x)$ are non-negative
and have invertible primitive functions. Then multichannel sampling \cite{Kleiss:1994qy}
extends on the importance-sampling prescription, by using the relative 
size of the integrals $I_i = \int g_i(x) \, \d x$ to each time pick a new
$g_i$ for the $x$ selection in \cref{eq:random-importance}.
The $y$ selection and the accept/reject works as before, since it is
easy to see that the weighted usage of the different $g_i(x)$ adds up
to $g(x)$.

In addition to the generic methods, it is also sometimes possible to find
special tricks. For instance, a single Gaussian $\exp(-x^2)$ is not
integrable, but the product of two is, by transforming to plane-polar
coordinates:
\begin{equation}
  e^{-(x^2 + y^2)} \, \d x \, \d y = e^{-r^2} \, r \d r \, \d \varphi
  \propto  e^{-r^2} \, \d r^2 \, \d \varphi ~,
\end{equation}
which gives
\begin{align}
  x & = \sqrt{- \ln R_1} \, \cos(2\pi R_2)  ~, \nonumber \\
  y & = \sqrt{- \ln R_1} \, \sin(2\pi R_2)  ~,  
\end{align}  
\ie two Gaussian-distributed numbers are obtained from two random ones.
Another trick is that a judicious choice of convolutions can be used to
show that $f(x) = x^{n-1} \, e^{-x} / (n -1)!$ can be obtained by
$x = - \sum_{i = 1}^n \ln R_i = - \ln \left( \prod_{i = 1}^n R_i \right)$.

\subsubsection{The veto algorithm}\index{Veto algorithm}
\label{subsubsec:vetoalgorithm}

A broad class of stochastic evolution algorithms, including ones
describing radioactive decays, parton showers, and also \pyt's modelling of
MPI, involve the generation of
\emph{ordered} sequences of state changes (transitions), where the
ordering parameter is typically a measure of time and/or
resolution scale. 

For probability distributions (and/or domains) that are complicated to
handle analytically, the veto  algorithm offers a convenient and
mathematically exact approach by which simple overestimates can be
used instead of the original functions that are then reimprinted via a
veto step. This circumvents the need for costly and delicate numerical
integrations and root finding, and the overestimating functions and
domains can be tailored to the problem at hand for maximum
efficiency. 

Before  describing the  algorithm itself, however, let us first clear up
a point of semantics. In the context of parton showers, the veto
algorithm is the main way by which Sudakov form factors (see below)
and related quantities are calculated. One therefore occasionally sees
the phrase ``Sudakov veto algorithm'', but this risks giving the
mistaken impression that Sudakov invented the veto algorithm. To avoid
this, the terms ``veto algorithm'' and ``Sudakov (form) factor'' are
kept separate in this work, with the former referring to the
broad numerical sampling method described in this section and the latter
being an (important) example of a physical quantity that can be
calculated with it. 

Consider a stochastic process that is ordered in some measure of evolution
scale. \Eg for nuclear decay, the ordering measure could be time (in
the rest frame of the decaying nucleus), while for \pyt's 
evolution algorithms, which are formulated in momentum space, the
ordering is normally done in a measure of  transverse momentum, from
high to low. This ensures that infrared and collinear divergences of
the corresponding transition  amplitudes are associated with vanishing
resolution scales, or equivalently with asymptotically late times in
the algorithmic sense.

Starting from a given initial value, $u$, for the evolution scale, the
probability for the  
\emph{next} transition (\eg a nuclear decay, or a shower branching) to
happen at a lower scale $t < u$, is given by  
\begin{equation} \label{eq:shower-distribution}
p(t|u) ~=~ f(t) \, \Pi(u,t)~,
\end{equation} 
where $f(t)$ is the probability (sometimes called the ``naive''
probability) for a transition to occur at the scale $t$ under the
implicit condition that the state still exists at $t$.
The latter is made explicit by the survival probability, $\Pi(u,t)
\in [0,1]$, which represents the probability that the state remains
unchanged over the interval $[u,t]$. Analogously to nuclear decay,
$\Pi(u,t)$ is given by a simple exponential of the integrated
naive transition rate, 
\begin{equation}
 \Pi(u,t)~=~\exp\left(-\int^u_t \d\tau f(\tau) \right)~,
\end{equation}
such that 
\begin{equation}
p(t|u) = \frac{\partial \Pi(u,\tau)}{\partial \tau}\big\vert_{\tau=t}~.
\label{eq:dSudakov}
\end{equation}
The survival probability $\Pi(u,t)$ is often referred to as the
Sudakov form factor. We note that the two are only strictly identical
for final-state showers, while for initial-state showers they are
related via ratios of parton distribution functions, and in the
context of MPI one can really only talk about a Sudakov-like factor.
In this section, only the survival probability itself, which we denote
by $\Pi(u,t)$, will be of interest.  

It is worth pointing out that the ordered probability density
$p(t|u)$ remains well-behaved 
and bounded by unity even if the integrated naive
transition rate exceeds unity. In fact, due to the aforementioned
collinear and soft singularities, $f(t)$ typically diverges for $t\to
0$, in which case the total probability for at least one-state change 
becomes 
\begin{equation}
  \int^u_0 \d\tau \, p(\tau|u) = 
  1 - \exp\left(F(0) - F(u)\right) \to 1~.
\end{equation}
That is, since $F(0) \to -\infty$ for a divergent kernel, the
probability for at least one-state change simply saturates at unity. This reflects the 
unitarity of the shower algorithm, which is also manifest in 
\cref{eq:dSudakov}. If the naive probability
does not diverge, or if the evolution is 
stopped at a finite cutoff $t_\mathrm{cut} > 0$, then there is a
non-zero probability, given by $\Pi(u,t_\mathrm{cut})$, to have no
state change at all. 

Starting from \cref{eq:shower-distribution}, probabilities for
two or more ordered transitions can easily be constructed as well,
\eg for two successive branchings with $t<u<v$: 
\begin{equation}
  P(t|u|v) = f_2(t) f_1(u) \, \Pi_2(v,u) \, \Pi_1(u,t)~,
\end{equation}
where $f_1(u)$ is the naive transition rate at the scale of the
``first'' transition, and that of the ``second'' transition is $f_2(t)$.
Note that we do not  assume $f_1 = f_2$ since the state undergoes a change
at the intermediate scale $\tau=u$ (and the phase space is generally
also different). This is also emphasized by the presence of two
separate survival probabilities with different subscripts instead of a
single combined $\Pi(v,t)$. 

\begin{figure}
\centering
\begin{tikzpicture}
\node[block] (sample) at (0, 3.4) {Sample $t$ from $g(t) \, \Pi_g(u,t)$};
\node[block] (accept) at (0, 1.7) {Accept trial with probability $\frac{f(t)}{g(t)}$};
\node[block] (done)   at (0, 0.0) {Done};
\node[block] (veto)   at (3.6, 2.5) {Set $u = t$};

\draw[blackarrow] (sample) to [out=270,in=90] (accept);
\draw[redarrow]   (accept) to [out=0,in=270](veto);
\draw[blackarrow] (veto) to [out=90,in=0](sample);
\draw[greenarrow] (accept) to [out=270,in=90](done);    
\end{tikzpicture}
\caption{Flowchart representation of the veto algorithm. 
The red and green arrows refer to rejection and acceptance of the
trial scale $t$ respectively.}
\label{alg:veto_algorithm}
\end{figure}
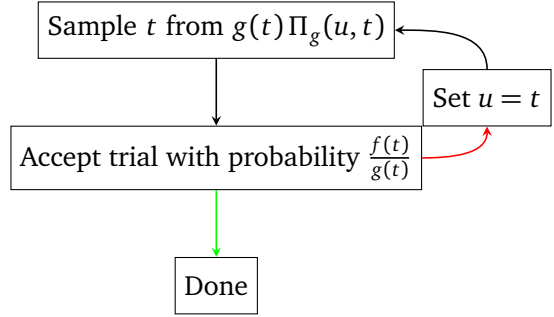

We now turn to how to actually sample from \cref{eq:shower-distribution}. 
The branching kernel $f(t)$ is typically not simple enough to allow 
for the use of inversion sampling as described in the previous section. 
Fortunately, the veto algorithm~\cite{Sjostrand:1987su,Sjostrand:2006za,Lonnblad:2012hz, 
  Platzer:2011dq, Kleiss:2016esx} (and its antecedents, see the
``thinning algorithm''~\cite{Lewis:1979,devroye2013non}) 
enables sampling from \cref{eq:shower-distribution} in a quite
efficient and flexible manner.  
This algorithm relies on the existence of an overestimating ``trial'' function
$g(t) \geq f(t)$ that is simple  
enough for samples to be drawn from \cref{eq:shower-distribution} directly, with $f$ replaced by $g$. 
A flowchart representation of the veto algorithm in its simplest form is shown in \cref{alg:veto_algorithm}.

To confirm that this algorithm produces \cref{eq:shower-distribution},
we follow along and write out its probability distribution $q(t|u)$ to find
\begin{align} \label{eq:veto-algorithm-dist}
q(t|u) = \int_0^u \d t' g(t') \, \Pi_g(u,t) \bigg[ \frac{f(t')}{g(t')} \delta\left(t - t'\right) 
+ \left(1 - \frac{f(t')}{g(t')}\right) p\left(t|t'\right) \bigg]\,,
\end{align} 
where the first term describes the probability to accept the proposed
trial scale $t'$, and the second 
term gives the probability to reject the trial scale.
Note that \cref{eq:veto-algorithm-dist} explicitly displays the Markovian nature 
of the veto algorithm, with every recursive step only
depending on the previous one. 
\Cref{eq:veto-algorithm-dist} may be solved by considering the differential equation
\begin{equation} \label{eq:veto-algorithm-diff}
\frac{\partial}{\partial u} q(t|u) =  f(u) \delta(t - u) - f(u) q(t|u),
\end{equation} 
which is found by application of Leibniz's rule for differentiation 
to \cref{eq:veto-algorithm-dist}.
We can find a solution by using an ansatz $q(t|u) = \hat{q}(t|u) e^{-F(u)}$, 
which after integration leads to 
\begin{equation} \label{eq:veto-algorithm-solution}
q(t|u) = f(t) \, \Pi_f(u,t) \Theta(t - \sigma) + q_0(t,\sigma).
\end{equation} 
The scale $\sigma$ in the step function
$\Theta(t - \sigma)$ and the function $q_0$ appear because information
is lost in converting \cref{eq:veto-algorithm-dist} to \cref{eq:veto-algorithm-diff},  
but they are easily understood by reconsidering the structure of the algorithm. 
Mathematically, no other scale $\sigma$ was introduced at any point, 
so \cref{eq:veto-algorithm-solution} cannot depend on it.
As a result, $\sigma$ must equal zero and the function $q_0$ must vanish, 
recovering \cref{eq:shower-distribution}.
In practice, however, the infrared cutoff on the shower evolution
\emph{does} introduce a scale $\sigma$. 
In that case, the algorithm shown in \cref{alg:veto_algorithm} 
is stopped whenever $t$ drops below $\sigma$. 
The function $q_0$ then represents the superfluous probability 
of sampling a scale below the cutoff, which is not associated with any
change of state.

Many extensions of the veto algorithm are possible and are often used, 
of which we only discuss a few.
Further details may be found in \citerefs{Sjostrand:1987su, Lonnblad:2012hz,
  Platzer:2011dq, Mrenna:2016sih, Kleiss:2016esx}. 

One can replace the acceptance probability $f(t)/g(t)$ by some other $r(t) \in [0,1]$ 
and compensate by modifying the event weight by a multiplicative factor 
$f(t)/g(t) r(t)$ in case the scale is accepted and 
$(1-f(t)/g(t))/(1-r(t))$ in case it is rejected.
Writing out the probability distribution again, we find
\begin{equation} \label{eq:veto-algorithm-weighted}
q(t|u) = \int_0^u \d t' g(t') \, \Pi_g(u,t') \bigg[ r(t') \delta\left(t - t'\right) \frac{f(t')}{g(t') r(t')}
+ \left(1 - r(t')\right) p\left(t|t'\right) \frac{1 - f(t')/g(t')}{1 - r(t')} \bigg],
\end{equation} 
where the weights appear as multiplicative factors. 
It is then straightforward to see that \cref{eq:veto-algorithm-weighted} 
reduces to \cref{eq:veto-algorithm-dist}.
This modification enables sampling from \cref{eq:shower-distribution} in 
cases where it is difficult to find a $g(t) \geq f(t)$, or even in cases where $f(t)$ may be negative.
However, in both cases events with negative weights will appear.

\index{Uncertainties!Parton showers@in Parton showers}\index{Weights}Applying \cref{eq:veto-algorithm-weighted},
shower uncertainties can be efficiently incorporated as event weights.
In that case, $r(t)$ represents the {\it baseline} acceptance
probability, while $f(t)$ is a modified branching kernel
that parameterizes the uncertainties through variations of the
renormalization scale, its non-singular components, or choice of
parton distribution function for initial state showers.
If $g(t)$ overestimates both the baseline and the modified branching
kernels, the event weights stay positive.   This is also the basis for
generating biased emissions of rare splittings.
More details can be found in \cref{sec:var}.

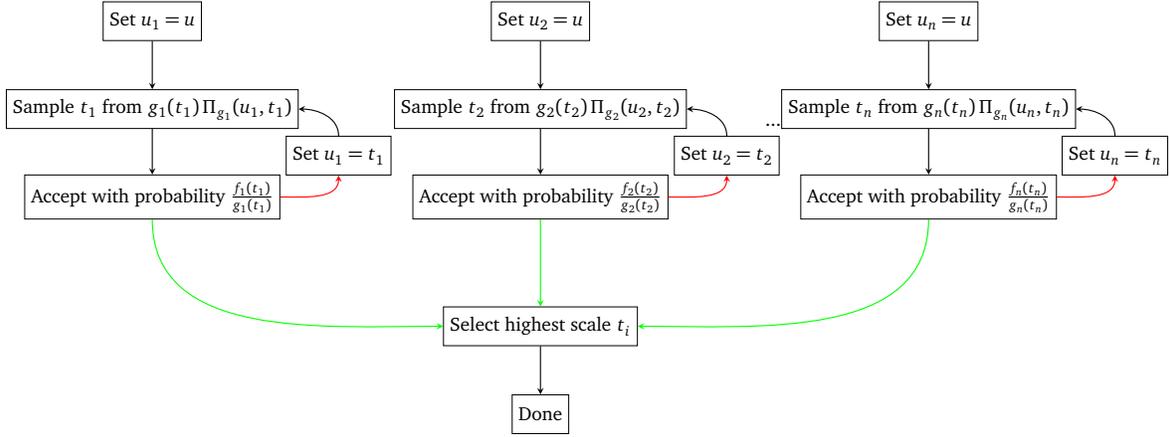
\begin{figure}
\centering
\resizebox{\columnwidth}{!}{
  \begin{tikzpicture}
  \node[block] (init1) at (0, 5.1) {Set $u_1 = u$};
  \node[block] (sample1) at (0, 3.4) {Sample $t_1$ from $g_1(t_1) \, \Pi_{g_1}(u_1,t_1)$};
  \node[block] (accept1) at (0, 1.7) {Accept with probability $\frac{f_1(t_1)}{g_1(t_1)}$};
  \node[block] (veto1) at (3.6, 2.5) {Set $u_1 = t_1$};

  \draw[blackarrow] (init1) to [out=270,in=90]   (sample1);
  \draw[blackarrow] (sample1) to [out=270,in=90] (accept1);
  \draw[redarrow]   (accept1) to [out=0,in=270]  (veto1);
  \draw[blackarrow] (veto1) to [out=90,in=0]     (sample1);

  \node[block] (init2) at (7.5, 5.1) {Set $u_2 = u$};
  \node[block] (sample2) at (7.5, 3.4) {Sample $t_2$ from $g_2(t_2) \, \Pi_{g_2}(u_2,t_2)$};
  \node[block] (accept2) at (7.5, 1.7) {Accept with probability $\frac{f_2(t_2)}{g_2(t_2)}$};
  \node[block] (veto2) at (11.1, 2.5) {Set $u_2 = t_2$};

  \draw[blackarrow] (init2) to [out=270,in=90]   (sample2);
  \draw[blackarrow] (sample2) to [out=270,in=90] (accept2);
  \draw[redarrow]   (accept2) to [out=0,in=270]  (veto2);
  \draw[blackarrow] (veto2) to [out=90,in=0]     (sample2);

  \node       (dots) at (12, 3.1) {... };
  \node[block] (initn) at (15, 5.1) {Set $u_n = u$};
  \node[block] (samplen) at (15, 3.4) {Sample $t_n$ from $g_n(t_n) \, \Pi_{g_n}(u_n,t_n)$};
  \node[block] (acceptn) at (15, 1.7) {Accept with probability $\frac{f_n(t_n)}{g_n(t_n)}$};
  \node[block] (veton) at (18.6, 2.5) {Set $u_n = t_n$};

  \draw[blackarrow] (initn) to [out=270,in=90]   (samplen);
  \draw[blackarrow] (samplen) to [out=270,in=90] (acceptn);
  \draw[redarrow]   (acceptn) to [out=0,in=270]  (veton);
  \draw[blackarrow] (veton) to [out=90,in=0]     (samplen);

  \node[block] (highest) at (7.5, -0.8) {Select highest scale $t_i$};
  \node[block] (done) at (7.5, -2.5) {Done};
  \draw[greenarrow] (accept1) to [out=270,in=180] (highest);
  \draw[greenarrow] (accept2) to [out=270,in=90]  (highest);
  \draw[greenarrow] (acceptn) to [out=270,in=0]   (highest);
  \draw[blackarrow] (highest) to [out=270,in=90]  (done);
  \end{tikzpicture}
}
\caption{Flowchart representation of the first competition veto algorithm.}
\label{alg:veto_algorithm_competition_1}
\end{figure}

We complete this section by discussing some variations of the 
veto algorithm in the context of competition between channels.
In most cases, multiple branching kernels $f_i(t)$ contribute 
to the total parton-shower probability distribution, which may then be written as 
\begin{equation} \label{eq:shower-distribution-competition}
\tilde{p}(t|u) = \tilde{f}(t) \, \Pi_{\tilde{f}}(t,u) \text{ where } \tilde{f}(t) = \sum_{i=1}^n f_i(t).
\end{equation} 
One way to handle competition is to apply the veto algorithm to all channels individually, 
and then select the channel with the highest scale $t_i$. 
A flowchart representation of this procedure is shown in \cref{alg:veto_algorithm_competition_1}.

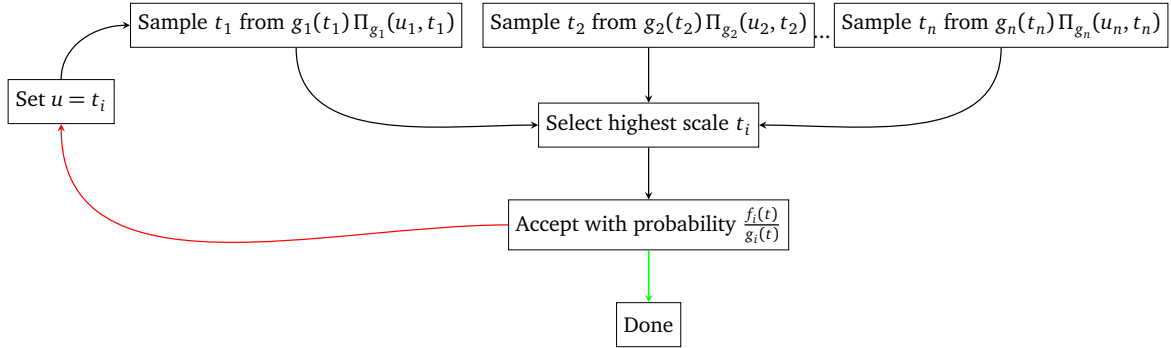
\begin{figure}
\centering
\resizebox{\columnwidth}{!}{
  \begin{tikzpicture}
  \node[block] (sample1) at (0, 3.4) {Sample $t_1$ from $g_1(t_1) \, \Pi_{g_1}(u_1,t_1)$};
  \node[block] (sample2) at (6, 3.4) {Sample $t_2$ from $g_2(t_2) \, \Pi_{g_2}(u_2,t_2)$};
  \node       (dots) at (8.95, 3.2) {... };
  \node[block] (samplen) at (12, 3.4) {Sample $t_n$ from $g_n(t_n) \, \Pi_{g_n}(u_n,t_n)$};

  \node[block] (veto) at (-4, 2.1) {Set $u = t_i$};

  \node[block] (highest) at (6, 1.7) {Select highest scale $t_i$};
  \node[block] (accept)  at (6, 0) {Accept with probability $\frac{f_i(t)}{g_i(t)}$};
  \node[block] (done)    at (6, -1.7) {Done};

  \draw[blackarrow] (sample1) to [out=270,in=180] (highest);
  \draw[blackarrow] (sample2) to [out=270,in=90]  (highest);
  \draw[blackarrow] (samplen) to [out=270,in=0]   (highest);
  \draw[blackarrow] (highest) to [out=270,in=90]  (accept);
  \draw[greenarrow] (accept) to  [out=270,in=90]  (done);
  \draw[redarrow]   (accept) to  [out=180,in=270] (veto);
  \draw[blackarrow] (veto)   to  [out=90,in=180]  (sample1);

  \end{tikzpicture}
}
\caption{Flowchart representation of the second-competition veto algorithm.
This algorithm is used to to interleave initial-state and final-state radiation 
with multiple parton interactions, and is more efficient when branching-kernel 
evaluation is expensive, such as for matrix-element corrections.}
\label{alg:veto_algorithm_competition_2}
\end{figure}

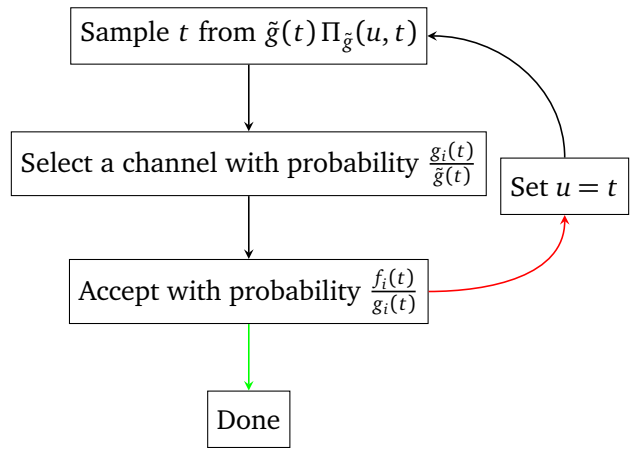
\begin{figure}
\centering
\begin{tikzpicture}
\node[block] (sample) at (0, 5.1) {Sample $t$ from $\tilde{g}(t) \, \Pi_{\tilde{g}}(u,t)$};
\node[block] (select) at (0, 3.4) {Select a channel with probability $\frac{g_i(t)}{\tilde{g}(t)}$};
\node[block] (accept) at (0, 1.7) {Accept with probability $\frac{f_i(t)}{g_i(t)}$};
\node[block] (done)   at (0, 0.0) {Done};
\node[block] (veto)   at (4.2, 3.1) {Set $u = t$};

\draw[blackarrow] (sample) to [out=270,in=90] (select);
\draw[blackarrow] (select) to [out=270,in=90] (accept);
\draw[redarrow]   (accept) to [out=0,in=270](veto);
\draw[blackarrow] (veto) to [out=90,in=0](sample);
\draw[greenarrow] (accept) to [out=270,in=90](done);    
\end{tikzpicture}
\caption{Flowchart representation of the third-competition veto algorithm.
Useful in situations where multiple channels have similar overestimates, 
like for quark-flavour selection in $\g \rightarrow \qqbar$ splittings and 
the efficient sampling of the large number of branchings in \vincia's electroweak shower.}
\label{alg:veto_algorithm_competition_3}
\end{figure}

It may be shown to yield \cref{eq:shower-distribution-competition} as follows:
\begin{align} \label{eq:veto-algorithm-competition-1}
&\left[\prod_{i=1}^{n}\int_{0}^{u} \d t_{i} f_{i}(t_{i}) \, \Pi_{f_i}(u,t_i) \right] 
\sum_{j=1}^{n} \Bigg[ \prod_{k\neq j} \Theta\left(t_{j}-t_{k}\right) \Bigg]\,
\delta\left(t-t_{j}\right) \nonumber \\
=& \sum_{i=1}^{n}\left[\prod_{j\neq i} \int_{0}^{t_{i}} \d t_{j} f(t_{j}) \, \Pi_{f_j}(u, t_j) \right] 
\int_{0}^{u} \d t_{i} f_{i}(t_{i}) \, \Pi_{f_i}(u, t_i) \,\delta\left(t-t_{i}\right) \nonumber \\
=& \sum_{i=1}^{n}f_{i}(t) \, \Pi_{f_i}(u, t) \prod_{j\neq i} \Pi_{f_j}(u, t) = \tilde{p}(t|u).
\end{align}
Equivalently, the result of \cref{eq:veto-algorithm-competition-1} may be used 
with overestimates $g_i(t_i)$ in place of $f_i(t_i)$, then selecting the highest 
scale before proceeding to the acceptance step. 
A flowchart representation of this procedure is shown in \cref{alg:veto_algorithm_competition_2}, and produces the same result. 
This algorithm is used to interleave initial-state and final-state radiation 
with multiple parton interactions. 
Furthermore, it is more efficient when branching-kernel evaluation is expensive, 
such as for matrix-element corrections. 
A third option is available, where instead a single scale is drawn according 
to the sum of overestimates $\tilde{g}(t)$ and a channel is selected with 
probability $g_i(t)/\tilde{g}(t)$ for the acceptance step.
A flowchart representation of this procedure is shown in \cref{alg:veto_algorithm_competition_3} 
and again produces \cref{eq:shower-distribution-competition}.
This procedure is used in a few specific places, in particular when the overestimates of 
several channels are very similar. 
Examples include quark-flavour selection in $\g \rightarrow \qqbar$ splittings and 
the efficient sampling of the large number of branchings in \vincia's \ac{EW} shower.
It is important to note that these algorithms may also be combined, 
for instance by grouping several channels for use with the algorithm depicted in
\cref{alg:veto_algorithm_competition_3}, 
and further combining them with the algorithms depicted in \cref{alg:veto_algorithm_competition_1} 
or~\cref{alg:veto_algorithm_competition_2}. 
In fact, the different shower models available in \pyt often use 
different procedures to optimize code structure and performance.

\subsubsection{Phase space (M-generator and RAMBO)}\index{Phase-space generation}
\label{subsection:phasespace}

One standard task is to distribute the momenta of final-state
particles uniformly according to \ac{LIPS}, on top
of which then later dynamical aspects can be added. 
(Non-uniform sampling methods, used \eg when resonances are present, are discussed separately, in \cref{subsection:processGenBasics}.) 
The relevant phase-space density is
\index{Phase-space generation!Lorentz Invariant@Lorentz Invariant
  Phase Space $\d\Phi_n$}\index{dPhin@$\d\Phi_n$}
\begin{equation}
  \d\Phi_n(P; p_1, p_2, \ldots, p_n) = \left(2\pi\right)^4 \,
  \delta^{(4)}\left(P - \sum_{i = 1}^{n} p_i\right) \;
  \prod_{i = 1}^{n} \frac{\d^3 p_i}{\left(2\pi\right)^3 \, 2p_i^0} ~.
  \label{eq:lips}
\end{equation}
where $P$ is the total four-momentum, and $p_i, i \geq 1$
are the $n$ different outgoing four-momenta. 
Usually the process is initially considered in the rest frame of the system, $P = (M;\mathbf{0})$, and later boosted to the relevant frame of the whole event.

A common special case is two-body final states, where the \ac{CM}-frame expression
reduces to
\begin{equation}
  \d\Phi_2= \frac{|\mathbf{p}|}{16 \pi^2 M} \, \d\Omega
  = \frac{|\mathbf{p}|}{16 \pi^2 M} \, \d\cos\theta \, \d\varphi ~.
\end{equation}
That is, the direction of one of the outgoing particles has to be
picked uniformly on the unit sphere, with the other moving out in the
opposite direction. The three-momentum length is
\begin{equation}
  |\mathbf{p}| = |\mathbf{p}_1| = |\mathbf{p}_2|
  =  \frac{\sqrt{\lambda\left(M^2, m_1^2, m_2^2\right)}}{2M} \,
\end{equation}
where the K\"all\'en $\lambda$ function can be written in a number of
equivalent ways
\begin{align}
  \lambda(a^2, b^2, c^2)
   & = a^4 + b^4 + c^4 - 2a^2 b^2 - 2a^2 c^2 - 2b^2 c^2 \nonumber \\
   & = (a^2 - b^2 - c^2)^2 - 4 b^2 c^2 \nonumber \\
   & = (a^2 - (b + c)^2) (a^2 - (b - c)^2) \nonumber \\
   & = (a + b + c) (a - b - c) (a - b + c) (a + b - c) ~.
\end{align}
The energies are given by
\begin{align}
  p_1^0 & = \sqrt{m_1^2 + \mathbf{p}^2}
        = \frac{M^2 + m_1^2 - m_2^2}{2M} ~,\nonumber \\
  p_2^0 & = \sqrt{m_2^2 + \mathbf{p}^2}
        = \frac{M^2 + m_2^2 - m_1^2}{2M} ~.
\end{align}

For three or more final particles, \pyt implements two different generic
methods, the older M-generator \cite{James:1968gu} and the newer RAMBO
\cite{Kleiss:1985gy} one, but approaches tailor made for the specific
situation are also common, \eg in parton showers. RAMBO is the best
choice for the case of massless products, whereas the situation is
less obvious once the masses constitute a significant fraction of the
full energy.

\index{Phase-space generation!M-generator}
The basic idea of the M-generator strategy is to view the full event as
arising from a sequence of fictitious two-body decays. Thus a four-body
decay $0 \to 1 + 2 + 3 + 4$, as an example, is viewed as a sequence
$0 \to 123 + 4 \to 12 + 3 + 4 \to 1 + 2 + 3 +4$, where $123$ and $12$
represent intermediate states. By a suitable insertion of a unit factor
\begin{equation}
  1 = \delta^{(4)}(p_{12} - p_1 - p_2) \, \d^4p_{12} \, 
  \delta(m_{12}^2 - p_{12}^2) \, \d m_{12}^2
  = \delta^{(4)}(p_{12} - p_1 - p_2) \, \frac{\d^3 p_{12}}{2p_{12}^0} \,
  \d m_{12}^2~, 
\end{equation}
and a similar one for $p_{123}$,
the four-body phase space can be reformulated as
\begin{equation}
  \d\Phi_4\left(P; p_1, p_2, p_3, p_4\right) \propto
  \d\Phi_2\left(P; p_{123}, p_4\right) \, \d m_{123}^2 \, \d\Phi_2\left(p_{123}; p_{12}, p_3\right)
  \, \d m_{12}^2 \, \d\Phi_2\left(p_{12}; p_1, p_2\right) ~. 
\end{equation}
The mass-dependent parts can be collected and simplified to 
\begin{equation}
\frac{\sqrt{\lambda(m_0^2, m_{123}^2, m_4^2)}}{m_0} \, \d m_{123} \,  
\frac{\sqrt{\lambda(m_{123}^2, m_{12}^2, m_3^2)}}{m_{123}} \, \d m_{12} \,  
\frac{\sqrt{\lambda(m_{12}^2, m_1^2, m_2^2)}}{m_{12}} ~. 
\end{equation}
The $(m_{123}, m_{12})$ phase space can easily be sampled within
allowed borders, but the rest of the expression then becomes a weight
that has to be taken into account by hit-and-miss Monte Carlo. This is
where the algorithm can be slow. Once the intermediate masses have been
selected, two-particle kinematics are constructed in a sequence of rest
frames for $1 + 2$,  $12 + 3$ and $123 + 4$, interleaved with Lorentz
boosts between them.

\index{Phase-space generation!RAMBO}\index{RAMBO}
The RAMBO algorithm provides an alternative sampling of $n$-body
phase space, which by construction has constant (uniform) weight for
arbitrary $n$ in the massless limit. The starting point is the following
identity for massless four-vectors, 
\begin{equation}
\int d^4 q \, \delta(q^2) \exp(-q^0) = \int_0^{\infty} \frac{q^0}{2} \exp(-q^0) 
\int d\Omega = 2\pi~.
\label{eq:flat-ps-component}
\end{equation} 
A four-momentum $q$ distributed according to the integrand of
the left-hand side of this identity can be generated via the steps
\begin{equation}
\qquad q^0 = -\log(R_1 R_2), \, \cos \theta = 2 R_3 - 1, \, \varphi = 2 \pi R_4~,
\end{equation}
\begin{equation}
  \implies ~q = (q^0, q^0 \sin \theta \sin \varphi, q^0 \sin \theta \cos \varphi, q^0 \cos \theta)~.
\end{equation} 
RAMBO repeats this process $n$ times to produce a set of momenta
$q_i$ that initially have $\sum_i q_i^\mu \equiv Q^\mu$. The final momenta $p_i$
are then constructed by applying a boost $\Lambda^\mu_{\, \nu}$ to the CM frame of $Q$ and scaling by an
overall factor $x = M/\sqrt{Q^2}$, so that $P^\mu \equiv \sum p^\mu_i =
x(\Lambda^\mu_{\, \nu} Q^\nu) = 
(M,{\bf 0})$.

To illustrate that this leads to momenta distributed according to \cref{eq:lips}, we may 
start from $n$ multiples of \cref{eq:flat-ps-component} and unitarily transform the momenta as 
\begin{align}
(2\pi)^n &= \prod_{i=1}^n \int d^4 q_i \, \delta(q_i^2) \exp(-q_i^0) \nonumber \\
&\times d^4 Q \, \delta^4 \left(Q - \sum_{i=1}^n q_i\right) 
dx \, \delta\left(x - \frac{M}{\sqrt{Q^2}}\right) \nonumber \\
&\times \prod_{i=1}^n d^4 p_i \, \delta^4(p_i - x \left(\Lambda q_i\right) )~.
\label{eq:massless-rambo}
\end{align} 
Identifying the phase-space measure \cref{eq:lips} in \cref{eq:massless-rambo} and 
integrating over the other variables then leads to 
\begin{equation}
\int \d\Phi_n(P; p_1, p_2, \ldots, p_n) = \left(\frac{\pi}{2}\right)^{n-1} 
\frac{M^{2n-4}}{(n-1)! (n-2)!}~,
\label{eq:massless-lips-weight}
\end{equation} 
which is indeed the $n$-body massless phase-space volume \cite{Kleiss:1985gy}.
RAMBO thus samples the massless phase space isotropically, with
constant weight given by \cref{eq:massless-lips-weight}.  

\index{Phase-space generation!Massive particles}
For massive particles, no equivalent general expression for the phase-space
volume exists. 
However, the massless RAMBO algorithm may be adapted to the massive case 
at the cost of introducing variable event weights, which translates to
a reduced efficiency at the unweighted level.
Starting from the massless momenta $p_i$, massive momenta $k_i$ are
obtained through 
\begin{equation}
\mathbf{k}_i = y \mathbf{p}_i, \, k_i^0 = \sqrt{|\mathbf{k}_i|^2 + m_i^2}.
\end{equation} 
The momenta $k_i$ are on-shell and preserve momentum conservation as
long as the  rescaling parameter $y$ is given by the solution of the equation
\begin{equation}
\sum_{i=1}^n \sqrt{y^2 |\mathbf{p}_i|^2 + m_i^2} = M.
\label{eq:massive-rambo-condition}
\end{equation} 
Since \cref{eq:massive-rambo-condition} is a monotonic function of
$y$ with a solution  
$0 \leq y \leq 1$, the value of $y$ may be determined easily using the
Newton--Raphson method. 
Through a similar procedure as the one followed in
\cref{eq:massless-rambo}, the  
event weight may be determined to be 
\begin{equation}
w = \left(\frac{\pi}{2}\right)^{n-1} 
\frac{1}{(n-1)! (n-2)!}
\left( \prod_{i=1}^n \frac{|\mathbf{k}_i|}{k_i^0}\right)
\left( \sum_{j=1}^n \frac{|\mathbf{k}_j|^2}{k_j^0} \right)^{-1}
\left(\sum_{j=1}^n |\mathbf{k}_j|\right)^{2n - 3}~,
\end{equation} 
which is bounded from above by the massless weight,
\cref{eq:massless-lips-weight}, so that the distribution can be
unweighted by accepting the generated massive phase-space point with the
probability
\begin{equation}
  P_\mathrm{accept} = \overbrace{\prod_{i=1}^n \frac{|{\bf k}_i|}{k_i^0}\left(\frac{
  \sum_{j} |\mathbf{k}_j|}{\sum_{j}
      \frac{|\mathbf{k}_j|^2}{k_j^0}}\right)}^{<1} \overbrace{
    \left(
    \frac{\sum_{j}|\mathbf{k}_j|}{M}\right)^{2n-4}}^{<1}~,
\end{equation} 
which (by construction) tends to unity in the massless limit.

%% file: introduction/process-generation.tex
\subsection{Process-generation basics}
\index{Phase-space generation}\label{subsection:processGenBasics}

Particle-physics cross sections can crudely be divided into
two categories: perturbative and non-perturbative. Both kinds of
processes play crucial roles in \pyt. The former can be computed
order by order in perturbation theory, \eg based on Feynman-diagram
rules. For the electroweak sector, the couplings are sufficiently
small that higher-order corrections should offer a rapidly converging
series. The exception is the enhanced emission of soft or collinear
photons, but this is a well-understood issue. For the strong sector,
on the other hand, 
the large $\alphas$ coupling leads to a slower convergence.
It can still work well for QCD processes involving large momentum
transfers. In the opposite limit, at low momentum transfers, the
perturbative coupling diverges and perturbation theory breaks
down. Therefore the 
total cross section in hadron-hadron collisions, which is dominated
by such low scales, can only be described in terms of effective,
phenomenological models. The same applies for its main components
--- elastic, diffractive and nondiffractive cross sections ---
which therefore are classified as non-perturbative processes. 

In the current section, the focus will be on perturbative processes,
introducing how these are 
defined and generated inside \pyt. Non-perturbative processes are
discussed separately in \cref{sec:soft}.  There are also
components that partly bridge the gap between the two, such as
multiparton interactions (MPIs), hard diffraction, and photoproduction
processes. These are also discussed in \cref{sec:soft}, along
with further aspects specific to simulating cross sections in
heavy-ion collisions.  

To begin the discussion of perturbative process generation, consider
a process $a + b \to f_n$, where $a$ and $b$ are 
two incoming particles that together create a final state $f$
consisting of $n$ particles. The differential cross section can then
be written as
\index{matrix element}
\begin{equation}
  \frac{\d\hat{\sigma}}{\d\Phi_n}
  = \frac{|\mathcal{M}|^2}{2\sqrt{\lambda(\hat{s}, m_a^2, m_b^2)}}
  \approx \frac{|\mathcal{M}|^2}{2\hat{s}} ~,
\end{equation}
 where
$\hat{s} = (p_a + p_b)^2$ is the squared invariant mass of the
collision system. Usually $m_a$ and $m_b$ are negligible in comparison
with $\sqrt{\hat{s}}$, and then the last expression is obtained. The
process-specific physics is encapsulated in the matrix element
$\mathcal{M}$, which we shall assume can be calculated perturbatively.
The $|\mathcal{M}|^2$ expression also has to be
averaged over incoming spin and colour configurations, and summed over
outgoing spin and colour configurations, where relevant.

In some rare cases $a$ and $b$ are the actual incoming beam
particles. Normally, however, $a$ and $b$ are constituents of the 
true beam particles, $A$ and $B$. Then one needs to introduce parton
distribution functions (PDFs), $f_a^A(x, Q^2)$ (and $f_b^B(x,Q^2)$),
that to leading order describe the probability to find a parton $a$
inside the particle $A$, with a fraction $x$ of the particle
four-momentum, if the hard-collision process probes the particle
at a (factorization) scale $Q^2$. The cross section then reads
\index{PDFs!Factorization theorem}
\index{Collinear factorization|see{PDFs}}
\index{Factorization theorem|see{PDFs}}
\begin{equation}
  \sigma = \int \d x_1 \, f_a^A(x_1, Q^2) \; \int \d x_2 \, f_b^B(x_2, Q^2)
  \; \int  \, \frac{\d\hat{\sigma}(\hat{s}, Q^2)}{\d\Phi_n} \, \d\Phi_n  ~,
  \label{eq:basics-sigma-nbody}
\end{equation}
where
\begin{equation}
  \hat{s} = x_1 x_2 s~~~~~\mbox{with}~~s = (p_A + p_B)^2~.
\end{equation}
The nature of the
PDFs varies depending on what kind of particle is concerned: hadrons,
nuclei, leptons, photons, or pomerons. They will therefore be discussed 
further in the respective beam context. The most commonly used and
best studied are the proton PDFs, \cf\cref{sec:hadronPDFs},
and for these we will omit the $A$ and $B$ superscripts. 

\subsubsection[$2\to2$ processes]{$\bf 2\to 2$ processes}
\index{Phase-space generation!2-body processes}%
\index{Kinematics!for 2to2 processes@for $2\to2$ processes}
\paragraph{Massless Kinematics:}
In a massless $2 \to 2$ subprocess $a(p_1) + b(p_2) \to c(p_3) + d(p_4)$
it is conventional to write the cross section in terms of the Mandelstam
variables
\begin{align}
  \hat{s} & =  (p_1 + p_2)^2 = (p_3 + p_4)^2 ~, \label{eq:sHat}  \\
  \hat{t} & =  (p_1 - p_3)^2 = (p_2 - p_4)^2
       =  -\frac{\hat{s}}{2} (1 - \cos\hat{\theta}) ~, \label{eq:tHat} \\
  \hat{u} & = (p_1 - p_4)^2 = (p_2 - p_3)^2
                =  -\frac{\hat{s}}{2} (1 + \cos\hat{\theta}) \label{eq:uHat}~,
\end{align}
where $\hat{\theta}$ is 
the scattering angle, defined as the polar angle of particle 3, in
the rest frame of the collision. Since $\d\Phi_2  
\propto \d\cos\hat{\theta} \propto \d\hat{t}$ 
(assuming a trivial flat $\varphi$ dependence, as is the case unless the
incoming beams are transversely polarized), it is common to recast 
\cref{eq:basics-sigma-nbody} accordingly. Furthermore,
$\d x_1 \, \d x_2 = \d \tau \, \d y$, where $\tau = x_1 x_2 = \hat{s}/s$
and $y = (1/2) \ln(x_1/x_2)$. It is also standard to use
$x f(x)$ rather than $f(x)$. In total this gives
\begin{equation}
  \sigma = \iiint \frac{\d\tau}{\tau} \, \d y \,  \d\hat{t} \,
  x_1f_a^A(x_1, Q^2) \, x_2f_b^B(x_2, Q^2) \,
  \frac{\d\hat{\sigma}(\hat{s}, \hat{t}, Q^2)}{\d\hat{t}} ~.
  \label{eq:basics-sigma-twobody}
\end{equation}
The $\hat{u}$ variable is redundant since $\hat{s} + \hat{t} + \hat{u} = 0$,
but often symmetry properties of matrix elements are apparent if it is used
judiciously. In a frame where $a$ and $b$ come in back-to-back, moving in
the $\pm z$ directions, $\hat{p}_\perp^2 = \hat{t}\hat{u}/\hat{s}$ is the squared
transverse momentum of the outgoing $c$ and $d$. A frequent choice is
to put $Q^2 = \hat{p}_\perp$ as the factorization scale.

The sampling ranges for each of the $(\tau, y, \hat{t})$ variables
depends on whether phase-space cuts are imposed at the
process-generation level, \cf
\cref{subsection:phasespacecuts}. Generically,
they are:
\begin{alignat}{2}
  \frac{\hat{s}_\mathrm{min}}{s} & < \tau && < \frac{\hat{s}_\mathrm{max}}{s}~,\\
  -\frac12 |\ln\tau| & < y(\tau) && < \frac12|\ln \tau|~, \\
  \sqrt{1 - \frac{4\hat{p}_{\perp\mathrm{max}}^2}{\tau s}} & < |z(\tau)| && <
 \sqrt{1 - \frac{4\hat{p}_{\perp\mathrm{min}}^2}{\tau s}}~,
\end{alignat}
where $\hat{t}$ has been replaced by $z = \cos\hat{\theta}$ via
\cref{eq:tHat}, and we emphasize that there are solutions for both
positive and negative $z$. 
The phase-space boundaries are set via the (user-specifiable)
parameters $\hat{m}_\mathrm{min,max}$, $\hat{p}_{\perp\mathrm{min,max}}$, and/or 
$\hat{Q}^2_\mathrm{min}$, \cf\cref{subsection:phasespacecuts}. To give
some examples:
\begin{itemize}
  \item For processes containing an $s$-channel resonance, it may be
desirable to only generate phase-space points within a specific range of
$\hat{m}$ values. Processes involving resonance production and decay
are discussed in more detail in \cref{sec:resonances}.
\item A restriction like $\hat{p}_\perp > \hat{p}_{\perp\mathrm{min}}$, implying
$\hat{s}_{\mathrm min} = \max(\hat{m}_\mathrm{min}^2,4\hat{p}_{\perp\min}^2)$, 
is mandatory for  matrix elements that diverge in the
$\hat{p}_\perp \to 0$ limit; this includes in particular
massless $t$-channel QCD processes. It may also be convenient for
studies focusing on the high-$p_\perp$ tail of ``hard'' $2\to 2$
processes.
\item The option to specify a $Q^2_\mathrm{min}$ value is
intended for $t$-channel DIS-type processes with distinguishable
final-state particles, \cf\cref{subsection:phasespacecuts}, in
which case $\hat{s}_\mathrm{min} \ge Q^2_\mathrm{min}$ and\\
$z(\tau) \le 1 - 2Q_\mathrm{min}^2/(\tau s)$.  
\end{itemize}

The selection of phase-space points $(\tau, y, z)$ is described
in detail in \citeone{Sjostrand:2006za}, and remains unchanged.
The basic strategy is to use multichannel sampling in each of the
three variables separately. Thus, for instance, the $\tau$ dependence is
modelled as a mix of sampling according to either $1/\tau$ or
$1/\tau^2$. The normalization factors of the two possibilities are
determined at initialization, and would depend on the process,
the choice of PDFs and $\alphas$, and the $p_{\perp\mathrm{min}}$ cut.
That way an upper envelope is found for the real cross-section
expression. The probability that a trial phase-space point is retained
is given by the ratio of the full differential cross section to the
multichannel overestimate, and the accepted events are assigned a
standard weight of unity. There is always the risk that the intended upper
estimate of the cross section is exceeded by the full expression in
some corner of phase space, even if it is not common. Such points are
associated with a weight correspondingly above unity.

The cross section for a process is obtained in parallel with the
generation of events, using the multidimensional generalization of
\cref{eq:random-integrate}. Thus the error decreases with the number
of events generated.

When several processes are to be generated simultaneously, an upper
envelope is found for each differential cross section separately. The
size of integrated envelopes, \ie the upper estimate of the respective
cross sections, is used as a relative weight when the next process is
selected. If the trial phase-space point is rejected then a new process
choice is made. That is, a larger overestimate will make a given
process more likely to be picked, but then afterwards also more likely
to be rejected. In the end, all processes are generated in proportion to
their correct integrated cross sections.

Generation in $(\tau, y, z)$ is only one possible choice. It has
the advantage that additional $\tau$ terms can be used for the sampling
of resonances in the cross section, \cf\cref{sec:resonances}.
For the generation of MPIs, however, it is essential to use
$\hat{p}_\perp^2$ rather than $\hat{t}$, \cf
\cref{subsection:mpi}. Then one may instead note that 
\begin{equation}
  \frac{\d x_1}{x_1} \, \frac{\d x_2}{x_2} \, \d\hat{t} =
  \frac{\d\tau}{\tau} \, \d y \, \d\hat{t} =
  \d y_3 \, \d y_4 \, \d\hat{p}_\perp^2 ~,
\end{equation}
where $y_3$ and $y_4$ are the rapidities of the two outgoing particles.

\paragraph{Massive Kinematics:}\index{Phase-space generation!Massive particles}

So far we have considered massless kinematics. It is quite common
to have cases where one or both of the outgoing particles are massive,
while the incoming ones still are assumed massless. In some cases,
such as elastic 
scattering, both incoming and outgoing masses need to be taken into
account. The fully general $\hat{t}$ expression is
\begin{equation}
  \hat{t} = - \, \frac{\hat{s}^2 - \hat{s} (m_1^2 + m_2^2 + m_3^3 + m_4^2)
   + (m_1^2 - m_2^2) (m_3^2 - m_4^2) - 
   \sqrt{\lambda(\hat{s}, m_1^2, m_2^2) \, \lambda(\hat{s}, m_3^2, m_4^2)}
   \, \cos\hat{\theta}}{2\hat{s}} ~,
\end{equation}
with $\hat{u}$ obtained by $m_3^2 \leftrightarrow m_4^2$ and
$\cos\hat{\theta} \to - \cos\hat{\theta}$, resulting in 
\begin{equation}
  \hat{s} + \hat{t} + \hat{u} = m_1^2 + m_2^2 + m_3^3 + m_4^2~.
\end{equation}
The limits
$\hat{t}_{\mathrm{min}} < \hat{t} < \hat{t}_{\mathrm{max}}$ (all negative or,
for $\hat{t}_{\mathrm{max}}$, zero) are obtained for
$\cos\hat{\theta} = \mp 1$. Often $\hat{t}_{\mathrm{max}}$ is close to zero
and a numerically safer recipe for it is obtained by noting that
\begin{equation}
  \hat{t}_{\mathrm{min}} \, \hat{t}_{\mathrm{max}} = (m_3^2 - m_1^2)(m_4^2 - m_2^2)
  + \frac{(m_1^2 + m_4^2 - m_2^2 - m_3^2)(m_1^2 m_4^2 - m_2^2m_3^2)}{\hat{s}} ~.
\end{equation}
If $m_1^2 = m_2^2 = 0$ then
$\hat{t}_{\mathrm{min}} \, \hat{t}_{\mathrm{max}} = m_3^2 m_4^2$ and
$\hat{p}_\perp^2 = (\hat{t}\hat{u} - m_3^2 m_4^2) / \hat{s}$.

\subsubsection[$2\to3$ processes]{$\bf 2\to 3$ processes}
\index{Phase-space generation!3-body processes}%
\index{Kinematics!for 2to3 processes@for $2\to3$ processes}

In pure $s$-channel $2 \to 3$ processes, say
(unpolarized) $\epem \to \gam^*/\Z \to \qqbar\g$, cross sections
factorize into production and decay steps, and the decay phase space is
easy to generate in terms of two energy variables and three angles.
Such decays are not coded as explicit hard processes,
however, but instead are handled during the parton-level shower
evolution. Three-body final states such as  $\epem \to \gam^*/\Z \to \qqbar\g$
are then reached via showering from
$\epem \to \gam^*/\Z \to \qqbar$ (\cf\cref{sec:resonances} on resonances
and \cref{sec:showers} on parton showers), with
matrix-element corrections applied to the extent available and
switched on, \cf\cref{section:matchmerge}.     

For $2\to 3$ hard processes that do not factorize into resonance
production and decay plus shower, it becomes much more messy to set up phase
space, since there are more possibilities for peaks in different places.
\pyt does not have a general-purpose machinery to handle generic cross
sections. Instead, the main assumption is
that such processes are 
provided via the Les Houches accord, \cf
\cref{subsection:lha}, from external programs that have their
own phase-space generators.

There are a few
internal $2 \to 3$ processes, however, for very specific tasks. These
are generated according to one of three different prescriptions, tailored
to the squared amplitudes for massless QCD $2\to 3$ processes,
\ac{VBF}, and  central diffractive processes,
respectively. These were developed separately and employ somewhat 
different notation in the code, here relabelled for clarity.
Note that all three assume a cylindrical symmetry with respect to
the collision axis. 

\index{Hard QCD processes!Phase-space sampling for 3-body processes}Massless QCD $a(p_1) + b(p_2)
\to \c(p_3) + d(p_4) + e(p_5)$ cross 
sections contain divergences when any of the final-state particles
become collinear to the beam, collinear to each other, and/or soft.
It is therefore important to choose a set of phase-space variables
that allows for the isolation of these singularities. The parameterization used
in \pyt is $(y_3, y_4, y_5, \pTs_3, \pTs_4, \varphi_3, \varphi_4)$.
The rapidity  
sampling here is simple and consistent, while the $\pT$ selection is not,
unfortunately. The $\pT_5$ is fixed opposite to the vector sum of the
other two, and in the first instance gets a different $\pT$ spectrum than
them. Notably, a requirement for all $\pT > p_{\perp\mathrm{min}}$ can be
imposed with full efficiency for two, but is inefficient for the third.
It is also important to avoid the collinear singularity by an
additional cut on $R = \sqrt{(\Delta y)^2 + (\Delta \varphi)^2}$
for all outgoing pairs.

\index{VBF}A process of special interest is vector-boson fusion to a Higgs boson,
$\Wp\Wm \to \H$ and $\Z\Z \to \H$ (and/or $\Wp\Wp \to
\mathrm{H}^{++}$ in some BSM scenarios).
Since the bosons are emitted from fermion 
lines this results in $2 \to 3$ processes of the character
$\f_1(p_1) + \f_2(p_2) \to \f_3(p_3) + \f_4(p_4) + \H(p_5)$. The variables chosen
in this case are $(\tau, y, y_5, \pTs_3, \pTs_4, \varphi_3, \varphi_4)$.
Here, special care is taken in the modelling of $\pTs_3$ and $\pTs_4$ which,
unlike the QCD cross sections, have no $\pT \to 0$ divergence but instead
are fairly flat out to the gauge-boson mass. Note that the physics of the
process here naturally singles out the Higgs $\pT$ as having a different
shape than the other two, again different from the QCD case. The same
machinery is also used for heavy-quark fusion to Higgs,
$\qqbar \to \QQbar\H$ and $\g\g \to \QQbar\H$, where top masses are
selected with a Breit--Wigner shape.

\index{Central diffraction}
Another special case is central diffraction, \eg 
$\p(p_1) + \p(p_2) \to \p(p_3) + \p(p_4) + X(p_5)$, where $X$ is the
central diffractive system. Here sampling in $t_1 = (p_1 - p_3)^2$
and $t_2 = (p_2 - p_4)^2$ is crucial to impose an exponential fall-off
in these variables. The energy fractions $x_1$ and $x_2$ taken from the
incoming proton defines $m_X^2 = p_5^2 = x_1 x_2 s$ in the collinear limit
$t_1, t_2 \to 0$. Away from it also the $\varphi_3$ and $\varphi_4$ angles
play a role, and one requires a more elaborate definition of $x_1$ and $x_2$.

\subsubsection{Processes involving resonances} \label{sec:resonances}
\index{Phase-space generation!Resonances}\index{Resonance decays}%
\index{Decays!Resonances@of Resonances}%

The term ``resonance'' has a specific meaning in \pyt and 
refers to particles whose decays are considered to be 
part of the hard process. This enables 
\pyt to modify the total calculated cross section depending on which decay
channels are open or closed (including effects of sequential decays,
such as $\t \to \b \W^+$ followed by $\W^+ \to \ele^+ \neu_e$),
and also provides a natural framework for incorporating
process-specific aspects such as spin correlations and/or finite-width
effects. Here, we focus aspects of
phase-space generation common to all processes involving resonances.
Details on cross-section considerations, process-specific
features, and some further sophistications are explained in
\cref{sec:hardRes}, while user implementations of ``semi-internal''
resonances is described in \cref{sec:semi-internal}, and the
handling of \ac{SLHA} decay tables is covered in \cref{sec:slha}.  

\index{meMode@\texttt{meMode}}We focus first on the \emph{simplest}
treatment available in \pyt, 
with partial widths and branching fractions fixed to their
on-shell values. Technically, in the code this corresponds to decay
channels that are assigned \texttt{meMode = 100}. Values of
\texttt{101}, \texttt{102}, and \texttt{103} additionally include
some simple kinematic threshold effects, also discussed below.
Lastly, we emphasize that the \emph{default} treatment often goes
further than this, with most decay modes of SM resonances (and some
BSM ones) assigned  
\texttt{meMode = 0} implying the use of dedicated matrix-element
expressions for branching fractions that can vary over a reasonably
broad resonance peak; this is covered separately in
\cref{sec:hardRes}; see further \cref{sec:semi-internal} for
user implementations of such expressions. 

\index{Breit-Wigner distribution}
Starting from a cross section computed in the zero-width
approximation, \ie for stable final-state resonances, the simplest
shape modelling available in \pyt is a relativistic Breit--Wigner
substitution of the type  
\begin{equation}
1 = \int \delta (m^2 - m_0^2) \mathrm{d}m^2 \to
\int_{m^2_\mathrm{min}}^{m^2_\mathrm{max}}
\frac{1}{\pi} \frac{m_0 \Gamma_0}
     {(m^2 - m_0^2)^2 + m_0^2 \Gamma_0^2}\mathrm{d}m^2~,
     \label{eq:basicBW}
\end{equation}
for each final-state resonance, where $m_0$ and $\Gamma_0$ are
the nominal (on-shell) mass and widths of the resonance, respectively, 
and $m$ is allowed to vary in a range
$m \in [m_\mathrm{min},m_\mathrm{max}]$ that can be specified
    individually for each resonance in \pyt's particle data table.
Note that choosing a small range will reduce the total cross sections
accordingly. 

The phase-space integral in \cref{eq:basics-sigma-twobody} is then
extended to include integrations over $m^2$ for each resonance, and
the sampled values\footnote{See \citeref[sec.~7.4.2]{Sjostrand:2006za}
  for details on the sampling procedure.} for these masses 
are used instead of the on-shell ones in the evaluation of
$\mathrm{d}\hat{\sigma}/\mathrm{d}\hat{t}$ and also in the relations
between kinematic variables such as between $\hat{t}$ and
$\cos\hat\theta$.  This offers 
a crude level of approximation to the expected mass dependence of the
full cross section, at least in the vicinity of the resonance(s) where
the resonant amplitudes can still be assumed to dominate over any
(non-resonant) background processes.  

A complication arises for processes that involve pair production of
the same kind of particle, such as $\t\tbar$, $\W^+\W^-$, or $\Z^0\Z^0$
production. For such processes, on-shell matrix elements are phrased
in terms of a single pole-mass value, while the procedure above produces two
different values, $m_3$ and $m_4$. For the specific case of
double-vector-boson production, \pyt uses 4-fermion matrix elements
that include the full mass dependence (as well as the full
$\gam^*/\Z$ interference). However, for more general processes
involving two of the same kind of resonance (such as $\t\tbar$
production), the choice made in \pyt is to use an average squared mass,      
\begin{equation}
  \bar{m}^2 = \frac{m_3^2 + m_4^2}{2} - \frac{(m_3^2 -
    m_4^2)^2}{4\hat{s}}~, \label{eq:s34Avg}
\end{equation}
which is defined so that $(\hat{s},m_3^2,m_4^2)$ and
$(\hat{s},\bar{m}^2,\bar{m}^2)$ correspond to the same CM-frame
three-momenta, 
\begin{equation}
   |{\bf p}^*(\hat{s},\bar{m}^2,\bar{m}^2)|^2 = |{\bf
     p}^*(\hat{s},m_3^2,m_4^2)|^2 =
   \frac{1}{4\hat{s}}\left(
\hat{s} - (m_3 + m_4)^2\right)\left(\hat{s} - (m_3 - m_4)^2\right)~.
   \end{equation}
Analogous modified values for the $\hat{t}$ and $\hat{u}$
variables are defined to correspond to the same CM-frame scattering angle, 
\begin{align}
  \bar{\hat{t}} & ~=~ \hat{t} - \frac{(m_3^2 - m_4^2)^2}{4\hat{s}}
  ~=~ -\frac12\left( \hat{s} - 2\bar{m}^2 - 2 |{\bf
    p}^*|\sqrt{\hat{s}}\cos\hat\theta \right)~,\\
  \bar{\hat{u}} & ~=~ \hat{u} - \frac{(m_3^2 - m_4^2)^2}{4\hat{s}}
  ~=~ -\frac12\left( \hat{s} - 2\bar{m}^2 + 2 |{\bf
    p}^*|\sqrt{\hat{s}}\cos\hat\theta \right) ~,\label{eq:uHatAvg}
\end{align}
and these variables ($\bar{m}$, $\bar{\hat{t}}$, and $\bar{\hat{u}}$)
are then used in the evaluation of the on-shell cross-section
formula. If in doubt whether full matrix elements or the
mass-symmetrized approximation represented by \crefrange{eq:s34Avg}{eq:uHatAvg} is used for a given process, the corresponding
\texttt{sigmaKin()} method can be inspected in the code (with $\bar{m}^2$ then
typically denoted \texttt{s34Avg}).   
We note that, when gauge bosons are involved, the procedure is not
guaranteed to be gauge invariant, nor positive definite, and
breakdowns should be expected if any resonance masses are far from their
on-shell values. The alternative would be to change to use full 4- or
6-body matrix elements instead (as already done for double-vector-boson
production), \eg by interfacing external hard-process generators,
\cf\cref{sec:externalGenerators}. 

Be aware that, if a decay mode has been assigned \texttt{meMode = 100}
and $m_\mathrm{min}$ is such that the decaying resonance can fluctuate
down in mass to below the nominal threshold for the given decay mode
(\ie, $m_\mathrm{min} < \sum_j m_j$ with $m_j$ the on-shell daughter
masses for the decay mode in question), it is assumed that at least
one of the daughters could also fluctuate  down to keep the channel
open. Otherwise the program will hit an impasse.

Alternatively, simple step functions $\Theta(m - \sum_j m_j)$ can be
applied to impose kinematic thresholds; this is done for decay
channels that are assigned \texttt{meMode = 101}.
A slightly more sophisticated alternative is to use a smooth threshold
factor,
\begin{equation}
  \beta = \sqrt{ \left(1 - \frac{m_1^2 + m_2^2}{m^2}\right)^2 - 4 \frac{m_1^2 m_2^2}{m^4}}
\end{equation}
for two-body decay modes, and
\begin{equation}
  \sqrt{1 - \frac{\sum_j m_j}{m}}
  \end{equation}
for multi-body ones, again with $m_j$ equal to the on-shell masses of
the decay products for the given mode. The former correctly encodes
the shrinking size of the phase space near threshold (but would still
miss any non-trivial matrix-element factors) while the latter is
 only a crude simplification. Two separate options exist
for this, depending on whether the stored on-shell branching fraction
should be considered to already include this factor (\texttt{meMode =
  103}) or whether it should be modified by it (\texttt{meMode =
  102}). In the former case  (\texttt{meMode =
  103}), the actual factor applied is the ratio of the above to the
corresponding value for $m=m_0$, with a safety limit imposed in case
that denominator turns out to be very small, to avoid
unintentionally large rescalings at large $m$.

Among the options discussed thus far, only the no-threshold one 
(\texttt{meMode = 100}) allows for purely off-shell decay modes, \ie
ones for which the on-shell daughter
mass values exceed $m_0$; as noted above one or more of the daughters
must then be able to fluctuate down in mass, or there will be
trouble. The remaining options (\texttt{meMode = 101} -- \texttt{103})
are all restricted to phase-space points satisfying $\sum m_j < m$,
with $m_j$ the on-shell daughter-mass values.

Currently, the only higher level of sophistication available in \pyt
is to go all-in and implement dedicated decay-rate calculations
specific to each given resonance and decay mode; this is obtained for
\texttt{meMode = 0}. As mentioned above, this is the default for most
SM resonance decays in \pyt as well as for some BSM ones, meaning that
such code exists in the program (in the form of 
process-specific \texttt{SigmaProcess::weightDecay()} methods and
resonance-specific \texttt{ResonanceWidths::calcWidth()} methods) and
is used by default. See further \cref{sec:hardRes}. 

Finally, note that both \pyt's simple shower as well as the \vincia
antenna shower allow for the insertion of resonance decays as $1\to n$ branchings
in the overall perturbative evolution, at decay-specific
perturbative scales. This is called interleaved resonance
decays~\cite{Brooks:2021kji} and is also further discussed in
\cref{sec:hardRes}.

%% file: physics/introduction.tex
The \pyt event generator is the product of a
physics development program in close touch with experimental reality.
The two have often gone hand in hand, by making it possible to check which
ideas work and which do not. Many of the concepts that today form
the accepted picture of high-energy collisions can be traced
directly to this undertaking, including string fragmentation, dipole showers,
multiparton interactions, colour reconnection, and more.
This part of the manual provides details on these and the other physics models
encoded in \pyt. Many of these models still evolve to handle new
experimental input, or to accommodate the progress in our theoretical
understanding of the standard model of particle physics.

The first section describes the physics processes --- sometimes denoted
``hard processes'' --- internal to \pyt. These processes are those
that can be calculated in leading-order perturbation theory in the
standard model or simple extensions.  While some of the calculations are
currently outclassed, and are more of interest as
a cross check or for quick studies, others, particularly the
treatment of jet production and BSM physics models, are still actively
used for comparisons with data. Further sections describe the core
of the \pyt engine: parton showers, multiparton interactions,
hadronization, and particle decays. An important complement to these
sections is the one on matching and merging, which documents the
methods for interfacing external calculations (that are more precise
in perturbation theory) with the \pyt engine.

Notable additions to the \pythia presentation here are the
descriptions of two parton-shower plugins --- \dire and \vincia --- and
the heavy-ion collision machinery.

%% file: physics/hard-proc.tex
\section{Internal process types}

\pythia includes a good selection of native hard processes at \ac{LO}. The hard processes are generated by sampling the allowed phase space using matrix elements, and convoluted with the PDFs, as a weight. Usually the multichannel sampling strategy results in accepted events with a common unit weight, but there are exceptions to this rule, so it is wise to be prepared for non-unit weights in event-analysis programs. In addition to the internal processes, there are several ways to feed in externally generated hard processes for showering and hadronization, including several options for matching and merging of higher-order processes. These options are discussed in detail in \cref{section:matchmerge} and \cref{section:intext}. This section classifies and lists internally defined hard processes and discusses about special features and appropriate settings for given processes. All internally defined processes are listed in \appref{sec:hardProcesses}, where also references to the cross-section formulae are given. If a process is included for both quark and lepton initial or final particles, the process is written with an ``f'' (denoting fermion) whereas a ``q'' is used when only quarks are expected. Charge-conjugate processes are always implicitly included.

\subsection{Hard QCD}\index{Hard QCD processes}\index{QCD processes}
\label{subsection:hardQCD}

\index{Hard Process! QCD}The internal QCD processes can be classified in three categories: (1) $2 \rightarrow 2$ scattering of light quarks and gluons, (2) production of heavy flavours (charm and bottom) in $2 \rightarrow 2$ processes, and (3) $2 \rightarrow 3$ processes involving light quarks and gluons. For hard processes suitable, process-dependent, phase-space cuts need to be applied to avoid soft and collinear singularities of perturbative QCD. In addition to these, soft QCD processes are included. These are discussed in \cref{subsection:sigmatotal} and include also a unitarized version of the hard $2 \rightarrow 2$ cross sections, where the divergences in the $\pT \rightarrow 0$ limit have been regulated with the screening parameter $\pTo$, see \cref{subsection:mpi} for more details. Normally the hard and soft QCD processes would not be used simultaneously, since typically they target different kinds of physics studies. If they are combined nevertheless, relevant phase-space cuts should be introduced to separate the regions handled by each machinery, to prevent double counting.

\subsubsection{Light quarks and gluons}

This subclass of processes contain all possible $2 \rightarrow 2$ scatterings of massless quarks and gluons. In total there are six possibilities:
\begin{itemize}
\item $\g \g \rightarrow \g \g $
\item $\g \g \rightarrow \q \qbar $
\item $\q \g \rightarrow \q \g $
\item $\q \q' \rightarrow \q \q' $ where incoming and outgoing flavours are the same
\item $\q \qbar \rightarrow \g \g $
\item $\q \qbar \rightarrow \q' \qbar' $
\end{itemize}
By default the light flavours include $\u$, $\d$ and $\s$ quarks but it is also possible to produce $\c$ and $\b$ quarks with the massless matrix elements.

\subsubsection{Heavy flavours}

These processes provide heavy-quark pair production where heavy flavours stand for $\c$ and $\b$. In LO, there are two possible processes each:
\begin{itemize}
\item $\g \g \rightarrow \c \cbar $
\item $\q \qbar \rightarrow \c \cbar $
\item $\g \g \rightarrow \b \bbar $
\item $\q \qbar \rightarrow \b \bbar $
\end{itemize}
Unlike the case of massless partons, the finite mass also makes the matrix element expressions finite in the $\pT \rightarrow 0$ limit, so there is no need to introduce phase-space cuts to avoid divergences. However, it is also possible to generate these processes within the regularized soft QCD framework, though the mass effects are then not accounted for in the matrix elements. When considering heavy-quark production, one should keep in mind that especially $\c$ quarks are abundantly produced in the parton shower at LHC energies~\cite{Norrbin:2000zc}. Therefore, to obtain \eg inclusive D-meson spectra, these processes should be combined with the light-parton processes above. This combination will also provide the total QCD jet cross section in LO. Notice also that the $\q \g \rightarrow \q \g $ process is available only in the massless approximation above. (The massive matrix element for this process incorrectly sets the incoming quark on mass shell, so it is not a better alternative.)

\subsubsection{Three-parton processes}

In addition to $2 \rightarrow 2$ QCD processes, LO expressions for processes with three final state partons are also included in \pyt. These contain only light partons, but if needed the massive quarks can be dealt with using the massless matrix elements. One should also keep in mind that, since three-jet events can be formed from two-parton final states in the parton showers, mixing these with the $2 \rightarrow 2$ QCD processes would lead to double counting. So far this section is partly incomplete, \eg colour flows are rather simple, so the purpose is rather to provide a way to check cross sections in specific kinematics where \eg all three jets need to be above a certain $\pT$. Included processes are listed below:
\begin{itemize}
\item $\g \g \rightarrow \g \g \g $
\item $\q \qbar \rightarrow \g \g \g $
\item $\q \g \rightarrow \q \g \g $
\item $\q \q' \rightarrow \q \q' \g $ where $\q$ and $\q'$ are different flavours
\item $\q \q \rightarrow \q \q \g $ where incoming and outgoing flavours are the same
\item $\q \qbar \rightarrow \q' \qbar' \g $ where $\q$ and $\q'$ are different flavours
\item $\q \qbar \rightarrow \q \qbar \g $ where incoming and outgoing flavours are the same
\item $\g \g \rightarrow \q \qbar \g $
\item $\q \g \rightarrow \q \q' \qbar' $ where $\q$ and $\q'$ are different flavours
\item $\q \g \rightarrow \q \q \qbar $ where incoming and outgoing flavours are the same
\end{itemize}

\subsection{Electroweak}\index{Electroweak processes}
\label{subsection:EWprocesses}

The internally defined electroweak (EW) processes contain prompt photon production, processes with photons in the initial state, and processes including electroweak bosons as an intermediate state or in the final state.

\subsubsection{Prompt photon production}\index{Prompt photons}

\index{Hard Process! Prompt photon}These processes include parton-initiated production that have one or two photons in the final state. The partonic cross sections are at LO in QCD for massless partons and contain only $2 \rightarrow 2$ processes. Thus, similarly as for light-flavour QCD, the expressions diverge in the $\pT \rightarrow 0$ limit and require a minimum $\pT$ cut to avoid QCD singularities. These processes are, however, also included in the eikonalized description of the soft QCD process class, where the divergences are regulated with the $\pT_{0}$. Therefore this event class should be preferred when low $\pT$ photons are considered. The available processes are
\begin{itemize}
\item $\q \g \rightarrow \q \gamma $
\item $\q \qbar \rightarrow \g \gamma $
\item $\g \g \rightarrow \g \gamma $
\item $\q \qbar \rightarrow \gamma \gamma $
\item $\g \g \rightarrow \gamma \gamma $
\end{itemize}
Notice that the processes with two gluons in the initial state are box graphs. By default, it is assumed that the five massless quarks may form the loop, such that the expressions should be valid in a region where $\pT$ is between the $\b$ and $\t$ quark masses. It is, however, possible to change the number of active flavours inside the loop if a different region is considered. In addition to the photons produced in the hard scattering (prompt photons), photons may also be formed in parton showers and hadron decays. Therefore QCD processes might be needed to obtain a realistic rates for photon production.

\subsubsection{Weak bosons}\index{Weak bosons}

\index{Hard Process! EW bosons}This section includes processes with standard model EW gauge bosons, $\gamma^*/\Z$ and $\Wpm$. The processes are classified into single and pair production, where the single production is associated with a parton and as an intermediate particle in $t$-channel exchange between two fermions.

As a highly-virtual photon $\gamma^*$ cannot be distinguished from a $\Z$ boson with equal quantum numbers, typically both contributions should be accounted for to include interference effects. It is, however, possible to consider these two components separately, without the interference. This applies for all of the following processes including neutral EW bosons.

\paragraph{Boson exchange, DIS}\index{DIS}

The EW boson $t$-channel exchange is mainly relevant in \ac{DIS} processes in lepton-hadron collisions but may also be applied for other types of collisions. There are two different contributions, one with neutral and one with charged EW bosons, namely
\begin{itemize}
\item $f f' \rightarrow f f'$ where a neutral $\gamma^*/\Z$ boson is exchanged so that the initial- and final-state fermion pair remains the same.
\item $f_1 f_2 \rightarrow f_3 f_4 $ where a charged $\Wpm$ boson is exchanged so that the initial- and final-state fermions are different. This includes charged current DIS with a charged-lepton beam and DIS with neutrino beams.
\end{itemize}
In $\pp$ collisions the factorization and renormalization scales are usually related to the transverse momentum, $\pT$, of the final-state particles. However, in DIS a more appropriate hard scale is usually the virtuality of the intermediate particle, $Q^2$. Therefore, when studying DIS with these processes, it is advised to set the renormalization and factorization scales appropriately, see \cref{subsection:couplingsscales} for details. Similarly, rather than having a phase-space cut on $\pT$ to avoid divergences, here it is more appropriate to set a minimum $Q^2$ value to make sure that the relevant phase space is covered. Furthermore, since the default parton shower distributes the emission recoils globally, it is not well suited for DIS studies where the scattered lepton is expected to stay intact. Instead it is recommended to use either the dipole-recoil option or the \dire shower, see \cref{sec:SimpleShower} and \cref{sec:Dire} for further details.

\paragraph{Single boson production}

Two different options are included for the single EW-boson production ($s$-channel) processes. In the first case the process is described as $2 \rightarrow 1$ scattering where the final state is either $\gamma^*/\Z$ or $\Wpm$:
\begin{itemize}
\item $\f \fbar \rightarrow \gamma^*/\Z$
\item $\f \f' \rightarrow \Wpm$
\end{itemize}
The decay products of the short-lived (or virtual) particles and their kinematics are then derived as described in \cref{subsection:processGenBasics}.

The other possibility is to consider single EW-boson production as a $2 \rightarrow 2$ process where the decay products can be predetermined. This is useful if only certain final states are studied and enables one to set phase-space cuts for the final state, \eg for the $\pT$ of the produced lepton. These overlap with the first class of single-boson processes so one should not mix these processes to avoid double counting. The possible processes are:
\begin{itemize}
\item $\f \fbar \rightarrow \gamma^* \rightarrow \f' \fbar'$
\item $\f \fbar \rightarrow \gamma^*/\Z \rightarrow \f' \fbar'$
\item $\f_1 \fbar_2 \rightarrow \Wpm \rightarrow \f_3 \fbar_4$
\end{itemize}
The difference between the first two is that in the first, the final state can be any of three possible lepton generations or five possible quark flavours, whereas in the second, the decay channels are set by the $\Z$-decay modes. In the former, only $\gamma^*$ exchange is included and the process is part of the MPI framework. In the latter, it is also possible to select between pure $\gamma^*$, $\Z$, and the full interference modes. For the last, $\Wpm$ production, the decay channels are always the same for $\Wp$ and $\Wm$. These are set for $\Wp$ and charge-conjugated channels are applied for $\Wm$. No quark-mass effects are included for the angular distribution of the decay products of the $\Wpm$.

\paragraph{Boson pair production}

These processes describe possible combinations of two EW-boson production, also including LO correlations for 4-lepton final states~\cite{Gunion:1985mc}.
\begin{itemize}
\item $\f \fbar' \rightarrow \gamma^*/\Z \, \gamma^*/\Z $
\item $\f \fbar' \rightarrow \Z \, \Wpm $
\item $\f \fbar \rightarrow \Wp \, \Wm $
\end{itemize}
Notice that for the second process, no contribution from a virtual photon is included. In addition, it is possible to produce EW bosons in the parton shower as described in \cref{sec:SimpleQEDEW} and \cref{sec:VinciaEW}. Therefore, a full EW-boson pair production might require a combination of different processes with some additional care to avoid possible double counting. 

\paragraph{Boson and parton production}

These processes produce events where an EW boson is produced in association with a parton, where the latter in this case refers either to a quark, gluon, photon, or lepton. The possible channels are:
\begin{itemize}
\item $\q \qbar \rightarrow \gamma^*/\Z \, \g $
\item $\q \g \rightarrow \gamma^*/\Z \, \q $
\item $\f \fbar \rightarrow \gamma^*/\Z \, \gamma $
\item $\f \gamma \rightarrow \gamma^*/\Z \, \f $
\item $\q \qbar \rightarrow \Wpm \, \g $
\item $\q \g \rightarrow \Wpm \, \q $
\item $\f \fbar \rightarrow \Wpm \, \gamma $
\item $\f \gamma \rightarrow \Wpm \, \f $
\end{itemize}
Again, there will be overlap with the single-boson production channels and the appropriate process depends on the final state and considered kinematics. For fully inclusive EW-boson production, the single-boson class is the relevant one, but for the high-$\pT$ tail these processes would provide more accurate kinematics. These processes should also be favoured when EW-boson production is studied with an associated high-$\pT$ jet or lepton.

\subsubsection{Photon collisions}\index{Photon-photon collisions}

\index{Hard Process! Photon collisions}Many modern PDF sets include perturbatively-generated photons as a constituent of protons. In addition, all electrically-charged particles accelerated to high energies may emit photons that act as initiators for hard processes. The difference between these two cases is that in the former case, the beam hadron will be resolved, whereas in the latter case, the beam particle will stay intact. The following two-photon initiated processes are included in \pythia and can be applied for resolved and unresolved beams:
\begin{itemize}
\item $\gamma \gamma \rightarrow \q \qbar $
\item $\gamma \gamma \rightarrow \c \cbar $
\item $\gamma \gamma \rightarrow \b \bbar $
\item $\gamma \gamma \rightarrow \eplus \eminus $
\item $\gamma \gamma \rightarrow \muplus \muminus $
\item $\gamma \gamma \rightarrow \tauplus \tauminus $
\end{itemize}

\subsubsection{Photon-parton scattering}\index{Photon-parton collisions}

A few processes with a photon and a parton as initiators have been included. These are mainly relevant for photoproduction in $\ep$ collisions but can also be applied to other collision types with beams that may provide photons and partons. Similarly as pure-QCD processes with light partons, these processes also contain collinear and soft divergences so a suitable phase-space cut, \eg on partonic $\pT$, must be applied to obtain finite cross sections. Unlike for the pure-QCD processes, no regularized description applicable at any $\pT$ is present for any of the photon-initiated processes. The included processes for photon-parton collisions are:
\begin{itemize}
\item $\g \gamma \rightarrow \q \qbar $
\item $\g \gamma \rightarrow \c \cbar $
\item $\g \gamma \rightarrow \b \bbar $
\item $\q \gamma \rightarrow \q \g $
\item $\q \gamma \rightarrow \q \gamma $
\end{itemize}
Here, the heavy-quark pair production processes contain the full mass dependence in the matrix elements. Similar to the case of pure-QCD processes, at high-enough collision energies heavy quarks, at least charm, may be produced via parton-shower emissions so several processes might need to be considered to obtain realistic heavy-quark production rates.

\subsection{Onia}\index{Quarkonium}
\label{subsection:onia}

\index{NRQCD}\index{Colour-octet onium production}Hard
processes involving charmonium and bottomonium are 
provided using 
\ac{NRQCD}~\cite{Bodwin:1994jh},
which includes both colour-singlet and colour-octet
contributions. The spectroscopic notation ${}^{2s+1}L_J$
specifies the necessary quantum numbers to define a state: spin $s$,
orbital angular momentum $L$, and total angular momentum $J$.
Processes are available for the ${}^3S_1$,
${}^3P_J$, and ${}^3D_J$ states containing \ccbar or \bbbar, given an arbitrary
radial excitation $n$, \eg any $\Upsilon(nS)$ for the ${}^3S_1$
$\b\bbar$ onia states.  Double
onium production is also available for double-${}^3S_1$ $\c\cbar$ and
$\b\bbar$ processes, but only with colour-singlet contributions
provided. Because of the long-standing discrepancy between
polarization in data and NRQCD predictions, only unpolarized processes
are provided, with isotropic decays, which can then be reweighted
accordingly by the user for a given polarization model.

Within NRQCD, the inclusive cross-section for a heavy onium state, $H$,
can be written as,
\begin{equation}
  \label{eq:onia_prod}
  \textrm{d}\sigma(\pp \to H + X) = \sum_{s,L,J}
  \textrm{d}\hat{\sigma}(\pp \to \Q\Qbar[{}^{2s+1}L_J] +
  X)\langle\mathcal{O}^H[{}^{2s+1}L_J]\rangle
\end{equation}
where the cross section is factorized into a sum of products between
short-distance matrix elements, $\mathrm{d}\hat{\sigma}$, and
long-distance matrix elements
$\langle\mathcal{O}^H[{}^{2s+1}L_J]\rangle$. The short-distance matrix
elements are calculated with perturbative QCD~\cite{Baier:1983va,
  Gastmans:1986qv, Cho:1995ce, Yuan:1998gr, Humpert:1983yj,
  Qiao:2002rh}, while the long-distance matrix elements are determined
from phenomenological fits to parameters~\cite{Nason:1999ta, Bargiotti:2007zz,
  Yuan:1998gr}. Default settings for these parameters are provided 
for the \Jpsi, $\psi(2S)$, $\chi_{c0}$, $\chi_{c1}$, $\chi_{c2}$,
$\psi(3770)$, $\Upsilon(1S)$, $\Upsilon(2S)$, $\Upsilon(3S)$,
$\chi_{b0}$, $\chi_{b1}$, and $\chi_{b2}$ states.

The sum in \cref{eq:onia_prod} for a given physical onium state
$\left|H[{}^{2s+1}L_J]\right\rangle$ is over the expansion of its
Fock states,
\begin{align}
  \left|H[{}^{2s+1}L_J]\right\rangle =~
  &\mathcal{O}(1)\left|\Q\Qbar[{}^{2s+1}L_J^{(1)}]\right\rangle +
  \mathcal{O}(v)\left|\Q\Qbar[{}^{2s+1}(L\pm1)_{J'}^{(8)}]\g\right\rangle
  \nonumber \\ & +
  \mathcal{O}(v^2)\left|\Q\Qbar[{}^{2s+1}(L\pm1)_{J'}^{(8)}]\g\g\right\rangle
  + \ldots
\end{align}
where the superscript $(1)$ indicates a colour-singlet state, $(8)$ a
colour-octet state, and the expansion is in the velocity $v$ of the
heavy-quark system. Consequently, a long-distance and short-distance
matrix element must be provided for each state in the expansion.

For the physical ${}^3S_1$ states the following processes are available.
\begin{multicols}{3}
  \begin{itemize}
  \item $\g\g \to \left|\c\cbar({}^3S_1)[{}^3S_1^{(1)}]\right\rangle \g$
  \item $\g\g \to \left|\c\cbar({}^3S_1)[{}^3S_1^{(1)}]\right\rangle \gamma$
  \item $\g\g \to \left|\c\cbar({}^3S_1)[{}^3S_1^{(8)}]\right\rangle \g$
  \item $\q\g \to \left|\c\cbar({}^3S_1)[{}^3S_1^{(8)}]\right\rangle \q$
  \item $\q\qbar \to \left|\c\cbar({}^3S_1)[{}^3S_1^{(8)}]\right\rangle \q$
  \item $\g\g \to \left|\c\cbar({}^3S_1)[{}^1S_0^{(8)}]\right\rangle \g$
  \item $\q\g \to \left|\c\cbar({}^3S_1)[{}^1S_0^{(8)}]\right\rangle \q$
  \item $\q\qbar \to \left|\c\cbar({}^3S_1)[{}^1S_0^{(8)}]\right\rangle \q$
  \item $\g\g \to \left|\c\cbar({}^3S_1)[{}^3P_J^{(8)}]\right\rangle \g$
  \item $\q\g \to \left|\c\cbar({}^3S_1)[{}^3P_J^{(8)}]\right\rangle \q$
  \item $\q\qbar \to \left|\c\cbar({}^3S_1)[{}^3P_J^{(8)}]\right\rangle \q$
  \end{itemize}
\end{multicols}
\noindent The ${}^3P_J^{(8)}$ Fock states are a summation of contributions
for $J = 0, 1, 2$. The ${}^3P_1^{(8)}$ and ${}^3P_2^{(8)}$
long-distance matrix elements are calculated from the ${}^3P_0^{(1)}$
long-distance matrix element.

The following processes are available for the physical ${}^3P_J$
states, again with $J = 0, 1, 2$.
\begin{multicols}{3}
  \begin{itemize}
  \item $\g\g \to \left|\Q\Qbar({}^3P_J)[{}^3P_J^{(1)}]\right\rangle \g$
  \item $\q\g \to \left|\Q\Qbar({}^3P_J)[{}^3P_J^{(1)}]\right\rangle \q$
  \item $\q\qbar \to \left|\Q\Qbar({}^3P_J)[{}^3P_J^{(1)}]\right\rangle \q$
  \item $\g\g \to \left|\Q\Qbar({}^3P_J)[{}^3S_1^{(8)}]\right\rangle \g$
  \item $\q\g \to \left|\Q\Qbar({}^3P_J)[{}^3S_1^{(8)}]\right\rangle \q$
  \item $\q\qbar \to \left|\Q\Qbar({}^3P_J)[{}^3S_1^{(8)}]\right\rangle \q$
  \end{itemize}
\end{multicols}
\noindent Similar to the ${}^3P_J^{(8)}$ states, the colour-singlet
${}^3P_1^{(1)}$ and ${}^3P_2^{(1)}$ long-distance matrix elements are
calculated from the ${}^3P_0^{(1)}$ long-distance matrix element.

For physical ${}^3D_J$ production, the following processes are provided:
\begin{multicols}{3}
  \begin{itemize}
  \item $\g\g \to \left|\Q\Qbar({}^3D_J)[{}^3D_J^{(1)}]\right\rangle \g$
  \item $\g\g \to \left|\Q\Qbar({}^3D_J)[{}^3P_J^{(8)}]\right\rangle \g$
  \item $\q\g \to \left|\Q\Qbar({}^3D_J)[{}^3P_J^{(8)}]\right\rangle \q$
  \item $\q\qbar \to \left|\Q\Qbar({}^3D_J)[{}^3P_J^{(8)}]\right\rangle \q$
  \end{itemize}
\end{multicols}
\noindent The colour-octet ${}^3P_J^{(8)}$ contributions are treated
in the same fashion as for the physical ${}^3S_1$ state. Finally,
double onium production is available for any arbitrary same-flavour
${}^3S_1$ configuration.
\begin{itemize}
\item $\g\g \to \left|\Q\Qbar({}^3S_1)[{}^3S_1^{(1)}]\right\rangle
  \left|\Q\Qbar({}^3S_1)[{}^3S_1^{(1)}]\right\rangle$
\item $\q\qbar \to \left|\Q\Qbar({}^3S_1)[{}^3S_1^{(1)}]\right\rangle
  \left|\Q\Qbar({}^3S_1)[{}^3S_1^{(1)}]\right\rangle$
\end{itemize}
The default configuration for double-onium production is to provide all possible
combinations of the same-flavour physical ${}^3S_1$ states.

Many of the short-distance matrix elements diverge as $\pT \to 0$, and
consequently must be regulated either with a hard cutoff or a smooth
damping factor.
Onium can be produced in a hard process, but also in multiparton interactions, except for double onium.
In a hard process, a hard cutoff in
$\pT_0$ is used, although it is also possible to implement smooth
damping through a user defined scheme, see
\cref{subsection:userhooks}.
In multiparton interactions, instead, a smooth
damping is performed with a cutoff scale $\pT_0$ for a given energy
$E_0$ with an evolution parameter. See
\cref{subsection:mpi} for more details. 

The colour-octet states are defined in the event record using a
non-standard numbering scheme, $\mathtt{99 n_q n_s n_r n_L n_J}$,
where $\mathtt{n_q}$ is the quark flavour of the state and
$\mathtt{n_s}$ is the colour-octet state type. Here, $\mathtt{0}$ is
${}^3S_1$, $\mathtt{1}$ is ${}^1S_0$, and $\mathtt{2}$ is
${}^3P_J$. All remaining numbers follow the standard \ac{PDG} numbering
scheme~\cite[sec.~45]{ParticleDataGroup:2020ssz}. As an example,
$\mathtt{9941003}$ is the ${}^1S_0^{(8)}$ 
$\c\cbar$ colour-octet state for the colour-singlet \Jpsi. After the
parton shower and hadronization, all colour-octet states are forced to
isotropically decay into their corresponding physical colour-singlet
state and a soft gluon. A user-configurable mass splitting is used to
set the mass of the colour-octet states for a given
colour-singlet. This determines the softness of the gluon emitted in
the octet to singlet transition.

Colour-octet states are allowed to evolve under the timelike QCD parton shower, see
\cref{sec:showers} for more details on parton
showers. 
This is meant to account for the competing effects of unbound $\Q\Qbar$
states that emit additional gluons to become a semi-bound state, and
semi-bound $\Q\Qbar$ states that are broken apart by additional gluon
radiation.
The combination is approximated by
allowing the colour-octet states to radiate in the parton shower with twice
the $\q \to \q \g$ splitting probability.
Both the probability of a colour-octet
state being considered in the parton shower and the pre-factor for the
splitting kernel can be configured.

This treatment of colour-octet production in the parton shower is a
simplification. The colour-octet state
can be treated as a gluon, and so a factor of $9/4$ rather than $2$
may be used.
Using a $\q \to \q \g$ splitting kernel, rather than $\g
\to \g \g$, is roughly equivalent to always following the path of the
harder gluon, resulting in harder final-state onia.
Additionally, soft
gluons producing heavy-quark pairs will have not have sufficient
phase space to produce hard semi-bound states. However, after the $\g
\to \Q \Qbar$ splitting, each heavy quark carries only approximately
half the onium energy, reducing the energy of the gluon emissions.
In
principle, these two effects between softer and harder gluon emissions
should approximately balance. However, comparisons to measurements of
prompt \Jpsi production in jets from \pp collisions indicate that this
treatment underestimates the local radiation surrounding
onia~\cite{Aaij:2017fak, Sirunyan:2019vlp}.

\subsection{Top production}\index{Top quark}\index{Single top}

\label{subsection:top}

\index{Hard Process! Top production}Standard model top production has now been part of standard measurements for over two decades and state-of-the-art experimental observations now make use of higher-order calculations.  However, we still maintain a minimum set of top-production processes that can be used either with a $K$-factor for quick testing or for designing searches for non-standard decay modes by modifying the top-decay table by hand.

Production processes available include:
\begin{itemize}
  \item $\g\g \rightarrow \t \tbar $
  \item $\q\q \rightarrow   \t \tbar $ 
   \item $\f \fbar \rightarrow   \t \tbar $ (via $t$-channel $\W$ or $s$-channel $Z/\gamma$ separately) 
   \item $\gamma \gamma  \rightarrow \t \tbar $
   \item $\g \gamma  \rightarrow \t \tbar $
   \item $\q \q^\prime \rightarrow   \q^{\prime \prime} \t $ (single top via $s$-channel $\W$)
\end{itemize}

It may be possible, for example, to test for the production of charged Higgses in top decays by adding the decay mode $\t \rightarrow \b \Hp$ to the decay table and using the BSM Higgs sector (see the next section for details of setting BSM Higgs parameters).

\subsection{Higgs}\index{Higgs bosons}
\label{subsection:Higgs}

\index{Hard Process! Higgs production}Pythia includes the capability of simulating production of standard
model or BSM Higgses via the \ac{2HDM}. 
The production processes for the SM Higgs include:
\begin{itemize}
  \item $\f \bar \f \rightarrow \H $
  \item $\g\g \rightarrow \H $ (via 1-loop)
  \item $\q\g \rightarrow \H \q $ (via 1-loop)
  \item $\gamma \gamma \rightarrow \H $ (via 1-loop)
  \item $\f \bar \f \rightarrow \Z \, \H $ (via $s$-channel Z)
  \item $\f \bar \f \rightarrow \Wpm \, \H $ (via $s$-channel W)
  \item $\f \bar \f \rightarrow  \H \f \bar \f $ (vector-boson fusion, $\Z\Z$ and $\Wpm \Wpm$ can be selected separately)
  \item $\f \bar \f \rightarrow \H \Q \bar \Q$ where $\Q = \b, \t$
\end{itemize}

\noindent BSM Higgses can be produced using \settingval{Higgs:useBSM}{on}.
To allow for CP-violating cases, the neutral Higgses are named $\H_1,
\H_2, \A_3$, which in the CP-conserving case refer to two scalar and one
pseudoscalar Higgs, respectively.  The neutral Higgses are ordered by
mass. All processes mentioned above for the SM Higgs are also available for BSM
ones by replacing $\H$ with the required BSM Higgs name.  Further processes
available for BSM Higgses are the pair-production processes.
\begin{itemize}
  \item $\f \bar \f \rightarrow \H_{1,2} \A_3 $
  \item $\f \bar \f \rightarrow \H^+ \H_{1,2} $
  \item $\f \bar \f \rightarrow \H^+ \H^- $
\end{itemize}
The couplings of each Higgs boson to SM fermions can be set
independently to account for all possible 2HDM structures.  Further
selection of the parity of each Higgs is also possible.  We refer the
user to the \htmlmanual~for a description of each parameter.

The decay of the Higgses is also calculated automatically based on
input parameters.  The decay table can be overwritten by the user,
either using the \pyt settings structure or using the SLHA interface
(see \cref{sec:slha}).  Since the LHC cross-section working group
recommends the usage of \ac{NLO} decay widths, we use a multiplicative
factor for all internally-calculated widths.  The factor is calculated
for $m_\H = 125$~GeV, but should be sensible for a range of masses.
Furthermore, the Breit--Wigner shape\index{Breit--Wigner for Higgses} of the Higgs resonance is
complicated due to a dependence on mass.  For resonance searches, it may
be useful to ``clip the wings'' of the Breit--Wigner shape using \setting{Higgs:clipWings} and \setting{Higgs:wingsFac} (what factor of width
beyond which to clip) parameters.

\subsection{Supersymmetry}\index{SUSY}\index{Supersymmetry|see{SUSY}}\index{MSSM|see{SUSY}}
\label{subsection:SUSY}

\index{Hard Process! Supersymmetry}The implementation of the \ac{MSSM} allows fully general, 
complex $6\times 6$ mixing in the squark
sector, and up to five neutral gauginos (corresponding to
next-to-minimal MSSM).  We also allow all four kinds of
R-parity violating couplings (one bi-linear and three tri-linear).
Users are expected to input parameters via an SLHA
file (see \cref{sec:slha}).  Typically, the Higgs sector of \ac{SUSY} is
identical to a type-2 2HDM model and can be generated via the Higgs
processes described above. \pyt is also capable of calculating
decay widths in the standard channels for all SUSY
particles.  However, if a decay table is provided in the SLHA file,
the internal calculation is turned off.  For very low-width particles,
the lifetime is set as the inverse of the total decay width.  All
particles with a decay width set to zero are set as stable.

Pair production of squarks ($\tilde q_i$), gluinos ($\tilde g$), and
gauginos ($\tilde \chi^0_j,\, \tilde \chi^\pm$), including pairs like
squark-gluino, squark-gaugino, and gluino-gaugino, are implemented with
EW contributions.  We also implement resonant production of squarks
via \ac{RPV} $\lambda^{\prime \prime}$ couplings,
with corresponding modification to showering and hadronization to
include the new colour structure.  Here follows a full list of the processes
available. 
\begin{itemize}
\item squark-pair production (including anti-squarks and EW interference)
  \begin{itemize}
  \item $\f \bar \f \rightarrow \tilde q_i \tilde q_j^{(*)}$
  \item $\g\g \rightarrow \tilde q_i \tilde q_j^{(*)}$
  \end{itemize}
\item gluino pair and gluino-squark production 
  \begin{itemize}
  \item $\q \bar \q \rightarrow \tilde g \tilde g$
  \item $\g\g \rightarrow \tilde g \tilde g$
  \item $\q_i\g \rightarrow \tilde q_i \tilde g$ (and charge conjugate)
  \end{itemize}
    
\item electroweak-gaugino pair production 
  \begin{itemize}  
  \item $\f \bar \f \rightarrow \tilde \chi^\pm \tilde \chi^\pm$
  \item $\f \bar \f \rightarrow \tilde \chi^0_i \tilde \chi^0_j$
  \item $\f \bar \f \rightarrow \tilde \chi^\pm \tilde \chi^0_j$
  \end{itemize}
  
\item gaugino-gluino and gaugino-squark production 
  \begin{itemize} 
  \item $\f \bar \f \rightarrow \tilde \g \tilde \chi^\mp$
  \item $\f \bar \f \rightarrow \tilde \g \tilde \chi^0_i$
  \item $\f \bar \f \rightarrow \tilde q_j^{(*)} \tilde \chi^\mp$
  \item $\f \bar \f \rightarrow \tilde q_j^{(*)} \tilde \chi^0_i$
  \end{itemize}

\item slepton or sneutrino-pair production
  \begin{itemize} 
  \item $\f \bar \f \rightarrow \tilde \ell_i \tilde \ell_j^{(*)}$
  \end{itemize}
  
\item resonant production of a squark via an R-parity violating process
    \begin{itemize}
    \item $\q_i \bar \q_j  \rightarrow  \tilde q_k$ (RPV)
    \end{itemize}
 \end{itemize}
    
Further selection of what processes to turn on is also possible by
specifying individual \ac{PDG} IDs of particles.  All supersymmetric
particles are given \ac{PDG} codes greater than $1000000$, with the
superpartners generally carrying the corresponding code to their SM
partner, \eg an up quark is $2$ and the two up squarks are named
$1000002$ and $2000002$.  The full list of \ac{PDG} codes is available in
the published review~\cite[sec.~45]{ParticleDataGroup:2020ssz}. 

\subsection{Hidden valley}
\label{subsection:HiddenValleyProcesses}
\index{Hidden valleys}\index{Dark photons}

\ac{HV} refers to a range of scenarios characterized by a gauge-symmetric dark sector with various possibilities of portals into the
``valley''. \pyt currently is the only general-purpose Monte Carlo
code that implements a HV scenario, including running of
gauge couplings, showering, and hadronization in the dark sector~\cite{Carloni:2010tw,Carloni:2011kk}.
There are multiple particle spectra and production modes available which
together can cover a wide range of phenomenology.

First, based on the rank N of the dark \su{N}, radiation to either dark
photons (\ie \uone) or dark gluons (\ie \su{N}) is implemented.  The
matter content is modelled of in two separate ways --- first via
partners of the SM fermions (named dark-u, dark-d, dark-e and so on)
that carry both the SM charges of their partner as well as fundamental
of the dark \su{N}, and second via a ``dark quark'' that carries only
the dark charge but does not carry any SM charges.  In the first case,
dark sector particles can be produced via normal SM gauge bosons and
radiate to both SM and dark bosons based on their mass and relative
strengths of the dark and SM couplings.  In the second case, we
implement an extra $Z^\prime$ portal to produce said quarks via a
kinetic mixing with the SM photon.  The spin of the dark-sector
particles (aside from the gauge bosons) can be set by the user to be
either scalars, fermions, or vectors.

Two kinds of models are available in \pyt, depending on the charge of the new fermions.  The first case is where the new fermions also carry some standard-model charge and can therefore be produced via one of the standard-model gauge bosons.  The radiation of the final state-fermions then includes both dark-sector radiation as well as SM radiation.  The processes that fall in this category include:
\begin{itemize} 
  \item $\g \g \rightarrow F_v \bar F_v$ via intermediate gluon where $F_v$ is the hidden sector quark --- either one of the quarks $U_v$, $D_v$, $S_v$, $C_v$, $B_v$, $T_v$, or generic quark $Q_v$ 
  \item $\q \bar \q \rightarrow F_v \bar F_v$ via intermediate gluon where $F_v$ is the hidden sector fermion --- either one of the quarks $U_v$, $D_v$, $S_v$, $C_v$, $B_v$, $T_v$, or generic quark $Q_v$ 
  \item $\f \bar \f \rightarrow F_v \bar F_v$ via intermediate Z or $\gamma^*$. $F_v$ includes all the quarks above, plus the ``leptons'' $E_v$, $\nu_{Ev}$, and similarly for $\mu$ and $\tau$ flavours
\end{itemize}

It is possible to simulate a hidden sector where the new fermions do not carry any SM charges, but in this case, we need a new portal, which \pyt assumes is a new vector $Z^\prime$.  This $Z^\prime$ is then expected to be able to decay to both SM fermions as well as dark-sector fermions.  Pair production of dark-sector fermions via this portal can be done using:
  \begin{itemize} 
   \item $\q \bar \q \rightarrow Z_v $ followed by $Z_v \rightarrow F_v \bar F_v$
   \end{itemize}

An important phenomenological effect is the running of the hidden-sector strong coupling which can make significant changes to the radiation pattern in the dark sector.  This is by default taken into account by using the one-loop beta function of \su{N} once the number of colours and flavours of new fermions is set. The running can also be turned off by the user to use a fixed-coupling value instead.  There is also an inherent ambiguity in the composition of the hadrons in the dark sector. \pyt allows the user to manually set the ratio of scalar to vector mesons as well as the parameters of the Lund symmetric fragmentation function or the dark sector (see \cref{sec:lund-model} for details of the fragmentation functions).  The decay table of the hidden mesons back into the standard model (should this be desirable), can be done by the user at run time using the standard particle data scheme that \pyt uses for all particles.

\subsection{Dark matter}\index{Dark matter}
\label{subsection:DarkMatter}
\index{Dark Matter}Multiple models for \ac{DM} are currently implemented
in \pyt. They may be separated into two different categories --- production via
$s$-channel mediator and production via pair production of mediators
(typically seen in co-annihilation or co-scattering scenarios of DM).
In all cases, the DM is assumed to be fermionic.  We provide the
possibility to produce dark matter with one associated jet for the
$s$-channel models (vector or axial-vector $Z^\prime$ and scalar or
pseudoscalar $A$).  For the mediator pair-production processes, all
mediators are produced via Drell--Yan production.

The \ac{PDG} provides some standard codes for common
DM particles and mediators, \cf\cite[sec.~45]{ParticleDataGroup:2020ssz}. Of these, the fermionic DM
code \texttt{52}, the 
$s$-channel scalar mediator (\texttt{54}) and vector mediator (\texttt{55}) are used in this
implementation.  The new mediators are either charged scalar (with PDG
code \texttt{56}), charged vector-like fermion (PDG code \texttt{57}), and doubly charged
fermion (PDG code \texttt{5}9).  The neutral partner that accompanies the
charged mediators is given the PDG code \texttt{58}.

The singlet model contains a scalar singlet with quantum numbers identical to 
a right-handed lepton. Therefore, it couples via a Yukawa-like
coupling to a SM right-handed lepton and the DM is a Dirac fermion.
Both the scalar and the DM are odd under a $Z_2$ symmetry to ensure
the stability of the DM.
\begin{align}
  \mathcal{L} = \partial_\mu \phi^* \partial^\mu \phi + \bar \chi (i \gamma^\nu \partial_\nu \chi) - m_\phi^2 |\phi|^2 - m \bar \chi \chi - (y_\ell \bar \ell \phi \chi + \mathrm{h.c.} ).
\end{align}

The fermionic mediators are based on models that have mixing between a
singlet and an $n$-plet vector fermion, both charged under a $Z_2$
symmetry for which all of SM particles are even.  The mixing between
the singlet and $n$-plet is then calculated based on the value of $n$.
The lightest neutral state is denoted as dark matter.  
\begin{align}
  \mathcal{L} =  \bar \chi (i \gamma^\nu \partial_\nu \chi) + \bar \psi (i \gamma^\nu D_\nu \psi) - m_1 \bar \chi \chi -m_2 \bar \psi \psi.
\end{align}

\noindent The mixing term depends on the representation of $\psi$.  For example, for a triplet case, we have
\begin{align}
\mathcal{L}_\mathrm{mix} = \frac{c}{\Lambda^2} \left( \bar \chi (\Phi^\dag_h \tau^a \Phi_h) \psi^a + \mathrm{h.c.} \right),
\end{align}
where $\Phi_h$ is the SM-Higgs doublet, $\psi$ is the triplet fermion, $\tau$ are the Pauli matrices and $\chi$ is the singlet fermion.

\index{Hard Process! Dark Matter}Production of DM can be studied in two ways --- either by directly producing the $s$-channel mediator, which then decays to DM, or by producing the charged partner of DM via Drell--Yan followed by the decay of the partner.  The production processes therefore are
  \begin{itemize} 
 	\item $\q \bar \q \rightarrow Z^\prime \rightarrow \bar \chi \chi$
	\item $\g \bar \g \rightarrow S \rightarrow \bar \chi \chi$, note that 1-loop $ \g \bar \g \rightarrow S$ via top-loop is included in this production.
	\item $\q \bar \q \rightarrow Z^\prime g$ (mono-jet)
	\item $\g \bar \g \rightarrow S g$ (also mono-jet, via 1-loop in production)
	\item $\q \bar \q \rightarrow Z^\prime H$, \ie mono-Higgs production (coupling of the SM to the new $Z^\prime$ has to be set by the user)
	\item $\f \bar \f \rightarrow \psi \bar \psi$ where $\psi = \tilde \ell^\pm$ (scalar with leptonic quantum numbers), $\chi^\pm$ (singly charged fermion), or $\chi^{\pm\pm}$ (doubly charged fermion), followed by decay of $\psi$ into DM (Drell--Yan for charged partners)
\end{itemize}	
Couplings of quarks and leptons to the mediators are assumed to be generation universal, however vector and axial-vector components (or equivalently scalar and pseudoscalar components for the scalar mediator) can be set individually for up type, down type, charged lepton, neutrino, and dark-matter fermions.  In case of $Z^\prime$, it is also possible to choose a kinetic-mixing parameter $\epsilon$ which then automatically sets the rest.

\subsection{Other exotica}
\label{subsection:otherexotica}
\index{Exotica}
\index{Hard Process! Exotica}
Finally, we mention other models of new physics that are implemented in \pyt, though they are perhaps not as popular as they once were.  
We refer the reader to the \htmlmanual~for the detailed descriptions of the model parameters and only provide a list here.

\begin{itemize}
\item \textbf{Fourth generation} includes production of fourth-generation quarks or leptons via the usual SM-mediated processes.
\item \textbf{New gauge boson $Z^\prime$, $W^\prime$ and horizontal gauge boson } production can be performed through $\f \fbar \rightarrow V$ production followed by decay.  For $Z^\prime$, full interference with SM $\gamma$, Z in the $s$-channel production is taken into account.  It is possible to have both universal and non-universal models where couplings to each generation should be set by hand by the user.
\item The \textbf{left-right symmetry} model includes a right handed \su{2} sector.  Along with the gauge bosons $W^\prime$ and $Z^\prime$, it also includes heavy right-handed neutrinos that can be used to study signatures of heavy neutral leptons.
\item \textbf{Leptoquark production} includes resonant single production or pair production of scalar leptoquarks via gluon-mediated diagrams. The flavour of the leptoquark should be set by the user by explicitly setting the decay table of the leptoquark.
\item \textbf{Compositeness} models include simple models of excited fermions and contact interactions that modify standard QCD dijet or Drell--Yan production of leptons.
\item \textbf{Extra dimensions} includes production of the graviton or the extra Kaluza--Klein gauge boson (\eg KK-gluon) of the Randall--Sundrum model.  Further processes include modification of SM dijet/dilepton production due to extra KK-bosons in the \TeV-scale or large extra dimension models.  Finally, \textbf{Unparticle} emission is modelled associated with a jet or photon.
\end{itemize}

\subsection{Couplings and scales for internal processes}
\label{subsection:couplingsscales}\index{alphaS@$\alphas$!in hard processes}
\index{Strong coupling|see{$\alphas$}}
\index{QCD coupling constant|see{$\alphas$}}

The perturbatively calculated cross sections for QCD and \ac{QED} processes depend directly on the value of the relevant coupling evaluated at the scale at which the hard scattering occurs. The scale dependence of the couplings arises due to the renormalization procedure required to obtain finite cross sections and can be calculated by solving the renormalization group equations of the applied theory.

In \pythia the running of the QCD coupling, $\alphas(Q^2)$, is implemented up to second order and applied at first order by default to match the precision of the internally-calculated cross sections. A fixed value can also be used, but the potential usage is limited to special cases and generally a running coupling should be applied for realistic cross-section estimates. The coefficients related to the value of the coupling are fixed by setting the $\alphas(Q^2)$ value at the mass of the $\Z$ boson.

Similarly, running of the QED coupling $\alphaem(Q^2)$ has been implemented in \pythia. This, however, runs much slower than the QCD one and only first-order running is implemented. An option to use a fixed value for $\alphaem(Q^2)$ is included, either by setting the value directly at the mass of the $\Z$ boson or by matching to its value at vanishing momentum transfer. In addition, it is possible to globally scale the cross sections with a $K$-factor if such behaviour is desired.

There are two relevant scales that needs to be set. The renormalization scale, $Q^{2}_{\mrm{ren}}$, arises from the renormalization procedure and defines at which scale the couplings are evaluated. The factorization scale $Q^{2}_{\mrm{fact}}$ arises from factorizing the short-distance phenomena (hard scattering) from the large-distance (soft) structure of hadrons. This scale determines at which $Q^2$ the PDFs of resolved beams are probed.

As the scale dependencies arise from an approximated description of QCD, there is some amount of freedom in the scale choices. The only solid guideline is that the scales should be related to the hardness of the scattering process and therefore the optimal choice depends on the type of the studied process. Multiple options for the scale choices have been implemented into \pythia, and all options are available for both $Q^{2}_{\mrm{ren}}$ and $Q^{2}_{\mrm{fact}}$.

\noindent For $2 \rightarrow 1$ processes two options exist:
\begin{itemize}
\item the squared invariant mass, $\hat{s}$, \ie the mass of the produced particle;
\item and a fixed scale.
\end{itemize}
For $2 \rightarrow 2$ a few more options are included:
\begin{itemize}
\item the smaller of the squared transverse masses of the outgoing particles, $\mrm{min}(m_{\perp,3}^2,m_{\perp,4}^2)$;
\item the geometric mean of the squared transverse masses of the outgoing particles, $m_{\perp,3} \cdot m_{\perp,4}$;
\item the arithmetic mean of the squared transverse masses of the outgoing particles,\\ $(m_{\perp,3}^2+m_{\perp,4}^2)/2$;
\item the squared invariant mass of the system, $\hat{s}$, relevant for $s$-channel processes;
\item the squared invariant momentum transfer $-\hat{t}$, relevant especially for DIS events as this coincides with the virtuality of the intermediate photon $Q^2$;
\item and a fixed scale.
\end{itemize}
For $2 \rightarrow 3$ processes the possible choices are:
\begin{itemize}
\item the smallest of the squared transverse mass of the outgoing particles, $\mrm{min}(m_{\perp,3}^2,m_{\perp,4}^2,m_{\perp,5}^2)$;
\item the geometric mean of the two smallest squared transverse masses of the outgoing particles, $\sqrt{m_{\perp,3}^2 \cdot m_{\perp,4}^2 \cdot m_{\perp,5}^2 / \mrm{max}(m_{\perp,3}^2, m_{\perp,4}^2, m_{\perp,5}^2)}$;
\item the geometric mean of the squared transverse masses of the outgoing particles,\\ $(m_{\perp,3}^2 \cdot m_{\perp,4}^2 \cdot m_{\perp,5}^2)^{1/3}$;
\item the arithmetic mean of the squared transverse masses of the outgoing particles,\\ $(m_{\perp,3}^2+m_{\perp,4}^2+m_{\perp,5}^2)/3$;
\item the squared invariant mass of the system, $\hat{s}$, relevant for $s$-channel processes;
\item and a fixed scale.
\end{itemize}

\index{VBF}\index{Vector-boson fusion|see{VBF}}\index{Weak-boson fusion|see{VBF}}For vector-boson-fusion (VBF) processes, such as $\f_1 \f_2 \rightarrow \f_3 \H \f_4 $, the virtualities of the intermediate bosons would not be accounted for with the above options and would likely underestimate the relevant scales. Therefore modified scale choices where, instead of the transverse mass of the final-state particle, a virtuality estimate $m_{\perp, Vi}^2 = m_{V}^2 + p_{\perp, i}^2$ can be used in the options above when relevant.

\index{Uncertainties!Hard@in Hard processes}Traditionally, the theoretical uncertainties related to the truncated pQCD expansion are estimated by varying the QCD scales by a factor of two or so. To enable such variations, options to multiply the scales determined by the options above by constant factors have been implemented. In a basic form, these variations will, however, require to generate a completely new set of events, so mapping out all possible uncertainties might become computationally demanding. Therefore, options to calculate weights for each event based on different scale variations have been implemented in \pythia for more efficient uncertainty estimation, see \cref{sec:using-weights} for details. Notice also that the couplings and scales can be set separately for MPIs and initial- and final-state showers.

\subsection{Handling of resonances and their decays}
\index{Decays!Resonances@of Resonances}
\index{Resonance decays}\index{Top quark}
\index{Higgs bosons}\index{Weak bosons} 
\label{sec:hardRes}

By default, the SM electroweak gauge
bosons, top quarks, the Higgs boson, and generally all BSM
particles are classified as resonances. Note that all of these have
on-shell masses above 20 GeV (with the exception of some hypothetical
weakly interacting and stable particles such as the gravitino, which
are also considered resonances). 

Importantly, neither hadrons nor any
particles that can be produced in hadron decays, such as
$\tauon$ leptons, are included in this category.
The decays of such particles are performed after hadronization, and
changing their decay channels will \emph{not} automatically affect
the reported cross section.
For example,
allowing only the decay $\Z\to \muon^+\muon^-$ will reduce the total
cross section reported by \pyt for hard processes like $\p\p \to \Z$ by
the appropriate branching fraction, while allowing only the
decay $\Jpsi \to \muon^+\muon^-$ will \emph{not} change the
cross section for $\g\g \to \Jpsi \g$. The reason for this is
that hadron and $\tauon$ decays involve multistep chains that cannot be
predicted beforehand: a hard process like $\g\g\to \g\g$ can develop a
shower with a $\g \to \b \bbar$ branching, where the $\b$ hadronizes to
a $\bar{\B}^0$ that oscillates to a $\B^0$ that decays to a $\Jpsi$.
Any bias at the hard-process level would not affect these other
production mechanisms and could thus be misleading.
Instead, the user must consider all relevant production sources and
perform their own careful bookkeeping.

Both types, ``resonances'' and ``unstable particles'', can have
Breit--Wigner distributed mass spectra (at least when generated by
internal \pyt processes); more on this below.
For the remainder of this
subsection we focus on the production and decay of those
particles that are classified as resonances, referring to
\cref{sec:particleDecays} for the treatment of hadron and $\tau$
decays.   

Note that the cross-section reduction factors to account for
decay modes that have been switched off are always evaluated at
initialization, for nominal masses. For instance, in the example
above, the $\Z\to\muon^+\muon^-$ reduction factor is evaluated at the
nominal $\Z$ mass, even when that factor is used, later on, say in the
description of the decay of a 125 GeV Higgs boson, where at least one
$\Z$ would be produced below this mass. We know of no 
case where this approximation has any serious consequences, however. 

Note also that, for the specific case of electroweak showers
(\cf\cref{sec:SimpleQEDEW} and \cref{sec:VinciaEW}),
the decays of any resonances that are produced by the shower (\ie not by
the hard process) are treated inclusively, ignoring any user
restrictions on which channels should be open or closed.
It is then up to the user to select the final states of interest and
reject the rest.  

Finally, a word of caution: the above logic implies that switching off
\emph{all} of the decay channels of a resonance will result in cross
sections evaluating to zero,
precluding \pyt from being able to generate any events.  Instead, to force a
resonance to be treated as stable for a given run, set
\settingval{NN:mayDecay}{false}, with \texttt{NN} being its particle ID code.  

\paragraph{Total and partial widths:}
For resonances, the partial widths to different decay
channels are typically perturbatively calculable, given the parameters of the
respective model. By default, during initialization \pyt therefore 
computes the hadronic widths of $\W$, $\Z$, $\t$, and SM Higgs bosons
at NLO in QCD, with 
\begin{align}
  \Gamma_{\V\to \q\qbar}^{\mathrm{NLO}} & =
  \left(1+\frac{\alpha_s(m^2_{V})}{\pi}\right)\Gamma_{\V\to
    \q\qbar}^\mathrm{LO}~, \nonumber\\ 
  \Gamma_{\t\to \b\W}^{\mathrm{NLO}} & =
  \left(1-\frac{5\alpha_s(m^2_\Z)}{2\pi}\right)\Gamma_{\t\to \b\W}^\mathrm{LO}~,
\label{eq:widNLO}
\end{align}
where $\V$ is a generic vector boson. For $\H^0$, the default
is a set of  channel-specific numerical NLO rescaling factors
recommended by the
LHCXSWG~\cite{LHCHiggsCrossSectionWorkingGroup:2011wcg}, with current
values given in \cref{tab:widHnlo}  
valid for a reasonable range around the nominal Higgs mass of $m_\H =
125$ GeV.
\begin{table}[t]
  \centering
  \begin{tabular}{lrrrrrrrrrr}
\toprule 
    SM $\H^0$ Decay Mode: & $\g\g$ & $\gam\gam$ & $\gam \Z$ & $\Z\Z$ &
$\W\W$ & $\b\bbar$ & 
    $\c\cbar$ & $\muon^+\muon^-$ & $\tauon^+\tauon^-$ \\
\midrule
NLO rescaling factor: & 1.47 & 0.88 & 0.95 & 1.10 & 1.09 &
1.11 & 0.98 & 0.974 & 0.992\\
\bottomrule
  \end{tabular}
  \caption{Numerical correction factors applied to the LO SM-Higgs
    decay partial width, based on LHCXSWG
    recommendations~\cite{LHCHiggsCrossSectionWorkingGroup:2011wcg}. Note
    that the strong coupling is fixed to 
    $\alpha_s = 0.12833$ in this context. \index{Higgs bosons}
  \label{tab:widHnlo}}
\end{table}
Note also that \pyt~8 computes the LO partial widths for
$\H^0 \to \gam\gam$ and $\H^0 \to \g\g$ using running quark-mass values
in the loop integrals (evaluated at $m_\H$); this gives a
non-negligible shift relative to \pyt~6 which used pole-mass values
in the same expressions.
For comparisons, the LHCXSWG
rescaling factors can optionally be replaced by simple $(1+
\alpha_s/\pi)$ correction for the decays to quarks, and for the
loop-induced decays the running mass values can be replaced by pole ones.

For BSM resonances, \pyt applies the $(1+\alpha_s/\pi)$ factor to all
integer-spin BSM particle decays to quark-antiquark pairs and to
semi-leptonic decays of right-handed neutrinos, while the
$(1 - 5\alpha_s/(2\pi))$ one is applied to $\t'\to \q\W$ decays. 

At the technical level, these decay-rate calculations are performed
by dedicated \texttt{calcWidth()}
methods in the derived \texttt{ResonanceWidths} class for the given
resonance. Note that this means that the tabulated
widths for these particles stored in the program's particle data table
are purely dummy values, overridden at initialization. To force a 
resonance with ID code \texttt{NN} to have a certain user-defined
width, $\Gamma$, set \settingval{NN:doForceWidth}{on} and
\settingval{NN:mWidth}{$\Gamma$}. Input of resonance widths via the \ac{SLHA}
interface is discussed separately below. 

\index{Breit--Wigner distribution}
\paragraph{Breit--Wigner modelling:}
We now turn to \pyt's modelling of resonance \emph{shapes}.
Note that this applies to resonances that are produced by \pyt
(\ie in \pyt's internal hard processes and/or in decays performed by
\pyt). For externally generated ones, \cf\cref{sec:externalGenerators}, it is the
responsibility of the external generator to model the shape of the 
produced resonances, though \pyt's modelling may still apply to any
resonances produced by subsequent decays of particles that are kept
stable in the external process. 

An important note in the
specific context that an external generator is responsible not only
for resonance production, but also for one or more of their decays is
that the total invariant mass of the resonance-decay products
(and hence the resonance shape) is only guaranteed to be preserved
during parton showering if an explicit resonance mother (with Les
Houches status code \texttt{2}) is present in the externally provided event
record. This is 
particularly relevant for any coloured resonances (such as top
quarks), for which the reconstructible resonance-mass distribution will
otherwise be impacted by unphysically large QCD recoil
effects to parton(s) outside the resonance-decay system. In principle,
the same issue exists for QED recoil effects in decays of
electrically charged resonances.  

\index{Decays!Resonances@of Resonances}
\index{weightDecay@\texttt{weightDecay()}}
\index{Resonance decays!calcWidth@\texttt{calcWidth()}}
\index{Resonance decays!weightDecay@\texttt{weightDecay()}}
\index{meMode@\texttt{meMode}}
The basics of phase-space generation and Breit--Wigner sampling in the 
context of processes involving resonances were covered in
\cref{sec:resonances}. As already mentioned there, decay-rate   
calculations specific to each given resonance and decay mode are the
default for most SM-resonance decays in \pyt as well as for some BSM
ones, via process-specific \texttt{SigmaProcess::\-weightDecay()} methods and
resonance-specific \texttt{ResonanceWidths::calcWidth()}
methods, enabled for decay channels assigned \settingval{meMode}{0}.
For
resonances that include such channels, \eqref{eq:basicBW} of
\cref{sec:resonances} is generalized to   
\begin{equation}
  \frac{1}{\pi} \frac{ m \sum_j \Gamma_j(m) }{(m^2 - m_0^2)^2 + m^2
    \Gamma^2_\mathrm{tot}(m)}~,
\end{equation}
where both the partial widths $\Gamma_j$ and the total width
$\Gamma_\mathrm{tot}$ are in principle allowed to depend on $m$. 
There are two main sources of $m$ dependence:
\begin{itemize}
\item  Running couplings in the relevant matrix elements.
  This also applies \eg to the NLO
normalizations given by \cref{eq:widNLO}, in which $\alpha_s(m_0^2)$ is
replaced by $\alpha_s(m^2)$. The SM-Higgs resonance is sufficiently
narrow that no appreciable running effects are expected, hence the
partial widths given in \cref{tab:widHnlo} are left unchanged. 
\item Threshold effects. For bosonic resonances ($\Z$, $\W$, $\H$, and
  particles that are 
trivially related to them such as $\Z'$, $\W'$, $\H^+$, and $\A$ bosons),
decays to same-flavour fermion pairs are 
associated with the following threshold factors:
\begin{equation}
  \Gamma(m) = \frac{m \Gamma_0}{m_0} \Theta(\hat{s} - 4m_{\f}^2) 
  \left\{
  \begin{array}{lcp{3cm}}
    \beta^3 & : & scalar \\ 
    \beta & : & pseudoscalar \\ 
    \beta(3-\beta^2)/2 & : & vector \\
    \beta^3 & : & axial-vector 
  \end{array}\right.~,
\end{equation}
where $\Gamma_0$ is the on-shell partial width and
$\beta = \sqrt{1-4m_{\f}^2/m^2}$ is the fermion velocity in the rest frame of the
decay. Resonances that have both vector and axial-vector (or both
scalar and pseudoscalar) couplings use appropriate mixtures of these
factors, and analogous but more complicated expressions are used for
decays into unequal masses \eg of the $\W^+$.
For other decays, the $m$ dependence is typically more
complicated.  
\end{itemize}
We refer to the corresponding implementations in the
\texttt{weightDecay()} and \texttt{calcWidth()} methods in the code,
which can be inspected for more details about the treatment of a given
process and/or decay mode, respectively.  

\paragraph{Decay angular distributions:}
In many cases, non-trivial angular distributions are
encoded in \pyt via process-specific LO matrix elements that include
the relevant decays. 
For example, for the hard process
$\f\fbar \to \W^+\W^-$ (with $\f$ denoting a generic fermion), \pyt generates the 
angular distributions for the two $\W$ decays at the same time,
using the full $\f\fbar \to \W^+\W^- \to 4$-fermion matrix elements.

 This allows for an accounting of the effects of spin
 correlations between the production and decay stages. Note, however,
 that only diagrams with the same resonant structure as the production
 process are included; interference with background processes is not
 accounted for by this method.
 
 \index{Top quark}\index{Weak bosons}\index{Higgs bosons}
 \index{Excited fermions}Using $V$ to denote a generic weak boson
 ($\W^\pm$ or $\Z^0$, with the 
latter typically including $\gam^*/\Z$ 
interference where relevant) and $\H$ to denote a generic neutral Higgs
boson, processes for which such matrix-element-corrected
resonance-decay distributions are generated by \pythia include:  
\begin{itemize} 
\item Decays of (unpolarized) top quarks: $\t\to \b\W^+ \to \b\, 2\f$. 
\item Electroweak decays of neutral Higgs bosons: $\H \to \V\V \to 4\f$ 
  and $\H \to \gam \Z \to \gam\, 2\f$, in both cases 
  allowing for generic (BSM) mixed-CP states. 
\item Electroweak resonant $s$-channel processes $2\f \to \V \to 2\f$. Note:
  this extends to BSM vector bosons such as $\V'$ and $\V_R$, and also
  includes the full 
  $\gam^*/\Z/\Z'$ interference for $\Z'$ ones. 
\item Electroweak resonant $2\to 4$  processes $2\f \to \V\V \to 4\f$ and 
  $2\f \to \H\V \to 4\f$. Also $2\f \to \V' \to \V\V \to 4\f$. 
\item $\W$ decays in $\f \fbar \to  \g/\gam\, \W \to \g/\gam\, 2\f$. 
\item BSM excited-graviton decays in $2\f \to \G^*$ and $\g\g\to \G^*$
  processes, \cf\cite{Park:2001vk}.
\item BSM compositeness excited-fermion decays in
  $2 \to \f^* \to \g/\gam\, \f$ and $2 \to \f^* \to \V\, \f$, with $\V$
  decaying isotropically for the latter.
\end{itemize}
A prominent example of a process that is absent from this list is
top-quark pair production, implying that internally generated $\t\tbar$
events in \pyt do not exhibit non-trivial correlations \emph{between}
the two top decays.  Note also that, for externally provided events
(\cf\cref{sec:externalGenerators}),
only the top- and Higgs-decay correlations in the two first points
above are applied. When interfacing external hard processes it is therefore
important to consider whether, and how, resonance decays are treated by
the external generator.

At the technical level, these process-specific angular distributions
are implemented via dedicated \texttt{weightDecay()} methods in the derived 
\texttt{SigmaProcess} class for the given hard process. 

\paragraph{Effects of PDFs on resonance
  shapes:}\index{PDFs!Effect on resonance shapes}
\index{Breit--Wigner distribution!Effect of PDFs} 
Often, the observable resonance shape results from a convolution
with non-trivial parton distribution functions. For hadrons, these tend
to be strongly peaked towards small $x$, with a typical asymptotic
behaviour roughly like $f(x) \propto 1/x$. When convoluted with the
Breit--Wigner shape, this tilts the overall resonance shape; the
parton-parton luminosity is higher in the low-mass tail than it is in
the high-mass tail.

If the low-mass enhancement is strong
enough, the wide tails of the Breit--Wigner can even lead to a
secondary peaking of the cross section towards very low masses. This
is obviously unphysical, as the resonant approximation is invalid that
far from the resonance, and non-resonant background processes would
anyway normally dominate in that region. The desire to cut away such
behaviour is one reason for the default choices made in \pyt for the
$m_\mathrm{min}$ limits in \cref{eq:basicBW}. For non-standard PDFs,
or when making user-defined modifications to the nominal mass and/or width
values (\eg for BSM particles), it is up to the user to check that sensible
$m_\mathrm{min}$ limits are imposed.

\paragraph{Interleaved resonance decays:}
\index{Decays!Resonances@of Resonances}
\index{Interleaved resonance decays}
\index{Resonance decays!Interleaving}

Rounding off the discussion of resonance production and decays, \pyt
also allows for interleaving resonance decays with the final-state
shower evolution, as described in \citeone{Brooks:2021kji}. Currently,
this is only done by default for the \vincia shower model, while it
exists as a non-default option for \pyt's simple showers. 

When interleaved resonance decays are enabled, resonance decays are
inserted into the final-state shower evolution as $1\to n$
branchings, at a scale which by default is given by the following
measure of the off-shellness of the resonance propagator, 
\begin{equation}
  Q_\mathrm{RES}^2 = \frac{ (m^2 - m_0^2)^2 }{m_0^2}~,
\end{equation}
with median value $\left< Q_\mathrm{RES} \right> = \Gamma$.
(A few alternative choices are also offered, including an option to
use a fixed scale $Q_\mathrm{RES}\equiv\Gamma$.)
As part of the resonance-decay branching process, a ``resonance
shower'' is also performed, in the region $m_0 > Q >
Q_\mathrm{RES}$. This shower stage only involves the decaying 
resonance and its decay products, with no recoils to any other
partons. Note that any nested resonance decays associated with intermediate
scales (\eg the $\W$ boson produced in a $\t\to \b\W$ decay) are also
performed during this stage, along with their
corresponding resonance showers, while any decays associated
with scales below $Q_\mathrm{RES}$ occur afterwards, sequentially. 

The main consequence is that resonances are prevented from
participating as emitters or recoilers for radiation at scales below
$Q_\mathrm{RES}$; only their decay products can do that.
We refer to \citeone{Brooks:2021kji} for further details.

\subsection{Parton distribution
  functions}\label{sec:hadronPDFs}\index{PDFs}\index{DGLAP}\index{Altarelli-Parisi|see{DGLAP}}
\index{Parton distribution functions|see{PDFs}}

Parton distribution functions provide number distributions of a parton flavour 
$i$ at a given momentum fraction $x$ when a hadron is probed at scale $Q^2$, and
are a necessary input for any hard process generation with hadron beams ~\cite{Kovarik:2019xvh}.
Here, we focus on PDFs for hadrons and nuclei --- PDFs for other types of beams
(including leptons, photons, and pomerons) are discussed separately in
\cref{sec:soft}. The scale evolution of the PDFs is provided by the
\ac{DGLAP} equations~\cite{Gribov:1972ri, Dokshitzer:1977sg, Altarelli:1977zs}
and usually these are derived in a global QCD analysis where the
non-perturbative input at an initial scale is fitted to a wide range of
experimental data. Further constraints are provided by the momentum- and
baryon-number sum rules. Nowadays, it is common that in addition to the best
fit, the PDF sets also provide error sets that can be used to quantify how the 
uncertainties in the applied data propagate into other
observables.\index{Uncertainties!PDF@from PDFs}

In the case of protons, the high-precision DIS data from HERA collider form the
backbone of the PDF analyses. On top of this, the modern PDF sets incorporate a
wealth of different LHC data to increase the kinematic reach of the analysis
and to obtain further constraints for the flavour dependence. With this,
in kinematic regions relevant for LHC studies, the proton structure is known
with a percent-level accuracy, except for in a few regions
like the very small-$x$ region.
\pyt comes with some 20 different proton PDF sets. There are a few 
(pre-HERA) sets that are out of date, \eg GRV94L and CTEQ5L, but are kept in
for historical reasons as some earlier tunes were based on these. In addition,
there are a few sets that include HERA data but did not have any from LHC data
(\eg CTEQ6L) which mainly differ from the older ones due different small-$x$
gluon behaviour. The more modern ones include several data sets from LHC
experiments which provide further constraints for gluon PDFs and flavour
separation between different quarks. Another recent development in the PDFs
is the inclusion of QED evolution that enables inclusion of photons as a part
of the hadron structure. The current default set is NNPDF2.3 QCD+QED LO, which does
contain some datasets from LHC, but not the most recent ones. It is important 
to note, however, that the default Monash tune is based on this PDF set, so
updating to a more recent PDF set would not lead to an improved description
unless a complete retuning of $\p\p$ parameters is performed. Many further sets 
are accessible via the LHAPDF interface, \cf\cref{sec:interface:lhapdf}.
This runs slightly slower than the built-in sets, but also offers further
facilities such as error bands around the central PDF member. Notice also
that there might be small differences between the internally defined PDFs sets
and the corresponding LHAPDF grids due to different interpolation routines
and different extrapolations beyond the provided interpolation grid.

The neutron PDF is obtained from the proton one by isospin conjugation.
This is not quite correct for some recent sets where the QCD evolution
is combined with a QED one, \ie where the quarks can radiate off photons,
but in practice it is good enough except for photon physics.

For pions, the main set is based on GRS~99~\cite{Gluck:1999xe}. 
This work makes the ansatz that valence, gluon, and sea PDFs are of
the form $N x^a (1-x)^b (1 + A \sqrt{x} + B x)$ at an initial scale
$Q_0^2 = 0.26\,\mathrm{GeV}^2$, with the parameters fitted to data.
By choosing a small $Q_0$, the distributions can be assigned a
valence-quark-like shape at that scale, and strange and heavier quarks
can be taken to vanish. An older set based on GRV~92~\cite{Gluck:1991ey}
is available, but is deprecated in favour of GRS~99. A similar PDF is
also available for the kaon~\cite{Gluck:1997ww}.

For other hadrons, rough estimates for PDFs have been made based on the form
above, with $A = B = 0$. No data is available, so the parameters $a$ and
$b$ have been chosen heuristically, based on the guiding principle that all
valence quarks should have roughly the same velocity for the hadron to stay
together over time, and thus heavier quarks must take a larger average
momentum fraction. The $N$ are fixed by the flavour and
momentum sum relations. For
details on this procedure, see \citeone{Sjostrand:2021dal}. These PDFs are
referred to as the SU21 sets, and are stored in the LHAPDF format and
distributed with \pythia. Specifically, the PDFs included this
way are available for the following hadrons: $\p$, $\pip$, $\kp$, $\phiz$, $\eta$,
$\D^0$, $\D_\s^+$, $\Jpsi$, $\B^+$, $\B_\s^0$, $\B_\c^+$, $\Upsilon$,
$\Sigma^+$, $\Xi^+$, $\Omega^-$, $\Sigma_\c^{++}$, $\Xi_\c^+$, $\Omega_\c^0$,
$\Sigma_\b^+$, $\Xi_\b^-$, and $\Omega_\b^-$.
The SU21 $\p$ and $\pion$ PDFs are less accurate than other available sets,
so they should not to be used in real studies, but are included for
completeness. Hadrons with the same quark contents as the ones above
are assumed to have the same PDFs. Furthermore, other cases can be defined
using isospin conjugation, since no QED effects are included
in the SU21 sets. Mixed cases
such as \piz and $\Sigma^0$ are assumed to have equal $\u$ and $\d$ contents,
which are given by the averages for the corresponding implemented PDF 
(\ie \pip and $\Sigma^+$, respectively). Using such rules, all normal
hadrons can be simulated, except for baryons with more than one
charm or bottom quark. One final technical point is that
in the SU21 LHAPDF files for flavour-diagonal mesons, the antiquark content
represents the sea, in order to make it possible to separate valence and sea
(\eg for \Jpsi, the \c column represents the charm content, while the \cbar
column represents charm sea). 

\index{PDFs!nuclear@for Nuclei}\index{Nuclear PDFs|see{PDFs for Nuclei}}Also, a
few nuclear PDF sets have been included internally. 
These can be used to estimate the leading nuclear effect for
inclusive high-$\pT$ observables, such as jet production, but for more
involved studies it is recommended to use the full heavy-ion
machinery, see \cref{sec:heavy-ion-collisions}. More nPDFs are
available as LHAPDF grids, but the advantage of the internally defined
sets is that any proton baseline PDF can be applied and, if needed,
the number of protons and neutrons can be redefined event-by-event.

A fair fraction of the internal PDFs are LO ones.
This ensures a sensible behaviour also for processes at low $x$ and/or $Q^2$
(discussed further below), but also some NLO and NNLO proton sets are
available, for the modelling of hard processes.
In this context, it can be mentioned that, at large $x$ and $Q^2$,
NLO corrections to the PDF shape are often more important than those for the
matrix elements, such that NLO PDFs and LO MEs can be a viable combination.
Further, in the large-$(x, Q^2)$ region where the behaviour is nowadays
rather well understood, PDFs do not risk turning negative. 

For showers and MPIs, the case is less clear; they both connect to low-$\pT$ scales around or below 1~GeV, and especially MPIs can probe
extremely small $x$ values, down to around $10^{-8}$ at LHC energies,
\cf\cref{subsection:mpi}.
In this region, all PDF components are poorly known, especially the
dominant gluonic one.  
In an LO description, the PDFs are required to be
non-negative, and HERA data in combination with Regge theory provide
some reasonable constraints on the low-$x$ behaviour. PDFs need not be
positive definite at higher orders, NLO or NNLO, since it is only the
convolution of NLO (NNLO) hard-process matrix elements with NLO (NNLO)
PDFs that should be non-negative, up to NNLO (N$^3$LO) terms. Actually,
at scales $\pT \sim 1$~GeV the whole perturbative expansion is poorly
convergent, since $\alphas$ is large.  Some recent PDFs attempt a resummed
description of the small-$x$ behaviour to restore a guaranteed PDF
positivity~\cite{Bertone:2018dse}. Nevertheless, in general, the
criteria for what constitutes an optimal or at least sensible PDF
choice for the hard process are not necessarily the same as for
showers and MPIs; for this reason, \pythia allows for the use of one PDF set
for the hard process and a different set for showers and MPIs. This can 
also be useful to preserve shower- and underlying-event tuning
properties while changing PDFs for the hard process. 

It is also possible to pick
different PDF sets for the two incoming beam particles, which may be
convenient as a technical trick but has no physics motivation when colliding
beams are the same.

\subsection{Phase-space cuts for hard processes}
\label{subsection:phasespacecuts}
\index{Kinematics!Phase-space cuts}
\index{Phase-space generation!Cuts}

Several different phase-space cuts have been implemented for the internal hard processes in \pythia. These serve two purposes: to properly set values that ensure the approximations in the theory description are valid, and to allow for more efficient event generation when only a certain part of the available phase space is considered. The principal example is the lower limit of the partonic $\pT$ of $2 \to 2$ processes, that needs to be set to a high enough value such that the divergent behaviour of the massless matrix elements in the $\pT \rightarrow 0 $ limit is avoided. Similarly, a suitable lower limit for $\pT$ should be applied when considering \eg jet production at higher values of $\pT$, to avoid the inefficiency otherwise associated with a rapidly dropping $\pT$ spectrum. (But also see comment at the end of this subsection.)

The number of implemented phase-space cuts for the hard scattering depends on the number of final-state particles of the process. For $2 \rightarrow 1$ only two options are included:
\begin{itemize}
\item the minimum invariant mass $m_{\mrm{min}}$
\item the maximum invariant mass $m_{\mrm{max}}$
\end{itemize}
If the value of the latter is lower than the value of the former, the invariant mass will be limited from above by the collision energy. The same cuts also apply to $2 \to 2$ and $2 \to 3$ processes.

For $2 \rightarrow 2$ processes some more options appear. The first three are related to invariant transverse momentum of the process: 
\begin{itemize}
\item the minimum transverse momentum $p_{\perp\mrm{min}}$
\item the maximum transverse momentum $p_{\perp\mrm{max}}$ 
\item an additional lower transverse-momentum cut $p_{\perp\mrm{diverge}}$
\end{itemize}
The latter is to prevent divergences in the $\pT \rightarrow 0$ limit for processes where a particle has a mass smaller than the set $p_{\perp\mrm{diverge}}$. In these cases, however, the larger of the $p_{\perp\mrm{min}}$ and $p_{\perp\mrm{diverge}}$ is always applied for the $\pT$ selection. The next set of cuts is related to limits of Breit--Wigner (BW) mass distributions. By default, the mass selection based on BW shapes is always applied for particles with a width above a certain threshold. There are two different thresholds that can be set:
\begin{itemize}
\item the minimum width of a resonance for which the Breit--Wigner shape can be deformed by the variation of the cross section across the peak;
\item and the minimum width of a resonance that is below the former threshold, for which a simplified treatment is applied instead, where a symmetric Breit--Wigner selection is decoupled from the hard-process cross section.
\end{itemize}
Notice that the allowed mass range of a given particle can be set by modifying the particle properties. In case of DIS, instead of $\pT$, the most relevant phase-space cut is the lower limit for the allowed virtuality of the intermediate photon:
\begin{itemize}
\item minimum $Q^2$ for $t$-channel processes with non-identical particles 
\end{itemize}
Notice that the cuts for $\pT$ will also be applied when a non-zero cut for $Q^2$ is applied.

For $2 \rightarrow 3$ processes that do not contain soft or collinear singularities, such as Higgs production in EW-boson fusion, the same cuts as in the $2 \rightarrow 2$ case can be applied. For QCD processes, where such singularities need to be accounted for, alternative cuts are defined. Also, since the outgoing partons are no longer back-to-back, cuts for individual partons can be used for a more detailed phase-space mapping:
\begin{itemize}
\item the minimum transverse momentum for the highest-$\pT$ parton
\item the maximum transverse momentum for the highest-$\pT$ parton
\item the minimum transverse momentum for the lowest-$\pT$ parton
\item the maximum transverse momentum for the lowest-$\pT$ parton
\item the minimum separation $R$ $(=\sqrt{(\Delta \eta)^2+(\Delta \phi)^2})$ between any two outgoing partons
\end{itemize}
The last one needs to have a high-enough value to avoid collinear divergences associated with the outgoing partons.

As described above, the phase-space cuts can be used to improve the sampling efficiency by focusing on a particular phase-space volume, \eg defined by cuts on partonic $\pT$. In some cases this might, however, require several runs that need to be combined later on. Similar improvement in efficiency can also be achieved by reweighting the cross section of the hard process with a suitable kinematic variable. In \pythia the events can easily be reweighted by $\pT^{-\alpha}$, where $\alpha$ is a power that could \eg approximate the $\pT$ dependence of the hard cross section. This allows for a more uniform filling of the phase space, even when the cross section itself drops rapidly. The downside is that when reweighting is applied, each event comes with a weight that needs to be accounted for \eg when filling histograms. In addition to this built-in reweighting of internally defined $2 \rightarrow 2$ hard processes, there are also more involved options for reweighting with different variables that can be enabled with the user hooks described in \cref{subsection:userhooks}.

An important aspect is that the described phase-space cuts are applied only for the hard scattering, \ie before any showering or hadronization. As the shower emissions will modify the four-momentum of outgoing partons; a jet formed from the final particles will have a somewhat different transverse momentum than the parton that originated the jet. Final-state radiation and hadronization can reduce the energy of the jet, whereas initial-state radiation and multiparton interactions may enhance it. Therefore a ``fiducial phase-space volume'' is needed, \ie hard processes must be generated in a larger volume than the volume of interest for final-state observables, at the unfortunate cost of generating many events that will be thrown away. The necessary amount of oversampling depends highly on the kinematics and beam configuration considered, so it needs to be checked case-by-case. For jet studies, this is usually done by plotting the hard-process $\pT$ associated with accepted jets or events. If a non-negligible fraction of events near the $p_{\perp\mrm{min}}$ scale are accepted then $p_{\perp\mrm{min}}$ is too high. 

\subsection{Second hard process}\index{Quarkonium}\index{MPI}
\label{subsection:secondhard}

The MPI framework in \pyt will generate a variable number of partonic $2 \rightarrow 2$ interactions in addition to the selected hard process itself. These, mainly QCD processes, will form the underlying event, typically consisting of rather soft particles. Occasionally, they may also contain a hard scattering but, due to power-law falloff of the relevant cross sections, such events are rare. There are, however, cases when the studied observable is such that more control over the kinematics of the second scattering can significantly improve the sampling efficiency (\eg of four-jet final states), or the second process is not included as a part of current MPI generation (\eg the production of an EW boson together with a jet). The machinery for a second hard process can be used in these situations. It can be viewed as an approach to generate so-called \ac{DPS} events, but with two key distinctions. First, the DPS framework as used for theoretical studies typically assumes that there are exactly two hard interactions in an event, while the second-hard setup allows there to be further MPIs just like when starting out from one hard interaction. Second, the MPI machinery uniquely fixes how two hard cross sections should be combined into a total, while this usually involves a free parameter in the DPS expressions.
 
The basic approach in the \pyt implementation for the generation of two hard processes in a single event is that, first, the two processes are selected completely independently and, afterwards, momentum conservation and the possible correlations in the PDFs are accounted for by the rejection of a fraction of the topologies. This makes the process sampling symmetric and thus the distinction between ``first'' and ``second'' is used only for bookkeeping. Furthermore, as long as there is some overlap in phase space of the two processes, any of the two can be the hardest one. In principle, this construction would allow the generation of any two internally (or externally) defined processes, but in practice there is no need for a very fine-grained control of both processes, and furthermore the combination of two rare processes would give a negligible cross section. Therefore a somewhat more limited set of second processes have been implemented. Still, the first process can be selected from the complete list of processes (\appref{sec:hardProcesses}), or even provided externally. The processes that can be enabled as a second hard one include:
\begin{itemize}
\item standard QCD $2 \rightarrow 2$ processes, \ie two-jet production
\item a prompt photon and a jet
\item two prompt photons
\item charmonium production, colour singlet and octet
\item bottomonium production, colour singlet and octet
\item $\gamma^*/\Z$ production with full interference
\item single $\Wpm$ production
\item production of a $\gamma^*/\Z$ and a parton
\item production of a $\Wpm$ and a parton
\item top-pair production
\item single-top production
\item bottom-pair production
\end{itemize}
Technically these can be combined freely, but some combinations would double count and therefore must be avoided. This includes a $\gamma^*/\Z/\Wpm$ together with a jet or on its own, and $\b \bbar$ production as part of the QCD $2 \rightarrow 2$ processes or on its own. Also, since the last one will include only $\b \bbar$ production explicitly in the hard scattering, the pairs produced by gluon splittings in the parton showers will not be present in that sample. Thus, depending on the kinematics, this might or might not be enough to give realistic cross-section estimates.

By default, the phase-space cuts, couplings, and scales for the second hard process are the same as for the primary scattering. It is, however, possible to set different cuts for the second one, and, due to fully symmetric treatment of the two processes, the cuts for the second process can be set higher or lower than for the primary one. The cuts that can be separately specified are the minimum and maximum values for the invariant mass and transverse momentum of the process.

It is instructive to consider some Poissonian statistics before showing how the cross sections of two processes should be combined. If the average number of subcollisions, $\langle n \rangle$, is known, the probability for $n$ of them to occur is given by
\begin{equation}
  P_n = \langle n \rangle^n \frac{\mrm{e}^{-\langle n \rangle}}{n !}
  \label{eq:secondhard:Poissonian}
\end{equation}
In case where $\langle n \rangle$ is small, as it is for hard processes, we can approximate $\mrm{e}^{-\langle n \rangle} = 1$. The probability for one event to happen is then $P_1 = \langle n \rangle$, and correspondingly for two such events we find $P_2 = \langle n \rangle^2 / 2 = P_1^2 / 2$. Now consider two independent event types $a$ and $b$, such that $\langle n \rangle = \langle n_a \rangle + \langle n_b \rangle = P_{1a} + P_{1b}$. The probability for any combination of two events $a$ and $b$ is then given by
\begin{equation}
P_2 = \frac{(P_{1a} + P_{1b})^2}{2} = \frac{P_{1a}^2 + 2P_{1a}P_{1b} + P_{1b}^2}{2}~.
\label{eq:secondhard:probtwo}
\end{equation}
From this it can be read off that a probability for having two different-type events comes with a factor 2 relative to the same-type cases. If modelled in terms of increasing time, or decreasing hard scale (say $\pT$), the mixed combination can occur in two ways, either where the event $a$ happens before $b$, or the other way around, which explains the factor of $2$.

The proper way to evaluate the resulting cross section thus depends on whether the two processes are the same, and on whether the phase-space regions overlap. The simplest case is when the two processes do not overlap, \ie either the phase-space regions are completely separated or the two process are different. An example of the latter would be a combination of processes where the first produces two jets and the second two photons. When the $a$ and $b$ cross sections are small fractions of the total non-diffractive cross sections $\sigma_{\mrm{ND}}$, naively the probabilities $P_{a,b} = \sigma_{a,b}/\sigma_{\mrm{ND}}$ enter multiplicatively. Thus their combined cross section is 
\begin{equation}
\sigma^{\mrm{naive}}_{2ab} = P_a \, P_b \, \sigma_{\mrm{ND}} =
\frac{\sigma_{1a} \sigma_{1b}}{\sigma_{\mrm{ND}}} ~. 
\end{equation}

This simplification neglects the dependence on collision geometry, however. The probability for a hard process is enhanced in central collisions, \ie for small impact parameter, while it is depleted in peripheral ones. This leads to a so-called ``trigger bias'' effect, where events containing a first hard process predominantly occur in central collisions, which thereby enhances the likelihood of a second hard process. In the context of traditional MPIs this is known as the ``pedestal effect'', where a selected high-$\pT$ process has more underlying-event activity than an average event, see more details in \cref{subsection:mpi:impactparameter}. When the colliding matter profiles have been specified, along with the parameters that set the $\langle n_{\mrm{MPI}} \rangle $, a correction factor $f_{\mrm{impact}}$ can be derived event-by-event within the MPI framework. Its average value gives a corrected combined cross section 
\begin{equation}
\sigma_{2ab} = \langle f_{\mrm{impact}} \rangle \frac{\sigma_{1a} \sigma_{1b}}{\sigma_{\mrm{ND}}} = \frac{\sigma_{1a} \sigma_{1b}}{\sigma_{\mrm{eff}}} ~.
\end{equation}
In the last step we introduce $\sigma_{\mrm{eff}}$, which is the conventional parameter that many experimental results are expressed in terms of, but here it is a prediction of the model.   

The cross section $\sigma_{2aa}$ of two identical processes follows the same pattern, except for the extra factor of $1/2$ that has already been explained. Often $a$ would itself be the sum of several subprocesses, \eg the six main classes of $2 \to 2$ QCD processes that contribute to two-jet production. If so, then a compensating factor of 2 will automatically occur for the mixed-subprocess configurations, in the same spirit as \cref{eq:secondhard:probtwo}.

The cross section calculation becomes somewhat more complicated in cases when there is partial overlap between the two processes. An example would be identical processes with different, but partly overlapping, cuts on $\pT$. In such cases it is useful to split the problem into two completely independent processes $a$ and $b$ and a common process $c$. The first (second) process can be selected according to $\sigma_a + \sigma_c$ ($\sigma_b + \sigma_c$). Half of the  events should be discarded if both processes are chosen as $c$, and the combined cross section should be reduced accordingly.   

So far it has been assumed that the generation of the two processes can be done independently, apart from the geometrical correction factor for the final cross sections. This obviously misses all possible correlations between the PDFs and, perhaps more importantly, may violate energy-momentum conservation. Part of the selected events will be discarded to account for these effects, even though each process would be acceptable on its own. The correlations in multiparton PDFs implemented in \pyt are described further in \cref{subsection:mpi:pdfrescaling}. The PDF reduction factor is obtained as the average of the two possible orderings, where either the second or first PDF is corrected for the parton taken out either by the first or second process.

\index{Uncertainties!Hard@in Hard processes}In the end, the cross sections provided by \pyt after the event generation do account for all these effects, including the correction factor $\langle f_{\mrm{impact}}\rangle$ and the PDF rescaling. The error estimates provided by \pyt are statistical ones and do not cover the potentially large model uncertainties, as usual. When the first process is provided externally, \pyt does not have the information whether there is an overlap between the first and the second process, and so will assume that this is not the case. The proper correction for an overlap then rests with the user.

%% file: physics/parton-showers.tex
\section{Parton showers}
\label{sec:showers}
\index{Parton showers}
\index{QCD showers|see{Parton showers}}

The most violent $\pp$ collisions at the LHC may have five to ten
easily separated jets. Zooming in on these, they display a
substructure of jets-inside-jets-inside-jets, associated with the
perturbative production of increasingly nearby partons. Such a
fractal nature is expected to continue down to the hadronization
scale, a bit below 1~GeV. At that scale, the event may contain up to
a hundred partons, even if the full partonic structure  is masked
by the subsequent non-perturbative hadronization process. There is no way to perform matrix-element calculations to describe
such complicated event topologies. Instead, the standard approach is to
start out from a matrix-element calculation with only a few well-separated partons, and then apply a parton shower to that.

Parton showers attempt to describe how a basic hard process is
dressed up by emissions at successively ``softer'' (longer-wavelength)
and/or more ``collinear'' (smaller-angle) resolution scales,
to give an approximate but realistic picture of the 
(sub)structure of the partonic state across the full range of
(perturbative) resolution scales.
Such a shower is constructed in a recursive manner,
from the large scale of the hard process down to a lower cutoff
at around the hadronization scale. In each step, the number of partons
is increased by one, or in very special cases two, and the random nature of the steps leads to
a large variability of final states. It is worth emphasizing that, 
although often thought of in the context of QCD, parton 
(or more generally particle) showers are in fact common to 
any quantum field theory with several (quasi-)massless particles. 
Thus, showers are present in QCD, QED, and the EW theory above 
the symmetry breaking scale and as such, dedicated modules 
describing all of these are part of \pythia.

One starting point is to study the ratio of two differential matrix
elements, $\d\sigma_{n+1}/\d\sigma_n$, where the numerator corresponds
to the emission of one more gluon in the final state. It then turns out
that this ratio is given by universal expressions, \ie independent of
which specific process is considered, if this gluon is either soft, or collinear
with one of the already existing partons. This means that
one can formulate a generic scheme that can be applied to any process
of interest. Such schemes started to be developed in the late 1970s.
A key ingredient has been the \ac{DGLAP} evolution equations~\cite{Gribov:1972ri,Altarelli:1977zs,Dokshitzer:1977sg}\index{DGLAP},
which describe
near-collinear emissions. Modern showers, like the three available
with \pythia, are based on many subsequent developments, intended to
make them cover the full phase space as well as possible. These aspects
are described later, but initially we introduce the simpler, classical
(collinear ``leading-log'') framework that helps in understanding the
overall picture. 

Historically, showers are split into two
kinds, \ac{ISR} and \ac{FSR}, which occur respectively before
or after the hard process. Alternatively, they may be referred to as
spacelike and timelike showers, respectively, since their representation
in terms of Feynman diagrams contain off-shell intermediate particles
that are either spacelike or timelike. The more virtual such a particle
is, the shorter it may exist. Therefore, the highest virtualities
occur in and closest to the hard interaction, and then showers with
decreasing virtualities stretch backwards (for ISR) or forwards (for FSR)
in time. LHC processes usually contain both ISR and FSR, and outside
the strictly collinear limits the distinction
can be blurred, just like
interfering Feynman graphs of a different nature may contribute to a
given final state. A decay $\gam^*/\Z \to \q\qbar$ is 
pure FSR, however, while its production $\q\qbar \to \gam^*/\Z$ can be
discussed in terms of ISR only, so these are often used as textbook
examples. (Conversely, ISR-FSR interference can be exemplified by
$t$-channel colour-singlet exchange, such as in deep inelastic
scattering or vector boson fusion.) 

\paragraph{FSR}\index{Final-state radiation|see{FSR}}\index{FSR}
Starting from $\gam^*/\Z \to \q\qbar$, either the
$\q$ or $\qbar$ may emit a $\g$, \eg $\q \to \q\g$. This produces a
$\q\qbar\g$ state, after which either of the three partons may branch, and so on.
The differential probability for a parton to branch can be written as
\begin{equation}
\d\mathcal{P}_a(z,Q^2) = \frac{\d Q^2}{Q^2} \, \frac{\alphas(Q^2)}{2\pi} \,
\sum_{b,c} P_{a \to bc}(z) \, \d z ~.
\label{eq:introshower:evforwfin}
\end{equation}
Here $a$ is the mother that splits into partons $b$ and $c$, where the
momentum-energy of the mother is split such that $b$ takes fraction
$z$ and $c$ takes $1 - z$. The $Q$ scale, used to order emissions in a
falling sequence, is a key distinguishing feature of different shower
implementations, and may be chosen \eg as mass, transverse momentum
or energy-weighted emission angle. That is, $z$ parameterizes the
longitudinal and $Q$ the transverse evolution of the shower. There is
also an azimuthal angle $\varphi$ that determines the orientation of the
decay plane; typically, and for the purpose of this brief
introduction, this is assumed to be distributed isotropically, though
we note that \pyt does allow for non-uniform distributions as well,
\eg to reflect gluon polarization effects. 

A key issue that distinguishes parton showers from so-called
analytic resummation approaches, is that the latter only maintain 
exact energy and momentum conservation in the strict soft and collinear limits
while showers do so over all of phase space. This difference
leads to the crucial aspect of \emph{recoil effects} in parton
showers, which will play an important role when we introduce dipole
showers later on.  
 
There are three different \ac{DGLAP} splitting kernels,
\begin{align}
P_{\q\to\q\g}(z) &= \frac{4}{3} \, \frac{1+z^2}{1-z} ~, 
\label{eq:introshower:dglapqqg}\\
P_{\g\to\g\g}(z) &= 3 \, \frac{\big(1-z(1-z)\big)^2}{z(1-z)} ~, 
\label{eq:introshower:dglapggg}\\
P_{\g\to\q\qbar}(z) &=  \frac{1}{2} \, \left(z^2 + (1-z)^2\right) ~.
\label{eq:introshower:dglapgqq}
\end{align}
These obey the trivial symmetry relations
$P_{a \to cb}(z) = P_{a \to bc}(1-z)$. The $P_{\g\to\q\qbar}$ kernel is normalized
for one quark flavour only, and has to be summed over all
kinematically allowed channels.

\index{QED showers}The same approach can also be used for other branchings, notably QED
ones, where $\alphas$ in \cref{eq:introshower:evforwfin} is replaced by
$\alphaem$ and the splitting kernels are
\begin{align}
P_{\f\to\f\gam}(z) &= e_{\f}^2 \, \frac{1+z^2}{1-z} ~, 
\label{eq:introshower:dglapffgam}\\
P_{\gam\to\f\fbar}(z) &= \Nc \, e_{\f}^2 \, \left(z^2 + (1-z)^2\right) ~,
\label{eq:introshower:dglapgamff}
\end{align}
where $\Nc = 3$ if $\f$ is a quark and $\Nc = 1$ if a charged lepton.

The DGLAP kernels are often written with additional terms that modify
the behaviour at $z = 1$ and 0, in order to conserve momentum-energy
and flavour in analytic calculations. This is not necessary in event
generators, partly because the 0 and 1 limits are never reached,
and partly because conservation issues are handled explicitly: parton
$a$ is removed at the same time as $b$ and $c$ are inserted in the list
of currently existing partons. 

The branching probability in \cref{eq:introshower:evforwfin} can be
integrated over the kinematically allowed $z$ range
\begin{equation}
\d\mathcal{P}_a(Q^2) = \frac{\d Q^2}{Q^2} \, \frac{\alphas(Q^2)}{2\pi} \,
\sum_{b,c} \int_{z_{\mathrm{min}}(Q^2)}^{z_{\mathrm{max}}(Q^2)} \,
P_{a \to bc}(z) \, \d z ~,
\label{eq:introshower:forwardsevol}
\end{equation}
to express the infinitesimal probability that $a$ branches in a $\d Q^2$
infinitesimal step. (Strictly speaking $|\d Q^2|$ since $Q^2$ is decreasing
in the evolution.) The probability for $a$ \emph{not} to branch in the
same step thus is $1 - \d\mathcal{P}_a(Q^2)$. By multiplication of the
no-emission probabilities (exponentiation), the probability for $a$ not
to branch between an initial scale $Q_1^2$ and a final lower $Q_2^2$
becomes the Sudakov factor~\cite{Sudakov:1954sw}
\index{Sudakov factor}
\begin{equation}
\Pi_a(Q_1^2, Q_2^2) = \exp \left( - \int_{Q_2^2}^{Q_1^2} \d\mathcal{P}_a(Q^2)
\right) ~.
\label{eq:introshower:sudakov}
\end{equation}
The differential probability for $a$ to evolve from a $Q_{\mathrm{max}}^2$
to a $Q^2$ and then branch at the latter scale, thus is
$\Pi_a(Q_{\mathrm{max}}^2, Q^2) \, \d\mathcal{P}_a(Q^2)$.
Note that the introduction of a Sudakov factor ensures that the total
probability for $a$ to branch cannot exceed unity, something that is
not guaranteed for $\d\mathcal{P}_a$ alone.

We observe that the Sudakov factor plays a crucial role in the selection
of a branching scale. The veto-algorithm technology in
\cref{subsubsec:vetoalgorithm} is eminently suited to handle
cases where the $Q^2$ and $z$ integrations cannot be done
analytically. The Sudakov factor also is closely related to virtual
corrections of matrix elements, \ie loop corrections. This will play
a key role for the matching and merging methods presented in the
next section.

\paragraph{ISR}\index{Initial-state radiation|see{ISR}}\index{ISR}
The ISR description starts out from the evolution
equation for \ac{PDF}s,
\begin{align}
\d f_b(x, Q^2) &= \frac{\d Q^2}{Q^2} \, \frac{\alphas(Q^2)}{2\pi} \,
\sum_{a} \int f_a(x', Q^2) \, \d x' \int P_{b/a}(z) \, \d z \,
\delta(x - x'z) \nonumber \\
&= \frac{\d Q^2}{Q^2} \, \frac{\alphas(Q^2)}{2\pi} \, \sum_{a} \int
\frac{\d z}{z} f_a\left(x' = \frac{x}{z}, Q^2\right) \, P_{b/a}(z) ~,
\label{eq:introshower:pdfevol}
\end{align}
where $f_i(x, Q^2)$ is the probability to find a parton $i$ inside a
hadron, with $i$ carrying a fraction $x$ of the full hadron momentum
if the hadron is probed at a scale $Q^2$.

As for FSR, the evolution is driven by branchings $a \to bc$ but,
where FSR is formulated in terms of the decay rate of $a$, ISR is
given in terms of the production rate of $b$. The simple splitting
kernels are easily related, $P_{b/a}(z) = P_{a \to bc}(z)$, except that
$P_{\g/\g}(z) = 2 P_{\g \to \g\g}(z)$, since two gluons are produced
for each gluon that decays. 

The evolution of PDFs starts at some low scale $Q_0^2$ and then proceeds
towards the $Q^2$ scale of the hard process, where they enter into the
cross-section expression. While \cref{eq:introshower:pdfevol} describes
the evolution of an inclusive distribution, an exclusive shower
formulation similar to the FSR one is possible, although more
complicated. A key problem is that the two incoming cascades, one from
each side of the event, may not end up as the colliding partons one is
interested in. For example, in $\g\g \to \H$ the two incoming gluons must have an
invariant mass that matches the Higgs mass.

\index{Backwards evolution}The solution to this problem is backwards
evolution~\cite{Sjostrand:1985xi}. 
In this method, the evolved PDFs are first used to select the hard process of
interest, say $\q\qbar \to \gam^*/\Z$ . Only afterwards are the incoming
showers then constructed backwards in time, from the high $Q^2$ scale
down to the low $Q_0^2$. To this end, we introduce
\begin{equation}
\d\mathcal{P}_b(x,Q^2) = \frac{\d f_b(x, Q^2)}{f_b(x, Q^2)}
= \frac{\d Q^2}{Q^2} \, \frac{\alphas(Q^2)}{2\pi} \,
\sum_{a} \int_{z_{\mathrm{min}}(Q^2)}^{z_{\mathrm{max}}(Q^2)} \d z \,
\frac{x' f_a(x', Q^2)}{x f_b(x,Q^2)} \, P_{b/a}(z) ~,
\label{eq:introshower:backwardsevol}
\end{equation}
where we have used that $z = x / x'$. Here $\d\mathcal{P}_b$ is
the probability that parton $b$ becomes associated with a branching
$a \to bc$ during the interval $\d Q^2$. A no-branching probability
$\Pi_b(x, Q_1^2, Q_2^2)$ can be defined in analogy with the Sudakov factor
\cref{eq:introshower:sudakov}. The corrected probability for a parton
$b$ that branches or interacts at $Q_{\mathrm{max}}^2$ to be assigned a
mother $a$ at $Q^2$ then is
$\Pi_b(x, Q_{\mathrm{max}}^2, Q^2) \, \d\mathcal{P}_b(x, Q^2)$. 
This $a$ in its turn must be evolved to yet lower $Q^2$ to find its
mother at an even higher $x$ value.

\paragraph{Recoils and dipoles}
An isolated parton cannot branch, if energy and momentum is to
be preserved. Take as an example
$\gam^*/\Z \to \q\qbar \to \q^*\qbar \to \q\qbar\g$, where $\q^*$
is the off-shell quark that branches as $\q^* \to \q \g$. Initially,
the $\q$ and $\qbar$ can split the energy equally, but the off-shell
$\q^*$ acquires a larger mass than $\qbar$, and so it must have a
larger energy while the $\qbar$ receives a smaller one. In this case we
would call $\q$ the radiator (or emitter) and $\qbar$ the recoiler,
but note that at the end, both may yield energy to create the $\g$.
Also, considering the existence of $\qbar \to \qbar\g$ branchings,
it may be simpler to say that it is the $\q\qbar$ pair that jointly
radiates the $\g$. Note that $\q$ and $\qbar$ have opposite and
compensating colours and thus form a colour dipole, hence the concept
of dipole radiation.

This picture generalizes to the subsequent emission of further
gluons~\cite{Gustafson:1986db,Gustafson:1987rq}. In the limit of
infinitely many colours, $\Nc \to \infty$~\cite{tHooft:1973alw},
the $\q\qbar\g$ system exactly splits into one $\q\g$ dipole and
one $\g\qbar$ dipole. These can radiate independently, and the recoil
is distributed within each dipole. It is still possible, but not necessary,
to split the radiation inside each dipole as being associated with
either dipole end. 

To allow dipole showers to operate, unique colour indices (in the
$\Nc \to \infty$ limit) are assigned to  all coloured
partons, both ones produced in the hard process and ones in the
subsequent shower evolution. For the extension to ISR, and to decays
like $\t \to \b\Wp$, one should note that the hole left behind
by a scattered or decayed colour parton can act like its anticolour. 

\paragraph{Formal basis of parton showers} In the previous discussion, we have developed the basic idea of parton showers, similarly to the historical development. We now want to turn to a more in-depth treatment about the formal basis of modern shower algorithms as the three implemented in \pythia.

We have seen above that parton showers build upon the factorization of (squared) amplitudes in soft and collinear limits. 
Technically, this means that whenever either two (or more) particles become collinear or one (or more) particle becomes soft, the full (squared) amplitude can be well approximated by the (squared) matrix element without the unresolved particle times a universal radiation function. It is the latter, which takes the effect of the soft or collinear radiation into account. 
This factorization is reminiscent of the perturbative physics of the hard process and occurs, because an intermediate, almost on-shell, propagator can be replaced by a polarization sum, such that the amplitude may be split into two independent pieces. Vital for the construction of showers is that this factorization is universal in the sense that it is process and multiplicity independent. This means that the same radiation functions can be used for different squared matrix elements and at any multiplicity, as long as only single-unresolved radiation is concerned. The latter comment serves to emphasize that at higher multiplicities also multiple-unresolved limits occur, in which, for instance, two particles become simultaneously soft or three particles become simultaneously collinear. For such configurations, it should be obvious that higher-order radiation functions are needed and the ones describing single-soft or (double-)collinear radiation are not sufficient.
\index{Strong ordering}At the same time, it is always possible to factorize phase-space integration measures into on-shell steps by introducing delta functions to factorize the decay system, and introducing recoiling systems to guarantee four-momentum conservation. Taking matrix-element and phase-space factorization together, it follows that cross sections can be factorized. This allows for iteration of the approximation, as long as the measure of ``softness'' or ``collinearity'' remains appropriate. In this context, the requirement of an appropriate measure leads to the notion of strong ordering, which means that radiation of soft particles is yet softer and radiation of collinear particles is yet more collinear. 
Although different possibilities to factorize matrix elements exist, all inherit that the approximation should recover the singularities of fixed-order results. On the one hand, DGLAP evolution is driven by collinear radiation; on the other hand, factorization-breaking (so-called non-global) logarithms are driven by soft radiation. These are the limits any parton shower resumming the leading, \ie largest, logarithms should recover.

\index{Evolution variable}\index{Shower evolution
  variable|see{Evolution variable}}\index{Shower ordering
  variable|see{Evolution variable}}Based on the above, we can start thinking about the construction of a shower model. It should be emphasized that the construction of showers is by no means unique. As the bare minimum, a shower algorithm must define the following.
\begin{enumerate}
	\item\label{enum:showers:MEfact} Radiation functions, \ie the matrix-element factorization.
	\item\label{enum:showers:PSfact} A phase-space factorization and recoil procedure.
	\item\label{enum:showers:OrderingVar} An ordering variable, \ie a measure of ``softness'' and/or ``collinearity''.
\end{enumerate}
For each of these points, different choices are possible and used, motivated by different desires to obtain certain objectives: simplicity, extendability, or simply to describe specific processes better at the cost of describing others worse. 

Using somewhat general language for now, we can denote the radiation functions by $K_{j/\tilde{i}\tilde{k}}$, describing the radiation of particle $j$ from the two parent particles $\tilde{i}$ and $\tilde{k}$, \ie the branching $\tilde{i} \tilde{k} \mapsto i j k$. Depending on the specifics of the shower algorithm, one of the two parents $\tilde{i}$ and $\tilde{k}$ may be distinguished as the ``emitter'' while the other, the ``recoiler'', only ensures four-momentum conservation, or both parents act as emitters and recoilers in an agnostic way. The former is how both \pyt's simple shower and \dire are structured, whereas the latter describes the antenna picture employed in the \vincia shower.
No matter which specific choice of radiation functions is made, the sum of terms must reproduce all single-unresolved limits of the full real-emission cross section,
\begin{equation}
	\deriv \sigma_{n+1} \xrightarrow{\text{single-unresolved}} \sum_j K_{j/\tilde{i}\tilde{k}} \, \deriv \Phi_{+1} \, \deriv \sigma_n =: K_{n\mapsto n+1} \, \deriv \Phi_{+1} \, \deriv \sigma_n  \, ,
\label{eq:showers:xSecFactorization}
\end{equation}
with the cross sections $\sigma$ defined as in \cref{eq:basics-sigma-nbody}. This factorization consists of two parts: the factorization of the squared matrix element and the factorization of the phase space.

\index{DGLAP}Specifically, in the case of two particles $i$ and $j$ becoming collinear, the $n+1$-particle matrix element factorizes into a product of the $n$-particle matrix element and the DGLAP splitting kernels \crefrange{eq:introshower:dglapqqg}{eq:introshower:dglapgamff}, 
\begin{equation}
	\abs{\ME_{n+1}}^2 \xrightarrow{i \parallel j}  \frac{8 \uppi \alpha}{2p_i\cdot p_j} \dglap{\tilde{i}}{ij}(z) \abs{\ME_{n}}^2 + \text{angular terms} \, .
\end{equation}
Generally, the collinear limit involves spin correlations between the factorized matrix element and the (spin-dependent) DGLAP kernels, here indicated by the additional ``angular terms''. These terms vanish upon azimuthal integration and are therefore not necessarily implemented in a parton-shower algorithm. It is, however, vital to account for these terms in so-called NLO subtraction schemes to ensure point-wise cancellation of singularities.
\index{Coherence}In the limit of a single gauge boson becoming soft, however, the emission of the soft boson can be described by a universal factor known as the soft eikonal. Different to the collinear limit, soft radiation is an intrinsically coherent phenomenon, meaning that the boson is emitted by the whole particle ensemble, introducing a sum over radiators:
\begin{equation}
	\abs{\ME_{n+1}}^2 \xrightarrow{E_j \to 0} 8 \uppi \alpha \sum\limits_{i<k} \colfac_{ik} \frac{2 p_i \cdot p_k}{(2p_i \cdot p_j)(2 p_j \cdot p_k)} \abs{\ME_{n}}^2 \, ,
\end{equation}
with charge factors $\colfac_{ik}$ depending on the charges of the radiators $i$ and $k$.
Especially in the case of QCD, these charge factors introduce intricate colour correlations for soft gluon emissions. It is because of these complications that most parton showers only consider the leading-colour limit, \ie neglect any contributions in the above sum that correspond to emissions from non-neighbouring partons.

Besides the factorization of matrix elements, in \cref{eq:showers:xSecFactorization} we used that the $n+1$-particle phase space exactly factorizes into a product of an $n$-particle phase space and the branching phase space $\deriv \Phi_{+1}$, obtained through a formal insertion of an intermediate off-shell particle with mass $m_{ij}^2 = (p_i+p_j)^2$, 
\begin{align}
	\deriv \Phi_{n+1}(q; p_1, \ldots, p_i, p_j, p_k, \ldots, p_{n+1}) &= \deriv \Phi_{n}(q; p_1, \ldots, p_{\tilde{i}}, p_{\tilde k}, \ldots, p_{n}) \nonumber \\
	&\qquad \times \abs{J(p_{\tilde{i}}, p_{\tilde{k}}; p_{ij}, p_k)} \, \frac{\deriv m_{ij}^2}{2 \pi} \, \deriv \Phi_2(p_{ij}; p_i, p_j)  \nonumber\\
	& \equiv \deriv \Phi_{n}(q;  p_1, \ldots, p_{\tilde{i}}, p_{\tilde{k}}, \ldots, p_{n}) \, \deriv \Phi_{+1}(p_i, p_j, p_k)   \, .
\end{align}
It must be emphasized that the $n$-particle phase-space measure is here written with on-shell momenta $p_{\tilde{i}}$ and $p_{\tilde{k}}$ instead of an off-shell intermediate momentum $p_{ij}$. This means we here assume an on-shell phase-space factorization, \ie that after each emission, all momenta are separately physical and momentum is conserved at each step in the shower,
\begin{equation}
	p_{\tilde{i}} + p_{\tilde{k}} = p_i + p_j + p_k \, .
\end{equation}
The change from the off-shell momenta $\{p_{ij}, p_k \}$ to the on-shell momenta $\{p_{\tilde{i}}, p_{\tilde{k}} \}$ is represented by the Jacobian $ \abs{J(p_{\tilde{i}}, p_{\tilde{k}}; p_{ij}, p_k)}$. 
Specific forms of kinematic mappings $\{p_{\tilde{i}}, p_{\tilde{k}} \} \mapsto$ $\{p_i, p_j, p_k \}$ (or ``recoil schemes'') are again shower specific.
Presently, however, all showers in \pythia employ an on-shell factorization as described here. While this might not generally be required, this is a key requirement for the matching and merging techniques utilized in \pythia, \cf\cref{section:matchmerge}.

The branching phase space $\deriv \Phi_{+1}$ accounts for the degrees of freedom entering through the emission of one particle from the $n$-particle configuration and can generally be expressed in terms of three ``shower variables'' $t$, $z$, and $\phi$,
\begin{equation}
	\deriv \Phi_{+1}(p_i, p_j, p_k) = \abs{J(t,z,\phi)} \, \deriv \Phi_{+1}(t, z, \phi) = \frac{1}{16 \pi^2} \abs{J(t,z,\phi)} \, \deriv t \, \deriv z \, \deriv \phi \, . \label{eq:showers:branchingPS}
\end{equation}
Usually, $t$ is interpreted as the ordering variable of the shower, $z$ as some kind of energy-sharing variable, and $\phi$ as the angle about the branching plane in the $i$-$j$-$k$ rest frame. However, different showers make different choices which may be more or less connected with this analogy.

Addressing point~\ref{enum:showers:OrderingVar} of the list above, it is instructive to start by noting that various choices of ordering variables are formally equivalent at the \ac{LL} level, as can be seen by comparing the differentials as they enter through the matrix-element and phase-space factorizations described above,
\begin{equation}
	\frac{\deriv t}{t} = \frac{\deriv p_{\perp,j}^2}{p_{\perp,j}^2} = \frac{\deriv m_{ij}^2}{m_{ij}^2} = \frac{\deriv \theta_{ij}^2}{\theta_{ij}^2} \, ,
\end{equation}
and by noting that in the collinear limit $p_{\perp,j}^2 \sim z (1-z) m_{ij}^2 \sim z^2(1-z)^2 E_j^2 \theta_{ij}^2$. 
It is straightforward to see that all these choices represent a certain measure of softness or collinearity, as required above. The requirement that this measure remains appropriate during the shower evolution then translates into strong ordering of emissions, \ie subsequent emissions evolve down in the ordering scale: $t_0 > t_1 > t_2 > \ldots t_n$.

Putting the above together, a no-branching
probability\index{No-branching probability|see{Sudakov factor}}, often
also called Sudakov factor\index{Sudakov factor}, can be defined:
\begin{equation}
	\Pi_n(t_n, t_{n+1}; \Phi_n) = \exp\left\{- \int\limits_{t_{n+1}}^{t_n} K_{n\mapsto n+1}(\Phi_{n},\Phi_{+1}(t', z', \phi')) \, \deriv \Phi_{+1}(t', z', \phi') \right\} \, .
\end{equation}
It describes the evolution from an $n$-particle state at scale $t_n$ to an $n+1$-particle state at scale $t_{n+1} < t_n$. By rewriting $K_{n\mapsto n+1}$ as the sum of radiation functions $K_{j/\tilde{i}\tilde{k}}$, $\Pi_{n}$ can written as the product of $\tilde{i}\tilde{k}\mapsto ijk$ no-branching probabilities:
\begin{align}
	\Pi_n(t_n, t_{n+1}; \Phi_n) & = \exp\left\{- \sum_j \int\limits_{t_{n+1}}^{t_n} \int\limits_{z_{\mrm{min}}}^{z_{\mrm{max}}} \int\limits_{0}^{2\pi} \frac{1}{16 \pi^2} \, K_{j/\tilde{i}\tilde{k}}(t', z', \phi') \, \abs{J(t',z',\phi')} \, \frac{\deriv \phi'}{2\pi} \, \deriv z'  \, \deriv t' \right\} \nonumber\\
	&= \prod_j \exp\left\{- \int\limits_{t_{n+1}}^{t_n} \int\limits_{z_{\mrm{min}}}^{z_{\mrm{max}}} \int\limits_{0}^{2\pi} \frac{1}{16 \pi^2} \, K_{j/\tilde{i}\tilde{k}}(t', z', \phi') \, \abs{J(t',z',\phi')} \, \frac{\deriv \phi'}{2\pi} \, \deriv z'  \, \deriv t' \right\} \label{eq:showers:noBranchingProbSingleEmission} \\
	&= \prod_j \, \Pi_{j/\tilde{i}\tilde{k}}(t_n, t_{n+1}; \Phi_n) \, . \nonumber
\end{align}
Written this way, it is also emphasized that each branching $\tilde{i}\tilde{k} \mapsto ijk$ comes with its own branching phase space and kinematic mapping. This is how the full no-branching probability $\Pi_{n\mapsto n+1}$ is implemented in shower algorithms in practice.

For the calculation of the expected value of an observable $O$, the no-branching probabilities enter to describe the shower evolution as a Markov chain,
\begin{equation}
	\avg{O}^\mrm{PS}_n = \int \frac{\deriv \sigma_n}{\deriv \Phi_n} \, \SC_n(t, O) \, \deriv \Phi_n \label{eq:showers:ExpValPS}
\end{equation}
which is generated by a ``shower operator'' $\SC_n(t,O)$, defined recursively as
\begin{equation}
	\SC_n(t, O) :=  \Pi_n(t,t_\mrm{c}; \Phi_n)O(\Phi_n) + \int\limits^t_{t_\mrm{c}} K_{n\mapsto  n+1} \, \Pi_n(t, t'; \Phi_n) \, \SC_{n+1}(t',O) \, \deriv \Phi_{+1}(t', z', \phi') \, . \label{eq:showers:showerOp}
\end{equation}
This shower operator makes the unitarity of the shower explicit. The first term implicitly accounts for all unresolved radiation and virtual corrections between the shower starting scale $t$ and the shower cutoff $t_\mrm{c}$\index{Cutoff scales}. The second term, on the other hand, describes the emission of a single particle, approximated by the sum of radiation functions $K_{n\mapsto n+1}$, and includes all unresolved and virtual corrections between the shower starting and cutoff scale.

\index{Backwards evolution}It is instructive to make the form of the no-branching probability \cref{eq:showers:noBranchingProbSingleEmission} more explicit for QCD showers. Implicitly, the radiation functions $K_{j/\tilde{i}\tilde{k}}$ above contain the strong-coupling constant, a colour factor, and, for ISR, a ratio of PDFs, 
\begin{equation}
	K_{j/\tilde{i}\tilde{k}}(t,z,\phi) = \gstrong^2(t)\, \RPDF(t,z) \, \colfac_{j/ \tilde{i}\tilde{k}}\,\bar{K}_{j/\tilde{i}\tilde{k}}(t,z,\phi)= 4\pi \alphas(t)\, \RPDF(t,z) \, \colfac_{j/ \tilde{i}\tilde{k}}\, \bar{K}_{j/\tilde{i}\tilde{k}}(t,z,\phi) \, ,
\end{equation}
where we have introduced the coupling-, PDF-, and colour-factor-stripped radiation function $\bar{K}_{j/\tilde{i}\tilde{k}}$, which depends solely on the branching kinematics. For FSR, the PDF ratio is equal to unity, $\RPDF = 1$, as the initial-state momenta do not change due to the branching. Differentially in the evolution variable $t$, the integral in the exponent of $\Pi_{j/\tilde{i}\tilde{k}}$ can thus be written as
\begin{align}
	\frac{\deriv \prob^\mrm{FSR}_{j/\tilde{i}\tilde{k}}(t)}{\deriv t} &= \frac{\alphas(t)}{2\pi} \frac{\colfac_{j/ \tilde{i}\tilde{k}}}{2} \int\limits_{z_{\mrm{min}}}^{z_{\mrm{max}}} \int\limits_{0}^{2\pi} \bar{K}_{j/\tilde{i}\tilde{k}}(t, z', \phi') \, \abs{J(t,z',\phi')} \, \frac{\deriv \phi'}{2\pi} \, \deriv z' \, , \label{eq:showers:branchingProbFSR} \\
	\frac{\deriv \prob^\mrm{ISR}_{j/\tilde{i}\tilde{k}}(t)}{\deriv t} &= \frac{\alphas(t)}{2\pi} \frac{\colfac_{j/ \tilde{i}\tilde{k}}}{2} \int\limits_{z_{\mrm{min}}}^{z_{\mrm{max}}} \int\limits_{0}^{2\pi} \RPDF(t,z) \, \bar{K}_{j/\tilde{i}\tilde{k}}(t, z', \phi') \, \abs{J(t,z',\phi')} \, \frac{\deriv \phi'}{2\pi} \, \deriv z' \label{eq:showers:branchingProbISR} \, ,
\end{align}
for FSR and ISR, respectively. Written this way, the connection to \cref{eq:introshower:forwardsevol,eq:introshower:backwardsevol}, respectively, is immediately evident.
It is worthwhile to point out here that typically different shower algorithms are inconsistent as to whether colour factors are included or excluded in radiation functions. Moreover, depending on whether a shower aims at describing the evolution of a single initial-state leg at a time or both at the same time, the PDF ratios $\RPDF$ have to include one PDF ratio,
\begin{equation}
	\RPDF(t, z) = \frac{x_i \fpdf_i(x_i, t)}{x_{\tilde{ij}} \fpdf_{\tilde{i}}(x_{\tilde{i}}, t)} \, , 
\end{equation}
or two PDF ratios if both initial-state particles are evolved at the same time, 
\begin{equation}
	\RPDF(t, z) = \frac{x_i \fpdf_i(x_i, t)}{x_{\tilde{i}} \fpdf_{\tilde{i}}(x_{\tilde{i}}, t)} \frac{x_k \fpdf_k(x_k, t)}{x_{\tilde{k}} \fpdf_{\tilde{k}}(x_{\tilde{k}}, t)} \, .
\end{equation}
The $x$-fractions $p_i = x_i P$, with $P$ the incoming hadron momentum, depend on the shower variables $t$ and  $z$. 

A similar analysis can be done in the cases of QED or EW showers, where the QCD coupling has to be replaced by the electromagnetic/electroweak coupling and QCD colour factors by the appropriate QED/EW charges.

\paragraph{Formal accuracy} Despite their success in describing wide classes of observables with often impressive agreement with experimental data, parton showers commonly work with a number of approximations. It is not an easy task to formally assess the accuracy of a given shower model, \ie to determine which exact terms of a perturbative series a shower includes. For a start, there are three expansions to be considered:
\begin{enumerate}
	\item the perturbative expansion in the coupling constant $\alpha^n(t)$, determining the accuracy of the hard process, \eg leading-order (LO), next-to-leading order (NLO), \etc;
	\item the perturbative expansion in large logarithms $\alpha^n(t) \log^m(t_\mrm{hard}/t)$, determining the accuracy of the resummation, \eg Leading Logarithmic (LL), \ac{NLL}, \etc;
	\item and for QCD showers, the expansion in the number of colours ($\Nc$), determining the accuracy of the colour factors in the resummation, \eg \ac{LC}, \ac{NLC}, \etc
\end{enumerate}
A baseline shower would for example start from a LO matrix element and (typically) generate the LL corrections arising from additional radiation under the LC assumption of planar colour flows. Such a shower could be assigned a LO+LL+LC accuracy. This can be expected from virtually all common shower models, although observables may exist for which a given shower does not correctly include the LL terms. It is more interesting, however, to determine if and for which observables showers reach sub-leading, \ie higher, accuracy than the LO+LL+LC minimum. 

Increasing the accuracy on the fixed-order side can be addressed by matching and merging methods, which are described in detail in \cref{section:matchmerge}. Matching and merging at LO and NLO have de-facto become state of the art for all showers and processes.

Assessing and increasing the logarithmic accuracy of showers has become a highly-active field, where no general solution has yet been developed. 
Different approaches to assess the logarithmic accuracy of showers have been developed in the recent past, such as ones based on comparison of analytic and numerical resummation~\cite{Hoeche:2017jsi,Baberuxki:2019ifp}, analytic examination of the logarithmic structure of showers~\cite{Dasgupta:2018nvj,Hamilton:2020rcu}, or numerical checks of logarithmic terms~\cite{Nagy:2020rmk,Nagy:2020dvz}. Moreover, for simple processes such as $\epem$ annihilation to jets, first shower models have been developed that can be shown to give NLL accuracy for a wider range of observables~\cite{Dasgupta:2020fwr,Forshaw:2020wrq}. Most common shower models currently only obtain a formal LL accuracy, with varying, observable-dependent subleading accuracy. 

\index{MECs}Lastly, the inclusion of sub-leading colour corrections in parton showers is an active field as well, with approaches based on matrix-element corrections~\cite{Giele:2011cb,Platzer:2012np,Platzer:2018pmd,Bellm:2018wwz,Hamilton:2020rcu,Holguin:2020joq}, sampling of colours~\cite{Isaacson:2018zdi,Hoeche:2020nsx}, quantum-probability density-matrix arguments~\cite{Nagy:2012bt,Nagy:2015hwa}, or amplitude-level evolution~\cite{Forshaw:2019ver,DeAngelis:2020rvq}. Sub-leading colour corrections are not in general universally applied in parton showers.

\index{Parton showers}
\paragraph{Showers in \pythia} There are three different shower
modules available in \pythia: the original/default simple shower, the
\vincia antenna shower, and \dire. These will be discussed in detail below in
\cref{sec:SimpleShower}, \cref{sec:Vincia}, and \cref{sec:Dire}, respectively.

\subsection{The simple shower}\label{sec:SimpleShower}
\index{Parton showers!Simple showers}
\index{Simple showers}
The ``simple shower'' is the oldest parton-shower
algorithm in \pyt~8 and is also the default shower model in \pythia.
It has its origin in the mass-ordered showers in \jetset/\pyt~\cite{%
Sjostrand:1985xi,Bengtsson:1986hr,Bengtsson:1986et,Norrbin:2000uu},
with the transition to $p_\perp$ ordering~\cite{Sjostrand:2004ef}
partly influenced by the Lund dipole picture~\cite{Gustafson:1987rq}
and partly by the desire to combine the ISR and FSR shower evolution
with MPI in a single interleaved
sequence~\cite{Sjostrand:2004ef}.

Over the years, significant revisions and extensions have been introduced,
many of them only available in recent \pyt versions. This includes:
\begin{itemize}
\item Full interleaving of ISR, FSR, and MPI~\cite{Corke:2010yf}.
\item Options for a dipole-style treatment of initial-final colour
flows~\cite{Cabouat:2017rzi}.
\item $\f \to \f \gamma$ and $\gamma \to \f\fbar$ splittings
(where $\f$ represents charged fermions).
\item Matrix element corrections for resonance decays and a few other processes~\cite{Bengtsson:1986et,Miu:1998ju,Norrbin:2000uu}.
\item Extensive facilities for matching and merging (\cf\cref{section:matchmerge}).
\item Reweighted shower branchings and uncertainty bands~\cite{Mrenna:2016sih}.
\item A flexible treatment of showers in baryon-number-violating processes~\cite{Desai:2011su}.
\item Weak showers~\cite{Christiansen:2014kba}.
\item Hidden-sector showers~\cite{Carloni:2010tw,Carloni:2011kk}.
\end{itemize}

The name ``simple'' shower here refers to the limited aim of a
consistent leading-logarithmic (and beyond) shower evolution,
with several known shortcomings~\cite{Dasgupta:2018nvj,Hamilton:2020rcu},
as opposed to the more sophisticated goals of the alternative
\vincia (\cf\cref{sec:Vincia}) and \dire (\cf\cref{sec:Dire})
shower options, also available in \pythia. It should be
emphasized that, by virtue of its longer history, many features are only
available in the simple shower, and that as such the naming might be
slightly misleading. As an example, the simple shower offers a much larger
selection of matching and merging methods than does \vincia or \dire.

The shower machinery consists of one algorithm for FSR and one for ISR.
These two are evolved together into one combined sequence of decreasing
$\pT$ scales. As an example, consider a partonic process
$a + b \to c + d$, where $a$ and $b$ are extracted from the beams
$A$ and $B$. It is then possible for $c$ and $d$ to undergo FSR
branchings, and for $a$ and $b$ backwards-evolution ISR ones.
Starting from some maximal scale $p_{\perp\mathrm{max}}$, downwards
evolution gives a possible branching $\pT$ scale for each of the four
partons. The one with largest $\pT$ is the winner that undergoes
a branching, leading to a new state of five partons. The selected $\pT$
value is taken as the new starting point for all five partons to evolve
further down in $\pT$, giving a new branching. This is applied iteratively
until some lower cutoff is reached and the evolution is stopped. Also, MPIs
will form part of this evolution, see \cref{subsubsec:soft:interleave}.

\subsubsection{Basic shower branchings}

The description of showers in the introduction of this section is
valid for the simple shower framework. Notably the branching
probabilities $\d \mathcal{P}_a$ of \cref{eq:introshower:forwardsevol}
and $\d \mathcal{P}_b$ of \cref{eq:introshower:backwardsevol} play a
central part, but with two key additions. 

One is that evolution is performed in terms of transverse momenta,
\ie the generic $Q^2$ scale in \cref{eq:introshower:forwardsevol}
and \cref{eq:introshower:backwardsevol} is replaced by a
$p_{\perp\mathrm{evol}}^2$. The use of transverse momentum as an evolution
variable has been shown to catch key coherence features and therefore
is a preferred choice~\cite{Gustafson:1986db,Gustafson:1987rq}. 

The other is that a dipole picture is being used, although with some
exceptions. In it each coloured parton has a unique anticolour partner,
and together the two form a dipole. Radiation is  split into one
contribution from each dipole end. When one end radiates, the other end
has to take a recoil such that total energy and momentum is preserved. 

\paragraph{Shower evolution}
\index{ISR!Kinematics}
\index{FSR!Kinematics}
\index{Kinematics!in shower branchings}
To understand basic kinematics in a branching $a \to b c$,
expressions become especially simple using light-cone (LC)
$p^{\pm} = E \pm p_z$, for which $p^+ p^- = m_{\perp}^2 = m^2 + \pTs$.
When $a$ moves along the $+z$ axis, with $p_b^+ = z_{\mathrm{LC}} p_a^+$
and $p_c^+ = (1-z_{\mathrm{LC}}) p_a^+$, $p^-$ conservation then gives
\begin{equation}
m_a^2 = \frac{m_b^2 + \pTs}{z_{\mathrm{LC}}} + \frac{m_c^2
+ \pTs}{1-z_{\mathrm{LC}}} ~,
\end{equation}
or equivalently
\begin{equation}
\pTs = z_{\mathrm{LC}} (1-z_{\mathrm{LC}}) m_a^2 - (1-z_{\mathrm{LC}}) m_b^2
- z m_c^2 = p_{\perp\mathrm{LC}}^2~.
\label{eq:simpleshower:pTlightcone}
\end{equation}
For a timelike branching $Q^2 = m_a^2$ and $m_b = m_c = 0$, assuming 
massless partons, so then
$p_{\perp\mathrm{LC}}^2 = z_{\mathrm{LC}}(1-z_{\mathrm{LC}}) Q^2$. For a
spacelike branching $Q^2 = - m_b^2$ and $m_a = m_c = 0$, where $b$ is
the parton that will enter the hard interaction, so instead
$p_{\perp\mathrm{LC}}^2\ = (1-z_{\mathrm{LC}}) Q^2$. We are inspired by
these relations to define abstract evolution variables
\index{Evolution variable!in Simple showers}
\begin{align}
p_{\perp\mathrm{evol}}^2 &=       z (1-z)Q^2 \quad \mathrm{for~FSR}~,
\label{eq:simpleshower:pTevolFSR}\\
p_{\perp\mathrm{evol}}^2 &=  \,\,\, (1-z)Q^2 \quad \mathrm{for~ISR}~,
\label{eq:simpleshower:pTevolISR}
\end{align}
in which to order the sequence of shower emissions. The $z_{\mathrm{LC}}$
definitions will be replaced by invariant-mass-based $z$ for the
final kinematics definitions, for better Lorentz invariance properties,
and as a consequence $p_{\perp\mathrm{evol}} \neq p_{\perp\mathrm{LC}}$. Further details on this are given later.

The evolution is now carried out, downwards
in $p_{\perp\mathrm{evol}}^2$ 
from some starting scale $p_{\perp\mathrm{evol,max}}^2$,
for FSR by parton $a$ branching to $b + c$, for ISR by parton $b$
being reconstructed as coming from the branching of an earlier $a$.
The branching probabilities of \cref{eq:introshower:forwardsevol}
and \cref{eq:introshower:backwardsevol}, with the addition of no-branching
probabilities $\Pi$, \cref{eq:introshower:sudakov}, gives
\begin{align}
\d\mathcal{P}_{\mathrm{FSR}} &=
\Pi_a(p_{\perp\mathrm{evol,max}}^2, p_{\perp\mathrm{evol}}^2) \,
\d\mathcal{P}_a(p_{\perp\mathrm{evol}}^2) ~, 
\label{eq:simpleshower:forwardsevol} \\
\d\mathcal{P}_{\mathrm{ISR}} &=
\Pi_b(x, p_{\perp\mathrm{evol,max}}^2, p_{\perp\mathrm{evol}}^2) \,
\d\mathcal{P}_b(x, p_{\perp\mathrm{evol}}^2) ~.
\label{eq:simpleshower:backwardsevol}
\end{align}
A $p_{\perp\mathrm{evol}}^2$ scale is selected for each existing dipole end,
and the end with the largest value is chosen to branch. 

The selection of a branching means that $p_{\perp\mathrm{evol}}^2$ and $z$
are fixed. From these, one can derive the virtuality of the evolving parton
\begin{align}
m_a^2 &= Q^2 = \frac{p_{\perp\mathrm{evol}}^2}{z (1-z)} \quad \mathrm{for~FSR}~,\\
-m_b^2 &= Q^2 = \frac{p_{\perp\mathrm{evol}}^2}{(1-z)Q^2} \quad \mathrm{for~ISR}~.
\end{align}
What now remains is to construct the kinematics of the branching. 
This works rather differently for FSR and for ISR, so the two cases are
presented separately.

\paragraph{FSR branching kinematics}
\index{Quark masses!in Simple showers}
Study the radiation inside a dipole, consisting of a radiator $a$ and
a recoiler $r$, in the dipole rest frame, with $a$ moving in the $+z$
direction, and with $m_{ar}^2 = (p_a +p_r)^2$.

For massless partons, the introduction of an off-shell $Q^2 = m_a^2$
increases $E_a$ from $m_{ar}/2$ to $( m_{ar}^2 + Q^2)/2 m_{ar}$, with $E_r$
reduced by the same amount, or in terms of four-momenta
\begin{equation}
p_{a'} =  p_a + \frac{Q^2}{m_{ar}^2} p_r ~, \qquad
p_{r'} = \left( 1 - \frac{Q^2}{m_{ar}^2} \right) p_r ~.
\label{eq:simpleshower:erfsrshift} 
\end{equation}
The two daughters share the energy according to $E_b = z E_a$ and
$E_c = (1-z) E_a$. With the modified $a$ still along the $+z$ axis,
the transverse momentum of the two daughters then becomes
\begin{equation}
p_{\perp b,c}^2 = \frac{z(1-z)(m_{ar}^2 + Q^2)^2 - m_{ar}^2 Q^2}%
{(m_{ar}^2 - Q^2)^2} \, Q^2 ~\leq~ z(1-z) Q^2 = p_{\perp\mathrm{evol}}^2~.
\end{equation}
The kinematics can now be completed, including a random $\varphi$
orientation of the $\pT$. Also, if the original dipole had to be boosted
and rotated to its rest frame, the new system should be transformed
back to the original frame. 

Colours are also assigned in the branching, such that the new
colour-dipole picture is set up. This is well defined in the
$\Nc \to \infty$ limit, except for $\g \to \g \g$ branchings.
Here a rewriting~\cite{Gustafson:1987rq},
\begin{equation}
P_{\g\to\g\g}(z) = 3 \, \frac{\big(1-z(1-z)\big)^2}{z(1-z)}
= \frac{3}{2} \, \frac{1+z^3}{1-z} + \frac{3}{2} \, \frac{1+(1-z)^3}{z}
\simeq 3 \, \frac{1+z^3}{1-z} ~,
\label{eq:simpleshower:dglapgggmod}
\end{equation}
allows the gluon that takes the (usually smaller) $1 - z$ fraction
to be the ``radiated'' gluon that connects the ``radiator'' gluon to the
recoiler.

Of note is that the light-cone sharing of momenta between daughters,
suggested initially, here is replaced by an energy sharing. It has the
advantage that $p_{\perp\mathrm{evol}}^2$ and this $z$ together exactly
match on to the singularity structure of matrix elements, such
as the textbook \mbox{$\gamma^*/\Z \to$ $\q(1) + \qbar(2) + \g(3)$} one,
when $\q\to\q\g$ and $\qbar\to\qbar\g$ radiation from the two dipole
ends is combined
\begin{equation}
\frac{\d p_{\perp\mathrm{evol},\q}^2}{p_{\perp\mathrm{evol},\q}^2} \, 
\frac{\d z_{\q}}{1 - z_{\q}} 
+ \frac{\d p_{\perp\mathrm{evol},\qbar}^2}{p_{\perp\mathrm{evol},\qbar}^2} \,
\frac{\d z_{\qbar}}{1 - z_{\qbar}}
= \frac{\d x_1 \, \d x_2}{(1 - x_2) x_3} 
+ \frac{\d x_1 \, \d x_2}{(1 - x_1) x_3} 
 = \frac{\d x_1 \, \d x_2}{(1 - x_1) (1 - x_2)} ~,
\label{eq:simpleshower:splitFSR}
\end{equation} 
with $x_i = 2 E_i/E_{\mathrm{tot}}$. Corrections to fully reproduce
several important matrix elements therefore are easily implemented.

Incidentally, note that
$1 - x_2 \propto \cos\theta_{\q\g}$ and
$1 - x_1 \propto \cos\theta_{\qbar\g}$, so \cref{eq:simpleshower:splitFSR}
provides a prescription for how radiation from the full dipole
smoothly can be split into radiation from the two ends as a function of
the gluon emission angle. This split also decides which of the two
original partons is the recoiler, the one that keeps its direction of
motion.

\index{Quark masses!in Simple showers}
The kinematics need to be modified when quark masses are included, with full
expressions in~\oneref{Norrbin:2000uu}. There are two key points, however.
First, if the branching parton $a$ has an on-shell mass $m_a$ and
off-shell mass $m_{a'}$, then \cref{eq:simpleshower:pTevolFSR} needs
to be modified to
\begin{equation}
p_{\perp\mathrm{evol}}^2 = z(1-z)Q^2 = z(1-z)(m_{a'}^2 - m_a^2) ~,
\label{eq:simpleshower:massFSR}
\end{equation}
to reproduce the singularities in matrix elements. Second, if the
daughters are initially assigned four-momenta $p_b^{(0)}$ and $p_c^{(0)}$
as if they were massless, then massive four-vectors can be constructed
as
\begin{align}
p_b &=  (1 - k_b) p_b^{(0)} + k_c p_c^{(0)} ~,\\
p_c &= (1 - k_c) p_c^{(0)} + k_b p_b^{(0)} ~,\\
k_{b,c} &= \frac{ m_a^2 - \sqrt{ (m_a^2 - m_b^2 - m_c^2)^2 - 4 m_b^2 m_c^2 }
  \pm (m_c^2 - m_b^2)}{2 m_a^2} ~.
\end{align}
The $\p_{\perp b,c}$ is also reduced in the process, by a factor
$1 - k_b - k_c$. 

\paragraph{ISR branching kinematics}
The handling of ISR branching kinematics is somewhat more complicated.
At any resolution scale $p_{\perp\mathrm{evol}}^2$ the ISR algorithm will
identify two initial partons, one from each incoming hadron, that are
the mothers of the respective incoming cascade to the hard interaction.
These partons should be set massless and collinear with the beams. 
When the resolution scale is reduced, using backwards evolution, either
of these two partons may turn out to be the daughter $b$ of a previous
branching $a \to b c$. The parton $r$ on the other side of the event
takes on the role of recoiler, needed for consistent reconstruction of
the kinematics when the parton $b$ previously considered massless now is
assigned a spacelike virtuality $m_b^2 = - Q^2$. This redefinition should
be performed in such a way that the invariant mass of the $b+r$ system is
unchanged, since this mass corresponds to the set of produced particles,
which in a case like $\g\g \to \H$ must not be modified. The system will
have to be rotated and boosted as a whole, however, to take into account
that $b$ not only acquires a virtuality but also a transverse momentum;
if previously $b$ was assumed to move along the event axis, now it is $a$
that should do so.
 
At any step of the cascade, the massless mothers suitably should have
four-momenta given by $p_i = x_i \, (\sqrt{s}/2) \, (1; 0, 0, \pm 1)$
in the rest frame of the two incoming beam particles, so that
$\hat{s} = x_1 x_2 s$. If this relation is to be preserved in the
$a \to b c$ branching, the $z = x_b/x_a$ should fulfil
$z = m_{br}^2/m_{ar}^2 = (p_b + p_r)^2/(p_a + p_r)^2$. 
This gives an explicit construction of the kinematics in the $a + r$ rest
frame, assuming $a$ is moving along the $+z$ axis and $c$ is massless:
\begin{align}
p_{a,r} &=  \frac{m_{ar}}{2} \left( 1 ; 0, 0, \pm 1 \right) ~, \\
p_b &=  \left( \frac{m_{ar}}{2} \, z ;
p_{\perp b,c} \cos\varphi, p_{\perp b,c} \sin\varphi, \frac{m_{ar}}{2}
\left(z + \frac{2Q^2}{m_{ar}^2} \right) \right) ~,\\
p_c &= \left( \frac{m_{ar}}{2} \, (1- z) ;
-p_{\perp b,c} \cos\varphi, -p_{\perp b,c} \sin\varphi,
\frac{m_{ar}}{2} \left(1 - z - \frac{2Q^2}{m_{ar}^2} \right) \right) ~,\\
p_{\perp b,c}^2 &= (1-z)Q^2 - \frac{Q^4}{m_{ar}^2}
< (1-z)Q^2 = p_{\perp\mathrm{evol}}^2 ~.
\label{eq:simpleshower:pTISRnoM}
\end{align}
 
For small $Q^2$ values the $p_{\perp b,c}^2$ and $p_{\perp\mathrm{evol}}^2$
measures agree well, but with increasing $Q^2$ the $p_{\perp b,c}^2$
will eventually turn over and decrease again (for fixed $z$ and $m_{ar}$).
Simple inspection shows that the maximum $p_{\perp b,c}^2$ occurs for
$p_{z c} = 0$ and that the decreasing $p_{\perp b,c}^2$ corresponds to
increasingly negative $p_{z c}$. The drop of $p_{\perp b,c}^2$ thus is
deceptive. Like for the FSR algorithm, $p_{\perp\mathrm{evol}}^2$ therefore
makes more sense than $p_{\perp b,c}^2$ as an evolution variable, despite
it not always having as simple a kinematic interpretation. One should note,
however, that emissions with negative $p_{z c}$ are more likely to come
from radiation off the other incoming parton, where it is collinearly
enhanced, so in practice the region of decreasing $p_{\perp b,c}^2$ is not
so important.

Quark-mass effects are less crucial for ISR: nothing heavier than charm
and bottom need be considered as beam constituents, unlike the multitude
of new massive particles one could imagine for FSR. Kinematics have to be
modified slightly if the outgoing parton $c$ is not massless, \eg in a
$\g \to \q \qbar$ branching. The main effect is a modified evolution $\pT$,
with \cref{eq:simpleshower:pTevolISR} replaced by
\begin{equation}
p_{\perp\mathrm{evol}}^2 = (1-z) (Q^2 + m_c^2) ~,
\end{equation}
and a reduced $\pT$ in the branching, replacing
\cref{eq:simpleshower:pTISRnoM} by
\begin{equation}
p_{\perp b,c}^2 = (1-z)Q^2 - \frac{Q^4}{m_{ar}^2} -
m_c^2 \left( z + \frac{Q^2}{m_{ar}^2} \right) =
Q^2 - z \, \frac{(Q^2 + m_c^2) (m_{br}^2 + Q^2)}{m_{br}^2} ~.
\end{equation}

\index{Backwards evolution!Heavy-quark thresholds}
\index{ISR!Heavy-quark thresholds}Charm and bottom
quarks raise another issue, namely what to do in the 
threshold region, \ie around the $Q_{\mathrm{thr}}^2$ scale where
$\g \to \c\cbar$ or $\g \to \b\bbar$ branchings are turned on in the
PDF evolution. Normally, it is assumed that these quark PDFs vanish below
$Q_{\mathrm{thr}}^2$ and then evolve above it as a massless quark would.
Initially, thus $f_{\q}(x, Q^2) \propto \ln(Q^2 / Q_{\mathrm{thr}}^2)$.
In backwards evolution of a $\c/\b$ quark, this leads to a diverging
$\d \mathcal{P}_b$ in \cref{eq:introshower:backwardsevol} for
$Q^2 \to Q_{\mathrm{thr}}^2$, and a vanishing no-branching probability.
While such a behaviour is possible to handle by evolving with gradually
smaller $Q^2$ steps as the threshold is approached, the chosen solution
is instead to rely on the known forwards-evolution PDF shape. Therefore,
once $p_{\perp\mathrm{evol}}^2 < f \, m_{\q}^2$, with $f$ a parameter of the
order of 2, a $p_{\perp\mathrm{evol}}^2$ is chosen logarithmically evenly
between $m_{\q}^2$ and $f \, m_{\q}^2$, and a $z$ flat in the allowed range.
Acceptance is based on the product of three factors, representing the
running of $\alphas$, the splitting kernel (including the mass term)
and the gluon density weight. At failure, a new $p_{\perp\mathrm{evol}}^2$
is chosen in the same range, \ie is not required to be lower since no
no-branching probability is involved. 

As for FSR, the choices of $p_{\perp\mathrm{evol}}^2$ and $z$ offers a
possibility to match onto the singularity structure of common matrix
elements, and thereby easily correct to matrix-element expressions.
Consider \eg $\q\qbar' \to \g\Wpm$~\cite{Miu:1998ju}. The
$\q\to \q\g$ branching gives a denominator $\hat{t}(\hat{t} + \hat{u})$
and $\qbar'\to \qbar'\g$ a denominator $\hat{u}(\hat{t} + \hat{u})$,
which combine to $\hat{t} \hat{u}$, in agreement with the matrix element.
This also illustrates how the full ISR radiation pattern can be subdivided
into contributions from the two sides.

\index{ISR!Rapidity ordering for Simple showers}
One special option in the ISR implementation, on by default, is the
possibility to order the emissions in rapidity, or equivalently in angle,
\ie to veto any trial emission that leads to unordered emitted partons~\cite{Corke:2010yf}. The backwards evolution is one towards smaller
$\pT$ and larger $x$ values, so angular ordering is already implicit
to first approximation, but the unordered emissions have a non-negligible
impact that appears to be detrimental for some distributions. There are
good arguments for a rapidity ordering to be a legitimate choice~\cite{Andersson:1995ju}, to provide a consistent separation between
ISR and FSR. But that was for a somewhat different algorithm, so this
option should more be seen as one possible variation beyond the basic
LL accuracy of the shower.

\paragraph{Strong coupling}
\index{alphaS@$\alphas$!in Simple showers}
By default a first-order running $\alphas(\p_{\perp\mathrm{evol}}^2)$ is
used, but alternatives are a fixed value or second-order running.  
Tuned $\alphas(m_{\Z}^2)$ values typically tend to come out somewhat
above the PDG $\overline{\mathrm{MS}}$ one~\cite{Zyla:2020zbs}. This
can be understood as absent higher-order effects, in splitting kernels
and shower kinematics, being absorbed into effective values.
Since these higher-order corrections differ between ISR and FSR, the
$\alphas(m_{\Z}^2)$ are also set separately for the two.

Furthermore, in the soft-gluon limit, it can be shown that the dominant
$\mathcal{O}(\alphas^2)$ splitting-function term, which generates
contributions starting from $\mathcal{O}(\alphas^2 \, \ln^2)$ at the
integrated level, can be absorbed into the LO splitting functions by
translating to the so-called \ac{CMW} (also known as MC) scheme~\cite{Catani:1990rr}.
This means that an $\overline{\mathrm{MS}}$ $\alphas(m_{\Z}^2) = 0.1185$
would translate into an MC $\alphas(m_{\Z}^2) = 0.126$. This goes some
of the way towards explaining the \pyt default $\alphas(m_{\Z}^2) = 0.1365$.
It is possible to switch on the usage of the CMW rescaling procedure 
to allow a lower input $\alphas(m_{\Z}^2)$, but physics is only mildly
modified by this. 

Another consequence of staying at leading order is that usage of LO
parton distributions is vastly to be preferred. If not, the description
of ISR branchings at low scales becomes quite unreliable, for physical
and technical reasons. The former are covered elsewhere, the latter 
are reflected in the need to have positive PDFs in 
\cref{eq:introshower:backwardsevol}, which is not guaranteed at NLO.

\paragraph{Shower cutoff}\index{Cutoff scales!in Simple showers}
\index{Primordial kT@Primordial $k_\perp$!Interplay with showers}
A lower cutoff scale $p_{\perp\mathrm{min}}$ is needed both
for ISR and FSR, but the two need not be same. The FSR one is related
to the transition from partons to hadrons, and LEP experience gives
us some understanding that too high a value does affect event shapes
detrimentally. The ISR case is less clear cut. Experimental signals,
such as the $\pT$ spectrum of $\Z$ bosons in $\pp/\ppbar$ collisions,
are affected by the non-trivial interplay with primordial $k_{\perp}$,
\cf\cref{sec:primordialKT}. 
A lower $p_{\perp\mathrm{min}}$ means more $\pT$ kicks to the $\Z$, 
but a shower initiator with a larger $x$, which means more dilution
of its $k_{\perp}$ in the cascade. One reasonable strategy therefore
is to assume the ISR is damped in the same way as MPIs are, \ie
the $\d \pTs / \pTs$ divergence is replaced by a
$\d \pTs / (p_{\perp 0}^2 + \pTs)$ one. Alternatively, it is also
possible to use a sharp cutoff.

\paragraph{Interleaving}\index{Interleaved evolution}
Multiparton interactions and ISR are in direct competition for
the beam-remnant momentum. Therefore, a combined downwards evolution
in $\pT$ of the two gives precedence to the harder parts of the
event activities. There is no corresponding competition requirement
for FSR to be interleaved, and FSR can also be viewed as occurring
after the other two components in time. Interleaving is allowed,
however, since it can be argued that a high-$\pT$ FSR occurs on
shorter time scales than a low-$\pT$ MPI, say. Backwards evolution
of ISR is also an example that physical time is not the only possible
ordering principle. Rather, one can work with conditional probabilities:
given the partonic picture at a specific $\pT$ resolution scale, what
possibilities are open for a modified picture at a slightly lower $\pT$
scale, either by MPI, ISR, or FSR? This is the default approach taken.

It is possible to switch off the interleaving, and consider FSR after
MPI and ISR. In that case it is also possible to allow FSR dipoles to
be formed between matching colour-anticolour pairs in two different
MPIs, whereas normally dipoles are local to each MPI separately.

\index{Decays!Resonances@of Resonances}
\index{Resonance decays!Interleaving}
\index{Resonance decays!Colour reconnections}
\index{Colour reconnections!Interplay with resonance decays}
Another ordering issue is when resonance decays and their showers
are considered. By default, this is done after 
the ISR/FSR/MPI evolution of the hard process, and also after the
handling of beam remnants and colour reconnections (CR). An option for
``early resonance decays'' allows for the resonance-decay to be handled
before remnants and CR; this does not alter the
perturbative evolution, but partons
from resonance decays can then participate in CR on an equal footing
with partons from the production process. The option for
``interleaved resonance decays''~\cite{Brooks:2021kji} moves the
resonance-decay handling 
even earlier, interleaving it with the ISR/FSR/MPI evolution of the
hard process, with a few different options for which value of the
perturbative evolution scale to associate to resonance decays,
the default being of order the width of the resonance. This
effectively represents an alternative treatment of finite-width
effects; it is not a big effect for the standard-model particles, none
of which has widths much larger than the shower cutoff, but could be
relevant for precision studies and/or in BSM scenarios. 

\subsubsection{The dipole evolution}

\begin{figure}[t!]
\includegraphics[width=0.4\linewidth]{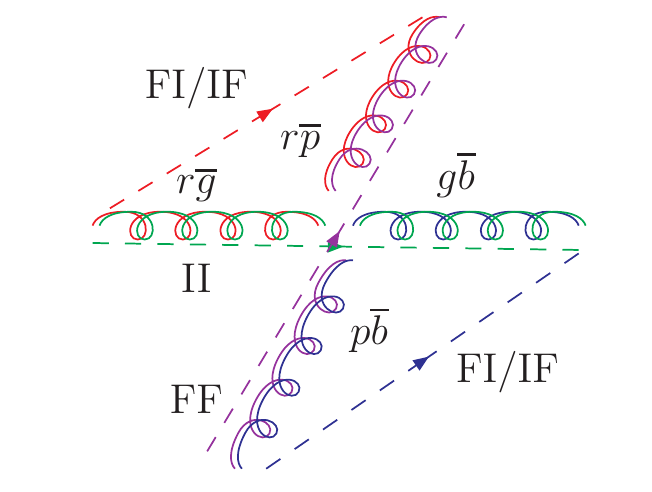}%
\includegraphics[width=0.38\textwidth]{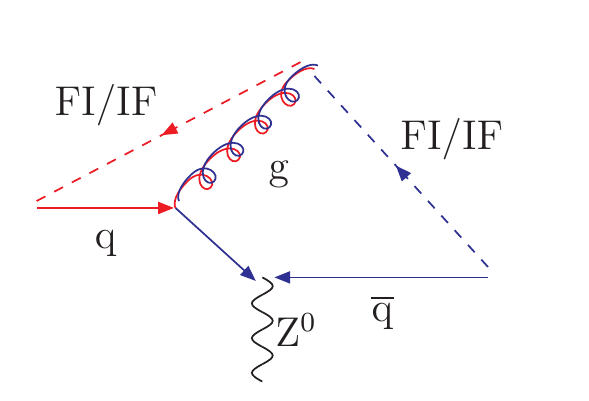}%
\includegraphics[width=0.25\textwidth]{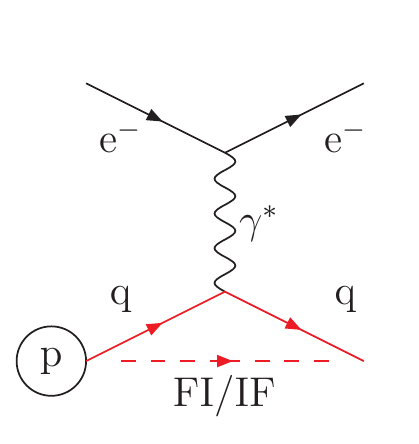}\\
\hspace*{0.16\linewidth}(a)\hspace*{0.30\linewidth}(b)%
\hspace*{0.30\linewidth}(c) 
\caption{
(a) Colour flow for the process 
$\g(r\overline{g}) + \g(g\overline{b}) \to \g(r\overline{p}) 
+ \g(p\overline{b})$. Here, the $\Nc \to \infty$ limit is used so that
$p$ stands for the new colour purple. The dashed lines represent the colour
lines stretching between the dipole ends. The type of dipole is indicated.
(b) $\q\qbar \to \Z\g$, again with colour lines and dipole types.
(c) Deeply inelastic scattering, again with colour lines and dipole types.}
\label{fig:simpleshower:dipoles}
\end{figure}

The previous subsection described the kinematics of a single branching.
The full evolution in an event requires some further consideration,
in particular related to the overall colour flow and the resulting set
of radiating dipoles. In hadronic collisions the dipole pattern can be
quite complicated. Consider the example of $\g\g \to \g\g$ scattering,
as shown in \cref{fig:simpleshower:dipoles}a, which is one of the six possible
colour topologies for this process in the $\Nc \to \infty$ limit.
Each radiation now is characterized by whether the radiator is in the
initial (I) or final (F) state, combined with the same classification for
the recoiler, so in general four different emission types need to be
considered.

\paragraph{Final-final radiation}
To begin with, consider the simple $\epem \to \gamma^*/\Z \to \q\qbar$ event.
The first emission of a gluon, to give $\q\qbar\g$, follows the pattern
already outlined. Now the $\Nc \to \infty$ limit is applied to split the
event into two dipoles $\q\g$ and $\g\qbar$. Each can be considered in
its respective rest frame, with the $p_{\perp\mathrm{evol}}$ scale of the
branching setting the upper limit for the continued evolution. In this
evolution, the full emission rate of $\g \to \g\g$ has to be split between
the two dipoles. Using \cref{eq:simpleshower:dglapgggmod}, the effective
splitting kernel becomes $P_{\g\to\g\g}(z) = (3/2) (1 + z^3)/(1 - z)$. 
Here, the emitter gluon takes the fraction $z$ and the emitted $1-z$,
where the latter is the one straddling the two new dipoles. The radiation
function from the $\q$ (or $\qbar$) and $\g$ ends of the dipole have
almost the same shape, the main difference being between the colour
factors $4/3$ \vs $3/2$, which are smoothly mixed around the middle of
the dipole, as already discussed for the angular dependence of
$\q\to \q\g$ \vs $\qbar\to \qbar\g$. There are known shortcomings with
this colour factor treatment~\cite{Gustafson:1992uh,Hamilton:2020rcu},
but these are of order $1/\Nc^2$ and are neglected here. On the kinematics
side, note that an emission in one dipole also affects the kinematics
of adjacent ones, by virtue of sharing one gluon with changed momentum.

\paragraph{\ac{II} radiation}
The ISR and FSR descriptions can be separated so long as
colour does not flow between the initial and the final state, as for
the first emission in $\q\qbar \to \Z$, which is pure II. But once a
gluon has been emitted, \cf\cref{fig:simpleshower:dipoles}b, the two
dipoles now bypass the $\Z$, and the $\Z$ does not receive any further
$\pT$ recoil during the subsequent evolution. This runs counter to
standard perturbation- and resummation-theory results, which is the reason
why traditionally ISR has only been handled as II dipoles. That is,
as shown in \cref{fig:simpleshower:dipoles}b, the emission of a second gluon
is handled as occurring from the (new) $\q\qbar$ dipole, with $\Z + \g$
together taking the recoil. Similarly, as shown in \cref{fig:simpleshower:dipoles}a,
the two IF dipole ends are replaced by doubling the strength of the
II dipole.

\paragraph{\ac{FI} radiation}
It should usually also be possible to replace the FI ends by FF ones,
with an arbitrary matching of the dipole ends, and such an option exists
for exploratory purposes, but is not the default. Instead, the incoming
colour-connected parton is designated as recoiler $r$. In a branching,
considered in the dipole rest frame, a fraction $Q^2 / m_{ar}^2$ of the
recoiler energy should be given from the recoiler to the emitter,
exactly as in \cref{eq:simpleshower:erfsrshift}. But the recoiler is
not a final-state particle, so the increase of $a$ momentum is not
compensated anywhere in the final state. Instead, the incoming parton
that the recoiler represents must have its momentum increased, not
decreased, by the same amount as the emitter. That is, its momentum
fraction $x$ needs to be scaled up as
\begin{equation} 
x_{r'} = \left( 1 + \frac{Q^2}{m_{ar}^2} \right) x_r ~.
\label{eq:simpleshower:newfraction}
\end{equation} 
Note that the direction along the incoming beam axis is not affected by
this rescaling, and that the kinematics construction therefore inevitably
comes to resemble that of Catani--Seymour dipoles~\cite{Catani:1996vz}.
The dipole mass $m_{ar}$ and the squared subcollision mass $\hat{s}$ are
increased in the process, the latter by the same factor as $x_r$. As with
ISR, the increased $x$ value leads to an extra PDF weight
\begin{equation} 
\frac{x_{r'} f_r(x_{r'},\pTs)}{x_r f_r(x_r,\pTs)} ~,
\label{eq:pdfratio}
\end{equation} 
in the emission and no-emission probabilities. This ensures a proper
damping of radiation in the $x_{r'} \to 1$ limit. The splitting of the
full dipole radiation pattern is not as well understood in this case as
for an FF dipole, however, but some order-of-magnitude estimates of how
the full dipole-emission rapidity range should be shared can be made~\cite{Corke:2010yf}. This suggests an extra damping factor like
$Q^2_{\mathrm{hard}}  / (Q^2 + Q^2_{\mathrm{hard}})$, where
$Q^2_{\mathrm{hard}}$ is the relevant hard scale of the process,
like $4 \pTs$ for QCD $2 \to 2$ processes, which is applied by default.

\paragraph{\ac{IF} radiation}
Finally, a non-default option exists, where IF dipole ends are treated
in their own right~\cite{Cabouat:2017rzi}. It then suffers from the
above-mentioned problems with $p_{\perp\mathrm{Z}}$ resummation, but it
enables handling \eg of deeply inelastic scattering (DIS),
\cf\cref{fig:simpleshower:dipoles}c, where II radiation is not an option
(using the $\eminus$ as recoiler would upset DIS kinematics), and
presumably offers a more realistic description \eg of weak-gauge-boson
fusion to a Higgs. The kinematics step from $b + r$, where $r$ is the
colour-connected recoiler in the final state, to $a + c + r'$, as a
consequence of the $a \to bc$ step, is easiest constructed in the
$b + r$ rest frame. There
\begin{align}
p_a &= \frac{1}{z} p_b ~,\\
p_c &= \left( \frac{1}{z} - 1 \right) p_b + p_{\mathrm{shift}} ~, \\
p_{r'} &=  p_r - p_{\mathrm{shift}} ~,\\
p_{\mathrm{shift}} &= \left( \frac{(2z - 1)Q^2}{2m_{br}} + z\frac{m_c^2}{m_{br}};
\pT \cos\varphi, \pT \sin\varphi, - \frac{Q^2}{2m_{br}} - z\frac{m_r^2}{m_{br}} 
\frac{Q^2 + m_c^2}{m_{br}^2 - m_r^2} \right) ~,\\
\pTs &=  \left((1 - z)(Q^2 + m_c^2) - m_c^2\right)
\left(1 - z\frac{Q^2 + m_c^2}{m_{br}^2 - m_r^2}\right)
- m_r^2\left(z\frac{Q^2 + m_c^2}{m_{br}^2 - m_r^2}\right)^2 ~.
\end{align}
The same set of rotations and boosts as used to recover the $b + r$
rest frame can then be inverted to bring $c$ and $r'$ back to the event
rest frame.

\paragraph{Special cases}
What remains is to combine IF and FI emissions consistently. In the
specific case of the first gluon emission from a DIS process, it
turns out that the IF-type branching $\q \to \q\g$ exactly reproduces 
the soft- and collinear-singularity structure of the $\gam^*\q \to \q\g$
matrix element on its own, with only a mild mismatch in the numerator
(which vanishes in the soft-gluon limit).
Therefore, it would be possible to leave aside FI emissions in this
case, and the same holds for $\g \to \g \g$ splittings, but not for
$\g \to \q\qbar$ ones. So in general. both IF and FI contributions have
to be used. One simplifying factor is that the incoming parton must always
be along the beam axis, so there will only be one common phase-space
mapping, unlike the case of FF or II dipoles. Nevertheless, the details
become technical and we refer to~\oneref{Cabouat:2017rzi} for further
discussion. One small comment, however: when an emission from
a $\q\g$ dipole is considered, the two ends radiate with different
colour charges, $4/3$ and $3/2$, respectively. The colour factors of the
two ends are then mixed in proportion to the $1/m^2$ values of the emitted
parton to the two dipole ends.

Another set of problems occurs in the decays of coloured resonances,
say $\t \to \b \W$. In this case the colour dipole is stretched between
the $\b$ and the hole left behind by the decayed $\t$. In order to
conserve momentum-energy, the $\b$ uses the $\W$ as a recoiler, and this
choice is unique. Once a gluon has been radiated, however, it is possible
to either still have the unmatched colour (inherited by the gluon)
recoiling against the $\W$, or to let it recoil against the $\b$
for this dipole as well. The former could give unphysical radiation patterns,
so the latter is chosen by default, although it is not perfect either. 
A more detailed discussion of this issue can be found in~\cite{Brooks:2019xso}. The same issue exists for a second emission of
QED radiation, \eg in $\Wp \to \eplus \nu_{\mathrm{e}}$, but is obviously
less significant there. 

\subsubsection{Matrix-element and other corrections}\label{sec:showers:simpleShowerCorrections}

In this subsection we give a survey of some methods used to make the
shower reproduce, or at least better approximate, known matrix-element
behaviours. The methods to match and merge external matrix-element
input to the showers are covered separately in \cref{section:matchmerge},
so here we mainly describe program elements internal to the simple shower.
Included are also some other ``correction'' aspects, that should offer
improvements to the shower, or at least provide increased understanding
by controlled variations.

\paragraph{Matrix-element corrections}\index{MECs}
One key capability is the first-order correction to resonance decays
$a \to bc$, where a gluon is emitted to give an $a \to bc\g$ final state.
The foremost example of this is
$\epem \to \gam^*/\Z \to \q\qbar \to \q\qbar\g$~\cite{Bengtsson:1986hr}. 
This works because \cref{eq:simpleshower:splitFSR} provides
a way that the parton shower exactly can reproduce the singularity
structure of the matrix element, \ie of the generic ratio
\begin{equation}
	\frac{1}{\sigma_{a \to bc}} \, \frac{\d\sigma_{a \to bc\g}}{\d x_1 \, \d x_2} ~.
\end{equation}
The $1 + z^2$ numerator of the splitting kernels also combines to an
expression that overestimates the numerator of the matrix elements,
\eg $x_1^2 + x_2^2$ for $\epem$ annihilation. In the veto-algorithmic downwards 
evolution of the shower, it is therefore trivial to use the ratio of the correct 
numerator to the shower-kernel numerator, as a probability that a trial emission 
will be retained. In fact, for the evolution down to the
first branching, it is as simple as putting the numerator equal to $2$
and correct down from that.

\index{Quark masses!in Simple showers}
This approach has then been extended to all combinations of colours and
spins for $a$, $b$ and $c$ that can occur within the SM and MSSM~\cite{Norrbin:2000uu}, and can be reused for other models where the same
colour and spin combinations occur. The inclusion of $b$ and $c$ masses
as in \cref{eq:simpleshower:massFSR} also reproduces the proper propagator 
poles $1/(m_{b'}^2 - m_b^2)$ and $1/(m_{c'}^2 - m_c^2)$ that are found in
the matrix elements, such that all correction factors are well behaved
over the whole phase space. Although the matrix elements are calculated for
a first emission only, they are reused in a suitably modified form to include 
mass effects also in subsequent steps. 

Similarly, there are a few processes where the first branching of an
ISR shower are corrected to the respective matrix element~\cite{Miu:1998ju},
based on a common singularity structure. These include
$\q\qbar \to \mathrm{V}\g$, $\q\g \to \mathrm{V}\q$,
$\f\fbar \to \mathrm{V}\gamma$, and $\f\gamma \to \mathrm{V}\f$,
where $\mathrm{V} = \gam^*/\Z/\Wpm/\mathrm{Z}'/\ldots$ is a colour-singlet vector boson. In the point-like-coupling approximation, also
Higgs production $\g\g \to \H$ and $\gamma \gamma \to \H$ is handled.

It should be feasible to include a matrix element correction to DIS
in the same fashion as already outlined, but this has not been done yet.
A generic and more detailed discussion of matrix-element corrections is given in
\cref{section:matchmerge}.

\paragraph{Power and wimpy showers}
\index{Power showers}
\index{Wimpy showers}
In the cases above, the ISR/FSR showers are allowed to cover the full
phase space, so-called \emph{power showers}~\cite{Plehn:2005cq}. We
have seen that they  
can reach the furthest corners no worse than being a factor two off,
which then could be fixed by modest reweighting. One guess is that this
would hold true also in other processes, where no matrix-element
correction factors have been implemented. But there are counterexamples.
Consider QCD jet production, say, starting out from
$2 \to 2$ partonic processes. Then a low-$\pT$ $2 \to 2$ process could
not be allowed to shower further partons at high $\pT$, or else such
high-$\pT$ production would be double counted and the whole perturbative
framework would be undermined. So the logical $p_{\perp\mathrm{evol,max}}$
shower starting scale is the $\pT$ scale of the $2 \to 2$ process,
\ie the factorization scale, giving \emph{wimpy showers}. Comparisons
with $2 \to 3$ matrix elements confirm that such a scale choice is close
to optimal~\cite{Corke:2010yf}.

In general, it is possible for the user to choose between power and wimpy
showers, even separately for ISR and FSR. The default option involves a
choice between the two based on the likelihood of double counting:
\begin{itemize}
\item If the final state of the hard process (not counting subsequent
resonance decays) contains at least one quark ($\u, \d, \s, \c ,\b$),
gluon, or photon then $p_{\perp\mathrm{evol,max}}$ is chosen to be the
factorization scale for internal processes and the \texttt{scale}
value for Les Houches input, \ie wimpy showers.
\item Else, emissions are allowed to go all the way up to the kinematic
limit, \ie power showers.
\end{itemize}
The reasoning is that in the former set of processes, the ISR emission
of yet another quark, gluon, or photon could lead to double counting,
while no such danger exists in the latter case.

In cases where more is known about the context of a particular
event sample, \eg when doing matching and merging, it is important
to make use of this knowledge to override the default behaviour.
One example is to start out with power showers but then implement
a user hook to reject those emissions that would double count the particular
cuts of the event sample.

\paragraph{Damped showers}\index{Power showers!Damped showers}
While there are processes where power or wimpy showers are appropriate,
there are also ones where the actual behaviour lies in between. It is
relevant to recall that the characteristic cross-section shape of a 
shower emission is $\d\pTs/\pTs$, while that of QCD $2 \to 2$ process
is $\d\pTs/p_{\perp}^4$. That is, the $\pT$ spectrum of a parton ought
to begin to drop faster around the scale where it goes from being a 
soft add-on to being a part of the core hard process. For top-pair
production $\g\g \to \t\tbar\g$, \eg the gluon emission can be
approximated by a shape
\begin{equation}
\frac{\d\mathcal{P}}{\d p_{\perp\g}^2} \propto \frac{1}{p_{\perp\g}^2} \,
\frac{k^2 M^2}{k^2 M^2 + p_{\perp\g}^2} ~,
\end{equation}
where $M^2$ is a reasonable scale to associate with the hard process
and $k^2$ is a fudge factor of order unity. This generalizes into the
possibility to use a power shower with an additional damping factor
$k^2 M^2/(k^2 M^2 + p_{\perp\mathrm{evol}}^2)$. Studies~\cite{Corke:2010zj}
show that this is a reasonable approach for coloured final states,
\eg for pairs of supersymmetric coloured particles, whereas a simple
power shower is more appropriate for pair production of uncoloured
particles. This can be understood as reduced emission by a destructive
interference between ISR and FSR  when colours flow from the initial
to the final state~\cite{Ellis:1986bv}, but only if there is such flow. 

\paragraph{Gluon splittings}\index{Gluon splittings@$\g\to\q\qbar$ splittings}
The pure $s$-channel nature of $\g \to \q\qbar$ splittings motivates the
introduction of an option with $\alphas(m_{\q\qbar}^2)$ rather than
$\alphas(p_{\perp\mathrm{evol}}^2)$, where $m_{\q\qbar}$ is the invariant mass
of the $\q\qbar$ pair. More importantly, the cuts on the allowed $z$
range during the FSR evolution imply that the branching rate is 
reduced relative to expectations from matrix elements. Therefore, for
this branching only, the default option is to weigh up the splitting
kernel inside the allowed $z$ range to give the correct integrated
matrix-element weight. Furthermore, this range is afterwards remapped
to cover the full range of decay angles, disregarding the normal $\pT$
ordering. This treatment is especially important for charm and bottom
quarks, where the mass is not negligible and mass corrections should
be reproduced both in rate and in angular distributions. As a final
twist, the matrix element for $\H \to \g\g \to \g\q\qbar$ does reproduce
the expected behaviour \eg from $\epem \to \gam^* \to \q\qbar$, but times
a factor $(1 - m_{\q\qbar}^2/m_{\H}^2)^3$. The default option uses this
factor, with the radiating dipole mass replacing the Higgs one, to
suppress high-mass branchings.

\paragraph{Dead cones}
\index{Dead cones|see{Quark masses}}
For topologies where a gluon recoils against a massive quark (or another
massive coloured particle) there are no suitable ME corrections
implemented into \pyt. When the dipole radiation pattern is split into
two ends, with a smooth transition between the two, this means that the
gluon end can radiate into the quark hemisphere as if the quark were
massless. The ``dead cone'' effect, that radiation collinear with a
massive quark is strongly suppressed, thereby is not fully respected.
(Unlike radiation from the quark end itself, where mass effects are
included.) By default, a further suppression is therefore introduced
for $\g \to \g \g$ branchings, derived as the massive/massless ratio
of the eikonal expression for dipole radiation, which eliminates radiation
collinear with the quark. The $\g \to \q \qbar$ branchings currently
are not affected; the absence of a soft singularity implies that there
is hardly any radiation into the recoiler hemisphere anyway. 

\paragraph{Global recoil}\index{Recoil schemes}
The default ISR and FSR showers differ, in that the former uses a
global recoil while the latter uses a dipole one. That is, the recoil
from an emission is carried by all final-state particles in ISR, 
but only by a single one in FSR. Then we introduced an option
where dipole recoil can be used for ISR. As it turns out, there
is also an option to obtain a global recoil in FSR. In such a scenario,
the radiation pattern is unrelated to colour correlations, which could
be seen as a disadvantage. It is convenient for some matching algorithms,
however, where a full analytic knowledge of the shower radiation pattern
is needed to avoid double counting, so it is by such user requests that 
the option is made available.

Technically, the radiation pattern is most conveniently represented
in the rest frame of the final state of the hard subprocess. Then, for
each parton at a time, the rest of the final state can be viewed as a
single effective parton. This ``parton'' has a fixed invariant mass
during the emission process, and takes the recoil without any changed
direction of motion. The momenta of the individual new recoilers are
then obtained by a simple common boost of the original ones. With the
whole subcollision mass as ``dipole'' mass, the phase space for
subsequent emissions is larger than for the normal dipole algorithm,
which leads to a too steep multiplication of soft gluons. Therefore, the
main application is for the first one or few emissions of the shower,
where a potential overestimate of the emission rate is to be corrected
by a matching to the relevant matrix elements. Thereafter, subsequent
emissions should be handled as before, \ie with dipoles spanned between
nearby partons. Several process-dependent settings are needed to use
this option.

\paragraph{Azimuthal asymmetries}\index{Azimuthal asymmetries}
Parton-shower branchings are assumed to occur isotropically in azimuthal
angle $\varphi$, in the rest frame of the respective dipole. The boost to
the overall CM frame then gives rise to the familiar ``string effect''~\cite{Andersson:1980vk,Azimov:1986sf} coherence phenomenon, where particle
production is enhanced in the region between two colour-connected partons.
But there are also azimuthal correlations arising from parton polarization~\cite{Webber:1986mc}. Notably, gluons tend to be plane polarized, with
the decay plane of $\g \to \g\g$ branchings favourably aligned with the
production plane, while $\g \to \q\qbar$ ones tend to be aligned orthogonal
to it. The former branching type is common but with small asymmetries,
while the opposite holds for the latter branching type, so that net
effects are small. They are included nevertheless, since they may have
some effect in charm and bottom production.

\paragraph{User hooks}\index{User hooks}
There are also other user hooks that can be used to modify the shower
evolution. The ones that allow an ISR or FSR emission to be vetoed play
a key role in matching and merging schemes and therefore are described in \cref{section:matchmerge}.

\subsubsection{QED, electroweak and other showers}\label{sec:SimpleQEDEW}

The simple shower includes several extensions beyond the QCD core
discussed so far. Characteristic is that these form part of the same
evolution in a common $p_{\perp\mathrm{evol}}$ scale, although with some
distinguishing features. 

\paragraph{QED shower}\index{QED showers!in Simple showers}
The most obvious extension is to QED. The required branching kernels
have been presented in \cref{eq:introshower:dglapffgam,eq:introshower:dglapgamff}.
In the evolution equations $\alphas(p_{\perp\mathrm{evol}}^2)$ is replaced
by $\alphaem(p_{\perp\mathrm{evol}}^2)$, but otherwise most that has been
written about $\q\to \q\g$ and $\g \to \q\qbar$ carries over. A dipole
language is used also for QED emissions, but the dipoles may be
different from the QCD ones. An example is
$\epem \to \gam^*/\Z \to \q\qbar \to \q\qbar\g$, where
the last stage contains two colour dipoles $\q\g$ and $\g\qbar$, but
only one charge dipole $\q\qbar$, since the gluon carries no electrical
charge. The complete multipole radiation pattern may be poorly
represented by a set of simple dipoles in cases with multiple charges,
since there is no confinement mechanism in QED to further a unique
dipole setup. In reality, few events contain multiple QED charges to
consider, and if so, often the event history suggests a reasonable
division, \eg when a new dipole arises from a $\gam \to \f \fbar$
branching. 

\index{Cutoff scales!in Simple showers}%
The lower cutoff on QED radiation in a hadron beam is not bound
to be the same as the QCD one, \ie since there is no issue of $\alphaem$
diverging at low scales. Nevertheless it is plausible to assume that the
QCD cutoff is related to the transition from quarks to hadrons, and thus
should be applied to all radiation. For radiation off a lepton, there is
no such restriction, and \pyt then by default sets
$p_{\perp\mathrm{evol,min}} = 10^{-6}$~GeV for FSR and $5\cdot10^{-4}$ for ISR. 
These values are fully sufficient to cover the emission of any photons
observable at a collider. They are also adjusted to be in a region where
kinematic reconstruction still works well in double precision. It has been
pointed out that they are not sufficiently low to generate the full
observable-photon spectrum when \pyt is applied to whatever processes
could give the highest-energy cosmic rays. \index{Cosmic rays}

The branching of a photon, $\gam \to \f \fbar$, does not fit well into
the dipole picture. The choice of a recoiler is based on the history
to the largest extent possible, \ie based on what the photon was produced in
association with. The photon branchings in part compete with the hard
processes involving $\gam^*/\Z$ production. In order to avoid overlap
it makes sense to correlate the maximum $\gam$ mass allowed in showers
with the minimum $\gam^*/\Z$ mass allowed in hard processes, by
default at 10~GeV. In addition, the shower contribution only contains
the pure $\gam^*$ contribution, \ie not the $\Z$ part, so the mass
spectrum above around $50~\GeV$ would not be well described.

\paragraph{Electroweak shower}
\index{Electroweak showers|see{Weak Showers}}
\index{Weak showers!in Simple showers}
The emission of $\Wpm$ and $\Z$ gauge bosons off fermions is an
integrated part of the ISR and FSR frameworks, and is fully
interleaved with QCD and QED emissions~\cite{Christiansen:2014kba}.
It is off by default, however, since it takes some time to generate
trial emissions, whereof very few result in real emissions unless
the fermion transverse momenta are much larger than the $\W/\Z$ masses.
These masses also have a considerable impact on the phase space of
emissions, which the shower is not set up to handle with a particularly
good accuracy. Therefore, the weak-shower emissions are always matched
to the matrix element for emissions off an $\f \fbar$ weak dipole, or some
other $2 \to 3$ matrix element that resembles the topology at hand. Even
if the match may not be perfect, at least the main features should be
caught that way. Notably, the correction procedure is used throughout
the shower evolution, not only for the emission closest to the hard
$2 \to 2$ process. Also, the angular distribution in the subsequent
$\mathrm{V} = \Wpm/\Z$ decay is matched to the matrix-element expression
for $\f \fbar \to \f \fbar \mathrm{V} \to \f \fbar \f' \fbar'$ (FSR) and
$\f \fbar \to \g^* \mathrm{V} \to \g^* \f' \fbar'$ (ISR). Afterwards, the
$\f' \fbar'$ system undergoes showers and hadronization just like any
$\Wpm/\Z$ decay products would. 

Special for the weak showers is that couplings are different for left-
and right-handed fermions. With incoming unpolarized beams this should
average out, at least so long as only one weak emission occurs. In the
case of several weak emissions off the same fermion, the correlation
between them will carry a memory of the fermion helicity. Such a memory
is retained for the affected dipole end. The flavour-changing character
of $\Wpm$ emissions also affects the tight relation between the
real-emission evolution and Sudakov factors, so-called Bloch--Nordsieck
violations. These effects are not expected to be large, but they are not
properly included. Another restriction is that there is no simulation of
the full $\gam^*/\Z$ interference: at low masses, the QED shower involves
a pure $\gam^*$ component, whereas the weak shower generates a pure $\Z$. 
Finally, it should be remembered that this is not a full (electro)weak shower,
which would also have required interactions among gauge bosons, and even
involved the Higgs boson. These interactions are included, \eg in the
\vincia EW shower, \cf\cref{sec:VinciaEW}.

\paragraph{Onia}\index{Onia|see{Quarkonium}}\index{Quarkonium}\index{Bottomonium|see{Quarkonium}}\index{Charmonium|see{Quarkonium}}
Hard production of charmonium and bottomonium can proceed either through
colour-singlet or colour-octet mechanisms. In the former case, the state
does not radiate and the onium is therefore produced in isolation, while
it is sensible to assume that a shower can evolve in the latter case,
giving an onium state embedded in some amount of jet activity.
Currently, both cases are initiated by $2 \to 2$ interactions directly
producing an onium state;  the alternative mechanism of producing
onia during the shower evolution itself~\cite{Ernstrom:1996am} is not
(yet) implemented. Emissions off an octet-onium state could easily break
up a semi-bound quark pair, but might also create a new semi-bound state,
and to some approximation these two effects should balance in the onium
production rate. The showering implemented here therefore should not be
viewed as an accurate description of the emission history step by step,
but rather as an effective approach to build up the onium environment.
The simulation of branchings is based on the assumption that the full
radiation is provided by an incoherent sum of radiation off the quark
and off the antiquark of the onium state. Thus, the splitting kernel is
taken to be the normal $\q\to \q \g$ one, multiplied by a factor of two.
A number of corrections to this picture could be imagined; since they
would come with opposite signs the assumption is that they cancel out. Further discussion is also included in \cref{subsection:onia}.

\paragraph{Baryon-number-violating decays}\index{Baryon number violation}
A complicated case for showering is baryon-number-violating decays,
\eg a neutralino decaying to three quarks. It is then not possible to
assign an ordinary dipole configuration. Instead half-strength dipoles
are constructed between each pair of quarks. That way the total emission
rate from each quark is at normal strength, and the recoil can be taken
by either of the other two quarks. Similar reduced-showering-rate dipoles
can be selected also in a few other cases.

\paragraph{Hidden Valley processes}
\index{Dark photons}\index{Parton showers!Dark photons}
\index{Hidden valleys}\index{Parton showers!Hidden valleys}
The Hidden Valley (HV) scenario, introduced in
\cref{subsection:HiddenValleyProcesses}, has been developed
specifically to allow the study of visible consequences of radiation
in a hidden sector, either by recoil effects or by leakage back into
standard-model particles. A key aspect therefore is that the normal
timelike showering machinery has been expanded with a third kind of
radiation, in addition to the QCD and QED(+EW) ones~\cite{Carloni:2010tw,Carloni:2011kk}. These three kinds are fully
interleaved, \ie evolution occurs in a common $\pT$-ordered sequence.
This radiation may be described either within a (possibly broken)
\uone or an unbroken \su{N} gauge group, but not both
simultaneously. Thus, one has either HV-photons or HV-gluons as
interaction carriers, where the latter are non-Abelian and may branch
into more HV-gluons. A set of 12 new particles mirrors the standard-model
flavour structure, and is charged under both the SM and the HV symmetry
groups, so that they can radiate both into the visible and invisible
sector. There is also a new massive particle with only HV charge, sitting
in the fundamental representation of the HV gauge group, denoted an
HV-quark.

HV particles are only produced in or after the hard process, so only
FSR needs to be considered. The HV radiation defines its own set of
dipoles, usually between opposite charges. Decays of massive particles
can give rise to the same kind of issues as for top decays, \ie that a
dipole properly involves the hole of the decaying particle. Matrix-element corrections are implemented for a number of decay processes,
with colour, spin, and mass effects included, as for SM processes.
These were calculated within the context of the particle content of the
MSSM, however, which does not include spin-1 particles with unit colour
charge. In such cases spin 0 is assumed instead. By experience, the main
effects come from mass and colour flow anyway, so this is not a bad
approximation. In the case of a broken \uone symmetry, the
HV-photon is massive, which requires some kinematics corrections
relative to ordinary QED radiation. If decays back to the SM occur,
\eg the HV-photon by mixing with the ordinary $\gam$, then also ordinary
showers are allowed. By default the coupling strength is fixed, but running
is allowed, given the gauge group and the contributing matter content.  

\subsubsection{Algorithms for automated shower variations and enhanced
  splittings}\label{sec:var}\index{Weights!in Simple showers}
Several variations of the simple shower are available in an automated fashion, to help construct
uncertainty bands for predictions~\cite{Mrenna:2016sih}. 
That is, weights are constructed and
associated with the shower evolution under different alternative
conditions, at the same time as the normal showers (with unit weight)
are evolved. The properties of an event only need to be analysed
once, but can then be filled in one histogram for each distinct
variation, with its associated event weight, and at the end these histograms can be combined to provide the uncertainty band. 
Variations can be set for the renormalization scale for ISR and FSR
QCD emissions (separately), for the inclusion of non-singular terms in
the ISR and FSR splitting kernels (separately), and for different PDF members.

The veto algorithm is used to generate parton-shower histories for the
physics parameters chosen at initialization as normal.
Using \cref{eq:veto-algorithm-weighted},
we can compute sets of weights (which we call variations) for each
event reflecting the changed probability for that event under
different possible choices of physics parameters. The number of variations
calculated is limited only by finite computing and memory resources.

While the proof of unitarity is more easily realized using
\cref{eq:veto-algorithm-weighted}, the algorithm is employed
discretely.
Thus the factors
\begin{equation}
\frac{f(t')}{g(t') r(t')}~\text{(acc) and ~} \frac{1-f(t')/g(t')}
{1-r(t')}~\text{(disc),}
\end{equation}
can be calculated at each
discrete step and book-kept during the shower to calculate an event
weight.
The factors (acc) account for the effect in accepted splittings, while the
factors (disc) preserve unitarity from the discarded trial
splittings. 

\paragraph{Parton-Shower Variations}\index{Uncertainties!Parton showers@in Parton showers}\label{sec:showerVariations}
Consider a parton shower based on the veto algorithm 
with a physical trial-accept probability, $P_\mrm{acc}$, given by
the ratio of a splitting kernel $P(t,z)$ and an oversampling kernel
$\hat{P}(t,z)$, 
and an alternative shower algorithm, defined by a
different physical trial-accept probability, $P'_\mrm{acc}$, given
by the ratio of an alternative radiation kernel $P'(t,z)$ to the same
oversampling kernel.  The difference between the radiation kernels 
could be different $\alphas$ scale
choices, different non-singular terms in the splitting kernels, and/or
different effective higher-order contributions to the splitting
kernels.
In the following, we assume that the $t$ and $z$ definitions
remain the same.
Translations between different $t$ choices are discussed
in~\oneref{Hartgring:2013jma} (and the resulting equations used in early
versions of \vincia to provide an uncertainty variation corresponding
to the difference between virtuality-ordered and $p_\perp$-ordered
showers) while exploring different $z$ definitions (and more
generally, different recoil strategies) would require a future
generalization. 
The algorithm to compute the probability of an event generated by $P'$
based on an event generated using $P$ is,
following~\citerefs{Giele:2011cb,Mrenna:2016sih} and suppressing the $z$ dependence for
clarity: 
\begin{enumerate}
\item Start the event evolution by setting all weights (nominal and
  uncertainty-variation ones) equal to the input weight of the event,
  $w'=w$. 
\item\label{step:niaveSplit} If the trial branching is accepted, multiply the alternative
  weight $w'$ by the relative ratio of accept probabilities,   
\begin{equation}
 R_\mrm{acc}'(t)~=~\frac{P'_\mrm{acc}(t)}{P_\mrm{acc}(t)}~=~ \frac{P'(t)}{P(t)}~.
\end{equation}
\item\label{step:noBranch} If the trial branching is rejected, multiply the alternative weight $w'$ by the relative ratio of discard probabilities, 
\begin{equation}
R_\mrm{disc}'(t)~=~\frac{P'_\mrm{disc}(t)}{P_\mrm{disc}(t)}~=~
 \frac{1-P'_\mrm{acc}(t)}{1-P_\mrm{acc}(t)}
~=~\frac{\hat{P}(t)-P'(t)}{\hat{P}(t)-P(t)}~.
\label{eq:Rdiscard}
\end{equation}
\item If desired, the detailed balance between the accept and discard probabilities could optionally be allowed to be broken by up to a non-singular term, $P'_\mrm{acc} \neq 1-P'_\mrm{disc}$, to represent uncertainties due to genuine (non-cancelling) higher-order corrections, which would modify the total cross sections. For the current implementation in \pyt, however, we do not consider this possibility further. 
\end{enumerate}
Step~\ref{step:niaveSplit} is responsible for adjusting the naive splitting probabilities, while step~\ref{step:noBranch} is responsible for adjusting the no-branching probabilities. The result is that the set of weights $w'$ represents a separately unitary event sample, with $\left<w'\right> = \left<w\right>$; \ie the samples integrate to the same total cross section. 
The relative discard-ratio, \cref{eq:Rdiscard}, contains the
difference $\hat{P}-P$ in the denominator.
If the trial overestimate, $\hat{P}$, is ``too efficient'' (meaning it
is very close to $P$),
the denominator can become close to singular, resulting in large and
possibly numerically unstable weights.
Algorithmically, what happens is that there are very few failed
trials, hence the modifications to the no-branching probability can have large fluctuations.
Technically, we address this by applying a ``headroom factor'' to the
trial functions when automated uncertainty variations
are requested, ensuring that there is always a non-negligible probability for trials to be discarded at the cost of computational
speed.

\index{Weights!in Simple showers}
The final event weight, $w'$, after the full shower evolution, is the
product of many such factors,
one $R'_\mrm{acc}$ for each accepted trial and one $R'_\mrm{disc}$ for each discarded one,
\begin{equation}
  w' ~= \prod_{i\in \mrm{accepted}}\frac{P'_{i,\mrm{acc}}}{P_{i,\mrm{acc}}} \prod_{j\in \mrm{discarded}} \frac{P'_{j,\mrm{disc}}}{P_{j,\mrm{disc}}}~.
\end{equation}
Given enough phase space for evolution, this factor can become
arbitrarily different from unity, representing that, \eg a very
active shower
history is exponentially more likely to occur in a shower with a large
value of $\alphas$ than in one with a small  value.
In principle, this \emph{is} both physically and mathematically
correct. In practice, however,
it is not desirable that branchings at low evolution scales in the
shower  should significantly alter the modified event weights.
Technically, we treat this by imposing a few limiting factors on the variations.

\paragraph{Renormalization-Scale Variations}

The first major class of variations we include are variations of the
shower renormalization scales. This can be done for both QED and QCD,
with the latter normally dominating the overall uncertainty. It is
worth noting, however, that for a coherent shower algorithm, a scale
choice of $p_\perp$ accompanied by the so-called CMW scale factor
absorbs the leading second-order corrections to the splitting
functions for soft-gluon emission.
A brute-force scale variation would destroy this agreement. We therefore provide an option to allow an explicit ${\cal O}(\alphas^2)$ compensating term to accompany each scale variation,  driving the effective scale choice back towards $p_\perp$ at the NLO level, while leaving the higher-order components of the scale variation untouched. 

Specifically, if the baseline gluon-emission density is
\begin{equation}
P(t,z) ~=~ \frac{\alphas(p_\perp)}{2\pi} \frac{P(z)}{t} ~,
\end{equation}
with $P(z)$ the DGLAP radiation kernel, then we may define a renormalization-scale variation, $\mu=p_\perp \to \mu'=kp_\perp$, with an NLO-compensating term (see, \eg \oneref{Hartgring:2013jma})
\begin{equation}
P'(t,z) ~=~ \frac{\alphas(k p_\perp)}{2\pi}\left(1 + \frac{\alphas}{2\pi}\beta_0 \ln k\right) \frac{P(z)}{t} ~,\label{eq:compensation}
\end{equation}
with $\beta_0=(11\Nc - 2n_F)/3$, $\Nc=3$, and $n_F$ the number of active flavours at the scale $\mu = p_\perp$. 
Note that, if there are any quark-mass thresholds in between $p_\perp$
and $kp_\perp$, then $\alphas(p_\perp)$ and $\alphas(kp_\perp)$ will
not be evaluated with the same $n_F$. Matching conditions are applied
in \pyt to make the running continuous across thresholds, so this
effect should be small for reasonable values of $k$. Nonetheless, one
could in principle add an additional term $\alphas/(2\pi)
\ln(m_q/(kp_\perp)) / 3$ to compensate for the different $\beta_0$
coefficients used in the region between the threshold and $kp_\perp$.
However, since the variation is numerically larger without that term,
and since the ambiguities associated with thresholds are anyway among
the uncertainties one could wish to explore, for the time being we
consider it more conservative to not include any such
terms.\index{Uncertainties!Parton showers@in Parton showers}

Note also that the scale and scheme of the $\alphas$ factor in the compensation term, inside the parenthesis in \cref{eq:compensation}, is not specified, as this amounts to an effect of a yet higher order, beyond NLO. To make the compensation as conservative as possible (and to avoid the risk of overcompensating), we choose the scale of the compensation term to be the largest local scale in the problem, namely the invariant mass of the emitting colour dipole
$m_\mrm{dip}$, thus making the correction term as numerically small (and hence as conservative) as possible, specifically $\mu_\mrm{max}=\max(m_\mrm{dip},kp_\perp)$. 
Furthermore, since this argument only pertains to the soft limit, our estimate of the compensation would be too optimistic if applied undiminished over all of phase space. To be more conservative, we therefore multiply the compensation term by an explicit factor $(1-\zeta)$, defined so as to vanish linearly outside the soft limit,
\begin{equation}
\zeta = \left\{ 
\begin{array}{ccl}
z&&\mbox{for splittings with a $1/z$ singularity}\\ 
1-z&&\mbox{for splittings with a $1/(1-z)$ singularity}\\
\min(z,1-z)&&\mbox{for splittings with a $1/(z(1-z))$ singularity}
\end{array}
\right.~.
\end{equation}
Combined, these arguments lead us to the following modified accept probability for a robust shower renormalization-scale variation compatible with the known second-order leading-singular structure:
\begin{equation}
P'(t,z)~=~ \frac{\alphas(k p_\perp)}{2\pi}\left(1 + (1-\zeta)\frac{\alphas(\mu_\mrm{max})}{2\pi}\beta_0 \ln k\right)\frac{P(z)}{t}~,
\end{equation}
hence 
\begin{equation}
R'_\mrm{acc}(t,z)~=~\frac{P_\mrm{acc}'(t,z)}{P_\mrm{acc}(t,z)} ~=~ \frac{\alphas(k p_\perp)}{\alphas(p_\perp)}\left(1 + (1-\zeta)\frac{\alphas(\mu_\mrm{max})}{2\pi}\beta_0 \ln k\right)~.
\end{equation}
The compensation term in the expressions above is only included for gluon emissions, not for $g\to q\bar{q}$ splittings. The latter are subjected to the full (uncompensated) variation, $\alphas(kp_\perp)/\alphas(p_\perp)$.

Finally, we impose an absolute limit on the allowed amount of $\alphas$ variation, by default
\begin{equation}
    |\Delta \alphas| \le 0.2~.
\end{equation}
This does not significantly restrict the range of variation for perturbative branchings (even when $\alphas \sim 0.5$, a full 40\% amount of variation is still allowed), but it does prevent branchings very near the cutoff from generating large changes to the event weights. Removing this bound would not significantly affect the perturbative physics uncertainties, but would cause much larger weight fluctuations (between events with and without some very soft branching near the end of the evolution), mandating much longer run times for the same statistical precision. 

At the technical level, the user decides whether to perform scale variations of ISR and FSR
independently, or whether to vary the respective $\alphas$ factors in a correlated manner. It 
is even possible to include both types of variations (independent and correlated), and compare 
the results obtained at the end of the run. From a practical point of view, the FSR $\alphas$ 
choice mainly influences the amount of broadening of the jets, while the ISR $\alphas$ choice 
influences resummed aspects such as the combined recoil given to a hard system (\eg a $Z$, 
$W$, or $H$ boson, or a $t\bar{t}$, dijet, or $\gamma+\mrm{jet}$ system) by ISR radiation and 
also how many extra jets are created from ISR. The latter of course also depends on whether and
how corrections from higher-order matrix elements are being accounted for. 
A few illustrations for the simple shower model can be found in~\oneref{Mrenna:2016sih}. 

\paragraph{Finite-Term Variations \label{sec:cNS}}\index{DGLAP}

All shower formalisms are based upon the universal nature of the singular infrared (soft and/or collinear) limits of QCD. In these limits, the exact form of the splitting functions are known (to a given order), regardless of whether we express them as DGLAP kernels, dipole/antenna functions, or by any other means. Away from these limits, however, in the physical phase space on which the kernels will be applied as approximations, there are in principle infinitely many different radiation functions to choose from, sharing the  same singular terms but having different non-singular ones. 
Attempting to evade this problem by setting the non-singular terms to zero would 
not only be arbitrary, it would also not be stable against reparameterizations of the radiation functions themselves. Thus, zero finite terms in a DGLAP parameterization does not translate to zero in a dipole one, nor does zero in one dipole parameterization correspond to zero in another, see \eg \citerefs{Giele:2011cb,LopezVillarejo:2011ap}. 

We also emphasize that finite terms are qualitatively different from renormalization-scale variations and produce quite different-looking uncertainty envelopes~\cite{Giele:2011cb}. The reason is that re\-normalization-scale variations are by construction proportional to the shower radiation functions. In regions far from the singular limits, the pole terms are highly suppressed and the shower radiation functions may not bear much resemblance to the matrix elements for the process at hand. In such regions, modest finite terms can therefore easily produce much larger variations than renormalization-scale changes. 

We therefore believe that an exhaustive exploration of parton-shower uncertainties should at least grant the \emph{capability} to perform finite-term variations, while the final decision whether and how to use them  can still be left up to the user. 
An observation of large finite-term uncertainties in the context of a physics study would be a direct indication of a need to incorporate further corrections from matrix elements, \eg via one of the many matching/merging strategies available in \pyt. This is because the matrix elements contain the correct finite terms for the  process at hand, thus nullifying the finite-term uncertainties at least in any phase-space regions populated by the matrix elements. 

To implement such variations in the context of a DGLAP approach, we do the following,
\begin{equation}
\frac{P(z)}{Q^2} \ dQ^2 \ \to\  \left(\frac{P(z)}{Q^2} + \frac{c}{m_\mrm{dip}^2}\right) dQ^2 = \left(P(z) + \frac{c \ Q^2}{m_\mrm{dip}^2}\right) \frac{dt}{t}~,
\end{equation}
where $m_\mrm{dip}$ is the invariant mass of the dipole in which the splitting occurs, $c$ is a dimensionless finite term of order unity, and in the last equality we used the identity $dQ^2/Q^2 = dt/t$ which holds for any $t=f(z)Q^2$, including in particular all the \pyt evolution variables. Note that, for gluon emission off timelike massive quarks, $Q^2$ should be the virtuality, or off-shellness of the massive quark, defined as $Q^2 = (p_b + p_g)^2 - m_b^2 = 2p_b\cdot p_g$~\cite{Norrbin:2000uu}, with $p_b$ the four-momentum of the massive quark and $p_g$ that of the emitted gluon.
Thus,
\begin{equation}
P'(t,z)~=~\frac{\alphas}{2\pi}\ {\cal C}\left(\frac{P(z) \ + \ c \ Q^2/m^2_\mrm{dip}}{t}\right)~,
\end{equation}
where $\cal C$ is the colour factor. The variation can therefore be obtained by introducing a spurious term proportional to $Q^2/m_\mrm{dip}^2$ in the splitting kernel used to compute the accept probability, hence
\begin{equation}
R'_\mrm{acc}~=~\frac{P'_{\mrm{acc}}}{P_\mrm{acc}}~=~1 + \frac{c \ Q^2/m_\mrm{dip}^2}{P(z)}~, 
\end{equation}
from which we also immediately confirm that the relative variation explicitly vanishes when $Q^2\to 0$ or $P(z)\to\infty$. 

To motivate a reasonable range of variations, we take the finite terms that different physical
matrix elements exhibit as a first indicator, and supplement that by considering the terms that
are induced by \pyt's \ac{MEC} for $Z$-boson
decays~\cite{Bengtsson:1986hr}. In particular, the study in~\oneref{LopezVillarejo:2011ap} found
order-unity differences (in dimensionless units) between different
physical processes and three different antenna-shower formalisms.
Therefore, here we also take variations of
order unity as the baseline for our recommendations.
A few illustrations for the simple shower model can be found in~\oneref{Mrenna:2016sih}. 

\paragraph{Veto Algorithm with Biased Kernels}\label{sec:biasedKernels}
\index{Enhanced splittings!in Simple showers}
A second important use case for modifying the veto algorithm is to evaluate the
fragmentation contributions to processes like photon and $B$-hadron
production, via splittings like $q\to q\gamma$ and $g\to b\bar{b}$,
respectively.
Since these processes are relatively rare ($\alphaem \ll
\alphas$ and $P_{g\to b\bar{b}} \ll P_{g\to gg}$), the generation of
adequate event samples
featuring these processes can suffer from substantial inefficiencies.
The method implemented in \pythia is described in \citerefs{Lonnblad:2012hz,Mrenna:2016sih}.
It is formally identical to the one presented for $q\to q\gamma$
branchings in \oneref{Hoeche:2009xc}. 

\index{Weights!in Simple showers}
Consider that we wish to enhance the rate of $g\to b\bar{b}$
splittings by a factor $E \gg 1$ 
until we have obtained at least one such splitting, after which we would normally 
want to let the probability to have a second $g\to b\bar{b}$ splitting in the same event drop 
back down to the normal level.   We can achieve this by first
increasing  the rate of trials for the corresponding
splitting function by a factor of $E$ by using a larger (biased) trial function (suppressing the dependence on both $t$ and $z$),
\begin{equation}
\hat{P}_\mrm{biased} = E \hat{P}~.
\end{equation}
We then keep the accept probability the same as normal, but reweight each accepted biased trial branching by the inverse of the biasing factor, 
\begin{equation}
P_\mrm{acc} ~=~\frac{P}{\hat{P}} ~~~;~~~R_\mrm{acc}~=~\frac{\hat{P}}{\hat{P}_\mrm{biased}}~=~\frac{1}{E}~,\label{eq:biasWeight}
\end{equation}
so that the product $R_\mrm{acc}P_\mrm{acc} \hat{P}_\mrm{biased} = P$
is the desired physical distribution.
For each discarded biased trial branching, we use the same technology as above to reweight the event,
\begin{equation}
R_\mrm{disc}~=~\frac{1-P_\mrm{acc}R_\mrm{acc}}{1-P_\mrm{acc}}~=~ 
\frac{\hat{P}}{\hat{P} - P}\left(1 - \frac{P}{\hat{P}_\mrm{biased}} \right) ~\stackrel{P\ll \hat{P}_\mrm{biased}}{\to}~ \frac{\hat{P}}{\hat{P} - P}~,
\label{eq:biasSudakov}
\end{equation}
where the last asymptote shows that the reweighting factor becomes
independent of the bias in the limit that the bias factor is very
large.
Nonetheless, the difference is important since it allows us to recover the physical no-branching probability. 
Currently enhancements of both ISR and FSR branchings have been
included, uniformly in phase space.

\subsection{The \vincia antenna shower}\label{sec:Vincia}
\index{Vincia@\vincia}\index{Parton showers!Vincia@\vincia}\index{Antenna showers|see{\vincia}}\index{Sector showers|see{\vincia}}\index{Ariadne@\ariadne}
The \vincia shower implements
an interleaved $\pT$-ordered evolution based on the so-called antenna
formalism. In event-generator contexts, this type of shower was
first pioneered by  the \ariadne
model~\cite{Gustafson:1987rq,Lonnblad:1992tz}, 
which was widely used \eg at LEP.
For completeness, we note that the objects we call ``antennae'' here were
actually called dipoles in that context, but today the term dipole has
taken on a different meaning, see, \eg \cref{sec:Dire} on \dire. 

Especially for FSR QCD radiation, \vincia shares
many features with \ariadne, including its
evolution-variable definition and its antenna-style $2\mapsto 3$ approach
to parton branchings in which both parents can acquire transverse recoil
and the soft eikonal remains unpartitioned. These latter two
properties are specific to antenna showers.

\index{Backwards evolution}For ISR, \vincia's treatment is quite
different from that of 
\ariadne, with \vincia extending the concept 
of (interleaved) backwards
evolution~\cite{Sjostrand:1985xi,Sjostrand:2004ef} to the antenna 
picture~\cite{Ritzmann:2012ca} via coherent II and IF 
antennae~\cite{Brooks:2020upa}, as well as
so-called \ac{RF} ones~\cite{Brooks:2019xso}. The
latter are relevant to the decays of coloured resonances, such as top
quarks. They come with their own, dedicated kinematic mapping which is 
constructed to preserve the invariant mass of the decaying resonance.
Since all of its building blocks are explicitly coherent
(at least at leading colour) and interleaved in a single common
sequence of decreasing $p_\perp$ values, \vincia should exhibit a
quite reliable description of soft coherence effects across
essentially all physical contexts.

This extends to QED, for which \vincia's default antenna
functions~\cite{Kleiss:2017iir,Skands:2020lkd} include fully coherent
(multipole) soft interference effects in addition to the collinear DGLAP
structures. We are not aware of any other multipole QED treatment 
that can be interleaved with the QCD evolution. (\eg
the YFS formalism~\cite{Yennie:1961ad} is constructed purely as an
``afterburner'', \ie not interleaved with the QCD shower, and
collinear logarithms can only be included order by order.)

A further difference with respect to \ariadne is that
\vincia's QCD and multipole QED showers are constructed as so-called
``sector'' antenna showers, in which the
phase space  
is divided into distinct (non-overlapping) colour and kinematics
sectors, each of which only receives contributions from one specific
antenna-branching kernel. This has a number of mainly technical
consequences which will be elaborated on further below, to do with
making the incorporation of higher-order corrections as
straightforward and efficient as possible. For the time being, 
\ariadne-style ``global'' showers also remain available as a
non-default option. 

\index{Quark masses!in Vincia@in \vincia}Effects of particle masses
are systematically included, both via mass  
corrections to the antenna functions such that all relevant
(quasi-)collinear limits are
reproduced~\cite{GehrmannDeRidder:2011dm,Brooks:2020upa}, and by the
use of exact massive phase-space factorizations.
The current default behaviour is to treat bottom and heavier
quarks, and muons and tau leptons, as massive in \vincia, though this
can be changed if desired. (Weak bosons are always treated
as massive.) A subtlety arises in the treatment of incoming
heavy-flavour quarks (and, potentially, muons). 
Kinematically, such partons are book kept as massless, similarly to the
choice made in \pyt's simple shower. The consequence is that the treatment 
of mass effects for initial-state partons in \vincia is less rigorous than for
final-state ones. One should also be aware that there can be a
non-trivial interplay with the flavour scheme employed by the chosen
PDF set.

As a complementary option to the multipole QED shower, \vincia also
includes a module for full-fledged electroweak
showers~\cite{Kleiss:2020rcg,Brooks:2021kji}. This option includes
the full set of EW-branching kernels, including both
Higgs couplings and gauge-boson self-interactions, tallying to more
than 1000 EW-antenna functions in total. The main limitation
is that only the relevant (quasi-)collinear limits are implemented,
not the full soft interference structure; thus, also the QED treatment
is effectively 
reduced to a DGLAP-style treatment when using this option. Note also
that the EW module is based explicitly on \vincia's underlying
formalism for helicity-dependent
showers~\cite{Larkoski:2013yi,Fischer:2017htu} (\eg to tell left- 
and right-handed weak bosons apart). This module therefore 
requires Born partons with assigned helicities, which is not 
the default in \pyt, and must be provided either via external LHEF
events with helicity information, or via \vincia's dedicated option
for hard-process helicity selection.
(The latter is based on \pyt's run-time interface to external
matrix-element libraries; see the program's \htmlmanual and example
programs for configuration and linking instructions.)  

\index{Interleaved resonance decays}
\index{Resonance decays!Interleaving}
\index{Top quark}\index{Weak bosons}A further feature that was
originally introduced with 
\vincia's electroweak-shower module, but is now applied independently of it, is 
a novel treatment of finite-width effects, called interleaved
resonance decays~\cite{Brooks:2021kji}, which 
are the default in \vincia. This means that decays of short-lived
resonances, such as top quarks, $W/Z$ bosons, or BSM particles, 
are inserted in the shower evolution 
at an evolution scale of order the off-shellness of the resonance,
instead of being treated sequentially, after the shower of the hard
process. This reflects the physical picture that short-lived
particles should not be able to radiate at frequencies lower than
the inverse of their lifetime; only their decay products can do
that. This can produce subtle changes in reconstructed
invariant-mass distributions, relative to conventional
(non-interleaved) resonance decays. 

\index{Interleaved evolution}
All of \vincia's shower modules are fully interleaved
with \pyt's treatment of multiparton interactions (MPI), in the
same manner as for the simple-shower model. 

\subsubsection{Common features}\label{sec:vinciacommon}
Some aspects are common to all of \vincia's shower modules.
This includes the definition of the evolution variable as well as
recoil schemes and phase-space factorizations. These common features
are discussed in this subsection before going into further detail
on each of the specific components of \vincia's shower implementations. 

\paragraph{Evolution variables}\index{Evolution variable!in Vincia@in
  \vincia}\index{Ordering variable|see{Evolution variable}}
All showers in \vincia, including the QED and EW ones, are evolved in
a Lorentz-invariant scaled notion of off-shellness, based on a
generalized version of the \ariadne definition of transverse
momentum. For a generic branching $IK \to i j k$,  
\begin{align}
	\pTs[j] &= \frac{\bar{q}_{ij}^2 \bar{q}_{jk}^2}{\sinv{\mrm{max}}}~,
             \label{eq:vincia-ordering-scales}
\end{align}
where the off-shellness for final-state partons is defined as
\begin{equation}
  \bar q_{ij}^2 = \displaystyle (p_i + p_j)^2 - \msq{I} = \displaystyle m^2_{ij} - \msq{I}~\qquad\mbox{$i$ is final}~,
\end{equation}
and that for initial-state partons is obtained via crossing (and an overall
sign change to make it positive), 
\begin{equation}
  \bar q_{ij}^2 = \displaystyle -(p_i - p_j)^2 +
  \msq{I}~\qquad\mbox{$i$ is initial}~. 
\end{equation}
These both involve the positive invariant $2p_i \cdot p_j$ but differ in the
signs of pre- \vs post-branching parton masses. This reflects the 
underlying crossing and sign change, combined with the propagator
structure of backwards evolution. For convenience, we define the
dimensionful invariant 
\begin{equation}
  s_{ij} \equiv 2p_i \cdot p_j~,
\end{equation}
regardless of whether particles $i$ and $j$ are massless or not.
The maximal antenna
invariant, $\sinv{\mrm{max}}$, is then defined by 
\begin{equation}
\sinv{\mrm{max}} = \begin{cases}
\sinv{IK} & \mrm{FF} \\[8pt]
\sinv{aj}+\sinv{jk} & \mrm{IF~\&~RF} \\[8pt]
\sinv{ab} & \mrm{II}
\end{cases}~,
\end{equation}
where initial-state partons are labelled with letters from the
beginning of the alphabet ($a$ and $b$) and final-state ones are
labelled by $i$, $j$, $k$, $\ldots$. Below, that labelling
convention will be used systematically to distinguish initial- and
final-state partons.  

We also define dimensionless (scaled) invariants and masses,  
\begin{equation}
  \yinv{ij} = \frac{\sinv{ij}}{\sinv{\mrm{max}}}~~~~;~~~~\mu_i^2 = \frac{m_i^2}{\sinv{\mrm{max}}}~.
\end{equation}
For massless kinematics, the scaled invariants
have very simple relations to the $z$ variables of \ac{DGLAP}-style
approaches. Thus, for final-final (FF) antennae, the CM energy
fractions are  
\begin{equation}
  x_k = \frac{2E_k}{\sqrt{s_{IK}}} = 1 - y_{ij}~,
\end{equation}
and similarly for the two other permutations of $(i,j,k)$. 
For initial-initial (II) antennae, the incoming legs are always
massless and the $y_{AB}$ invariant can be identified with the $z$ variable,
since  
\begin{equation}
  y_{AB} = \frac{s_{AB}}{s_{ab}} = \frac{x_A x_B}{x_a x_b} = z_a z_b~,
\end{equation}
where we have emphasized that, for antenna-II branchings, in general both $x$
values change, with 
\begin{equation}
z_a = \frac{x_A}{x_a} = \sqrt{\yinv{AB}\frac{1-\yinv{jb}}{1-\yinv{aj}}}
~~~\mbox{and}~~~
z_b = \frac{x_B}{x_b} = \sqrt{\yinv{AB}\frac{1-\yinv{aj}}{1-\yinv{jb}}}~.
\label{eq:vinciaIIz}
\end{equation}
There is still the constraint that, in the $a$-collinear limit $z_b\to
1$ and \textit{vice versa} (for massless $j$). 

For initial-final (IF) ones, the exact relations are more involved but
again the collinear limits can be examined via 
\begin{equation}
  z_a = 1-y_{jk}~~~~;~~~~z_k = 1-y_{aj}~.
\end{equation}

\begin{figure}[t!]\centering
\hspace*{-0.9cm}\includegraphics[width=1.08\textwidth]{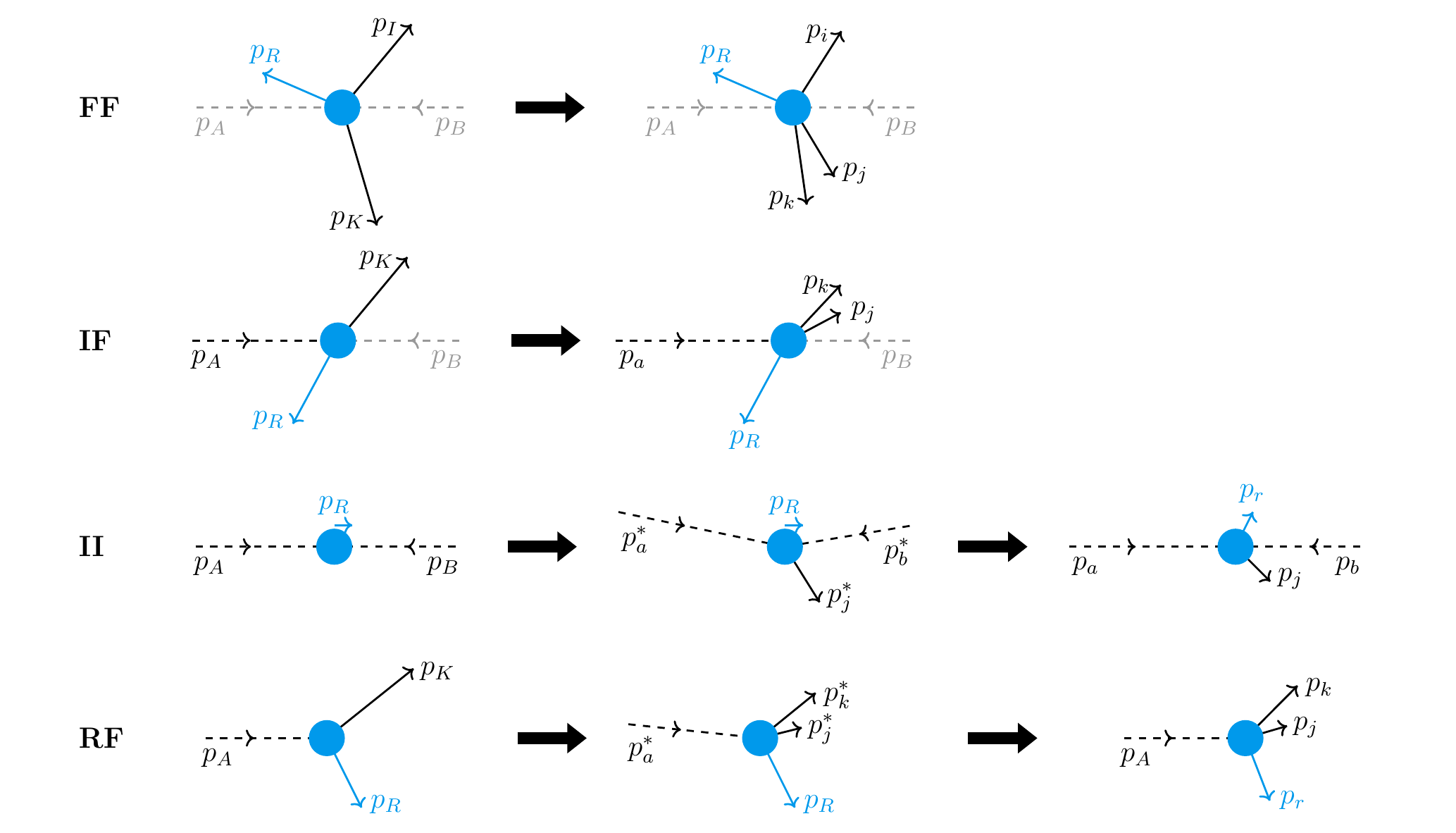}\hspace*{-0.9cm}
	\caption{Illustration of \vincia's kinematic maps, for
          final-final (FF), initial-final (IF), initial-initial (II),
          and resonance-final (RF) branchings. Dashed lines represent
          initial-state momenta, non-participating legs are shaded
          grey, and the set of final-state spectators ($R$) is
          shown in cyan. For II and
          RF branchings, the frame reinterpretation done in the last
          step imparts a 
        collective recoil to the final-state spectators, $p_R \to p_r$.}  
	\label{fig:ps:vinciaRecoils}
\end{figure}

\paragraph{Recoil schemes}
\index{Recoil schemes}\index{Final-state radiation (FSR)!Kinematics}
\index{Kinematics!in shower branchings}
In the antenna formalism, the branching recoil is shared between both
antenna parents for FSR emissions in an on-shell kinematics map along
the lines of \citerefs{Kosower:1997zr,Giele:2007di,GehrmannDeRidder:2011dm} including
full mass dependence. This is illustrated in the top row in \cref{fig:ps:vinciaRecoils}.
In the collinear limits, any transverse momentum is fully absorbed
within the collinear pair, and the anti-collinear parent recoils
purely longitudinally; therefore, in these limits the map agrees with
the conventional dipole ones,
\cf\citerefs{Catani:1996jh,Catani:1996vz,Catani:2002hc}. 
This means that the post-branching momenta are constructed as 
\begin{align}
	p_i^\mu &= \left(E_i,0,0,\vert \vec p_i \vert\right) \, , \\[1mm]
	p_j^\mu &= \left(E_j,-\vert \vec p_j \vert \sin \theta_{ij},0,\vert \vec p_j \vert \cos \theta_{ij}\right) \, , \\[1mm]
	p_k^\mu &= \left(E_k,\vert \vec p_k \vert \sin \theta_{ik},0,\vert \vec p_k \vert\cos \theta_{ik}\right) \, ,
\end{align}
in the rest frame of the parent $I$-$K$ antenna. Here, the energies are given by
\begin{equation}
	E_i = \frac{\sinv{ij}+\sinv{ik}+2m_i^2}{2m_{IK}} \, ,\qquad
	E_j = \frac{\sinv{ij}+\sinv{jk}+2m_j^2}{2m_{IK}} \, ,\qquad
	E_k = \frac{\sinv{ik}+\sinv{jk}+2m_k^2}{2m_{IK}} \, ,
\end{equation}
and the angles by
\begin{equation}
	\cos \theta_{ij} = 
	\frac{2E_iE_j - \sinv{ij}}{2\vert \vec p_i \vert \vert \vec p_j \vert} \, ,
	\qquad
	\cos \theta_{ik}  =  \frac{2E_iE_k - \sinv{ik}}{2\vert \vec p_i \vert \vert \vec p_k \vert} \, .
\end{equation}
Subsequently, the branching plane is rotated by an angle $\phi$, uniformly sampled in $[0,2\uppi]$, in the $x$-$y$ plane, and by an angle $\psi$ between the mother parton $I$ and the daughter parton $i$, which establishes the relative orientation of the post-branching partons with respect to the pre-branching ones. 
As the choice of $\psi$ is not unique away from the collinear limits, \vincia implements a few different options for $\psi$, \cf\citerefs{Giele:2007di,GehrmannDeRidder:2011dm}. In any of the choices, $\psi\to 0$ ensures that parton $i$ recoils purely longitudinally in the $K$-collinear limit and $\psi \to \pi - \theta_{ik}$ ensures that $k$ recoils purely longitudinally in the $I$-collinear limit.

\index{ISR!Kinematics}
For initial-state radiation, one (or both) parents must remain collinear to the beam axis, and instead the hard system can now acquire transverse recoil. This makes it more complicated to define a truly antenna-like recoil scheme, and \vincia's choices~\cite{Fischer:2016vfv,Verheyen:2020yor,Brooks:2020upa} are more similar to dipole treatments such as the ones in~\citerefs{Schumann:2007mg,Dinsdale:2007mf,Platzer:2009jq,Hoche:2015sya}.
In the case of IF antennae, this amounts to constructing the post-branching momenta as
\begin{align}
  x_a &= x_A / y_{AK}
  ~~~~~~(\implies
  p_a^\mu = \frac{1}{\yinv{AK}}p_A^\mu) \, , \\[1mm]
	p_j^\mu &= \frac{(\yinv{ak}+\mu_j^2-\mu_k^2)+(\yinv{ak}-\yinv{aj})\mu^2_K-\yinv{AK}\yinv{ak}}{\yinv{AK}}\,p_A^\mu + \yinv{aj}\, p_K^\mu + \sqrt{\Gamma_{ajk}}\,q_{\perp\mathrm{max}}^\mu \, , \\[1mm]
	p_k^\mu &= \frac{(\yinv{aj}-\mu_j^2+\mu_k^2)+(\yinv{aj}-\yinv{ak})\mu_K^2-\yinv{AK}\yinv{aj}}{\yinv{AK}}\,p_A^\mu + \yinv{ak}\, p_K^\mu - \sqrt{\Gamma_{ajk}}\,q_{\perp\mathrm{max}}^\mu \, ,
\end{align}
in the $A$-$K$ rest frame, as illustrated in the second row in
\cref{fig:ps:vinciaRecoils}. In this context,
$\Gamma_{ajk} = \yinv{aj} \yinv{jk} \yinv{ak}$ and $q_{\perp\mathrm{max}}^\mu$
denotes the transverse component in terms of a spacelike four-vector
that is perpendicular to $p_A$ and $p_K$ and obeys
$q_{\perp\mathrm{max}}^2 = -(\sinv{aj}+\sinv{ak})$. 

For II antennae, both initial-state particles are evolved at the same
time and therefore both momentum fractions change simultaneously~\cite{Daleo:2006xa,Ritzmann:2012ca}, \cf the third row in
\cref{fig:ps:vinciaRecoils} for an illustration. (Note that
this is different to ``dipole'' kinematics, in which only one of the
incoming $x$ fractions can change in each branching.)
Consequently, the post-branching momenta are constructed as 
\begin{align}
p_B^\mu \, , \\
x_a &= x_A / z_a
~~~~~~(\implies
p_a^\mu = p_A^\mu/z_a)
\, , \\[1mm]
x_b &= x_B / z_b
~~~~~~(\implies
p_b^\mu =p_B^\mu/z_b)
\, ,\\[1mm]
p_j^\mu &= y_{jb} p_a^\mu + y_{aj} p_b^\mu
+ \sqrt{\yinv{aj}\yinv{jb}-\mu_j^2}\,q_{\perp\max}^\mu
\, , \\[1mm]
p_r^\mu &= p_a^\mu +p_b^\mu - p_j^\mu,
\end{align}
where the $z_{a,b}$ fractions are defined in \cref{eq:vinciaIIz},
$q_{\perp\mrm{max}}^\mu$ is again a 
spacelike four-vector perpendicular to $p_A$ and $p_B$ with
$q_{\perp\mathrm{max}}^2 = -\sinv{ab}$, and 
$r$ denotes the recoiling spectator system whose combined invariant
mass and rapidity are both unchanged by the branching: $p_r^2 = p_R^2$ and
$y_r = y_R$.

In both IF and II antennae, all momenta are rotated about the
branching plane by a uniformly distributed angle $\phi \in
[0,2\uppi]$. 

\index{FSR!Kinematics}
For RF antennae~\cite{Brooks:2019xso}, the invariant mass of
the resonance must be kept fixed, $p_a^2 = p_A^2 = m_A^2$. The
post-branching kinematics are therefore constructed in the resonance
rest frame with the $z$-axis defined along $p_K$, so that 
\begin{align}
	p_A^\mu &= p_a^\mu = \left( m_A , 0, 0, 0 \right)\,, \\
	p_k^\mu &= \left( E_k , 0, 0, \sqrt{E^2_j - m^2_k} \right)\,, \\
	p_j^\mu &= \left(E_j , \sqrt{E^2_j - m^2_j} \sin \theta_{jk}, 0, \sqrt{E^2_j - m^2_j}\cos{\theta_{jk}} \right)\, , \\
	p_r^\mu &= \left(m_A - E_k - E_j , -\sqrt{E^2_j - m^2_j} \sin \theta_{jk}, 0, -\sqrt{E^2_j - m^2_k} -\sqrt{E^2_j - m^2_j}\cos{\theta_{jk}} \right)\, ,
\end{align}
where $r$ denotes the remainder of the resonance decay system and 
\begin{equation}
	E_j = \frac{\sinv{aj}}{2m_a}, \quad E_k = \frac{\sinv{ak}}{2m_a}, \quad \cos \theta_{jk} = \frac{2 E_b E_g -\sinv{jk}}{2 \sqrt{(E_k^2 -m_k^2)(E_j^2-m_j^2)}}\, .
\end{equation}
These momenta are rotated about the $y$ axis such that the set of recoilers are along -$z$, so that only $j$ and $k$ receive transverse recoil. 
Again, the momenta are subsequently rotated by a uniformly sampled
angle $\phi \in [0,2\uppi]$ about the $z$ axis. The original
orientation of $p_K$ with respect to $z$ is then recovered in a final
step. This map is illustrated in the bottom row of
\cref{fig:ps:vinciaRecoils}. 

\paragraph{Helicity Dependence}
All of \vincia's QCD and EW (but not QED) antenna functions
\index{Antenna functions} are implemented with full helicity
dependence, \ie decomposed 
into distinct terms  for each set of contributing helicities.
This facilitates helicity-dependent showering and matching, given a
polarized Born state~\cite{Fischer:2017htu}. For brevity, the QCD antenna
functions shown below are averaged over pre-branching helicities
and summed over post-branching ones; see~\oneref{Brooks:2020upa} for
details on their individual helicity components. 

\paragraph{Biased branchings and uncertainty weights}
\index{Enhanced splittings!in Vincia@in \vincia}
\index{Weights!in Vincia@in \vincia}
Just as for the simple shower, \vincia contains several options for
artificially increasing (or suppressing) the probabilities for
different branching types to occur, accompanied by non-unity event
weights to compensate for how over- or under-represented each generated
event becomes in the resulting sample.
This can be especially useful to enhance the rate of rare
splittings, such as $g\to b\bar{b}$. The general procedure is
described in \cref{sec:biasedKernels} and
follows the formalism presented in~\oneref{Mrenna:2016sih}.

As a relatively minor extension, \vincia also allows for ``enhancement'' factors
smaller than unity, which then act to suppress the corresponding
branchings. The intended use case is to focus on
Sudakov-suppressed regions of phase space.
In the algorithms described in~\citerefs{Lonnblad:2012hz,Mrenna:2016sih},
enhancement factors smaller than unity are not guaranteed to produce
positive weights. In \vincia's implementation, this issue
is sidestepped by letting trial branchings be enhanced by a factor
$\max(1,E)$,
\begin{equation}
  \hat{P}_\mathrm{biased} = \max(1,E) \hat{P}~,
\end{equation}
where $\hat{P}$ is the unenhanced trial-generation probability density
and $E$ is the enhancement (or suppression) factor. 
Thus for $E<1$ the trial probability is not modified. Conversely, each
trial branching is accepted with a probability
\begin{equation}
  P_\mathrm{acc} = \frac{\min(1,E) P}{\hat{P}}~, 
\end{equation}
where $P/\hat{P}$ is the unbiased accept probability.
The reweighting factor for an accepted trial branching
remains $R_\mathrm{acc} = 1/E$ (\cf\cref{sec:biasedKernels}),
while the reweighting factor for a discarded one generalizes to  
\begin{equation}
  R_\mathrm{disc} = \frac{\hat{P}_\mathrm{biased} - P}{\hat{P}_\mathrm{biased} - EP}~.
\end{equation}
Thus, in \vincia's version of the enhancement algorithm, both
$R_\mathrm{acc}$ and  $R_\mathrm{disc}$ are positive definite for any
$E > 0$ and $P<\hat{P}$. 

Although automated shower-variation weights were a signature feature
of early versions of \vincia~\cite{Giele:2011cb},
such variations have not yet been incorporated into the current \vincia
implementation in \pythia but remain planned for a future revision. 
See the program's \htmlmanual for updates. 

\subsubsection{QCD showers}\label{sec:vinciaqcd}
In their present incarnation, \vincia's QCD showers are fully
developed within the so-called sector 
framework~\cite{Kosower:1997zr, Kosower:2003bh,
  Larkoski:2009ah, Larkoski:2011fd, LopezVillarejo:2011ap,
  Larkoski:2013yi, Brooks:2020upa}, in which only a single branching
contributes per phase-space point. This is enforced by dividing the
phase space into sectors according to a decomposition of unity as
given by the following sum of Heaviside step functions, 
\begin{equation} \label{eq:vincia-sector-step}
	1 = \sum\limits_j \thetasct_j(\pTs[j], \zeta_j, \phi_j) = \sum\limits_j \theta\left(\min\limits_k \{\qsqres{k}\} - \qsqres{j}\right) \, .
\end{equation}
\index{Sector resolution variable}To discriminate between the different
sectors, a ``sector resolution'' 
variable is used, which we define to
be~\cite{LopezVillarejo:2011ap} 
\begin{equation} \label{eq:vincia-sector-resolution}
	\qsqres{j} = \begin{cases}
		\displaystyle\pTs[j] & \mrm{if~} j \mrm{~a~gluon}\\[8pt]
		\displaystyle\bar q_{ij}^2 \sqrt{\frac{\bar q_{jk}^2}{\sinv{\mrm{max}}}} & \mrm{if~} (i,j) \mrm{~a~quark-antiquark~pair}
	\end{cases} \, ,
\end{equation} 
with $\pTs[j]$ and $\bar{q}_{ij}$ as defined in
\cref{eq:vincia-ordering-scales}. The asymmetric choice for
quark-antiquark pairs accounts for the fact that in gluon splittings
with an arbitrary colour-connected recoiler $X_I \g_K \mapsto X_i \q_j
\qbar_k$, there is no singularity associated to the $i$-$j$-collinear
limit~\cite{LopezVillarejo:2011ap}. 

The shower evolution is given by the exponentiation of leading-order
antenna functions\index{Antenna functions}~\cite{Campbell:1998nn,Kosower:2003bh,GehrmannDeRidder:2004tv,GehrmannDeRidder:2005hi,GehrmannDeRidder:2005aw,GehrmannDeRidder:2005cm,Daleo:2006xa,Larkoski:2009ah,Larkoski:2011fd},
specifically sector-antenna ones defined by the ratio of
colour-ordered squared amplitudes, 
\begin{equation}
\antfun{j/IK} = \gstrong^2 \colfac_{j/IK} \antfunbar{j/IK} = \frac{\abs{\ME_{ijk}(q; p_i, p_j, p_k)}^2}{\abs{\ME_{IK}(q; p_I, p_K)}^2} \, ,
\label{eq:vincia-barred-antenna}
\end{equation} 
and the coupling- and colour-factor-stripped antenna function
$\antfunbar{j/IK}$. Antenna functions for quark-antiquark,
quark-gluon, and gluon-gluon parents can be derived from off-shell
photon decays $\gamma \to \q \qbar$~\cite{GehrmannDeRidder:2004tv},
neutralino decays $\tilde{\chi}^0 \to \tilde{\gp{g}} \g$~\cite{GehrmannDeRidder:2005hi}, and Higgs decays $\H \to \g \g$~\cite{GehrmannDeRidder:2005aw}, respectively. An antenna function
derived in this way will include the full single-unresolved
singularity structure of all colour dipoles in the given
colour-ordered amplitude.

When multiple colour dipoles are present in the three-particle state used to derive the function, these can be divided into sub-antenna functions,
\begin{equation}
  \begin{array}{lcl}
	\antfun{\g/\q \g}(p_i,p_j,p_k) & = & \subantfunqgemit(p_i,p_j,p_k) + \subantfunqgemit(p_i,p_k,p_j) ~, \\
	\antfun{\g/\g \g}(p_i,p_j,p_k) & = &
        \subantfunggemit(p_i,p_j,p_k) + \subantfunggemit(p_i,p_k,p_j)
        + \subantfunggemit(p_j,p_i,p_k)~.
  \end{array}
  \label{eq:defSubAntFuns}
\end{equation}
\index{Gluon splittings@$\g\to\q\qbar$ splittings}Such functions build the basis for so-called \emph{global} antenna showers, in which every antenna radiates over all of its branching phase space, and only the sum of all antennae recovers the full single-unresolved singularity structure. Denoting these by $\subantfun{j/IK}$, the specific choices for global final-state antenna functions in \vincia are
\begin{align}
	\subantfunqqemit(\sinv{IK}; \yinv{ij}, \yinv{jk}, \musq{i}, \musq{k}) &= \frac{\gstrong^2 \colfac_{\g/ \q\qbar} }{\sinv{IK}}\left[ \frac{(1 - \yinv{ij})^2 + (1 - \yinv{jk})^2}{\yinv{ij} \yinv{jk}} + 1 - \frac{2\musq{i}}{\yinv{ij}^2} - \frac{2\musq{k}}{\yinv{jk}^2}\right] \, , \label{eq:VinciaQQemitGl}\\
	\subantfunqgemit(\sinv{IK}; \yinv{ij}, \yinv{jk}, \musq{i}) &= \frac{\gstrong^2 \colfac_{\g/\q\g} }{\sinv{IK}}\left[ \frac{(1 - \yinv{ij})^3 + (1 - \yinv{jk})^2}{\yinv{ij} \yinv{jk}} + 2 - \yinv{ij} - 
	\frac{\yinv{jk}}{2} - \frac{2\musq{i}}{\yinv{ij}^2}\right] \, , \label{eq:VinciaQGemitGl}\\
	\subantfunggemit(\sinv{IK}; \yinv{ij}, \yinv{jk}) &= \frac{\gstrong^2 \colfac_{\g/\g\g}}{\sinv{IK}}\left[ \frac{(1 - \yinv{ij})^3 + (1 - \yinv{jk})^3}{\yinv{ij} \yinv{jk}} + 3 - \frac{3}{2} \yinv{ij} - \frac{3}{2}\yinv{jk} \right] \, , \label{eq:VinciaGGemitGl}\\
	\subantfungxsplit(\sinv{IK}; \yinv{ij}, \yinv{jk}, \yinv{ik}, \musq{Q}) &= \frac{\gstrong^2 \colfac_{\q/X\g}}{2 \sinv{IK}}\frac{1}{\yinv{ij} + 2 \musq{Q}}\left[\yinv{ik}^2 + \yinv{jk}^2 + \frac{2 \musq{Q}}{\yinv{ij} + 2 \musq{Q}}\right] \label{eq:VinciaGXsplitGl}\, .
\end{align}
They only differ from the ones given in \citerefs{Campbell:1998nn,Kosower:2003bh,GehrmannDeRidder:2004tv,GehrmannDeRidder:2005hi,GehrmannDeRidder:2005aw,GehrmannDeRidder:2005cm,Daleo:2006xa,Larkoski:2009ah,Larkoski:2011fd}
by non-singular terms. Below, we show how \vincia's sector-antenna
functions, $A^\mathrm{sct}_{j/IK}$, are constructed from these
building blocks. 

\paragraph{Single-unresolved limits}
\index{Antenna functions}\index{DGLAP}
In the sector shower formalism, there is only a single branching kernel that contributes per phase-space point. In order to capture the correct leading-logarithmic structure of QCD matrix elements, it is therefore vital that sector-antenna functions fully incorporate all single-unresolved limits of a given antenna/dipole. 
This means, that a single sector-antenna function has to reproduce the full eikonal in the soft-gluon limit,
\begin{equation}
\sctantfun{j/IK}(\sinv{IK};\yinv{ij},\yinv{jk},\musq{i}, \musq{k}) \xrightarrow{g_j \text{ soft}} \gstrong^2  \colfac_{j/IK} \left[\frac{2\sinv{ik}}{\sinv{ij}\sinv{jk}} - \frac{2\msq{i}}{\sinv{ij}^2} - \frac{2\msq{k}}{\sinv{jk}^2}\right] \, ,
\end{equation}
while reproducing the full massive DGLAP splitting kernel $\dglap{I}{ij}(z, \musq{i})$ (or $\dglap{I}{ij}(z, \musq{i})/z$ for initial-state partons) in any (quasi-)collinear limit,
\begin{equation}
\sctantfun{j/IK}(\sinv{IK};\yinv{ij},\yinv{jk}, \musq{i}, \musq{k}) \xrightarrow{i\parallel j} \gstrong^2 \colfac_{j/IK} \frac{\dglap{I}{ij}(z, \musq{i})}{\sij} \, .
\end{equation}
This differs from conventional (non-sector)
parton-shower algorithms, in which the soft and/or collinear singularity
structures are partial fractioned onto different branching
kernels. \Eg in DGLAP-based and dipole approaches, the soft eikonal
is partial fractioned onto two separate kernels (which are associated
with two different recoil maps), and the same is true of
gluon-collinear singularities in both dipole and global-antenna
showers. In the sector-antenna formalism, each antenna function
reproduces both the full eikonal and the full DGLAP kernel in the
respective limits, and double counting is avoided by allowing only one
such antenna function to contribute per phase-space point. 

\paragraph{FSR antenna functions} 
\index{Antenna functions}
In \vincia, final-final (FF) sector-antenna
functions for gluon emissions are
constructed from their global counterparts, \crefrange{eq:VinciaQQemitGl}{eq:VinciaGXsplitGl}, by symmetrizing over colour-connected
gluons in the following way,
\begin{align}
	\sctantfun{g/IK}(\sinv{IK}; \yinv{ij}, \yinv{jk}, \musq{i}, \musq{k}) &= \subantfun{g/IK}(\sinv{IK}; \yinv{ij}, \yinv{jk}, \musq{i}, \musq{k}) \nonumber\\
	& \qquad + \delta_{I\g}\, \subantfun{\g/IK}(\sinv{IK}; \yinv{ij}, 1-\yinv{jk}, \musq{i}, \musq{k}) \label{eq:VinciaEmitSct}\\ 
	& \qquad + \delta_{K\g} \, \subantfun{\g/IK}(\sinv{IK}; 1-\yinv{ij}, \yinv{jk}, \musq{i}, \musq{k})\, , \nonumber
\end{align}
where $\delta_{I\g} = 1$ if $I$ is a gluon and zero otherwise, and 
similarly for $K$.
Note that the symmetrization is done on the CM energy fraction of the
relevant gluon(s), as in $(y_{jk} = 1 - x_i) \to (1-y_{jk} = x_i)$
in the symmetrization for $I \to ij$, instead of via explicit
permutations of the $i$ and $j$ momenta as in
\cref{eq:defSubAntFuns}, which would correspond to $y_{jk} \to
(y_{ik} = 1 - y_{jk} - y_{ij})$.  
This slight difference (which vanishes in the relevant collinear
limit $y_{ij} \to 0$) is to ensure finiteness in phase-space regions
close to the ``hard'' boundary $\yinv{ik}$. Although the $\yinv{ik} =
0$ region will never belong to the $j$-emission sector, this damping
of the singularity is important as it allows for the sampling of the sector-antenna function over all of phase space with a post-hoc imposed sector veto. Additionally, it ensures numerical stability whenever sector boundaries become close to the $\yinv{ik}$-singular region.

For gluon-splitting sector-antenna functions, an equivalent procedure yields 
\begin{equation}
	\sctantfungxsplit(\sinv{IK}; \yinv{ij}, \yinv{jk}, \yinv{ik}, \musq{Q}) = 2\, \subantfungxsplit(\sinv{IK}; \yinv{ij}, \yinv{jk}, \yinv{ik}, \musq{Q}) \, . \label{eq:VinciaSplitSct}
\end{equation}
Antenna functions for final-state partons that are
colour-connected to incoming ones, as in initial-final (IF) or
resonance-final (RF) colour flows, are discussed below. 

\paragraph{ISR antenna functions}\index{Antenna functions}
As for final-state radiation, sector-antenna functions involving 
initial-state partons can be obtained by symmetrizing corresponding
global ones over final-state gluons.
The reason initial-state legs do not need to be symmetrized 
is that there is no sector for ``emission into the initial state''.
(Analogously, while jet algorithms may decide to cluster final-state partons
either with each other or with the beam, the beam itself is hard by
definition and cannot be clustered away.)

This means that, even in the global-antenna approach, beam-collinear
singularities do not need to be partial-fractioned.  
Hence, for II antennae, there is no difference
between global and sector-antenna functions; while for initial-final
gluon emissions, antenna functions with two final-state gluons are
symmetrized as follows,  
\begin{align}
	\antfun{\g/AK}^{\mrm{sct,IF}}(\sinv{AK}; \yinv{aj}, \yinv{jk}, \musq{a}, \musq{k}) &= \antfun{\g/AK}^{\mrm{sct,IF}}(\sinv{AK}; \yinv{aj}, \yinv{jk}, \musq{a}, \musq{k}) \nonumber \\
		&\qquad + \delta_{\g K} \, \antfun{\g/AK}^{\mrm{sct,IF}}(\sinv{AK}; 1-\yinv{aj}+\yinv{jk}, \yinv{jk}, \musq{a}, \musq{k}) \, .
\end{align}
Finiteness close to the spurious $\yinv{ak} \to 0$ singularity is here again ensured by adding $\yinv{jk}$ to the symmetrized argument. Initial-final antenna functions describing final-state gluon splittings are obtained in exactly the same way as in \cref{eq:VinciaSplitSct}.

Global initial-final and initial-initial antenna functions are
obtained from \crefrange{eq:VinciaQQemitGl}{eq:VinciaGXsplitGl}
by crossing partons from the final state into the initial state.
For initial-initial antennae, the crossing $(I,K,i,k) \to
(-A,-B,-a,-b)$ implies:
\begin{align} \label{eq:vincia-II-crossing}
	\yinv{ij} = \frac{\sinv{ij}}{\sinv{IK}} \quad    & \to \quad \frac{-\sinv{aj}}{\sinv{AB}} = -\frac{\yinv{aj}}{\yinv{AB}} \, , \nonumber\\
	\yinv{jk} = \frac{\sinv{jk}}{\sinv{IK}} \quad  & \to \quad \frac{-\sinv{jb}}{\sinv{AB}} = -\frac{\yinv{jb}}{\yinv{AB}} \, , \\
	\yinv{ik} = \frac{\sinv{ik}}{\sinv{IK}} \quad  & \to \quad~ \frac{\sinv{ab}}{\sinv{AB}} = \frac{1}{\yinv{AB}} \, , \nonumber
\end{align}
while for initial-final antennae, the crossing $(I,i) \to (-A,-a)$ yields:
\begin{align} \label{eq:vincia-IF-crossing}
	\yinv{ij} = \frac{\sinv{ij}}{\sinv{IK}} \quad   & \to \quad  \frac{-\sinv{aj}}{-\sinv{AK}} = \frac{\yinv{aj}}{\yinv{AK}} \, , \nonumber \\
	\yinv{jk} = \frac{\sinv{jk}}{\sinv{IK}} \quad & \to \quad \frac{\sinv{jk}}{-\sinv{AK}} = -\frac{\yinv{jk}}{\yinv{AK}} \, , \\
	\yinv{ik} = \frac{\sinv{ik}}{\sinv{IK}}\quad  & \to \quad \frac{-\sinv{ak}}{-\sinv{AK}} = \frac{\yinv{ak}}{\yinv{AK}} \, . \nonumber
\end{align}
The RF antenna functions are identical to the IF ones.
The full set of \vincia antenna functions, including their helicity contributions, can be found in~\oneref{Brooks:2020upa}.\index{Antenna functions}

\paragraph{The strong coupling}\index{alphaS@$\alphas$!in Vincia@in \vincia}

\vincia offers the same basic options for the strong coupling as the
simple shower does, with up to 2-loop running matched across flavour
thresholds and an option to use the CMW scheme. However, whereas the
main tuneable parameters in the simple shower are the effective values of
$\alphas^\mathrm{ISR}(M^2_Z)$ and $\alphas^\mathrm{FSR}(M^2_Z)$ (which
may then be interpreted as being given in a renormalization scheme not
necessarily identical to $\overline{\mathrm{MS}}$), in \vincia one
instead specifies a single common value for
$\alphas^\mathrm{\overline{MS}}(M^2_Z)$ --- normally just set to agree
with a reasonable global average value such as that given by the 
PDG~\cite{Zyla:2020zbs} --- with different effective values 
for different branching types obtained via user-specifiable
renormalization-scale prefactors,  
\begin{equation}
  \alphas^\mathrm{\overline{MS}}(M^2_Z) \to \left\{
  \begin{array}{lcp{5cm}}
    \alphas(\,k_\mrm{E}^\mrm{F} \pTs[j] + \mu_0^2 \,) & \mbox{for} & FF and RF gluon emissions~, \\[1.5mm]
    \alphas(\,k_\mrm{S}^\mrm{F} \pTs[j] + \mu_0^2\,) & \mbox{for} & final-state
    gluon splittings~, \\[1.5mm]
    \alphas(\,k_\mrm{E}^\mrm{I} \pTs[j] + \mu_0^2\,) & \mbox{for} & II and IF gluon emissions~, \\[1.5mm]
    \alphas(\,k_\mrm{S}^\mrm{I} \pTs[j] + \mu_0^2\,) & \mbox{for} & initial-state gluon splittings~, \\[1.5mm]
    \alphas(\,k_\mrm{C}^\mrm{I} \pTs[j] + \mu_0^2\,) & \mbox{for} & initial-state
   gluon conversions~, 
  \end{array} \right.
\end{equation}
where the scheme can be either $\overline{\mathrm{MS}}$ or CMW and
$\mu_0 \sim {\cal O}(\Lambda_\mathrm{QCD})$ is a fixed scale that 
forces the effective coupling to asymptote to
$\alphas(\mu_0^2)$  for $\pT[j]\to 0$. A maximum value can also be
specified beyond which $\alphas$ is not allowed to grow, effectively
freezing the coupling at that value in the infrared. 

\paragraph{Evolution equations} The differential branching probability as implemented by the sector shower is given as the sum of individual $IK \mapsto ijk$ antenna branching probabilities,
\begin{equation}
	\frac{\deriv \prob}{\deriv \pTs[j]} = \sum\limits_j \frac{\deriv \prob_{j/IK}}{\deriv \pTs[j]} \, ,
\end{equation}
which can be written in terms of the shower evolution variable $\pTs[j]$
and an arbitrary complementary phase-space variable $\zeta$ as
\begin{equation}
	\frac{\deriv \prob_{j/IK}}{\deriv \pTs[j]} = \frac{\alphas(\pTs[j])}{4\uppi}\, \colfac_{j/IK}\, \int\limits_{\zeta_{\mrm{min}}(\pTs[j])}^{\zeta_{\mrm{max}}(\pTs[j])} \int\limits_{0}^{2\uppi}  \sctantfunbar{j/IK}(\pTs[j], \zeta)\, \RPDF \, \FPS \, \abs{J(\pTs[j], \zeta)} \, \thetasct(\pTs[j], \zeta, \phi)\, \frac{\deriv \phi}{2\uppi} \, \deriv \zeta \, .
\label{eq:vincia:branchingProbQCD}
\end{equation}
Here, the Jacobian $J(\pTs[j],\zeta)$ accounts for the change to the shower
variables $(\yinv{ij}, \yinv{jk}) \mapsto (\pTs[j], \zeta)$, for which
different choices are implemented in \vincia, depending on the
branching type, \cf\citeref[section 2.5]{Brooks:2020upa}. Note that,
since the starting point is an exact phase-space factorization and the  
Jacobian factor $J(\pTs[j],\zeta)$ accounts for the mapping to shower
variables, there is no physical dependence on the choice of $\zeta$ in
\vincia; it only affects how simple or complicated the trial integrals
become, and the efficiency with which trial branchings can be generated. 
The phase-space factor
\begin{equation}
\FPS = \begin{cases}
\displaystyle\sinv{IK} \frac{\sinv{IK}}{\sqrt{\kallen{\sinv{IK}}{\msq{i}}{\msq{k}}}} & \mrm{FF} \\[8pt]
\displaystyle\frac{\sinv{AK}+\msq{j}+\msq{k}-\msq{K}}{(1-\yinv{jk})^3} \frac{\sinv{AK}+\msq{j}+\msq{k}-\msq{K}}{\sqrt{\kallen{\msq{A}}{\msq{AK}}{\msq{K}}}} & \mrm{RF} \\[8pt]
\displaystyle \frac{\sinv{AK}}{1-\yinv{jk}} & \mrm{IF} \\[8pt]
\displaystyle \frac{\sinv{AB}}{1-\yinv{aj}-\yinv{jb}} & \mrm{II} 
\end{cases}
\end{equation}
accounts for the relative size of the post-branching phase space to the pre-branching phase space.
For ISR, a PDF ratio is included for every initial-state parton,
\begin{equation}
	\RPDF = \begin{cases}
		\displaystyle 1 & \mrm{FF~\&~RF} \\[8pt]
		\displaystyle \frac{\fpdf_a(x_a, \pTs[j])}{\fpdf_A(x_A, \pTs[j])} & \mrm{IF} \\[8pt]
		\displaystyle \frac{\fpdf_a(x_a, \pTs[j])}{\fpdf_A(x_A, \pTs[j])}\frac{\fpdf_b(x_b, \pTs[j])}{\fpdf_B(x_B, \pTs[j])} & \mrm{II}
	\end{cases}~.
\end{equation}

Two things should be noted in
\cref{eq:vincia:branchingProbQCD}. First, the colour factor
$\colfac_{j/IK}$ is normalized such that the integral prefactor is
always $1/4\uppi$ (as opposed to $1/2\uppi$), with the specific
\vincia choices being 
\begin{align}
	\colfac_{\g/ \q\qbar} &= 2 \CF = \frac{8}{3} \, , \\
	\colfac_{\g/ \q\g} &= \frac{1}{2}\left(2\CF+\CA\right) = \frac{17}{6}\, , \\
	\colfac_{\g/\g\g} &= \CA = 3\, , \\
	\colfac_{\q/X\g} &= 2 \TR = 1\, ,
\end{align}
where an interpolation between $2\CF$ and $\CA$,
\begin{equation}
	\colfac_{\g/ \q\g} = \frac{(1-\yinv{ij})2\CF + (1-\yinv{jk})\CA}{2-\yinv{ij}-\yinv{jk}}
\end{equation}
is available for $\q \g$ antennae.
Second, the azimuthal integration is made explicit although the antenna functions have no azimuthal dependence. This is to emphasize a potentially non-trivial azimuthal dependence of the sector veto $\thetasct$.

\paragraph{Matching, merging, and matrix-element corrections}
A unique property of \vincia's sector-based approach to parton showers
is that there is only a very small number of ``shower
histories'' leading to each distinct parton configuration. For gluon
emissions, \vincia's sector shower is entirely bijective, \ie there is
only a single unique shower history leading from the Born to any given
Born+$n$-gluon parton configuration. For $g\to q\bar{q}$ splittings,
one has to sum over all possible same-flavour quark-antiquark pairings, 
but the number of contributing histories for a given phase-space point
is still drastically reduced relative to conventional, non-sectorized,
showers. We say that the sector shower is ``maximally
bijective'', and this provides an optimal framework for
high-multiplicity matching and merging, discussed further in
\cref{sec:VinciaMatchMerge}.   

\paragraph{Infrared cutoffs}\index{Shower cutoff scale|see{Cutoff scales}}%
\index{Cutoff scales!in Vincia@in \vincia}
For $\pT$ scales below 1~GeV or so, perturbative
approximations become increasingly inaccurate as $\alphas(\pT)$
shoots towards divergence at $\Lambda_\mathrm{QCD} \sim 200 -
300$~MeV. Like for the simple shower model,
\vincia's perturbative shower evolution is therefore also halted some
distance above $\Lambda_\mathrm{QCD}$, at which point the parton
system is handed to \pyt's string-fragmentation model for
hadronization. In \vincia, the precise scale at which the shower is
stopped can be set independently for FF, IF, and II antennae. 

The shower cutoff for FF antennae in \vincia is analogous to the FSR cutoff in
the simple-shower model. It can be regarded as
the effective factorization scale between the perturbative and
non-perturbative parts of the overall fragmentation description. It
therefore has the interpretation as the scale at which the
parameters of the non-perturbative hadronization modelling are defined.  
Ideally, the hadronization parameters should ``run'' with the shower
cutoff, but since the relevant running equations are not known, in practice
the hadronization parameters simply have to be retuned for each new
value of the cutoff. 
In other words, the FF-cutoff value can be considered part of the
fragmentation tuning. In general, one would seek not to leave too much of
a gap between the lowest \pT scales generated by shower branchings
(down to the cutoff) and the highest  \pT scales generated by string
breaks (with typical size set by the fragmentation \pT width,
\cf\cref{sec:had-flavour}).  

\index{Primordial kT@Primordial $k_\perp$!Interplay with showers}
For II antennae, the cutoff can be regarded as an effective
colour-screening resolution scale, or a lowest scale for which partons
inside hadrons can be said to be well represented by plane waves. This
could possibly be tied to the physics of parton saturation, though no explicit
such connection is made here. The practical considerations
are similar as for the ISR shower cutoff in the simple-shower model,
striking a balance between \pT kicks generated by the shower and
contributions from so-called ``primordial $k_\perp$'',
\cf\cref{sec:primordialKT}. 

For IF antennae, the fact that \vincia's default recoil strategy is
fully local, and does not impart \pT recoil to any partons outside
of the $2\to 3$ branching itself, leads to some pathologies. In particular,
each IF branching \emph{dilutes} the  primordial $k_\perp$, and does
not add any perturbative \pT of its own, to the hard system. The
non-smooth interplay between the II and IF recoil strategies can make
it challenging to describe the soft peak of experimental signals such
as the Drell--Yan \pT spectrum, and can produce seemingly
counter-intuitive scaling with the value of the IF cutoff.  

\subsubsection{QED
  showers}\label{sec:VinciaQED}\index{QED showers!in Vincia@in \vincia}\index{Coherence}
The \vincia shower offers a number of options for the inclusion of
electromagnetic and weak corrections. They all share common
features, like the  
phase-space factorizations and ordering scale, with the QCD shower, \cf\cref{sec:vinciacommon}. In this section, we describe the first (and
default) option, which is a pure QED shower that incorporates a fully
coherent multipole treatment of the simulation of photon radiation off
systems of charged fermions, vectors, and scalar particles, as well as
photon splittings to pairs of charged
fermions~\cite{Kleiss:2017iir,Skands:2020lkd}. We also include
a simpler and somewhat faster alternative, in which the full
multipole sum is replaced by individual dipole terms according to a
principle of maximal screening, analogous to how QED is handled
in the simple-shower model.  

The basic building block for \vincia's treatment of photon radiation
is the photon-emission antenna function for a single pair of
final-state charged particles $i$ and $k$,  
\begin{align}
	\sctantfun{\gamma/IK}(\sinv{IK}; \yinv{ij}, \yinv{jk}, \musq{i}, \musq{k}) = 2 g^2 \frac{ Q_I Q_K}{\sinv{IK}} \bigg[ &2\frac{\yinv{ik}}{\yinv{ij}\yinv{jk}} - 2\frac{\musq{i}}{\yinv{ij}^2} - 2\frac{\musq{k}}{\yinv{jk}^2} + \delta_{I \f}\frac{\yinv{ij}}{\yinv{jk}} + \delta_{K \f}\frac{\yinv{ij}}{\yinv{jk}} \nonumber \\
	& + \delta_{I \W} \frac{4}{3} \yinv{ij} \left( \frac{\yinv{jk}}{\yinv{IK} - \yinv{jk}} + \frac{\yinv{jk}(\yinv{IK} - \yinv{jk})}{\yinv{IK}^2} \right) \nonumber \\
	& + \delta_{K \W} \frac{4}{3} \yinv{jk} \left( \frac{\yinv{ij}}{\yinv{IK} - \yinv{ij}} + \frac{\yinv{ij}(\yinv{IK} - \yinv{ij})}{\yinv{IK}^2} \right) \bigg]~,
\end{align}
where the Kronecker deltas ensure the correct collinear terms are incorporated in the cases where $I$ and $K$ are fermions or \W bosons. 
The factors $Q_I$ and $Q_K$ represent the relative electromagnetic charges of $I$ and $K$, respectively. 
The II and IF antennae may be found by crossing symmetry following \cref{eq:vincia-II-crossing,eq:vincia-IF-crossing}.
For notational convenience we define
\begin{equation} \label{eq:vincia-barred-qed-antenna}
\sctantfun{\gamma/IK}(\sinv{IK}; \yinv{ij}, \yinv{jk}, \musq{i}, \musq{k}) = g^2 Q_I Q_K \sctantfunbar{\gamma/IK}(\sinv{IK}; \yinv{ij}, \yinv{jk}, \musq{i}, \musq{k}),
\end{equation} 
similar to the QCD equivalent \cref{eq:vincia-barred-antenna}.

While possibly counter intuitive, the definition of a coherent QED
shower using \cref{eq:vincia-barred-qed-antenna} is not as
straightforward as for its (leading-colour) QCD shower counterpart. 
The reason is the absence of an equivalent of the  leading-colour
approximation, which in QCD allows one to discard the majority of the
soft eikonal contributions that are subleading in colour.  
Conversely, in QED no eikonal is subleading to any other, and full coherence can only be accomplished by the inclusion of all of them simultaneously. 
\vincia's most sophisticated photon emission algorithm accomplishes this by the definition of a single branching kernel 
\begin{equation}
\sctantfunbar{\gamma/\text{coh}} = \sum_{\{IK\}} \sigma_I Q_I \sigma_K Q_K \sctantfunbar{\gamma/IK}(\sinv{IK}; \yinv{ij}, \yinv{jk}, \musq{i}, \musq{k})~,
\end{equation} 
where $\{IK\}$ runs over all pairs of charged particles, and $\sigma_I$ and $\sigma_K$ are sign factors that have $\sigma_I=1$ for final-state particles and $\sigma_I=-1$ for initial-state particles.
This branching kernel includes all soft multipole terms, as well as the correct collinear limits~\cite{Kleiss:2017iir}, but its singular structure is highly complex.
The coherent algorithm is able to sample it by sectorizing the phase space according to 
\begin{align}
	\frac{\deriv \prob_{j,\text{coh}}}{\deriv \pTs[j]} = \sum\limits_{\{ik\}} \int\limits_{\zeta_{\mrm{min}}(\pTs[j])}^{\zeta_{\mrm{max}}(\pTs[j])} \int\limits_{0}^{2\uppi} &\frac{\alphaem(\pTs[j])}{4\uppi}\, \sctantfunbar{\gamma/\text{coh}}(p_{\perp}^2, \zeta) \nonumber \\
	&\times \RPDF \, \FPS \, \abs{J(\pTs[j], \zeta)} \, \thetasct_{ik}(\pTs[j], \zeta, \phi)\, \frac{\deriv \phi}{2\uppi} \, \deriv \zeta \, ,
	\label{eq:vincia-qed-coherent-prob}
\end{align}
where $\thetasct_{ik}(\pTs[j], \zeta, \phi)$ is given by \cref{eq:vincia-sector-step}, but with a sum over charged-particle pairs, and the sector resolution is the same as that of a gluon emission as given by \cref{eq:vincia-sector-resolution}.
This procedure ensures the soft and collinear singularities are correctly regularized by the transverse momenta of the photon with respect to all pairs of charged particles.
The coherent emission algorithm is the default choice, but in some specific high-multiplicity cases it may be slow due to the large number of sectors that need to be sampled. 

As a backup, a faster, unsectorized alternative is implemented that rephrases the photon emission probability as 
\begin{align}
	\frac{\deriv \prob_{\text{pair}}}{\deriv \pTs[j]} = \sum\limits_{[IK]} \int\limits_{\zeta_{\mrm{min}}(\pTs[j])}^{\zeta_{\mrm{max}}(\pTs[j])} \int\limits_{0}^{2\uppi} \frac{\alphaem(\pTs[j])}{4\uppi}\, Q_{[IK]}^2 \sctantfunbar{\gamma/IK}(\pTs[j], \zeta) \RPDF \, \FPS \, \abs{J(\pTs[j], \zeta)} \, \frac{\deriv \phi}{2\uppi} \, \deriv \zeta \, ,
	\label{eq:vincia-qed-pairing-prob}
\end{align}
where $[IK]$ now runs over all pairings of charged particles with identical but opposite charge $Q_{[IK]}$.
The factors $\sigma_I$ and $\sigma_K$ have been absorbed into the definition of $Q_{[IK]}$, meaning that a final-state charged particle may be paired with a same-sign initial-state particle.
That is, every charged particle now only appears once, and pairings are constructed to minimize the sum of dipole-antenna invariant masses as per the principle of maximum screening~\cite{Skands:2020lkd}.
The task of pairing particles under such a constraint in
$\mathcal{O}(n^3)$ time complexity is accomplished using the Hungarian
algorithm~\cite{Kuhn55thehungarian,Munkre1957s,Jonker1987}. 
While this algorithm is generally faster, it only approximates the complete multipole structure. 
Furthermore, it may not always be possible to pair up all charges.
For instance, in a $\Wp \rightarrow \u \dbar$ decay, no pairings are possible at all.
In such cases, as many charges as possible are paired up, and the fully coherent algorithm is used for the remainder. 

The QED shower also includes photon splittings to charged fermions, which use antennae that are kinematically identical to their gluon-splitting counterparts.
Furthermore, while photon radiation off quarks is cut off at a scale
of order the hadronization scale, leptonic photon radiation continues
to much lower scales and has its own cutoff scale. 
Since the system of leptons is not necessarily charge conserving by
itself, which is a requirement for the above algorithms, the pool of
charges is supplemented with the colour-neutral strings that enter the
hadronization stage. 
When acting as the recoiler of a lepton, the antenna function is
replaced by a dipole function that only contains the singular limits
relevant to the lepton.

QED radiation off charged hadrons and/or in hadron decays, is not
present in the current implementation but may be included in future
work; see the program's \htmlmanual for updates. 

\subsubsection{EW showers}\label{sec:VinciaEW}
\index{Weak showers!in Vincia@in \vincia}
As an alternative to the coherent QED shower described above, \vincia
also offers the option to interleave the QCD shower with a
full-fledged EW shower, in which all possible branchings from the EW
sector are incorporated, albeit only in a collinear approximation
without any attempt at incorporating soft-interference
effects~\cite{Kleiss:2020rcg,Brooks:2021kji}. For each given
application, one must therefore choose whether weak-shower corrections
are more important than QED coherence effects for the study at hand,
the default choice being the coherent-QED one. 

When enabled, \vincia's EW option includes not only the branchings 
that are also available in the simple shower (heavy vector-boson
emissions off initial- and final-state fermions) but
also final-state triple vector-boson branchings, Higgs emissions, and
decay-like splittings. Like the QED module, the EW one also shares the
common features of the QCD shower, allowing for a sensible
interleaving of the two.  However, it is important to be aware that
the EW shower relies on the helicity-dependent evolution described in
\cref{sec:vinciacommon}, which must therefore also be enabled.
The resulting intermediate states of
definite helicity are vital in the EW sector due to its chiral nature.  
Helicity-dependent antenna functions are present for all EW
branchings, capturing their associated quasi-collinear limits. 
Due to the rich physics landscape of the EW sector of the SM and the
many different helicity combinations, there are hundreds of distinct polarized
collinear-splitting kernels. The antenna functions are therefore not
included here, but they may be found in~\oneref{Brooks:2021kji}. 
Note that, since the EW shower does not incorporate soft-interference effects, the antenna functions are more like dipole
functions, only including the single quasi-collinear limit of the
branching particle, while the other just functions as a
recoiler. 

A number of features unique to the EW sector are incorporated. 
For example, in a shower sequence like $\eminus \rightarrow \eminus \, \Z/\gam \rightarrow \eminus \Wp \Wm$, the interference between the $\Z$ and the $\gam$ can be of $\mathcal{O}(1)$~\cite{Chen:2016wkt,Kleiss:2020rcg}. 
A full treatment of this effect may for instance be accomplished by the evolution of density matrices, which can quickly become prohibitively expensive. 
\index{Weights!in Vincia@in \vincia}
\vincia instead implements a simplified approach, in which the emission probability is corrected at first order by an event weight.
This weight is computed using quasi-collinear amplitude-level branching amplitudes using the spinor-helicity formalism. 

These same amplitudes are also used to determine recoilers for the quasi-collinear branchings of the EW shower.
Unlike in the QCD sector, where recoilers are typically chosen to be a colour-connected parton, no such mechanism is available in the EW sector.
Furthermore, because the EW shower only models quasi-collinear branchings without soft-interference effects, the choice of recoiler is formally arbitrary.
One can however select the recoiler probabilistically such that the kinematic effects of recoil on previous branchings is minimized~\cite{Kleiss:2020rcg}.

Another peculiar feature of the EW sector is the fact that branchings like $\t \rightarrow \b \Wp$ and $\Z \rightarrow \q \qbar$ appear both as shower branchings and as resonance decays. 
For off-shellness scales $Q^2 = m^2 - m_0^2 \sim \Gamma^2$, the physics is best described by a Breit--Wigner distribution, while for scales above the electroweak scale $Q_{\text{EW}}^2$, the EW shower is most accurate. 
In the intermediate region a matching procedure is required.
When the EW shower produces a heavy resonance like one of the EW gauge bosons, a top quark, or a Higgs boson, its mass is sampled from a helicity-dependent Breit--Wigner distribution (see~\oneref{Brooks:2021kji} for details)
\begin{equation} \label{eq:vincia-EW-BW}
	\text{BW}(Q^2) \propto \frac{m_0 \Gamma(m)}{Q^4 + m_0^2 \Gamma(m)^2}~.
\end{equation}
This procedure mirrors the treatment of resonances that are part of
the hard process as described in \cref{sec:resonances}, which can
also be branched by the EW shower. 
The shower is matched to the Breit--Wigner distribution by applying a suppression factor $Q^4/(Q^2 + Q^2_{\text{EW}})^2$, and the resonance is decayed when the evolution scale reaches the sampled resonance off-shellness without generating an EW branching.
In that case, if the EW shower produced the resonance, the decay is distributed according to the appropriate helicity-dependent $1\rightarrow 2$ matrix element.
If instead the resonance was part of the hard process, the decay has already been generated and is inserted.

Finally, double counting issues appear with the inclusion of EW branchings in the shower. 
For instance, the state $\pp \rightarrow \V \V \j$ may be reached by starting from $\pp \rightarrow \V \V$ and performing an initial-state QCD emission, or from $\pp \rightarrow \V \j$ and performing an EW emission.
To avoid double counting such phase-space points, \vincia implements an overlap veto procedure that can be used when overlapping matrix elements are enabled. 
It is based on the $k_{T}$ jet algorithm~\cite{Catani:1993hr} distance measures, generalized to account for the massive states that appear in the EW sector, given by
\begin{align}
	d_{iB} &= k_{T,i}^2, \nonumber \\
	d_{ij} &= \min(k_{T,i}^2, k_{T,j}^2) \frac{\Delta_{ij}}{R} + |m_i^2 + m_j^2 - m_I^2|.
\end{align}
The distance between the beam and final-state particle $i$ is measured by $d_{iB}$, while $d_{ij}$ measures the distance between two final-state particles $i$ and $j$. 
If, for example, a gluon is emitted by the QCD shower, the distances with respect to its colour-connected partons are computed. 
Furthermore, the distances of all possible $2 \rightarrow 1$ EW clusterings of the state after the gluon emission are also evaluated.
If one of these distances is smaller than the QCD ones, then the current phase-space point should be populated by the EW shower rather than the QCD shower, and the gluon emission is vetoed.
This procedure ensures no double counting occurs and the QCD and EW showers populate the regions of phase space they are most accurate in. 

\subsection{The \dire shower}\label{sec:Dire}
\index{Dire@\dire}
\index{Parton showers!Dire@\dire}

The \dire parton shower, introduced in \oneref{Hoche:2015sya}, offers another alternative showering model. It aims to combine aspects traditionally associated with $2\rightarrow 3$ dipole (antenna) showers with features of ``conventional'' $1\rightarrow 2$ parton showers. The goal of this hybrid is to inherit the modelling of soft-emission effects from dipole showers, while keeping an explicit association of splittings with specific collinear directions. This should, in principle, allow for an uncomplicated comparison to the ingredients in QCD factorization theorems. The physics aspects of \dire have been developed in a series of articles~\cite{Hoche:2015sya,Hoche:2017iem,Hoche:2017hno,Dulat:2018vuy,Prestel:2019neg,Andersen:2020sjs,Gellersen:2021caw}, and we refer the reader to these publications for details. Below, we will summarize the most important choices, virtues and current limitations.

\subsubsection{Phase-space coverage and ordering}\label{sec:direphasespace}
\index{Quark masses!in Dire@in \dire}
\dire employs an exact factorization of the single- and double-emission phase
spaces. The single-emission phase space is adapted from \citerefs{Dittmaier:1999mb,Catani:2002hc}, and allows for any combination of masses in $2\rightarrow 3$ branchings. The construction of post-emission momenta through the mapping $\mathrm{M}^{(1)}$ can be sketched by
\begin{equation}
\mathop{\mathlarger{\mathrm{M}^{(1)}}}
\left(\vcenter{\hbox{\includegraphics[width=.2\textwidth]{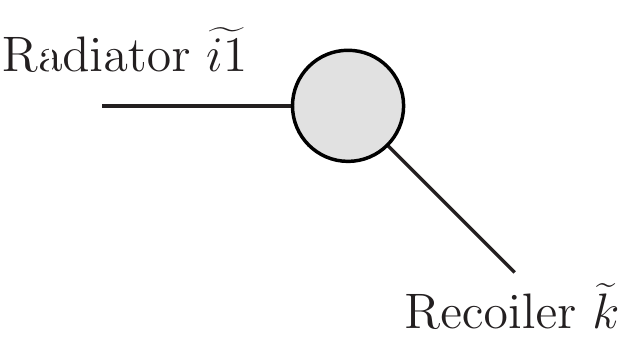}}}~\oplus~t^{(1)},z^{(1)},\phi^{(1)}\right) 
~~=~~
\vcenter{\hbox{\includegraphics[width=.2\textwidth]{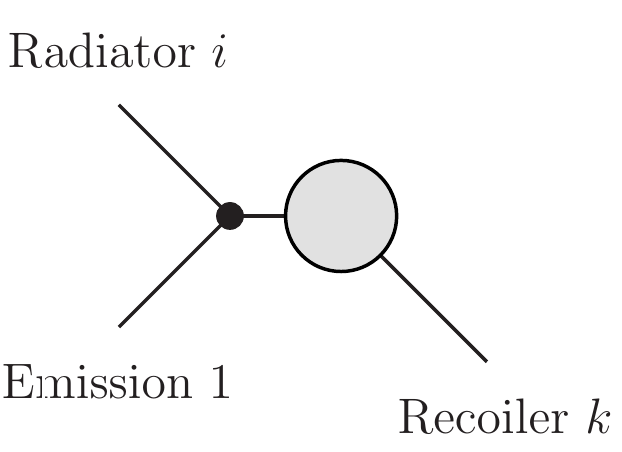}}}~,
\end{equation}
where $t^{(1)}$ is the evolution variable, $z^{(1)}$ a momentum-sharing variable, and $\phi^{(1)}$ an azimuthal angle. Note that under the mapping $\mathrm{M}^{(1)}$, the direction of the recoiler is not affected by the branching. Only its longitudinal momentum components change. This deliberate \emph{choice} ensures that the collinear direction defined by the recoiler, and consequently its mapping onto factorization theorems, remains intact. A caveat to this approach --- related to initial-state emissions --- is discussed below.

The double-emission phase space --- relevant for NLO parton evolution~\cite{Hoche:2017iem} --- may similarly be illustrated by a map $\mathrm{M}^{(2)}$
\begin{equation}
\mathop{\mathlarger{\mathrm{M}^{(2)}}}
\left(\vcenter{\hbox{\includegraphics[width=.2\textwidth]{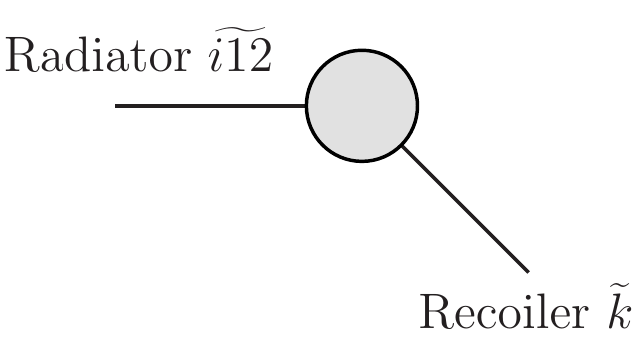}}}~\oplus~t^{(12)},z^{(12)},\phi^{(12)},s_{12},x,\phi'\right) 
~~=~~
\vcenter{\hbox{\includegraphics[width=.2\textwidth]{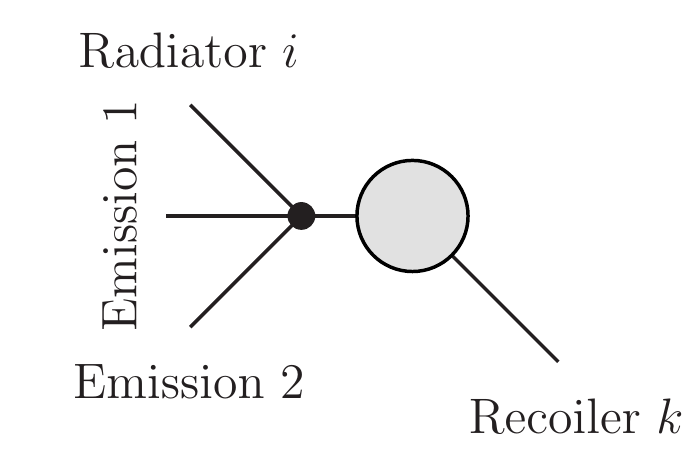}}}~.
\end{equation}
Here,  $t^{(12)}$ is the evolution variable assigned to the emission of the system $(12)$, $z^{(12)}$ ($\phi^{(12)}$) a momentum-sharing (azimuthal angle) variable, and $s_{12}$ the virtuality of the system $(12)$, while $x$ and $\phi'$ are related to the momentum sharing (azimuthal angle) between emissions $1$ and $2$. Again, the direction of the ``recoiler'' is preserved.

The momentum mappings $\mathrm{M}^{(1)}$ and $\mathrm{M}^{(2)}$ are (re)arranged to ensure that the phase-space coverage is fully symmetric between radiation from the ``radiator'' or the ``recoiler'', \ie given a fixed post-branching phase-space point and fixed branching variables, an identical pre-branching phase-space point is produced, independent of assigning the emissions to the radiator or recoiler. However, it should be noted that the momentum-sharing variables $z$ are \emph{not symmetric} under exchange of the ``emission'' for one of the other involved particles, since the limit $z\rightarrow 1$ is associated with a soft emission. 
\dire separates the generation of post-branching momenta into four distinct cases:
\begin{itemize}
\item[]{\bf FF}, \ie emission from a final-state particle, using a final-state recoiler:\\
This case has the fewest kinematic constraints, but the richest set of combinations of possible masses. Thus, the mapping is constructed to ensure that regions of phase space in which mass-corrected transition rates would lead to negative contributions are outside the physical phase-space boundaries. 
\item[]{\bf FI}, \ie emission from a final-state particle, using an initial-state recoiler:\\
This case again has few kinematic constraints, after the choice of keeping the recoiler direction intact. \dire will, if not instructed otherwise, treat incoming particles as massless for the purpose of phase-space generation. This means that in this configuration, negative transition rates due to mass corrections may occur. This is handled by a weighted shower algorithm.\index{Weights!in Dire@in \dire}
\item[]{\bf IF}, \ie emission from an initial-state particle, using a final-state recoiler:\\
This case has several kinematic constraints that need to be considered. In fact, the system is over-constrained if both the initial-state particle and the final-state recoiler should retain their directions. In this case, the transverse momentum generated in the branching can be balanced by extending the set of particles that may change their momentum. \dire offers the possibility to employ a ``global''-recoil strategy, in which the transverse momentum of the splitting is balanced by all final-state particles within the decaying system\footnote{Here, ``decaying system'' refers to the particle content of a single $2\rightarrow n$ scattering, in case several such scatterings exist due to the inclusion of multiparton interactions.}. It is also possible to instruct \dire to relax the condition that the final-state recoiler retains its direction. In this ``local'' strategy, the system of particles that change their momentum does not need to be extended. 
\item[]{\bf II}, \ie emission from an initial-state particle, using an initial-state recoiler:\\
This case also has several kinematic constraints, and is over constrained, since both initial-state particles should retain their directions. Here, no attempt is made to construct a ``local''-recoil strategy. Instead, the transverse momentum of the branching is collectively balanced by all final-state particles within the decaying system.
\end{itemize}
In general, the construction of post-branching momenta is subject to many choices. The choices above have foremost been guided by providing a simple procedure and Jacobian factors, such that analytic integrations of the emission patterns are as straight-forward as possible. This helps when improving the evolution with next-to-leading order corrections.

\index{Evolution variable!in Dire@in \dire}
The evolution variables in \dire are chosen to lead to a symmetric phase-space sampling and simple phase-space boundaries. Soft transverse momenta fulfil these criteria, if defined by
\begin{equation}
t^{(a)} \propto \frac{(p_ip_a) (p_kp_a)}{Q^2} \propto p_a^+p_a^-~,
\end{equation}
where $Q^2$ is a maximal scale, $p_a$ may be a sum of one or two emission momenta, and $p_i$ and $p_k$ are the radiator and recoiler post-branching momenta. Thus, in all of the four cases above, and for both single-emission and double-emission contributions, \dire employs soft transverse momentum as ordering variable, see~\citerefs{Hoche:2015sya,Hoche:2017iem} for details.

\subsubsection{Transition rates}\label{sec:direkernels}\index{DGLAP}

The \dire parton shower aims to model configurations containing soft particles or collinear configurations with high fidelity. As in a traditional parton shower, separate transition rates are used in each collinear direction. It might be helpful to explain this choice with an example. Imagine a dipole stretched between a quark and a gluon. The primary contributions to radiation collinear to the quark should be proportional to the colour factor $\CF$, while the radiation pattern collinear to the gluon should, up to small corrections, be proportional to $\CA$. Similarly, higher-order corrections to the radiation pattern in either region differ. 

\index{Coherence}However, simultaneously radiating from both dipole ``ends'' with the full rate expected in the collinear limit (given by the DGLAP splitting functions) will naively lead to an incorrect pattern in the soft limit. This problem is circumvented by replacing collinear-soft parts of DGLAP kernels by an improved description. The latter may be obtained by distributing the correct soft radiation pattern among all coherently radiating particles,
\begin{nonfloatfig}
\begin{tikzpicture}[remember picture]
\node (eikonal) [nobox] {\Large$\frac{(p_ip_k)}{(p_ip_a) (p_kp_a)} =$};
\node (propagator) [nobox, right=of eikonal,yshift=-0.275cm] {$\underbrace{\frac{1}{(p_ip_a)}}_{}$};
\node (partial) [nobox, right=of propagator] {$\underbrace{\frac{(p_ip_k)}{(p_ip_a) + (p_kp_a)}}_{}$};
\node (rest) [nobox, right=of partial,yshift=0.3cm] {$+~ (i\leftrightarrow k) $};
\node (kernel) [nobox, below=of partial] {\Large$\frac{2(1-z)}{(1-z)^2 + t/Q^2}$};
\path[every node/.style={font=\sffamily\small},->,>=stealth]
    (partial) edge [out=south, in=north]  (kernel);
\node (1byT) [nobox, below=of propagator] {\Large$ \frac{1 }{ t }$};
\path[every node/.style={font=\sffamily\small},->,>=stealth]
    (propagator) edge [out=south, in=north]  (1byT);
\node (dglap) [nobox, right=of kernel,xshift=3cm] {\Large$\frac{2}{1-z}$};
\path[every node/.style={font=\sffamily\small},->,>=stealth]
    (kernel) edge [out=east, in=west]  (dglap);
\node (description1) [description, below=of partial, yshift=0.8cm, xshift=1.7cm] {rewrite in $t$ and $z$};
\node (description2) [description, below=of propagator, yshift=0.8cm, xshift=-2.1cm] {combine with Jacobian};
\node (description3) [description, below=of dglap, yshift=1.1cm, xshift=-2.4cm] {collinear limit $t\rightarrow 0$};
\end{tikzpicture}
\end{nonfloatfig}
\noindent
\index{Quark masses!in Dire@in \dire}
The resulting soft-collinear pattern is supplemented with hard-collinear terms~\cite{Catani:1996vz}. As an extension of~\oneref{Catani:1996vz}, the $1/z$-terms present in DGLAP kernels are also shifted $\frac{1}{z}\rightarrow\frac{z}{z^2+t/Q^2}$ to ensure that sum rules for the splitting kernels are maintained. Finally, mass-dependent corrections based on~\oneref{Catani:2002hc} are added. The exact splitting kernels used in \dire are
listed in~\oneref{Hoche:2015sya}. The above chain of reasoning is used for all branching types in \dire.

\paragraph{QCD}
The above reasoning directly applies to QCD branchings at leading order, \ie when increasing the multiplicity by one particle, and while not including explicit virtual corrections to single-parton emission. At leading order:
\begin{itemize}
\item Dipoles are formed from radiator-recoiler pairs connected by a colour flow in the $\Nc\rightarrow\infty$ limit. At the point of compiling this manual, the fixed-colour corrections discussed in~\oneref{Gellersen:2021caw} have not been included in the core \pyt code.
\item Colour factors due to colour-charge correlators in the $\Nc\rightarrow\infty$ limit are given by:
\begin{enumerate}
\item gluon-radiation off (anti)quarks $\propto \CF$,
\item gluon-radiation off gluons $\propto \CA/\#($possible recoilers$) = \CA/2$,
\item and gluon branching to quark pairs $\propto \TR$.
\end{enumerate}
\item \index{alphaS@$\alphas$!in Dire@in \dire}Coupling-factors $\alpha_s$ for all QCD splittings are evaluated with dynamic arguments, with the preferred scheme being $\alpha_s(t)$. However, it should be noted that the emergence of the running coupling is driven by soft-gluon emissions, and thus, it is \textit{a priori} not obvious if the evaluation $\alpha_s(t)$ extends also to hard-collinear configurations. Thus, the user may instruct \pyt to use different arguments to evaluate $\alpha_s$: the running coupling may be evaluated using the ``collinear transverse momenta'' ${\bf k}_{\perp}^2$ defined as evolution variables in~\oneref{Schumann:2007mg}, \ie  $\alpha_s({\bf k}_{\perp}^2)$, or it may be evaluated using the strict definition of the (inverse) eikonal term, \ie $\alpha_s\left( \frac{(p_ip_a) (p_kp_a)} {(p_ip_k)}\right)$. 
\end{itemize}
The usage of a running coupling effectively includes ``universal'' virtual corrections to the emission rates. For inclusive soft-gluon emission, it is possible to include further next-to-leading order corrections rescaling the soft-gluon emission rate:
\begin{align}
&\frac{\alpha_s}{2\pi}\left[
\frac{2(1-z)}{(1-z)^2 + t/Q^2}
+ \left(\textnormal{hard-collinear terms}\right)
\right]
\nonumber\\
&\qquad \rightarrow~
\frac{\alpha_s}{2\pi} \left( 1 + K  \frac{\alpha_s}{2\pi}\right)
\frac{2(1-z)}{(1-z)^2 + t/Q^2}
~+~
\frac{\alpha_s}{2\pi}\,\cdot \left( \textnormal{hard-collinear terms}\right)~.
\end{align}
This may be considered as a conservative implementation of the conventional CMW (or MC) scheme~\cite{Catani:1990rr}. Note that different strategies for the evaluation of running couplings will induce \emph{different} higher-order corrections. Without better modelling, none of these \textit{ad hoc} choices are completely satisfactory.

Higher-order corrections to QCD evolution have been known for a long time. \dire implements several aspects of QCD evolution at next-to-leading order:
\begin{itemize}
\item Inclusive branching rates can be augmented with hard-collinear corrections by employing NLO DGLAP splitting functions. These improvements are available for both initial-state and final-state branchings. The benefit of such corrections is mainly in a more consistent treatment of PDF evolution in backwards initial-state evolution, since the latter relies on the parton shower distributing emissions according to the rates used to evolve externally pre-tabulated PDFs from low to high scales.    
\item Correlated triple-collinear emissions, \ie branchings of the form $1\oplus 1\rightarrow 3 \oplus 1 $, have been included to yield NLO DGLAP (initial- or final-state) evolution in the collinear limit from fully differential double-emission matrix elements.
\item Correlated double-soft emissions and explicit real-virtual corrections, \ie branchings of the form $2 \rightarrow 4 $ and $2\rightarrow 3$ at 1-loop, can be employed for final-state branchings. The inclusion of such NLO corrections is mainly a reduction of the renormalization-scale uncertainty of the parton shower. At the point of writing this manual, the consistent combination of triple-collinear and double-soft NLO corrections outlined in~\oneref{Gellersen:2021eci} has not been included in a public \pyt release.
\end{itemize}

\paragraph{QED}\index{QED showers!in Dire@in \dire}

The description of QED in \dire~\cite{Prestel:2019neg,Gellersen:2021caw} follows a very similar structure to that of QCD branchings, and is inspired by~\oneref{Schonherr:2017qcj}. 

\begin{itemize}
\item Dipoles are formed from all electrically charged radiator-recoiler pairs, much like the fixed-colour QCD dipole assignments discussed in~\oneref{Isaacson:2018zdi,Gellersen:2021caw}.
\item Charge-factors due to electric-charge correlators are determined from
\begin{alignat}{2}
Q^2 &= -\frac{\eta_{\tilde{i1}}\eta_{\tilde{k}} Q_{\tilde{i1}} Q_{\tilde{k}}  }{  Q_{\tilde{i1}}^2 } ~ &&\textnormal{(photon emission)} \nonumber\\
Q^2 &= \frac{1}{ \#\mathrm{recoilers}}&&\textnormal{(photon splitting)}~,
\end{alignat}
where $Q_{\tilde{i1}}$ and $Q_{\tilde{k}}$ are the charges of the radiator and recoiler, respectively, and $\eta_i=+1(-1)$ if $i$ is a final- or initial-state particle. These correlators multiply the splitting functions in place of the QCD colour factors, and may readily lead to negative contributions to the transition rates. Thus, a weighted shower algorithm is crucial for the QED modelling in \dire.\index{Weights!in Dire@in \dire}
\item Coupling-factors $\alpha_\mathrm{em}$ for all QED splittings are evaluated in the Thompson limit, \ie no running QED coupling is employed in the shower.
\end{itemize}

\paragraph{Kinetically mixed dark photons}
\index{Dark photons}\index{Parton showers!Dark photons}

\dire further implements kinetically mixed dark photon interactions, featuring dark photon emission from and decay into standard-model particles. These transitions are handled analogously to QED interactions, except that the dark photon may be massive. The decay width of the dark photon is currently ignored for both dark photon emission and decay. 

\paragraph{Electroweak effects}
\index{Weak showers!in Dire@in \dire}

Finally, \dire allows for electroweak-boson radiation and fermionic weak-boson decay, using a simplistic model similar to the ideas of~\citerefs{Christiansen:2014kba,Krauss:2014yaa}. Electroweak effects are mainly included because of the necessity for consistent matrix-element merging at LHC energies: to avoid an overly QCD-evolution biased scale setting for vector-boson plus jets configurations that exhibit giant $K$-factors~\cite{Rubin:2010xp}, the inclusion of parton-shower histories containing electroweak clusterings are mandatory for some showers~\cite{Schalicke:2005nv,Christiansen:2015jpa}. Thus, \dire implements electroweak evolution using:
\begin{itemize}
\item Transition rates are determined from partial-fractioned massive dipole kernels~\cite{Catani:2002hc}.
\item Dipoles are formed from all pairs of particles that may emit the same electroweak vector bosons. Electroweak-boson decays employ the same recoiler selection as vector-boson radiation, much like the QED case~\cite{Prestel:2019neg,Gellersen:2021caw}.
\item Coupling factors are calculated under the assumption of chirality-summed evolution, \cf\oneref{Krauss:2014yaa}. The coupling value is kept fixed, \ie no running coupling is employed.
\end{itemize}
This overly simplified model may appropriately handle electroweak history effects in the context of multi-jet merging, especially for $\Wpm$-boson plus multi-jet configurations at the LHC~\cite{Andersen:2020sjs}. Beyond this, studies of weak-boson radiation with \dire are discouraged.

\subsubsection{Weight handling aspects}\label{sec:diretech}
\index{Event weights|see{Weights}}
\index{Weights!in Dire@in \dire}
\index{Enhanced splittings!in Dire@in \dire}

The transition rates outlined in \cref{sec:direkernels} may be relatively complex. This results in some technical requirements that need to be met to produce a sound simulation since:
\begin{itemize}
\item it may not be possible to find efficient overestimates of complex transition kernels, such as for correlated double emission;
\item and it may not be possible to guarantee positivity, \eg due to mass effects, electric charge correlators, or higher-order corrections.
\end{itemize}
Both of these points (as well as the automated renormalization scale variations available in \dire) may be addressed with the help of a weighted veto algorithm, which was discussed in \cref{subsubsec:vetoalgorithm}. \dire employs this method more heavily than the rest of \pyt. Thus, the relevant features and extensions beyond the literature will be discussed here. 

The core realization of weighted shower algorithms is that acceptance
rates in the veto algorithm may be factored into a contribution that
is ``unweighted'' via the veto algorithm, and an event-by-event
``weight'' factor that encapsulates the effect of sign changes or
underestimations of the transition rate. To preserve inclusive cross
sections, this is naturally complemented by event-by-event weights
that augment the rejection rate of the parton shower. Once acceptance
and rejection weights have been introduced, it also becomes possible
to only partially unweight the use of overestimates through the veto
algorithm, and correct for the partial unweighting by amending the
event weight. This allows for an enhancement of certain transitions beyond their natural rate, leading to an improved statistical error, at the expense of a larger weight variance. Finally, the veto algorithm may be implemented in a series of distinct accept-reject steps. Each such step can be upgraded to incorporate complex rates\footnote{For example, the algorithm of~\oneref{Gellersen:2021caw} relies on a three-step (un)weighting.}. At present, \dire employs a weighted shower when choosing a branching according to splitting kernels, and another, stacked weighted shower to incorporate matrix-element corrections. A weighted approach to the latter is necessary as matrix-element corrections may induce sign changes, or because matrix elements are underestimated by the (sum of all possible) splitting kernels.

The weighted shower algorithm in \dire rests on the realization that the rate of producing one transition in the shower after $n$ rejections (through the veto algorithm) is given by
\begin{align}
\mathcal{P}(t) &= g(t)\frac{f(t)}{g(t)} \exp\left(-\int^{t_n}_{t} d\bar{t} g(\bar{t})\right) \prod_{i=1}^n \left[\frac{g(t_i) - f(t_i)}{g(t_i)}\right] g(t_i)\exp\left(-\int^{t_0}_{t_i} d\bar{t} g(\bar{t})\right)~.
\end{align}
This equation is the formal requirement for the validity of the veto algorithm, and does not strictly constrain the transition rates $f(t)$ by the ``overestimate'' $g(t)$ through $0<f(t)<g(t)$. The numerical implementation of the equation does, however, require sensible acceptance \emph{probabilities}. This may be achieved by introducing an auxiliary function $h(t)$ that guarantees acceptance probabilities $0<f(t)/h(t)<1$, see \eg also the description at the end of \cref{subsubsec:vetoalgorithm}. Rewriting in terms of this auxiliary function leads to
\begin{align}
\mathcal{P}(t) &=
\Bigg[g(t)\frac{f(t)}{h(t)}
\exp\left(-\int^{t_n}_{t} d\bar{t} g(\bar{t})\right)
\prod_{i=1}^n \left[ 1- \frac{f(t_i)}{h(t_i)}\right] g(t_i)
\exp\left(-\int^{t_0}_{t_i} d\bar{t} g(\bar{t})\right)\Bigg] \nonumber\\
&\otimes \frac{h(t)}{g(t)} \prod_{i=1}^n \frac{h(t_i)}{g(t_i)} \frac{ g(t_i) - f(t_i)}{ h(t_i) - f(t_i) }~.
\end{align}
The second line may be interpreted as a corrective factor due to disconnecting the sampling and rejection distributions. It does not have to be bounded, and is implemented as an event weight. It is important to note that
the acceptance rates $f(t)/h(t)$ only need to be bounded point-wise in $t$, \ie they may be adjusted depending on the value of $f(t)$. In particular, \dire uses
\begin{equation}
h(t)=
\begin{cases}
\mathrm{sign}[f(t)] g(t) & \textnormal{if}~ g(t) > |f(t)| \\
k f(t) & \textnormal{if}~ g(t) < |f(t)| \quad(\textnormal{with}~ k \sim 1.1)~.
\end{cases}
\end{equation}
An artificially enhanced sampling may be achieved by 
shifting $g\rightarrow g' = C g$, while keeping all rejection steps (\ie the definition of $h(t)$) fixed to their original values. The compensating event weight will then be shifted to
\begin{align}
\frac{1}{C}\frac{h(t)}{g(t)} \prod_{i=1}^n \frac{h(t_i)}{g(t_i)} \frac{ g(t_i) - f(t_i)/C}{ h(t_i) - f(t_i) } ~.
\end{align}
\index{Uncertainties!Parton showers@in Parton showers}
Once the weighted shower is in place, parton-shower variations may be included by keeping track of multiple weights of the form
\begin{align}
w^{[k]} =
\frac{1}{C}\frac{h(t)}{g(t)} \prod_{i=1}^n \frac{h(t_i)}{g(t_i)} \frac{ g(t_i) - f^{[k]}(t_i)/C}{ h(t_i) - f(t_i) } ~,
\end{align}
where $f^{[k]}(t_i)$ is the value of the varied transition kernel. \dire allows for renormalization-scale variations in the argument of running-coupling evaluations, as well as variations of parton distribution functions.

Finally, \dire stacks weighted-shower steps, especially to allow the incorporation
of matrix-element corrections. The possibility for stacking relies on two 
realizations: weighted-shower induced event weights are multiplicative, and 
after applying the event weight of previous steps, the shower rate will be 
correctly determined from the full splitting rate. Thus, for subsequent weighted 
shower steps, the full splitting kernel that would be obtained after applying the
weight is the new sampling rate --- or ``overestimate'' --- for the next, stacked, 
weighted-shower step. 

\dire currently stacks two weighted-shower 
steps. Ignoring, for the sake of a simple presentation, splitting enhancements and variations, then the first weighted shower (used to exponentiate complicated splitting kernels) yields 
a weight
\begin{align}
w_{1} =
\frac{h_1(t)}{g_1(t)} \prod_{i=1}^n \frac{h_1(t_i)}{g_1(t_i)} \frac{ g_1(t_i) - f_1(t_i)}{ h(t_i) - f(t_i) }~,
\end{align}
while the stacked weighted shower (used within the context of matrix element corrections) further induces the weight
\begin{align}
w_{2} =
\frac{h_2(t')}{f_1(t')} \prod_{i=1}^m \frac{h_2(t_i')}{f_1(t_i')} \frac{ f_1(t_i') - f_2(t_i')}{ h_2(t_i') - f_2(t_i') }~,
\end{align}
with
\begin{align}
f_2(t_i') = f_1(t_i') \otimes\textnormal{ME-correction}
~\textnormal{and}~
h_2(t)=
\begin{cases}
\mathrm{sign}[f_2(t)] f_1(t) & \textnormal{if}~ |f_1(t)| > |f_2(t)| \\
k_2 f_2(t) & \textnormal{if}~ |f_1(t)| < |f_2(t)| 
\end{cases} ~,
\end{align}
where $k_2 \sim 1.5$. Since matrix-element corrections are applied only
once a viable splitting has been selected, and the corresponding phase-space point
generated. Thus, the set of all $t'$ is different, and smaller, than the set of all $t$.
Currently, \dire does not implement enhancements or variations in the stacked 
weighted-shower step, since there does not seem to be a strong need for such 
complications. Variations may, in the future, be used to embed uncertainties
due to the underlying Lagrangian entering the matrix elements.

Note that only the product of all event weights is required. Thus, the stacked algorithm is identical to the original weighted-shower algorithm from an outside perspective.

%% file: physics/match-merge.tex
\section{Matching and merging}\index{Matching and merging}
\label{section:matchmerge}

Matching and merging methods aim to augment the event generator with (multiple) 
calculations performed within fixed-order perturbation theory. This is
rather straight forward for individual (and simple) hard-scattering calculations, which may 
be treated as the ``hard process'' from which further event generation steps start.
When the fixed-order calculation includes virtual and/or real corrections, a
consistent treatment quickly becomes more complex, such that dedicated schemes
of combining external calculations with the event generator need to be
developed. 

Naive parton showers aim to reproduce the effect of many collinear or soft emissions, and thus require improvements when describing observables that depend on well-separated hard particles. Fixed-order perturbative calculations furnish, on the other hand, an appropriate description of events with a handful of well-separated particles, but may fail in the collinear and soft limits. 
At high-energy colliders, observables typically exhibit effects of both approximations. On top of that, both the bulk cross sections (of low jet multiplicity) and tails (depending on the correct rate of high jet multiplicities) are often equally important. Methods to perform a matching or a merging of the fixed-order calculations with parton showers aim to combine the strengths of both approaches. 

Before going into the details of matching and merging methods, it is useful to discuss some aspects of fixed-order calculations. Higher-order calculations require the calculation of virtual and real corrections. The latter introduce additional final-state particles, so that the (next-to-)$^k$order prediction for an observable $O$ is
\begin{align}
\langle O \rangle &=
  \int \deriv\Phi_n \sum\limits_{i=0}^k \frac{\deriv \sigma_n^{(i)}}{\deriv \Phi_n}O(\Phi_n)
+ \int \deriv\Phi_{n+1} \sum\limits_{i=0}^{k-1} \frac{\deriv \sigma_{n+1}^{(i)}}{\deriv \Phi_{n+1}}O(\Phi_{n+1})
+ \int \deriv\Phi_{n+1} \sum\limits_{i=0}^{k-2} \frac{\deriv \sigma_{n+2}^{(i)}}{\deriv \Phi_{n+2}}O(\Phi_{n+2})\nonumber\\
&+ \dots
+ \int \deriv\Phi_{n+k} \frac{\deriv \sigma_{n+k}^{(0)}}{\deriv \Phi_{n+k}}O(\Phi_{n+k})\quad,
\end{align}
where the superscript $(i)$ determines the loop order. The symbols $\deriv\Phi_n$, $\deriv\Phi_{n+1}$, \ldots, $\deriv\Phi_{n+k}$ refer to the $n$-, $(n+1)$-, \ldots, $(n+k)$-particle phase-space measures defined in \cref{eq:lips}.
We will refer to a ``matching method'' as a method to combine complete higher-order corrections (\ie all terms for the order $k$) to a single inclusive process. A ``merging method'' combines several calculations for a lowest-multiplicity base process and related processes with additional well-separated jets (\ie up to a certain multiplicity $n+m$, but possibly omitting some higher-$(i)$ terms) with parton showering. The goals of these two approaches are often overlapping. Next-to-leading order matching methods aim to include the NLO prediction
\begin{equation}
\langle O \rangle_{\mathrm{NLO,in}} =
  \int \deriv\Phi_n \left( \frac{\deriv \sigma_n^{(0)}}{\deriv \Phi_n} +  \frac{\deriv \sigma_n^{(1)}}{\deriv \Phi_n}\right) O(\Phi_n)
+ \int \deriv\Phi_{n+1} \frac{\deriv \sigma_{n+1}^{(0)}}{\deriv \Phi_{n+1}}O(\Phi_{n+1})~,
\end{equation} 
while a leading-order merging combines the calculation
\begin{align}
\langle O \rangle_{\mathrm{LO,in}} & =
  \int \deriv\Phi_n \frac{\deriv \sigma_n^{(0)}}{\deriv \Phi_n}O(\Phi_n)\Theta(Q(\Phi_n) - \QMS)
+ \int \deriv\Phi_{n+1} \frac{\deriv \sigma_{n+1}^{(0)}}{\deriv \Phi_{n+1}}O(\Phi_{n+1})\Theta(Q(\Phi_{n+1}) - \QMS)\nonumber\\
& + \dots
+ \int \deriv\Phi_{n+m} \frac{\deriv \sigma_{n+m}^{(0)}}{\deriv \Phi_{n+k}}O(\Phi_{n+k})\Theta(Q(\Phi_{n+m}) - \QMS)
\end{align}
with the parton shower. Here, $\QMS$ denotes the so-called merging scale, which separates the hard (fixed-order) region $Q > \QMS$ from the soft/collinear (resummation) region $Q\leq\QMS$. This scale is in principle arbitrary, and merging algorithms should not develop a strong dependence on the exact choice, as long as it amounts to a reasonably small scale.

Before any combination is attempted, it is important to remember that virtual and real corrections are separately infrared divergent, and only their sum is free of singularities. This means that the individual contributions need to be regularized carefully, making the (unweighted) generation of events challenging. Matching and merging can help with this, as explained below. Furthermore, fixed-order calculations are \emph{inclusive}, meaning that a calculation for the process $a b\rightarrow c + X$ includes real-emission corrections implicitly, as part of $X$. For example, a leading-order calculation for $pp\rightarrow e^+e^- + X$ implicitly includes the process $pp\rightarrow e^+e^-g$. Fixed-order calculations for different processes can thus not simply be added --- they first have to be reorganized as \emph{exclusive} cross sections. At fixed order, this is achieved by including all relevant virtual corrections. The parton shower employs Sudakov factors or no-emission probabilities to produce exclusive all-order cross sections --- a reminder that Sudakov factors resum virtual corrections to all orders. High-multiplicity fixed-order calculations and showered low-multiplicity predictions may overlap as well. 

Thus, various sources of overlap between calculations should be handled when combining fixed-order calculations with parton showers. Matching and merging methods typically employ a mix of subtraction, phase-space division and (emission or event) vetoes for this task. The subtractions that are required in matched or merged calculations can occur at fixed perturbative order or at all perturbative orders. 

The aim of additional fixed-order subtractions is to remove the parton-shower approximations of real and/or virtual corrections from the fixed-order calculation, such that the resulting remnant can be showered without introducing overlaps. At next-to-leading order, this leads to (NLO) matching formulas that schematically have the form
\begin{align}
\label{eq:nlops0}
\langle O \rangle_{\mathrm{NLO+PS}} &=
  \int \deriv\Phi_n \left( \frac{\deriv \sigma_n^{(0)}}{\deriv \Phi_n} +  \frac{\deriv \sigma_n^{(1)}}{\deriv \Phi_n} - \frac{\deriv \sigma_n^{\mrm{PS}(1)}}{\deriv \Phi_n} \right) \SC(O,\Phi_n)\nonumber\\
&+ \int \deriv\Phi_{n+1} \left(\frac{\deriv \sigma_{n+1}^{(0)}}{\deriv \Phi_{n+1}} -  \frac{\deriv \sigma_{n+1}^{\mrm{PS}(0)}}{\deriv \Phi_{n+1}} \right)\SC(O,\Phi_{n+1})~,
\end{align}
where $\SC$ is the shower operator defined in \cref{eq:showers:showerOp}.
This also shows that these ``parton-shower subtractions'' are typically mandatory to allow for fixed-order event generation, since this would allow for the generation of the bracketed terms in \cref{eq:nlops0} as individual event samples. Additionally, this highlights that matrix-element-corrected parton showering can lead to simple NLO matching methods. Matrix-element corrections guarantee that the first emission in the parton shower is distributed according to the full tree-level rate by improving the splitting kernel \eqref{eq:showers:xSecFactorization},
\begin{equation}
	K_{j/\tilde{i}\tilde{k}} \to \Pmec K_{j/\tilde{i}\tilde{k}} \quad \mrm{with}\quad \Pmec = \frac{\abs{\ME_{n+1}^{(0)}}^2}{\sum\limits_j K_{j/\tilde{i}\tilde{k}} \abs{\ME_{n}^{(0)}}^2} \, .
\label{eq:matchmerge:defMEC}
\end{equation}
It is straightforward to see that in the sum over all branchings this reproduces the full $n+1$-particle matrix element,
\begin{equation}
	\sum\limits_j \Pmec K_{j/\tilde{i}\tilde{k}}\, \abs{\ME_n^{(0)}}^2 = \abs{\ME_{n+1}^{(0)}}^2 \, .
\end{equation}
\index{Matrix-element corrections|see{MECs}}\index{MECs}In practice, the correction factor $\Pmec$ is implemented via an additional multiplicative factor in the accept probability of the shower veto algorithm, \cf\cref{subsubsec:vetoalgorithm}. It is worth noting that matrix-element-correction methods historically appeared well before generic NLO matching methods~\cite{Bengtsson:1986hr,Seymour:1994df,Seymour:1994we,Andre:1997vh}. MECs identified
\begin{align}
 \frac{\deriv \sigma_{n+1}^{\mrm{PS}(0)}}{\deriv \Phi_{n+1}} &= \frac{\deriv \sigma_{n+1}^{(0)}}{\deriv \Phi_{n+1}} \\
 \frac{\deriv \sigma_{n}^{\mrm{PS}(1)}}{\deriv \Phi_{n}} &= -\int  \deriv\Phi_{1} \frac{\deriv \sigma_{n+1}^{(0)}}{\deriv \Phi_{n+1}} ~,
\end{align} 
leading to the \powheg matching prescription\index{powheg@\powheg!}
\begin{equation}
\label{eq:nlops1}
\langle O \rangle_{\mathrm{NLO+PS}} =
  \int \deriv\Phi_n \left( \frac{\deriv \sigma_n^{(0)}}{\deriv \Phi_n} +  \frac{\deriv \sigma_n^{(1)}}{\deriv \Phi_n} + \int  \deriv\Phi_{1}\left.\frac{\deriv \sigma_{n+1}^{(0)}}{\deriv \Phi_{n+1}}\right|_{\Phi_n} \right) \SC_\mathrm{MEC}(O,\Phi_n)~.
\end{equation}
Since the matrix-element-corrected parton shower $\SC_\mathrm{MEC}$ would now produce real-emission events, it is not possible to combine this calculation naively with further pre-calculated multiparton fixed-order predictions. It would, however, be possible to add new tree-level samples if the contributions are also subtracted in the overall result:
\begin{align}
\label{eq:lops0}
\langle O \rangle_{\mathrm{STACKED}} &=
  \int \deriv\Phi_n \left( \frac{\deriv \sigma_n^{(0)}}{\deriv \Phi_n} +  \frac{\deriv \sigma_n^{(1)}}{\deriv \Phi_n} + \int  \deriv\Phi_{1}\left.\frac{\deriv \sigma_{n+1}^{(0)}}{\deriv \Phi_{n+1}}\right|_{\Phi_n} \right) O(\Phi_n)\nonumber\\
&- \int \deriv\Phi_{n} \deriv\Phi_{1} \left. \frac{\deriv \sigma_{n+1}^{(0)}}{\deriv \Phi_{n+1}} \right|_{\Phi_n} \Theta(Q(\Phi_{n+1}) - \QMS) O(\Phi_{n}) \nonumber\\
&+ \int \deriv\Phi_{n+1} \frac{\deriv \sigma_{n+1}^{(0)}}{\deriv \Phi_{n+1}}\Theta(Q(\Phi_{n+1}) - \QMS)O(\Phi_{n+1})
\end{align}
This somewhat academic exercise of ``stacking'' fixed-order calculations can be cast into a more familiar form
related to the parton shower. For that, we introduce the ``parton-shower weight''
of an $n$-parton state $w^\mathrm{PS}_n$ as the exact parton-shower rate
of the $n$-parton state, excluding the product of naive splitting probabilities.
Thus, 
\begin{equation}
\label{eq:pswt}
w^\mathrm{PS}_n = f_{0}(x(\Phi_{0}),\mu_{f\mathrm{PS}}^2) \prod_{i=0}^{n-1} \Pi_i(t_i, t_{i+1}; \Phi_i) \frac{\alpha_s(t_{i+1})}{2\pi} \frac{f_{i+1}(x(\Phi_{i+1}),t_{i+1})}{f_{i}(x(\Phi_{i}),t_{i+1})}
\end{equation}
Similarly, we may collect all coupling and PDF factors used at fixed order into
the fixed-order weight
\begin{equation}
\label{eq:fowt}
w^\mathrm{FO}_n = \left(\frac{\alpha_s(\mu_r^2)}{2\pi}\right)^n f_{n}(x(\Phi_{n}),\mu_f^2)
\end{equation}
Applying the ratio of the former to the latter weight, to an $n$-parton fixed-order
calculation introduces appropriate parton-shower higher orders. With this,
we may instead add and subtract all-order contributions, leading to
\begin{align}
\label{eq:lops1}
\langle O \rangle_{\mathrm{MERGED}} &=
  \int \deriv\Phi_n \left( \frac{\deriv \sigma_n^{(0)}}{\deriv \Phi_n} +  \frac{\deriv \sigma_n^{(1)}}{\deriv \Phi_n} + \int  \deriv\Phi_{1}\left.\frac{\deriv \sigma_{n+1}^{(0)}}{\deriv \Phi_{n+1}}\right|_{\Phi_n} \right) \frac{w^\mathrm{PS}_{n}}{w^\mathrm{FO}_{n}} O(\Phi_n)\nonumber\\
&- \int \deriv\Phi_{n} \deriv\Phi_{1} \frac{w^\mathrm{PS}_{n+1}}{w^\mathrm{FO}_{n+1}} \left. \frac{\deriv \sigma_{n+1}^{(0)}}{\deriv \Phi_{n+1}} \right|_{\Phi_n} \Theta(Q(\Phi_{n+1}) - \QMS) O(\Phi_{n}) \nonumber\\
&+ \int \deriv\Phi_{n+1} \frac{w^\mathrm{PS}_{n+1}}{w^\mathrm{FO}_{n+1}} \frac{\deriv \sigma_{n+1}^{(0)}}{\deriv \Phi_{n+1}} \Theta(Q(\Phi_{n+1}) - \QMS)O(\Phi_{n+1})\nonumber\\
&\approx
  \int \deriv\Phi_n \left( \frac{\deriv \sigma_n^{(0)}}{\deriv \Phi_n} +  \frac{\deriv \sigma_n^{(1)}}{\deriv \Phi_n} + \int  \deriv\Phi_{1}\left.\frac{\deriv \sigma_{n+1}^{(0)}}{\deriv \Phi_{n+1}}\right|_{\Phi_n} \right) \frac{w^\mathrm{PS}_{n}}{w^\mathrm{FO}_{n}} \Pi_n(t_n, t_{cut}; \Phi_n; > \QMS ) O(\Phi_n)\nonumber\\
&+ \int \deriv\Phi_{n+1}  \frac{w^\mathrm{PS}_{n+1}}{w^\mathrm{FO}_{n+1}} \frac{\deriv \sigma_{n+1}^{(0)}}{\deriv \Phi_{n+1}} \Theta(Q(\Phi_{n+1}) - \QMS)O(\Phi_{n+1})~.
\end{align}
Here, the additional argument ``$> \QMS$'' in the no-emission probability indicates that only emissions leading to states with $Q(\Phi_{n+1}) >\QMS$ should be considered --- leading to what is sometimes called the ``vetoed shower'' no-emission probability.
The lines after the approximate equality would be fully equivalent to the lines before if the shower correctly reproduced the rate $ \deriv \sigma_{n+1}^{(0)} /\deriv \Phi_{n+1}$.

This equation leads to an interesting interpretation: the inclusion of a no-emission probability $\Pi_n$ on top of the fixed-order $n$-parton cross section \emph{is producing a subtraction} that allows it to be combined the $(n+1)$-parton event samples. Event samples that are made exclusive with the help of no-emission probabilities can be added without further complication. This realization is the basis of merging methods, which extend the argument to the combination of several multiparton calculations. If the no-emission probabilities are approximated by jets after the complete evolution sequence, then the merging procedure can become independent of the shower details. This is the case for MLM jet matching\index{MLM jet matching}. In the \ac{CKKW-L}\index{CKKW(-L) merging} method, the second equation and the exact (partonic) no-emission probabilities of the parton shower are used to calculate the
rescalings $w^\mathrm{PS}_{n+1}/w^\mathrm{FO}_{n+1}$. Incidentally, such ratios are often called the ``CKKW-L weight'' or ``merging weight''. Unitarized merging methods retain the explicit add-subtract structure to guarantee the correct inclusive cross sections even if the parton shower does not accurately reproduce the (higher-order) emission pattern.

\index{MECs}As of today, a broad spectrum of matching and merging techniques has been developed. Historically, the first method were matrix-element corrections (MECs)~\cite{Bengtsson:1986hr}, where the shower kernel itself is corrected to the full matrix element after the first emission. This method has later been extended to include higher orders as well~\cite{Giele:2007di,Giele:2011cb,LopezVillarejo:2011ap,Fischer:2016vfv,Fischer:2017yja}.
For NLO matching, two general schemes exist, namely \mcatnlo\index{madgraph@\mg5amc}\index{NLO matching}\index{MCatNLO@\mcatnlo}~\cite{Frixione:2002ik} and \powheg~\cite{Nason:2004rx,Frixione:2007vw,Hoche:2010pf}, with the former being automated in the \madgraphamc~\cite{Alwall:2014hca} and \sherpa~\cite{Hoeche:2011fd} event-generation frameworks and the latter available through the \powhegbox\index{powheg@\powheg} program~\cite{Alioli:2010xd}.
Well-established tree-level merging methods are MLM~\cite{Mangano:2001xp,Mangano:2006rw} and CKKW~\cite{Catani:2001cc,Hamilton:2009ne}, which utilize a simple
jet-matching algorithm and analytic Sudakov factors,
respectively. The CKKW-L method~\cite{Lonnblad:2001iq,Lonnblad:2011xx,Brooks:2020bhi} and METS~\cite{Hoeche:2009rj} extend the CKKW merging scheme to use numerical
no-branching probabilities, generated by trial showers.

\index{UMEPS merging}The \ac{UMEPS} scheme~\cite{Lonnblad:2012ng,Platzer:2012bs} improves the unitarity of tree-level merging and thereby addresses the issue that these change the inclusive cross section of the event samples.
At NLO, multiple refinements of the aforementioned LO merging schemes exist. \index{NL3 merging}\index{UNLOPS merging}The NL3 technique~\cite{Lavesson:2008ah} extends CKKW-L to NLO, just as UNLOPS~\cite{Lonnblad:2012ix,Bellm:2017ktr} does the same with UMEPS. At the same time, UNLOPS may be viewed as the unitarity-improved version of NL3.
The MENLOPS scheme~\cite{Hamilton:2010wh,Hoche:2010kg} combines an NLO calculation for the lowest multiplicity with LO calculations for higher multiplicities in the METS scheme, while the full extension to NLO is treated in the MEPS@NLO scheme~\cite{Gehrmann:2012yg,Hoeche:2012yf}. The FxFx method~\cite{Frederix:2012ps} combines \mcatnlo matching with MLM merging. The MiNLO scheme~\cite{Hamilton:2012np,Frederix:2015fyz} may be regarded as a scale-setting-improved NLO extension of the CKKW algorithm, with analytically calculated NLL Sudakov factors between the clustered states.

Before describing the matching and merging methods available in \pythia, it should be emphasized that (NLO) matching and merging methods introduce the stated fixed-order accuracy only up to the matched (merged) multiplicities. That is, an NLO-matched $n$-jet event sample has NLO accuracy only for $n$-jet observables, while $(n+1)$-jet observables will have LO accuracy, and $(n+2)$-jet as well as higher-multiplicity observables have parton-shower accuracy only.
Similarly, a merged event sample with up to $n$ jets at (N)LO accuracy, has (N)LO accuracy for $m$-jet observables with $m\leq n$. In the case of NLO merging, $(n+1)$-jet observables will have LO accuracy, while they will have parton-shower accuracy for LO merging. In either case, $m$-jet observables with $m > n+1$ have only parton-shower accuracy.
Another question is the accuracy of the inclusive cross section. In NLO matching schemes, the inclusive cross section is accurate to NLO by design. In merging schemes, the inclusive cross section of $n$-jet cross sections are individually retained only if the merging scheme is constructed to be unitary, such as UMEPS or UNLOPS. In non-unitary merging schemes, inclusive cross sections are changed by the inclusion of higher-multiplicity event samples.

\subsection{\pyt methods for leading-order multi-jet merging}\label{sec:PythiaLOMerging}
\index{Matching and merging!Leading order methods}

As discussed above, \pyt offers a variety of leading-order merging schemes. This allows for a test of the robustness of merged predictions, beyond assessing the uncertainty due to scheme-specific parameters. The main task for a leading-order merging scheme is to produce an inclusive event sample that provides a simultaneous tree-level prediction for observables depending on any number of jets $\leq n$. This entails removing any overlap between the tree-level calculations for $\leq n$ partons. The second main task is to provide a smooth transition between the ``well-separated region'' ($Q(\Phi_{n}) > \QMS$) described by (reweighted) tree-level results, and the ``soft/collinear region'' ($Q(\Phi_{n}) < \QMS$) described by the parton-shower radiation pattern. Internal merging schemes in \pyt also ensure that the self-consistency of the \pyt event-generation chain is intact.   

CKKW-L multi-jet merging is the oldest tree-level merging scheme implemented in \pyt. It allows both standard-model and BSM core processes\footnote{Note, though, that no attempt is made at diagram removal or subtraction for colour-changed BSM resonances that decay into SM quarks.}, to which multiple several colour-charged partons or $\W$ bosons are added. Lepton and hadron-collider processes are accepted. The resulting tree-level samples are combined with each other and the default parton shower by employing the merging formula
\begin{align}
\label{eq:ckkwl}
\langle O \rangle_{\mathrm{CKKW-L}} &=
  \sum\limits_{n=0}^{N-1}\int \deriv\Phi_n \frac{\deriv \sigma_n^{(0)}}{\deriv \Phi_n}  \frac{w^\mathrm{PS}_{n}}{w^\mathrm{FO}_{n}} \Pi_n(t_n, t_{cut}; \Phi_n; > \QMS ) \Theta(Q(\Phi_{n}) - \QMS) ~\SC(O,\Phi_n; < \QMS)\nonumber\\
&+ \int \deriv\Phi_{N} \frac{w^\mathrm{PS}_{N}}{w^\mathrm{FO}_{N}} \frac{\deriv \sigma_{N}^{(0)}}{\deriv \Phi_{N}} \Theta(Q(\Phi_{N}) - \QMS)  ~\SC(O,\Phi_N) ~,
\end{align}
where the showers $\SC(O,\Phi_n; < \QMS)$ of all but the highest-multiplicity event sample fill in emissions below the merging scale $\QMS$. The main challenge of CKKW-L merging lies in the correct calculation of the weights $w^\mathrm{PS}_{n}$. Their calculation in \pyt ensures that the value of the weights is identical to the all-order weight the shower would had attached to the state $\Phi_n$, had it produced the state internally. This requires the construction of all possible parton-shower histories, and an admixture of the (history-dependent) weight factors identical to that of the shower~\cite{Lonnblad:2011xx}.
 
A theoretical drawback of CKKW-L is that inclusive jet cross sections change upon inclusion of higher-multiplicity tree-level samples. The size of the change is determined by the value of the merging scale $ \QMS$, leading to unacceptable changes for $\QMS$ of $\mathcal{O}(1\mrm{GeV})$. For low merging scales, the exact ``subtract what you add'' strategy highlighted in \cref{eq:lops1} has to be employed. For this purpose, the UMEPS method has been introduced in \pyt. The UMEPS implementation can handle the same process classes as the CKKW-L scheme. The UMEPS merging formula reads
\begin{align}
\label{eq:umeps}
\langle O \rangle_{\mathrm{UMEPS}}
&=
  \sum\limits_{n=0}^{N-1} \int \deriv\Phi_n \Bigg[ \frac{\deriv \sigma_n^{(0)}}{\deriv \Phi_n}  \frac{w^\mathrm{PS}_{n}}{w^\mathrm{FO}_{n}} \Theta(Q(\Phi_{n}) - \QMS) \nonumber\\
&\phantom{\sum\limits_{n=0}^{N-1} \int \deriv\Phi_n}
- \int \deriv\Phi_1 \frac{\deriv \sigma_{n+1}^{(0)}}{\deriv \Phi_{n+1} } \frac{w^\mathrm{PS}_{n+1}}{w^\mathrm{FO}_{n+1}} \Theta(Q(\Phi_{n+1}) - \QMS)  \Bigg] ~\SC(O,\Phi_n; < \QMS)\nonumber\\
&+ \int \deriv\Phi_{N} \frac{w^\mathrm{PS}_{N}}{w^\mathrm{FO}_{N}} \frac{\deriv \sigma_{N}^{(0)}}{\deriv \Phi_{N}} \Theta(Q(\Phi_{N}) - \QMS)  ~\SC(O,\Phi_N) ~,
\end{align}
The subtractive samples in this formula are produced with the help of the parton-shower history employed to calculate the weights $w^\mathrm{PS}_{n}$~\cite{Lonnblad:2012ng}.

As a final remark on leading-order merging, it should be noted that \pyt offers detailed semi-internal \texttt{UserHooks} utilities for MLM jet matching~\cite{mlm}. This early  approach to combining fixed-order matrix-element
calculations with parton showers approximates the parton-shower no-emission probabilities necessary to remove overlap between samples with a pragmatic event-veto procedure: the parton ensemble at fixed order is stored, and compared to jets after showering the ensemble. If each jet directions overlaps with one parton direction, the event is retained. The rate of rejected events mimics the application of no-emission probabilities. This convenient approach has the benefit of simplicity and computational efficiency, at the expense of sacrificing a formal understanding of the result.   

\subsection{\pyt methods for matching}\label{sec:PythiaMatching}\index{powheg@\powheg}
\index{Matching and merging!Next-to-leading order methods}

The \powheg NLO matching formalism as given in \cref{eq:nlops1} provides an elegant and universal method for the combination of NLO calculations and parton showers. It is universal, as it does not depend on the exact implementation of the parton shower to be matched. This is, because in addition to the NLO-corrected Born-level event, the first emission is generated according to a matrix-element corrected no-branching probability
\begin{equation}
	\bar{\Pi}(Q_0^2,Q_1^2) = \exp\left\{ -\int\limits_{Q_1^2}^{Q_0^2} \, \deriv \Phi_{+1} \, \frac{\deriv \sigma^{(0)}_{n+1}}{\deriv \sigma^{(0)}_n} \right\} \, ,
\label{eq:matchmerge:PowhegSudakov}
\end{equation}
which is independent of the branching kernels used by the shower. In principle, an NLO-matched prediction could therefore be obtained with any given shower by starting the shower evolution at the scale of the first \powheg emission. In practice, the ordering variable of the shower $t$ will disagree with the ordering variable $Q^2$ used in the \powheg formalism. To avoid over- or under-counting emissions, it is hence preferable to start the shower at the phase-space maximum (\ie using a ``power shower'', \cf\cref{sec:showers:simpleShowerCorrections}) and vetoing emissions that are harder than the \powheg emission according to the \powheg ordering variable. This method was outlined in \citeone{Corke:2010zj} and since then \pythia provides the relevant \powheg hook to supply consistent showering of \powhegbox events with the simple showers.
It is worth noting that this procedure leads to the somewhat awkward situation that the first, \ie hardest, jet is produced from the kinematics and colour configuration of the Born+1-jet state. To circumvent this, a more complete treatment would involve clustering the first emission and running a vetoed truncated shower off the actual Born state for the first emission. This is currently not available in \pythia.

It might potentially be regarded as an inelegance that the \powheg no-branching probability \cref{eq:matchmerge:PowhegSudakov} exponentiates the full matrix element, including process-specific non-singular terms, and that the value of $Q_0^2$ is typically given by the (hadronic) phase-space limit, and not a scale that more adequately defines the transition between ``hard'' and ``soft'' radiation. These concerns are avoided in the (historically first developed) \mcatnlo method, in which the real-radiation matrix element is separated into an infrared-singular (``soft'') and infrared-regular (``hard'') part,
\begin{equation}
	\deriv \sigma_{n+1}^{(0)} = \deriv \sigma_{n+1}^{\mrm{S}(0)} + \deriv \sigma_{n+1}^{\mrm{H}(0)} \, .
\end{equation}
Therefore, only the singular part in the no-branching probability is retained,
\begin{equation}
	\bar{\Pi}^{\mrm{S}}(Q_0^2,Q_1^2) = \exp\left\{ -\int\limits_{Q_1^2}^{Q_0^2} \, \deriv \Phi_{+1} \, \frac{\deriv \sigma^{\mrm{S}(0)}_{n+1}}{\deriv \sigma^{(0)}_n} \right\} \, ,
	\label{eq:matchmerge:MCatNLOSudakov}
\end{equation}
so that the \mcatnlo matched expectation value of an observable $O$ reads
\begin{align}
	\avg{O}_{\mrm{NLO+PS}}^{(\mcatnlo)} &= \int \, \deriv \Phi_n \, \Bigg[\frac{\deriv \sigma^{(0)}_{n}}{\deriv \Phi_{n}} + \frac{\deriv \sigma^{(1)}_n}{\deriv \Phi_n} + \int\, \deriv \Phi_{+1}\, \frac{\deriv \sigma^{\mrm{PS}(0)}_{n+1}}{\deriv \Phi_{n+1}} \nonumber\\
&\phantom{\int \, \deriv \Phi_n \,}
+ \int \deriv \Phi_{+1} \, \left(\frac{\deriv \sigma^{\mrm{S}(0)}_{n+1}}{\deriv \Phi_{n+1}} - \frac{\deriv \sigma^{\mrm{PS}(0)}_{n+1}}{\deriv\Phi_{n+1}} \right) \Bigg] \SC(O,\Phi_n) \nonumber\\
&+ \int \, \deriv \, \Phi_{n+1} \frac{\deriv \sigma^{\mrm{H}(0)}_{n+1}}{\deriv \Phi_{n+1}} \,. 
\label{eq:matchmerge:MCatNLOMaster}
\end{align}
As evident from \cref{eq:matchmerge:MCatNLOMaster}, the \mcatnlo method requires stringent consistency between the NLO calculation and the shower. Different to the \powheg method, it therefore does not provide a universal scheme but needs to be implemented explicitly for each shower.
To facilitate a simple implementation of the \mcatnlo technique for \pyt's simple shower, \pythia provides a global-recoil scheme, \cf\cref{sec:showers:simpleShowerCorrections}. A publicly available parton-level event generator supporting the generation of \mcatnlo events for \pyt's simple shower is \madgraphamc.
Caution is, however, advised, as the global-recoil scheme might not be a suitable choice for each and every process. 

Different to the \powheg formalism, the \mcatnlo scheme explicitly employs negative-weighted events (in fact \powheg was designed to remove negative weights from \mcatnlo). While not posing a problem technically, negative weights reduce the efficiency of any simulation, simply because they have to compensated for in histograms with positive-weighted events, of which more are needed to obtain the same statistics as for simulations with a strict probability interpretation.
The fraction of negative-weighted events can be reduced by the \mcatnlo-$\Delta$ formalism~\cite{Frederix:2020trv}, which regulates the divergent structure of real-emission matrix elements via shower-generated no-branching probabilities. In addition, the \mcatnlo-$\Delta$ prescription introduces an independent shower starting scale for each colour line in the hard process. Such multi-scale treatment is also required in the \powheg formalism, if resonance-aware matching is pursued, \eg when using the \powhegboxres generator~\cite{Jezo:2015aia}. In both cases, \pythia offers the necessary machinery to deal with multiple scale definitions~\cite{FerrarioRavasio:2019vmq} through \texttt{UserHooks} (see \cref{subsection:userhooks}).
While \pythia offers in-house implementations for \mcatnlo and \powheg matching as alluded to above, other matching schemes can conveniently be implemented via user hooks, \cf\cref{subsection:userhooks}. A prominent example is the NNLO+PS matching framework \geneva~\cite{Alioli:2012fc,Alioli:2013hqa,Alioli:2015toa,Alioli:2016wqt}.\index{Geneva@\geneva}

\subsection{\pyt methods for NLO multi-jet merging}\label{sec:PythiaNLOMerging}
\index{Matching and merging!Next-to-leading order methods}

A number of schemes to combine several matched NLO (QCD) calculations with each other and parton showering are available in \pyt. As was the case for tree-level merging (\cf\cref{sec:PythiaLOMerging}), this allows for an exploration --- through comparison within the same code base --- of the benefits and limitations of various approaches, as well as NLO merged predictions more generally.

\index{NL3 merging}Historically, the first two NLO merging
schemes embedded in \pyt were NL$^3$ (an extension of the CKKW-L
approach to NLO) and UNLOPS\index{UNLOPS merging},
the extension of UMEPS to NLO accuracy. Beyond the theoretical and computational challenges already present at leading order, NLO merging needs to ensure that the application of all-order weights $w^\mathrm{PS}_{n}/w^\mathrm{FO}_{n}$ does not invalidate the NLO precision of the input samples due to overlapping virtual corrections. This can be achieved by subtracting the first-order expansion of the shower weights attached to tree-level events. Thus, the main complication in calculation is in generating terms in the shower expansion~\cite{unlops}. \pyt uses trial showering to generate the expansion of no-emission probabilities, and analytic results to calculate the expansion of running-coupling and PDF factors. Once these technicalities are under control, the NLO extension of CKKW-L implements the matching formula
\begin{align}
\label{eq:nl3}
\langle O \rangle_{\mathrm{NL}^3} &=
  \sum\limits_{n=0}^{N}\int \deriv\Phi_n \frac{\deriv \sigma_n^{(0)}}{\deriv \Phi_n}  \frac{w^\mathrm{PS}_{n}}{w^\mathrm{FO}_{n}} \Pi_n(t_n, t_{cut}; \Phi_n; > \QMS ) \Theta(Q(\Phi_{n}) - \QMS) ~\SC(O,\Phi_n; < \QMS)\nonumber\\
&+
\sum\limits_{n=0}^{N}\int \deriv\Phi_n
\Bigg( \frac{\deriv \sigma_n^{(1)}}{\deriv \Phi_n} + \int  \deriv\Phi_{1}\left.\frac{\deriv \sigma_{n+1}^{(0)}}{\deriv \Phi_{n+1}}\right|_{\Phi_n} \nonumber\\
&\phantom{\sum\limits_{n=0}^{N}\int \deriv\Phi_n}
- 
\frac{\deriv \sigma_n^{(0)}}{\deriv \Phi_n}  \left[\frac{w^\mathrm{PS}_{n}}{w^\mathrm{FO}_{n}} \Pi_n(t_n, t_{cut}; \Phi_n; > \QMS)\right]_{\mathcal{O}(\alpha_s)} 
\Bigg)\, \SC(O,\Phi_n; < \QMS)\nonumber\\
&+
\int \deriv\Phi_{N+1}\frac{\deriv \sigma_{N+1}^{(0)}}{\deriv \Phi_{N+1}} \frac{w^\mathrm{PS}_{N+1}}{w^\mathrm{FO}_{N+1}}  \Theta(Q(\Phi_{N+1}) - \QMS)  ~\SC(O,\Phi_{N+1})
~.
\end{align}
where square bracket with subscript $\mathcal{O}(\alpha_s)$ indicate that the $\mathcal{O}(\alpha_s)$ term of the expansion of the bracketed term is required. The first and last line of \cref{eq:nl3} are identical to the CKKW-L result. The second line
incorporates the inclusive NLO correction, and the subtraction of double counting of virtual corrections. As was the case for CKKW-L, the calculation of all necessary terms in \cref{eq:nl3} employs parton shower histories.

However, the NL$^3$ formula \cref{eq:nl3} has the same theoretical drawback as the CKKW-L results: inclusive $n$-parton rates are changed when including corrections to $m>n$ parton rates. The size of the effect is determined by $\QMS$, and can easily be of a similar numerical size as NLO corrections for moderately small $\QMS$. Thus, \pyt also extends the UMEPS method, which corrects this behaviour, to NLO accuracy. The resulting UNLOPS merging formula can be found in~\citeone{unlops}. UNLOPS is the preferred NLO merging scheme in \pyt. 

Before moving on, it is interesting to observe that any reweighting of the second line in \cref{eq:nl3} with higher-order terms will neither affect the NLO fixed-order nor the shower accuracy of the combined calculation. Thus, variants of NLO merged methods can be obtained from reweighting these contributions. This freedom, and the resulting uncertainty, is exposed in \pyt by offering three different variants of UNLOPS~\cite{Gellersen:2020tdj}. Sensible uncertainties from NLO merged calculations should include the envelope of these variants as ``scheme uncertainty''.

As part of its semi-internal implementation of MLM jet matching, \pyt also offers semi-internal utilities to combine input samples produced for FxFx merging~\cite{Frederix:2012ps}. This scheme extends the MLM method to NLO QCD accuracy, and handles the overlap between events of different multiplicity before showering in a hybrid scheme: fixed-order events are reweighted with analytic Sudakov form factors to produce results that are additive before showering. The overlap after showering is addressed in a pragmatic way, following along the lines of MLM jet matching. The resulting scheme is computationally efficient, especially since Sudakov form factors can directly be integrated into the fixed-order calculation code. However, this efficiency comes at the price of an unclear all-order structure of the prediction. Nevertheless, the scheme has found a large user base.
 
\subsection{Matching and Merging in
  \vincia}\label{sec:VinciaMatchMerge}
\index{Matching and merging!Leading order methods}
\index{Sector merging}

The unique ``maximally bijective'' property of \vincia's sector
antenna 
showers, \cf\cref{sec:Vincia}, make them well suited for
incorporating corrections from fixed-order matrix elements,
especially at high multiplicities. The focus so far has been on QCD corrections,
although the sector nature of \vincia's coherent QED shower should
make adaptations to include QED corrections as well fairly
straightforward. 

At leading order, a dedicated CKKW-L merging scheme has been
implemented which exploits the properties of \vincia's sector
showers. This is discussed below in \cref{sec:matchmerge:vinciaCKKWL}. 
Details on the general \pyt CKKW-L implementation can be found in
\cref{sec:PythiaLOMerging}. 

Next-to-leading order matching in the antenna framework is so far not
generally available, but \vincia's QCD showers, including the
resonance-final one, can be combined with NLO QCD calculations by
shower-independent matching methods, such as \powheg. This is
described in \cref{sec:matchmerge:vinciaPOWHEG}.

Although \vincia currently has no built-in NLO matching schemes, a generalization of the scheme developed in \citerefs{Hartgring:2013jma,Li:2016yez} may become available in the future.
In a similar vein, \vincia does not offer the merging of multiple NLO-matched samples in the current version.
Schemes extending the ones described in \cref{sec:PythiaNLOMerging} to sector showers may, however, be implemented in the future.

\vincia's NNLO matching approach presented in \citeone{Campbell:2021svd}
is not yet part of the public \pythia releases. We expect it to become
available in a future release, once it has been applied and validated
for a larger class of processes. 

A signature feature in early developments of \vincia, iterated
matrix-element corrections~\cite{Giele:2011cb,Lopez-Villarejo:2011pwr,Larkoski:2013yi,Fischer:2017yja} 
have not yet been made available in \pythia. Plans are underway to
do so, building on the new matrix-element interfaces
described in \cref{sec:mg5amc}. Note also that \vincia's weak
shower, described in \cref{sec:VinciaEW}, is
currently not sectorized and hence full-fledged 
EW merging would presumably require some work. Finally, note
that dedicated matching and merging strategies for \vincia's 
interleaved treatment of resonance decays have not yet been worked
out. See the program's \htmlmanual for updates.

\subsubsection{Leading-order merging}\label{sec:matchmerge:vinciaCKKWL}
\index{CKKW(-L) merging}
Tree-level merging with \vincia is done in the CKKW-L scheme~\cite{Catani:2001cc,Lonnblad:2001iq,Lonnblad:2011xx} according to \cref{eq:ckkwl}, properly extended to sector showers~\cite{Brooks:2020bhi}. The phase-space sectorization particularly facilitates the merging at very high multiplicities and offers increased control over highly-complex final states.
Most of the merging method is identical to the \pyt implementation described in \cref{sec:PythiaLOMerging}, with the difference only in the construction of shower histories needed for the Sudakov reweighting. The settings relevant to \vincia's CKKW-L implementation can be found in \cref{sec:standalone:matchmerge}.

In the default CKKW-L scheme, all possible shower histories are constructed and the one maximizing the branching probability is chosen, \cf\cref{sec:PythiaLOMerging}. In the sector-shower CKKW-L implementation, however, the construction of all possible histories is replaced by a deterministic inversion of the shower evolution. This is possible because \vincia's sector showers generate branchings only if these correspond to the minimal sector-resolution variable, \cf\cref{sec:vinciaqcd}. The sector-resolution variable can then be used to exactly invert any branching. The only subtlety in this algorithm stems from the treatment of multiple quark pairs, for which all possible quark-antiquark clusterings must be taken into account. To this end, the same procedure as in the default CKKW-L method is utilized and a shower history is constructed for all viable permutations of colour strings between quark pairs, and the one maximizing the branching probability is picked. 
This algorithm replaces the shower history tree by maximally a few linear history branches, which not only positively affects the CPU time needed for the computation, but more importantly reduces the prohibiting scaling in memory allocation intrinsic to the default CKKW-L algorithm.

\subsubsection{NLO
  matching}\label{sec:matchmerge:vinciaPOWHEG}\index{powheg@\powheg}
\index{Matching and merging!Next-to-leading order methods}

If an NLO-matched calculation with \vincia is desirable, the \powheg method~\cite{Nason:2004rx,Frixione:2007vw} with externally matched NLO event samples, as \eg produced by the \powhegbox program~\cite{Alioli:2010xd,Jezo:2015aia}, can be used~\cite{Brooks:2019xso}. To this end, the difference in the \powhegbox and \vincia evolution variables are properly accounted for by increasing the shower starting scale and vetoing emissions above the \powheg scale~\cite{Corke:2010zj}. The usage of \powhegbox with \vincia is described in detail in \citeref[appendix A]{Hoche:2021mkv}.

\subsection{Matching and Merging in \dire}\label{sec:DireMatchMerge}

At the time of compiling this manual, the matching and merging machinery with \dire have not been validated within \pythia. Previous versions of \dire + \pyt~8.2 included CKKW-L tree-level merging~\cite{Prestel:2019neg}, UNLOPS NLO merging~\cite{Andersen:2020sjs}, iterated matrix-element corrections within the MOPS approach~\cite{Prestel:2019neg,Gellersen:2021caw}, and TOMTE N3LO+PS matching~\cite{Prestel:2021vww}. We expect these previous developments to become accessible and validated in \pythia in the future.

%% file: physics/soft-proc.tex
\section{Soft and beam-specific processes}
\label{sec:soft}
\index{PDFs}
Hadrons and nuclei are composite objects, mainly made out of quarks and
gluons. This requires the introduction of parton distribution functions
(PDFs) $f_a^A(x, Q^2)$, expressing the probability density that parton
$a$ exists inside particle $A$ with a momentum fraction $x$ if the particle
is probed at a resolution scale $Q^2$. Given such PDFs, hard collisions
between the constituent partons can be described by perturbation theory,
see \cref{subsection:processGenBasics}. But in the limit $\pT \to 0$
the cross section for perturbative QCD scattering diverges, and traditional
perturbation theory breaks down.

\index{Pomeron}\index{Reggeon}\index{Regge--Gribov theory}The
alternative offered already since before the advent of QCD is 
so-called Regge--Gribov theory~\cite{Gribov:1968fc,Collins:1977jy,
Forshaw:1997dc,Donnachie:2002en,Barone:2002cv},  
wherein an effective field theory is formulated in terms of the
exchange of reggeon (\Reg\index{Reggeon}) and pomeron (\Pom\index{Pomeron}) objects
between the colliding hadrons, with propagators and vertex-coupling
strengths, the latter both to hadrons and among themselves.
A reggeon\index{Reggeon} (pomeron\index{Pomeron}) contribution
represents the resummed effect 
from the exchange of (an infinite set of) mesons (glueballs) with a
common set of flavour quantum numbers, but ordered in a linear
relationship (a ``trajectory'') between increasing orbital angular
momentum $L$ and increasing $m^2$. This language can be used to motivate
expressions for total, elastic, and diffractive cross sections, even if
today this is done in a pragmatic spirit, where not fully consistent
adjustments of parameters can be made to better fit data. 

Leptons are fundamental particles, unlike hadrons, and it would
seem like traditional perturbation theory can always be applied.
But a charged lepton is surrounded by a cloud of virtual photons,
and these carry some of the total momentum. It therefore becomes
necessary to introduce PDFs also to describe the distribution of
a lepton and photons inside the whole charged lepton, as a function
of $Q^2$. Either of these two components can then
collide with constituents of the other beam. The photon, in its turn,
can fluctuate further into a lepton or quark pair, and the latter
again can have a non-perturbative behaviour. This requires a similar
approach for photon interactions as for hadron ones, in fact with even
more complexity. Since hadrons and nuclei also can contain or be
surrounded by photons, by coupling to the charge of individual quarks
or to the hadron as a whole, similar issues can arise in hadronic
collisions.

Also fluxes of $\Wpm$ and $\Z$ bosons can be defined, and have
been used in the past, both for leptons and for protons. The large
weak-gauge-boson masses suppress the rate in the $\pT \to 0$
limit, however, and so their contributions are better handled as
propagators in Feynman graphs, like the top quark and the Higgs boson.
This also implies that neutrinos can be considered point-like for our
purposes.

Heavy-ion collisions introduce further physics aspects, relative to
hadronic collisions. Some of these are reasonably well understood,
such as the role of the initial geometry, where models for the
distribution of nucleons inside a nucleus can be used to find the
``wounded'' nucleons, \ie those that interact. But most of the
subsequent physics is still open to interpretation, and different
approaches exist. One such is the \angantyr model, presented here.

\subsection{Total and semi-inclusive cross
  sections}\index{Diffraction}\index{Elastic scattering}\index{Single
  diffraction}\index{Double diffraction}
\index{Central diffraction}\index{Total cross sections}
\index{Diffraction!Single|see{Single diffraction}}
\index{Diffraction!Double|see{Double diffraction}}
\index{Diffraction!Central|see{Central diffraction}}
\label{subsection:sigmatotal}

Here we introduce the components of the total hadron-hadron
cross section, and how these vary as a function of the collision energy.
The intention is not to go into the modelling of the collision
processes as such, which is the main topic for subsequent subsections,
but some such information is necessary when the differential cross
sections are the basic building blocks, and the integrated ones only a
consequence thereof. See also the \htmlmanual, under the heading ``Total Cross Sections''.

Throughout this section, we will discuss collisions
between two high-energy  
hadrons $A$ and $B$ at a squared CM energy $s = \ECMs$. By high energy, we 
mean roughly $\ECM > 10$~\GeV, where the perturbative model is valid. 
Low-energy non-perturbative processes are discussed in 
\cref{subsubsection:lowenergyprocesses}. The (strong-interaction) total 
cross section for the collision of two high-energy hadrons is conveniently 
subdivided into several components, typically 
\begin{align}
\sigma_{\mrm{tot}}^{AB}(s)
& = \sigma_{\mrm{el}}^{AB}(s) + \sigma_{\mrm{inel}}^{AB}(s) \nonumber \\
& = \sigma_{\mrm{el}}^{AB}(s) + \sigma_{\mrm{sd}(XB)}^{AB}(s)
+ \sigma_{\mrm{sd}(AX)}^{AB}(s) + \sigma_{\mrm{dd}}^{AB}(s)
+ \sigma_{\mrm{cd}}^{AB}(s) + \sigma_{\mrm{nd}}^{AB}(s) ~.
\label{eq:soft:sigmasplit}      
\end{align}
The components are:
\begin{itemize}
\item Elastic scattering (el) $AB \to AB$ where the hadrons are
scattered through an angle but are otherwise unharmed. Everything
else, where the final state is not $AB$, is collectively called
inelastic. 
\item Single diffraction (sd) where either of the incoming hadrons
becomes an excited system, while the other remains intact, $AB \to XB$
or $AB \to AX$. Here, $X$ represents the excited system that eventually
will produce two or more hadrons.
\item Double diffraction (dd) where both hadrons are excited,
$AB \to X_1X_2$, but remain as separate objects.
\item Central diffraction (cd), where both hadrons survive but lose
energy to a new central system, $AB \to AXB$.
\item Non-diffractive interactions (nd), or more formally inelastic
non-diffractive ones, where both hadrons break up and form a common
system, $AB \to X$, that is not (easily) subdivided into separate
subsystems, unlike diffraction.  
\end{itemize}
In principle, one could imagine higher diffractive topologies,
say $AB \to X_1 X_2 X_3$, but these are expected to be small and are
neglected here. For applications at low energies we will also introduce
annihilation and resonance contributions.

The dividing line between these different components is unclear,
notably between diffractive and non-diffractive events. Single- or double-
diffractive systems $X$ predominantly have low masses, and thus only
produce a few particles at either end of the full rapidity range.
In between, there is a large rapidity gap, \ie a region in rapidity
space without any particle production. That is unlike the non-diffractive
events, where particle production is assumed to span the whole available
rapidity range. But, since particles are discrete objects randomly
produced, there will be a falling distribution of increasing gap sizes
also in non-diffractive events. In contrast, the falling
tail of large-mass diffractive systems can leave no obvious gap in a
diffractive event. We therefore need to distinguish the theoretical
modelling of cross-section classes and event properties presented here
from the experimental-detector and analysis-dependent classification
of observed events.

\begin{figure}[t!]
\index{Pomeron}\index{Pomeron!Cut}\index{Cut pomeron}\includegraphics[width=\linewidth]{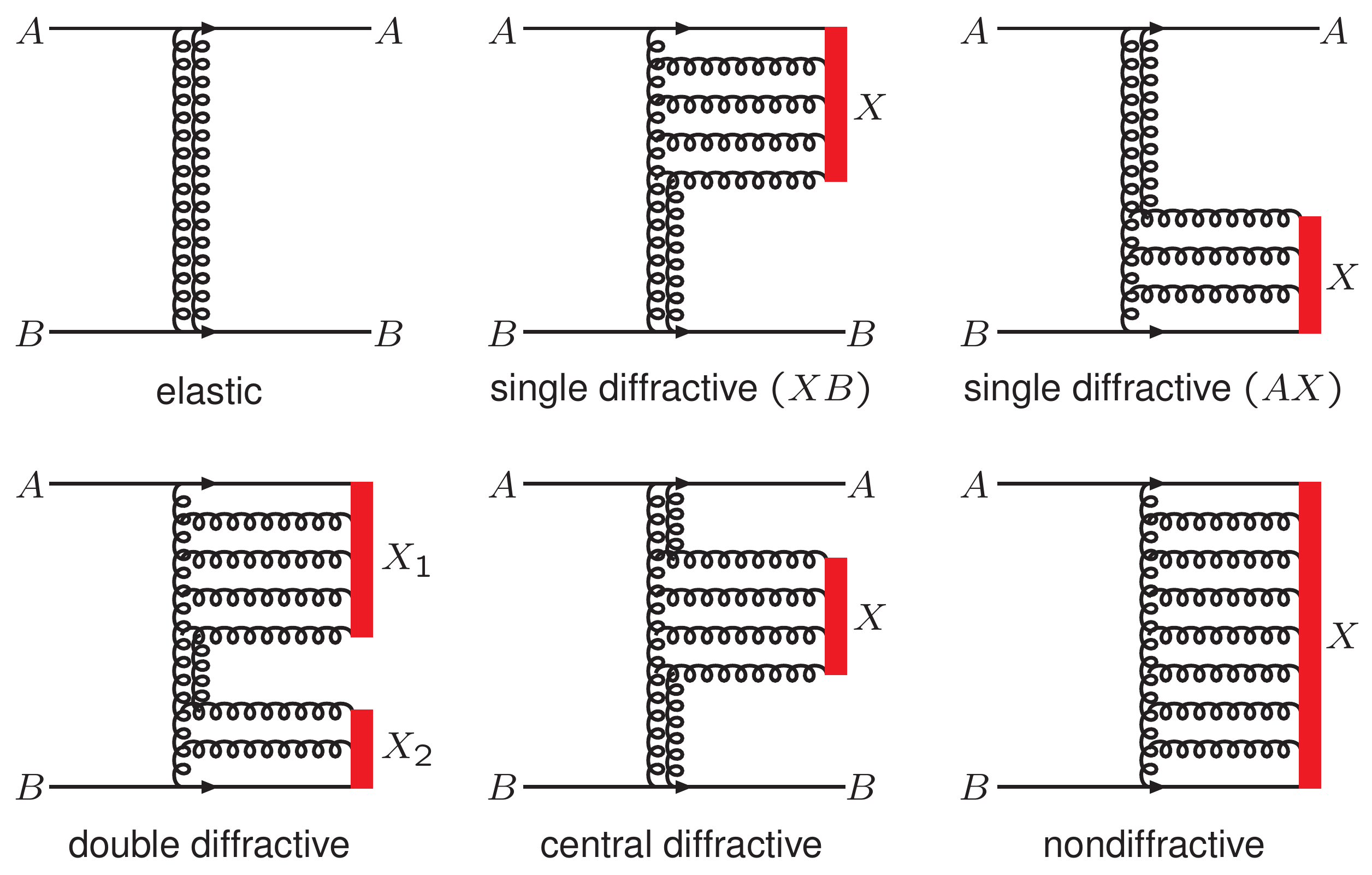}
\caption{\label{fig:soft:sigmatypes}
Schematic Feynman-diagram-style illustration of the six event classes
in \cref{eq:soft:sigmasplit}. A pair of parallel vertical gluons represent
a net colour-singlet exchange, a pomeron, while a single vertical gluon
gives a colour-octet exchange, a cut pomeron. The vertical axis can be
viewed as representing the rapidity range spanned between $A$ and $B$,
where horizontal gluons are regions with possible partonic final-state
presence. The red bars represent the final regions where strings will be
drawn and produce hadrons, whereas the regions without them are
rapidity gaps.}
\end{figure}

\index{Pomeron}\index{Pomeron!Cut}\index{Cut pomeron}In modern
nomenclature, where a pomeron is viewed as shorthand for a 
two-gluon system in a colour-singlet state, the different event classes
can be illustrated in terms of the colour and momentum-energy transfers
between the two incoming hadrons, see \cref{fig:soft:sigmatypes}.
An elastic scattering requires a
pomeron (or reggeon) to be exchanged, so that the scattered hadrons
remain colour singlets, but with (modestly) changed outgoing momenta.
If only one gluon is exchanged, a so-called cut pomeron, then the colour
transfer turns the $A$ and $B$ hadrons into colour-octet objects, which
means they will be connected by colour strings that can fragment into
hadrons over the whole rapidity range, \ie this gives a non-diffractive
event. Single diffraction, \eg $AB \to AX$, can be viewed as a two-step
process. First the emission of a \Pom from $A$, carrying a fraction $\xi$
of the $A$ momentum. And second the collision between the
\Pom and $B$, giving rise to a system with $M_X^2 = \xi s$. 
For the first step a pomeron flux $f_{\Pom}^A(\xi, t)$ can be introduced
in analogy with conventional PDFs, while the second step can be viewed
as a non-diffractive-type $\Pom B$ subcollision, at least for large $M_X$.
Double and central diffraction can be described in a similar manner.

The hadronic cross sections that will be discussed late are
for reasonably high hadron-hadron CM energies, say $\ECM > 10$~GeV,
corresponding to a fixed-target proton-proton beam energy of
$E_{\mrm{beam}} \gtrsim 50$~GeV. Separate
from this, low-energy cross sections will be discussed in the context
of hadronic rescattering, \cref{subsubsection:lowenergyprocesses}.
To a large extent the same language can be used, but at low energies
the contribution from exclusive resonances, $AB \to R \to AB$ or
$AB \to R \to CD$, can give rise to rapid fluctuations in the cross
section as a function of $s$.

\subsubsection{Proton total cross sections}

\index{Odderon}\index{Pomeron}\index{Reggeon}As already mentioned, pomeron and reggeon contributions
play a large 
role in the modelling of cross sections. Both are expected to give an
$s^{\delta}$ energy dependence, where $\delta \geq 0$ for pomerons and
$\delta < 0$ for reggeons, such that high-energy cross sections are
dominated by the pomeron term. The pomeron contribution is even,
\ie the same for $AB$ and $A\overline{B}$, while reggeons can be even
or odd, the latter giving opposite-sign contributions for the two
processes. A hypothetical odderon contribution would be odd, as the
name indicates, and have $\delta \geq 0$ like the pomeron, so that
$\sigma^{A\overline{B}}(s) - \sigma^{AB}(s)$ would not vanish in the
$s \to \infty$ limit. Its existence is supported by recent TOTEM data~\cite{Antchev:2017yns}, but is still not included in most models.

The simple $s^{\delta}$ behaviour is obtained for the exchange of a
single object, whereas multiple exchanges can come in with opposite
signs and damp the rise of cross sections. The Froissart bound~\cite{Froissart:1961ux} shows that cross sections cannot rise faster
than $\ln^2 s$ asymptotically, but that limit is far off. Other bounds
are more relevant, for instance that diffractive cross sections cannot
become larger than the total one~\cite{Schuler:1993wr}. 

The most important hadronic cross sections are the \pp and \ppbar ones.
Here, four different $\sigma_{\mrm{tot}}(s)$ parameterizations are
available for high-energy collisions in \pyt, plus one placeholder,
see further the overview in~\citeone{Rasmussen:2018dgo}. They are roughly
ordered in increasing number of free parameters tuned to data,
with numbers corresponding to the options of the \setting{SigmaTotal:mode} switch.
\begin{enumerate}
\item[0.] A zero option allows the user to set any value at the
currently studied energy, \ie it does not model any energy dependence.
\item The \ac{DL} form~\cite{Donnachie:1992ny},
with one pomeron and one reggeon term,
\begin{equation}
\sigma_{\mrm{tot}}^{AB}(s) = X^{AB} \, s^{0.0808} + Y^{AB} \, s^{-0.4525} ~,
\label{eq:soft:DonLan}
\end{equation}
with $s$ in units of GeV and $\sigma$ in mb. The coefficients 
$X^{\ppbar} = X^{\pp}$, as discussed above. There is no such symmetry
for the $Y^{AB}$, which can be viewed as having one even and one odd
reggeon, but with the same power.
\item The \ac{MBR} parameterization~\cite{Ciesielski:2012mc}, which uses two different expressions.
Below 1.8~TeV the form is given by one pomeron and two reggeon
terms, whereof one odd and one even, with different $\delta$.
Above it a common Froissart-inspired form like
$a \ln^2 s + b \ln s + c$ is used.
\item The ABMST model~\cite{Appleby:2016ask} includes a soft and a
hard pomeron, \ie lower or higher $\delta > 0$, an even and an odd
reggeon, plus terms for two-pomeron and triple-gluon exchange.
\item The COMPAS/RPP parameterization~\cite{Patrignani:2016xqp} contains
a total of six even and six odd terms, including pomeron, odderon,
reggeon, and double-exchange ones.
\end{enumerate}
The relevant cross section parameterizations are hard coded in options
1 -- 4, and cannot easily be changed. 

\subsubsection{Proton elastic cross sections}\index{Elastic scattering}
\label{subsection:sigmaelastic}

Elastic cross sections are related to total ones via the optical
theorem:
\begin{equation}
  \frac{\d\sigma_{\mrm{el}}}{\d t}(s, t=0) = \frac{1 + \rho^2}{16\pi} \,
  \sigma_{\mrm{tot}}^2(s) ~,
\end{equation}
where $t$ represents the squared momentum transfer between the initial
and final proton on the same side of the event.  
For detailed modelling, a suitable starting point is the 
elastic scattering amplitude $T(s,t)$, from which one derives
$(\d\sigma_{\mrm{el}}/\d t)(s,t) \propto |T(s,t)|^2$,
$\sigma_{\mrm{tot}} \propto \mrm{Im}\, T(s,0)$, and
$\rho = \mrm{Re}\, T(s,0) / \mrm{Im}\, T(s,0)$.
Typically $\rho$ is close to 0 and can be neglected to first approximation.
The total elastic cross section is obtained by integration over $t$.

For the simple Regge-theory-motivated ansatz, that
$(\d\sigma_{\mrm{el}}/\d t)(s,t) \propto \exp(B_{\mrm{el}}t)$,
one obtains
\begin{equation}
  \sigma_{\mrm{el}}(s) = \frac{1 + \rho^2}{16\pi} 
  \frac{\sigma_{\mrm{tot}}^2}{B_{\mrm{el}}} ~,
  \label{eq:soft:el:opticalReg}
\end{equation}
where, to a very good approximation at high energies, the $t$
integration range has been extended to $[-\infty, 0]$. The ansatz
also assumes that
\begin{equation}
  B_{\mrm{el}}^{AB}(s) = 2 b_A + 2 b_B + 2 \alpha' \ln (s/s_0) ~,
  \label{eq:soft:el:Bel}
\end{equation}  
where $b_{A,B}$ come from the respective hadronic form factors,
with $b = 2.3~\mrm{GeV}^{-2}$ for unflavoured baryons and 1.4~GeV$^{-2}$
for mesons, $\alpha' = \d L / \d m^2 = 0.25$~GeV$^{-2}$ is the slope of the
pomeron trajectory, and $s_0 = 1 / \alpha' = 4~\mrm{GeV}^2$ is a
typical hadronic scale~\cite{Donnachie:1985iz,Schuler:1993wr}.

In detail, the total-cross-section models above, as selected by
\setting{SigmaTotal:mode}, also handle elastic scattering as follows.
\begin{enumerate}
\item[0.] It is possible for the user to set their own $\sigma_{\mrm{el}}$,
$B_{\mrm{el}}$, and $\rho$ at the current energy.  
\item \index{Schuler-Sjostrand model@Schuler-Sj\"ostrand model}The original DL
model was extended to \ac{SaS}/DL~\cite{Schuler:1993wr} 
by the simple Regge-theory ansatz above, but with the difference that
the $\ln s$ dependence in \cref{eq:soft:el:Bel} is replaced by
an $s^{0.0808}$ term to ensure that $\sigma_{\mrm{el}}$ does not
grow faster than $\sigma_{\mrm{tot}}$ asymptotically. There is no
modelling of $\rho$, but a value can be set by hand.
\item In MBR the ratio $\sigma_{\mrm{el}}(s)/\sigma_{\mrm{tot}}(s)$
is parameterized, separately below and above 1.8~TeV, and separately
for \pp and \ppbar below it. Then \cref{eq:soft:el:opticalReg}
is used to derive a $B_{\mrm{el}}(s)$ slope, with $\rho = 0$.
\item In ABMST the fundamental building block is a complex scattering
amplitude $T(s,t)$, containing the six terms of the total cross
section, each with a separate $t$ dependence, usually, but not always,
of an exponential character. From this, both total and elastic cross
sections are derived, including the $\rho$ parameter.
\item Also the COMPAS/RPP parameterization starts out from a complex
$T(s,t)$, with the same comments as for ABMST, except that the number
of terms now is larger.
\end{enumerate}
There are no further free parameters in the code, beyond the ones
mentioned above.

So far, only strong interactions have been considered. But, since
protons are charged particles, there are also electromagnetic (EM)
interactions. These are given by the traditional Coulomb scattering
cross section, $\d\sigma_{\mrm{el}}/\d t \propto \alphaem^2/t^2$, which
blows up in the $t \to 0$ limit, \ie for vanishing scattering angle.
Therefore, it is always necessary to specify a minimal angle or
equivalently a $t_{\mrm{max}} < 0$. There are two aspects that make
it possible to disregard the EM contributions at the LHC, except for special
runs. Firstly, the EM contribution exceeds the strong one only
below a $|t|$ of order $10^{-3}$~GeV$^2$, which corresponds to extremely
small scattering angles. Secondly, owing to this, inelastic EM collisions
are completely negligible. By default, Coulomb corrections therefore
are not taken into account, but can be switched on.

What complicates the issue is that the elastic scattering amplitude
\begin{equation}
T(s,t) = T_n(s,t) + e^{i\alphaem\phi(t)} \, T_c(s,t) ~,
\end{equation}  
contains a phase factor in front of the Coulomb $T_{\mrm{c}}$
amplitude, relative to the definition of the real part of the
nuclear/QCD $T_{\mrm{n}}$ amplitude. Three different expressions are
used, one for SaS/DL, and also for MBR and
\setting{SigmaTotal:mode}{0}, and one each for ABMST and
COMPAS/RPP. Although written in slightly different ways, they
give almost identical results.

\subsubsection{Proton diffractive cross sections}\index{Diffraction}
\label{subsection:sigmadiffractive}\index{Single diffraction}
\index{Double diffraction}\index{Central diffraction}
\index{Pomeron}

Diffractive cross sections are differential in several variables: 
for single diffraction in $t$ and $M_X$; for double diffraction in
$t$, $M_{X_1}$, and $M_{X_2}$; and for central diffraction in $t_1$,
$t_2$, and $M_X$. Here, $M_X$ represents the mass of the respective 
diffractive system. Alternatively the scaled variable $\xi = M_X^2/s$
is often used, but it is less intuitive when modelling contributions
from low-mass resonances. The fundamental objects are the differential
expressions, and the integrated rates in general do not have simple
closed forms. Within Regge theory it is possible to relate the
differential expressions to the ones for total and elastic ones,
with minor extensions. Specifically, single diffraction is modelled
with triple-Regge exchange graphs that involve the same pomeron (or
reggeon) propagators as before, but requires the introduction of
triple-Regge couplings. If only pomerons are considered, as could be
reasonable at high energies, then mass spectra will behave roughly
like $\d M_X^2 / M_X^2$ and $t$ spectra roughly like $\exp(Bt)$,
where $B = B(s, M_X^2)$ depends on the process considered.

In reality it is more complicated, for a number of reasons.
At low masses the experimental $M_X$ spectrum is not smooth, but reflects
the presence of well defined $\mrm{N}$ and $\Delta$ resonance states.
At high masses phase-space restrictions kick in, \eg in the allowed
$t$ range, as well as a wish to minimize the overlap between
diffractive and non-diffractive event topologies. In addition to
the pomeron also reggeons should be considered, in various combinations,  
contributing to different mass distribution shapes and CM energy
dependencies. Some terms increase faster with CM energy than the
total cross section itself, implying that the description has to break
down at some point. The solution to this is likely to involve the
possibility of multiple exchanges of both a diffractive and non-diffractive
nature, leading to a competition between the two~\cite{Aurenche:1991pk}.

Three different diffractive models are implemented~\cite{Rasmussen:2018dgo}, matching the first three descriptions of
total and elastic cross sections, plus again an additional placeholder,
enumerated in the \setting{SigmaDiffractive:mode} switch in the same way as in \setting{SigmaTotal:mode}.
It is possible to combine the choice of total plus elastic and diffractive
models freely.
\begin{enumerate}
\item[0.] One can set user defined single, double and central
diffractive cross sections for the current energy. In this option
there is a choice between seven possible $M_X^2$ spectra, with
related $t$ shapes.
\item \index{Schuler-Sjostrand model@Schuler-Sj\"ostrand model}The SaS
  model is based on pomeron contributions only, 
\ie is of the form $(\d M_X^2 / M_X^2) \, \exp(Bt)$ to first
approximation. At low masses a smooth enhancement is added, to provide
a simple smeared-out further contribution from resonances.
At large masses the rate is suppressed to reduce the rate
of diffractive events with small rapidity gaps. The rise of the
diffractive cross section with energy is given by integration.
It turns out, however, that the initial ansatz gives a steeper rise
than data, so energy-dependent damping factors have been introduced. 
Central diffraction is a rather recent addition, not included
in many commonly used tunes, and therefore not on by default.
The $B$ slope is similar in spirit to \cref{eq:soft:el:Bel},
but without any form factor contribution for protons that break up,
and the logarithmic term is related to the rapidity gap size,
\eg $\ln(s/M_X^2)$ for single diffraction.
\item In the MBR model the single-, double- and central-diffractive
cross sections are given as ratios of two integrals, one being
the Regge cross section and the other a renormalized flux.
These are matched so that the increase of diffractive cross sections
is kept at an acceptably low rate. The differential distributions in 
$M_X^2$ and $t$ are given by somewhat lengthier expressions than in
SaS, but qualitatively similar.
\item The ABMST model is the most sophisticated one, in terms of number
of components considered. The single-diffractive description is split
into two parts, for high- and low-mass diffraction. The former includes
$\Pom\Pom\Pom$, $\Pom\Pom\Reg$, $\Reg\Reg\Pom$ and $\Reg\Reg\Reg$ graphs,
plus pion exchange, each with a characteristic mass, $t$, and energy 
dependence. Four resonances are modelled in the low-mass regime, along
with a background from the high-mass regime and a contact term
matching the two regimes smoothly. The resonances are excited states
of the proton, each a unit of angular momentum higher than the
previous one. Taken together, the ABMST model gives a very good
description of data at lower energies. Unfortunately the energy
dependence of some terms is too steep, such that single diffraction
at the LHC is overestimated by about a factor of two, and results at
100~TeV would be completely unphysical. A few different options have
therefore been included in the \pyt implementation to damp this
rise~\cite{Rasmussen:2018dgo}. Another problem is that ABMST does not
model double diffraction. One expects an approximate relationship
\cite{Barone:2002cv}
\begin{equation}
  \frac{\d^3\sigma_{\mrm{dd}}}{\d M_{X_1}^2 \, \d M_{X_2}^2 \, \d t}
  \approx \frac{\d^2\sigma_{\mrm{sd}}}{\d M_{X_1}^2 \, \d t} \,
  \frac{\d^2\sigma_{\mrm{sd}}}{\d M_{X_2}^2 \, \d t} \,
  \left( \frac{\d\sigma_{\mrm{el}}}{\d t} \right)^{-1} ~,
\end{equation}  
however, and this has been used to extend the modelling. Also central
diffraction can be introduced by a similar ansatz. 
\item The COMPAS/RPP parameterization does not extend to diffraction,
so there is no such option.  
\end{enumerate}
Each of the models above contain a wide selection of modifiable
parameters, specific to that diffraction model. Both the integrated and
the differential cross sections can be modified, notably affecting
the dependence on CM energy and the shape of the $M_X$ spectra.

In summary, the modelling of diffraction is highly nontrivial, and at
a more primitive stage than that of total and elastic cross sections.
There also exist alternative starting points to the Regge formalism
we have worked with here, notably the Good--Walker one~\cite{Good:1960ba}.
In it, it is assumed that the interaction eigenstates do not agree with
the mass ones. That is, an incoming proton can be viewed as a coherent
sum of interaction eigenstates. During the collision process, parts
of these eigenstates are absorbed to give rise to non-diffractive events.
The remaining parts of the incoming wave function can be projected
back on to a spectrum of possible masses for the outgoing object,
including one component corresponding to elastic scattering. Actually
the ``diffraction'' name comes from the close analogue with optics,
where an opaque disk put in a beam of light absorbs part of the light
but also generates a quantum mechanical diffraction pattern in the
remaining light. Such a picture implies that diffractive and elastic
collisions are more peripheral than non-diffractive ones. The same also
holds in the Regge-language-related MPI framework to be discussed
in the next subsection, so even of the models seemingly are unrelated,
there are many common traits.

\subsubsection{Other cross sections}
\label{subsection:sigmaother}

Except for the absence of Coulomb elastic scattering, collisions
involving (anti)neutrons are assumed to have the same cross sections
as (anti)protons in \pyt, and this similarity appears supported by
data~\cite{Donnachie:1992ny}. Therefore all of the models above can
be used for $\p\n$, $\p\nbar$, $\n\n$ and $\n\nbar$.

For other hadron combinations, the only alternative beyond the user-defined
option is an extension of the SaS/DL setup. It encompasses the following
collision types.
\begin{itemize}
\item Combinations where $\sigma_{\mrm{tot}}(s)$ were fitted by
DL~\cite{Donnachie:1992ny}: $\pip\p$, $\pim\p$, $\kp\p$, $\km\p$,
and $\gamma\p$.
\item \index{Schuler-Sjostrand model@Schuler-Sj\"ostrand model}SaS extensions~\cite{Schuler:1996en}: $\rhoz\p$, $\phiz\p$,
$\Jpsi\p$, $\rhoz\rhoz$, $\rhoz\phiz$, $\rhoz\Jpsi$, $\phiz\phiz$,
$\phiz\Jpsi$, and $\Jpsi\Jpsi$. Particles with identical flavour
content are assumed to give identical cross sections. The prime
example is $\piz$, $\rhoz$, and $\omega$. The emphasis on the
interactions of vector mesons is related to the SaS modelling
of $\gamma\p$ and $\gamma\gamma$ physics, where an important aspect
is that a real photon can fluctuate into a vector meson like $\rhoz$,
$\omega$, $\phiz$, or $\Jpsi$, and interact as such most of the time
(\ac{VMD}, see \cref{subsection:leptonhadron}
and \cref{subsection:photonphoton}). Total $\gamma\gamma$ cross
sections are also provided.
\item Later extensions in the SaS/DL spirit~\cite{Sjostrand:2021dal}: 
$\kz\p$, $\eta\p$, $\eta'\p$, $\D^{0,+}\p$, $\D_{\mathrm{s}}^+\p$,
$\B^{0,+}\p$, $\B_{\mathrm{s}}^0\p$, $\B_{\mathrm{c}}^+\p$, $\Upsilon\p$,
$\Lambda\p$, $\Xi\p$, $\Omega\p$, $\Lambda_{\mathrm{c}}\p$,
$\Xi_{\mathrm{c}}\p$, $\Omega_{\mathrm{c}}\p$,  $\Lambda_{\mathrm{b}}\p$,
$\Xi_{\mathrm{b}}\p$, and $\Omega_{\mathrm{c}}\p$. Isospin symmetry is
used to equate the cross sections of closely related particles,
\eg $\n$ with $\p$ and $\Sigma^{+, 0, -, *+, *0, *-}$ with $\Lambda$.
For the baryon-baryon processes, the corresponding baryon-antibaryon
ones are also implemented. The purpose of these extensions is to
allow the tracing of the evolution of cascades in matter, also
in collisions with nuclei, meaning that essentially all hadronic
collisions with protons or neutrons have to be included.
\item As a final case, a $\sigma^{\Pom\p}_{\mrm{tot}}(s)$ is defined
for pomerons, but more for model studies of diffractive systems at
a given mass than for comparisons with data. 
\end{itemize}
In summary, by suitably mapping a particle onto one of equivalent
flavour content, the possibilities above cover a fair fraction
of all possible hadron collisions. The main exceptions are those
involving baryons with more than one charm or bottom quark, and
(most) collisions between two short-lived particles. Should the need
arise, further extensions along the same lines would be possible for
these cases.

\index{Uncertainties}It should be clear from the onset that the
accuracy expected for these  
cross sections cannot compare with the $\p\p$ and $\pbar\p$ ones.
As a rule of thumb, the rarer the particle, the more uncertain
the assumptions that went into deriving related cross sections.
For many applications, notably the evolution of a cascade in matter,
it is the average collision rates that count rather than the
individual ones, however, one may assume that it should still
work out reasonably well.

The starting point in all these total cross sections is the
pomeron plus reggeon ansatz of \cref{eq:soft:DonLan}. The $X^{AB}$ pomeron
term strength appears to obey the \ac{AQM} rule~\cite{Levin:1965mi,Lipkin:1973nt}, \ie be proportional to the number
$n_i$ of valence quarks in each hadron, but with a reduced contribution
for strange and heavier quarks. Thus we have made the ansatz that
$X^{AB} \propto n_{\mathrm{eff}}^A \, n_{\mathrm{eff}}^B$, with
\begin{equation}
  n_\textrm{eff} = n_\d + n_\u + 0.6 n_\s + 0.2 n_\c + 0.07 n_\b ~.
  \label{eq:soft:neff}
\end{equation}
The prefactors for heavier quarks have been assumed roughly inversely
proportional to their respective constituent quark masses,
\index{Constituent quark masses|see{Quark masses}}\index{Quark masses!Constituent quark masses} which could
be viewed as a reflection of a reduced size of their spatial
wave functions.

The modelling of the  $Y^{AB}$ reggeon factors is considerably less
systematic, since typically several reggeon trajectories may contribute.
The mix of charge-even and charge-odd contributions gives
$Y^{\overline{A}B} \neq Y^{AB}$, while $X^{\overline{A}B} = X^{AB}$.
For baryon collisions $Y^{\overline{A}B} > Y^{AB}$, which can be
viewed as a reflection of possible contributions from $\q\qbar$
annihilation graphs. This is supported by the observation that
$Y^{\phi\p} \approx 0$, consistent with the OZI rule~\cite{Okubo:1963fa,Zweig:1981pd,Iizuka:1966fk},
and we assume that this suppression of couplings between light $\u/\d$
quarks and $\s$ quarks extends to $\c$ and $\b$. Thus, for baryons, the
reggeon $Y^{AB}$ and $Y^{\overline{A}B}$ values are assumed proportional to
the number of $\u/\d$ quarks only, scaled separately from the
$Y^{\p\p}$ and $Y^{\pbar\p}$ reference values. Thereby baryons with the
same flavour content, or only differing by the relative composition of
$\u$ and $\d$ quarks, are taken to be equivalent, \ie
$\sigma^{\Lambda\p}(s) = \sigma^{\Sigma^+\p}(s) =
\sigma^{\Sigma^0\p}(s) = \sigma^{\Sigma^-\p}(s)$.
Another simplification is that $\Dbar/\Bbar$ mesons are assigned
the same cross sections as the respective $\D/\B$, taken to be some
average. 

The $B_{\mrm{el}}^{AB}$ slope for hadronic collisions is defined as in
\cref{eq:soft:el:Bel}, with a universal $\alpha'$ but $b_{A,B}$
taken to be 1.4 for mesons and 2.3 for baryons, except that mesons
made only out of $\c$ and $\b$ quarks are assumed to be more tightly
bound and thus have lower values, in the spirit of the AQM factors.
Given this, and assuming $\rho \approx 0$, the integrated elastic
cross sections are given by \cref{eq:soft:el:opticalReg}.
\index{VMD}For photon
interactions, only the VMD part is assumed to undergo ``elastic''
scatterings, so the fractions of fluctuations to $\rhoz$, $\omega$,
$\phiz$, or $\Jpsi$ are combined with the expressions for these
respective states to scatter elastically.

\index{Diffraction!Schuler-Sjostrand@Schuler-Sj\"ostrand|see{Schuler-Sj\"ostrand model}}
\index{Schuler-Sjostrand model@Schuler-Sj\"ostrand model}Also
diffractive cross sections are calculated 
using the SaS ansatz. 
The differential formulae are integrated numerically for each relevant
collision process and the result suitably parameterized, including a
special threshold-region ansatz~\cite{Sjostrand:2020gyg}. If the
hadronic form factor from pomeron-driven interactions is written as
$\beta_{A\Pom}(t) = \beta_{A\Pom}(0) \, \exp(b_A t)$ then, with suitable
normalization, $X^{AB} = \beta_{A\Pom}(0) \, \beta_{B\Pom}(0)$.
Thus we can define $\beta_{\p\Pom}(0) = \sqrt{X^{\p\p}}$
and other $\beta_{A\Pom}(0) = X^{A\p}/\beta_{\p\Pom}(0)$. These numbers
enter in the prefactor of single diffractive cross sections, \eg
$\sigma_{AB \to AX} \propto \beta_{A\Pom}^2(0) \, \beta_{B\Pom}(0) =
X^{AB} \, \beta_{A\Pom}(0)$.
This relation comes about since the $A$ side scatters (semi)elastically,
while the $B$ side description is an inclusive one, \cf the optical
theorem. In double diffraction $AB \to X_1X_2$ neither side is elastic
and the rate is directly proportional to $X^{AB}$.
For photons again only the VMD parts undergo diffractive scatterings.

The descriptions mentioned so far are intended for cross
sections at high energies. Specifically, the original DL ansatz was
tuned to data down to 6~GeV. At low energies, different descriptions
are used, as outlined in the next subsection, most of which are not
intended to be used much above 10~GeV. In cases where the full energy
range from threshold upwards needs to be used, a smooth interpolation
is therefore applied between the low- and high-energy descriptions.
More precisely, the transition is linear in the range between
\begin{align}
E_{\mathrm{CM}}^{\mathrm{begin}} &= E_{\mathrm{min}} + \max(0., m_A - m_{\p})
+ \max(0., m_B - m_{\p}) ~~\mathrm{and}\\
E_{\mathrm{CM}}^{\mathrm{end}} &= E_{\mathrm{CM}}^{\mathrm{begin}}
+ \Delta E ~,
\end{align}
where $E_{\mathrm{min}}$ is 6~GeV and $\Delta E$  is 8~GeV by default.

\subsubsection{Low-energy processes}
\label{subsubsection:lowenergyprocesses}
\index{Low-energy processes}

At low energies (below $\sim10~\GeV$), the perturbative framework described
in this section breaks down. In modern high-energy physics, experimental
beam energies lie far above this threshold,
but processes at these energies still have applications for example 
in hadronic rescattering (see \cref{subsection:hadronicrescattering}). 
\pyt provides a framework for simulating such low-energy collisions. This
framework can be used explicitly by enabling \setting{LowEnergyQCD:*} processes, 
and is used implicitly inside \pyt when rescattering is turned on. 
The following gives a summary of the available low-energy processes:

\index{Elastic scattering}\paragraph{Elastic scattering} $AB \to AB$ is implemented similarly to elastic
scattering at high energies, except the cross section is calculated
differently, as described below.

\index{Diffraction}\paragraph{Diffractive scattering} (both single and double) is also
similar to how it is implemented at high energies.
Central diffractive ($AXB$) has a very small cross section
at low energies, and is thus not implemented. In addition, at low energies the
diffractive system can sometimes be viewed as a resonance excitation, for
example $\pp \to \p\gp{\Delta}^+$. In \pythia, these excitation processes are
implemented only for nucleon-nucleon interactions.

\paragraph{Non-diffractive scattering} Works similarly in principle to
high-energy non-diffractive interactions, but with extra steps to ensure
the process does not reduce to an elastic scattering at energies very close
to the threshold.

\paragraph{Annihilation processes} Baryon-antibaryon interactions
where one or two quarks annihilate.

\paragraph{Resonance formation} A meson interacting with a baryon or
another meson can form a resonance particle, \eg $\p\pip \to \Delta^{++}$
or $\pip\pim \to \rhoz$.

\noindent 
While several of these processes correspond to similar high-energy processes,
their cross sections are in most cases calculated differently, as perturbative
calculations cannot be used at these energies. Only a short overview of how
the cross sections are calculated is given here, and the reader is referred to
\citeone{Sjostrand:2020gyg} for further details.

When \ac{PDG} data is available\footnote{
\url{https://pdg.lbl.gov/2018/hadronic-xsections/hadron.html}~\cite{Tanabashi:2018oca}}, total cross sections are
calculated using parameterizations or by fitting to data. The
$HPR_1R_2$ parameterization is used when available, as is the case for \eg
nucleon-nucleon interactions~\cite{Tanabashi:2018oca}. For
baryon-antibaryon interactions, a parameterization due to UrQMD is used~\cite{Bass:1998ca}. \pion\pion and \pion\gp{K} interactions use
 parameterizations by Pelaez \etal~\cite{GarciaMartin:2011cn,Pelaez:2019eqa,Pelaez:2016tgi}.
For other processes involving mesons, if the pair can form resonances,
the total cross section is calculated by summing the contributions from each
resonance, possibly also adding an elastic contribution.
While these cases describe the most common processes, there
is also a large set that is not covered. For these remaining
processes, the total cross section is calculated using the
additive quark model (AQM)~\cite{Levin:1965mi,Lipkin:1973nt} with small
modifications introduced to also include charm and bottom quarks
\cite{Sjostrand:2020gyg}. Specifically, the total AQM cross section is given by
\begin{equation} 
  \sigma_{\mrm{AQM}}^{AB} = (40\text{ mb}) \frac{n_{\textrm{eff}}^A}{3}
\frac{n_{\textrm{eff}}^B}{3}~,
\end{equation}
where $n_{\textrm{eff}}$ is the ``effective'' number of quarks in each
hadron, given by \cref{eq:soft:neff}. With this, low-energy
processes are available for all possible hadron-hadron types.

In our description, we define elastic interactions as processes where the
hadrons exchange momenta without ever changing their types, \eg through a
pomeron exchange. We do not include for example ``pseudo-elastic'' scattering
through a resonance, $AB \to R \to AB$. Note that this distinction usually
cannot be made experimentally, so one often considers a process
elastic as long as the outgoing hadrons are of the same type as the incoming
ones. For nucleon-nucleon and nucleon-pion interactions, the
elastic cross section is found by fitting to data below 5~\GeV~\cite{Tanabashi:2018oca}, and by using the CERN/HERA parameterization
above 5~\GeV~\cite{Montanet:1994xu}. Elastic cross sections
for baryon-antibaryon interactions are calculated using another
parameterization by UrQMD~\cite{Bass:1998ca}, and for \pion\pion and
\pion\gp{K}, we use parameterizations by Pelaez \etal~\cite{GarciaMartin:2011cn,Pelaez:2019eqa,Pelaez:2016tgi}. Other cross sections
are given by an elastic AQM-style parameterization. The angular distribution
of the outgoing hadrons is the same as for the high-energy case
(\cref{subsection:sigmaelastic}).

\index{Schuler-Sjostrand model@Schuler-Sj\"ostrand model}
Diffractive cross sections are calculated using the SaS model~\cite{Schuler:1993wr,Schuler:1996en}, with two modifications. First, the
basic model is designed to deal with processes involving only \p, \pbar,
\pion, \rhomeson, \gp\omega, and \gp\phi hadrons. In the low-energy framework,
the generic case is calculated by replacing each incoming hadron by the most
similar among these particles (\eg treating each baryon as a proton), then
rescaling the calculated cross section by the appropriate AQM factor. The
second modification is due to the fact that the basic SaS model is intended
for collision energies above 10~\GeV. This is compensated for by multiplying
by an \textit{ad hoc} factor below 10~\GeV. At low energies, diffractive processes can
lead to the formation of explicit resonances, \eg $\pp \to \p\gp{\Delta}^+$.
This is implemented in \pythia only for nucleon-nucleon interactions, using
the description by UrQMD~\cite{Bass:1998ca}.

Non-diffractive cross sections are calculated by subtracting all other partial
cross sections from the total cross section. One important difference between
non-diffractive interactions at low and high energies is that at low energies,
the hadronization process might sometimes produce a hadron pair that is the
same as the incoming pair, essentially reducing the interaction to an elastic
process ($AB \to X_1 X_2 \to AB$). This is a problem in cases where the calculated
elastic cross section has already been adjusted to fit data. Several steps are
taken to ensure that this does not give unexpected contributions to the
elastic cross section, and are outlined in \cref{sec:small-mass}.

Annihilation processes in our framework refer to baryon-antibaryon 
interactions where one or two quark-antiquark pairs are annihilated.
Strings are drawn between the remaining quark-antiquark pairs, and
hadronize to form outgoing hadrons. The cross section for annihilation
in \ppbar is given by a parameterization by Koch and Dover~\cite{Koch:1989zt},
\begin{equation}
\sigma_\text{ann} = 120 \frac{s_0}{s} \left( \frac{A^2 s_0}{(s - s_0)^2 + A^2 s_0}
+ 0.6 \right)~,
\end{equation}
where $s_0 = 4 m_\p^2$ and $A = 0.05$~GeV. The cross section for other
\gp{B}\gpbar{B} interactions is found by rescaling this value by an AQM
factor. The only exception is when the quark-contents make annihilation
impossible, \eg like in a $\gp{\Delta}^{++} + \gpbar{\Sigma}^-$ interaction,
in which case the annihilation cross section is set to zero.

Finally, resonance production refers to processes where the two hadrons
combine to form one resonance particle. The cross section for the process
$AB \to R$ is given by a non-relativistic Breit--Wigner~\cite{Tanabashi:2018oca},
\begin{equation}
\sigma_{AB \to R} = \frac{\pi}{p_\mrm{CM}^2} \, \frac{(2S_R+1)}{(2S_A+1)(2S_B+1)} 
  \frac{\Gamma_{R \to AB} \Gamma_R}{(m_R - \sqrt{s})^2 + \frac14 \Gamma_R^2} ~,
  \label{eq:resonanceFormation}
\end{equation}
where $S$ is the spin of each particle, $p_\mrm{CM}$ is the CM momentum of the
incoming particles, $\Gamma_{R \to AB}$ is the mass-dependent partial
width of the decay $R \to AB$, and $\Gamma_R$ is the mass-dependent
total width of $R$. The full list of implemented resonances is given
in~\citeone{Sjostrand:2020gyg}. For \pion\pion and \pion\gp{K} where the total
cross section is calculated using the parameterization by Pelaez \etal,
the partial cross sections are rescaled to ensure their sum equals the
total cross section.

\subsection{Multiparton interactions basics}
\label{subsection:mpi}

\index{MPI}
Hadrons are composite objects. A proton consists of three valence
quarks, plus countless gluons and sea quarks. When two hadrons
collide there is a possibility for several parton pairs to collide
--- multiparton interactions (\ac{MPI}s). Processes with exactly two
parton pairs, double parton scattering (DPS), was proposed in the
early days of QCD, but then viewed as a rare perturbative process~\cite{Landshoff:1978fq,Goebel:1979mi}. Regge--Gribov theory, on the
other hand, allowed for events with multiple cut pomerons, \ie
several ``strings'' crossing from one rapidity end of the event to
the other, each generating its sequence of low-$\pT$ hadrons~\cite{Abramovsky:1973fm}. The \pyt philosophy for the first time
introduced a merger and extension of these two approaches~\cite{Sjostrand:1985vv}. In it, semiperturbative MPIs both generate
multiple minijets, that contribute to the $\pT$ flow, and multiple
colour connections between the beam remnants, that leads to events
with higher multiplicity. This picture is now generally accepted
in its essentials. An overview of MPI theory and phenomenology
can be found in~\citeone{Bartalini:2017jkk}, with the \pyt perspective
described in~\citeone{Sjostrand:2017cdm}, with many further references.
See also the \htmlmanual under the ``Multiparton Interactions'' heading.
 
\subsubsection{The perturbative cross section}
\begin{figure}[t!]
\includegraphics[width=0.5\linewidth]{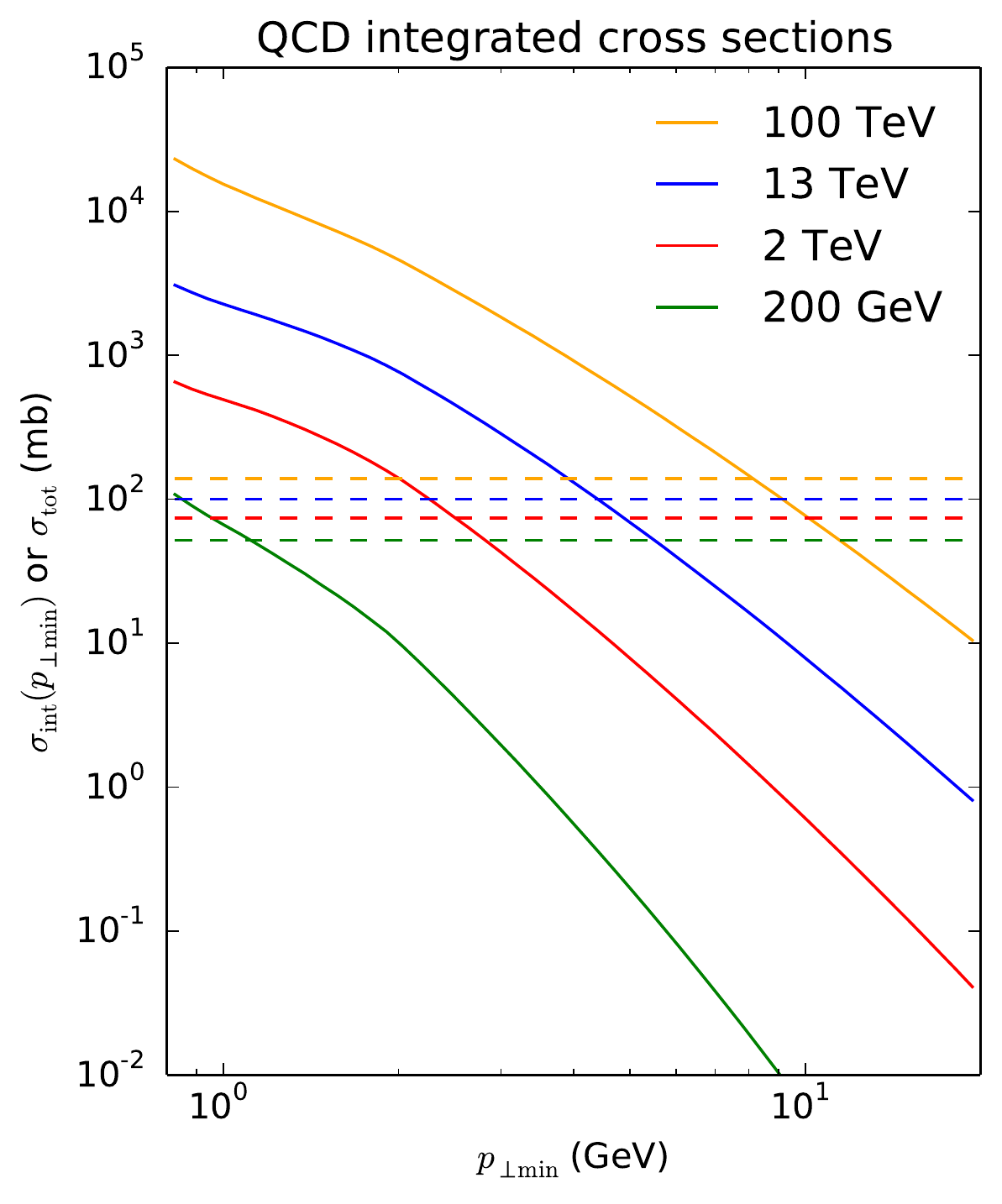}%
\includegraphics[width=0.5\linewidth]{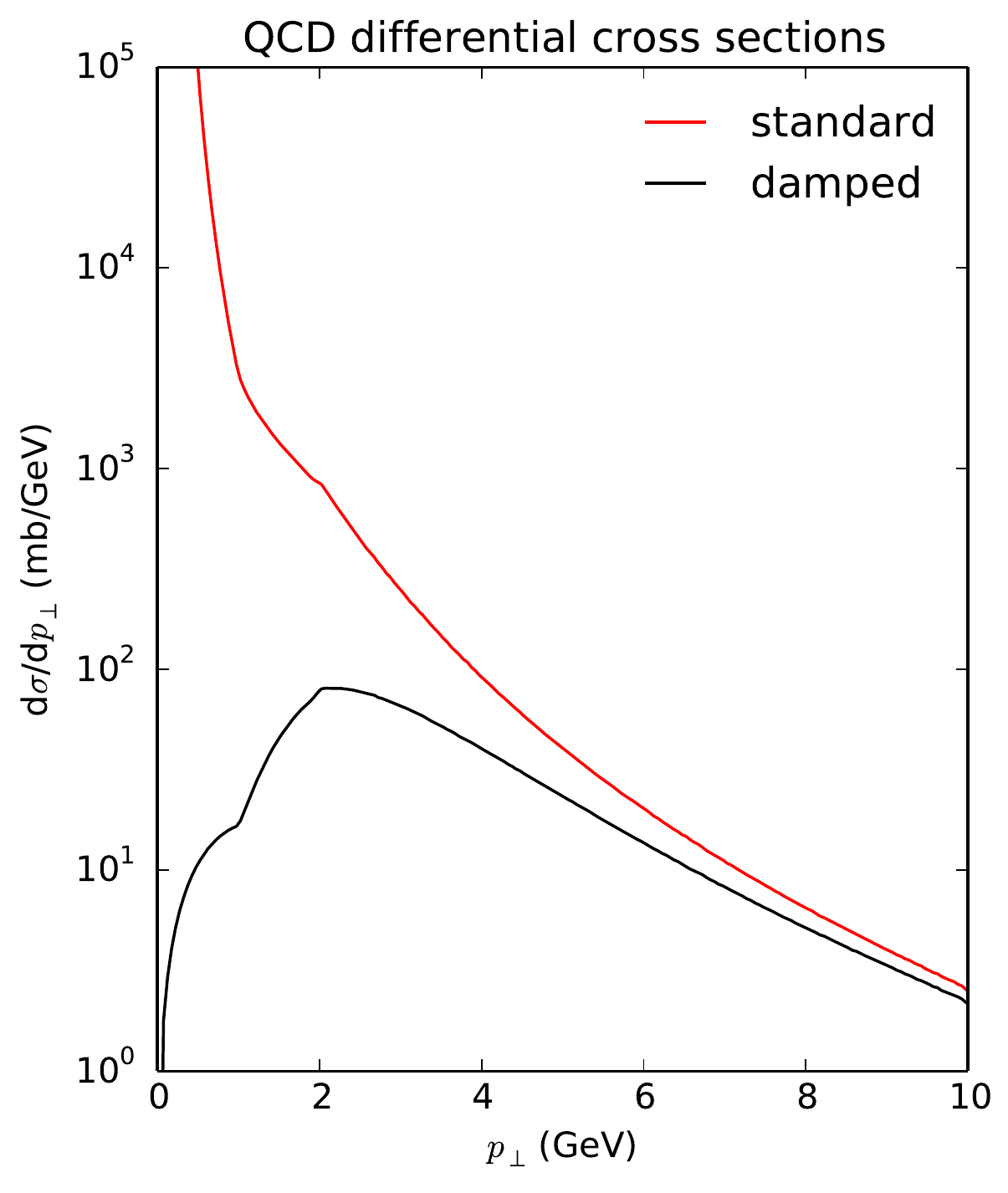}\\
\null\hfill (a) \hfill\hfill (b) \hfill\null
\caption{\label{fig:mpi:sigmaintdiff}
(a) Integrated standard $2 \to 2$ QCD cross section as a function of
the lower cutoff $\pTmin$ for $\pp$ collisions at 200~GeV, 2~TeV,
13~TeV and 100~TeV, respectively. Horizontal dashed lines give the
total cross section at their respective energy.
(b) Differential $2 \to 2$ QCD cross section at 13 TeV, as obtained in
standard perturbation theory, and after multiplication by the damping
factor \cref{eq:soft:MPI:ptdamp}. Minor breaks in slopes come from
transitions, notably the freeze of PDFs below 1~GeV. Results have been
obtained for the default \pythia setup, and details depend \eg on the
choice of PDF set.}
\end{figure}

\index{alphaS@$\alphas$!in MPI}
The $\pT$-differential perturbative QCD $2 \to 2$ cross section
can, to leading order, be written as
\begin{equation}
\frac{\d\sigma}{\d \pTs} = \sum_{i,j,k} \iiint f_i(x_1, Q^2) \,
f_j(x_2, Q^2) \, \frac{\d\hat{\sigma}_{ij}^k}{\d\hat{t}} \,
\delta \left( \pTs - \frac{\hat{t}\hat{u}}{\hat{s}} \right)
\, \d x_1 \, \d x_2 \, \d \hat{t} ~,
\label{eq:soft:MPI:dsigma}
\end{equation}
with $Q^2 = \pTs$ as factorization and renormalization scale,
partons assumed massless, and $k$ running over processes with the same
initial state but different final states
(\cf\cref{eq:basics-sigma-nbody} and \cref{eq:basics-sigma-twobody}).
The partonic cross section $\d\hat{\sigma}/\d\hat{t}$ is dominated by
$t$-channel gluon exchange, \ie $\q \q' \to \q \q'$, $\q \g \to \q \g$
and $\g \g \to \g \g$. (Including those $u$-channel graphs that easily
can be relabelled into $t$-channel ones.) This cross section has an
approximate behaviour
\begin{equation}
\frac{\d\hat{\sigma}}{\d\hat{t}} \propto \frac{\alphas^2(Q^2)}{\hat{t}^2}
~~ \Rightarrow ~~ \frac{\d\hat{\sigma}}{\d\pTs} \propto
\frac{\alphas^2(\pTs)}{p_{\perp}^4} ~.
\label{eq:soft:MPI:dsigmahat}
\end{equation}
Evidently this cross section is divergent in the limit $\pT \to 0$,
as shown in \cref{fig:mpi:sigmaintdiff}. The integrated $2 \to 2$ cross section
above some $\pTmin$ scale, $\sigma_{\mrm{int}}(\pTmin)$, is increasing
with the $\pp$ collision energy. But, taking $\pTmin = 1$~GeV as a scale 
where perturbation theory would be expected to hold, already at a 
collision energy of 200~GeV, the $\sigma_{\mrm{int}}$ value exceeds the
total \pp cross section $\sigma_{\mrm{tot}}$ at this energy.

A further aspect is that $\sigma_{\mrm{tot}}$ is subdivided into different
components, as already discussed, and the $2 \to 2$ partonic interactions
primarily occur within the non-diffractive one, which is what we will
assume next. They are absent in elastic scattering and low-mass
diffraction, while they can occur in high-mass diffraction. This is a
small fraction of the total cross section, however, so to first
approximation we may neglect it. Later on we will correct the picture. 

Putting it together, one finds that $\sigma_{\mrm{int}}(\pTmin)$ is
around 60~mb for $\pTmin \approx 5$~\GeV at LHC energies, which is
also the order of the non-diffractive \pp cross section
$\sigma_{\mrm{nd}}$. Going to lower $\pTmin$ scales the cross section
rapidly explodes, $\sigma_{\mrm{int}}(2~\GeV) \approx 1000~\mrm{mb} 
\approx 15\, \sigma_{\mrm{nd}}$. In the context of MPIs, this is not
as bad as it may sound, since we may interpret the ratio
$\sigma_{\mrm{int}}(\pTmin) /\sigma_{\mrm{nd}}$ as the average number
of MPIs above the $\pTmin$ scale that occur in a non-diffractive
collision. Nevertheless, an infinity of MPIs in the $\pTmin \to 0$
limit is not attractive. 

A damping of the cross section at low $\pT$ can be viewed as a consequence
of colour screening: in the $\pT \to 0$ limit a hypothetical exchanged
gluon would not resolve individual partons but only (attempt to) couple to
the vanishing net colour charge of the hadron. By contrast, traditional
perturbation theory is based on the assumption of asymptotically free
incoming and outgoing partons. To be specific, a multiplicative damping
factor
\begin{equation}
\left( \frac{\alphas(\pTos + \pTs)}{\alphas(\pTs)} \,
\frac{\pTs}{\pTos + \pTs} \right)^2 ~.
\label{eq:soft:MPI:ptdamp}
\end{equation}
is introduced, with $\pTo$ a free parameter. This means a
modification to \cref{eq:soft:MPI:dsigmahat}
\begin{equation}
\frac{\d\hat{\sigma}}{\d\pTs} \sim
\frac{\alphas^2(\pTs)}{p_{\perp}^4} \longrightarrow
\frac{\alphas^2(\pTos + \pTs)}{(\pTos + \pTs)^2}  ~,
\end{equation}
which is finite in the limit $\pT \to 0$, \cf\cref{fig:mpi:sigmaintdiff}b.

\index{Uncertainties!MPI@in MPI}
The $\pTo$ value is not provided from first principles, although
suggestions have been made to equate it with the saturation scale
$Q_s$ in colour glass condensate models~\cite{McLerran:1993ni,Gelis:2010nm}.
Fits to $\pp/\ppbar$ data give a result that increases with energy,
by default like
\begin{equation}
\pTo(\ECM) = (2.28~\GeV) \left( \frac{\ECM}{7~\TeV}
\right)^{0.215} \, , \label{eq:pToDef}
\end{equation}
but alternatively a logarithmic rise could be assumed. It should
be noted that results are sensitive to the choice of PDF set, and
especially to the low-$x$ behaviour of the gluon distribution at small
$Q^2$. The numbers are for the default 
NNPDF2.3 QCD+QED LO $\alphas(M_{\Z}) = 0.130$ set~\cite{Ball:2013hta}.
The choice of an LO PDF is deliberate, since the description of partonic
collisions is also an LO one, but in particular since NLO PDFs tend
to become unphysical at small $x$ and $Q^2$. This is why \pyt offers
the possibility to use two different sets of PDFs, one for the hard
processes, where these kinematic regions are not accessed, and one
for MPIs and showers, where often they are.

\begin{figure}[t]
    \centering
    \includegraphics*[width=0.85\textwidth]{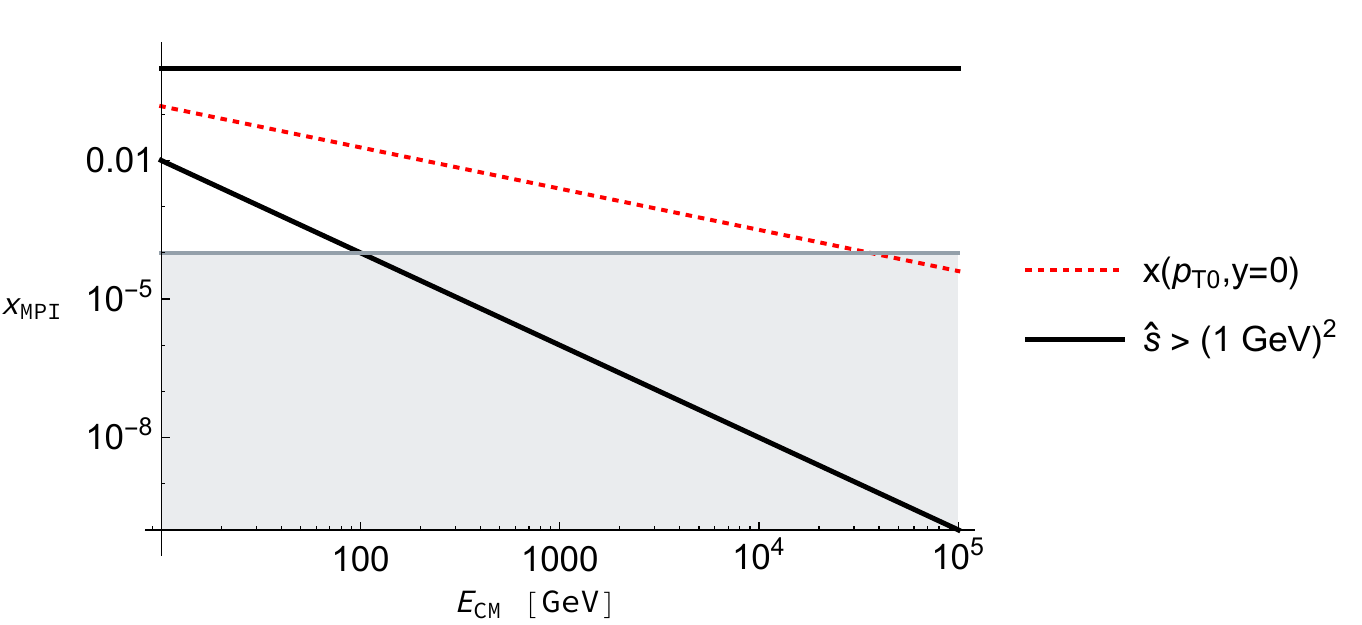}
    \caption{Range of $x$ values accessible to MPI in \pyt, for
      $10\,\mathrm{GeV}<\ECM< 100\,\mathrm{TeV}$. Scatterings at $\hat{p}_\perp \sim \pTo$
      will involve at least one $x$ fraction below the 
      red dashed line.
      Grey shading highlights the low-$x$ extrapolation region $x <
      10^{-4}$ in which current PDFs are uncertain by a factor two or more.
      \label{fig:xranges}}
\end{figure}

\index{PDFs}
\index{Uncertainties!PDF@from PDFs}
\index{Uncertainties!MPI@in MPI}
The range of $x$ values that can be accessed by MPI in
\pyt is illustrated by the thick black lines in \cref{fig:xranges}, for 
hadronic CM energies ranging from 10~GeV (at the left-hand edge of the plot) to 
100~TeV (at the right-hand edge). The shaded area 
emphasizes the region of low $x \le 10^{-4}$ in which current PDFs are
uncertain by a factor two or more.
The red dashed line indicates the
solution to $x^2s = 4p_{\perp 0}^2$, for the default form of $\pTo(\ECM)$
given by \cref{eq:pToDef}. Any partonic collision with $\hat{p}_\perp
\sim \pTo$ will involve at least one $x$ value below this line.
Thus, especially at LHC energies and beyond, it is important to keep in mind
that the effective MPI cross section (and hence any observables
derived from it) around $\hat{p}_\perp\sim \pTo$ really depends on the
\emph{combination} of $\pTo$ and the shape of the low-$x$ PDF
parameterization. Since the latter can change 
drastically between different PDF sets, any ``tuned'' values of $\pTo$
should be considered valid only for the PDF set they were obtained with.

\subsubsection{The impact-parameter model}\index{Impact parameter}
\label{subsection:mpi:impactparameter}

A hadron is characterized not only by its longitudinal structure,
as encoded in the PDFs, but also by its transverse one. That is, the
``impact parameter'' plane overlap of partons in the two hadrons
influences the possible collisions. The hadrons are Lorentz contracted
to pancake shapes in high-energy collisions, such as the LHC, so the
partons can be considered as frozen during the short collision time.

As a first approximation we will assume a common spatial distribution
$\rho(\mathbf{x}) \, \d^3x = \rho(r) \, \d^3x$ for all parton types
and momenta in a hadron. In the collision between two hadrons, passing by
at an impact parameter $b$, the overlap between the two distributions
is then given by
\begin{align}
\widetilde{\mathcal{O}}(b) & = \iint \d^3x \, \d t \,
\rho_{\mrm{boosted}}\left( x - \frac{b}{2}, y, z - vt \right) \,
\rho_{\mrm{boosted}}\left( x + \frac{b}{2}, y, z + vt \right)
\nonumber \\
& \propto \iint \d^3x \, \d t \, \rho ( x, y, z) \,
\rho ( x, y, z - \sqrt{b^2 + t^2} )  ~,
\label{eq:soft:MPI:convolute}
\end{align}
where the second line is obtained by suitable scale changes.

A few different $\rho$ distributions have studied and made available 
as options. Using Gaussian distributions is especially convenient, since the
convolution then becomes trivial. However, a single Gaussian does not give a
good enough description of the data, and a better description is obtained
with a sum of two Gaussians, with a small core region embedded in a
larger hadron. This can be viewed as a manifestation of the  
``hot spot'' concept~\cite{Gribov:1984tu,Mueller:1985wy}, wherein
partons may tend to cluster in a few small regions, typically associated
with the three valence quarks, as a consequence of partons cascading
from them. Another alternative, that is currently the default, is a
one-parameter shape 
\begin{equation}
\widetilde{\mathcal{O}}(b) \propto \exp\left( -b^d \right) ~,
\end{equation}
where $d < 2$ gives more fluctuations than a Gaussian and $d > 2$
less. The default value is $d = 1.85$, \ie slightly more peaked 
than a Gaussian. Note that the expression is for the overlap, not for
the individual hadrons, for which no related simple analytic form
is available.
            
It is now assumed that the interaction rate, to first approximation,
is proportional to the overlap
\begin{equation}
\langle \widetilde{n}_{\mrm{MPI}}(b) \rangle
= k \, \widetilde{\mathcal{O}}(b)~.
\end{equation}
Interactions are assumed to occur independently of each other for a
given $b$, to first approximation, which leads to a Poissonian number
distribution. Zero interactions means that the hadrons pass each other
without interacting. The $\widetilde{n}_{\mrm{MPI}}(b) \geq 1$
interaction probability therefore is
\begin{equation}
\mathcal{P}_{\mrm{int}}(b) =
 1 - \exp \left( - \langle \widetilde{n}_{\mrm{MPI}}(b) \rangle \right)
= 1 - \exp \left( - k \, \widetilde{\mathcal{O}}(b) \right) ~.
\label{eq:soft:MPI:bprob}
\end{equation}
We notice that $k \widetilde{\mathcal{O}}(b)$ is essentially the same as
the eikonal $\Omega(s,b) = 2\, \mrm{Im}\chi(s,b)$ of optical models~\cite{Glauber:1959xx,Chou:1968bc,Bourrely:1984gi,LHeureux:1985qwr},
but split into one piece $\widetilde{\mathcal{O}}(b)$ that is purely
geometrical and one $k = k(s)$ that carries the information on the
parton-parton interaction cross section. 

Simple algebra shows that the average number of interactions in events,
\ie hadronic passes with $n_{\mrm{MPI}} \geq 1$, is given by
\begin{equation}
\langle n \rangle = \frac{\int k \, \widetilde{\mathcal{O}}(b) \,
\mrm{d}^2 b}{\int \mathcal{P}_{\mrm{int}}(b) \, \mrm{d}^2 b}
= k \langle \widetilde{\mathcal{O}} \rangle 
= \frac{1}{\sigma_{\mrm{nd}}} \int_0^{\s/4} \frac{\d\sigma}{\d\pTs}
\, \d\pTs   ~,
\label{eq:soft:MPI:nAvg}
\end{equation}
which fixes the absolute value of $k$ (numerically). We have also
taken the occasion to introduce
$\langle \widetilde{\mathcal{O}} \rangle$ as the average overlap. Hence
$\widetilde{\mathcal{O}}(b) / \langle \widetilde{\mathcal{O}} \rangle$
represents the enhancement at small $b$ and depletion at large $b$.

So far, we have assumed the transverse $b$-space profile to be decoupled
from the longitudinal $x$ one. This is not the expected behaviour,
because low-$x$ partons in a hadron should diffuse out towards larger
$r$ during the evolution down from higher-$x$ ones~\cite{Frankfurt:2005mc}.
Additionally, if $r = 0$ is defined as the centre of energy of a hadron,
then by definition a parton with $x \to 1$ also implies $r \to 0$. In
this spirit, there is a non-default \pyt option  with correlated $x$
and $r$~\cite{Corke:2011yy}. It does not explicitly trace the evolution
of cascades in $x$, but assumes that the $r$ distribution of partons at
any $x$ can be described by a simple Gaussian, but with an $x$-dependent
width:
\begin{equation}
  \rho(r, x) \propto \frac{1}{a^3(x)} \, \exp
                     \left( - \frac{r^2}{a^2(x)} \right)
~~\mrm{with}~~
  a(x) = a_0 \left( 1 + a_1 \ln \frac{1}{x} \right) ~,
\end{equation}
where $a_0$ and $a_1$ are free parameters to be determined. The overlap
is then given by
\begin{equation}
\widetilde{\mathcal{O}}(b, x_1, x_2) = \frac{1}{\pi} \,
  \frac{1}{a^2(x_1) + a^2(x_2)} \,
  \exp \left( - \frac{b^2}{a^2(x_1) + a^2(x_2)} \right) ~.
\end{equation}
In principle one could argue that also a third length scale should be
included, related to the transverse distance the exchanged propagator
particle, normally a gluon, could travel. This distance should be made
dependent on the $\pT$ scale of the interaction. For simplicity,
this further complication is not considered but, a finite effective radius is allowed also for $x \to 1$.
The generation of events is more complicated with an $x$-dependent
overlap, but largely involves the same basic principles. Until now,
there is no evidence that this option provides a better description
of data than the default, unfortunately.

\subsubsection{The generation sequence}

To introduce the MPI generation algorithm, leave aside the
impact-parameter issue for a moment. The probability to have an MPI
at a given $\pT$ in a non-diffractive event is then given by
$(1 /\sigma_{\mrm{nd}}) \d\sigma/\d\pT$. If interactions occur
independently of each other, the number of MPIs
would be distributed according to a Poissonian, with the zero
suppressed. There are a few ways to generate such a Poissonian.

The \pyt approach is inspired by the parton-shower paradigm.
The generation of consecutive MPIs is formulated as an evolution
downwards in $\pT$, resulting in a sequence of $n$ interactions
with $\sqrt{s}/ 2 > p_{\perp 1} > p_{\perp 2} > \cdots > p_{\perp n} > 0$.
The probability distribution for $p_{\perp 1}$ becomes
\begin{equation}
\frac{\d\mathcal{P}}{\d p_{\perp 1}} =
\frac{1}{\sigma_{\mrm{nd}}} \frac{\d\sigma}{\d p_{\perp 1}} \,
\exp \left( - \int_{p_{\perp 1}}^{\sqrt{s}/2} \frac{1}{\sigma_{\mrm{nd}}}
\frac{\d\sigma}{\d p'_{\perp}} \, \d p'_{\perp} \right) ~.
\label{eq:soft:MPI:phardest}
\end{equation}
Here the naive probability is corrected by an exponential factor
expressing that there must not be any interaction in the range
between $\sqrt{s}/2$ and $p_{\perp 1}$ for $p_{\perp 1}$ to be the hardest
interaction. The procedure can be iterated, to give
\begin{equation}
\frac{\d\mathcal{P}}{\d p_{\perp i}} =
\frac{1}{\sigma_{\mrm{nd}}} \frac{\d\sigma}{\d p_{\perp i}} \,
\exp \left( - \int_{p_{\perp i}}^{p_{\perp i - 1}} \frac{1}{\sigma_{\mrm{nd}}}
\frac{\d\sigma}{\d p'_{\perp}} \, \d p'_{\perp} \right) ~.
\label{eq:soft:MPI:pnext}
\end{equation}
The exponential factors resemble Sudakov form factors of parton showers~\cite{Sudakov:1954sw}, or virtual corrections of ``uncut pomerons'' in the
Regge--Gribov framework, and fills the same function of ensuring that
probabilities are bounded by unity. We will use the Sudakov terminology
to stress this similarity. Summing up the probability for a scattering
at a given $\pT$ scale to happen at any step of the generation chain
gives back $(1 / \sigma_{\mrm{nd}}) \, \d\sigma / \d \pT$,
and the number of interactions above any $\pT$ is a Poissonian with
an average of $\sigma_{\mrm{int}}(\pT) / \sigma_{\mrm{nd}}$,
as it should. The downwards evolution in $\pT$ is handled
by using the veto algorithm, like for showers. If no MPIs are generated
in the evolution, a sequence is rejected and a new try made.

When the impact-parameter variability is to be included as well,
\cref{eq:soft:MPI:phardest} generalizes to
\begin{equation}
\frac{\d\mathcal{P}}{\d^2 b \, \d p_{\perp 1}} =
\frac{\widetilde{\mathcal{O}}(b)}{\langle \widetilde{\mathcal{O}} \rangle}
\, \frac{1}{\sigma_{\mrm{nd}}} \frac{\d\sigma}{\d p_{\perp 1}}
\, \exp \left( -
\frac{\widetilde{\mathcal{O}}(b)}{\langle \widetilde{\mathcal{O}} \rangle}
\int_{p_{\perp 1}}^{\sqrt{s}/2} \frac{1}{\sigma_{\mrm{nd}}}
\frac{\d\sigma}{\d p'_{\perp}} \, \d p'_{\perp} \right) ~.
\label{eq:soft:MPI:phardestwithb}
\end{equation}
This expression can be integrated over $p_{\perp 1}$ to give
\cref{eq:soft:MPI:bprob}. Once $b$ has been chosen, the selection
is similar to that in \cref{eq:soft:MPI:phardest}, except that there
is now a factor
$\widetilde{\mathcal{O}}(b)/\langle \widetilde{\mathcal{O}} \rangle$
multiplying the rate. The same factor enters in the extension of
\cref{eq:soft:MPI:pnext}, for the continued evolution, to
\begin{equation}
\frac{\d\mathcal{P}}{\d p_{\perp i}} =
\frac{\widetilde{\mathcal{O}}(b)}{\langle \widetilde{\mathcal{O}} \rangle} \,
\frac{1}{\sigma_{\mrm{nd}}} \frac{\d\sigma}{\d p_{\perp i}} \,
\exp \left( -
\frac{\widetilde{\mathcal{O}}(b)}{\langle \widetilde{\mathcal{O}} \rangle} \,
\int_{p_{\perp i}}^{p_{\perp i - 1}} \frac{1}{\sigma_{\mrm{nd}}}
\frac{\d\sigma}{\d p'_{\perp}} \, \d p'_{\perp} \right) ~.
\label{eq:soft:MPI:pnextwithb}
\end{equation}

The usefulness of the doubly differential expression in
\cref{eq:soft:MPI:phardestwithb} is not so apparent in the generation
of an inclusive non-diffractive event sample, where $p_{\perp 1}$ can be
integrated out before selecting $b$. But it gives important insights,
especially since the MPI machinery is also intended to be used to
generate the underlying event associated with other processes. Assume
\eg that we want to produce a hard jet sample, \ie $p_{\perp 1} > \pTmin$.
For a large $\pTmin$ the steep fall of $\d\sigma/\d\pT$ ensures that the
argument of the exponent is tiny, and so the exponent itself is close to
unity and can be neglected. The $b$ and $p_{\perp 1}$ expressions then
factorize. The former variable is selected proportional to  
$\widetilde{\mathcal{O}}(b)$, while the latter is selected according to
the conventional differential cross section. Since
$\widetilde{\mathcal{O}}(b)$ is more peaked
at small $b$ than $\mathcal{P}_{\mrm{int}}(b)$, it means that hard
processes are selected at more central $b$ values than inclusive
non-diffractive events. The physics is quite clear: the probability to
obtain a hard collision is proportional to the full parton-parton
collision rate, $\langle \widetilde{n}_{\mrm{MPI}}(b) \rangle \propto%
\widetilde{\mathcal{O}}(b)$, and so it is strongly peaked at small $b$,
while already a single MPI is enough to obtain a non-diffractive event,
and so that probability saturates at unity in $\mathcal{P}_{\mrm{int}}(b)$. 
The consequence of picking a smaller $b$ in hard processes is that the
selection rate for subsequent MPIs, \cref{eq:soft:MPI:pnextwithb},
also is larger, thus giving a higher level of underlying activity than
that of the full non-diffractive event sample, the ``pedestal effect''. 

While the expression in \cref{eq:soft:MPI:phardestwithb} provides
for interpolation between hard and soft events, it is important to
note that only the non-diffractive processes, \ie the ones where the
hardest interaction is selected by the MPI machinery, involve the full
correlation. If one studies a hard process, be it hard QCD jets or
something else, then in \pyt the selection of process kinematics is
done with no reference to MPIs. It is only if and when, after
the MPI machinery is invoked, that the $\pT$ scale of the hard process
is used to select a $b$ value that takes into account the Sudakov factor.

Therefore, in the study of hard QCD jets, one should not pick such a
low $\pTmin$ that the Sudakov factor deviates appreciably from unity.
In practice, this means that one should have $\pTmin$ at least above
20~GeV at LHC energies. If one wants to study jets below that scale,
one can as well start out from the full non-diffractive sample. When a
hard process is fed into the MPI machinery, however, $b$ is chosen
according to \cref{eq:soft:MPI:phardestwithb} in full, \ie including
the Sudakov. That is, if by mistake one were to generate LHC jets at or
below 10~GeV, the Sudakov would not be used in the $\pT$ selection, and thus 
the cross section would be overestimated, but it would be used in the
$b$ selection, and thereby provide the correct underlying-event activity.

So far, we have only considered $2 \to 2$ QCD processes in the MPI
framework, but the list can be extended also to other ones. By default
\pyt allows other $2 \to 2$ processes to be included in the Sudakov
factor, and thereby also in the MPI generation: jet pairs via $s$-channel
$\gamma^*$ or $t$-channel $\gamma^*/\Z/\Wpm$ exchange, events with one
or two photons, or charmonium or bottomonium recoiling against a jet.
Needless to say, these cross sections are much lower than the standard
QCD ones, and therefore do not make much of a difference, but nevertheless
help provide a richer non-diffractive or underlying-event structure.

Another issue is what upper limit to set for the selection of $p_{\perp 2}$.
If studying QCD jets, the ordering  $p_{\perp 1} > p_{\perp 2}$ is obvious;
anything else would not reproduce the inclusive scattering cross
section. But, if the hard process is single $\Z$ production, say,
then this is not part of the MPI machinery, and so there is no
double counting involved by allowing the underlying events to contain
jets up to the kinematic limit. (The exception is if weak showers are
switched on; then a hard QCD jet can emit a softer $\Z$, and so such
topologies could be double counted.) A few options are available, but the
default strategy in \pyt is to split events into two types. If the final
state of the hard process contains only ($\d, \u, \s, \c$ or $\b$) quarks,
gluons, and photons then $\pTmax$ is chosen to be the factorization
scale for internal processes, and the \texttt{scale} value for external
Les Houches input. If not, interactions are allowed to go all the way
up to the kinematic limit.

\subsubsection{Momentum and flavour conservation}
\label{subsection:mpi:pdfrescaling}
\index{PDFs!MPI@for MPI}
\index{Multiparton interactions|see{MPI}}

As formulated so far, the same PDFs are used for all MPIs. This would
allow more momentum to be taken out of a beam than there is, and also
favour the repeated collisions of valence quarks that have already reacted.
It is here that the ordering of the emissions becomes important. Standard
PDFs can indeed be used for the first emission, which is the hardest one
and therefore the one most visible and the one that standard PDFs have
been tuned to describe. For subsequent emissions, the PDFs can gradually
be modified to take into account the effects of the previous ones. An
obvious modification is to rescale the $x$ scale such that PDFs do not
extend to higher values than left by the previous ones, \ie
\begin{equation}
x_i < x_{i,\mrm{max}} \equiv X_i = 1 - \sum_{j = 1}^{i - 1} x_j ~, 
\end{equation}
but we will also want to consider flavour aspects. The beauty is that
these successive modifications, that gradually let the PDFs diverge from
the conventional ones, occur at falling $\pT$ scales, where individual
MPIs become less easily studied, so imperfections do not give large
effects. The consecutive reduction of remaining momentum also means that
the $n_{\mrm{MPI}}$ distribution, for a fixed $b$, will fall off
faster than the assumed Poissonian. What does not change, fortunately,
is the fraction of $n_{\mrm{MPI}} = 0$ events that have to be thrown
away, because that is entirely determined by whether a first MPI can
be generated with standard PDFs or not.

To extend the PDF framework, to include not only a simple $x$ rescaling
but also flavour counting, it is assumed that quark distributions
can be split into a valence and a sea part. In cases where this is
not explicit in the PDF parameterizations, it is assumed that the sea
is flavour-antiflavour symmetric, so that one can write \eg
\begin{equation}
u(x,Q^2) = u_{\mrm{val}}(x,Q^2) + u_{\mrm{sea}}(x,Q^2) =
u_{\mrm{val}}(x,Q^2) + \overline{u}(x,Q^2) ~.
\end{equation}
The parameterized $u(x, Q^2)$ and $\overline{u}(x, Q^2)$ distributions
can then be used to find the relative probability for a kicked-out
$u$ quark to be either valence or sea.

For valence quarks two effects should be considered. One is the
reduction in content by previous MPIs: if a $\u$ valence quark has
been kicked out of a proton then only one remains, and if two then
none remain. In addition, the constraint from momentum conservation
should be included. Together this gives
\begin{equation}
u_{i,\mrm{val}}(x,Q^2)
= \frac{N_{u,\mrm{val,remain}}}{N_{u,\mrm{val,original}}} \,
\frac{1}{X_i} \, u_{\mrm{val}} \left(\frac{x}{X_i},Q^2\right) ~,
\label{eq:soft:MPI:valfrescale}
\end{equation}
for the $\u$ quark in the $i$'th MPI, and similarly for the $d$. The
$1/X_i$ prefactor ensures that the $\u_i$ integrates to the remaining
number of valence quarks. The momentum sum is also preserved, except
for the downwards rescaling for each kicked-out valence quark. The
latter is compensated by a uniform scaling up of the gluon and
sea PDFs.

When a sea quark (or antiquark) $\q_{\mrm{sea}}$ is kicked out of a
hadron, it must leave behind a corresponding antisea parton in the
beam remnant, by flavour conservation, which can then participate in
another interaction. We can call this a companion antiquark,
$\qbar_{\mrm{cmp}}$. In the perturbative approximation the pair comes
from a gluon branching $\g \to \q_{\mrm{sea}} + \qbar_{\mrm{cmp}}$.
This branching often would not be in the perturbative regime, but we
choose to make a perturbative ansatz, and also to neglect subsequent
perturbative evolution of the $q_{\mrm{cmp}}$ distribution. Even if
approximate, this procedure should catch the key feature that a sea
quark and its companion should not be expected too far apart in $x$.
Given a selected $x_{\mrm{sea}}$, the distribution in
$x = x_{\mrm{cmp}} = y - x_{\mrm{sea}}$ then is
\begin{align}
q_{\mrm{cmp}}(x;x_{\mrm{sea}}) & =  C \int_0^1
g(y) \, P_{\g \to \q_{\mrm{sea}}\qbar_{\mrm{cmp}}}(z)
\, \delta(x_{\mrm{sea}} - zy) \, \mrm{d}z
\nonumber \\
& =  C~\frac{g(x_{\mrm{sea}}+x)}{x_{\mrm{sea}}+x} \,
P_{\g \to \q_{\mrm{sea}}\qbar_{\mrm{cmp}}}
\left(\frac{x_{\mrm{sea}}}{x_{\mrm{sea}}+x}\right) ~.
\label{eq:sea_comp}
\end{align}
Here $P_{\g \to \q\qbar}(z)$ is the standard DGLAP branching kernel,
$g(y)$ an approximate gluon PDF, and $C$ gives an overall normalization
of the companion distribution to unity. Furthermore, an $X_i$ rescaling
is necessary as for valence quarks. The addition of a companion quark
does break the momentum sum rule, this times upwards, and so is
compensated by a scaling down of the gluon and sea PDFs.

In summary, in the downwards evolution, the kinematic limit is
respected by a rescaling of $x$. In addition, the number of remaining
valence quarks and new companion quarks is properly normalized. Finally,
the momentum sum is preserved by a scaling of gluon and (non-companion)
sea quarks. It is interesting to note that the joint PDFs for the first
two MPIs behave rather similarly to the Gaunt--Stirling DPS PDFs~\cite{Gaunt:2009re}, whereas the \pyt approach currently is the only
one that explicitly offers triple parton distributions and beyond.

\subsubsection{Interleaved and intertwined evolution}
\label{subsubsec:soft:interleave}\index{Interleaved evolution}

So far we have only considered an MPI as a $2 \to 2$ process, but it
should be associated with ISR and FSR showers. In particular, ISR needs
to take momentum from the beams, and can also change the ``original''
flavour taken out of the beam during the backwards evolution. This
implies a more intricate competition between the MPI systems than
already outlined. If all MPIs are first considered, then their number
will be maximized, whereas there may be little room left for ISR. If
instead ISR is added to each MPI before proceeding to the next, then
there will be less room left for MPIs. 

Time ordering does not give any clear guidance what is the correct
procedure. Incoming high-energy hadrons can be viewed as flat pancakes,
such that all MPIs happen simultaneously at the collision moment,
while ISR stretches backwards in time from it, and FSR forwards.
But we have no clean way of separating the hard interactions themselves
from the virtual ISR cascades that ``already'' exist in the colliding
hadrons.

Instead we choose the same guiding principle as we did when we originally
decided to consider MPIs ordered in $\pT$: it is most important
to get the hardest part of the story ``right'', and then one has to live
with an increasing level of approximation for the softer steps.
Since also showers are ordered in (some kind of) $\pT$, it is
meaningful to choose $\pT$ as common evolution scale. Thus
the scheme is characterized by one master formula
\begin{align}
\frac{\mrm{d} \mathcal{P}}{\mrm{d} \pT}&=
\left( \frac{ \mrm{d}\mathcal{P}_{\mrm{MPI}}}{\mrm{d} \pT}  +
\sum   \frac{ \mrm{d}\mathcal{P}_{\mrm{ISR}}}{\mrm{d} \pT}  +
\sum   \frac{ \mrm{d}\mathcal{P}_{\mrm{FSR}}}{\mrm{d} \pT} \right)
\nonumber \\
 & \times \exp \left( - \int_{\pT}^{\pTmax}
\left( \frac{ \mrm{d}\mathcal{P}_{\mrm{MPI}}}{\mrm{d} p_{\perp}'}  +
\sum   \frac{ \mrm{d}\mathcal{P}_{\mrm{ISR}}}{\mrm{d} p_{\perp}'}  +
\sum   \frac{ \mrm{d}\mathcal{P}_{\mrm{FSR}}}{\mrm{d} p_{\perp}'}
\right) \mrm{d} p_{\perp}' \right)
\label{eq:soft:MPI:combinedevol}
\end{align}
that probabilistically determines what the next step will be.
Here the ISR sum runs over all incoming partons, two per
already produced MPI, the FSR sum runs over all outgoing partons
(or dipoles), and $\pTmax$ is the $\pT$
of the previous step. Starting from the hardest interaction,
\cref{eq:soft:MPI:combinedevol} can be used repeatedly to construct a
complete parton-level event. The flavour and momentum used by previous
MPIs or shower branchings are book kept in accordance with the principles
outlined previously, with a few straightforward extensions.
For ISR, \eg the $x$ and flavour of the own MPI does not count as used up.

MPIs are not only related to each other by overall momentum and flavour
conservation issues, but may be directly interacting with each other.
Two such examples are joined interactions and partonic rescattering.

In the former, two partons participating in two separate MPIs may turn
out to have a common ancestor when the backwards ISR evolution traces
their prehistory. The joined interactions are well known in the context
of the forwards evolution of multiparton densities~\cite{Konishi:1979cb,Kirschner:1979im}. It can approximately be
turned into a backwards evolution probability for a branching $a \to bc$
\begin{equation}
\mrm{d} \mathcal{P}_{bc}(x_b, x_c, Q^2) \simeq
\frac{\mrm{d} Q^2}{Q^2} \, \frac{\alpha_{\mrm{s}}}{2 \pi} \
\frac{x_a f_a(x_a, Q^2)}{x_b f_b(x_b, Q^2) \, x_c f_c(x_c, Q^2)} \,
z (1-z) P_{a \to bc}(z) ~,
\label{eq:soft:MPI:fcorrevol}
\end{equation}
with $x_a = x_b + x_c$ and $z = x_b/(x_b + x_c)$. The main approximation
is that the two-parton differential distribution has been been factorized
as $f^{(2)}_{bc}(x_b, x_c, Q^2) \simeq f_b(x_b, Q^2) \, f_c(x_c, Q^2)$,
to put the equation in terms of more familiar quantities.

Just like for the other processes considered, a form factor is given by
integration over the relevant $Q^2$ range and exponentiation. Associating
$Q \simeq \pT$, joined interactions can be included as a fourth term
in \cref{eq:soft:MPI:combinedevol}. But technical complications arise
when the kinematics of joined branchings are reconstructed,
notably in transverse momentum,  and the code to overcome these was
never written. One reason is that already the evolution itself showed that
joined-interaction effects are small and tend to occur at low $\pT$
values~\cite{Sjostrand:2004ef}.

The second intertwining possibility is rescattering, \ie that a parton
from one incoming hadron consecutively scatters against two or more partons
from the other hadron. The simplest case, $3 \to 3$, \ie one rescattering,
has been well studied~\cite{Paver:1982yp,Paver:1983hi,Paver:1984ux}.
The conclusion is that it should be less important than two separated
$2 \to 2$ processes: $3 \to 3$ and $2 \times (2 \to 2)$ contain the same
number of vertices and propagators, but the latter wins by involving one
parton density more. The exception could be large $\pT$ and $x$ values,
but there $2 \to n, n \geq 3$  QCD radiation anyway is expected to be the
dominant source of multi-jet events.

For rescattering, a detailed implementation is available as an option
in \pyt~\cite{Corke:2009tk}, as follows. In order to allow a rescattering
then a scattered parton has to be put back into the PDF, but now as a
$\delta$ function. A hadron can therefore be characterized by a new PDF
\begin{equation}
f(x,Q^2) \rightarrow
f_{\mrm{rescaled}}(x,Q^2) + \sum_{i} \delta(x - x_i)
= f_{\mrm{un}}(x,Q^2) + f_{\delta}(x,Q^2) ~,
\end{equation}
where $f_{\mrm{un}}$ represents the unscattered part of the hadron and
$f_{\delta}$ the scattered one. The scattered partons have the same
$x$ values as originally picked, in the approximation that small-angle
$t$-channel gluon exchange dominates, but more generally there will be
shifts. The sum over delta functions runs over all partons that are
available to rescatter, including outgoing states from hard or MPI
processes and partons from ISR or FSR branchings. All the partons of
this disturbed hadron can scatter, and so there is the possibility for
an already extracted parton to scatter again. With the PDF written in
this way, the MPI scattering rate can be seen as a sum of four terms,
depending on whether the $f_{\mrm{un}}$ or the $f_{\delta}$ is involved
on either incoming side. Unfortunately, like for the joined interactions
above, the kinematics become quite messy, specifically the
propagation of recoils between systems that are partly intertwined but
also partly separate.  

A third and more dramatic intertwining possibility is that the
perturbative cascades grip into each other. An example is the ``swing''
mechanism, whereby two dipoles in the initial state can reconnect
colours, which is a key aspect of the \dipsy generator~\cite{Avsar:2006jy,Bierlich:2014xba}. An implementation exists
in a branch of \pythia~\cite{Bierlich:2019wld}, but not yet in the
public version.

\subsubsection{Spatial parton vertices}\index{Impact parameter}

While setting spatial production vertices of unstable hadrons and leptons
is a standard task (see \cref{sec:hadron-lifetimes}), the corresponding
task for parton vertices in MPIs (as well as for beam remnants and parton
shower) is not. The main issue to tackle is, that as the MPI and shower
models are formulated in momentum space only, no obviously correct
correlation with an impact-parameter picture exists. The plan is to further develop
such an integrated framework, based on matching with dipole calculations
on proton Fock states in impact-parameter space~\cite{Bierlich:2019wld}, but
as such information is needed for string interactions 
(\cref{sec:string-interactions}) and hadronic rescattering 
(\cref{subsection:hadronicrescattering}), a basic framework is in
place already now. 

The basic framework includes four choices for the $\p\p$ overlap region,
from which vertices are sampled randomly. For all model choices, vertices
of ISR and FSR partons are smeared relative to their mother by a Gaussian
distribution, with a width of $\sigma_{\mathrm{v}}/k_\perp$, where $k_\perp$ is the
transverse momentum of the produced parton, and $\sigma_{\mathrm{v}}$ is
a parameter to be set by the user.

The four possible choices for the overlap region are:
\begin{itemize}
	\item The proton profile is a Lorentz-contracted ball of uniform density. 
	This gives an almond-shaped overlap region, similar to heavy-ion collisions,
	favouring MPIs being displaced perpendicular to the collision plane. This option
	somewhat collides with impact-parameter selection in the MPI model, as it does
	not allow any interactions of the impact parameter to be larger than twice the 
	hadron radius.
	\item The proton profile is a Lorentz-contracted three-dimensional Gaussian
	(motivated by the proton mass distribution), easily reduced to a two-dimensional
	one, as the $z$ can be integrated out. The overlap region is taken as the product
	of the two displaced Gaussians, which is in itself a Gaussian.
	\item A variation of the above Gaussian scheme, but elongated by a factor 
	$\sqrt{(1+\epsilon)/(1-\epsilon)}$, where $\epsilon$ is a parameter determining
	whether production should be favoured in the collision frame or out of the 
	collision frame.
	\item Another variation of the Gaussian scheme, but with a modulation factor
	$1 + \epsilon + \cos(2\phi)$, and $\phi$ defined with respect to the collision
	plane. 
\end{itemize}

It should be noted that the models for spatial parton vertices are at a very early stage of
development, and subject to change in the future.

\subsubsection{Other MPI aspects}

There are several topics that concern MPIs, that will be described
separately. One such is the issue of colour flow. The colours within
each MPI, and its associated ISR and FSR, are initially assigned in the
$\Nc \to \infty$ limit. This implies that each parton taken out of a
hadron, to go into a MPI, leaves its corresponding unique anticolours
behind in the beam remnant. With many MPIs involved this gives an
unrealistically complicated remnant, and so there is a machinery that
attempts to associate an initial-parton colour from one MPI with an initial-parton
 anticolour from another MPI. Remaining colour lines attach to the
remnant partons, see further the beam remnants description. This still
allows colour lines to be drawn criss-cross in the event. Colour
reconnection (CR) is a mechanism whereby these colour lines may be
reconnected, typically in such a way that the total string length is
reduced, further described in \cref{sec:colRec}.

In this section we have reasoned around MPIs in the non-diffractive
component in hadron-hadron collisions, which is the prime, but not
the only, application of the MPI framework. One extension is that
photons have a resolved component, where they behave more-or-less
like hadrons, and undergo MPIs in a similar manner. Another is that
diffraction may be viewed as involving the collision of pomerons with
hadrons or with each other, and that also pomerons can be associated
with a hadronic structure that allows MPIs to occur. These aspects
will be discussed further in their respective context.

A standard task for \pyt is to generate one predetermined hard process
and then add underlying-event activity to that, which means that most
of the time the additional MPI activity will be too soft to give
explicitly visible jets. This means that generation efficiency will be
low if one is interested in studies of double parton scattering.
But, there is a possibility to request two hard scatterings in an event,
each of a given type and within given kinematic ranges. While
one of the two processes can be selected from the full range of
possibilities, the other must be chosen from a list of a dozen
process groups. This is not a fundamental limitation, but covers all
that we could see a possible application for, and if need be the list
could be extended. Furthermore, as a non-standard extension to the
Les Houches Accord, it is also possible to feed in external events
with two hard processes for further handling in \pyt. See \cref{subsection:secondhard} for further details.

Since MPIs play such a key role for hadronic event properties, it is
important to tune them as well as possible to describe minimum bias,
\ie predominantly non-diffractive, and underlying events alike.
A number of settings and parameters are available to that end.
Of special interest is the $\pTo$ parameter, that directly influences
important properties such as the multiplicity distribution.
Finally, it is worth mentioning that the MPI component normally
is the most time-consuming task of the \pyt initialization step.
In order to prepare the Monte Carlo sampling of the differential
cross section, it is necessary to find an upper envelope of it in
the $(x_1, x_2, \hat{t})$ phase space. This envelope is based on
multichannel sampling, where the relative importance of the
channels should be optimized to allow a reasonably high sampling
efficiency. The MPI cross section itself also needs to be integrated,
as part of the $p_{\perp}$-evolution formalism. The initialization
of non-diffractive events therefore may take around a second,
\ie almost two orders more than it takes to generate an LHC event
afterwards. If one had to repeat the MPI initialization for each
new event, this step would form a bottleneck. That would be the case
in diffractive events, where the mass of the diffractive system varies
from one event to the next. To this end, diffraction is initialized
for a number of logarithmically evenly spaced mass values, and then
parameters for intermediate masses are obtained by interpolation.
If the incoming beams have varying energies, also non-diffractive
events can be set up for a range of collision energies.
Thus initialization may take tens of seconds for the full set of
inelastic processes, while the subsequent interpolation time is
negligible compared to the event-generation one. If furthermore \pyt
is initialized for multiple hadron types, the time needed becomes
proportionately longer. An option therefore exists to save the MPI
initialization data to a file, for reuse in subsequent runs,
see \cref{sec:standalone:variableBeams}.

\subsection{Beam remnants}
\index{Beam remnants}\index{Remnants|see{Beam remnants}}
\label{subsection:beamRemnants}

What is left of a beam particle, after the partons initiating hard
interactions and MPI have been removed from it (and showered), is
called the beam remnant.
By definition, the remnant itself does not participate in any momentum
exchanges at scales larger than ${\cal O}(1\,\mathrm{GeV})$. Hence, in
\pyt, it is regarded as a purely non-perturbative object, which does
not undergo a parton shower.

The general strategy is to add the minimal number of partons required
to conserve the beam particle's quantum numbers (flavour, colour and
baryon number), taking into account which valence and sea flavours
have been scattered out of it. The remaining beam-particle momentum is
then shared amongst those partons, as described below.
Note that what is relevant to determining the
remnant structure is not which partons initiated Born-level processes
(or MPI) at the respective hard-process factorization scale(s), but
instead the ones \emph{after} initial-state radiation, at $Q \sim
Q_\mathrm{cutoff} \sim {\cal O}(1\,\mathrm{GeV})$.
For brevity, we henceforth refer to these low-scale partons as
``initiator partons''. 

By default, also some ``intrinsic transverse momentum'' is added for the 
initiators and the remnant partons. Final momentum conservation is then 
ensured by rescaling the sampled momenta of the remnant partons appropriately. 
The procedure is discussed in more detail in \citeone{Sjostrand:2004pf} and 
outlined below for hadron beams and the more specialized cases of lepton and 
photon beams will be discussed in \crefrange{subsection:leptonlepton}{subsection:photonphoton}.
 
\subsubsection{Flavour structure}

The first step in beam remnant generation is to determine the number
and flavours of the remnant partons. This begins by including the
remaining valence quarks. For baryons, if two or more valence quarks
are present, a randomly selected pair of these is turned into a
diquark state. In this case, relative probabilities for different
diquark spins are derived within the context of the non-relativistic
\su{6} model, \ie flavour $\su{3}_\mathrm{uds}$
times spin \su{2}. For instance, a $ud$ diquark in a proton
remnant is 3/4 spin-0 and 1/4 spin-1, while a $uu$ diquark always has
spin-1. If the initiator was a gluon, then the remnant is a
colour-octet object, which is split into a triplet and an antitriplet,
again using \su{6} to determine relative weights. For a
proton remnant, $P(u+ud_0)=\frac12$, $P(u+ud_1) = \frac16$, and $P(d +
uu_1)=\frac13$.

Otherwise, the valence flavours are unambiguous assuming that valence content 
has been fixed beforehand. As sea quarks are created in pairs, for all
sea quarks  
that have taken out from the beam particle a companion quark with an opposite 
flavour and colour is added if such have not been already found during partonic 
evolution. If no other remnants are needed, a gluon (photon) is added to carry 
the momentum of the hadron (lepton) beam, otherwise no gluons are added as 
remnant partons unless required to balance for the colour structure. 
\index{DIS}For DIS events, it is also possible to collapse two remnant 
partons directly into a colour-singlet hadron.

\subsubsection{Colour structure}

\index{Uncertainties!Hadron@in Hadronization}Since the incoming hadrons (or,
more generally, the incoming beam 
particles) are colourless, the combined set of initiator
and beam-remnant partons must be colourless, too. In the very simplest
cases, such as when the remnant consists of a single triplet and/or
antitriplet colour, there is no ambiguity. But when there are several 
such charges, the assignment of colour flow in the remnant (roughly,
which remnant-parton colours to associate with which initiator-parton
colours) is inherently ambiguous, and there is no first-principles
solution. \pyt contains two distinct models that address this
ambiguity, called  
``old'' and ``new'', based on the time they were developed and
implemented. Currently, the ``old'' one is the default. 

The old model~\cite{Sjostrand:2004pf} is motivated by the way colour
flow is treated in parton showers, and extends this to the
beam remnant, as follows. Starting from the simplest
representation of the colour structure of the valence quarks in the 
incoming beam particle (a quark-antiquark pair for a meson and a three-quark
junction structure for a baryon, simplified to a quark-diquark
structure when possible), initiator gluons are
attached in random order to one of the valence quarks (selected
at random if there are several), and quark-antiquark pairs are added
as if they came from gluon splittings. Thus this model captures the
qualitative behaviour that is expected from leading-colour QCD.  

The new model~\cite{Christiansen:2015yqa} is motivated by \su(3) colour
algebra, and essentially extends the QCD-based colour-reconnection
model to the beam remnant, as follows. 
First, the set of initiator partons is considered. An \su{3} product
determines the possible overall multiplets that can be formed from
those partons. If one assumes they are uncorrelated, the naive
probability for the set to be in any of those multiplets would be
given simply by state counting. A free parameter allows for the application of an (exponential) weighting factor favouring small multiplets over larger
ones. This is intended as a way to mimic correlations due to
possible saturation effects which are not otherwise explicitly
represented in \pyt. Having selected a multiplet for the set of initiator
partons, the beam-remnant colour configuration has to be the inverse
of that, to conserve the colour-singlet nature of the beam particle.
The minimum amount of gluons are added to the beam remnant in order to
obtain this colour configuration.  

\subsubsection{Primordial $k_\perp$}\index{Primordial kT@Primordial $k_\perp$}
\label{sec:primordialKT}

As the hard processes and parton showers in \pyt are based on collinear factorization, only the longitudinal momenta are generated during the perturbative treatment. However, some transverse momentum of non-perturbative origin due to Fermi motion of partons inside a hadron is expected. Furthermore, studies on \Z-boson transverse-momentum distributions have indicated that a significant amount of partonic $\pT$ is required to reproduce these distributions in hadron-hadron collisions. In \pyt such partonic transverse momentum is modelled with primordial $k_\perp$ that acts as a proxy for non-perturbative and possibly perturbative initial $\pT$.

In \pyt the primordial $k_\perp$ is generated from a two-dimensional Gaussian distribution. For hard-process initiators the width of the Gaussian is parameterized as
\begin{equation}
\sigma(Q) = \frac{\sigma_{\mathrm{soft}} Q_{1/2} + \sigma_{\mathrm{hard}}}
{(Q_{1/2} + Q)} \frac{m} {(m + m_{1/2} y_{\mathrm{damp}})},
\label{primKTwidth}
\end{equation}
where $Q$ is the renormalization scale for the hardest process and $\pT$ for subsequent MPIs and $m$ the mass ($\sqrt{\hat{s}}$) of the system. The $Q$-dependent factor provides an interpolation between a soft scale set by parameter $\sigma_{\mathrm{soft}}$ and a hard scale, set by $\sigma_{\mathrm{hard}}$, and $Q_{1/2}$ controls the midpoint between these two. The $m$-dependent factor on the right-hand side in turn provides damping for small-mass and/or large-rapidity systems. Such damping is introduced due to purely technical reasons so the controlling parameters $m_{1/2}$ and $y_{\mathrm{damp}} = (\frac{E}{m})^{r_{\mathrm{red}}}$, where $r_{\mathrm{red}}$ controls the of amount rapidity damping, should not have much influence on related observables. For the remnant partons not directly connected to any hard process, the width of the $k_\perp$-distribution is fixed by an another parameter $\sigma_{\mathrm{remn}}$ and does not depend on any scale related to hard scattering or MPIs. After sampling the $k_\perp$ for each parton in the beam it is inevitable that the total transverse momentum of the beam becomes non-zero. To retrieve the original beam $\pT$, the $k_\perp$ of all partons will be rescaled with a common factor in such a way that the net four-momentum of the beam particles will be preserved.

\subsubsection{Longitudinal momentum}

In addition to the transverse momentum, the remnant partons should also carry the remaining longitudinal momentum of the beam particle, $X$. As a first step, a momentum fraction $x < X$ is sampled for each remnant parton. In case of valence quarks, the value is sampled according to $(1-x)^a/\sqrt{x}$, where the power $a$ can be adjusted for each parton flavour. Such a distribution approximates the valence quark PDFs around the initial scale ${\cal O}(1\,\mathrm{GeV})$ at which the remnants are constructed. For the remaining companion quarks, the momentum fraction is sampled from the distribution defined in \cref{eq:sea_comp} which takes into account that the sea quarks are always created in pairs, by definition, from gluon splittings. Gluons (and photons) are only added as remnants if no valence or companion quarks are remaining in the beam. As only one of these will be added as a remnant, it will carry all the remaining beam particle's momentum $X$.

After the initial momentum fractions have been sampled for each remnant parton, these have be to rescaled to make sure that the total four-momentum is conserved in each event. As now both the initiator and the remnant partons carry also transverse momentum, the longitudinal-momentum fraction of the remnants cannot be simply rescaled with $X$ but some momentum have to be shared between the two beams to balance the event, for details see \citeref[section 4.4]{Sjostrand:2004pf}. In some special cases, such as DIS processes, only one remnant is required and no such balancing can be done. To account for momentum conservation, the final-state parton momenta are then boosted and rotated in such a way that the total four-momentum is conserved for the sampled remnant configuration.

\subsection{Hadron-hadron collisions}
\label{subsection:hadronhadron}

In \cref{subsection:sigmatotal} we introduced the main event types in
hadron-hadron collisions, and how their total and differential cross
sections are parameterized in \pythia.
Elastic-scattering events are trivial to model, given the $\d\sigma/\d t$
cross-section expression; there are just two hadrons coming in and the
same two coming out, with a momentum transfer $t$ and a randomly-selected $\varphi$ angle. See \cref{subsection:sigmaelastic} for the
various options available for proton elastic-scattering cross sections, and
\cref{subsection:sigmaother} for the less sophisticated expressions
used for other hadrons.
 The subsequent test on MPIs and beam
remnants are mainly concerned the non-diffractive
component. It has the largest cross section, and especially it is the one
where the bulk of hard processes occur, which makes it the most studied
one experimentally. In this section we provide some further comments
on this event class in \cref{sec:soft:mb}, but in particular describe
additional aspects in the description of diffraction in
\cref{sec:soft:diff,sec:soft:hardDiff}.

\subsubsection{Minimum-bias and related inclusive
  processes}\label{sec:soft:mb}\index{Minimum bias}

The inelastic non-diffractive event type is often also called
\ac{MB}. Strictly speaking, however, MB refers to the
smallest possible trigger bias that allows for the identification of non-empty events
in a given experimental context. Depending on the detector acceptance,
MB will typically also include contributions from processes that \pyt
labels as diffractive. Thus, if the aim is to simulate an inclusive
sample of ``minimum bias'' events, usually both diffractive and
non-diffractive events must be included, and then subjected to the
appropriate experimental trigger requirements.
\index{Zero bias}\index{Pileup}

Other, related, experimental terms are zero bias (\eg based
on a bunch-crossing timing trigger, including some \textit{a priori} unknown
fraction of genuinely empty events), pileup (essentially also zero
bias except in cases where pileup contamination may affect trigger
variables such as calorimeter energies), inelastic $\ge N$ events
(inelastic events with at least $N$ particles in some given fiducial
region), and non-single-diffractive events (typically a
``double-sided'' MB trigger). 

Related to this, note that the distinction between diffractive and
non-diffractive processes is not without ambiguity. In experimental
contexts, diffraction may be defined in terms of observable ``rapidity
gaps'' with no particle production detected in specific region(s) of
the detector, while in theoretical contexts processes that are
classified as diffractive typically produce a whole spectrum of gaps
with small ones suppressed but not excluded, see \cref{sec:soft:diff}.
Conversely, events that are modelled as non-diffractive in origin may
produce large rapidity gaps, due to fluctuations in the fragmentation
process and/or if colour reconnections --- see \cref{sec:colRec}
--- are allowed to produce such gaps, and in the transition region
there could even be quantum interference between the two categories
(not modelled by \pyt). Thus, for any given application it is important
to phrase experimental measurements in terms of clearly defined physical
observables, and consider which MC processes are going to be able to
contribute to those.

Usually hard processes, such as jet or gauge-boson production, are
assumed to occur within the non-diffractive event class. This is not quite
true, since it is possible also for diffractive topologies to contribute
to hard cross sections, see \cref{sec:soft:hardDiff}. That contribution
typically is of the order of a per cent when modelled or measured
experimentally, however, and is neglected by default. This means that the
full parton distribution functions (PDFs) are associated with the
non-diffractive component. They are used not only for the hard process
itself but also for the associated MPI, ISR and FSR activity. See
further \cref{sec:hadronPDFs}

\subsubsection{Diffractive processes}\index{Diffraction}
\label{sec:soft:diff}

\index{Diffraction!Ingelman--Schlein}
\index{Ingelman--Schlein}\index{Pomeron}Diffractive event topologies
are illustrated in 
\cref{fig:soft:sigmatypes} on \pgref{fig:soft:sigmatypes}, 
and the differential cross sections are described in
\cref{subsection:sigmadiffractive}. The choice of diffractive mass(es)
and $t$ values sets the overall kinematics of the events, but does not
describe the hadronization of the diffractive system. To this end, the
Ingelman--Schlein approach is used~\cite{Ingelman:1984ns}, with details
as described further in~\citeone{Rasmussen:2018dgo}. In this approach, a
pomeron is viewed as a physical particle, akin to a glueball state, with
an internal structure and notably with PDFs. Similarly, a reggeon is
viewed as a mesonic state, but for the practical handling the two are
not distinguished. Single diffraction therefore contains a pomeron-proton
subcollision, double diffraction two such, and central diffraction a
pomeron-pomeron subcollision. Each such subcollision is assumed to produce
particles as in a normal inelastic non-diffractive hadron-hadron collision. 

At high energies the modelling on the perturbative level is then given by
the MPI machinery, augmented by ISR and FSR. There are a few issues that
need to be clarified, however. Notably the MPI collision rate involves a
combination of the pomeron-inside-proton flux with the parton-inside-pomeron
PDF. What is measured, \eg at HERA, is the convolution of the two, where
the absolute normalization of each individually is not known.
Historically, the flux normalization was specified, such that then the 
pomeron PDF does not have to obey the momentum sum rule. This may seem
odd, but is in line with some theoretical arguments that the pomeron
is not a real particle and therefore is not bound by such constraints.
There are a dozen different pomeron PDFs that come with \pyt (plus three
special-purpose ones), and most of these have a momentum sum of the order
of 0.5. It is possible to scale them by a factor, to restore unit
normalization. Whether that is done or not, the rescaling of remaining
momentum for subsequent MPIs is done as for a normal hadron, however.
That is, the normalization matters for the rate of MPI production, 
but not for the handling of those MPIs that do occur.

\index{Uncertainties!MPI@in MPI}Further, the ordinary non-diffractive MPI rate is
related not only to PDFs 
but also to the normalization with the non-diffractive total cross section
$\sigma_{\mathrm{nd}}$, \cf\cref{eq:soft:MPI:nAvg} and other MPI expressions.
This is an unknown number from first principles, and with the same
pomeron-flux-normalization uncertainty as the PDFs, so effectively it
can be used to compensate for a non-unit momentum sum. The default value
is 10~mb at a collision CM energy of 100~GeV, where it has been tuned
(with default PDFs \etc) to produce about the same average charge
multiplicity as ordinary $\p\p$ non-diffractive collisions at the same
energy. This value could be energy-dependent, \cf the pomeron term in
\cref{eq:soft:DonLan}, but currently the default is a constant value.

\index{Impact parameter}Diffraction tends to be peripheral, \ie occur at high-to-intermediate
impact parameter for the two protons. That aspect is implicit in the
modelling of diffractive cross sections. For the simulation of the
pomeron-proton subcollision itself, however, it is rather the
impact-parameter distribution of that particular subsystem that should
be modelled. That is, it also involves the transverse coordinate-space
shape of a pomeron wave function. The outcome of the convolution with a
proton wave function could be a different shape than for non-diffractive
events, and therefore it can be set separately. The default is a simple
Gaussian, for lack of any relevant data. The $p_{\perp 0}$ scale is
assumed the same as in non-diffractive events at the same collision
energy, but also that is an assumption that could be questioned.

The diffractive mass spectrum extends down to the $\Delta^+$ mass for
$\p\p$ collisions, and obviously a perturbative MPI description would
not make sense at such low energies. Instead a separate low-mass
description has been implemented. Up to 1~GeV above the hadron mass, the
diffractive system is allowed to decay isotropically into a two-hadron
state. Above that, a diffractively-excited hadron is modelled as if either
a valence quark or a gluon is kicked out from it, along the collision
axis with some ``primordial $k_{\perp}$'' smearing,
\cf\cref{sec:primordialKT}.

\index{Primordial kT@Primordial $k_\perp$}
In the former case 
this produces a simple string to the leftover remnant, in the latter it
gives a hairpin arrangement where a string is stretched from one quark
in the remnant, via the gluon, back to the rest of the remnant. The latter
topology ought to dominate at higher mass $M_X$ of the diffractive system.
Therefore an approximate behaviour like
\begin{equation}
\frac{P_q}{P_g} = \frac{N}{M_X^p}
\end{equation}
is assumed, with $N$ ($= 5$ by default) and $p$ ($= 1$) as free parameters,
and $M_X$ in GeV.

There is a smooth transition between the low-mass non-perturbative and
the high-mass perturbative descriptions. The probability for applying
the latter is given by~\cite{Navin:2010kk}
\begin{equation}
P_{\mrm{pert}} = 1 - \exp\left( -
\frac{\max(0, M_X - m_{\mrm{min}})}{m_{\mrm{width}}}\right) ~,
\end{equation}
with $m_{\mrm{min}}$ and $m_{\mrm{width}}$ free parameters, both by
default 10~GeV. Note how $P_{\mrm{pert}}$ vanishes when below 
$m_{\mrm{min}}$. 

\subsubsection{Hard diffraction}
\index{Hard diffraction|see{Diffraction}}
\index{Diffraction!Hard diffraction}
\index{Diffraction!Gap survival}
\index{Rapidity gap survival|see{Diffraction}}
\index{Pomeron}
\index{Ingelman--Schlein}
\index{Diffraction!Ingelman--Schlein}
\label{sec:soft:hardDiff}

The model for hard diffraction is somewhat different from the soft (low- and high-mass) diffraction and it can be applied to any hard process, including \eg high-\pT jets and EW bosons. The starting point is again the Ingelman--Schlein picture where these interactions are mediated by a pomeron whose internal structure is given by the diffractive PDFs. It has been observed, however, that this factorization-based approach is broken as the predictions based on the diffractive PDFs determined in diffractive DIS overshoot the hard diffractive data in hadron-hadron collisions roughly by an order of magnitude~\cite{Affolder:2000vb, Aad:2015xis}. In the \pyt framework this can be naturally explained by having several non-diffractive partonic interactions, MPIs, in the same hadron-hadron collisions on top of the diffractive process. These may then produce particles that fill up the rapidity gap used to select the diffractive events leading to seemingly a suppressed diffractive cross section. The details of this \emph{dynamical rapidity gap survival model} are presented in~\citeone{Rasmussen:2015qgr}, together with several data comparisons, and are briefly outlined below.

After a hard process and its kinematics are sampled, the events of diffractive origin are first selected based on relative magnitude of the diffractive, $f^{\mrm{p}, \mrm{D}}_{i}$, and non-diffractive, $f^{\mrm{p},\mrm{ND}}_{i}$, PDFs which together form the inclusive (the usual) hadronic PDFs
\begin{equation}
f_{i}^{\mrm{p}}(x,Q^2) = f^{\mrm{p}, \mrm{ND}}_{i}(x,Q^2) + f^{\mrm{p}, \mrm{D}}_{i}(x,Q^2) ~.
\label{eq:soft:hadron:diffraction:PDFdcomp}
\end{equation}
The diffractive part, in turn, can be defined as a convolution between the pomeron flux $f^{\mrm{p}}_{\Pom}$ and pomeron PDF $f^{\Pom}_{i}$:
\begin{equation}
f_{i}^{\mrm{p}}(x,Q^2) = \int_x^1 \frac{\mrm{d}x_{\Pom}}{x_{\Pom}} f_{\Pom}^{\mrm{p}}(x_{\Pom}) \, f_{i}^{\Pom}(x/x_{\Pom},Q^2) ~,
\label{eq:soft:hadron:diffraction:diffPDF}
\end{equation}
which can be considered as parton-in-pomeron-in-proton PDFs, typically
determined using diffractive DIS data from HERA. After this tentative
selection of diffractive events corresponding to the Ingelman--Schlein
approach, the pomeron kinematics are sampled and the event is
processed further. The essence of the \pyt model is then to perform a
full parton-level evolution for the original hadron-hadron system and
to check whether any MPIs, that would render the event to a
non-diffractive one, has occurred. This allows for the generation of a sample where only events without such additional interactions remain and the rapidity gap has survived. It is also possible not to perform such a check and obtain the purely factorization-based result that serves as a baseline for the expected cross-section suppression. Notice, however, that MPIs in the pomeron-hadron system are still allowed as these would not fill up the rapidity gap between the excited hadron and the pomeron remnants. Remarkably, this model relies solely on the MPI model in \pyt and does not require any further parameters tuned to data. Yet, it can qualitatively explain the order-of-magnitude difference between the purely factorization-based predictions and Tevatron and LHC data, and reproduces the latest CMS data for diffractive dijets~\cite{CMS:2018udy} with a good precision. Only single diffraction is currently implemented, and if both beams have been found to emit pomerons, the diffractive side is selected randomly with equal probabilities. It is possible to consider pomeron emissions from one side only which can be useful for non-symmetric collisions.

\subsection{Lepton-lepton collisions}
\label{subsection:leptonlepton}

Lepton colliders have a reputation for providing the cleanest collisions
possible, with $\epem \to$ $\Z \to \f\fbar$ at LEP/SLC providing a prime
example, where $\Z$ properties could be studied in minute detail.
At lower energies, charm and beauty factories have advanced our
understanding of the standard model, \eg the weak unitarity triangle(s).
The key argument for future lepton colliders often is precision Higgs
physics. Nevertheless, lepton colliders also have their challenges,
as will be discussed in this section.

\subsubsection{Bremsstrahlung and lepton PDFs}\index{PDFs!Leptons@for Leptons} 

A lepton is surrounded by a cloud of virtual photons. In a collision,
such as $\epem$ annihilation, some of those photons survive in the
final state as so-called bremsstrahlung, mainly travelling near the
incoming lepton directions, and the annihilation energy is reduced
correspondingly. Similarly to the traditional PDF evolution in $Q^2$
of a hadron, one can here start from a low-$Q^2$
$f_{\electron}^{\electron}(x, Q_0^2) = \delta(x - 1)$
and evolve it with a splitting kernel
\begin{equation}
\d\mathcal{P}_{\electron \to \electron \gamma} = \frac{\d Q^2}{Q^2} \,
\frac{\alphaem}{2\pi} \, \frac{1 + z^2}{1 - z} \, \d z ~,
\label{eq:soft:lepton:evol}
\end{equation}
in close analogy with $\q \to \q \g$. The resummed effects of multiple
photon emissions are described in \pyt by an NLO expression
\cite{Kleiss:1989de} of the approximate shape
\begin{equation}
f_{\electron}^{\electron}(x,Q^2) \approx \frac{\beta}{2} (1-x)^{\beta/2-1} ~; ~~~~
\beta = \frac{2 \alphaem}{\pi} \left( \ln \frac{Q^2}{m_{\electron}^2} - 1
\right) ~.
\end{equation}
The form is divergent but integrable for $x \to 1$, \ie the electron
tends to keep most of the energy. To handle the numerical precision
problems for $x$ very close to unity, where 64-bit double precision would
not be sufficient, the (electron) parton distribution is set to zero for
$x > 1-10^{-10}$, and is rescaled upwards in the range
$1-10^{-7} < x < 1-10^{-10}$, in such a way that the total area under the
parton distribution is preserved:
\begin{equation}
\left( f_{\electron}^{\electron}(x,Q^2) \right)_{\mrm{mod}} =
\left\{ \begin{array}{ll}
 f_{\electron}^{\electron}(x,Q^2) & 0 \leq x \leq 1-10^{-7} \\[2mm]
 \frac{\displaystyle 1000^{\beta/2}}{\displaystyle 1000^{\beta/2}-1}
\, f_{\electron}^{\electron}(x,Q^2) &  1-10^{-7} < x < 1-10^{-10}  \\[4mm]
0 & x > 1-10^{-10} \, ~.
\end{array} \right.
\end{equation}

Turning to the photon flux, the evolution equation
\cref{eq:soft:lepton:evol} is deceptive in that it appears to treat
the electron and photon on equal footing. But, there is no resummation
of the photon spectrum, as there is for the one-and-only electron,
only an increasing number of photons as the
evolution continues. The typical kinematics is also different.
When we consider $f_{\electron}^{\electron}(x, Q^2)$, it is for an
annihilating $\epm$, where $m_{\electron}^2 \ll Q^2 \sim s$, and the
radiated energy manifests itself in terms of massless photons.
For the $f_{\gamma}^{\electron}(x, Q^2)$, it is instead the electron that
has to be on mass shell, a requirement that leads to a non-trivial
$Q^2_{\mrm{min}}$, and the photon that is virtual. This gives a PDF like
\begin{equation}
f_{\gamma}^{\electron}(x,Q^2) =
\frac{\alphaem}{2 \pi} \, \frac{1+(1-x)^2}{x}
\, \ln \left( \frac{Q^2}{Q^2_{\mrm{min}}} \right) ~,~~~~
Q^2_{\mrm{min}} \approx \frac{m_{\electron}^2 x^2}{1-x} ~,
\label{eq:soft:lepton:PDF}
\end{equation}
which obviously should vanish if $Q^2 \leq Q^2_{\mrm{min}}$.
In typical physics applications, it is conventional to set
$Q^2 = Q_{\mrm{max}}^2 \sim 1~\mrm{GeV}^2$ to define a beam of quasi-real
photons, that then can lead to $\gamma\p$ and $\gamma\gamma$ collisions.
A photon more virtual than that would rather be considered as the
propagator of a deep inelastic scattering event, and one would not use
PDF language to describe it. See further \cref{subsection:leptonhadron}
and \cref{subsection:photonphoton}.

The above equations for an electron beam can easily be extended to a
muon one, simply by replacing $m_{\electron}$ by $m_{\mu}$, and
similarly for $\tau$. Neutrinos do not couple to photons and so
there is no need to introduce a substructure for them.

Returning to the issue of $\epem$ annihilation, the effects of
bremsstrahlung are more easily illustrated if only one photon emission
is considered, but from either side, in which case
\begin{equation}
\frac{\d \sigma}{\d x_{\gamma}} = \frac{\alphaem}{\pi} \,
\left( \ln\frac{s}{m_{\electron}^2} -1 \right) \,
\frac{1 + (1-x_{\gamma})^2}{x_{\gamma}} \, \sigma_0 (\hat{s}) ~,
\end{equation}
where $x_{\gamma}$ is the photon energy fraction of the beam energy,
$\hat{s} = (1-x_{\gamma}) s$ is the squared reduced hadronic CM
energy, and $\sigma_0$ is the ordinary annihilation cross section at
the reduced energy. For $\epem \to \gamma^* \to \f\fbar$, where
$\sigma_0(\hat{s}) \propto 1/\hat{s} \propto 1/(1-x_{\gamma})$,
the bremsstrahlung spectrum thus is singular both for
$x_{\gamma} \to 0$ and $x_{\gamma} \to 1$. The former is a true
singularity, corresponding to infinitely soft photons, that
fortunately also carry away infinitely little energy from the electron.
The latter is cut off by the mass threshold for $\f\fbar$ production.

If instead the $\epem$ collider is running on a peak in the cross
section, like the $\Z$ one at LEP~1, and neglecting interference with
$\gamma^*$ for simplicity, then $\sigma_0(\hat{s}) < \sigma_0(s)$.
While the soft-photon singularity remains, any non-negligible photon
energy will push the $\Z$ propagator further off-shell, which leads to
a suppression of such photon emissions and of the total $\Z$ cross
section.

The situation is even more extreme for charm and beauty factories
when they run on a narrow $\psi$ or $\Upsilon$ state, where the net
effect is a loss of cross section. \pyt does not simulate such
emissions, however, or indeed the production of onium states by
$\epem$ colliders.

Finally, note that leptons can be polarized both transversely and
longitudinally, the former by plane polarization in circular rings
and the latter by spin rotation thereof. This can lead to non-trivial
effects on cross sections, since the standard model distinguishes
between left- and right-handed fermions, and is therefore expected to
be a main staple at future linear colliders. While \pyt~6.4 encoded
spin-dependent cross sections for a few common processes, none of these
have been ported to \pythia. If Les-Houches event input is used,
such effects can be taken into account already at that level, and
will not affect the continued handling of the event by \pyt.

\subsubsection{Beamstrahlung}\index{Beamstrahlung}

At potential future linear $\epem$ colliders, the beams will be
so tightly collimated that the electrical field of one beam will
significantly deflect the individual $\epm$ of the other. This
acceleration of charges leads to the emission of photons ---
beamstrahlung. Like bremsstrahlung, it gives a reduced collision
energy, a disadvantage that has to be balanced against the gains 
of a higher luminosity. Beamstrahlung emits real photons and keeps
the electrons real as well, so there is no $Q^2$ dependence but only
an $x$ one. The $f_{\electron}^{\electron}(x)$ spectrum is highly
dependent on the beam parameters, and varies \eg between the front
and the tail of a bunch. It is therefore in the realm of machine
physicists to provide relevant spectra, \eg with the
\textsc{Guinea-Pig} program~\cite{Schulte:1999tx}.
Simplified parameterizations are found in the \textsc{Circe} program ~\cite{Ohl:1996fi}. 

For $\epem$ annihilation, the beamstrahlung and
bremsstrahlung effects must be convoluted. Relevant code for
handling such a convolution does not (yet) exist in \pythia.
In case of need, a temporary solution is to split the energy remaining
after beamstrahlung, but before bremsstrahlung, into small bins that
are generated separately and combined in proportion to their
respective cross section. This requires an initialization for each bin,
but this is not such a big overhead since the MPI bottleneck is absent
in $\epem$ annihilation.

\subsubsection{Processes}

\pyt contains many processes initiated by a fermion-antifermion pair,
and these can almost all be used both for hadron and lepton
colliders. The list includes electroweak processes, top production,
Higgs physics, new gauge bosons, supersymmetry, and so on.

Most prominent is $\epem \to \gamma^*/\Z \to \f\fbar$. It has been the
main staple of all lepton colliders so far, possibly with the
exception of LEP~2. In addition to precision electroweak physics, it
has allowed the study of FSR and hadronization under the cleanest
conditions that we can hope for. The simplest $\gamma^*/\Z \to \q\qbar$
process produces a single string between the $\q$ and $\qbar$ endpoints.
One order up, $\gamma^*/\Z \to \q\qbar\g$ offers access both to $\alphas$ 
and to tests of string topologies, specifically to confirm that
a string is drawn from the $\q$ via the $\g$ to the $\qbar$.
With four-jet events, mainly $\gamma^*/\Z \to \q\qbar\g\g$, the
non-Abelian nature of QCD could be established. Taken together, the
measured particle composition can be used to tune flavour parameters,
measured jet rates and correlations to tune showers, and measured
particle spectra to tune longitudinal and transverse fragmentation
properties.

For LEP~2, instead $\Wp\Wm$ pair production was the most prominent
process, although $\gamma\gamma$ physics contributed at an even higher
rate. Apart from electroweak physics, of note is that
$\epem \to \Wp\Wm \to \q_1\qbar_2\q_3\qbar_4$ offers a test bed for
colour reconnections, further described in \cref{sec:colRec}.

\subsection{Lepton-hadron collisions}\index{DIS}
\label{subsection:leptonhadron}

In lepton-hadron collisions the events are often classified in terms of virtuality of the intermediate photon, $Q^2$. Events where the virtuality is large, or the mass of exchanged EW boson is large, and the target hadron breaks up are referred to as deep inelastic scattering (DIS). At low virtualities ($Q^2 \lesssim 1~\mrm{GeV}^2$), the events are in the photoproduction region where the photons can either interact directly as unresolved particles or fluctuate into a hadronic state with equal quantum numbers. In \pythia these two event classes are handled in separate frameworks and the special features of the former class are discussed below. The photoproduction framework is, in turn, introduced in \cref{subsection:photonphoton}. 

\subsubsection{Parton distribution functions and structure functions}\index{PDFs}\index{PDFs!Structure functions}

In DIS the intermediate boson scatters off a parton in the target hadron in a relatively clean scattering process where the kinematics characterizing the scattering can be related to the four-momentum of the outgoing lepton. Therefore, such collisions can be used to study the structure of the hadron and the initial-state QCD dynamics. Let $P$ denote the four-momentum of the incoming hadron, $k$ the incoming lepton and $k'$ the scattered lepton. Then it is possible to define the following Lorentz-invariant quantities
\index{Kinematics!for DIS}
\begin{align}
Q^2 &= -q^2 = -(k - k')^2 \nonumber \\
W^2 &= (P + q)^2  \nonumber \\
x &= \frac{Q^2}{2\,P \cdot q}  \nonumber \\
y &= \frac{P \cdot q}{P \cdot k}~,
\label{eq:LeptonHadron:DIS:kinematics}
\end{align}
purely based on measured energy and scattering angle of the scattered lepton. In fully inclusive events, where the hadronic final state is integrated out, it is possible then to write down the cross section of such a scattering process in terms of these quantities without making further assumptions on the proton structure
\begin{equation}
\frac{\mrm{d}^2\sigma}{\mrm{d}x \mrm{d}y} = N^l \left( y^2 \,x \,F_1^l(x,Q^2) + (1 - y) \,F_2^l(x,Q^2) \mp (y - \frac{y^2}{2}) x \,F_3^l(x,Q^2) \right)~.
\end{equation}
The coupling factor $N^l$ is different for neutral- and charged-current DIS and the sign of the last term depends on  whether the incoming lepton $l$ is charged or neutral (neutrino) and if it is a particle or an antiparticle. The structure functions $F_i^l(x,Q^2)$ represent the partonic structure of the hadron. In the leading-order parton model~\cite{Feynman:1969ej, Bjorken:1969ja} the structure functions are simply proportional to the sum of the parton distributions $f(x,Q^2)$ but do depend also on beam lepton type. The $x$ can be interpreted as the momentum-energy fraction of the parton with respect to the hadron momentum $P$ and the $Q^2$ dependency arise from the QCD corrections at higher orders. The goal of \pythia is, however, to provide fully exclusive events for which the relevant treatment is described next.

\subsubsection{Deep inelastic scattering}\index{DIS}
The DIS framework describes processes where the scattered lepton emits a highly-virtual (point-like) photon or a massive gauge boson that interacts with the constituents of the target hadron breaking it up. As there currently are no models for intermediate photon virtualities ($Q^2 \sim 1~\GeV^2$) where high-virtuality, point-like, and low-virtuality hadronic processes contribute to cross sections, the DIS framework provides a reliable description only at sufficiently large $Q^2$ where the scattering is purely mediated by a point-like particle. As the resolved-photon contribution fade continuously (roughly as $\sim 1/Q^2$), it is impossible to set a hard cut for such a region. In most applications, however, a limit of $Q^2> 5~\GeV^2$ has turned out to be sufficient to ensure negligible contributions from the hadronic fluctuations. As the model for intermediate virtualities implemented in \pyt~6 was based on several parameterizations mimicking the physical picture and turned out to be somewhat fragile with restricted predictability, we have decided to develop a completely new model for such processes that will be implemented in a future \pythia release. 

\paragraph{Hard processes}
In the LO DIS implemented in \pyt, the incoming lepton scatters off a quark from the target hadron by exchanging an EW boson. As described in \cref{subsection:EWprocesses}, this includes both neutral- and charged-current processes with charged leptons and neutrinos, and the interference between a virtual photon and the $\Z$ boson can be accounted for. The DIS-optimized scale-setting options are listed in \cref{subsection:couplingsscales} and the relevant phase-space cuts in \cref{subsection:phasespacecuts}. It is also possible to provide the hard process as an input from an ME generator in the LHE format. However, no matching of higher-order processes and the default parton shower has been implemented. As the hard-processes are set up in the collinear approximation, no off-shellness is allowed for the initial lepton line. Thus no radiation should be allowed for the initial-state lepton and no PDFs for the lepton used. The phase-space sampling for DIS is inherited from generic massless $2 \rightarrow 2$ scattering where the initiators are assumed massless but the final-state particles can have finite masses. This is not ideal for DIS, however, since the invariants typically studied in DIS, Bjorken $x$ and $Q^2$, are often derived from the four-momentum of the scattered lepton. Due to a mismatch in masses, these variables might then not match the internally sampled values which can lead to unphysical configurations such as $x>1$ when invariants are derived from the scattered lepton. To fix the issue, a new phase-space sampling optimized for $t$-channel exchange of bosons with (potentially varying) masses will be implemented. Heavy-quark pairs can be produced in two different ways: if the lepton scatters off a heavy quark, a companion will be added by ISR, or the heavy-quark pair can be formed from a gluon splitting by FSR. Similarly, DIS events with more than one jet can be formed via PS emissions but no explicit hard dijet processes have been included. The showers do, however, include matrix-element corrections for the first emissions. As the DIS process is a scattering of a single point-like particle, no MPIs are allowed.

\paragraph{Parton showers}\index{Parton showers!DIS}
Both initial- and final-state radiation from deep-inelastic-scattering processes require a careful treatment of the branching kinematics. If emissions from the hadronic system disturb (via recoil) the lepton line, or \textit{vice versa}, then both the $x$ and the $Q^2$ distribution are affected by showering. In this case, an intricate recalculation of the hard-scattering cross section after each parton-shower emission is required, making the strategy sub optimal\footnote{Similar concerns apply to any scattering via $t$-channel colour-singlet exchange, \eg to Higgs production in vector-boson fusion.}. The natural resolution is to ensure that any recoil due to the branching process is contained either in the hadronic or the leptonic system. For the hadronic system, this is most easily achieved by employing a ``local dipole recoil'' strategy, in which the kinematic recoil is absorbed by a colour-connected partner. Such a strategy is employed by the \dire \cref{sec:Dire} and \vincia \cref{sec:Vincia} showers, and is an option for the simple-shower methods~\cite{Cabouat:2017rzi}. To model the QED evolution of the leptonic line, this approach is insufficient, however, and more complex strategies are necessary, or the conservation of $x$ and $Q^2$ may need to be relaxed, \eg for charged-current DIS events, where electric charge flows from the lepton to the hadron system. The latter is the case in the \dire shower.

Another important aspect of modelling radiation in DIS events is the phase-space boundary for emissions that involve incoming partons. The factorization scale (\ie typically $Q^2$) gives a natural phase-space boundary when using backward initial-state evolution~\cite{Sjostrand:1985xi}. However, the kinematic boundary is more accurately given by the invariant mass of the radiating dipole or the invariant mass of the hadronic system ($W^2$), which are often, and especially for low-$x$ values, significantly larger than $Q^2$. A natural resolution of this issue is to 
keep the tight $Q^2$ constraint for the shower, and use (tree-level) merging to supplement the missing phase-space regions. Another approach is to abandon the DGLAP-based initial-state evolution~\cite{Kharraziha:1997dn}. In lieu of the latter, hard initial-state emission in the partons shower models of \pyt should be considered with caution.

As the DIS events are rather clean, they offer a very good environment to study parton-shower dynamics. For example, since the parton shower produces $\pT$ kicks for the initiator via emissions, it can be thought to resemble perturbative evolution of transverse-momentum dependent (TMD) PDFs~\cite{Collins:1984kg,Angeles-Martinez:2015sea}. Thus one should obtain reasonable $\pT$ distributions in \ac{SIDIS} from the parton-shower enabled \pyt simulations. 

\paragraph{Hadronization}\index{DIS}
The hadronization of DIS events is analogous to that of hadron-hadron scattering systems. The scattered lepton does not partake in hadronization and since no multiparton interactions are included in DIS events, no colour reconnection model is employed. At present, no DIS data has been used in the tuning of the hadronization model. The study of spin polarizations and the higher-dimensional structure of the hadron are typically important aspects of DIS analysis. In this context, it should be noted that the \pythia hadronization model does not by default consider polarization, though external tools to model such effects have been proposed~\cite{Kerbizi:2021pzn}.

\subsection{Photon-hadron and photon-photon collisions}
\label{subsection:photonphoton}

The possibility to turn a charged lepton into a photon using laser back scattering has been studied, but has not been realized in the current or foreseen colliders. Thus photon-induced collisions are usually studied in colliders with charged beam particles that may emit photons when accelerated to high energies. The shape of the photon flux and the virtuality spectra are, however, different for different beam types but, given an appropriate flux, the photon-induced processes can be treated in a single framework regardless of the original beam configuration. Here we focus on low-virtuality (quasi-real) photons and introduce the current simulation framework in \pythia for processes involving such effective beams.

\subsubsection{Parton distribution functions of resolved photons}
\index{Resolved photons}
\index{PDFs!Resolved photons@for Resolved photons}
\index{DGLAP!Resolved photons@for Resolved photons}

In total there are three separate contributions for processes with low-virtuality photons: a photon can interact either as an unresolved particle, it can split perturbatively into quark-antiquark pair, or it can fluctuate into hadronic state non-perturbatively. The two latter contributions, where the partonic constituents act as initiators for hard scattering, can be described with DGLAP-evolved PDFs. As in the case of hadrons, the evolution equation for resolved photons do include a hadron-like component where a non-perturbative ansatz is evolved according the usual QCD DGLAP kernels. In addition to this, however, the evolution equation contains also a point-like component which feeds in more quark-antiquark pairs with increasing evolution scale that may evolve further by QCD splittings. The full evolution equation for resolved photons is
\begin{equation}
\frac{\partial f_{i}^{\gamma}(x, Q^2)}{\partial\log(Q^2)} =
  \frac{\alphaem(Q^2)}{2\pi}e_i^2 P_{i\gamma}(x)
  + \frac{\alphas(Q^2)}{2\pi} \sum_j\int_{x}^1\frac{\d z}{z}
  P_{ij}(z)f_{j}^{\gamma}(\frac{x}{z}, Q^2)~,
\label{eq:soft:photon:evolution}
\end{equation}
where the $\gamma \rightarrow \q \qbar$ splitting kernel in LO is $P_{i\gamma}(x) = 3(x^2 + (1-x)^2)$ and $Q^2$ is the factorization scale at which the partonic structure is probed. As in the case of proton PDFs, the parameters related to the non-perturbative ansatz at the initial scale are determined in a global QCD analysis comparing to experimental data. In \pythia, the default set for the resolved photon PDFs is from the CJKL analysis~\cite{Cornet:2002iy} which conveniently provides the hadron-like and point-like parts separately, which can be used for finer classification of the events with resolved photons. No dependence on the photon virtuality is included in these PDFs, but all photons are taken as real with zero virtuality, which is the case also for the LO photon-initiated cross sections currently implemented in \pythia.

\subsubsection{Photoproduction}
\index{Photoproduction}

Photoproduction typically refers to processes where a beam lepton emits a low-virtuality (quasi-real) photon that then collides with a hadron from the other beam. The following describes some special features of such collisions. These are not unique to \ep colliders -- similar processes can take place also in \epem, \pp, \pA, and \AA collisions as will be discussed in the following.

\begin{figure}[htb]
\centering
\includegraphics*[width=0.4\textwidth]{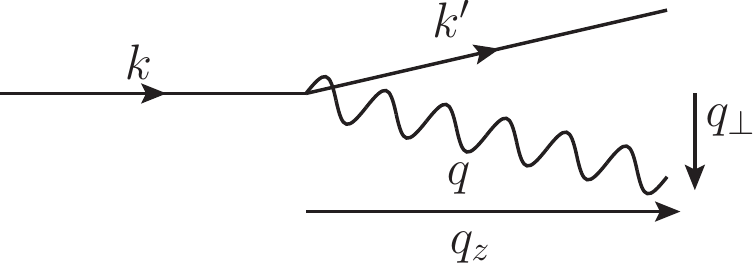}
\caption{Kinematics of a photon emission. \label{fig:soft:photon:kinematics}}
\end{figure}

\paragraph{Photon flux and kinematic limits}\index{Kinematics!for photoproduction}
When the emitted photons are quasi-real and almost collinear with the
beam leptons, the cross section calculations can be simplified by
factorizing the photon flux from the hard perturbatively calculated
part. In case of lepton beams, the flux of quasi-real photons can be
obtained from the well-known Weizs\"acker--Williams~\cite{vonWeizsacker:1934nji, Williams:1934ad} or
\index{Equivalent photon approximation}\index{Weizs\"acker-Williams}\ac{EPA}. The flux differential in photon virtuality $Q^2$ is
\begin{equation}
f_{\gamma}^l(x_{\gamma},Q^2) = \frac{\alphaem}{2\pi} \frac{\mathrm{d}Q^2}{Q^2} \frac{1+(1-x_{\gamma})^2}{x_{\gamma}}~,
\label{eq:soft:photon:EPAflux}
\end{equation}
where $x_{\gamma}$ is the momentum fraction carried by the (almost) collinear photon with respect to the parent lepton. Integration from the minimum allowed virtuality yields the photon-in-lepton PDF in \cref{eq:soft:lepton:PDF}. In photoproduction, the upper limit $Q^2_{\mrm{max}}$ is typically of the order $1~\GeV^2$, depending on the considered experimental setup and detector acceptance. The lower limit is restricted by the requirement of physical kinematics (on-shell leptons) for the $1 \rightarrow 2$ splitting and depends on $x_{\gamma}$, the mass of the lepton, $m_l$, and the energy of the beam in the CM frame, $E$
\begin{equation}
Q^2_{\mathrm{min}}(x_{\gamma}) = \frac{2 m_l^2 x_{\gamma}^2}{1 - x_{\gamma} - m_l^2/E^2 + \sqrt{1 - m_l^2/E^2} \sqrt{ (1 - x_{\gamma})^2 - m_l^2/E^2} } \approx \frac{m_{l}^2 x_{\gamma}^2}{1-x_{\gamma}}~.
\end{equation}
From a similar consideration, one can find the kinematically allowed upper limit for $x_{\gamma}$
\begin{equation}
x_{\gamma}^{\mathrm{max}} = \frac{2 \, \left( 1 - \frac{Q^2_{\mathrm{max}}}{ 4 E^2} - \frac{m_l^2}{E^2}\right) } { 1 + \sqrt{\left(1 + \frac{4 m_l^2}{Q^2_{\mathrm{max}}} \right) \left(1 - \frac{m_l^2}{E^2}\right)}}~,
\end{equation}
which typically is very close to unity. The lower limit of $x_{\gamma}$ can be derived from the minimum considered $W$ of the photon-hadron system. Similarly, as for hadron-hadron collisions, this should be large enough to justify the perturbative treatment that \pyt is largely based on. After the values for $x_{\gamma}$ and $Q^2$ have been sampled from the allowed phase space, the full kinematics for the intermediate photon can be derived. The transverse and longitudinal momentum, $q_{\perp}$ and $q_z$ as shown in \cref{fig:soft:photon:kinematics}, can be calculated from 
\begin{align}
q_{\perp} &= \sqrt{\frac{ \left( 1 - x_{\gamma} - \frac{ Q^2 }{ 4 E^2} \right) Q^2 - \left( x_{\gamma}^2 + \frac{Q^2}{E^2} \right) m_l^2 } { 1 - \frac{m_l^2}{E^2}}} \label{eq:soft:photon:qt} \\
q_{z} &= \frac{E (x_{\gamma} + \frac{Q^2}{2 E^2})} { \sqrt{1 - \frac{m_l^2}{E^2}}}~.
\end{align}
The azimuthal angle is sampled from a flat distribution and the scattered lepton four-momentum can be obtained simply from $k'=k-q$. It is also possible to provide the photon flux externally in \pythia, but the sampling has been optimized for the form in \cref{eq:soft:photon:EPAflux}. The kinematics and the allowed phase-space region are independent from the applied flux.

\paragraph{Direct and resolved photons}
If the (quasi-)real photon is the initiator of the hard scattering, \ie an unresolved (or direct) photon, the photon flux acts essentially as a PDF and can be directly applied for sampling of the process kinematics. If the photon has fluctuated into a hadronic state, for which the partonic structure is given by the resolved photon PDFs described above, these PDFs have to be convoluted with the flux to define so-called parton-in-photon-in-lepton PDFs
\begin{equation}
f_{i}^{\gamma}(x,Q^2) = \int_x^1 \frac{\mrm{d}x_{\gamma}}{x_{\gamma}} f_{\gamma}^\mrm{p}(x_{\gamma}) \, f_{i}^{\gamma}(x/x_{\gamma},Q^2)~,
\label{eq:soft:photon:fluxPDF}
\end{equation}
where the photon virtuality has been integrated out and $Q^2$ refers to the factorization scale at which the resolved photon is probed. Here, it is also assumed that the PDFs are independent of the photon virtuality, though alternatives containing such information exist, see \eg \citeone{Schuler:1996fc}. The flux is also used to sample the intermediate photon kinematics required to reconstruct the full event including the remnants of the resolved photon and the kinematics of the scattered lepton. In \pythia both of these contributions, direct and resolved, are included and can be generated simultaneously to obtain the correct mixture of the possible contributions for a given process at considered kinematics.
  
\begin{figure}[htb]
  \begin{center}
    \includegraphics[width=0.5\textwidth]{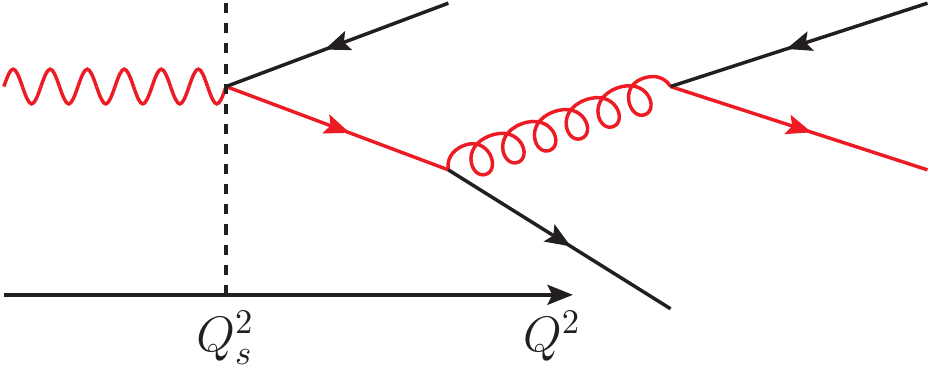}
  \caption{Backwards evolution of a point-like photon that collapses into unresolved photon at the scale $Q_s$. The hard-process initiator whose splittings are traced back in ISR is highlighted with red colour. \label{fig:soft:photon:remnant}}
  \end{center}
\end{figure}

\paragraph{ISR with photon beams}
\index{ISR!Resolved photons}
For direct photons, no ISR splittings have been implemented as in these cases the effect from additional QED emissions is typically small. For the resolved photons, however, some additional care needs to be taken when generating ISR due to the extra term in the PDF evolution, see \cref{eq:soft:photon:evolution}, compared to purely hadronic beam particles. As this term feeds in quark-antiquark pairs when evolving forwards with DGLAP, in backwards evolution, relevant for the ISR, this will collapse partons back into the original unresolved photon as illustrated in \cref{fig:soft:photon:remnant}. If such splittings are found during the PS evolution, one can think of these processes being of point-like origin and if not, the partons have originated from the hadron-like part of the PDFs. This dynamical selection of these two contributions have then further implications for beam remnants and MPIs as discussed below. This is also one of the key differences between the old \pyt~6 implementation where such selection was done already when sampling the hard scattering, and no MPIs were allowed for the point-like contribution at any scale.

\paragraph{MPIs with resolved photons} 
\index{MPI!Resolved photons}
Similarly as with resolved hadron beams, the resolved photons may also experience several partonic interactions in each collision. These MPIs are modelled in the same way as for hadrons as described in \cref{subsection:mpi}, but some aspects require further attention. The first one follows from the ISR generation discussed above. If the photon has collapsed back to an unresolved state, it can not have further MPIs below the scale at which this splitting has occurred, in \cref{fig:soft:photon:remnant} this scale is denoted with $Q_s$. Such an ordering is possible thanks to the interleaved evolution of PS and MPIs, see \cref{eq:soft:MPI:combinedevol}. Another potential difference is related to the screening parameter in semi-hard cross sections from which the MPI probabilities are calculated from. Since the partonic and spatial structure of resolved photons are quite different compared to protons, it would be expected that the value of this parameter should be separately tuned for collisions involving resolved photons. Indeed, first comparisons to HERA data~\cite{Helenius:2018bai} indicate that a somewhat larger screening parameter yielding a lower MPI probability is preferred but the constraints are still rather sparse and would benefit from further measurements of low-$\pT$ hadron production. Also, the impact-parameter profile could be modified but this would require more experimental data sensitive to MPIs. 

\paragraph{Remnants}\index{Beam remnants!for resolved photons}
Since the PDFs for resolved photons contain both a hadron- and a
point-like part, the remnant construction also needs to be adjusted to
handle both cases. The main difference to a purely hadronic state is
that since the point-like contribution is of a perturbative nature, the collapse back to a pure photon state should also be handled
perturbatively, namely with the parton showers. Unlike in \pyt~6 the
distinction into a point-like and hadron-like part is not done when the
(semi-)hard scattering is selected, but the term corresponding to
$\gamma \rightarrow q \bar{q}$ splitting in ISR algorithm will select
the cases where the initiator has originated from a perturbative
photon splitting. In cases where there are no MPIs in the event, ending
up in such a configuration means that there is no need to add any
non-perturbative remnants, as the necessary partons have been added
perturbatively by the parton shower as illustrated in
\cref{fig:soft:photon:remnant}. If the ISR generation will not end in a $\gamma \rightarrow q \bar{q}$ splitting, the resolved photon
is taken to be hadron-like, and the remnants will be constructed
similarly as for any hadrons. In this case, the valence flavour is
sampled based on relative weights derived from the PDFs. The remnant
construction becomes more complicated if the initiator is found to be
of a point-like origin but the beam photon has encountered additional
MPIs before (at scale $Q^2 > Q_s^2$) the resolved state is collapsed
into an unresolved one. Then there are several initiators kicked out
from the beam, so a single companion cannot make the beam
configuration flavour and colour neutral.
\index{Primordial kT@Primordial $k_\perp$} In this case the remnant is again constructed as for any hadron, but the primordial $k_{\perp}$ for the initiator of the hardest process and its companion are derived from the scale $Q^2_s$ at which the $\gamma \rightarrow q \bar{q}$ branching collapsing the photon to an unresolved state has occurred. 

\paragraph{Hard processes and diffraction}
\index{Diffraction!for resolved photons}
\index{Diffraction!Gap survival}
The hard diffraction with dynamical rapidity-gap survival model introduced in \cref{sec:soft:hardDiff} has been implemented also for photoproduction. For direct photons, the no-MPI requirement has zero effect since there are no MPIs with unresolved photons. However, as MPIs can still occur with resolved photons, some suppression is expected also for hard-parton initiated processes. Indeed, there are indications that diffractive dijet photoproduction cross sections are suppressed compared to pQCD predictions based on diffractive PDFs for the target proton. The observed suppression factor depends on the applied kinematic cuts and varies between 0.5--0.9 in different analyses~\cite{Chekanov:2007rh,Andreev:2015cwa}. The milder suppression compared to hadron-hadron collisions at the Tevatron and LHC is explained by the presence of the direct component and the smaller invariant mass of the photon-proton system at HERA kinematics which both reduce MPI probability compared to hadronic collisions at higher energies. As demonstrated in~\citeone{Helenius:2019gbd}, the MPI-based model in \pythia provides a reasonable description for the various HERA data.

\paragraph{Soft QCD processes}
\index{Schuler-Sjostrand model@Schuler-Sj\"ostrand model}
\index{VMD}\index{Vector-meson dominance|see{VMD}}
Apart from the non-diffractive low-$\pT$ $2 \rightarrow 2$ scatterings that are generated with the regulated cross section from the MPI framework using photon PDFs, the soft processes with real photons are modelled according to the vector meson dominance (VMD) model. In this model the photon is described as a linear combination of different vector-meson states with prefactors derived from experimental data. In \pythia, the values are taken from the analysis presented in \citeone{Bauer:1977iq} that have also been used in an SaS fit~\cite{Schuler:1996en} for total and elastic cross sections applied here. The included vector meson states are $\rhoz$, $\omega$, $\phiz$, and $\Jpsi$ but $\Upsilon$ is currently neglected. In the VMD model for elastic and diffractive processes, the incoming photon will first transform into a vector-meson state sampled according to relative weights. Then, the interaction is handled similarly as for any other hadron-hadron case described in \cref{subsubsection:lowenergyprocesses}. The elastic scattering process in photoproduction is often referred to as exclusive vector-meson production for which there are nowadays a good amount of data from HERA experiments, see \eg \citerefs{Derrick:1995vq, Derrick:1996yt, Derrick:1996af, Breitweg:1997rg, Chekanov:2002xi, Aktas:2005xu, Alexa:2013xxa}. The SaS parameterization tends to provide a good description for low-mass vector-meson production, \eg in case of $\rhoz$, but underestimates higher-mass states such as the $\Jpsi$ by a large margin. This indicates the need for further, possibly pQCD based, modelling for high-scale elastic processes.

\subsubsection{Photon-photon collisions}

Similarly as photoproduction in \ep collisions, the charged-lepton beams in \epem collisions may emit photons that can interact with each other leading to effective photon-photon collisions. If both of the photons have a low virtuality, there are a number of possible combinations that must be accounted for. In the most complex case, where both photons are resolved, the collisions are generated in a similar manner as in hadron-hadron collisions, including parton showers for the initial and final state, beam remnants, and, in particular, MPIs with the same special features as with photoproduction as discussed earlier. If one photon is unresolved and other resolved, the interactions are somewhat simpler, since the unresolved photon scatters off a parton from a resolved photon. In this case, no MPIs can take place and ISR and beam remnants are generated only for the hadron side. Both photons can also interact as unresolved particles when all particles are produced from the outgoing particles through FSR and hadronization, which are relevant also for other possible contributions.

\paragraph{Kinematics}\index{Kinematics!for photon-photon collisions}
The initial phase-space sampling assumes that the incoming photons are collinear with respect to the beam particles. However, as kinematically allowed photons emitted from massive (on-shell) particles will always have a finite virtuality, they will also possess some transverse momentum given by \cref{eq:soft:photon:qt}. The direction of this $q_{\perp}$ is not a priori known and is sampled only after the hard process kinematics are determined. Thus the final invariant mass of the photon-photon system, $W_{\gamma\gamma}$, will depend on the virtualities of the photons and their relative azimuthal angle, $\Delta \phi = \phi_1 - \phi_2$. The resulting $W$ can again be derived from the kinematics, giving
\begin{equation}
W_{\gamma\gamma}^2 = 2 E_1 E_2 x_{\gamma 1} x_{\gamma 2} - Q^2_1 - Q^2_2 + 2 q_{z 1} q_{z 2} - 2 q_{\perp 1} q_{\perp 2} \cos (\Delta \phi)~,
\label{eq:Wgmgm}
\end{equation}
where $x_i$ are the momentum fractions of the photons with respect to the beam leptons whose CM energies are $E_i$. To account for the possibly modified $W^2$ ($=\hat{s}$ for the direct-direct case), the cross section and relevant kinematic variables are recalculated after the virtualities and the direction of the photons are sampled. Typically the changes in the cross section and kinematics are negligible, but are needed in order to preserve the four-momentum of the event. An exception is, however, $2\rightarrow 1$ processes where it is important to keep the mass of the intermediate particle intact, a prime example being Higgs-boson production, where the photon momentum fractions are modified instead. 

\paragraph{Possible final states}
There are many topics that can be studied in photon-photon collisions and the relative importance of direct and resolved contributions varies by the process and considered kinematics. For example, Higgs production in $\gamma\gamma$ collisions is dominated by the direct-direct contribution but for QCD processes, such as jets or heavy quarks that contribute to the background of Higgs studies, the resolved photons may also have a significant contribution. Another interesting phenomenon is the MPIs in a photon-photon system which can be studied with low-$\pT$ hadrons that arise almost completely from resolved-resolved interactions. Also QED processes, such as dilepton production, can be considered to calibrate the photon fluxes as they are not sensitive to QCD effects.

\subsubsection{Ultra-peripheral collisions}
\label{sec:ultra-peripheral}
\index{Ultra-peripheral collisions}

As briefly mentioned earlier, other charged beam particles, including protons and heavy nuclei, may also emit photons that interact with the other beam or photons emitted by the other beam. When the beam particles do not interact hadronically but stay intact and emit photons that give rise to a hard interaction, the events are referred to as \ac{UPCs}. Due to the requirement of beam particles with finite size not breaking up, the emitted photons have always a small virtuality and can therefore be handled with the photoproduction framework introduced above. The photon-induced processes where the beam hadron break ups can be simulated by using a PDF set that includes perturbatively generated photons from DGLAP evolution with the usual \pyt model for hadron-hadron collisions.

\index{Equivalent photon approximation!protons}
The key difference between photon fluxes from hadrons and charged leptons is that the finite size of the emitting particle needs to be accounted for. For protons, a good approximation is obtained with the electric dipole form factor, giving a $Q^2$-differential flux of the form
\begin{equation}
f_{\gamma}^{\mrm{p}}(x_{\gamma},Q^2) = \frac{\alphaem}{2\pi} \frac{1+(1-x_{\gamma})^2}{x_{\gamma}} \frac{\mathrm{d}Q^2}{Q^2} \frac{1}{(1+Q^2/Q^2_0)^4},
\label{eq:EPAfluxProton}
\end{equation}
where $Q^2_0 = 0.71~\GeV^2$. Integrating over the possible virtualities will provide the flux derived in \citeone{Drees:1988pp}. Another flux has been implemented for protons that is based on work by Budnev \etal (see \citeone{Budnev:1974de}). The downside in the latter is that since only a virtuality-integrated form is provided, there is not enough information to sample the full kinematics of the intermediate photon and the virtuality sampling needs to be turned off. Therefore, this flux is not suited to study observables sensitive to the transverse momentum of the intermediate photon as the $q_{\perp}$ is set to zero.

\index{Equivalent photon approximation!heavy nuclei}
\index{Impact parameter}For heavy nuclei, it is possible to use form factors and derive the photon flux in a similar manner as for protons. Usually it is more convenient to work in the impact-parameter space since the heavy nuclei have a well-defined size and therefore it is possible to remove events where hadronic interactions dominate the particle production by rejecting events with small impact parameter. As shown in \citeone{Jackson:1998nia}, it is possible to derive an analytic form for the flux differential in the impact parameter by assuming a point-like charge distribution. In fact, this provides a good approximation for the flux with a more realistic density profile when considering the region outside of the nucleus relevant for UPCs. Integrating this from the minimum allowed impact-parameter value $b_{\mrm{min}}$ gives
\begin{equation}
f_{\gamma}^A(x_{\gamma}) = \frac{\alphaem Z^2}{\pi\, x_{\gamma}} \left[ 2 \xi K_1(\xi) K_0(\xi)  - \xi^2 \left(K^2_1(\xi) - K^2_0(\xi) \right) \right]~,
\label{eq:EPAfluxNuclei}
\end{equation}
where $Z$ is the electric charge (number of protons) of the nuclei $A$, $\xi = b_{\mrm{min}} x_{\gamma} m_{N}$. As the nuclear beams are typically defined in terms of per-nucleon energy, $m_{N}$ here also refers to average nucleon mass. A suitable value for $b_{\mrm{min}}$ is given by the sum of the radii of the colliding nuclei. Such a flux is included in \pythia but can only be enabled by providing this as a pointer to the \texttt{Pythia} object with a dedicated method. The shape and magnitude of this flux is very different from the flux for charged leptons, and therefore the phase-space sampling must be re-optimized for efficient event generation. A suitable over-estimate is included, but the parameters may have to be re-adjusted for different beam configurations. When using this flux, the virtuality sampling has to be disabled since the allowed virtualities have been essentially integrated over when converting to impact-parameter space by Fourier transform from the momentum space.

The current framework can already be applied to many processes studied
in UPCs but have a few limitations as well. In proton-proton collisions it is possible to study both photon-photon and photon-proton collisions with fully reconstructed kinematics, when a 
$Q^2$-dependent flux is used. This includes all hard processes 
initiated by photons or partons and also soft QCD processes apart from 
central- and double-diffractive events. These allows for the study of minimum-bias photon-proton collisions, inclusive and diffractive jet production, and photon-initiated dilepton production with all different contributions, to name a few. In case of heavy ions, the palette is somewhat more limited due to a $Q^2$-independent photon flux and lack of model for photon-nucleus collisions, which will be addressed in future releases. In \pA collisions, where the flux from the heavy nucleus is amplified by the $Z^2$ factor so that $\gamma \p$ component dominates the cross sections, almost all the same final states can be studied as in proton-proton collisions apart from observables highly sensitive to transverse momentum of the intermediate photon. For QCD observables, the effect from neglected $Q^2$ dependence will be washed out by the QCD radiation. In \AA collisions subsequent photon-nucleon interactions are not modelled, but high-$\pT$ observables and direct-photon dominated processes can be generated with reasonable accuracy. Photon-photon interactions can also be considered, with the only limitation being the neglected $Q^2$ dependence in the kinematics that again has an effect for the $q_{\perp}$-dependent observables, \eg the acoplanarity of dilepton pairs produced by two direct photons.

\subsection{Heavy ion collisions}

\label{sec:heavy-ion-collisions}

The \ac{HI} collider physics community has traditionally not
had very close ties to the rest of the \ac{HEP} community.  This has also
been reflected in the event generator community, where the authors of
HI event generators, although they some times make use of \eg the
string fragmentation in \pyt, did not interact much with the authors
of the main general purpose event generators for \pp, \ep, and \epem collisions.
However, with the arrival of the LHC, the situation has changed. Not
only are HI and particle physicists now part of the same
collaborations, the physics questions being asked are also starting to
converge, and typical observables studied in HI collisions are being
applied to \pp, and \textit{vice versa}. It should therefore not come as a
surprise that \pythia now also has some HI functionality implemented.

There are several ways to study HI collisions in \pythia. In \cref{sec:ultra-peripheral} we described how to
study ultra-peripheral HI collisions, and there is also the
possibility to use nuclear PDFs to study some observables. Here, we will concentrate on the modelling of complete exclusive
hadronic final states using the so-called \Angantyr model~\cite{Bierlich:2018xfw}, which is the default way of handling HI
collisions in \pythia.

\subsubsection{Wounded nucleons}
\label{sec:heavy-ion-collisions-wounded}

The \Angantyr model in \pythia can be said to be the successor of the old
\fritiof program~\cite{Andersson:1986gw} which used string
fragmentation to generate final states in HI collisions, and was based
on the so-called wounded nucleon model\index{wounded nucleons}~\cite{Bialas:1976ed}. The basic
assumption in the wounded nucleon model is that each nucleon that
participates in a HI collisions contributes to the multiplicity of the
full final state, according to a multiplicity function $W(y)$ which
has a triangular form in rapidity
\begin{equation}
  \label{eq:wounded-mult-fn}
  W(y)\propto \frac{1}{2}\left(1 + \frac{y}{y_{\max}}\right)~,
\end{equation}
where $y_{\max}$ is the rapidity of the nucleon in the collision
rest frame. This would yield the following simple form of the rapidity
distribution in an \AA, for a given number of wounded nucleons (or
\emph{participants}), $N_{\mrm{part},p},N_{\mrm{part},t}$ in the projectile and
target nuclei respectively,
\begin{equation}
  \label{eq:wounded-mult-tot}
  N(y)=N_{\mrm{part},p}W(y) + N_{\mrm{part},t}W(-y)~.
\end{equation}

\fritiof, in its simplest form, used the fact that the distribution of
particles of a hadronizing string is flat in rapidity.  For each
wounded nucleon, a string was stretched out to an endpoint randomly
positioned uniformly in rapidity, which then on average reproduces the
form in \cref{eq:wounded-mult-fn}. Despite the simplistic nature
of the model, \fritiof was able to provide a fairly good description of
collider data at the energies available in the 1980s. In fact, even
\pp collisions (with $N_{\mrm{part},p}=N_{\mrm{part},t}=1$) were reasonably
described.

With the energies achievable at RHIC and LHC, the basically
non-perturbative \fritiof model falls short of reproducing data, and
the \Angantyr model was developed to address these shortcomings.

\subsubsection{The \Angantyr model}

\index{Angantyr@\textsc{Angantyr}}In comparison to \fritiof, the \Angantyr model introduces two major
new ingredients. First, rather than wounded nucleons only resulting in a
string stretched out and being hadronized, a full diffractive
excitation is generated using the full multiparton interaction
machinery of \pyt where these are described in terms of a
pomeron-proton collision. In addition, a more advanced version of the
Glauber simulation is used where special attention is given to the
fluctuations in the nucleon wave functions, making it possible to
differentiate between different types of \ac{$NN$}
subcollisions.

Starting with the new Glauber modelling, we rely on the Good--Walker
formalism~\cite{Good:1960ba}\index{Good--Walker} to connect the different types of
$NN$ semi-inclusive cross sections with fluctuations in
the wave functions~\cite{Bierlich:2016smv}. 

For a projectile particle with an internal substructure, it is
possible that the mass eigenstates differ from the elastic scattering
eigenstates. We denote the mass eigenstates $\Psi_i$, with the
projectile in the ground state (\eg a nucleon) denoted $\Psi_0$,
while $\Phi_l$ are the eigenstates to the scattering amplitude $T$,
with $T \Phi_l = t_l \Phi_l$. The mass eigenstates are linear
combinations of the scattering eigenstates,
$\Psi_i = \sum_l c_{il} \Phi_l$. The scattering can be treated as a
measurement, where the projectile selects one of the eigenvalues
$t_l$, with probability $|c_{0l}|^2$.

The elastic amplitude for the ground state projectile is then given by
$\langle \Psi_0|T|\Psi_0\rangle = \sum_l |c_{0l}|^2 t_l \equiv \langle
T \rangle$, where $\langle T \rangle$ is the expectation value for the
amplitude $T$ for the projectile. The elastic cross section is then
given by
\begin{equation}
d\sigma_{\mrm{el}}/d^2 b =\langle T(b) \rangle^2 ~. 
\end{equation}
Working in impact-parameter space, the amplitude depends on $b$, and the
total diffractive-scattering cross section, $\sigma_{\mrm{diff}}$,
is the sum of transitions to all states $\Phi_l$:
\begin{equation}
d\sigma_{\mrm{diff}}/d^2 b = \sum_l \langle \Psi_0|T|\Phi_l\rangle \langle \Phi_l
|T|\Psi_0\rangle = \langle \Psi_0| T^2| \Psi_0 \rangle~, 
\end{equation}
where we have used the fact that the $\Phi_l$ form a complete set of
states. Subtracting the elastic cross section, we then obtain the cross
section for diffractive excitation, which thus is given by the
fluctuations in the scattering amplitude:
\begin{equation}
d\sigma_{diff-\mrm{tot}}/d^2 b = \langle T^2 \rangle - \langle T \rangle^2 ~.
\end{equation}

In a $NN$ collision, both the projectile and the target are
fluctuating, leading to single-diffractive excitation of the
projectile or the target, as well as to double diffraction. The
different $NN$ cross sections are then given by
\begin{align}
  \label{eq:ang-nn-xsecs}
  d\sigma_{\mrm{tot}}/d^2b &= \llangle 2T(\mathbf{b})\rrangle_{p,t}\nonumber\\
  d\sigma_{\mrm{abs}}/d^2b &= \llangle 2T(\mathbf{b})-
                          T^2(\mathbf{b})\rrangle_{p,t}\nonumber\\
  d\sigma_{\mrm{el}}/d^2b &= \llangle\llangle T(\mathbf{b})\rrangle_t^2\rrangle_p -
                          \llangle T(\mathbf{b})\rrangle^2_{p,t}\nonumber\\
  d\sigma_{\mrm{DD}}/d^2b &= \llangle T^2(\mathbf{b})\rrangle_{p,t} -
                         \llangle\llangle T(\mathbf{b})\rrangle_p^2\rrangle_t -
                         \llangle\llangle T(\mathbf{b})\rrangle_t^2\rrangle_p +
                         \llangle T(\mathbf{b})\rrangle^2_{p,t}~.
\end{align}
Here $\llangle\cdots\rrangle_p$ and $\llangle\cdots\rrangle_t$ are
averages over projectile and target states respectively, and
subscripts $Dt$, $Dp$, and $DD$ stand for single-diffractive excitation
of the target, the projectile, and double diffraction,
respectively. The absorptive or
\emph{non-diffractive inelastic} cross section is given by $\sigma_{\mrm{abs}}$. We note that the diffractive
excitation is directly related to fluctuations in the nucleon
wave function.

In \Angantyr, we use these cross sections in the Glauber modelling\index{Glauber modelling} to
determine not only which nucleons have been wounded, but also to
differentiate if they were non-diffractively scattered or only
diffractively excited. The fluctuations are by default modelled using
a varying radius of the nucleons, according to a Gamma function,
\begin{equation}
  \label{eq:ang-nucl-fluct}
  P(r)=\frac{r^{k-1}e^{-r/r_0}}{\Gamma(k)r_0^k} ~,
\end{equation}
and in addition, introducing a varying \emph{opacity} of the elastic
amplitude, which depends on the radii of the projectile and target
nucleons, $r_p$ and $r_t$,
\begin{equation}
  \label{eq:ang-var-amp}
  T(\mathbf{b},r_p,r_t)=T_0(r_p+r_t)
  \Theta\left(\sqrt{\frac{(r_p+r_t)^2}{2 T_0(r_p+r_t)}}-b\right) ~,
\end{equation}
where
\begin{equation}
  \label{eq:varyT0}
  T_0(r_p+r_t)=\left(1-\exp\left(-\pi(r_p+r_t)^2/\sigma_t\right)\right)^\alpha ~.
\end{equation}
We then obtain the differential semi-inclusive cross sections in
\cref{eq:ang-nn-xsecs} using
$\llangle\cdots\rrangle_i=\int dr_i P(r_i) (\cdots)$, which gives \eg
\begin{equation}
  \label{eq:ang-avex}
  \llangle\llangle T(\mathbf{b})\rrangle_p^2\rrangle_t=
  \int P(r_t)\left(\int P(r_p)T(\mathbf{b},r_p,r_t)dr_p\right)^2 dr_t ~.
\end{equation}

Three parameters ($k$, $r_0$ and $\sigma_t$) depend on the
$NN$ collision energy, and need to be determined. By default this is
done in the \Angantyr initialization by fitting the integrated total
and semi-inclusive $NN$ cross sections to the parameterization in \pythia
(see \cref{subsection:sigmatotal}), using a simple genetic
algorithm. If needed, the parameters can be specified by the user to avoid
the somewhat time-consuming fitting procedure.

The Glauber calculation works as follows. First the 3D positions of
the nucleons in the nuclei are modelled using a Woods--Saxon
parameterization (by default the parameterizations with a \emph{hard
  core} from~\citeone{Broniowski:2007nz,Rybczynski:2013yba} is
used). Then, an impact parameter between the nuclei is generated
according to a user-specified importance sampling (by default a 2D
Gaussian). For each nucleon we then sample the wave function according
to \cref{eq:ang-nucl-fluct}. This gives us the probability that a
projectile nucleon, $i$, scatters non-diffractively with target
nucleon, $j$, as
\begin{equation}
  \label{eq:ang-prob-nd-scat}
  2T(\mathbf{b},r_i,r_j)-T^2(\mathbf{b}_{ij},r_i,r_j) ~,
\end{equation}
where $\mathbf{b}_{ij}$ is the impact parameter between the nucleons.
But, we also want to obtain the probability of diffractive excitation,
which involves the fluctuations. We do this by generating an
additional radius, $r'$, for each nucleon, thus \emph{sampling} the
fluctuations. In this way we obtain four statistically equivalent $NN$
collisions and we can ensure that on the average we obtain the correct
integrated non-diffractive and diffractive excitation cross sections, by shuffling the probabilities between the four combinations so for each the probability never exceeds unity, as explained
in~\citeone{Bierlich:2018xfw}. It should be noted that this trick does
not allow us to determine the correct amount of elastic scattering, but these scattering are of less importance in a Glauber calculation.

In the end of the Glauber modelling, we have a long list of all
potential $NN$ subcollisions with an assigned type of
interaction. These will now tell us how many, and of which kind of $NN$
events we will generate using the normal \pp\ minimum-bias framework
in \pythia, to be merged together into a full HI collision
event. The way this is done is as follows.
\begin{itemize}
\item Order all non-diffractive subcollisions in the $NN$ impact
  parameter, $b_{ij}$, and iterate with increasing $b_{ij}$.
\item If none of the nucleons has been involved in a non-diffractive
  subcollision with smaller $b_{ij}$, generate a (primary)
  non-diffractive subevent.
\item If one of the nucleons has been involved in a previous
  subevent, generate a single-diffraction $NN$ event corresponding to
  the diffractive excitation of the other nucleon (using a special
  modification as explained in~\citeone{Bierlich:2018xfw}) and merge this
  with the corresponding previous subevent.
\item If both of the nucleons are already in a generated subevent, do
  nothing.
\end{itemize}
When we merge a single diffraction subevent, we only add the
diffractively excited subsystem, removing the elastically scattered
nucleon. We also take some longitudinal momentum from the remnants of
the primary event to ensure momentum-energy conservation.

In a similar way, we go through all double- and single-diffractive subcollisions, and add these to the full HI event. In the end,
we take all non-interacting nucleons and collect them into projectile
and target nucleus remnants, which each end up as a single entry in
the event record with PDG-ID codes of the form $100\mathit{ZZZAAA}9$, depending
on the number of neutrons and protons, which in the PDG standard
corresponds to a highly-excited nucleus.

It should be noted that all subevents above are generated on the parton
level, which allows us to hadronize them together. This
enables us the option to perform string shoving and rope formation (see
\cref{sec:shoving,sec:ropes}) on the full HI
partonic state.

The main use of the \Angantyr model is to generate minimum-bias events. It
is however, also possible to generate specific hard processes in HI
events. If a hard process is specified by the user, the Glauber
modelling will proceed as before, but (at least) one of the
non-diffractive primary $NN$ events will be replaced by a specific
hard interaction event, and at the same the event will be reweighted by
a factor given by
\begin{equation}
  \label{eq:5}
  N_{\mrm{ND}} \sigma_{\mrm{hard}}/\sigma_{\mrm{ND}}~,
\end{equation}
where $N_{\mrm{ND}}$ is the number of non-diffractive
subcollisions. Note that for the specified hard processes, \Angantyr
treats \pp, \p\n, \n\p, and \n\n\ subcollisions separately, which is
not the case for the minimum bias, where isospin symmetry is assumed.

By default, \pythia will automatically initialize the \Angantyr
machinery as soon as one of the beams is specified to be a nucleus
(using the PDG ID of the form $100\mathit{ZZZAAAI}$, where $I$ indicates the
excitation level). It is possible to use the \Angantyr machinery also
for minimum-bias \pp collisions, by setting \settingval{HeavyIon:mode}{2}.

Finally, it should be noted that only the most commonly used nuclei are
defined by default in \pythia, but a user can easily define further nuclei. Note
also that the beam energy of a nucleus is specified by giving the energy
per nucleon, following the convention of the field.

%% file: physics/hadronization.tex
\section{Hadronization}
Hadronization (often also referred to as fragmentation)
\index{Hadronization}\index{Fragmentation| see {hadronization}, } 
is the process of turning the final outgoing, coloured partons into 
colourless hadrons. This transition is non-perturbative, and must be handled by
models. In \pyt it is based on the Lund string model~\cite{Andersson:1983jt,Sjostrand:1984ic}, which is also historically
the core of the \jetset/\pyt programs. Even though the core methods for string 
hadronization are identical to previous versions of \pyt, the past years have seen significant 
activity in the area of fragmentation dynamics, guided by the discovery of 
heavy-ion-like effects in hadronic collisions. In \pyt, these efforts have culminated
in a multitude of models modifying the original Lund strings in the presence of other 
strings in an event. 

\subsection{The Lund String model}
\begin{figure}
  \begin{center}
    \includegraphics[width=0.35\textwidth]{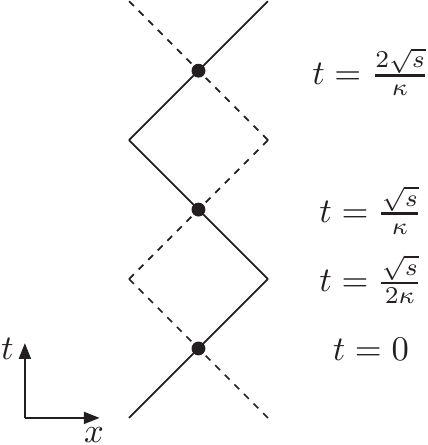} \hfill
    \includegraphics[width=0.45\textwidth,trim=0 300 0 0,clip]{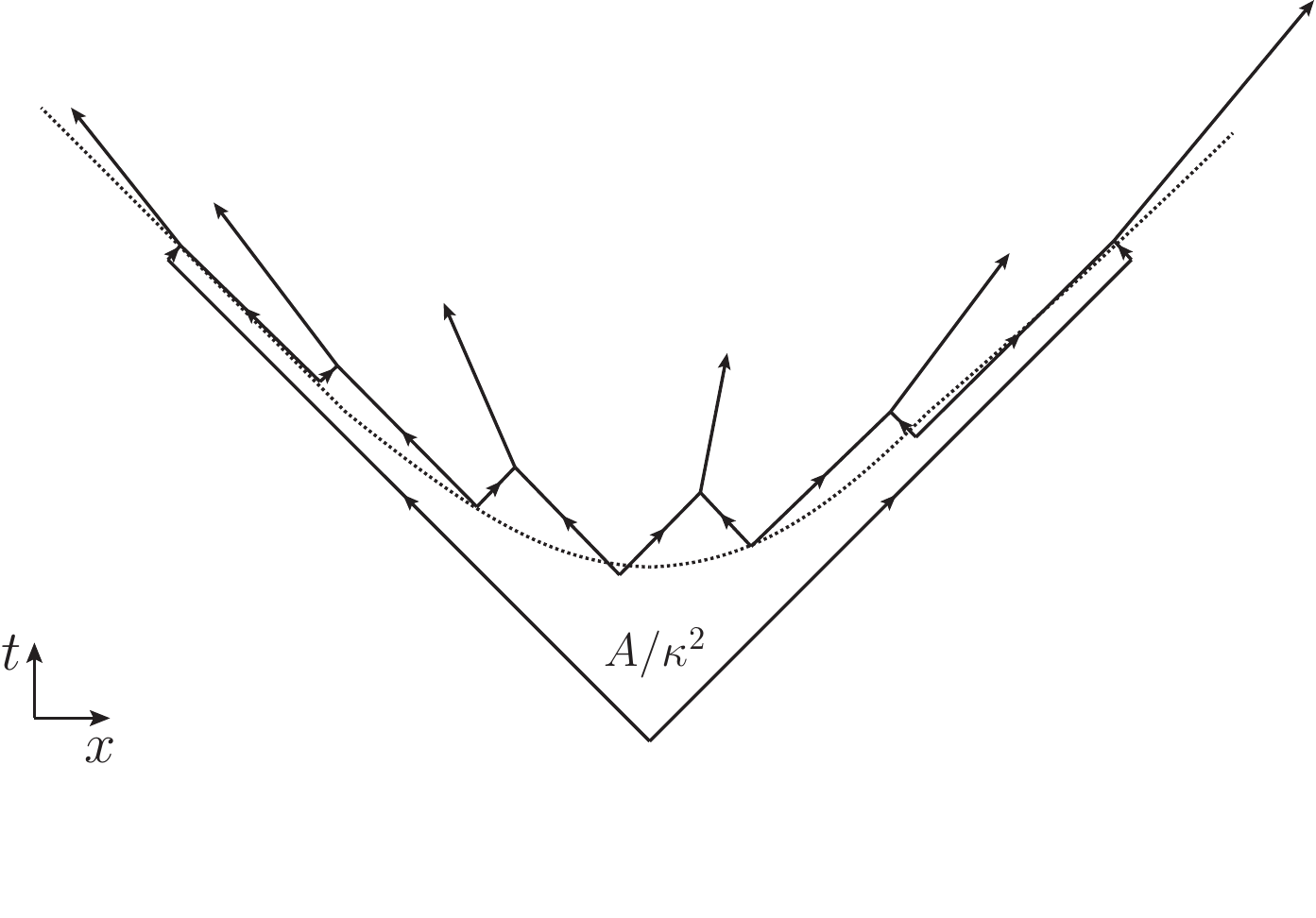} \\ 
    \null\hfill (a) \hfill\hfill (b) \hfill\null
    
  \end{center}
  \caption{\label{fig:string-model} (a) The yoyo picture of a meson, at several steps in time as explained in the text. (b) A quark-antiquark string breaking into hadrons. The original pair is moving outwards along light-like trajectories. New \qqbar pairs are produced around a hyperbola in $(x,t)$, and combine into hadrons.}
\end{figure}

\index{Lund model}\index{String model}
\label{sec:lund-model}
Results from lattice QCD support viewing the confining force field between a colour and an 
anti-colour charge, such as a \qqbar pair, as a flux tube with potential energy increasing 
linearly with the distance between the charge and the anti-charge. As the partons move apart,
energy is transferred from the partons at the ends of the string to the string itself, 
by $\kappa \approx 1$ $\GeV/\fm$. This directly gives rise to the so-called ``yoyo'' modes of single
\qqbar dipoles in 1+1 dimensions\footnote{The following convention for spatial coordinates is used. When 
discussing the 1+1 dimensional string, $x$ is taken as the spatial coordinate. When we move on to discuss 3+1
dimensional strings, the coordinate $z$ is chosen to be the coordinate along the string axis, as this will often
coincide with the coordinate along the beam axis, which is normally denoted $z$.}, as illustrated 
in \cref{fig:string-model}~(a). In the figure, an evolution
starts at time $t = 0$, where all the energy is stored in the ends, and none in the string,
\begin{equation}
	\label{eq:string-yoyo1}
	(E,p_x)_{\qqbar} = \frac{1}{2}(\sqrt{s},\pm\sqrt{s}),\quad E_{\mrm{string}} = 0~.
\end{equation}
The string reaches its maximal extension at time $t=\sqrt{s}/2\kappa$. Here, all energy has been transferred from the end-points to the string:
\begin{equation}
	(E,p_x)_{\qqbar} = (0,0),\quad E_{\mrm{string}} = \sqrt{s} ~.
\end{equation}
At time $t=\sqrt{s}/\kappa$, the string ends are back at their starting point, but with their momenta swapped compared to \cref{eq:string-yoyo1}, and finally at $t=2\sqrt{s}/\kappa$, the string has been through a full period.

In the string picture, yoyo modes like this are identified as mesons, with flavour determined by their quark content 
(see \cref{sec:had-flavour}). Longer strings will break into hadrons, with new \qqbar pairs breaking up the original 
string. Aligning the string axis of the original string with the $x$ axis, this process is depicted in 
\cref{fig:string-model}~(b). The \qqbar pairs are produced around a hyperbola, and joins together to form the hadrons, depicted as arrows. A hadron produced on the string is then characterized by two adjacent vertices ($i$ and $i-1$), with
space-time coordinates ($x$ and $t$) correlated through the hadron mass ($m$):
\begin{equation}
	m^2_i/\kappa^2 = (x_i - x_{i-1})^2 - (t_i - t_{i-1})^2 ~.
\end{equation}

\index{String breaks}In general, a string will break into a state with $n$ hadrons, which in the model is given by the probability~\cite{Andersson:1985qr}:

\begin{equation}
	\label{eq:had-momentum}
	\mrm{d}\mathcal{P} \propto \prod_{i=1}^n \left[Nd^2p_i \delta(p^2_i - m^2)\right]
    \delta^{(2)} \!\left(\sum p_i - P_{\mathrm{tot}}\right)\exp\left(-bA\right) ~,
\end{equation}
where $A$ is the area covered by the string before breakup in units of $\kappa$, as shown in \cref{fig:string-model}~(b), and $b$ is a parameter.
If the string breaking is imagined as an iterative process, the consistency constraint that the same result should be obtained (on average) by
fragmenting from the left or the right, one obtains the distribution of momentum fraction ($z$) of remaining light-cone momentum taken by
each hadron as
\begin{equation}
f(z) \propto \frac{(1-z)^a}{z}\exp\left(-\frac{bm^2}{z}\right) ~,
\label{eq:string:lsff}
\end{equation}
where $a$ is a new parameter related to $N$ and $b$ in \cref{eq:had-momentum}. Once transverse momenta are introduced, the
substitution $m^2 \rightarrow m^2_\perp$ is performed, with the
``transverse mass'' defined by
\index{Transverse mass}
\index{mT@$m_\perp$|see{Transverse mass}}
\begin{equation}
m_\perp^2 = m^2 + p_\perp^2 ~.
\end{equation}
The resulting form of \cref{eq:string:lsff} is known as the
\textit{Lund symmetric fragmentation function}. 
\index{Lund symmetric fragmentation function}
This simple picture of a \qqbar system can be extended to topologies including gluons, without introducing new parameters, by viewing the gluon
as a kink on the string in the $\Nc \rightarrow \infty$ limit, with separate colour and anti-colour indices. A string can as such stretch from
\eg the quark end through a number of gluons, and end in the antiquark end~\cite{Sjostrand:1984ic}.

\index{String breaks}While the default behaviour of \pyt is to always use \cref{eq:string:lsff} with given values for the parameters $a$ and $b$, the $a$ parameter can in principle be different for each flavour. This possibility is implemented for \s quarks and diquarks. Going from an old flavour $i$ to a new flavour $j$, the fragmentation function would thus be modified as:
\begin{equation}
	f(z) \propto \frac{z^{a_i}}{z} \left(\frac{1-z}{z}\right)^{a_j} \exp\left(-\frac{b m^2_\perp}{z}\right) ~.
\end{equation}
Finally, the Bowler modification~\cite{Bowler:1981sb} done in the Artru--Mennesier model~\cite{Artru:1974hr} allows for massive endpoint quarks with mass $m_Q$. This modified the symmetric fragmentation function, as the areas swept out by massive endpoint quarks is reduced compared to massless ones. Though using this modification is a break with the Lund-string philosophy, it is available as an option, where an effective $a$ term for a discrete mass spectrum~\cite{Morris:1987pt} is used:
\begin{equation}
	f(z) \propto \frac{1}{z^{1+r_Qbm^2_Q}}z^{a_\alpha} \left(\frac{1-z}{z}\right)^{a_\beta} \exp\left(-\frac{bm^2_\perp}{z}\right) ~.
\end{equation}
A common use case is to enable the Bowler modification for fragmentation for heavy quarks, as it can describe the somewhat harder spectrum better.

The derivation of \cref{eq:string:lsff} also gives the probability distribution in proper time ($\tau$) of \qqbar breakup vertices, \ie a quantity that
can be interpreted as (input to) a hadron production time. In terms of $\Gamma = (\kappa \tau)^2$ it is:
\begin{equation}
	\mathcal{P}(\Gamma) \mrm{d}\Gamma \propto \Gamma^a \exp(-b\Gamma)\mrm{d}\Gamma ~.
\end{equation}
From this distribution it is possible to calculate the average breakup time of a \qqbar string:
\begin{equation}
	\langle \tau^2 \rangle = \frac{1+a}{b\kappa^2} ~.
\end{equation}
Default \pyt values for $a$ and $b$ give $\langle \tau^2 \rangle \approx 2$~fm.
The $\Gamma_i$ values can be defined recursively
\begin{equation}
\Gamma_i = (1 - z) \left(\Gamma_{i-1} + \frac{m_{\perp}^2}{z} \right) ~,
\label{eq:string:gammanext}
\end{equation}
with $\Gamma_0 = 0$.

\subsubsection{Selection of flavour and transverse momentum}
\label{sec:had-flavour}
\index{Light-flavour hadrons}\index{String breaks}
In the previous section, the \qqbar pairs in the string breaking were treated as massless and without transverse momenta. If the quark and antiquark
has a transverse mass, they can no longer be produced in a single vertex, but must tunnel through a forbidden region of size $m_\perp/\kappa$. The
tunnelling probability can be calculated in the WKB approximation, giving \cite{Andersson:1983jt}:
\begin{equation}
	\label{eq:q-mass-and-pt}
	\frac{1}{\kappa}\frac{\mrm{d}\mathcal{P}}{\mrm{d}^2\pT} \propto \exp(-\pi m^2_\perp/\kappa) = \exp(-\pi m^2/\kappa) \exp(-\pi p^2_\perp/\kappa) ~.
\end{equation}
Here $m_\perp$ is the transverse mass of the quark, and the factorization of the result allows separation of the generation of $m$ and $p_\perp$.

\index{Strangeness suppression}The relative production of light quarks of different mass, and thus of different flavour, could in principle be obtained directly by inserting \u, \d and \s quark 
masses in \cref{eq:q-mass-and-pt}. It is, however, not obvious what quark masses to use. Current quark masses lead to too little strangeness suppression, 
and constituent quark masses \index{Quark masses!Constituent quark masses} lead to too much. Instead, the suppression is viewed as a free parameter, and tuned to LEP data. The current default \s suppression 
relative to \u or \d types is $0.217$, which does not imply unreasonable effective quark masses in \cref{eq:q-mass-and-pt}. Heavier quark flavours are suppressed too 
heavily to be produced in string breakings, for any reasonable value of their masses.

The generation of $\pT$ by \cref{eq:q-mass-and-pt}, can be implemented by giving the quark and antiquark Gaussian $\pT$-kicks with $\sigma^2 = \kappa/\pi \approx (0.25~\GeV)^2$. 
Fits to data have this number higher, around $\sigma = 0.35$~\GeV, implying that a large fraction of the $\pT$ kick comes from another
source, such as soft gluon radiation below the parton shower cutoff.

\begin{figure}
	\begin{center}
	\includegraphics[width=0.7\textwidth]{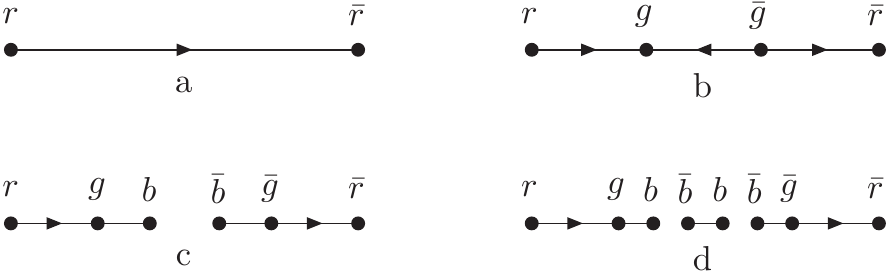}
	\end{center}
	\caption{\label{fig:popcorn}Illustration of step-wise popcorn production of a baryon-antibaryon pair, with a meson in between. In frame a), a string is spanned between a red-antired ($r\bar{r}$) \qqbar pair, with colour flow indicated by the arrow. In frame b), a green-antigreen ($g\bar{g}$) \qqbar pair has appeared as a vacuum fluctuation between them, reversing the colour flow in the central part of the string. In frame c), an additional pair is produced, breaking the string, and in frame d) another breakup produces a meson between the baryon and anti-baryon. Figure from \citeone{Bierlich:2014xba}.}
\end{figure}

Besides production of the normal light-quark species, other hadron types can be produced through the same mechanism with a few modifications. Excited mesons are allowed by letting quarks and antiquarks combine to a total spin of either 0 or 1. Considering only pseudoscalar and vector multiplets, the expectation of the relative rate is $1 : 3$, while data -- at least in the case of $\pi : \rho$ -- prefers a ratio about 1. This difference between prediction and data can be explained as a result of differences in the hadronic wave function~\cite{Andersson:1983ia,Bierlich:2022vdf}, but this comes at the expense of many free parameters, which have to be tuned to data. \index{Baryons}\index{Popcorn}Baryons can be produced using \cref{eq:q-mass-and-pt} as well, by allowing diquark-antidiquark string breakings~\cite{Andersson:1981ce}. Compared to the production of \s quarks, this process will be suppressed by a larger (effective) diquark mass. In such an approach, the produced baryon-antibaryon pair will be neighbours along the string, and share two flavours. This simple picture is modified by considering an underlying step-wise mechanism for baryon production, first suggested by Casher, Neuberger and Nussinov~\cite{Casher:1978wy}, and realized in the ``popcorn'' model~\cite{Andersson:1984af} in \pyt. In the popcorn model, diquarks are generated by first producing a \qqbar pair as a vacuum fluctuation on the string, without breaking it. By producing more new \qqbar pairs in between, meson production between the baryon-antibaryon pair is allowed. The whole process is illustrated in \cref{fig:popcorn}. In principle, several mesons can be produced in between a baryon-antibaryon pair through the popcorn mechanism, but currently only the simplest case of a single meson is implemented in \pyt.

While this explanation above suffices for an introduction of the physics behind the model, there are many important implementation details to be faced when going from a ``physics level'' description of the Lund string
to the actual implementation in \pyt, which must be able to handle arbitrarily complicated configurations of partons. In the next subsections we outline several of the more specialized features in \pyt string fragmentation, and the thought behind the implementation. While some are completely new models on top of the old hadronization framework, others remain the same as even the oldest version of the \jetset and \pyt~6.3 programs \index{Jetset@\jetset}. Those specific parts of the discussion are therefore largely carried over from the \pyt~6.4 manual~\cite{Sjostrand:2006za}.

\subsubsection{Joining two jets in \qqbar events}

Keeping with the simple picture of a single \qqbar pair, the iterative procedure obtained by successive application of \cref{eq:string:lsff}, is only valid when the remaining mass of the system, after fragmenting off a hadron, is large. If the algorithm implementing \cref{eq:string:lsff} were to start from one end, and create hadrons successively until the other end is reached, the mass of the last hadron would be fully constrained by four-momentum conservation, and would therefore be off-shell.

The practical route taken in \pyt, is to randomly fragment off hadrons from either the \q or \qbar end in each step, with $z$ taken to be either the positive or negative light-cone momentum respectively. To wit, if the step is on the \q side, $z$ is the remaining $E + p_z$ fraction, and if the step is on the \qbar side, $z$ is the remaining $E - p_z$ fraction. Once the mass of the remaining system has dropped below a certain value, with some smearing to avoid an unphysical sharp cutoff, the remaining system is fragmented into two ``final'' hadrons, and the chain ends.

\subsubsection{Fragmentation of systems with gluons}

Most of the preceding discussion has involved the simple system of a single string spanned between a \qqbar pair. While sufficient to explain the basic features of the model and implementation, it is far from covering the complexity in hadronization of multiparton systems. A Lorentz covariant algorithm exists, however, and in this section the machinery employed for this task is outlined, noting that the complete machinery is complicated, and covered in detail in \citerefs{Sjostrand:1984iu,Sjostrand:1984ic}.

The basis of the algorithm is to divide multiparton systems to be fragmented into smaller string pieces, spanned between individual partons. Consider a long string spanned between a \qqbar pair (labelled 1 and $n$ in the following), with a number of gluons in between (labelled $2,...,n-1$). Such a string will contain $n-1$ separate pieces. The kinematics of those pieces are, as for simple \qqbar strings, determined by the four-momenta of the endpoint partons. In the case of gluons, the four-momentum is shared between the two neighbouring string pieces, each taking half. It must furthermore be assumed that endpoint (anti-)quarks are massless, for the fragmentation algorithm to work. In practise this is done by attaching a fictitious string piece with a massless (anti-)quark to the string end, replacing the massive quark. This string piece in a later step becomes part of the massive hadron produced from the massive quark. 

In summary, we have therefore $n-1$ string pieces defined by adjacent\footnote{It is possible to have string regions spanned by non-adjacent pairs as well, created when a gluon loses all its energy to the string. These regions form an integral part of the formalism, and help ensure that string fragmentation is rather insensitive to soft and collinear gluon emissions in the parton-shower stage.} four-momentum pairs $(j,k)$, with the parton going towards the \q end further indexed with a $+$ and the parton going towards the \qbar end with a $-$. In general, a hadron is now created by taking a step from a region $(j_1,k_1)$ to $(j_2,k_2)$. A step may be taken within just a single region, or between two different regions. The resulting hadron four-momentum can be written as
\begin{equation}
p = \sum_{j=j_1}^{j_2} x_+^{(j)} p_+^{(j)} + \sum_{k=k_1}^{k_2} x_-^{(k)} p_-^{(k)} +
      p_{x1} \hat{e}_x^{(j_1 k_1)} + p_{y1} \hat{e}_y^{(j_1 k_1)} +
      p_{x2} \hat{e}_x^{(j_2 k_2)} + p_{y2} \hat{e}_y^{(j_2 k_2)} ~,
\label{eq:string:psum}
\end{equation}
where the four-momentum fraction of $p^{i}_\pm$ taken by the hadron is denoted $x^{i}_\pm$, and $(p_x,p_y)$ are the transverse momenta produced at the string breaks according to \cref{eq:q-mass-and-pt} with $(\hat{e}_x,\hat{e}_y)$ spacelike unit four-vectors normal to the string direction in the respective region.

The only remaining degree of freedom is $z$, to be determined by \cref{eq:string:lsff}. The interpretation of $z$ is, however, only well-defined for a step in the initial string regions. But via \cref{eq:string:gammanext} a $z$ value can be translated into a new $\Gamma = (\kappa \tau)^2$ value, and $\Gamma$ is well defined across region boundaries. Together with the $p^2 = m^2$ constraint on \cref{eq:string:psum} this is sufficient to find the relevant $x_+^{(j_2)}$ and $x_-^{(k_2)}$ values of the next breakup vertex.

\subsubsection{Hadron vertices}\index{Hadron production vertices}
\label{sec:hadron-vertices}
While the production vertices of hadrons are impossible to detect experimentally,
calculating them still has applications in other parts of the simulation, most 
notably hadronic rescattering. In this section we describe the space-time
picture for $\qqbar$ pairs, based on methods developed in \citeone{Ferreres-Sole:2018vgo}. 

From the linear potential $V(r) = \kappa r$, the equations of motion are
\begin{equation}
\left| \frac{\d p_{z,\q/\qbar}}{\d t} \right| =
\left| \frac{\d p_{z,\q/\qbar}}{\d z} \right| =
\left| \frac{\d E_{\q/\qbar}}{\d t} \right| =
\left| \frac{\d E_{\q/\qbar}}{\d z} \right| = \kappa ~.
\end{equation}
The sign on each derivative is negative if the distance between the quark is
increasing, and positive if the distance is decreasing. 
After sampling $E_{h_i}$ and $p_{h_i}$ for each hadron, these equations lead
to simple relations between the space-time and momentum-energy pictures, 
$z_{i-1} - z_i = E_{h_i}/\kappa$ and $t_{i-1} - t_i = p_{h_i}/\kappa$, 
where $z_i$ and $t_i$ denote the space-time coordinates of the $i$th breakup
point (note that $z_{i-1} > z_i$ since points are enumerated from right to left).
\index{Uncertainties!Hadron@in Hadronization}In the massless approximation, the endpoints are given by
$z_{0,n} = t_{0,n} = \pm \sqrt{s}/2\kappa$. This specifies the breakup points,
but there is still some ambiguity as to where the hadron itself is produced.
The default in \pythia is the midpoint between the two breakup points, but it is
also possible to specify an early or late production vertex at the point
where the light-cones from the two quark-antiquark pairs intersect.

A complete knowledge of both the space-time and momentum-energy pictures violates
the Heisenberg uncertainty principle. This is compensated for in part by
introducing smearing factors for the production vertices, but outgoing 
hadrons are still treated as having a precise location and momentum. Despite
not being a perfectly realistic model, there is no clear systematic bias in
this procedure, and any inaccuracies associated with this violation are
expected to average out.

There are several further complications to these process. One is more
complicated topologies such as those involving gluons or junctions. Another
is the fact that the massless approximation is poor for heavy $\qqbar$
pairs. For massive quarks, rather than moving along their light-cones, the 
quarks move along hyperbolas $E^2 - p_z^2 = m^2 + \pTs = m_{\perp}^2$. Both these
issues are addressed in more detail in \citeone{Ferreres-Sole:2018vgo}.

\subsubsection{Junction topologies}
\label{sec:junctions}%
\index{Junctions}%
\index{String junctions|see{Junctions}}%
\index{Baryon number violation}
\index{Baryons}
\index{Baryons!Junction baryons}

Junction topologies in their simplest form arise when three massless quarks
in a colour-singlet state move out from a common production vertex, a
textbook example of which is given by a baryon-number-violating
super-symmetric decay $\chi^0 \to \q\q\q$. 
In that case it is assumed that each of them pull out a string piece,
a ``leg'', to give a Y-shaped topology, where the three legs meet in a
common vertex, the junction. This junction is the carrier of the baryon
number of the system: the fragmentation of the three legs from the
quark ends inwards will each result in a remaining quark near to the
junction, and these three will form a baryon around it.

\begin{figure}[t]
  \centering
  \includegraphics*[width=\textwidth]{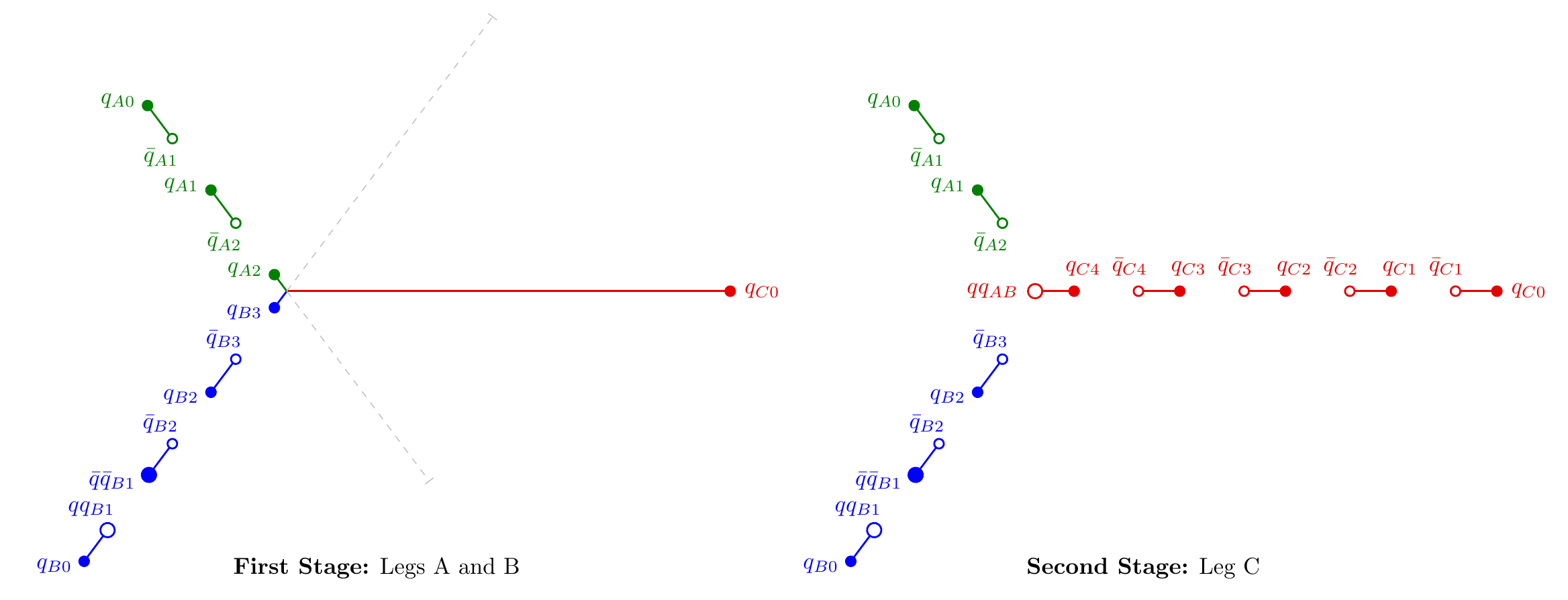}
  \caption{Illustration of the two main stages of junction
    fragmentation. (left) First, the junction rest frame (JRF) is
    identified, in which the pull directions of the
    legs are at $120^\circ$ to each other. (If no
    solution is found, the CM of the parton system
    is used instead.) The two lowest-energy legs ($A$ and $B$) in this
    frame are then  fragmented from their respective endpoints
    inwards, towards a 
    fictitious other end which is assigned equal energy and opposite direction,
    here illustrated by grey dashed lines. This fragmentation
    stops when any further hadrons would be likely to have negative
    rapidities along the respective string
    axes. (right) The two leftover quark endpoints from the
    previous stage ($q_{A2}$ and $q_{B3}$) are combined into a
    diquark (${qq}_{AB}$) that is then used as endpoint for a
    conventional fragmentation along the last leg,  alternating
    randomly between fragmentation from the $q_C$ end and
    the $qq_{AB}$ end as usual.
    \label{fig:junctionFragmentation}\index{Junctions}}
\end{figure}
\index{Junction rest frame}
The junction will be at rest in a frame where the pull of the three
legs balance each other, which is when the angle between each quark pair
is $120^{\circ}$. It is therefore convenient to handle the hadronization
in such a frame. There is no first-principles description of junction-string fragmentation. Instead the process is split into a few steps,
to make use of the existing string machinery in a credible manner~\cite{Sjostrand:2002ip}, illustrated in
\cref{fig:junctionFragmentation}. First, the two lowest-energy legs
are considered separately, each as if it were a $\q\qbar$ string, with
a fictitious $\qbar$ in the opposite direction to the $\q$.
All fragmentation is from
the $\q$ end of the respective system, however, and keeps on going until
almost all the original $\q$ energy is used up, resulting in the
situation illustrated in the left-hand pane of
\cref{fig:junctionFragmentation}. At that stage the remaining
unmatched two quarks ($q_{A2}$ and $q_{B3}$ in the figure) are
combined into a diquark, carrying the unspent 
energy and momentum. 
This diquark now forms one end of the remaining
string out to the third quark, which can be fragmented as a normal string
system, illustrated in the right-hand pane of
\cref{fig:junctionFragmentation}.
One criterion that the procedure works, \eg that the fragmentation
of the two first legs is stopped at about the right remaining energy, is
that the junction baryon is formed with a low momentum and with minimal
directional bias in the junction rest frame. Additional checks are
also made to ensure that the final string mass is above the threshold for
string fragmentation. Otherwise, repeated attempts are made, starting over
with the first two strings.  

Unfortunately real-life applications introduce a number of complications.
One such is that the pull is more complicated when the endpoints are not
massless. Then, in a fraction of the events, there is no analytic solution.
Typically this happens when a massive quark is almost at rest in the
configurations that come closest to balance, and an approximate balance
along these lines may be obtained. An even more complicated case is when
a leg is stretched via a number of intermediate gluons between
the junction and the endpoint quark, as would be a natural consequence
of parton-shower evolution in the $\chi^0 \to \q\q\q$ decay. Then the
initial motion of the junction is set by the gluon nearest to it.
But often this gluon has low energy and, once that is lost to the
drawn-out string, it is the direction of the next-nearest gluon that sets
a new net pull. Thus, there is no frame where the junction remains at rest
throughout the whole fragmentation process. An effective average pull is
then defined for each of the three legs, as a weighted sum of the
respective parton momenta, where the weight drops exponentially as the
energy sum of partons closer to the junction increases, \cf\citeone{Sjostrand:2002ip}.

The absence of an exact solution for the junction rest frame leads to an
approximate iterative procedure being used. One of the more common
sources of \pyt warnings is that this procedure does not converge.
If no fix can be found any other way, then ultimately the centre-of-mass
frame of the system is taken as the junction rest frame.

Junction fragmentation is not only a topic for exotic physics, 
but very much part of ordinary QCD hadronic physics. It appears if two
valence quarks are kicked out of a baryon beam by the MPI machinery.
Since these interactions typically involve colour exchange, two of the
ends will stretch to partons from the other incoming beam, unless
colour reconnection gives another result. The fragmentation follows the
already outlined procedure, which can lead to the beam baryon number
being transported in to the central region of the event,
\cf\citeone{Sjostrand:2004pf}. 

Also antijunctions may exist, where the colour lines from three antiquarks
meet, and such antijunctions carry a negative baryon number. A string
system may contain both a junction and an antijunction, or even multiple
of such. The simplest such topology is when one leg connects a junction
to an antijunction, leaving two other junction legs to quarks and two
antijunction legs to antiquarks. It is here assumed that the total string
length (see \cref{sec:colRec}) is smaller for such a topology than
for having two simple $\q\qbar$ strings, or else the junction pair would
annihilate to give the simple string
topology, \cf\cite{Sjostrand:2002ip,Sjostrand:2004pf}. Conversely, when the string
length can be reduced, more-or-less parallel $\q\qbar$ strings may colour
reconnect into junction-antijunction systems, see further
\cref{sec:colRec:QCD-based}.

To reduce the complexity of multijunction fragmentation, each system
is split up into smaller ones that only contain (at most) one junction
or antijunction each. Consider \eg a junction-antijunction topology.
If the leg connecting the two contains at least one gluon, it can be
split up by a replacement $\g \to \q\qbar$. If not, a small amount of
energy can be shuffled from the regular $\q$ and $\qbar$ legs into some
energy (and momentum) for this connecting leg, so that it can be split.

Another subtlety concerns what spin state to choose for the diquark
that is formed at the end of the fragmentation of the two first legs,
the one labelled $qq_{AB}$ in \cref{fig:junctionFragmentation}, which
we will call the junction diquark. For conventional (non-junction)
fragmentation, 
empirically one finds that $S=1$ diquark states are heavily
suppressed, interpreted as due to significantly higher masses
and smaller binding energies. However, unlike in conventional string
breaks, where diquark-antidiquark pairs are formed together in a
\emph{single} coherent tunnelling process (modulo fluctuations such as in the
popcorn scenarios), the junction diquark is formed by combining
the leftovers from two \emph{separate} string breaks; \pyt~8 therefore
allows for the $S=1$ suppression factor for junction diquarks to be
set independently of that for conventional diquarks. Moreover,
analogously to in the meson sector, it can be set independently for
$b$-, $c$-, $s$-, and light-flavoured junction diquarks, where the
label always refers to the heaviest of the two constituents.  

It is also worth emphasizing that, within the context of the
current \pyt modelling, junctions represent the sole 
mechanism for producing baryons containing multiple heavy flavours,
such as $\Xi_{cc}$, $\Omega_{cc}$, $\Omega_{ccc}$, and their
$b$-flavoured relatives. Note, however, that this will still be quite
rare; since heavy flavours cannot be produced by string breaks, they
can only appear as endpoints, say $q_{A0}$ and $q_{B0}$ in
\cref{fig:junctionFragmentation}. The only possibility to form a
double-heavy-flavoured baryon involving these is if there is too
little energy in \emph{both} legs $A$ and $B$ for any other 
string breaks to occur, so that $q_{A0}$ and $q_{B0}$ are combined
directly into the junction diquark, which is then
doubly-heavy flavoured. We note that, so far, no dedicated emphasis 
has been placed on developing the heavy-quark aspects of junction
fragmentation, though that may change with experimental interest.
Predictions should therefore be regarded as tentative.  

In summary, the full machinery for junction hadronization is convoluted
and not without weaknesses, but overall it serves its purpose, and finds
use in several physics contexts.

\subsubsection{Small-mass systems}
\label{sec:small-mass}\index{Cluster fragmentation}

If the invariant mass of the \qqbar system is small, a few complications
to the fragmentation process can arise. For example, for an \ssbar
system at 0.9~GeV, the string cannot fragment as there is
not enough energy to form an outgoing \kaon\kaonbar pair,
nor can the quarks enter a ``yoyo motion'' as there
is no hadron with compatible mass and flavour content. Furthermore,
even if the string can fragment, at low energies the available phase
space might be so small that the fragmentation algorithm becomes
very inefficient. These situations can occur for instance towards the end
of a parton shower by $\g \to \q\qbar$ branchings
or during hadronic rescattering, and are handled using approaches
inspired by cluster fragmentation~\cite{Norrbin:1998bw}.

To improve the efficiency of the algorithm, the first step is to assume that
the string will break at only a single point, and a few attempts are made to
find possible outgoing two-hadron states. If these attempts fail, next the
algorithm tries to form a single hadron from the endpoints, then put that
hadron on-shell by transferring momentum to or from a neighbouring string
system. If no momentum rearrangement is possible, further attempts are made to
find possible two-hadron states, but now only the lightest possible hadrons
for the given flavour content are considered. If this still does not work,
the string may collapse to the lightest possible hadron given the
endpoints, and produce one additional \piz. Finally, if this is not possible
either, the last resort is to collapse the string to the lightest
possible hadron, and transfer momentum with a neighbouring parton or hadron.

String systems are handled in order of increasing mass relative to the
two-body threshold, so normally other systems are still unfragmented
when addressing this kind of issue. Especially in (low-energy) hadronic
rescattering there may two low-energy strings. Then, when the first string
is handled, its collapse may reduce the mass of the other string. 
In this case, that system may also collapse to a single hadron, which is
put on-shell by transferring momentum with
a hadron from the previously fragmented string.

\subsection{Colour reconnections}
\index{Colour reconnections}\index{Leading colour}
\label{sec:colRec}

In \pyt (and other event generators), a simplified set of rules
for colour flow is used to determine between which partons confining
potentials should arise. In the context of the string model, this
determines a unique string topology which sets the stage for the
subsequent hadronization.  

Specifically, all perturbative processes (including MPIs, ISR and FSR)
are handled in a \emph{leading colour} (LC) limit in which the
probability for any two random colours to both be the same
vanishes. Formally, this is done by taking the limit
$\Nc \to \infty$ with $\alpha_s \Nc$ kept fixed~\cite{tHooft:1973alw}
so that QCD amplitudes retain their $\Nc=3$ normalizations.
This accomplishes two things: 1) it eliminates colour-interference effects  
which are suppressed by powers of $1/\Nc^2\to 0$, and
2) it allows for a particularly simple representation of gluons in
colour space, as direct products of a colour and an
anticolour, since the weight of the singlet in 
$\Nc\otimes \bar{N}_\mrm{c} = (\Nc^2 - 1) \oplus 1$ vanishes as $\Nc\to
\infty$.

In the LC limit, Feynman-diagram amplitudes in colour space are 
represented by products of independent ``colour lines''. Each of
these expresses conservation of a distinct colour charge, and is 
represented by a $\delta_{ij}$ connection between partons carrying
colours $i$ and $j$ (suitably crossed). We call this an LC dipole
connection. Due to the orthogonality of the basis states
and the lack of interference in this limit, each such line translates
directly to a coherent colour-singlet structure at the colour-summed
amplitude-squared level, which is confining at large
distances. Thus, each LC dipole emerging from the perturbative stages
of the event evolution can be uniquely mapped to a string piece
(discussed further, \eg in~\cite{Gustafson:1986db}). We use the term
``colour \textit{re}connection'' (CR) to refer to any scenario that results
in changes relative to this map in defining the starting configuration of
hadronizing strings in an event.

\begin{figure}[t]
  \centering
  \includegraphics*[width=0.5\textwidth]{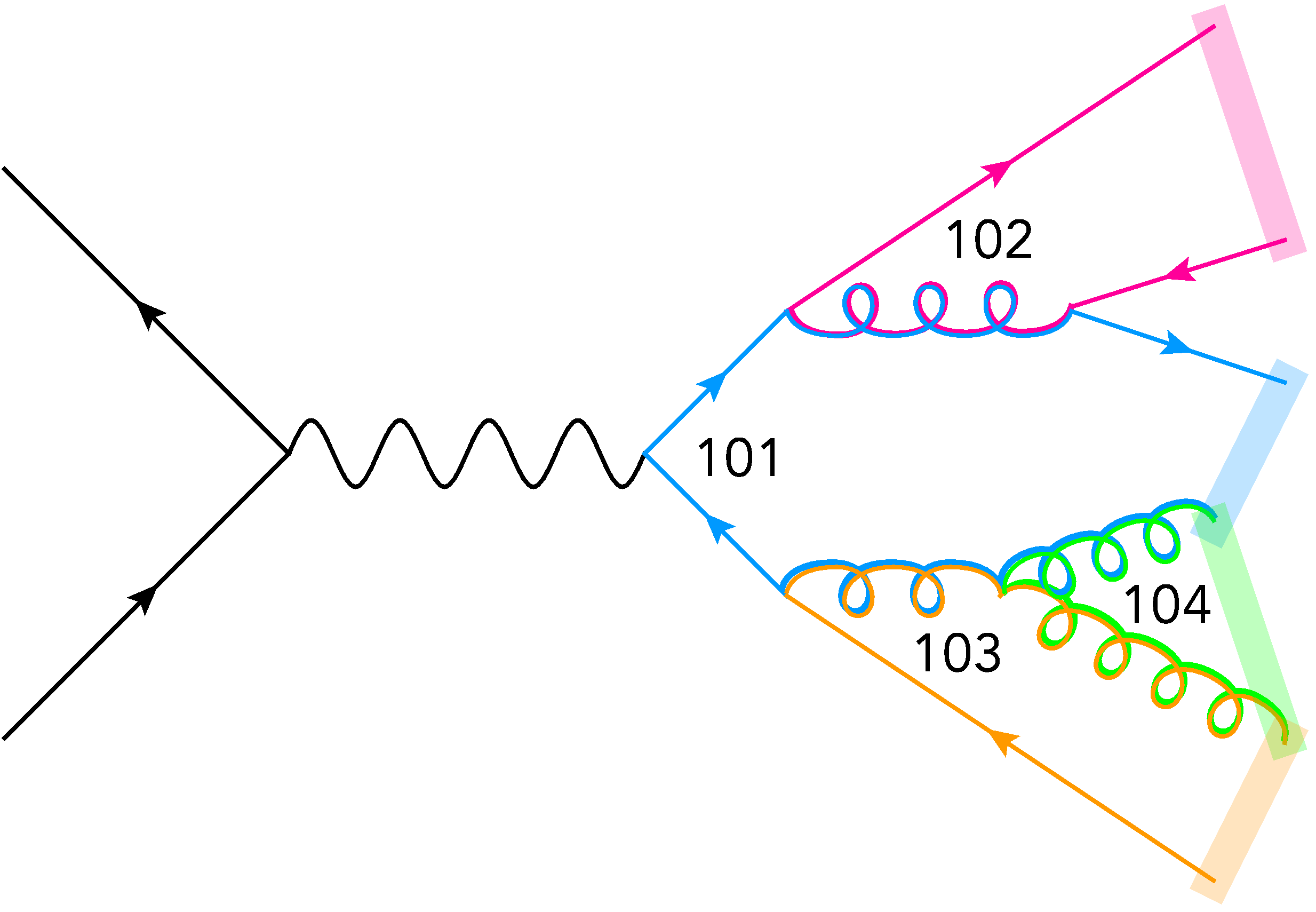}
  \caption{Illustration of LC colour flow in a simple 
    $e^+e^- \to q\bar{q}\,\otimes\,\mathrm{shower}$ event.
    The shaded regions represent the resulting unique LC
    string topology. 
  \label{fig:colFlow1}}
\end{figure}

A simple illustration of the map between LC dipoles and string pieces,
for an  $e^+e^- \to$ $\gamma^*/Z \to$ $q \bar{q} q' \bar{q}' gg$ event, is
shown in \cref{fig:colFlow1}.
Matching colour (and anticolour) charges are represented by
Les Houches colour (and anticolour)
tags~\cite{Boos:2001cv,Alwall:2006yp} numbered from 101 -- 104 in 
this example and indicated by coloured lines in the diagram.
In keeping with the
 $\Nc\to \infty$ nature of the LC limit, the number of different tags is not
limited to three, and each new tag is distinct from all
others. This produces a unique set of colour connections which can be
traced to form the LC string topology (shaded regions).

In hadronic collisions, the structure of the beam remnants is also to
be modelled, after MPIs have extracted multiple coloured objects
from them. Here it is useful define rules on how to equate some of these
colours and anticolours with each other, so as to keep the total
colour charge of a remnant within reasonable bounds. Note that
this would still classify as ``colour connection'', insofar as it is the initial
assignment of remnant colours, although the consequences propagate in
from the remnants to the central perturbative interactions. This is
discussed further in the section on beam remnants,
\cref{subsection:beamRemnants}. 
As used in this section, the term CR applies to models that go
beyond this, \ie that allow for departures from the simple colour rules
discussed above and/or address ambiguities that are left unresolved by them.
CR may be classified as one example of a broader palette of string interactions,
with other examples presented in \cref{sec:string-interactions}.

\index{Junctions}Note that, occasionally, ``junction'' structures (see
\cref{sec:junctions}) may also be present. Unlike dipole-type
$\delta_{ij}$ connections, junctions (and antijunctions) represent
$\epsilon_{ijk}$ structures in colour space; these are explicit
$\Nc=3$ structures which have no analogy in the $\Nc\to \infty$
limit. In \pyt, they can appear in the initial state in proton
beams~\cite{Sjostrand:2004pf}, 
in hard BSM processes (or decays) with baryon number
violation~\cite{Sjostrand:2002ip,Desai:2011su}, and/or as a product of colour
reconnections in the final state (in pairs of junctions and
antijunctions to conserve overall baryon
number)~\cite{Christiansen:2015yqa}. Due to the added technical
complexity of dealing with junction structures, the latter possibility is,
however, so far only invoked by the QCD-based CR model,
\cf\cref{sec:colRec:QCD-based}.   

Several different scenarios are included in \pyt, as described in the
following subsections, each with its own motivations and underpinnings.
The unifying feature is that these models act only by reassigning colours,
with no explicit momentum exchanges between the involved partons.
The decisions whether and how to reassign still can depend both on
momentum-energy and on space-time relations between partons. Also,
the changes at the level of produced hadrons still can be dramatic,
due to the changed lengths and orientations of the resulting hadronizing
strings. 

Historically, CR was first discussed in the context of charmonium 
production~\cite{Fritzsch:1977ay,Ali:1978kn,Fritzsch:1979zt}, 
notably in weak $\gp{B}$ decay to $\Jpsi$, \eg
$\gpbar{B}^0  = \b \dbar \to \Wm \c \dbar \to 
\s \cbar \c \dbar \to \Jpsi \kzbar$. In such decays 
the $\c$ and $\cbar$ belong to two separate colour singlets, 
but ones that overlap in space-time, with the possibility of soft 
gluon exchange to create the new singlets. 

The first large-scale application of CR was in the \textsc{Pythia}
MPI model of hadronic collisions~\cite{Sjostrand:1987su}, notably to
explain the increasing mean transverse momentum $\langle p_{\perp} \rangle$
with increasing charged multiplicity $n_{\mathrm{ch}}$ observed at CERN's
S$\p\pbar$S collider~\cite{Albajar:1989an}. If all MPIs draw out strings
and fragment in the same manner, $\langle \pT \rangle(n_{\mathrm{ch}})$
would be essentially flat. CR was therefore introduced in such a way that
the total string length is reduced. Each further MPI then on the average 
increases $n_{\mathrm{ch}}$ less than the previous one, while giving
the same $p_{\perp}$ from (mini)jet production, resulting in an increasing
$\langle p_{\perp} \rangle(n_{\mathrm{ch}})$.

LEP~2 offered a good opportunity to search for CR effects. Specifically, 
in a process $\epem \to$ $\Wp \Wm \to$ $\q_1 \qbar_2 \q_3 \qbar_4$,
CR could lead to the formation of alternative ``flipped'' singlets 
$\q_1 \qbar_4$ and $\q_3 \qbar_2$, and correspondingly for more complicated 
string topologies, formed when parton showers are included. Such CR would be
suppressed at the perturbative level, since it would force some $\Wpm$
propagators off the mass shell~\cite{Sjostrand:1993hi}. This suppression
would not apply in the soft region. Based on a combination of results
from all four LEP collaborations, the no-CR null hypothesis is excluded
at a $99.5\%$ CL~\cite{Schael:2013ita}. Within the SK~I scenario, described
below, the best description is obtained for $\sim$50\% of the 189~GeV
$\Wp\Wm$ events being reconnected, in qualitative agreement with
predictions.

\index{Top quark!Mass}\index{Quark masses!Top quark mass}More
recently, Tevatron~\cite{CDF:2016vzt} and 
LHC~\cite{ATLAS:2018fwq,CMS:2018tye} measurements of the top-quark mass in 
hadronic top-quark decays brought CR effects on precision
observables to the fore again, with several new models geared 
towards the increased complexity of hadron collisions produced
first in \pyt~6~\cite{Sandhoff:2005jh,Skands:2007zg,Skands:2010ak}
and later in \pyt~8~\cite{Argyropoulos:2014zoa,Christiansen:2015yqa}.
Hadronic reconstruction of the top-quark mass remains an important
impetus for further explorations of CR model space and for the
development of systematic and exhaustive ways to constrain modelling
ambiguities and parameters experimentally.
\index{Uncertainties!Hadron@in Hadronization}

The importance of colour algebra versus dynamics differs widely
between models. Taking the simple $\Wp \Wm$ case above, there is a
$1/9$ probability that $\q_1 \qbar_4$ and $\q_3\qbar_2$ are singlets
purely by colour algebra. But such accidental singlets do not stop
$\q_1 \qbar_2$ and $\q_3 \qbar_4$ from still being singlets as well;
so nevertheless, a dynamics principle would be needed to decide which
singlet set takes precedence when it is time to hadronize.
Furthermore, once parton showers are included, the number of colour
charges in an event increases, and the possibilities for CR with it.
In the extreme limit, a string may be viewed as a chain of
(non-perturbative) gluons infinitesimally closely spaced, such that
the string constantly flips colour, so there would be no
suppression of CR for lack of nearby matching colours.

\index{String length@String-length measure $\lambda$}
\index{Lambda measure@$\lambda$ measure of string length}
In several of the models below the
concept of a ``string-length'' 
$\lambda$ plays a prominent dynamics role. It is a measure of how many
hadrons of some reference hadronic mass $m_0$ there are room for (in
phase space), if the hadrons are evenly spaced in rapidity along the string. 
For a simple $\q\qbar$ string of mass $m_{\q\qbar}$ one possible definition is
$\lambda = \ln( m_{\q\qbar}^2 / m_0^2)$. In principle, $\lambda$ is
well defined also for more complicated string topologies~\cite{Andersson:1985qr}, but in practice its construction is too
complicated. Instead, approximate expressions are used, like
\begin{equation}
\lambda \approx 
\sum_{i = 0}^{n} \ln \left( 1 + \frac{ m_{i,i+1}^2}{m_0^2} \right) ~,
~~~~ m_{i,i+1}^2 = (\epsilon_i p_i + \epsilon_{i+1} p_{i+1})^2 ~,
~\epsilon_q = 1 ~, ~ \epsilon_g = \frac{1}{2}~,
\label{eq:colRec:intro:lambda}
\end{equation}
for a string $\q_0 \g_1 \g_2 \cdots \g_n \qbar_{n+1}$, where
$\epsilon_g = 1/2$ because gluon momenta are shared between two string
pieces. The addition of 1 is to ensure that a low-mass section does not
give a negative contribution, and is not always used. More generally,
if low masses are common, it probably signals that there is a larger underlying issue, \eg having too low a cut-off for shower evolution.

Loosely speaking, $\lambda$ can be viewed as the ``free energy'' of
a string system, available for particle production. It
provides a useful momentum-space measure of the worldsheet
area that a given string system will span, on average, before string
breaks occur. Since the classical (Nambu--Goto) string action is
proportional to (the negative of) that area, it is generally
assumed that, other things being the same, nature prefers a low
string length.

It should be noted that such a principle does not apply
to the perturbative stage of an event, where the hard interaction and
MPIs signal the transition from a state of small $\lambda$ (partons
confined in the incoming protons) to a state of significantly higher
$\lambda$. The principle of string-length minimization rather refers
to longer time scales, when strings begin to be pulled out between the
partons moving from the central collision.

Similarly, some general considerations of the space-time picture are necessary. One is that the spatial evolution of showers need
not be traced. That is, parton showers occur at time scales sufficiently
shorter than hadronization ones so that, to first approximation, all the
final partons can be viewed as emerging from a common vertex. Furthermore,
while the branching of a low-mass high-energy parton can be significantly
displaced, the daughters will tend to be sufficiently close, by any distance
measure, such that CR is unlikely to break them apart. Another issue is how the
lifetime of intermediate resonances compares with the CR time. The $\W$,
$\Z$, and $\t$  have intermediate decay time scales, about an order of
magnitude shorter than typical hadronization times. (Whereas the $\H$ is
much more long lived.) But the two would become more comparable if time
is added for the decay products to expand and begin interacting with the
environment given by the rest of the $\pp$ collision. Ideally, the
situation should therefore be simulated dynamically, where different time
orderings are possible outcomes, but that would be fraught with
uncertainties and is typically not done. Instead, a more common option
is to allow only early or only late resonance decays, \ie before or after
hadronization. In early decays, all partons can reconnect, while in late
decays the resonance decay products cannot.

\subsubsection{The MPI-based model}\index{Colour reconnections!MPI-based model}

The first CR model implemented in \pyt~8, and currently still the default,
attempts to reduce $\lambda$ by a complete merge of the partons of
separate MPI systems. The probability for two MPIs to be reconnected
this way is a function of the lower $\pT$ scale of the two, of the form
\begin{equation}
P_{\mathrm{rec}}(\pT) =
\frac{(R_{\mathrm{rec}}\, \pTo)^2}{(R_{\mathrm{rec}}\, \pTo)^2+\pTs} ~,
\label{eq:colRec:mpibased:recProb}
\end{equation}
where $\pTo$ is the parameter introduced in \cref{eq:soft:MPI:ptdamp}
to damp the $\pT \to 0$ infinity of the QCD $2 \to 2$ cross section,
and $R_{\mathrm{rec}}$ is a phenomenological parameter. An $R_{\mathrm{rec}}$
of order unity would seem reasonable; empirically somewhat larger values
are found. The reconnection probability is chosen to be higher for soft
systems, reflecting that the latter are described by more extended wave
functions, thus having a higher probability to overlap and interact with
other systems.

Now consider an event containing $n$ MPIs, which have been generated in 
order of falling $\pT$, $p_{\perp 1} > p_{\perp 2} > \ldots > p_{\perp n}$.
The reconnections are then done in a two-step procedure, as follows.

First, the MPI systems are tested for reconnection in sequence of 
increasing $\pT$, \ie starting with system $n$. 
For an arbitrary $m$, $2 \leq m \leq n$, the reconnection probability
$P_m = P_{\mathrm{rec}}(p_{\perp m})$ is used to decide whether system $m$
should be merged with $m-1$ or not. If not, the same relative probability
holds for a merger with $m - 2$, and so on to the top. That is, there is
no explicit dependence on the higher $\pT$ scale, but implicitly there is
via the survival probability of not already having been merged with a
lower-$\pT$ system. In total, the probability for $m$ not to merge
therefore is $(1 - P_m)^{m-1}$. Note that mergings may cascade:
if $m$ is merged with $l$, $1 < l < m$, then $l$ in its turn may be
merged with an even-higher-$\pT$ system $k$, $1 \leq k < l$, and then
also $m$ counts as merged with $k$.

Second, once it has been decided which systems should be reconnected, the
actual merging is carried out in the opposite direction. That is, first
the hardest system is studied, and all colour dipoles $(i,k)$ in it are
found, as usual in the $\Nc \to \infty$ limit. This includes those to the
beam remnants, as defined by the holes of the incoming partons. Then
consecutively, each softer system to be merged with it is considered in 
order of decreasing $\pT$. For each such system, the gluons $j$ are
inserted, in order of decreasing gluon $p_T$, into the dipole $(i,k)$
that minimizes the increase in the $\lambda$ measure for the harder
system
\begin{equation}
\Delta\lambda = \lambda_{j;ik} \equiv \lambda_{ij}+\lambda_{jk}-\lambda_{ik}
=\ln\frac{(p_i\cdot p_j)(p_j\cdot p_k)}{(p_i\cdot p_k) m_0^2} ~.
\end{equation}
Note that the first term of \cref{eq:colRec:intro:lambda} is not required
here, since an $E_k \to 0$ (for fixed relative angles) would affect all
$\lambda_{j;ik}$ the same way and thus not alter the choice of the winning
$(i,k)$ dipole. Although gluons dominate, MPIs may also contain quarks. 
Those $\q\qbar$ pairs that originate from the splitting of a gluon
can be inserted into the higher-$\pT$ system by the same criterion
as would have been used for such a gluon. The (few) other quarks 
are not affected by the CR procedure, but remain for the beam-remnant
handling to address.

The CR procedure is carried out before resonance decays are considered
by default, \ie the late decay option introduced above. It is possible
to switch to early decays, however.
  
\subsubsection{QCD-based colour reconnections}
\index{Colour reconnections!QCD-based model}
\label{sec:colRec:QCD-based}

As discussed in the introduction to \cref{sec:colRec},
during the perturbative stages of the event evolution, LC colour flow
is used to keep track of which partons are colour connected to each
other. In the LC limit, each colour tag is matched by only a single
unique anticolour tag in the event (or a combination of two colour
tags, if junctions are present). At the perturbative level, these 
connections represent LC dipoles/antennae, and they are one-to-one
mapped to string pieces at the non-perturbative stage, enforcing
colour confinement. 

Beyond the LC limit however, there should be a finite probability
also for LC-unconnected partons to ``accidentally'' find themselves in a
colour-singlet state, or in some other coherent state with a lower
total colour charge than the scalar sum of their individual
charges. This follows from the \su{2}~colour-algebra rules:
\begin{align}
  {\bf 3} \otimes \overline{{\bf 3}} & = {\bf 8} \oplus \textcolor{red}{\bf 1} \label{eq:33bar}\\
  {\bf 3} \otimes {\bf 3} & = {\bf 6} \oplus
  \textcolor{red}{\overline{\bf 3}}  \label{eq:33}\\
  {\bf 3} \otimes {\bf 8} & = {\bf 15} \oplus
  \textcolor{red}{\overline{\bf 6}} \oplus
  \textcolor{red}{{\bf 3}} \label{eq:38}\\
  {\bf 8} \otimes {\bf 8} & = {\bf 27} \oplus
  \textcolor{red}{{\bf 10}} \oplus
  \textcolor{red}{\overline{\bf 10}} \oplus
  \textcolor{red}{{\bf 8}} \oplus
  \textcolor{red}{{\bf 8}} \oplus
  \textcolor{red}{{\bf 1}}~, \label{eq:88}
\end{align}
where the representations that correspond to a coherent addition of
charges (with lower total charge) are highlighted in red. In the LC limit,
colour-unconnected quark-antiquark pairs are never allowed to form a
singlet; they are always in an overall octet state, while 
quark-quark, quark-gluon, and gluon-gluon ones are in sextet,
quindecuplet, and vigintiseptet states, respectively.

The starting point for the QCD-based CR
scheme~\cite{Christiansen:2015yqa} is that slightly simplified 
versions of \crefrange{eq:33bar}{eq:88} can be used to compute
probabilities for LC-unconnected partons to \emph{stochastically}
enter into coherent states with one another. This does not invalidate the LC
colour topology, but it does allow for (potentially many) \emph{other}
viable mappings of the same parton system to different string
configurations. Optionally, configurations that involve (re)connections
between systems with large relative boosts can be excluded if
deemed to be in conflict with causality, as discussed further below.
The model then chooses between the remaining allowed configurations by 
selecting the one that minimizes the $\lambda$ measure,
\cref{eq:colRec:intro:lambda}.
In principle, one could allow fluctuations around this, but that is
not currently done in the model.

A characteristic feature of this model is that it provides a
qualitatively new mechanism for the creation of baryon-antibaryon
pairs, in addition to the conventional mechanism of string breaks to
diquark-antidiquark pairs. 
The $\overline{\bf 3}$ in \cref{eq:33}, the $\overline{\bf 6}$ in
\cref{eq:38}, and the two decuplets in 
\cref{eq:88} represent colour states that involve colour-epsilon
structures. \index{Junctions}In the context of the string
model, these map to string junctions (and antijunctions), 
around which baryons will form,
\cf\cref{sec:junctions} and \citeone{Sjostrand:2002ip}.
As a consequence, the effective
\index{Baryons!Baryon-to-meson ratio}baryon-to-meson ratio increases with the
amount of CR in this model, 
and hence more active events (\eg with many MPI) will generally
exhibit higher baryon fractions. Note that  
colour conservation implies that the model always creates equal
numbers of baryons and antibaryons; these pairs can, however, be well
separated in phase space, contrary to the more localized nature of the
conventional diquark string breaks. Moreover, the model also allows for 
the formation of doubly-heavy-flavour baryons such as $\Xi_{bc}$,
\index{Baryons!Multiply-heavy-flavour}
a possibility that does not occur within the conventional diquark-type
string breaks. In the current formulation of the model,
however, no special attention has been devoted to questions specific to
heavy quarks, hence this aspect should be considered to be
associated with substantial uncertainty.
\index{Uncertainties!Hadron@in Hadronization}

At the technical level, the model approximates the QCD probabilities 
expressed by \crefrange{eq:33bar}{eq:88} by randomly
assigning an index between 0--8 to each Les-Houches colour tag,
subject to the
requirement that gluons must have different colour and anticolour
indices. Any parton pairs with matching colour and anticolour indices
are then considered to be in relative singlets and are candidates
for dipole-type string pieces. (This mimics the representation
first proposed in~\cite{Lonnblad:1995yk}.)   
Stochastically, this reproduces the $\frac19$
probability of \cref{eq:33bar} exactly.

The algorithm starts from the LC topology and
considers each index group in turn, working its way down from
high to low dipole invariant masses, at each step considering all
allowed possibilities and executing a swap if that lowers the total
$\lambda$ measure.
Note that $q\bar{q}$ pairs originating directly from $g\to q\bar{q}$
branchings are also excluded from having the same index. Consequently,
\index{Quarkonium}quarkonium
formation from such pairs is not expected 
in this model in its current formulation.

If junction-type reconnections are enabled, the algorithm then
works its way through three separate groups of indices:
 [0,3,6], [1,4,7], and [2,5,8] (chosen so that they are trivial to
 separate using the modulo 3 operation). 
Within each of these groups, any partons carrying two different
colour indices (say, 0 and 3) are allowed to add coherently to the overall
anticolour of the third (say, -6) and enter into corresponding
junction-type reconnections if that reduces the $\lambda$ measure.
This enables a decent ($\frac29$) approximation to the probability for
junction-type reconnections, but does underestimate the true QCD group weights
somewhat, see~\cite{Christiansen:2015yqa}. 
This procedure (dipole-style reconnections followed by junction
reconnections) is iterated until no more favourable
reconnections are identified. 

In addition to the colour rules, the dipoles also need to be
causally connected in order to perform a reconnection. 
The definition
of causally connected dipoles is not exact, and several different
options are available. 
All the time-dilation modes introduce a tunable
parameter, which provides a handle on the overall amount of CR.

When the two strings are allowed to reconnect, they will reconnect
if it lowers the total string length, as defined by an approximation to
the $\lambda$ measure. Several options for different approximations
are available. The $\lambda$ measure is not well understood, especially
for junction structures, and a tunable parameter allows for the enhancement or
suppression of junction-type connections to
dipole ones. This affects how many baryons are generated by the
model. See also the description of junction fragmentation in
\cref{sec:junctions}.

Although the main objective of the model is to treat reconnections involving
large invariant masses, there is of course a tail towards small
masses as well. For very low masses $< {\cal O}(1\,\mathrm{GeV})$,
string fragmentation becomes technically complicated (as each
hadron needs to straddle several gluon ``kinks''), especially when
junctions are involved, and also the approximations made in the $\lambda$
measure are not particularly reliable. Therefore, reconnections
involving string pieces with masses below $m_0$, \cf\cref{eq:colRec:intro:lambda},
are excluded from participating in the CR framework. (Technically,
partons making up such low-mass systems are treated collectively as a
single pseudo-particle for the purpose of reconnections.) 

\subsubsection{The gluon-move scheme}
\index{Colour reconnections!Gluon-move model}

In the effort to determine the top mass as accurately as possible, 
CR is one of the major sources of systematic error. To better understand
the situation, a range of new models were developed and implemented in
\citeone{Argyropoulos:2014zoa}. Many of these are crude straw-man models,
or applicable only to top decay. They are therefore not integrated as
standard options, but may be obtained by using the
\texttt{ColourReconnectionHooks.h} plugin; see \texttt{main29.cc}
for an example.

In the late resonance decays approach it is possible to allow separate
CR models for the underlying event and for the top decay products.
Then two collections of gluons are constructed, one containing the gluons
radiated from the top decay products and the other containing the gluons
from the rest of the event. Iterating over the former in random order,
one forces a random fraction of the gluon from the top to exchange colours
with a gluon from the rest of the event. The latter gluon can be picked
according to one of five different criteria,
\emph{(i)} at random,
\emph{(ii)} giving the smallest invariant mass,
\emph{(iii)} giving the largest invariant mass,
\emph{(iv)} giving the smallest (with sign) $\Delta\lambda$ value, or
\emph{(v)} as \emph{(iv)} but only if $\Delta\lambda < 0$.  

For early resonance decays, three possible operations were implemented,
swap, move, and flip. The latter two are implemented in the main body of
\pyt.

The swap model is similar to option \textit{(iv)} above. A random fraction
of all final-state gluons are chosen for possible reconnection. For each
such gluon pair $j$ and $m$, on dipoles $(i,k)$ and $(l,n)$ respectively,
one calculates the difference $\Delta\lambda$ resulting
from a swap of the two gluon colours
\begin{equation}
\Delta \lambda(j,m) = \lambda_{m;ik} + \lambda_{j;ln}
- \left(\lambda_{j;ik} + \lambda_{m;ln}\right) 
= \lambda_{im} + \lambda_{mk} + \lambda_{lj} + \lambda_{jn} -
\left( \lambda_{ij} + \lambda_{jk} + \lambda_{lm} + \lambda_{mn}  \right) ~.
\end{equation}
A reconnection is performed if
$\min_{j,m}\Delta\lambda(j,m)\leq\Delta\lambda_{\mathrm{cut}}$,
where $\Delta\lambda_{\mathrm{cut}}\leq 0$ is a tunable parameter that
expresses a CR strength. The procedure is repeated until no allowed swaps
remain.

The closely related move model works as follows. Again a random fraction
of all final-state gluons are singled out. Starting from each such gluon
$j$ on a final-state dipole $(i,k)$, the change in the string length
$\Delta\lambda$ that would result from moving the gluon to any
other final-state dipole $(l,n)$ is calculated using
\begin{equation}
\Delta\lambda(j, ik \to ln) = \lambda_{j;ln} - \lambda_{j;ik}=
\lambda_{lj} + \lambda_{jn} + \lambda_{ik} - 
\left( \lambda_{ij} + \lambda_{jk} + \lambda_{ln} \right) ~.
\end{equation}
Now the minimum is found
$\Delta\lambda_{\mathrm{min}} = \min_{j,l,n}\Delta\lambda(j, ik \to ln)$,
and the move carried out if
$\Delta\lambda_{\mathrm{min}} \leq \Delta\lambda_{\mathrm{cut}}$.
This is then repeated as long as the latter criterion is fulfilled.

There is some fine print. If a colour-singlet subsystem consists of
two gluons only, then it is not allowed to move any of them, since that
would result in a colour-singlet gluon. Also, at most as many moves
are made as there are gluons, which normally should be enough. A specific
gluon may be moved more than once, however. Finally, a gluon directly
connected to a junction cannot be moved, and also no gluon can be
inserted between it and the junction. This is entirely for practical
reasons, but should not be a problem, since junctions are rare in
this model. 

Neither the swap nor move methods reconnect quarks. That is, if a $\q\qbar$ pair
start out at the opposite ends of a string then so they will remain.
The gluons found along this string can change, and in the move model
even the number of such gluons, but the endpoints do not. To lift this
restriction, a flip step can be added subsequent to the swap or move one.
The basic idea here is to flip two string pieces, $(i,k)$ and $(l,n)$,
and instead connect them as $(i,n)$ and $(l,k)$. For any two separate colour-singlet subsystems one finds
\begin{equation}
\Delta\lambda_{\mathrm{min}} = \min_{i,k,l,n}
\left[ \lambda_{in} + \lambda_{lk}- (\lambda_{ik} + \lambda_{ln}) \right] ~.
\end{equation}
The system pair with smallest $\Delta\lambda_{\mathrm{min}}$ is selected for
a flip, as long as
$\Delta\lambda_{\mathrm{min}} \leq \Delta\lambda_{\mathrm{cut}}$.    
Singlet systems that have undergone one flip are not allowed any further
ones. As a minor variation, junction topologies are either excluded or
included among the allowed flip possibilities. It is also possible to
switch on/off move and flip separately.

\subsubsection{The SK models}\index{Colour reconnections!SK models}

The SK~I and SK~II models~\cite{Sjostrand:1993hi,Sjostrand:1993rb}
were specifically developed for
$\epem \to \Wp \Wm \to \q_1 \qbar_2 \q_3 \qbar_4$ at
LEP~2, and work (almost) equally well for an $\gamma^*/\Z \, \gamma^*/\Z$
intermediate state. They are not intended to handle hadronic collisions,
however, except in special contexts. The prime example is Higgs decays
of the same character as above, $\H \to \Wp \Wm / \Z \Z$, since the
Higgs is so long lived that its decay is decoupled from the rest of
the event~\cite{Christiansen:2015yca}. 

The labels I and II refer to the colour-confinement strings being
modelled either by analogy with type~I or type~II superconductors.
In the former model the strings are viewed as transversely extended
``bags''~\cite{Chodos:1974je}. The likelihood of reconnection is then
related to the integrated space-time overlap of string pieces from the
$\Wp$ with those from the $\Wm$. In the latter model, instead, strings
are assumed to be analogous with vortex lines, where all the topological
information is stored in a thin-core region. Reconnection, therefore, only
can occur when these cores pass through each other.

The imagined time sequence is the following. The $\Wp$ and $\Wm$ fly
apart from their common production vertex and decay at some distance.
Around each of these decay vertices, a perturbative parton shower
evolves from an original $\q\qbar$ pair. The typical distance that a
virtual parton (of mass $m \sim 10$~GeV, say, so that it can produce
a separate jet in the hadronic final state) travels before branching
is comparable with the average $\Wp\Wm$ separation, but shorter than the 
fragmentation time. Each $\W$ can therefore effectively be viewed as
instantaneously decaying into a string spanned between the partons,
from a quark end via a number of intermediate gluons to the antiquark
end. The strings expand, both transversely and longitudinally, at a
speed limited by that of light. They eventually fragment into hadrons
and disappear. Before that time, however, the string(s) from the $\Wp$
and the one(s) from the $\Wm$ may overlap. If so, there is some
probability for a colour reconnection to occur in the overlap region.

In scenario~I, the reconnection probability is proportional to the 
space-time volume over which the $\Wp$ and $\Wm$ strings overlap, with
saturation at unit probability. This probability is calculated as follows.
In the rest frame of a string piece expanding along the $\pm z$ direction,
the colour field strength is assumed to be given by
\begin{equation}
\Omega(\mathbf{x},t) = 
\exp \left\{ - (x^2 + y^2)/2r_{\mathrm{had}}^2 \right\}
\; \theta(t - |\mathbf{x}|) \;
\exp \left\{ - (t^2 - z^2)/\tau_{\mathrm{frag}}^2 \right\} ~,
\end{equation}
where $\mathbf{x} = (x, y, z)$.
The first factor gives a Gaussian falloff in the transverse 
directions, with a string width $r_{\mathbf{had}} \approx 0.5$~fm
of typical hadronic dimensions. The time retardation factor 
$\theta(t - |\mathbf{x}|)$ ensures that information on the decay of
the $\W$ spreads outwards with the speed of light. The last factor 
gives the probability that the string has not yet fragmented at 
a given proper time along the string axis, with 
$\tau_{\mathrm{frag}} \approx 1.5$~fm. For a string piece \eg 
from the $\Wp$ decay, this field strength has to be appropriately
rotated, boosted, and displaced to the $\W$ decay vertex. 
In addition, since the $\Wp$ string can be made up of many pieces, 
the string field strength $\Omega_{\mathrm{max}}^+(\mathbf{x},t)$ 
is defined as the maximum of all the contributing $\Omega^+$'s
in the relevant point. The probability for a reconnection to occur 
is now given by   
\begin{equation}
\mathcal{P}_{\mathrm{recon}} =  1 - \exp \left( - k_{\mathrm{I}} 
\int \d^3 \mathbf{x} \, \mathrm{d} t \; 
\Omega_{\mathrm{max}}^+(\mathbf{x},t) \, 
\Omega_{\mathrm{max}}^-(\mathbf{x},t) \right) ~,
\end{equation} 
where $k_{\mathrm{I}}$ is a free parameter. The integration cannot be
done analytically, but is approximated by Monte-Carlo methods.
Exponentiation has been applied to saturate the probability at unity.
If a reconnection occurs, however, the space-time point for this
reconnection is selected according to the differential probability 
$\Omega_{\mathrm{max}}^+(\mathbf{x},t) \, \Omega_{\mathrm{max}}^-(\mathbf{x},t)$
without any saturation. This defines the string pieces involved, 
and the new colour singlets are obtained by a flip as described above
(dipoles $(i,k) + (l,n) \to (i,n) + (l,k)$). 

In scenario~II it is assumed that reconnections can only take place
when the core regions of two string pieces cross each other. This means
that the transverse extent of strings can be neglected, which leads to
considerable  simplifications compared with the previous scenario.
The position of a string piece at time $t$ is described by a
one-parameter set $\mathbf{x}(t,\alpha)$, where $0 \leq \alpha \leq 1$
is used to denote the position along the string. To find whether two
string pieces $(i,k)$ and $(l,n)$ from the $\Wp$ and $\Wm$ decays cross,
it is sufficient to solve the equation system
$\mathbf{x}_{(i,k)}^+(t , \alpha^+) = \mathbf{x}_{(l,n)}^-(t , \alpha^-)$
and to check that this (unique) solution is in the physically allowed
domain. As an example, if there is no shower activity, so that the
event only consists of the two $\q_1\qbar_2$ and $\q_3\qbar_4$ strings,
it is easy to see that these are moving apart from each other already
from their creation and will never meet. A solution will nevertheless
be found, but with negative $t$ and possibly either or both of the
$\alpha^{\pm}$ outside their allowed range. Further, it is required
that neither string piece has had time to fragment, which gives
two extra suppression factors of the form
$\exp \{ - \tau^2/\tau_{\mathrm{frag}}^2 \}$, with $\tau$ the proper
lifetime of each string piece at the point of crossing, \ie as in
scenario~I. If there are several string crossings, only the one that
occurs first is retained. Reconnection is done with a flip, as in
scenario~I.

In models~I and II the string length is not tested, so it may increase.
The geometry of the process still tends to favour a reduced $\lambda$.
For the model variants I$'$ and II$'$, a reduced $\lambda$ is imposed
as an additional requirement on allowed reconnections. 

\subsubsection{Other CR models}

It is relevant to remember that many more CR models have been proposed,
and several implemented in past \pyt versions. Some of these could be
resuscitated using the existing colour-reconnection user hook, or an
expanded version thereof, should the need arise. 

\index{Colour annealing}\index{Colour reconnections!Colour annealing}In \pyt~6.4, several colour-annealing
scenarios were available 
\cite{Sandhoff:2005jh,Wicke:2008iz}, again primarily intended to be
useful for top-mass uncertainty studies in hadronic collisions. They
start from the assumption that, at hadronization time, no information
from the perturbative colour history of the event is relevant, so all
existing colour tags are erased. Instead, what determines how hadronizing
strings form between the partons is a minimization of the total potential
energy stored in these strings, as represented by the $\lambda$ measure.
The minimization is achieved by an iterative procedure, which
unfortunately can be quite time consuming. The scenarios differ by
details such as whether closed gluon loops are suppressed or not, 
or whether only free colour triplets are allowed to initiate string
pieces.

Also in \pyt~6.4, the GH model~\cite{Gustafson:1994cd} offered a simpler
option for $\Wp\Wm$ events, based on colour factors and string length
reduction, without any space-time picture.  

\index{Ariadne@\ariadne}\index{Dipole swing}\index{Colour reconnections!Dipole swing}In the \ariadne program
for $\epem$ and $\epm\p$, CR was introduced 
based on $\lambda$ minimization~\cite{Lonnblad:1995yk}, but CR could
occur after each new parton-shower emission, and thereby affect
the continued shower evolution. A similar idea is the dipole-swing
mechanism for the initial-state evolution of incoming hadrons~\cite{Avsar:2006jy,Bierlich:2019wld}. 

\index{Diffraction!Gap survival}\index{Diffraction!Ingelman--Schlein}\index{Ingelman--Schlein}\index{Pomeron}When rapidity gaps were found
in HERA DIS  
events, one early alternative 
to the Ingelman--Schlein pomeron picture~\cite{Ingelman:1984ns} was that the gaps were a consequence of CR~\cite{Buchmuller:1995qa,Edin:1995gi,Edin:1996mw,Pasechnik:2010cm}.
The Uppsala group has subsequently expanded this soft colour interactions
approach to encompass also hadronic events, for topics such as
diffraction and other rapidity gaps~\cite{Enberg:2001vq}, and
charmonium production~\cite{Edin:1997zb}. One important difference
relative to many of the models above is the frequent use of an
``area law''~\cite{Rathsman:1998tp} rather than the $\lambda$ measure.
The area that a string motion sweeps out is related to its $m^2$.
For a string consisting of several pieces, the total area is defined as
$A = \sum_{i = 0}^n m_{i, i+1}^2$, with masses calculated as in
\cref{eq:colRec:intro:lambda}. The probability of a reconnection is
then $P = R_0 \left[ 1 - \exp(-b \, \Delta A) \right]
=  R_0 \left[1 - \exp(-b (A_{\mathrm{old}} - A_{\mathrm{new}})) \right]$.
The $R_0 \approx 1/\Nc^2$ is an assumed colour-factor suppression,
and $b$ is the same as in \cref{eq:string:lsff}. Note that, had
$A$ been defined as the product of masses rather than a sum, then 
$\ln A$ would have been closely related to $\lambda$, and in particular
a $\Delta \lambda$ and a $\Delta A$ scan would find the same optimal
reconnection region, but that is not the case now. The related code
is available in some earlier \pyt versions. 

CR has also been studied in the context of other generators,
such as \textsc{Herwig}~\cite{Gieseke:2012ft,Gieseke:2018gff,Bellm:2019wrh}
and \textsc{Sherpa}~\cite{Khoze:2010by}. It is not possible to address CR
in equivalent terms for cluster as for string fragmentation, so there is 
no straight correspondence, but some basic ideas nevertheless are shared.

\subsection{String interactions and collective effects}
\index{Collective effects}
\index{String interactions}
\index{Hadronization!String interactions}
\index{Quark-gluon plasma|see{QGP}}\index{QGP}
\label{sec:string-interactions}

Heavy-ion collision experiments have for decades studied the possible creation of
a \ac{QGP} in high-energy collisions of heavy nuclei. Monte-Carlo simulations
of physics processes involving QGP creation, is mostly carried out in designated generators or
generator frameworks such as \jewel~\cite{Zapp:2013vla} or \jetscape~\cite{Cao:2017zih} (both
of which in fact use \pyt as a hard-process generator). Another approach is to segment 
individual events into ``core'' and ``corona'' parts~\cite{Werner:2007bf}, where the former
are treated as QGP, and the latter in vacuum. This is the case for EPOS-LHC~\cite{Pierog:2013ria},
which is an independent framework, as well as for other approaches built on top of \pyt ~\cite{Kanakubo:2019ogh}.

\pyt has thus, historically, played on a different field than generators focused on the special
observables obtained in heavy-ion collisions. Instead, \pyt is often used as a generator supplying an initial state, with a focus 
on the hard process, parton shower, and hadronization as in lepton collisions, with no QGP produced or assumed.
This clear division of tasks was questioned by data from LHC. First in 2010, with the discovery of long-range
azimuthal correlations of final-state hadrons in high multiplicity \pp collisions, referred to as ``the near-side ridge''~\cite{Khachatryan:2010gv}, and later by observations of enhanced production of strange and multi-strange
final-state hadrons, incompatible with model fits to LEP data ~\cite{Khachatryan:2011tm,Aaij:2012ut,Aamodt:2011zza,Abelev:2012jp}. The latter culminated in the observation
that not only is the observed strangeness production incompatible with model fits to LEP data, strange/non-strange
ratios also increase with multiplicity, and the increase smoothly connects \pp with \pA and \AA collision systems~\cite{ALICE:2017jyt}. This clearly meant that \pyt could no longer assume that effects traditionally ascribed
to QGP formation are only present in heavy-ion collisions. While CR models can account qualitatively for some of the
observed effects~\cite{Ortiz:2013yxa,Bierlich:2015rha}, they are wholly unsuitable in others~\cite{Bierlich:2018lbp}. 
Instead of introducing QGP formation into \pyt, as
the approaches cited above in some sense have already done, the route taken is to expand the Lund string model to
its furthest consequence, by allowing interactions between strings in densely populated regions of space. Whether
interactions between strings are indeed responsible for all collective effects observed in \pp and heavy-ion
collisions, is still unknown. The models introduced here should thus clearly be understood as one possibility
among several others, however unified by the underlying assumption that QGP is not formed. Furthermore, they
are all work in progress at the time of writing, and subject to change. There is no clear demarcation between
what constitutes a model of colour reconnection, as introduced above, and models of string interactions. In 
this manual we have drawn the line between models operating in momentum space (the colour reconnection models)
and models operating in real space.

\subsubsection{String shoving}
\index{Hadronization!Close-packing effects}\index{String shoving}
\label{sec:shoving}
In the original formulation of the Lund string model, strings are treated as massless relativistic strings, which presupposes 
that strings have no transverse extensions. In collisions with many strings occupying the available spatial volume, this
approximation breaks down, and strings are allowed to interact with mainly repulsive forces. The realization of this picture 
is denoted the ``string shoving model''. While similar ideas were explored analytically already in 1988~\cite{Abramovsky:1988zh},
the modern version of the string shoving model is formulated to take into account input from lattice QCD, and is based
more firmly on the correspondence with a superconductor. This model is rather new at the time of writing~\cite{Bierlich:2018xfw} and is still being extended with further consequences being explored~\cite{Bierlich:2019ixq,Bierlich:2020naj}.
The model contains three basic physics components: 1) the string shape, 2) the string transverse width, and 3) the interaction force
between two strings.

The transverse shape of the colour-electric field of the flux tube (the string shape) is determined with input from lattice 
QCD~\cite{Cea:2014uja}, and can be well described by a Gaussian:

\begin{equation}
	\label{eq:string-e-field}
	E(\rho) = N\exp\left(-\frac{\rho^2}{2R}\right) ~, 
\end{equation}
where $N$ is a normalization factor, $\rho$ is the radial coordinate, and $R$ is the string equilibrium radius. The normalization
constant is determined by assuming that the field energy per unit length $\int \mathrm{d}^2\rho E^2/2$ is a constant fraction ($g$)
of the string tension. This gives $N^2 = 2g\kappa/(\pi R^2)$.
The strings expand from their time of formation with infinitesimal width, until they either attain the maximum width $R$, or until the string's 
fragmentation proper time, $\tau_{\mathrm{had}}$, has been reached. 
While the equilibrium width of a string can be argued either by lattice considerations or from models, the number is associated with such
large uncertainty, that it is in practice kept as a free parameter of the model, with reasonable values between around 0.5 and 1.5~fm. 
The same holds for the parameter $g$.
The string repulsion force can then be calculated from the energy of the colour-electric field of two overlapping, parallel strings
$\int \mathrm{d}^2 \rho (\vec{E}_1 + \vec{E}_2)^2/2$. If strings are separated from each other by the transverse distance $d_\perp$,
the interaction energy becomes $2g\kappa \exp(-d^2_\perp/(4R^2))$, which gives the interaction force per unit length:

\begin{equation}
	\label{eq:gaussian-profile}
	f(d_\perp) = \frac{g \kappa d_\perp}{R^2} \exp\left(-\frac{d_\perp^2}{4R^2} \right) ~.
\end{equation}

The above treatment leading to \cref{eq:gaussian-profile}, is made in terms of Abelian fields. As such, anti-parallel strings would
attract each other rather than repel. In a non-Abelian theory like QCD, the picture is more complex, leading to repulsion being the 
dominant mechanism. As an example, consider the case of oppositely oriented triplet fields. One obtains an octet field with probability 8/9, 
which still leads to a repulsion, and a singlet field with probability 1/9, leading to attraction. Since singlets correspond to the total
attenuation of fields, it can further be assumed that singlets are already handled by colour-reconnection mechanisms~\cite{Bierlich:2020naj,Bierlich:2020naj}.

The technical implementation is concerned with two further questions, namely calculating $d_\perp$ for two given string pieces, and distributing
the resulting pushes in the event. For the former, a suitable Lorentz frame is defined, where a string pair always lies in parallel planes, called 
the parallel frame~\cite{Bierlich:2020naj}. One can then boost a pair of strings from the lab frame to the parallel frame, where the string topology is 
specified with an opening angle between the two partons in the string ends and a skewness angle between the two strings --- both of which are constrained by 
momentum-energy conservation. The angles can be expressed in terms of pseudorapidity and invariant masses $s_{ij}$ for the string formed by partons $i$ and $j$: 

\begin{equation}
	\label{eq:string-angles}
	\cosh (\eta) = \frac{s_{14}}{4p_{\perp1} p_{\perp4}} + \frac{s_{13}}{4p_{\perp1} p_{\perp3}},\mrm {~and~}  \cos (\phi) = \frac{s_{14}}{4p_{\perp1} p_{\perp4}} -  \frac{s_{13}}{4p_{\perp1} p_{\perp3}} ~.
\end{equation}
Furthermore, the strings now evolve and interact in the proper time in the parallel frame. Calculating this interaction for every possible string 
pair is, among other aspects, a computational challenge, and to curb the possibility of running into being an extreme computing resource-consuming program,
we for now neglect end-string effects which, for example, have been studied in \citeone{Sjostrand:1984ic}.
 
The shoving force is distributed to the outgoing hadrons formed after string fragmentation, taking into account that the total applied push is 
a result of a time evolution. The integrated push $\Delta$ p$_\perp$ is:

\begin{equation}
	\label{eq:shoving-pT}
	\Delta p_\perp = \int \mathrm{d}t \int \mathrm{d}z f(d_\perp (t)) ~,
\end{equation}
where the integration limits in $z$ are time dependent. Since the time ordering of pushes is important, $\Delta p_\perp$ is split up into
several (fixed) small pushes $\delta p_\perp$, according to a probability distribution $P(t)$. The total push is then:

\begin{equation}
	\Delta p_\perp = \int \mathrm{d}t P(t) \delta p_\perp \text{,~~~~~with~~~~~} P(t) = \frac{1}{\delta p_\perp} \int \mathrm{d}z f(d_\perp) ~,
\end{equation}
when $\delta p_\perp$ is small. The pushes can then be ordered in time (in the parallel frame) using the veto algorithm. The resulting
procedure corresponds to a time evolution with dynamical time stepping, where steps are large when pushes are small and \textit{vice versa}.

\begin{figure}[h!]
	\begin{center}
		\includegraphics[width=0.7\textwidth,trim={0.4cm 1.5cm 0.4cm 1.5cm},clip]{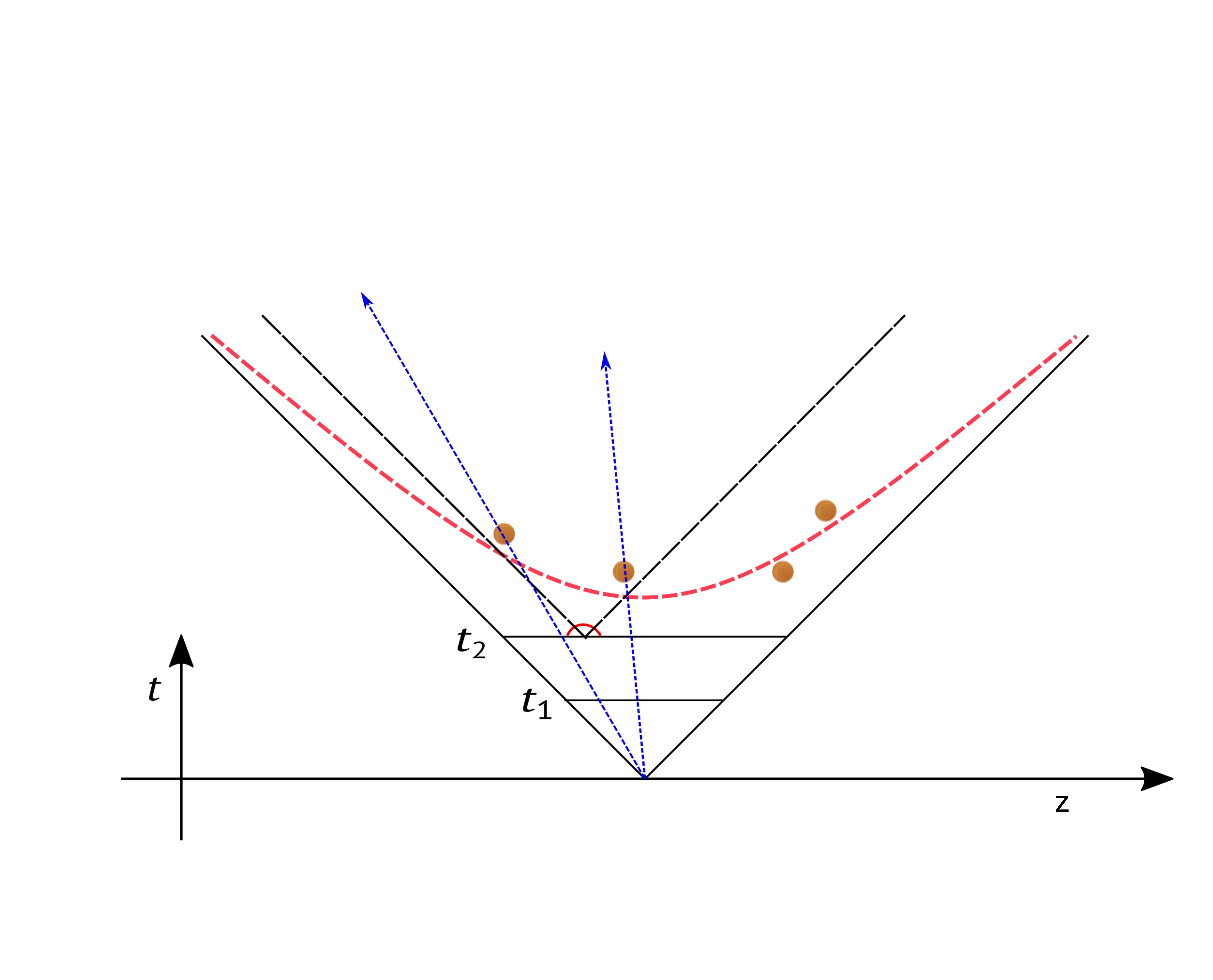}
		\caption{\label{fig:dist-of-pushes}Space-time diagram of a Lund string showing the trajectory of hadrons when they receive their share of $\Delta$p$_\perp$ resulting from string shoving interactions. The blue lines show the initial pseudorapidity lines for the hadrons formed, the red line implies a $\delta$p$_\perp$ generated from shoving, and the red dashed line shows the $\tau_{had}$.}
	\end{center}
\end{figure}

In $t,z$ space, this would look like hadrons flying out along the direction of their original pseudorapidity, even after the pushes are applied, 
spreading out in a light-cone that extends in such a way that it encloses all the hadrons which 
receive a share of this generated $\Delta p_\perp$. This distribution of pushes is performed as shown in \cref{fig:dist-of-pushes}.  
  
\subsubsection{Rope hadronization}
\index{Hadronization!Ropes}\index{Ropes}\index{Colour ropes|see{Ropes}}
\label{sec:ropes}
A simple string drawn between a quark and an antiquark is an $\mathbf{SU(3)}$ triplet (or anti-triplet depending on 
direction of colour flow). When several strings overlap with each other at hadronization time, the rope-hadronization model posits that end-point colour charges will act together coherently to form a stronger
field --- a rope. This possibility was noted in the classic paper by Biro, Nielsen and Knoll from
1984~\cite{Biro:1984cf}.

The new, stronger field is an \su{3} multiplet. According to lattice calculations~\cite{Bali:2000un},
the energy density (and thus the string tension) scales as the second Casimir operator ($C_2$) of the rope multiplet.
When a rope is formed by ordinary triplet and anti-triplet strings, the net colour charge is obtained from the
addition of random coloured triplets and anti-triplets~\cite{Biro:1984cf,Jeon:2004rk,Bierlich:2014xba}. A 
resulting multiplet is uniquely characterized by two quantum numbers $p$ and $q$, with a specific state corresponding
to $p$ coherent triplets and $q$ coherent anti-triplets (a normal triplet string is thus characterized as $\{p,q\} = \{1,0\}$).
The multiplicity of a multiplet is given by:
\begin{equation}
\label{eq:rope-multiplicity}
	2N = (p+1)(q+1)(p+q+2) ~.
\end{equation}
This allows for an iterative addition of multiplets. Starting from a given multiplet $\{p,q\}$, adding a triplet gives
the three possible multiplets~\cite{Bierlich:2014xba}:
\begin{equation}
	\{p+1,q\}, \{p-1,q-1\} \mrm{,~~and~~}\{p,q-1\} ~,
\end{equation}
with weights given by \cref{eq:rope-multiplicity}. The anti-triplet case is given directly from symmetry. Once it
is established which triplets and anti-triplets are overlapping in an event, a random walk procedure can be carried
out to find $p$ and $q$ for the rope multiplet. Since the energy density of the rope is proportional to $C_2$, the 
relative tension of the multiplet to the triplet can be calculated directly as:
\begin{equation}
	\label{eq:rope-tension}
	\frac{C_2(\{p,q\})}{C_2\{1,0\}} = \frac{1}{4}(p^2 + pq + q^2 + 3p + 3q) ~.
\end{equation}
When the rope breaks up, it does so in a step-wise manner, one string at a time. By considering the change in available field energy in the transition
$\{p,q\} \rightarrow \{p-1,q\}$, neglecting the contribution from the vacuum pressure to the total energy, the energy available in a single string breaking 
becomes the \emph{effective string tension} $\tilde{\kappa}$:
\begin{equation}
	\tilde{\kappa} = (2p - 1)\kappa ~.
\end{equation}

While the string tension $\kappa$ does not enter explicitly\footnote{The string tension does enter explicitly into the vertex positions in \cref{sec:hadron-vertices}, but the effect of rope formation has so far not been taken into account for hadron vertices.} into the \pyt implementation of string hadronization, it does enter implicitly
through the parameters of \cref{eq:string:lsff,eq:q-mass-and-pt}. From the implicit dependence on $\kappa$, transformation
rules for all parameters can be defined, given the assumption that the \pyt default values of all parameters correspond to 
$\tilde{\kappa} = \kappa$, as they are tuned to LEP data~\cite{Skands:2014pea} where there are no overlapping strings, and thus $p = 1$
and $q = 0$. The most important affected parameters~\cite{Bierlich:2014xba}, are: those involved in suppression of strangeness ($\rho$), diquark production ($\xi$),
diquark with strange-quark content relative to diquarks without strange quarks ($x$), the suppression of spin-1 diquarks relative to
spin-0 diquarks, and the width of the transverse momentum distribution in string breakings ($\sigma$). Letting $h = \tilde{\kappa}/\kappa$, the transformation rules
for $\rho$, $x$ and $y$ are similar:
\begin{equation}
	\rho \mapsto \tilde{\rho} = \rho^{1/h}\mrm{,~} x \mapsto \tilde{x} = x^{1/h}\mrm{,~~and~~} y \mapsto \tilde{y} = y^{1/h} ~,
\end{equation}

while $\sigma \mapsto \tilde{\sigma} = \sigma^{1/h}$. The $\xi$ parameter is more complicated, and transforms like:
\begin{equation}
	\xi \mapsto \tilde{\xi} = \tilde{\alpha}\beta \left(\frac{\xi}{\alpha\beta}\right)^{1/h} ~,
\end{equation}
with $\alpha$ depending on all the above parameters, and $\beta$ a free parameter.

\subsubsection{The thermal model}
\index{Thermal string breaks}\index{Hadronization!Thermal string breaks}\index{String breaks}

The thermal model~\cite{Fischer:2016zzs}, available as a non-default
option, can partly be viewed as an alternative to the rope model, sharing
similar objectives. Not all details have been fully developed, so its
main purpose is for exploration. One motivation for it is that hadronic
$\pT$ spectra in low-energy collisions are reasonably well described by
an exponential fit
\begin{equation}
	\frac{\d\sigma}{\d^2\pT} = N \exp(-m_{\perp\mathrm{had}}/T) ~~\mathrm{with}~~
	m_{\perp\mathrm{had}} = \sqrt{m_{\mathrm{had}}^2 + \pTs} ~,
\end{equation}
where $N$ and $T$ are (approximately) common for all hadron types.
Another motivation is that local quantum-mechanical fluctuations in
the string transverse profile translate into a fluctuating string
tension $\kappa$, which can broaden the Gaussian $\pT$ into an
exponential-like form~\cite{Bialas:1999zg}. (Compare with fluctuations
in the proton size, which are commonly advocated and used \eg in the
\Angantyr modelling of cross sections~\cite{Bierlich:2018xfw}.)
While traditionally $T$ is associated with a temperature, in such
a scenario it would rather be derived from $\kappa$.

The thermal model is implemented as follows. In each string break the
$\q$ and $\qbar$ receive opposite and compensating $\pT$ values,
chosen such that the $\pT$ sum of two
adjacent string breaks precisely gives an $\exp(-\pT/T)$ spectrum.
Starting from a known flavour in one string break, the next flavour
and the resulting intermediate hadron is chosen among all possibilities
according to a relative weight $\exp(-m_{\perp\mathrm{had}}/T)$. Assuming
the production of two hadrons with different masses $m_1$ and $m_2$,
this approach then implies the same production rate for
$\pT \gg m_1,m_2$, but more suppression of the heavier hadron at
low $\pT$. Thus, there is less production of heavier states, but they
come with a larger $\langle \pT \rangle$, which is as intended.

\index{Baryons!Thermal@in Thermal model}There is some fine print, like that each particle should be weighted
by the number of spin states, that flavour-diagonal mesons can mix,
that baryons need $\mathbf{SU(6)}$ symmetry factors, that baryons
receive a free overall normalization factor with respect to mesons,
and so on. The number of flavour-related free parameters still is
significantly reduced relative to the ordinary string fragmentation.

Overall the particle composition comes out reasonably well, with some
excess of the heavier baryons. This is in contrast to the normal
string fragmentation, where it is difficult to produce enough of these
particles. The larger $\langle \pT \rangle$ for heavier particles
also improves agreement with  $\pp$ data, but resonance decays act to
dilute the effects, so $\langle \pT \rangle(m_{\mathrm{had}})$ still
does not rise quite fast enough.

\index{Strangeness enhancement}\index{Close-packing of
  strings}\index{Hadronization!Close-packing effects}
Another issue is what happens when several strings are close packed.
In the rope model, this leads to a higher $\kappa$, in quantised
steps. An alternative is to assume a continuously increasing $\kappa$
as each string is squeezed into a decreasing effective area.
Such an option is implemented as part of the thermal model, but can also
be applied to the default Gaussian one. In this approach, the $T$ or
$\kappa$ parameter is rescaled by a power of the effective number of
strings in the neighbourhood of a new hadron. Therefore a trial average
step along the string is made before a new hadron is produced, giving
a likely hadron rapidity and $\pT$. Then, one may count the number of
strings crossing that rapidity, as a simple measure of string density.
A smooth damping is applied for particles produced at larger $\pT$,
which are likely to be produced in minijets sticking out from the
denser-populated central region. The close-packing enhancement can be
used \eg to increase strangeness production in high-multiplicity $\pp$
events, similar to the rope model, but it has not been as extensively
compared with data. 

\subsection{Hadronic rescattering}
\index{Collective effects}\index{Hadronization!Hadronic rescattering}\index{Hadronic Rescattering}
\label{subsection:hadronicrescattering}
After hadrons have been produced, outgoing hadrons can interact in secondary
collisions. This rescattering can be relevant when studying collective
effects, but can lead to a significant slowdown of \pyt, and is therefore not
on by default. It is enabled by setting \settingval{HadronLevel:Rescatter}{on}.
Here, we will outline the rescattering algorithm, then 
summarize some notable effects of rescattering of which the average user should
be aware. A more detailed discussion of the rescattering framework is
given in \citeone{Sjostrand:2020gyg} in the context of \pp collisions, while
\citeone{Bierlich:2021poz} discusses physics results for \pA and \AA collisions.

There are two aspects to the rescattering algorithm:
first, describing how two hadrons interact with each other in their rest
frame; and second, describing the evolution of the event as a whole.

Consider two hadrons in their rest frame, with CM energy $\sqrt{s}$ and impact
parameter $b$. We assume that the probability of an interaction occurring is
a function of $b$ and the total cross section $\sigma_{\mathrm{tot}}$. The cross
section generally depends on $\sqrt{s}$ and the specific hadron species, as
described in \cref{subsubsection:lowenergyprocesses}. There is no solid
theory for how $P$ depends of $b$, so two different models are implemented
in \pythia. The default is a Gaussian dependency,
\begin{equation}
P(b) = P_0 e^{-b^2/b_0^2} ~,
\label{eq:bGaussianEdge}
\end{equation}
where $P_0$ is referred to as the opacity, a free parameter that is 0.9 by 
default, and the characteristic length scale is
\begin{equation}
b_0 = \sqrt{\frac{\sigma_\mathrm{tot}}{P_0 \pi}} ~.
\end{equation}
The alternative model is a disk model,
\begin{equation}
P(b) = P_0 \Theta(b - b_0) ~,
\end{equation}
where $\Theta$ is the Heaviside step function. For $P_0 = 1$, this corresponds
to the black-disk model used by most existing hadronic rescattering frameworks.
The two models are normalized such that if $b$ is chosen uniformly on a disk
with radius much larger than $b_0$, then both models will give the same
interaction probability. In practice, rescattering is more likely in dense
regions where $b$ tends to be biased towards lower values, so the narrower
distributions like the black disk will lead to more rescattering activity.
If it is determined that the hadrons should interact, the interaction time is 
defined as the time of closest approach in their rest frame.

The algorithm for performing rescattering for the whole event proceeds as follows:
\begin{enumerate}
\item Start with an event right after hadronization.
\item For each hadron pair, test whether they could interact, using the
probability $P$ defined above.
\item If a pair could potentially interact, record the interaction time for
that pair in a time-ordered list.\label{step:record}
\item Choose the earliest interaction in the list where participants have
not already interacted, and simulate the collision. Which process to
simulate is chosen with probabilities proportional to the partial cross 
sections for each process.
\item Check whether the newly produced hadrons can interact with existing ones,
and if so, add the interaction times for those pairs to that list. 
\item Continue picking interactions from the list until there are no more 
potential rescatterings.
\end{enumerate}
Short-lived hadrons can also decay during the rescattering phase. To model
this, the decay times of those hadrons are recorded in the list together with 
rescattering interaction times, and the decay occurs when it is chosen in 
step~\ref{step:record} above, if it has not already rescattered.

Enabling rescattering has a few consequences for the shape of events.
First, rescattering increases charged multiplicity, since
only processes with two incoming hadrons are allowed,
but inelastic processes can produce more than two outgoing ones.
For \pyt~8.307 with default parameters and \pp at 13 TeV, this can be compensated
by setting \settingval{MultipartonInteraction:pT0Ref}{2.345}. For beams
such as \pPb and \PbPb, other values might restore the multiplicity, but a
more thorough retune is necessary in order to simultaneously obtain the correct
multiplicity in all three cases. In such a retune, it would also be relevant
to include other effects such as ropes (\cref{sec:ropes}) and
shoving (\cref{sec:shoving}). For now, the user is advised to assume that
rescattering will change the charged multiplicity.

\index{Baryons!Rescattering}Similarly, hadron composition will change. Baryon number in particular is
reduced in rescattering through annihilation processes. For example, the
process $\ppbar \to \pip\pim\piz$ is possible, but not the reverse. Another
way the composition changes is through resonance production, \eg 
$\pion\kaon \to \kaon^*$, but be aware that this resonance production is not
easily detectable in experiment; for a resonance production followed by
a decay, $\pion\kaon \to \kaon^* \to \pion\kaon$, the invariant mass of the
outgoing system is the same as for the incoming one. In other words, this
process produces a $\kaon^*$ that is visible in the event record, without
necessarily changing the observable $\pi\kaon$ mass spectrum.

A particular consequence of the increased multiplicity is that each hadron
will on average have lower $\pT$, which could affect \eg spectra that are
sensitive to $\pT$ cuts. At the same time, the mean $\pT$ for particular
hadron species may increase. This is the case for example with protons,
which will move slower than pions with similar $\pT$, and will therefore
receive a push from behind. This phenomenon is referred to as the ``pion wind''.

Rescattering has been shown to give rise to some collective effects, in
particular azimuthal flow in \PbPb collisions. \angantyr with rescattering
provides a good description of elliptic flow coefficients at large
multiplicities, and a more modest contribution at low multiplicities. It can
also lead to some jet modifications, but with the aforementioned $\pT$ shift,
it is not clear how to interpret these modifications.
See \citeone{Bierlich:2021poz} for further discussion.

\subsection{Bose--Einstein effects}\index{Hadronization!Bose--Einstein effects}
\index{Bose--Einstein effects|see{Hadronization}}
\index{Hanbury-Brown--Twiss|see{Bose--Einstein effects}}
\index{HBT|see{Bose--Einstein effects}}
\label{sec:bose-einstein}
Ideally, coloured partons could be formed into colourless final-state
hadrons using amplitude-based quantum mechanics, but because these
transitions are non-perturbative, phenomenological models of
hadronization are employed. Due to the probabilistic nature of these
hadronization models, coherence in final-state particles cannot be
directly described. A classic example of such final state coherence is
Bose--Einstein effects, where correlations arise between identical
bosons in an event from the symmetrization of the production
amplitude. While these correlations are expected to have a negligible
impact for most measurements in \pp collisions, Bose--Einstein
effects\footnote{Within the heavy-ion and astrophysics communities
  these effects are oftentimes discussed in the context of
  Hanbury-Brown--Twiss interferometry~\cite{HanburyBrown:1956bqd}.}
have been observed in minimum bias \pp and \ppbar
data~\cite{Neumeister:1991bq, Aad:2015sja, Khachatryan:2011hi,
  Aaij:2017oqu}, as well as \epem data~\cite{Acton:1991xb,
  Acton:1992kc, Decamp:1991md, Abreu:1994we}. Additionally, some
precision measurements such as $W$-mass determination using hadronic
final states may be sensitive to Bose--Einstein
effects~\cite{Lonnblad:1995mr}.

Assuming a geometric picture with a Gaussian distribution of
production vertices in space-time, two-particle correlations of
identical bosons are enhanced by a unitless factor of,
\begin{equation}
  f_2(Q) = 1 + \lambda e^{-Q^2 Q_0^{-2}} ~,
  \label{equ:f2}
\end{equation}
with respect to a final state with no coherence
effects~\cite{Gyulassy:1979yi}. Here, $Q^2$ is $(p_i - p_j)^2$ where
$p_i$ and $p_j$ are the four-momenta of the two particles, $\lambda$
is the incoherence parameter, and $Q_0$ is a reference $Q$ related to
the radius of the particle source as $r \equiv \hbar/Q_0$. The
incoherence parameter is limited between $0$ where there is no effect,
and $1$ with a maximal effect.

For a high multiplicity event with multiple two-particle correlations,
the event weight can be naively approximated as the product of
$f_2(Q)$ for each particle pair. Note, this is a slight overestimate
of the event weight for most event topologies. These event weights
cannot modify the overall normalization, as this would increase the
cross section for the final state. If the weights are normalized to
unity, the total cross section is not modified, but the multiplicity
distribution will be shifted to higher multiplicities. Neither of
these behaviours is desirable, as both cross sections and multiplicity
distributions are already well described without Bose--Einstein
effects.

Instead, in \pyt Bose--Einstein effects are introduced by shifting the
momenta inside particle pairs. Assuming the distribution of $Q$ for
particle pairs is given by flat phase space, then solving,
\begin{equation}
  \int_0^Q \d{q}\, \frac{q^2}{\sqrt{q^2 + 4m^2}} =
    \int_0^{Q'} \d{q}\, f_2(q)\frac{q^2}{\sqrt{q^2 + 4m^2}} ~,
\end{equation}
for $Q'$ determines the new $Q$ value needed to produce an enhancement
of $f_2(Q)$ for that particle pair with individual particle mass
$m$. The three-momentum for the two particles can then be shifted by,
\begin{equation}
  \Delta \vec{p}_{i,j} = c(\vec{p}_i - \vec{p}_j) ~,
\end{equation}
where $\vec{p}_i^{\,\prime} = \vec{p}_i + \Delta \vec{p}_{i,j}$
$\vec{p}_j^{\,\prime} = \vec{p}_j - \Delta \vec{p}_{i,j}$, and the
constant coefficient $c$ is determined from setting $Q^{\prime2} =
(p_i^\prime - p_j^\prime)^2$. Because events may have more than one
particle pair, the total shift for a given particle is then,
\begin{equation}
  \vec{p}_i^{\,\prime} = \vec{p}_i + \sum_{j \neq i} \Delta \vec{p}_{i,j} ~,
\end{equation}
where the sign of $\Delta \vec{p}_{i,j}$ is such that the
three-momenta of the event is conserved.

Effectively, this shifting of momentum corresponds to pulling particle
pairs closer together, and while three-momenta is conserved throughout
this process, energy conservation is violated and the total energy of
the event is reduced. The form of $f_2(Q)$ from \cref{equ:f2} arises
from integrating the pair symmetrization term $1 + \cos(\Delta x \cdot
\Delta p)$ over a Gaussian distribution of production vertices in
space-time. Consequently, any source distribution other than a
Gaussian will result in an oscillatory behaviour of $f_2(Q)$, with
alternating values of $f_2(Q) > 1$, where $Q' < Q$ results in the pair
pulled together, and $f_2(Q) < 1$, where $Q' > Q$ results in the pair
separated apart. With the appropriate damping of this oscillatory
behaviour, for a given particle configuration, a form of $f_2(Q)$ can
be found where both conservation of three-momenta and energy is
achieved. Some pairs at low $Q$ are pulled together, reducing the net
energy, while other pairs at middle $Q$ are separated apart,
increasing the net energy.

To achieve this behaviour, a form of $f_2(Q)$ is selected to have one
oscillation before damping. The ansatz of the $\mathrm{BE}_{32}$
algorithm~\cite{Lonnblad:1997kk},
\begin{equation}
  f_2(Q) = \left[1 + \lambda e^{-Q^2 Q_0^{-2}} \right] \left[1 +
    \alpha \lambda e^{-Q^2 Q_0^{-2}/9} \left(1 - e^{-Q^2
      Q_0^{-2}/4}\right)\right]
\end{equation}
is chosen for $\alpha < 0$, where the new second factor effectively
models the initial minimum of the oscillation as a smeared
Gaussian~\cite{Sjostrand:2006za}. This form does not have any deep
physical meaning, but provides the necessary first oscillation while
maintaining the initial Gaussian distribution form, and has the
limiting behaviour of $f_2(0) = 1 + \lambda$. The factor $\alpha$ is
iteratively determined per event after calculating all relevant
$p_i'$, such that energy is still conserved even after three-momentum
shifting is performed for each relevant particle. Consequently, at
least two particle pairs must be present for Bose--Einstein effects to
be introduced.

Bose--Einstein correlations are performed after hadronization but prior
to particle decays, and are not included by default. Effects may be
switched on or off for different particle groupings: pions with \piz,
\pip, and \pim pairs; kaons with \kshort, \klong, \kp, and \km pairs;
and eta mesons with $\eta$ and $\eta'$ pairs. Many of these particle
species are produced not only from primary hadronization, but also
from the decay of short-lived particles. Consequently, a configurable
minimum decay width can be set so that any particles with a larger
width are decayed prior to the application of Bose--Einstein
effects. The default minimum decay width is set at $0.02~\GeV$ so that
both \rhomeson and $\gp{K}^*$ mesons are decayed before correlations
are introduced. Both the shifted and unshifted versions of particles
are kept in the event record for bookkeeping purposes. All shifted
particles are assigned a status of $\mathtt{99}$ and are set as the
children of their unshifted entry.

\subsection{Deuteron production}\index{Hadronization!Deuteron production}
\index{Deuteron production|see{Hadronization}}

The deuteron, \D, is a bound proton and neutron state, which, similar
to Bose--Einstein effects (see \cref{sec:bose-einstein}), must be formed after
hadronization. Understanding deuteron production in the context of
collider-based experiments can help efforts in modelling nuclei
formation and reduce prediction uncertainties when searching for
possible dark-matter induced excesses in cosmic ray flux
ratios~\cite{Donato:1999gy}. In heavy-ion physics, formation of loosely
bound systems are often used to determine the chemical freeze-out
temperature in statistical hadronization models~\cite{Andronic:2010qu}.
In \pythia, two deuteron formation models are
available, the coalescence model~\cite{Schwarzschild:1963zz, Kapusta:1980zz} and the 
more sophisticated Dal--Raklev model~\cite{Dal:2015sha}. Both models are implemented
through the same configurable framework, with the Dal--Raklev model set as the default
configuration. All deuteron production is switched off unless
explicitly requested by the user. Note that while the discussion here
is for the deuteron, anti-deuteron production is also performed
following the exact same method, but with all particles swapped to
antiparticles.

\index{Coalescence}\index{Hadronization!Coalescence}In the coalescence
model, all 
possible~\p~and~\n~pair combinations are 
determined. For each pair the magnitude of their three-momenta
difference,
\begin{equation}
  k = \sqrt{(\vec{p}_i - \vec{p}_j)^2}
\end{equation}
is calculated in the rest frame of the pair, where $\vec{p}_i$ and
$\vec{p}_j$ are the two three-momenta of the pair. If $k$ is less than
some cutoff value $c_0$, the pair is bound into a deuteron,
otherwise the nucleons remain unmodified. Spatial separation, in
addition to momentum separation, could also be considered, although
this has not been implemented in any of the models described here. The
ordering of testing pairs for binding is randomized, and after each
successful binding, any remaining pairs containing one of the bound
nucleons are no longer considered for binding. For the coalescence
model, this implies that the binding cross section is flat as a
function of $k$. If there are two unique pairs each with $k < c_0$,
both pairs have an equal probability of being bound, even if one $k$
is smaller than the other.

After a nucleon pair is selected for binding, a deuteron is formed. In
principle, the three-momentum of this deuteron could be calculated as
$\vec{p}_i + \vec{p}_j$, and while three-momentum for the event would
be conserved, energy would not. Instead, an isotropic decay into a
deuteron and photon is performed in the rest frame of the
pair. Because the primary process for deuteron formation at the low
momentum differences of the coalescence model is radiative capture,
$\p \n \to \gamma \D$, this provides a reasonable approximation of the
process and conserves both energy and momentum. While spin
correlations could be considered, these typically are negligible after
boosting the deuteron and photon into the laboratory frame.

\index{Dal--Raklev model}The Dal--Raklev model expands upon the
coalescence model by considering 
the following formation channels, other than just $\p \n \to \D
\gamma$.
\begin{multicols}{4}
  \begin{itemize}
   \item $\p\n \to \gamma \D$
   \item $\p\n \to \piz \D$ 
   \item $\p\n \to \pim \pip \D$
   \item $\p\n \to \piz \piz \D$
   \item $\p\p \to \pip \D$
   \item $\n\n \to \pim \D$
   \item $\p\p \to \pip \piz \D$ 
   \item $\n\n \to \pim \piz \D$
  \end{itemize}
\end{multicols}
\noindent Channels can be removed, modified, or added. Each channel
must have a two-body initial state and an $n$-body final state where
$n > 1$ and at least one of the outgoing particles is a
deuteron. For each of these channels the kinematics of the final state
are determined by an isotropic decay in the rest frame of the initial
state pair. Additionally, the binding cross section is no longer
considered as a uniform distribution up to a cutoff parameter, but is
instead determined from fits of differential nucleon-scattering data from a 
number of experiments~\cite{Dal:2015sha}.

Four cross-section parameterizations are available. For each channel,
one of the following parameterizations must be selected, and the
necessary coefficients $c_i$ provided.
\begin{enumerate}
\item The coalescence model parameterization as previously described
  is a step function with two parameters, the cutoff parameter $c_0$
  and a normalization parameter $c_1$. The normalization allows
  channels using this parameterization to be used in combination with
  other channels.
  \begin{equation}
    \frac{\d{\sigma(k)}}{\d{k}} = c_1 \Theta(c_0 - k)
  \end{equation}
\item The $\p\n \to \gamma\D$ cross-section differential in $k$ can
  be parameterized by a polynomial below a cutoff of $c_0$, and with
  an exponential above. Due to Runge's phenomenon, the polynomial is
  fixed to its value at $k = 0.1~\GeV$ for $k < 0.1~\GeV$.
  \begin{equation}
    \frac{\d{\sigma(k)}}{\d{k}} = 
    \begin{cases}
      \d{\sigma(0.1~\GeV)}/\d{k} & \text{for } k < 0.1~\GeV \\
      \sum_{i = 1}^{12} c_i k^{i - 2} & \text{for } 0.1~\GeV \geq k < c_0 \\
      e^{-c_{13}k - c_{14}k^2} & \text{otherwise} \\
    \end{cases}
  \end{equation}
\item The two-body final states with a pion and deuteron are
  parameterized using a cross section differential in $q$, the
  momentum magnitude of the pion in the nucleon-pair rest frame,
  divided by the mass of the pion. Because the final state is
  two-body, the pion momentum magnitude is already known \textit{a
    priori}.
  \begin{equation}
    \frac{\d{\sigma(q)}}{\d{q}} = \frac{c_0 q^{c_1}}{(c_2 - e^{c_3
        q})^2 + c_4}
  \end{equation}
  In the default Dal--Raklev model, the $\p\n \to \piz \D$, $\p\n \to
  \pip \D$, and $\p\n \to \pim \D$ channels use this parameterization.
\item The cross sections for the three-body final states with pions
  are differential in $k$ and are parameterized with,
  \begin{equation}
    \frac{\d{\sigma(k)}}{\d{k}} = \sum_{i = 0} \frac{c_{5i} k^{c_{5i +
          1}}}{(c_{5i + 2} - e^{c_{5i + 3}k})^2 + c_{5i + 4}} ~,
  \end{equation}
  where the number of coefficients is variable but must be a multiple
  of $5$. The default $\p\n \to \pim \pip \D$, $\p\n \to \piz \piz
  \D$, $\p\n \to \pip \piz \D$, and $\n\n \to \pim \piz \D$ channels
  use this parameterization.
\end{enumerate}

Not only does the shape of the differential cross sections matter, but
also the normalization, as this determines the relative rates between
the channels. In the Dal--Raklev model the $\gamma\D$ channel
dominates at low $k$. For $k > 1~\GeV$ the $\pion\D$ channels dominate
except between roughly $1$ and $2~\GeV$ where the $\pion\pion\D$
channels dominate. An additional unitless normalization scale can be
configured to increase or decrease the total deuteron production
cross section. A number of fits for this normalization constant have
been made using various data sets from the LHC, with the default
normalization set from differential $7~\TeV$ ALICE
data~\cite{Dal:2015sha}.

%% file: physics/particle-decays.tex
\section{Particles and decays}\index{Decays!Hadrons@of Hadrons}
\label{sec:particleDecays}

There are several ways to classify unstable particles, and in \pyt at least three classifications are useful:
\begin{itemize}
  \item by lifetime, specifically for coloured particles whether above
    or below the hadronization time;
  \item by if the partial and total widths of a particle are
    perturbatively calculable, such as for the $\gp{\mu}$,
    $\gamma^*/\Z$, \Wpm, top, Higgs bosons, and most BSM particles, or
    not, such as for hadrons;
  \item by if a particle is part of the hard process, and cannot
    be produced anywhere else, such as in parton showers or
    hadronization, \eg \Z, \Wpm, top, and Higgs bosons.
\end{itemize}
These classifications are necessary to understand how particles are
technically treated.

In \pyt a distinction is made between the following technical
representations of particle states: resonances with an average
lifetime shorter than the hadronization scale; particles with an
average lifetime comparable to or longer than the hadronization scale;
and partons that carry colour charge and must be hadronized. By
default, any state with a nominal mass above $20~\GeV$ is considered
as a resonance, \eg $\gamma^*/\Z$, \Wpm, top, Higgs bosons, and most
BSM states such as sfermions and gauginos. However, some light
hypothetical weakly interacting or stable states, such as the
gravitino, are also by default considered as resonances to ensure a
full treatment of angular correlations in their decays. All remaining
states without colour charge, primarily leptons and hadrons, are
treated as particles, while quarks and gluons are considered as
partons. There are some exceptions like colour-octet onia, which are
treated as both partons carrying colour charge, and particles that
decay after hadronization.

Resonance states are sequentially decayed during the hard process, see \cref{sec:hardRes} for details. For
example, in the hard process $\g\g \to \H \to \Z[\to
  \gp{\mu^+}\gp{\mu^-}] \Z[\to \gp{\mu^+}\gp{\mu^-}]$ the decay
of the Higgs into \Z bosons is first performed, followed by the decays
of the $\gam^*\Z$ resonances into muon pairs. The cross sections
calculated for hard processes with resonances depend upon the
available decay channels of the resonances; closing decay channels
will reduce the cross section for the process while opening decay
channels will increase the cross section. Consequently, when using the
cross section calculated for a resonance produced in a hard process,
the available branching fractions of the resonance are already
included in the cross section. In most cases, angular correlations are
included in the decay of the resonance. In some cases, mixed decays of
the resonances are needed, \eg $\g\g \to \H \to \Z[\to
  \gp{\mu^+}\gp{\mu^-}] \Z[\to \eplus \eminus]$. In this
example, both the muon and electron channels could be left
open. However, in some cases this might result in inefficient
generation of the required final state. Consequently, a special class
is available in \pyt, \texttt{ResonanceDecayFilterHooks}, which can be
used to select specific final states from the resonance decays.

Lighter states such as the \Jpsi, which can be produced by the hard
processes of \cref{subsection:onia}, are technically treated by
\pyt as particles and not resonances. This is because the \Jpsi can be
produced in both hadronization and particle decays, where the
cross section of these \Jpsi production mechanisms is not known
\textit{a priori}. The reduced cross section of the \Jpsi due to
closed decay channels can only be determined after the generation of
full events, including \Jpsi production from the hard process,
hadronization, and particle decays. Similarly, states that are only
produced in the hadronization and particle decays, \eg the \rhomeson,
are also considered particles and not resonances. An important
exception to the treatment of resonances is the production of weak
bosons in the parton shower, see \cref{sec:SimpleQEDEW}. Here,
while closing the decay channels of the weak bosons will modify the
hard-process cross section, the decays of the weak bosons in the
parton shower will still remain inclusive. The decay channels of the
weak bosons in the parton shower can be selected using the special IDs
\texttt{93} and \texttt{94} for the \Z and \W, respectively. However,
changing these decay channels will not affect the hard-process
cross section and must be book kept carefully by the user.

\subsection{Particle properties}

\index{Particle properties}
For all states, a number of properties must be defined. Each state is
uniquely identified by its \ac{PDG} ID, or when a PDG convention is not
available, a \pyt specific numbering convention, \ie for BSM and colour-octet onium states. For each state a human readable name is stored, as
well as an antiparticle name when relevant. The quantum numbers for
each state must be defined: the spin, electric charge, and colour
charge. Note that the spin information is duplicated for hadrons,
where the spin can also be determined from the PDG ID. The
experimentally observable properties of the state are also specified
including a nominal mass, a nominal width, allowed limits of this
width, and a nominal proper lifetime. Additionally, a number of decay
related options can be specified including whether the state may
decay, if the width is perturbatively calculable, and if the width
should be forced to be rescaled. Each state may also have a list of
decay channels which determine how the state is decayed. Each channel
is configured with a flag specifying if the channel is available for
the particle/antiparticle state, a branching ratio, a mode specifying
a possible matrix element for the decay, and a list of the decay
products.

\subsubsection{Masses}
\label{prop:masses}

The default masses for most particles in \pyt are taken from
experimental observation as summarized by the \ac{PDG}~\cite{ParticleDataGroup:2012pjm}. There are three exceptions:
quarks and diquarks, unobserved or poorly studied hadrons, and
hypothetical BSM particles. For hypothetical particles, \eg BSM Higgs
bosons, hidden valley hadrons, or fourth generation fermions,
reasonable defaults are provided, see \cref{sec:slha} for details.

Due to ill-defined quark masses, two types of quark masses are used in
\pyt, kinematic and running. The kinematic masses are those defined in
the \pyt particle database, and are used when generating phase
space. For example, in the process $\g \g \to \c \cbar$, the kinematic
mass of the \c quark is used. Similarly, the $\g \to \q \qbar$ splittings of
the parton shower use these kinematic masses. While these quark masses
can be changed, their default values have been carefully chosen
following a number of considerations~\cite{Norrbin:1998bw}. Modifying
these default values can lead to unintended consequences across all
aspects of \pyt including hard process generation, the parton shower,
hadronization, and even particle decays. Consequently, care should be
taken when changing these quark masses from their default values.

\index{Running quark masses|see{Quark masses}}%
\index{Quark masses!Running quark masses}Running quark masses are used when calculating mass-dependent
couplings, which include couplings of the quarks to SM and BSM Higgs
bosons. The running masses for the quarks are calculated at one loop
in the \msbar normalization scheme using,
\begin{equation}
  m(Q) = m_0 \left(\frac{\ln(Q_0 / \Lambda)} {\ln(Q /
    \Lambda)}\right)^{12/(33 - n_\f)} ~,
\end{equation}
where $m_0$ is the input mass at the reference scale $Q_0$, and $n_\f$
is the number of active flavours in calculating $\alphas$. For the
light quarks, $Q_0$ is set at $2~\GeV$, while for the \c, \b , and \t
quarks $Q_0$ is set at $m_0$. The input masses can be configured with
the parameters \setting{ParticleData:mXRun}, where \texttt{X} is the 
quark name, \ie either \texttt{d, u, s c, b}, or \texttt{t}, to be put
equal to the \msbar mass of the quark. The reference value of $\alphas$ used in
calculating $\Lambda$ is defined at the scale $m_{\Z}$ and set with the
parameter \setting{ParticleData:alphaSvalueMRun}.

The default masses of unobserved hadrons and diquarks have been set
using the constituent mass \index{Quark masses!Constituent quark masses} model from \pyt~6~\cite{DeRujula:1975qlm,
  Sjostrand:2006za}, which considers the spin-spin couplings of the
quark combinations. The semi-empirical formula for a hadron mass is
given by,
\begin{equation}
  m = m_0 + \sum_i m_i + k \, m_{\d}^2 \sum_{i<j} \frac{
    \mathcal{S}_{ij}}{m_i \, m_j} ~,
  \label{prop:mass}
\end{equation}
where the terms $m_0$ and $k$ are determined from known hadron masses
and depend upon the multiplet of the hadron, $m_i$ are the constituent
quark masses, and $\mathcal{S}_{ij}$ are the spin-spin interactions
for each quark-pair combination. The constituent quark masses are taken
from \pyt~6 as $0.325~\GeV$ for the \u and \d, $0.5~\GeV$ for the \s,
$1.6~\GeV$ for the \c, and $5~\GeV$ for the \b. Since the \t does not
form narrow bound states, the \t constituent mass is not needed.

For mesons and diquarks, there is only one quark pair, given by $\q_1$
and $\q_2$. For diquarks and meson multiplets with orbital angular
momentum $L= 0$, the spin-spin term for $S = 0$ states is
$\mathcal{S}_{12} = -3$, while for the $S = 1$ states this term is
$\mathcal{S}_{12} = 1$. For both pseudoscalar and vector mesons, $m_0$
is set to $0~\GeV$ and $k$ is fitted to be $0.16~\GeV$. For the
excited multiplets with $L = 1$, the spin-spin terms vanish with $k$
set to $0~\GeV$ and $m_0$ fitted to be $0.45~\GeV$, $0.5~\GeV$,
$0.55~\GeV$, and $0.6~\GeV$ for scalars, $S = 0$ axial-vectors, $S =
1$ axial-vectors, and tensors, respectively. The masses of diquarks
are calculated using the same $k$ value as for baryons, $0.048~\GeV$,
and $m_0 = 0.077~\GeV$ which is two-thirds the baryon $m_0$ value.

There are three possible combinations for baryons, and the spin-spin
terms depend not only upon the spin of the baryon, but also the quark
composition. For $S = \frac{1}{2}$ baryons the spin-spin term is given
by,
\begin{equation}
  \sum_{i<j} \frac{\mathcal{S}_{ij}}{m_i \, m_j} = \frac{1}{m_1 m_2} -
  \frac{2}{m_1 m_3} - \frac{2}{m_2 m_3} ~,
\end{equation}
if there are either two identical flavours, $\q_1$ and $\q_2$, or all
the quark flavours are different and the two lighter quarks are in an
anti-symmetric spin state. For this anti-symmetric case $\q_3$ is the
heaviest quark, while $\q_1$ and $\q_2$ are the two lighter
quarks. When all the quarks are all different flavours and the light
quark pair is symmetric, the spin-spin term is given by
\begin{equation}
  \sum_{i<j} \frac{\mathcal{S}_{ij}}{m_i \, m_j} = -\frac{3}{m_2 m_3} ~,
\end{equation}
where $\q_2$ and $\q_3$ are the two lighter quarks when relevant. For
the $S = \frac{3}{2}$ baryons, the spin-spin term is given by,
\begin{equation}
  \sum_{i<j} \frac{\mathcal{S}_{ij}}{m_i \, m_j} = \frac{1}{m_1 m_2} +
  \frac{1}{m_1 m_3} + \frac{1}{m_2 m_3} ~,
\end{equation}
where the ordering of the quarks does not matter. For all baryons, the
fitted parameters are set as $m_0 = 0.11~\GeV$ and $k =
0.048~\GeV$. The default masses for a number of baryons in \pyt are
calculated using these factors and \cref{prop:mass}. These baryons
include the double and triple-heavy $\Xi$ and $\Omega$ baryons.

\subsubsection{Widths}\index{Breit--Wigner distribution}
\label{prop:width}

Widths are relevant for sampling the masses of both resonances and
particles. For particles, widths are fixed when sampling a particle
mass except for the case of hadronic rescattering, see
\cref{subsection:hadronicrescattering,dec:vwidth} for
further details. The parameter \setting{ParticleData:modeBreitWigner}
determines what type of distribution is used to select particle
masses. Note that this parameter is set for all particle species; it
is not possible to choose different mass shapes on a
species-by-species basis. For a value of \texttt{0} the fixed on-shell
particle mass is used, while for \texttt{1} a non-relativistic
Breit--Wigner is used,
\begin{equation}
  {\cal P}(m) \, \d m \propto \frac{1}{(m - m_0)^2 + \Gamma^2/4}
  \, \d m  ~.
  \label{prop:nrBW}
\end{equation}
By setting a value of \texttt{2} a mass dependent width can be
included,
\begin{equation}
  \Gamma(m) = \Gamma_0 \sqrt{\frac{m^2 - m_\mathrm{thr}^2}{m_0^2 -
      m_\mathrm{thr}^2}} ~,
  \label{prop:gamma}
\end{equation}
where $m$ is the selected mass, $m_0$ is the on-shell mass, and
$m_\mathrm{thr}$ is the average threshold mass. The threshold mass is
the sum of the on-shell masses for the decay products, and is
consequently channel dependent. However, to decouple mass selection
and decay, the mass threshold is taken as the average mass threshold
for all decay channels, weighted by branching fraction.

A relativistic Breit--Wigner can also be selected,
\begin{equation}
  {\cal P}(m^2) \, \d m^2 \propto \frac{1}{(m^2 - m_0^2)^2 + m_0^2
    \Gamma^2} \, \d m^2 ~,
  \label{prop:rBW}
\end{equation}
with the option \texttt{3}, where a fixed $\Gamma$ is used just as for
option \texttt{1}. The relativistic Breit--Wigner can also be used
with the mass dependent width of \cref{prop:gamma} with option
\texttt{4}. For all mass selection options, relativistic or otherwise,
the mass distribution is truncated via the \setting{NN::mMin} and
\setting{NN:mMax} parameters set for each particle species; here, \texttt{NN} is the given particle species ID. The default
mass shape in \pyt is option \texttt{4}, a mass-dependent relativistic
Breit--Wigner.

For particles with broad mass distributions that are not treated as
resonances, the mass selection models above can distort the particle
branching ratios. Regardless of the selected mass of a particle, all
decay channels, even those with an on-shell mass threshold above the
selected mass, are considered. Only after the masses for the decay
products are sampled, are channels eliminated if not kinematically
available. In this way, decay channels can remain open if there are
downward fluctuations in the selected masses of the decay
products. However, if the mass distribution for a particle is
truncated at a lower mass, decay channels with lower mass thresholds
may be enhanced. A good example is the $\rho^0$ which, as a broad low-mass resonance, has any number of non-perturbative and threshold
effects. The mass distribution for the $\rho^0$ is limited by the rare
$\eplus \eminus$ decay channel. However, truncating the mass
distribution at this mass threshold can result in oversampling the
lighter decay channels. Consequently, the mass threshold for the
$\pion \pion$ channel is used instead for the default low-mass
truncation of the $\rho^0$ mass distribution.

\index{meMode@\texttt{meMode}}
For resonances, widths are sampled using the relativistic Breit--Wigner
of \cref{prop:rBW}, but with a number of options available for
determining the partial widths of the resonance at a given mass. This
calculation method can be set by the user with the \setting{NN:meMode}
parameter when defining the decay channels for a particle. The default
value is \texttt{0}, where the partial width is calculated
perturbatively for the resonance if already available in \pyt. If a
given width is not available via a perturbative calculation, then this
width is set to zero. However, a number of alternative partial width
calculations are available.
\begin{itemize}
  \item \settingval{NN:meMode}{100}: The partial width is defined as
    the branching fraction for that decay channel, multiplied by the
    total width. This method results in mass-independent widths and
    does not account for mass-threshold effects, which may result in
    issues when a resonance is produced significantly off-shell with a
    mass below the on-shell mass. When this occurs, it is possible
    that no decay channels remain kinematically open, and the
    resonance can no longer be decayed. However, it is also possible
    that downward mass fluctuations may occur in the masses of the
    decay products, allowing some channels. Consequently, all decay
    channels are considered whenever a resonance is decayed, even if
    the on-shell masses of the products kinematically exclude the
    channel.
  \item \settingval{NN:meMode}{101}: The partial widths are calculated
    in the same fashion as for \settingval{NN:meMode}{100}, but are
    now set to zero if the sum of the on-shell masses for the decay
    products is not kinematically allowed at the mass for which the
    partial width is being calculated. Consequently, the total width
    becomes mass dependent through the introduction of step functions
    at the kinematic limits for each decay channel.
  \item \settingval{NN:meMode}{102}: This method builds upon the method of
    \settingval{NN:meMode}{101} but uses a smooth threshold factor, rather
    than a step function. For two-body decays the partial width is
    multiplied by the factor,
    \begin{equation}
      \beta = \sqrt{(1 - m_1^2/m^2 - m_2^2/m^2)^2 - 4m_1^2 m_2^2/m^4} ~,
    \end{equation}
    where $m_i$ are the masses of the decay products and $m$ is the
    selected mass of the decaying resonance. While this correctly
    includes the phase-space suppression for an isotropic two-body
    decay, any channel specific modifications due to the matrix
    element for the decay are not included. For higher multiplicity
    decays, a less sophisticated factor of,
    \begin{equation}
      \beta = \sqrt{1 - \sum_i m_i / m} ~,
    \end{equation}
    is used which roughly approximates the phase-space
    suppression. For this method, the branching ratio for each decay
    channel should be provided without a phase-space suppression
    factor, otherwise phase-space suppression for that channel will be
    double counted. When using this method, the branching fractions
    for the resonance as calculated by \pyt will not match those
    provided by the user.
  \item \settingval{NN:meMode}{103}: The phase-space suppression of
    \settingval{NN:meMode}{102} is used, but the branching fraction for
    the channel is divided by the $\beta$ factor calculated at the
    on-shell mass of the resonance, $\beta_0^{-1}$. Consequently, the
    branching fractions no longer need to be adjusted for phase-space
    suppression, and the branching fractions calculated by \pyt will
    match those provided by the user. However, in some cases
    $\beta_0^{-1}$ can be very large if a channel is very near
    threshold for the nominal mass of the resonance. The parameter
    \setting{ResonanceWidths:\-minThreshold} defines the minimum
    allowed $\beta_0$ and limits the correction for resonance masses
    well above the on-shell mass.
\end{itemize}
Note that it is possible to mix and match partial width calculation
methods for a given resonance, \ie some decay channels may have their
partial width calculated perturbatively, while the methods outlined
above are used for others.

\subsubsection{Lifetimes}
\label{sec:hadron-lifetimes}

While the lifetime of a particle is inversely related to its width,
decoupling the lifetime and width of a particle is oftentimes useful
for practical purposes. Consequently, both the lifetime and the width
of a particle species can be specified independently in \pyt. The
lifetime is given as the nominal proper lifetime multiplied by the
speed of light, $c\tau_0$, and has units of millimetres. For particles
with a non-zero lifetime, a lifetime is sampled according to an
exponential decay,
\begin{equation}
  {\cal P}(\tau) \, \d \tau \propto \exp(- \tau / \tau_0 ) \, \d \tau ~,
\end{equation}
where the $\tau_0$ used here is not calculated from the width, but
rather specified independently. When the hadronic-rescattering
framework is enabled and the independently provided $\tau_0$ is zero,
the nominal proper lifetime is automatically calculated using the
width, if the particle species has at least a single available decay
channel. See \cref{subsection:hadronicrescattering} for
details. Similarly, missing lifetimes are calculated when vertex
positions and rapid hadron decays are enabled in the
hadronization. For resonances, $\tau_0$ is automatically determined
from the calculated width of the resonance. However, in some cases
very long lifetimes are necessary, which could result in such narrow
widths that the calculation of the cross section becomes numerically
unstable. Here, the width and lifetime for a resonance can be made
independent by setting the flag \settingval{NN:tauCalc}{false} for
that resonance. This can be particularly useful when scanning lifetime
space for BSM resonances.
 
After the lifetime for a particle or resonance is selected, the decay
vertex position is calculated as,
\begin{equation}
  x_\textrm{dec} = x_\textrm{pro} + \tau \, \frac{p}{m} ~,
  \label{pro:vrt}
\end{equation}
where $m$ is the mass of the particle, $p$ the momentum, and
$x_\textrm{pro}$ the production-vertex position that may be either
the primary interaction point or from some previous decay. This
treatment of the decay vertex assumes all particles travel without
interaction, including no magnetic fields or interactions with
detector materials. Consequently, decay chains can be stopped to allow
the subsequent decays of the particles to be handed to a detector
simulation. A number of criteria for stopping decays is
available. Particles with a specified minimum nominal lifetime can be
stopped from decaying with the flag \setting{ParticleDecays:limitTau0}. Similarly,
particles with a selected lifetime greater than a configurable minimum
lifetime can be set stable with the \setting{ParticleDecays:limitTau} flag. Particles
can also be limited from decaying geometrically, either within a
sphere with \setting{ParticleDecays:limitRadius}, or within a cylinder with
\setting{ParticleDecays:limitCylinder}.
 
\subsection{Decays}\index{Hadron decays}
\label{ss:partdecays}\index{Particle decays}
 
Particle decays might at first appear to be one of the simpler
components of \pyt, given the clear factorization between the
production and decay of particles. The masses, widths, and decay
channels for most particles can be set directly to experimentally
observed values, and typically do not require sophisticated
calculations. Once this information is provided, a particle can be
decayed by randomly selecting a decay channel with a weight
proportional to its branching fraction, and then distributing the
products of the selected channel according to phase space. However,
there are a number of complications which require modifications to
this initial approach.

The technical generation of phase space for decays with more than
three products can be non-trivial to perform efficiently, and requires
the use of specialized algorithms such as the M-generator or RAMBO,
which are introduced in \cref{subsection:phasespace}. After
phase-space generation, a matrix-element weight can be applied to
ensure the correct kinematic distribution, given the nature of the
decay. For particles with non-zero spin, spin effects can change the
kinematic distribution not only for a single decay, but also between
correlated decays. Finally, additional photons need to be
probabilistically included in radiative decays.

All these complications assume the decay channel is exclusive, \ie the
number and type of decay products is fixed. For many decays, such as
those of charm and bottom hadrons, this is not the case. A full list
of the available decays are provided in \cref{dec:summary}. About
$40\%$ of decay channels in \pyt have dedicated matrix elements,
corresponding to $50\%$ of decays when weighted by branching
fraction. The remainder of this section describes these decays.

\begin{table}
  \caption{Available matrix element modes for particle decays. Here,
    $V$ is a vector meson, $P$ a pseudoscalar, $H$ a generic hadron,
    $X$ any non-partonic initial state, and $A$ and $B$ any
    non-partonic final states.\label{dec:summary}\index{meMode@\texttt{meMode}}}
  \begin{center}
    \begin{tabular}{lp{0.5\textwidth}cc}
      \toprule
      force & process & eq. & \texttt{meMode} \\
      \midrule
      any
      & $X \to \q\qbar$ or $X \to \g\g$
      & none
      & 91 \\
      any
      & $X \to \q\qbar A$, where $A$ is a colour singlet
      & none
      & 93 \\
      any
      & $X \to \q\qbar \ldots$
      & none
      & 42 - 80 \\
      any
      & $H \to A B$
      & (\ref{dec:H2AB})
      & 3 - 7 \\
      strong
      & $V \to \pi^+\pi^-\pi^0$, where $V$ is an isoscalar
      & (\ref{dec:V2PPP})
      & 1 \\
      strong
      & $P \to P V[\to P P]$
      & (\ref{dec:P2PV})
      & 2 \\
      strong
      & $P \to \gam V[\to P P]$
      & (\ref{dec:P2gmV})
      & 2 \\
      strong
      & $V \to \g \g \g$ or $V \to \gam \g \g$
      & (\ref{dec:V2ggg})
      & 92 \\
      EM
      & $H \to A \gam^*[\to \ell^+ \ell^-]$
      & (\ref{dec:H2gmll-mass}), (\ref{dec:H2gmll-angle})
      & 11 \\
      EM
      & $H \to \q\qbar \gam^*[\to \ell^+ \ell^-] \to A \ell^+ \ell^-$
      & (\ref{dec:H2gmll-mass}), (\ref{dec:H2gmll-angle})
      & 11 \\
      EM
      & $H \to A B \ldots \gam^*[\ell^+ \ell^-]$
      & (\ref{dec:H2gmll-mass}), (\ref{dec:H2gmll-angle})
      & 12 \\
      EM
      & $H \to \q\qbar \gam^*[\to \ell^+ \ell^-] \to A B \ell^+ \ell^-$
      & (\ref{dec:H2gmll-mass}), (\ref{dec:H2gmll-angle})
      & 12 \\
      EM
      & $H \to \gam^*[\to \ell^+ \ell^-] \gam^*[\to \ell^+ \ell^-]$
      & (\ref{dec:H2gmll-mass}), (\ref{dec:H2gmll-angle})
      & 13 \\
      weak
      & $H \to \gp{\bar{\nu}_\ell} \gp{\ell^-} A$
      & (\ref{dec:ff2ff})
      & 22/23 \\
      weak
      & $H \to \gp{\bar{\nu}_\ell} \gp{\ell^-} \q \qbar$
      & (\ref{dec:ff2ff})
      & 22/23 \\
      weak
      & $X \to \q\qbar A$, where $A$ is a colour singlet
      & (\ref{dec:ff2ff})
      & 94 \\
      weak
      & $H \to \gp{\bar{\nu}_\ell} \gp{\ell^-} A B \ldots$
      & (\ref{dec:H2nulAB})
      & 22/23 \\
      weak
      & $H \to \q \qbar \q \qbar$
      & (\ref{dec:H2qqqq})
      & 22, 23 \\
      weak
      & $\ell^- \to \gp{\nu_\ell} A \ldots$
      & (\ref{dec:H2qqqq})
      & 21 \\
      weak
      & $\ell^- \to \gp{\bar{\nu}_\ell} \gp{\ell^-} \gp{\ell^+} \gp{\nu_\ell}$
      & (\ref{dec:ff2ff})
      & 22/23 \\
      weak
      & $H \to \gam \q \qbar$
      & (\ref{dec:H2gmqq})
      & 31 \\
      \bottomrule
    \end{tabular}
  \end{center}
\end{table}

\subsubsection{Hadron decays with parton showers}
\index{Decays!Hadrons@of Hadrons}
\index{Parton showers!Hadron decays@in Hadron decays}
\index{meMode@\texttt{meMode}}
\label{dec:parton}

The decays of many particles are not known in an exclusive hadronic
form but instead, the relative rates between exclusive partonic
channels is known. Consequently, it is necessary to evolve these
exclusive partonic decays into final state hadrons. In \pyt there are
two mechanisms for this evolution. In the first method, the partons
are passed to the timelike parton shower of \cref{sec:showers},
followed by the hadronization of \cref{sec:lund-model}. This method
is used for $\b\bbar$ states, and typically the parton shower does not
significantly modify the decay. By default, the partons produced in
the decay are distributed uniformly in phase space, with the notable
exception of \settingval{NN:meMode}{92} detailed in \cref{dec:strong} and
\settingval{NN:meMode}{94} detailed in \cref{dec:weak}.

A number of parton and colour configurations are available for this
type of inclusive decay via the parton shower as follows. Here, $c_i$
is used to indicate a colour index and $\bar{c}_i$ anti-colour index.
\begin{itemize}
\item $\q\qbar$: The quark carries $c_1$ the antiquark
  $\bar{c}_1$. This type of decay is set with \settingval{NN:meMode}{91}. Examples of decays using this matrix element mode are
  $\gp{\Upsilon} \to \q \qbar$. Hidden valley hadrons also heavily
  utilize this decay.
\item $\g\g$: The first gluon carries $c_1$ and $\bar{c}_2$ while the
  second gluon carries $c_2$ and $\bar{c}_1$. This decay is also
  specified by \settingval{NN:meMode}{91} and is primarily used for
  quarkonia, \eg $\gp{\eta_b} \to \g \g$.
\item $\g\g\g$: The first gluon carries $c_1$ and $\bar{c}_2$, the
  second $c_2$ and $\bar{c}_3$, and the third $c_3$ and
  $\bar{c}_1$. This configuration is intended for the decays of
  quarkonia, \eg $\gp{Upsilon} \to \g\g\g$, and set with
  \settingval{NN:meMode}{92}.
\item $\g\g\gam$: The first gluon carries $c_1$ and $\bar{c}_2$ and
  the second $c_2$ and $\bar{c}_1$. This decay is also intended for
  quarkonium decays, \eg $\gp{\Upsilon} \to \gam \g \g$ and is set
  with \settingval{NN:meMode}{92}.
\item $\q\qbar X$: This is the same as the colour-singlet $\q\qbar$
  decay mode, except with an additional colour singlet $X$, and is
  selected with \settingval{NN:meMode}{93} for flat phase space, and
  \settingval{NN:meMode}{94} for a weak decay.
\end{itemize}
For all of these decays, the ordering of the partons as passed to \pyt
does not matter.

\subsubsection{Inclusive hadron decays}\index{Decays!Hadrons@of Hadrons}
\label{dec:inclusive}\index{meMode@\texttt{meMode}}

The second method for inclusive hadronic decays is to first determine
hadrons from the partons and then distribute these hadrons in phase
space. This method is used primarily for multibody decays of hadrons
such as the \D and \B mesons, where only a few channels are known
experimentally. The flavours for a channel can then be dynamically
built from the initial partonic content of a weak decay. For this type of
decay, either one or two parton pairs can be specified in the decay,
in addition to any non-parton particles. Here, a parton is either a
quark or diquark. The number of final particles is determined from a
Poisson distribution with a mean of,
\begin{equation}
  \label{dec:lambda}
  \lambda = \frac{n_\textrm{known} + n_\textrm{spec}}{2} +
  \frac{n_\textrm{partons}}{4} + \rho_\textrm{mult} \ln
  (m_\textrm{diff} / m_\textrm{mult})
\end{equation}
where $n_\textrm{known}$ is the number of non-partonic particles in
the specified decay, $n_\textrm{spec}$ is the number of spectator
partons, and $n_\textrm{partons}$ is the number of partons. Here, the spectator partons are those partons that do not participate in the partonic weak decay. The mass
$m_\mathrm{diff}$ is the difference between the decaying particle mass
and the sum of the nominal decay-product masses. A reference mass
$m_\mathrm{mult}$ is set by the parameter
\setting{ParticleDecays:multRefMass} and can be used to tune the
average decay multiplicity. An additional factor,
$\rho_\textrm{mult}$, also determines the average decay multiplicity
and is set via the parameter \setting{ParticleDecays:multIncrease} for
all relevant matrix-element modes except \settingval{NN:meMode}{23} where
\setting{ParticleDecays:\-multIncreaseWeak} is used instead. See
\cref{dec:weak} for further details.

The method for selecting the final hadrons is as follows.
\begin{enumerate}
\item\label{dec:mult} The multiplicity is selected according to
  \cref{dec:lambda} and is required to be less than $10$. A minimum
  multiplicity can be required by setting the \setting{NN:meMode}
  between \texttt{42} and \texttt{50}, where the minimum multiplicity
  is given by \texttt{meMode - 40}. Alternatively, the multiplicity
  can be fixed by setting the \setting{NN:meMode} between \texttt{62}
  and \texttt{70}. Here the multiplicity is calculated as
  \texttt{meMode - 60}.
\item The number of hadrons to form is the difference between the
  selected multiplicity from the previous step, and the number of
  non-parton particles in the decay.
\item One of the partons is selected at random and a new parton and
  hadron is formed, following the flavour selection of
  \cref{sec:had-flavour}.
\item The previous step is repeated until the number of remaining
  hadrons to select is the same as the number of parton pairs.
\item The remaining parton pairs are formed into hadrons.
\item If there are two pairs, they may be reshuffled, as determined by
  the probability \setting{ParticleDecays:\-colRearrange}, \ie for a
  value of \texttt{0} the pairs will never be reshuffled but for a
  value of \texttt{1} they will always be reshuffled.
\item\label{dec:mass} If the mass of the final decay products is less
  than the decaying particle, the hadron selection is kept, otherwise
  the process begins again with step~\ref{dec:mult}.
\end{enumerate}
This model is very similar to the hadronization model, but the momenta
of the hadrons is now just determined with phase space. For most
decays this approximation is valid as the decay-product momenta should
be very low and on average reproduce the correct kinematic
behaviour. While the flavour selection is the same as for
hadronization, the mass constraint of step~\ref{dec:mass} will
typically bias decays to the lighter pseudoscalar mesons, particularly
for high multiplicity decays.

For these types of inclusive decays, the special particle ID
\texttt{82} can be used to randomly select a light flavour pair, \ie
$\u\ubar$, $\d\dbar$, or $\s\sbar$. The suppression of selecting an
$\s\sbar$ pair with respect to $\u\ubar$ and $\d\dbar$ is configured
by the parameter \setting{StringFlav:probStoUD} which is also used in
the flavour selection of the hadronization algorithm of
\cref{sec:had-flavour}. When specifying decays with this ID, the
channel should be given as an \texttt{82 -82} pair, where the ordering
does not matter. A similar ID is \texttt{83} which is the same as
\texttt{82}, but intended for decays that proceed through a gluon
loop. Since this loop will increase the average multiplicity of the
decay, \eqref{dec:lambda} is modified by adding an additional constant
specified by the parameter \setting{ParticleDecays:multGoffset}. The
primary decay of the $\Jpsi$ into three gluons, as well as many of the
other onium states, use this special ID.

For some particles, exclusive decays must be specified in addition to
inclusive decays. Matrix-element modes are provided in \pyt to prevent
double counting the exclusive decays in the inclusive decays. An
\setting{NN:meMode} between \texttt{52} and \texttt{60} reproduces the
same behaviour as an \setting{NN:meMode} between \texttt{42} and
\texttt{50}, but will exclude any generated final state that matches a
non-partonic decay channel. An example of such a decay is $\gp{\eta_c}
\to \q \qbar$. Similarly, if \setting{NN:meMode} is between \texttt{72}
and \texttt{80}, the behaviour for \texttt{meModes} between
\texttt{42} and \texttt{50} is reproduced, but again excluding any
generated final state that matches a non-partonic decay channel.

\subsubsection{Variable-width hadrons}\index{Decays!Variable widths}
\label{dec:vwidth}\index{meMode@\texttt{meMode}}

For standard particle decays, the probability used to select a decay
channel is calculated using a fixed branching ratio, independent of
the decaying particle mass. The hadronic rescattering
framework (\cf\cref{subsection:hadronicrescattering}), however, includes
mass-dependent partial widths for two-body decays of hadrons. For
hadrons included in the rescattering framework, decay channels are
picked using these partial widths. The partial width for the decay of
a hadron resonance $H$ into particles $A$ and $B$, $H \to A B$, is
given by,
\begin{equation}
  \label{dec:H2AB}
  \Gamma_{H \to AB}(m) = \Gamma_0 \frac{m_0}{m}
  \frac{\Phi(2l + 1,m)}{\Phi(2l + 1,m_0)}
  \frac{1.2}{1.0 + 0.2\frac{\Phi(2l, m)}{\Phi(2l, m_0)}} ~.
\end{equation}
Here, $\Gamma_0$ is the nominal partial width of the decaying hadron
at its nominal mass $m_0$, set from experiment. The angular momentum
of the two-body decay is given by $l$. In \pyt, this angular momentum
is specified by the user as $l = \mathtt{meMode} - 3$. At high masses
the final multiplicative factor regulates the partial width. Similar
to resonance production, see \cref{prop:width}, these partial
widths define not only the branching fractions of the hadron but also
production.

The phase space is given by
\begin{equation}
  \Phi(l, m) = \int \deriv m_A \int \deriv m_B\,
  q^l(m, m_A, m_B) \mathit{BW}(m_A) \mathit{BW}(m_B) ~,
\end{equation}
where $q(m, m_A, m_B)$ is the magnitude of the $A$ and $B$ momentum in
the centre-of-mass frame,
\begin{equation}
  q(m, m_A, m_B) = \frac{\sqrt{(m^2 - (m_A + m_B)^2)
      (m^2 - (m_A - m_B)^2)}}{2m} ~.
\end{equation}
Finally, the mass distribution for each of the two decay products is
given by a Breit--Wigner,
\begin{equation}
  \mathit{BW}(m) = \frac{1}{2\pi} \frac{\Gamma(m)}{(m^2 - m_0^2)^2 +
    \frac{1}{4}\Gamma^2(m)} ~.
\end{equation}
While this mass distribution does include a mass-dependent width,
phase-space considerations ensure these mass-dependent widths can be
evaluated recursively from the lowest mass particle to the
highest. Note that performing decays with variable partial widths only
affects the branching ratios of the decay channels, and not the
angular distribution of the decay products. By default, a number of
hadrons are decayed using variable partial widths in \pyt. This
includes many of the excited mesons as well as a number of the
baryons. For technical reasons, variable partial width decays are
never performed for the \rhomeson or $\gp{f}_2$ mesons.

\subsubsection{Strong decays}\index{Strong decays}
\label{dec:strong}\index{meMode@\texttt{meMode}}

Most decays proceeding via the strong force in \pyt are modelled with
pure phase space. However, there are four special cases that are
generated according to matrix elements: isoscalar vector mesons
decaying into pseudoscalar mesons, pseudoscalar mesons decaying into a
pseudoscalar and vector mesons, pseudoscalar mesons decaying into a
photon and vector meson, and vector mesons decaying into a three gluon
final state.

The $\gp{\omega}$ meson decays predominantly into a three-pion final
state of $\pip\pim\piz$. This decay can be modelled using the isobar
model~\cite{Herndon:1973yn}, where the decay proceeds via the
intermediate $\rhoz\piz$ or $\rhomeson^\pm\pion^\mp$ state. The matrix
element for this decay is given by
\begin{equation}
  \label{dec:V2PPP}
  |\ME|^2 \propto \big[(m_1 m_2 m_3)^2 - (m_1 p_2 p_3)^2 - (m_2 p_1
    p_3)^2 - (m_3 p_1 p_2)^2 + 2 (p_1 p_2)(p_1 p_3) (p_2 p_3)\big]
  |\mathcal{F}|^2 ~,
\end{equation}
where $m_i$ and $p_i$ are the mass and momentum of decay product
$i$. Here, $\pip$ corresponds to $i = 1$, $\pim$ to $i = 2$, and
$\piz$ to $i = 3$. The function $\mathcal{F}$ includes possible
final-state interactions of the pions, and depends upon the full
kinematics of the decay. When no final-state interactions are present,
$\mathcal{F} = 1$, which corresponds to $P$-wave distributed phase
space. In \pyt, this assumption of no final-state interactions is
made. However, there is experimental evidence that final-state
interactions could play an important role in this
decay~\cite{BESIII:2018yvu}.

The $\gp{\phi}$ meson is also an isoscalar like the $\gp{\omega}$
meson and has a non-negligible branching to the $\rhoz\piz$ and
$\rhomeson^\pm\pion^\mp$ channels, where the larger $\gp{\phi}$ mass
provides sufficient phase space for a \rhomeson resonance. However, a
contact $\pip\pim\piz$ decay, without the \rhomeson-resonance
structure, is also possible~\cite{Rudaz:1984bz}, and is described by
the same matrix element as for the $\gp{\omega}$ meson. The $\rhoz$
itself can also decay into a $\pip\pim\piz$ final state described by
this matrix element, although this decay channel is heavily suppressed
due to phase space. For both the $\gp{\phi}$ and $\rhoz$ mesons, no
final-state interactions are considered in these decay channels. The
matrix element of \cref{dec:V2PPP} can be selected by setting
\settingval{NN:meMode}{1}.

In the decay chain $P_0 \to P_1 V_2[\to P_3 P_4]$, where $P$ is a
pseudoscalar meson and $V$ a vector meson, the decay products $P_3$
and $P_4$ are distributed in the rest frame of $V_2$ according to
$\cos^2\theta$, where $\theta$ is the angle between $P_0$ and
$P_3$. The corresponding matrix element, is given by
\begin{equation}
  \label{dec:P2PV}
  |\ME|^2 \propto (p_0 p_2) (p_2 p_3) - m_2^2 (p_0 p_3) ~,
\end{equation}
where again $i$ specifies the particle in the decay chain, $m_i$ is
the mass of that particle, and $p_i$ is the momentum. Similarly, for
the decay chain $P_0 \to \gamma V_2[\to P_3 P_4]$, the distribution of
$P_3$ and $P_4$ is now given by $\sin^2\theta$ in the rest frame of
$V_2$. The matrix element for this decay is,
\begin{equation}
  \label{dec:P2gmV}
  |\ME|^2 \propto m_2^2 \big[2 (p_2 p_3)(p_0 p_2)(p_0 p_3)
    - m^2(p_2 p_3)^2 - m_2^2(p_0 p_3)^2 - m_3^2(p_0 p_2)^2
    + (m m_2 m_3)^2\big] ~.
\end{equation}
While these two matrix elements are relevant for all appropriately
produced vector-meson decays into a pseudoscalar-meson pair, in
practice the relevant vector-meson decay channels are: $\rhomeson \to
\pion \pion$, $\omega \to \pip \pim$, $\gp{K}^* \to \kaon \pion$,
$\gp{\phi} \to \kaon \kaon$, $\gp{\phi} \to \pip \pim$, and $\D^* \to
\D \pion$. Note that when the vector meson is not produced in the
decay chain $P_0 \to P_1/\gamma V_2$, these matrix elements are not
used. As an example, in the decay chain $\D \to \pion \kaon^*[\to
  \kaon \pion]$, the decay products of the $\kaon^*$ are distributed
according to \cref{dec:P2PV}. To use these matrix elements,
\settingval{NN:meMode}{2} must be set.

For the decays of vector-like onium states into a partonic final state
of gluons, $V_0 \to \g_0 \g_1 \g_2$, or gluons and a photon, $V_0 \to
\gam_0 \g_1 \g_2$, the matrix element,
\begin{equation}
  \label{dec:V2ggg}
  |\ME|^2 \propto \left(\frac{1 - x_1}{x_2 x_3}\right)^2 +
  \left(\frac{1 - x_2}{x_1 x_3}\right)^2 +
  \left(\frac{1 - x_3}{x_1 x_2}\right)^2 ~,
\end{equation}
is used. Here, $x_i$ is twice the energy of particle $i$ divided by the mass
of the decayer in the rest frame of the decayer, $2E_i/m$. For the two
gluon and photon decay, the two-gluon system is required to have a
minimum mass configured by the parameter
\setting{StringFragmentation:stopMass} to ensure that the system can
properly hadronize. This matrix element is set using \texttt{meMode =
  92} as is done for the partonic decays $\gp{\Upsilon} \to \g \g \g$
and $\gp{\Upsilon} \to \gam \g \g$. Because \cref{dec:V2ggg} is
symmetric, ordering of the decay products when configuring \pyt does
not matter.

\subsubsection{Electromagnetic decays}
\index{Electromagnetic decays (of hadrons)}
\label{dec:em}\index{meMode@\texttt{meMode}}

The electromagnetic decay $\piz \to \gam^*[\to \eplus \eminus] \gam$
can be generated with a factorized approach. To begin, the $\gam^*$
mass is selected, using the decay matrix element integrated over the
solid angle, but still dependent upon the $\gam^*$ mass, $m_1$.
\begin{equation}
  \label{dec:H2gmll-mass}
  |\ME|^2 \propto \frac{1}{m_1^2}
  \left(1 + \frac{2m_2^2}{m_1^2} \right)
  \sqrt{1 - \frac{4m_2^2}{m_1^2}}
  \left(1 - \frac{m_1^2}{(m - m_{\max})^2} \right)^3
  \frac{1}{(m_{\rhoz}^2 - m_1^2)^2 + m_{\rhoz}^2 \Gamma_{\rhoz}^2 } ~.
\end{equation}
The subscript $i$ is $0$ for the \piz, $1$ for the virtual $\gam^*$,
$2$ for the \eplus, $3$ for the \eminus, and $4$ for the real $\gam$;
the mass for each particle is given by $m_i$ and $m_{\max}$ is the
maximum mass of the off-shell photon, \ie $m_{\max} = m_4 = 0$ for
this decay channel of the \piz. \index{VMD}The final factor of this
expression is 
the \ac{VMD} propagator for the \rhoz, where $m_{\rhoz}$ is the mass of the
\rhoz, and $\Gamma_{\rhoz}$ the width. This propagator is negligible for
any decaying particle with a mass far from the \rhoz mass, which
includes the case of the \piz. Next, after the $\gam^*$ mass is
selected, the two-body decay of $\piz \to \gam^* \gam$ is
performed. Finally, the angular distribution of the \epem pair is
generated according to,
\begin{equation}
  \label{dec:H2gmll-angle}
  |\ME|^2 \propto (m_1^2 - 2 m_2^2) \big[(q p_2)^2 + (q p_3)^2 \big] +
  4m_2^2 \big[(q p_2) (q p_3) + (q p_2)^2 + (q p_3)^2 \big] ~,
\end{equation}
where $p_i$ is the momentum of the corresponding particle with index
$i$, and $q = p_0 - p_1$. For efficiency and simplicity, this angular
distribution is generated in the rest frame of the decaying particle,
which if highly boosted, can result in minor numerical induced
violations in momentum-energy conservation. Consequently, the momentum
of the final lepton is calculated as $p_3 = p_1 - p_2$ in the
laboratory frame.

The matrix element for this decay channel is also valid for similar
processes where a lepton pair, $\gp{\ell}^+ \gp{\ell}^-$, is produced
via an off-shell photon. Such decay channels include $\gp{\eta} \to
\gp{\ell}^+ \gp{\ell}^- \gam$, $\gp{\omega} \to \gp{\ell}^+
\gp{\ell}^- \piz$, $\gp{\phi} \to \gp{\ell}^+ \gp{\ell}^- \gp{\eta}$,
$\gp{B} \to \gp{\ell}^+ \gp{\ell}^- \kaon/\kaon^*$, $\gp{B}_\s^0 \to
\gp{\ell}^+ \gp{\ell}^- \phi$, and $\gp{\Sigma}^0 \to \gp{\ell}^+
\gp{\ell}^- \gp{\Lambda}^0$. This matrix element can also be used for
the final state $\gp{\ell}^+ \gp{\ell}^- \q \qbar$. In this particular
case, the $\q\qbar$ is converted into a single hadron, following the
inclusive decay selection of \cref{dec:inclusive} but with the
multiplicity of the decay fixed to three. In all the cases described
above, the matrix element for these decay channels is set with
\settingval{NN:meMode}{11}.

The form of \eqref{dec:H2gmll-mass} and \eqref{dec:H2gmll-angle} are
also approximately valid for decay channels with the final state
$\gam^*[\gp{\ell}^+ \gp{\ell}^-] A B \ldots$, where there are two or
more decay products in addition to the lepton pair. Such decays
include $\gp{\eta} \to \gp{\ell}^+ \gp{\ell}^- \pip \pim$, $\kshort
\to \gp{\ell}^+ \gp{\ell}^- \pip \pim$, $\gp{B}^0 \to \gp{\ell}^+
\gp{\ell}^- \piz \piz$, and $\gp{B}^+ \to \gp{\ell}^+ \gp{\ell}^- \u
\sbar$. For this type of decay channel, \cref{dec:H2gmll-mass} is
still used to select the mass, but with $m_{\max} = m_A + m_B +
\ldots$, and \cref{dec:H2gmll-angle} is used without
modification. The phase-space generation, after selecting $m_1$, is
now performed as a decay with multiplicity $n - 1 > 2$, where $n$ is
the final multiplicity of the decay. Setting \settingval{NN:meMode}{12}
selects this matrix element. If $A$ and $B$ are replaced with a
$q\bar{q}$ final state, the system is collapsed down into two hadrons,
with the flavour selection again performed using the inclusive decay
algorithm but with the multiplicity fixed at four.

Finally, these matrix elements are also used to approximate
$\gam^*[\gp{\ell}^+ \gp{\ell}^-] \gam^*[\gp{\ell}^+ \gp{\ell}^-]$
decay channels. Following the same numbering convention, the mass of
the first off-shell photon, $m_1$ is selected using
\cref{dec:H2gmll-mass} where $m_{\max} = m_5 + m_6$, \ie twice the
mass of the second lepton flavour. Then, the mass of the second
off-shell photon, $m_3$, is selected again with
\cref{dec:H2gmll-mass} but using indexing $i - 2$ and setting
$m_{\max} = m_2 + m_3$. After performing the two-body decay of the
$\gam^* \gam^*$ system, the angular distributions for the two lepton
pairs are generated independently using \cref{dec:H2gmll-angle}. This
type of decay channel is specified with \settingval{NN:meMode}{13} and can
be used for decays such as $\piz \to \eplus \eminus \eplus
\eminus$. The technical implementation for all decays using
\cref{dec:H2gmll-mass,dec:H2gmll-angle} require that the
lepton pair should always be set as the final two decay products when
defining these decay channels.

\subsubsection{Weak decays}\index{Weak decays}
\label{dec:weak}\index{meMode@\texttt{meMode}}

The helicity averaged matrix element for the $t$-channel weak
scattering of fermions, $\f_0 \f_1 \to \f_2 \f_3$, is,
\begin{equation}
  \label{dec:ff2ff}
  |\ME|^2 \propto (p_0 p_1)(p_2 p_\textrm{rem})~,
\end{equation}
where $p_i$ is the momentum of particle with index $i$ and
$p_\textrm{rem} = \sum_{i = 3} p_i$ is the sum of the remaining
momenta, which here is just $p_3$. By crossing symmetry, this matrix
element can also be used for weak decays. An example is the fully
leptonic decay of the $\gp{\tau}$ lepton, $\gp{\tau^-} \to
\gp{\bar{\nu}_\ell} \gp{\ell^-} \gp{\bar{\nu}_\tau}$. The particle
ordering determines the corresponding $i$ for each particle in
\cref{dec:ff2ff}, \eg $i = 1$ for the anti-neutrino and $i = 2$ for
the charged lepton. This matrix element can also be used to
approximate semi-leptonic decays of \D and \B mesons, \eg $\D^0 \to
\gp{\ell^+} \gp{\nu_\ell} \pim$ or $\B^0 \to \gp{\nu_\ell} \gp{\ell^+}
\pim$, where the final fermion pair is collapsed into a single
hadron. In this example, the ordering of the neutrino and charged
anti-lepton is swapped between the two decays. This is because for
\D-meson decays, the partonic $\f_0$ is a \c quark, while for the
\B-meson this is a \b antiquark. Similarly, this matrix element can
be used for the semi-leptonic decays of baryons, \eg $\n \to
\gp{\bar{\nu}} \eminus \p$, or the leptonic decays of charged leptons,
\eg $\gp{\mu}^- \to \gp{\bar{\nu}_e} \eminus \gp{\nu}_\mu$. When not
using the sophisticated $\gp{\tau}$ decays of \cref{dec:tau}, this
helicity averaged matrix element can also be used for the leptonic
decays of the $\gp{\tau}$.

Semi-leptonic decays can also be specified with their partonic
content, \eg $\D^0 \to \gp{\ell^+} \gp{\nu_\ell} \d \ubar$ or $\B^0
\to \gp{\nu_\ell} \gp{\ell^+} \d \ubar$, where the ordering of the
quarks does not matter. Similarly, baryon decays of this nature like
$\gp{\Xi_c^0} \to \eplus \gp{\nu_e} \s \gp{(sd)_0}$, can be decayed
using this matrix element where one of the partons is a diquark, \ie
$\gp{(sd)_0}$. When partonic content is specified, the parton system
is collapsed to a single hadron following the flavour-selection rules
of \cref{sec:had-flavour}. The matrix element of \cref{dec:ff2ff}
is used for all the decays described above by setting either
\settingval{NN:meMode}{22} or \settingval{NN:meMode}{23}. For these types of
decays there is no difference between these two matrix-element
modes. The only technical requirement for these decays is that the
first two particles of the decay are the neutrino/charged-lepton pair,
followed by either a hadron or a parton pair, where ordering of the partons does not
matter. In some cases it is convenient to use the special particle ID
\texttt{81} to act as a place holder for the spectator quark or
diquark, which is then automatically replaced with the correct
spectator flavour. For baryons, an ambiguity can arise in this
selection where the spin of the diquark cannot always be determined
uniquely. For the example decay of the $\gp{\Xi_c^0}$ given here, the
spectator flavour can either be $\gp{(sd)_0}$ or $\gp{(sd)_1}$, while
the automatic flavour will always select the $\gp{(sd)_0}$ diquark.

In some cases, semi-leptonic decays with more than one final-state
hadron are needed, \eg $\D^0 \to \eplus \gp{\nu_e} \kz \pim$. The
additional hadrons can be physically interpreted as being produced
from the fragmentation of the spectator parton, resulting in hadrons
with a significantly softer momentum than the hadron containing the
spectator quarks. This softer momentum is modelled by taking the
product of \cref{dec:ff2ff} and an exponential damping factor,
\begin{equation}
  \label{dec:H2nulAB}
  |\ME_\textrm{damp}|^2 \propto |\ME|^2 \prod_{i = 4}
  e^{-|p_i|^2/\sigma_\textrm{soft}^2} ~,
\end{equation}
which is calculated in the rest frame of the decay, where the product
is taken over all hadrons following the spectator hadron with momentum
magnitude $|p_i|$. Here, $|\ME|^2$ is calculated with
\cref{dec:ff2ff} and $\sigma_\textrm{soft}$ is the damping term which
can be configured by the user with the parameter
\setting{ParticleDecays:sigmaSoft}. A single damping parameter is used
for all decays and is expected to fall within the range 0.2 -- 2,
where a smaller value increases the damping. For semi-leptonic decays
with two or more final state hadrons, this matrix element can be used
by setting \settingval{NN:meMode}{22} or \settingval{NN:meMode}{23}. Again, there
is no difference between these two matrix-element modes for decays of
this type. As before, the ordering of the decay as passed to \pyt
matters. The neutrino/charged-lepton pair must be specified first, in
the correct order as discussed above, followed by the hadron
containing the spectator quark, followed by any remaining hadrons,
which will then have their momentum damped.

The matrix element for weak decays into purely hadronic final states,
where the decay is defined only by partonic content, is approximated
by \pyt. An example of this class of decay is $\B^0 \to \u \dbar \cbar
\d$ which will result in a final state with a \D meson. The partonic
content should be set as $\q_1 \q_2 \q_3 \q_4$ where $\q_1$ and $\q_2$
are colour connected, and either $\q_3$ or $\q_4$ is the spectator
quark/diquark. The special particle code \texttt{81} can be used here
to automatically determine the spectator flavour. Just like for the
partonic semi-leptonic decays, the final two partons are collapsed
into a single hadron following the flavour-selection rules of
\cref{sec:had-flavour}. The first two partons are then fragmented
into multiple hadrons, following the method of
\cref{dec:inclusive}.

When \settingval{NN:meMode}{22} is used, the mean number of final particles
in the decay is calculated with \cref{dec:lambda} using the
$\rho_\textrm{mult}$ parameter
\setting{ParticleDecays:multIncrease}. When \settingval{NN:meMode}{23} is
used instead, the mean number of final particles is calculated using
\setting{ParticleDecays:\-multIncreaseWeak}. The former parameter is
intended, although not required, to be smaller than the latter, since
in weak decays only the mass of the off-shell \W boson is available to
the fragmenting partonic system, and not the entire parent
mass. Additionally, for \settingval{NN:meMode}{23} a minimum of three final
particles are required in the decay after flavour selection. After the
final particles are determined for each decay, the matrix element,
\begin{equation}
  \label{dec:H2qqqq}
  |\ME|^2 \propto \frac{2 E_1}{m}\left(3 - \frac{4E_1}{m}\right) ~,
\end{equation}
is used where $m$ is the mass of the decaying hadron and $E_1$ is
the energy of the hadron containing the spectator quark, in the rest
frame of the decay. This matrix element can also be used for hadronic
$\gp{\tau}$ decays when the sophisticated treatment is not needed by
specifying \settingval{NN:meMode}{21}. Here, the first decay product should
always be the $\gp{\nu_\tau}$, which increases the energy of the
neutrino with respect to flat phase space.

Partonic radiative decays via the weak force are roughly approximated
with the matrix element,
\begin{equation}
  \label{dec:H2gmqq}
  |\ME|^2 \propto \left(\frac{2 E_1}{m}\right)^3
\end{equation}
where $m$ is the mass of the decaying hadron and $E_1$ is the energy
of the photon in the rest frame of the decay. Effectively, this
increases the photon energy with respect to flat phase space. The
partonic content for these decays should be set as a photon, the
spectator quark, and the flavour-changing quark, \eg $\B^0 \to \d \sbar
\gam$ where $\d$ is the spectator quark. Unlike the previous weak
decays, where the spectator system is collapsed to a single hadron,
the spectator system is fragmented into multiple hadrons following the
inclusive selection of \cref{dec:inclusive}. However, the
multiplicity for the decay is selected with a geometric distribution,
\begin{equation}
  P(n) = \left(1 - \frac{1}{2}\right)^{n - 1} \frac{1}{2} ~,
\end{equation}
rather than a Poisson distribution, where a minimum multiplicity of
$2$ and a maximum multiplicity of $10$ is required. This type of decay
is specified by setting \settingval{NN:meMode}{31}, and the decay products
can be assigned in an arbitrary order.

In all the decays above, the matrix element is applied to the final
particles of the decay, not the partonic content. In some cases it is
useful to apply the matrix element to the partonic content of the
decay, and then perform a full parton shower followed by
hadronization, using the parton-shower method of
\cref{dec:parton}. Specifying \settingval{NN:meMode}{94} does this, where
the matrix element of \cref{dec:ff2ff} is used to distribute the
phase space of the partons from the decay.

In addition to the weak decays described above, \B systems may mix prior to decay. This mixing is controlled by the flag \setting{ParticleDecays:mixB} and has a probability of,
\begin{equation}
  {\cal P} = \sin^2\left(\frac{x \tau}{2 \tau_0}\right) ~,
\end{equation}
where $\tau$ is the selected lifetime of the particle, and $\tau_0$
the nominal proper lifetime. The mixing parameter $x$ is set with
\setting{ParticleDecays:xBdMix} and \setting{ParticleDecays:xBsMix} for
the $\B_\d$ and $\B_\s$ systems, respectively.

\subsubsection{Helicity decays}\index{Decays!Helicity density formalism}
\label{dec:helicity}

A generic helicity-density formalism is available in \pyt which can be
used for $\gp{\tau}$ decays as well as muon decays in lepton-flavour
violating production. External tools have also used this framework for
heavy-neutral-lepton decays. The weight for an $n$-body decay of an
arbitrary particle is given by,
\begin{equation}
  \mathcal{W} = 
  \rho_{\lambda_0\lambda_0'}
  \ME_{\lambda_0;\lambda_1 \ldots \lambda_n}
  \ME_{\lambda_0';\lambda_1' \ldots \lambda_n'}^*
  \prod_{i = 1,n} \mathcal{D}_{\lambda_i\lambda_i'}^{(i)} ~.
  \label{dec:weight}
\end{equation}
The decaying particle is given index $0$ and the decay products are
assigned indices $i$ through $n$. The helicity for each particle is
given by $\lambda_i$ and summations are performed over each repeated
helicity index. The helicity density matrix for the decaying particle
is given by $\rho$, while the decay matrix for each decay product is
given by $\mathcal{D}$. The helicity matrix element for the decay is
\ME and depends upon the helicity of the decaying particle as well as
the decay products.

For a particle produced from a $2 \to n$ hard process, the helicity-density matrix for an outgoing particle with index $i$ is given
by,
\begin{equation}
  \rho_{\lambda_i \lambda_i'}^{(i)} = 
  \rho_{\kappa_1\kappa_1'}^{(1)}\rho_{\kappa_2\kappa_2'}^{(2)}
  \ME_{\kappa_1\kappa_2; \lambda_1 \ldots \lambda_n}
  \ME_{\kappa_1'\kappa_2'; \lambda_1' \ldots \lambda_n'}^*
  \prod_{j \neq i} \mathcal{D}_{\lambda_j\lambda_j'}^{(j)} ~,
  \label{dec:rho2}
\end{equation}
where $\rho^{(1,2)}$ are the helicity-density matrices for the
incoming particles, \ME is the helicity matrix element for the
process, and $\kappa_{1,2}$ are the helicities of the incoming
particles. For incoming two-helicity-state beam particles with a known
longitudinal polarization $\mathcal{P}_z$ the helicity-density matrix
is diagonal with elements $(1\pm\mathcal{P}_z)/2$.

Before any particles are decayed in a given sequence, all decay
matrices in \cref{dec:weight} and \cref{dec:rho2}, $\mathcal{D}$,
are initialized to the identity matrix. In a $2 \to n$ process, a
first outgoing particle is randomly selected and decayed using a
helicity-density matrix determined with \eqref{dec:rho2}. The decay
matrix for this first decay is calculated as
\begin{equation}
  \mathcal{D}_{\lambda_0\lambda_0'}^{(0)} = 
  \ME_{\lambda_0;\lambda_1 \ldots \lambda_n}
  \ME_{\lambda_0';\lambda_1' \ldots \lambda_n'}^* \prod_{i
    = 1,n} \mathcal{D}_{\lambda_i\lambda_i'}^{(i)} ~.
  \label{dec:D}
\end{equation}
After the full decay tree for this first particle is determined, the
remaining particles for the $2 \to n$ process are then randomly
selected and decayed using the helicity-density matrix of
\eqref{dec:rho2} with the updated decay matrices for the already
decayed outgoing particles.

When a particle from the hard process is selected for decay, the full
decay tree of that particle is performed. A single branch of the decay
tree is followed until a final stable particle is reached. The helicity-density matrices for particles produced from decays are calculated
with,
\begin{equation}
  \rho_{\lambda_i \lambda_i'}^{(i)} = 
  \rho_{\lambda_0\lambda_0'}^{(0)}
  \ME_{\lambda_0;\lambda_1 \ldots \lambda_n}
  \ME_{\lambda_0';\lambda_1' \ldots \lambda_n'}^* \prod_{j
    \neq i} \mathcal{D}_{\lambda_j\lambda_j'}^{(j)} ~,
  \label{dec:rho1}
\end{equation}
where the $\rho^{(0)}$ is the helicity-density matrix of the parent
particle. The algorithm then calculates the decay matrix for the last
particle decayed with \cref{dec:D} and the next undecayed branch of
the decay tree is traversed until all branches of the decay tree have
been decayed. In this way the decays of the outgoing particles from
the hard process are correlated. As implemented in \pyt, this full
recursion is not necessary since the implemented $\gp{\tau}$ decays
are typically provided with stable final-state particles.

\index{SPINUP@\texttt{SPINUP}}The hard-process generation of \pyt uses
unpolarized matrix elements 
to generate the phase space of the hard process, and so dedicated $2
\to n$ helicity matrix elements are needed to determine the helicity
density matrix after phase-space generation. For $\gp{\tau}$ decays a
number of helicity matrix elements are available. Correlated decays
from \gam, \Z, $\gp{Z'^0}$, $\gam/\Z/\gp{Z'^0}$, neutral Higgs bosons,
and $t$-channel $\gam \gam \to \ell \ell$ production are
provided. Single $\gp{\tau}$ decays from $\W$, $\W'$, charged Higgs
bosons, and $\B/\D$ decays are also provided. For all these production
mechanisms the relevant parameters that can be configured for the
unpolarized production mechanisms are also used in the helicity matrix
elements. This includes the axial and vector couplings for the new
gauge bosons, as well as the parity of the Higgs bosons. When a
particle used in the helicity decay framework is provided from outside
of \pyt, the \texttt{SPINUP} digit is interpreted as the helicity of
the particle in the laboratory frame. A number of options can be
configured to fine tune the helicity treatment of $\gp{\tau}$ decays
in \pyt.

\subsubsection{Tau decays}\index{Tau decays@$\tau$ decays}
\index{Decays!Taus@of $\tau$ leptons}
\label{dec:tau}

\index{Tauola@\textsc{Tauola}}While unpolarized simplified models of
$\gp{\tau}$ decays are 
available in \pyt, see \cref{dec:weak}, dedicated models which use
the helicity-density framework are available. These models are based
on those provided in \textsc{Tauola}~\cite{Jadach:1993hs}, and are
available for all decay channels with branching fractions greater than
$0.04\%$, including up to six-body tau decays. The general helicity-density matrix for these decays used in \cref{dec:weight} is given by
\begin{equation}
  \ME \propto \bar{u}_{\nu_\tau} \gamma_\mu (1 - \gamma^5) u_\tau J^\mu ~.
\end{equation}
where only the current $J^\mu$ needs to be specified. Here, $u$ and
$\bar{u}$ are Dirac spinors, $\gamma^\mu$ are the Dirac matrices, and
the Weyl basis as adopted in \textsc{Helas}~\cite{Murayama:1992gi} is
used throughout.

\begin{table}
  \caption{Summary of available $\gp{\tau^-}$ decay models in
    \pyt. The $\gp{\nu_\tau}$ is omitted from the decay products for
    brevity and charge conjugation is implied for $\gp{\tau^+}$
    decays.\label{tab:dec:tau}\index{meMode@\texttt{meMode}}}
  \begin{center}
    \begin{tabular}{c|cc|l}
      \toprule
      \multicolumn{1}{c}{mult.}
      & \multicolumn{1}{c}{ref.}
      & \multicolumn{1}{c}{\texttt{meMode}}
      & \multicolumn{1}{c}{decays} \\
      \midrule[\heavyrulewidth]
      $2$
      &
      & 1521
      & $\pi^-$, $K^-$ \\
      \midrule
      \multirow{3}{*}{$3$}
      & 
      & 1531
      & $e^- \gp{\bar{\nu_e}}$, $\mu^- \gp{\bar{\nu_\mu}}$ \\
      & \cite{Kuhn:1990ad}
      & 1532
      & $\pi^0 \pi^-$, $K^0 K^-$, $\eta K^-$ \\
      & \cite{Finkemeier:1996dh}
      & 1533
      & $\pi^- \bar{K}^0$, $\pi^0 K^-$ \\
      \midrule
      \multirow{6}{*}{$4$}
      & \cite{CLEO:1999rzk}
      & 1541
      & $\pi^0 \pi^0 \pi^-$, $\pi^- \pi^- \pi^+$ \\
      & \multirow{2}{*}{\cite{Finkemeier:1995sr}}
      & \multirow{2}{*}{1542}
      & $K^- \pi^- K^+$, $K^0 \pi^- \bar{K}^0$, $K_S^0 \pi^- K_S^0$,
         $K_L^0 \pi^- K_L^0$, $K_S^0 \pi^- K_L^0$, \\
      &
      &
      & $K^- \pi^0 K^0$, $\pi^0 \pi^0 K^-$, $K^- \pi^- \pi^+$,
        $\pi^- \bar{K}^0 \pi^0$ \\
      & \multirow{2}{*}{\cite{Decker:1992kj}}
      & \multirow{2}{*}{1543}
      & $\pi^0 \pi^0 \pi^+$, $\pi^- \pi^- \pi^+$, $K^- \pi^- K^+$,
        $K^0 \pi^- \bar{K}^0$, $K^- \pi^0 K^0$, \\
      &
      &
      & $\pi^0 \pi^0 K^-$, $K^- \pi^- \pi^+$, $\pi^- \bar{K}^0 \pi^0$,
        $\pi^- \pi^0 \eta$ \\
      & \cite{Jadach:1993hs}
      & 1544
      & $\gamma \pi^0 \pi^-$ \\
      \midrule
      $5$
      & \cite{Bondar:2002mw}
      & 1551
      & $\pi^0 \pi^- \pi^- \pi^+$, $\pi^0 \pi^0 \pi^0 \pi^-$ \\
      \midrule
      $6$
      & \cite{Kuhn:2006nw}
      & 1561
      & $\pi^0 \pi^0 \pi^- \pi^- \pi^+$, $\pi^0 \pi^0 \pi^0 \pi^0 \pi^-$,
        $\pi^- \pi^- \pi^- \pi^+ \pi^+$ \\
      \bottomrule
    \end{tabular}
  \end{center}
\end{table}

Here, a brief description of the available $\gp{\tau}$ decays is provided; more
details can be found in \citeone{Ilten:2013yed} with a summary given
in \cref{tab:dec:tau}. Note that the ordering of the particles matters,
and whenever numerical indices are used, $0$ is the decaying
$\gp{\tau^-}$ while $\gp{nu_\tau}$ has index $1$.  For two-body decays
into a neutrino and pseudoscalar meson, $\gp{tau^-} \to \gp{\nu_\tau}
P$, the hadronic current is given by,
\begin{equation}
  J^\mu \propto p_2^\mu ~.
\end{equation}
The current for the fully leptonic three-body decay $\gp{\tau^-} \to
\gp{\nu_\tau} \gp{\ell^-} \gp{\bar{\nu_\ell}}$ is
\begin{equation}
  J^\mu = \bar{u}_2 \gamma^\mu (1 - \gamma^5) v_3 ~.
\end{equation}
Three-body decays with hadronic states can proceed via vector and scalar
currents,
\begin{align*}
  J^\mu \propto~ & \frac{c_v}{\sum_i {w_v}_i} \bigg( (p_3 -
  p_2)^\mu\sum_i {w_v}_i
  BW_p(m_2, m_3, s_2, {m_v}_i, {\Gamma_v}_i) \\
  & - s_1 (p_2 + p_3)^\mu \sum_i \frac{{w_v}_i
    BW_p(m_2, m_3, s_2, {m_v}_i, {\Gamma_v}_i)}{{m_v}_i^2}
  \bigg) \\ 
  & + \frac{c_s }{\sum_j {w_s}_j} (p_2 + p_3)^\mu \sum_j {w_s}_j
  BW_s(m_2, m_3, s_2, {m_s}_j,
  {\Gamma_s}_j) ~,
\end{align*}
where ${w_s,v}_i$ are complex weights for each vector and scalar
current, $c_{s,v}$ are the scalar and vector couplings, and $BW_p$ is
a $P$-wave Breit--Wigner. The final state determines the relevant
couplings and weights to use. The general form of the hadronic current
for four-body decays is given by,
\begin{align*}
  J^\mu \propto &\left(g^{\mu\nu} - \frac{q^\mu q^\nu}{s_1}\right)
  \bigg((F_3 - F_2) p_2 + (F_1 - F_3) p_3 + (F_2 - F_1) p_4
  \bigg)^\mu \\
  & + F_4 q^\mu + iF_5 \epsilon^\mu(p_2, p_3, p_4) ~,
\end{align*}
where each $F_i$ is a model specific form factor and $\epsilon$ is the
permutation operator.

The hadronic current for the decay $\gp{\tau^-} \to \gp{\nu_\tau} \gam
\piz, \pim$ is given by~\cite{Jadach:1993hs}
\begin{align*}
  J^\mu \propto~ & F(s_1, \vec{m}_\rho, \vec{\Gamma}_\rho, \vec{w}_\rho) 
  F(0, \vec{m}_\rho, \vec{G}_\rho, \vec{w}_\rho) 
  F(s_4, \vec{m}_\omega, \vec{G}_\omega, \vec{w}_\omega) \\
  & \bigg(
  \varepsilon_2^\mu\left(m_{\pi^-}^2{p_4}_\nu p_2^\nu -
    {p_3}_\nu p_2^\nu({p_4}_\nu p_3^\nu - {p_4}_\nu p_2^\nu)\right) \\ &
  - {p_3}^\mu \left(({p_3}_\nu \varepsilon_2^\nu)({p_4}_\nu p_2^\nu)
    - ({p_4}_\nu \varepsilon_2^\nu)({p_3}_\nu p_2^\nu) \right) \\ &
  - {p_2}^\mu \left(({p_3}_\nu \varepsilon_2^\nu)({p_4}_\nu p_3^\nu)
    - ({p_4}_\nu \varepsilon_2^\nu)(m_{\pi^-}^2 + {p_3}_\nu p_2^\nu) \right)
  \bigg) ~,
\end{align*}
where $F$ is a sum over the possible vector currents including
$\gp{\rho}$ and $\gp{\omega}$ resonances. The five-body decays depend
on sub-currents for each allowed resonance~\cite{Bondar:2002mw,
  Golonka:2003xt},
\begin{align*}
  J^\mu_{\pi^0\pi^0\pi^0\pi^-} & \propto J^\mu_{0,a_1 \rightarrow
    \rho
    \pi} + J^\mu_{0,a_1 \rightarrow \sigma \pi} \\
  J^\mu_{\pi^0\pi^-\pi^-\pi^+} & \propto J^\mu_{-,a_1 \rightarrow
    \rho \pi} + J^\mu_{-,a_1 \rightarrow \sigma \pi} +
  J^\mu_{-,\omega
    \rightarrow \rho \pi} ~,
\end{align*}
and are based in the Novosibirsk model. The six-body decay
model~\cite{Kuhn:2006nw} can be written as a summation of $a$ and
$b$-type currents,
\begin{equation}
  J^\mu \propto \sum J_a^\mu + \sum J_b^\mu ~,
\end{equation}
where each term is one of the possible final state permutations. The
$a$-type currents proceed through a $\gp{a_1} \to \gp{\omega}
\gp{\rho}$ resonance structure, while the $b$-type proceed via a
$\gp{a_1} \to \gp{\sigma} \gp{a_1}[\to \gp{\rho} \gp{\pi}]$ structure.

%% file: using-pythia/introduction.tex
\pythia provides comprehensive choices for modelling all kinds of physics effects in collision experiments, 
as can be seen from the bulk of this manual. It is often not necessary to know the details of all components 
to start using the program to calculate useful quantities, however.  The descriptions provided in this section
 should allow a new user to set up and use \pyt for most standard model and new physics processes, using 
default settings for showers, MPIs, and hadronization that have been tested to 
work at the LEP and LHC experiments. By extension, it should also be useful in many other contexts. 
All settings corresponding to particular physics models or to changing the ``tunes'' (\ie parameter 
fitting for showers, MPIs, and hadronization) are documented in the \texttt{HTML} \htmlmanual, which is also distributed 
in the \texttt{share/Pythia8/htmldoc/} directory of the released source code. One can begin browsing from 
the \texttt{Welcome.html} home page of that directory.

\pyt is under constant and active development. Therefore, any specific detail of this article can become 
obsolete soon after it is released. We therefore urge users seeking specific information to:
\begin{itemize}
        \item Make sure to read the most recent version of this manuscript, in conjunction with the most recent code version. Some information, which may have been correct when the manuscript was obtained, may be outdated when being read.
        \item Consult the \texttt{HTML} manual, which always contains specific settings and reasonable defaults for all physics processes, as well as suggestions for analyses. It also contains a detailed change-log documenting updates between code versions.
        \item Use the examples distributed with the working version of \pyt for inspiration. Examples are kept up to date, and should always correspond to the program version downloaded.
\end{itemize}
 
Past and present code versions, documentation, some relevant presentations, and more can be found at the \pyt website: 
\begin{center}
        \url{https://www.pythia.org/}
\end{center}
It is continuously kept up to date.

In \cref{sec:using-stand-alone} we will describe the logic behind using \pyt as a library to write a stand-alone analysis, and in section \cref{section:intext} we describe interfacing to external programs.

%% file: using-pythia/stand-alone.tex
\section{Using \pyt stand-alone}

\label{sec:using-stand-alone}
The default way of using \pyt, is to use it as a \cpp library, and write ``\texttt{main}'' programs performing the desired simulation tasks. This can be done completely stand-alone, as \pyt in principle contains everything needed for a complete physics analysis. Several such example \texttt{main}'s are shipped with \pyt in the \texttt{examples/} sub-directory. In the following we will describe and exemplify how such user code can be written, and then go on to give more advanced use cases, covering deeper interactions with the simulation than allowed from an example \texttt{main}.

\subsection{Installation}\index{Installing Pythia@Installing \pyt}
The latest version of \pythia (as well as older versions) can be downloaded from \url{https://www.pythia.org/} as a gzipped tarball \texttt{pythia83XX.tgz}.  On Unix, Linux, or MacOS systems this can be unzipped with 
\begin{codebox}
tar -xvfz pythia8307.tgz
\end{codebox}
\noindent (On Windows systems, we recommend to install a virtual
machine running Linux, \cf\eg,
\href{https://ubuntu.com/tutorials/how-to-run-ubuntu-desktop-on-a-virtual-machine-using-virtualbox}{this
  tutorial}.) The simplest installation can be made using the
standard commands 
\begin{codebox}
./configure \\
make
\end{codebox}
\noindent Configuration options (especially for linking against external libraries) can be found by typing
\begin{codebox}
./configure ----help
\end{codebox}
\noindent Details can also be found in the \texttt{README} file distributed with \pythia.  If an install location is specified with \texttt{----prefix}, then \texttt{make install} will copy libraries, headers, and shared documentation to that location in the standard Unix/Linux hierarchy. Details of the configuration can be accessed either via the generated \texttt{Makefile.inc} file or the \texttt{pythia8-config} script in the \texttt{bin} directory.

Most users would then change to the \texttt{examples/} sub-directory, find a suitable example to use as a template for their analysis, modify the desired parts, and compile and run the examples (say, \texttt{main01}) by:
\begin{codebox}
make main01 \\
./main01
\end{codebox}
\noindent It is, however, also possible to compile and run \pyt programs outside the \texttt{examples/} directory. Three environment variables could be potentially useful, providing the paths to the compiled libraries, and to the settings and particle properties databases,
\begin{codebox}
\begin{verbatim}
PYTHIA8PATH = <set to head Pythia directory>
PYTHIA8DATA = $PYTHIA8PATH/share/Pythia8/xmldoc
LD_LIBRARY_PATH = $PYTHIA8PATH/lib:$LD_LIBRARY_PATH
\end{verbatim}
\end{codebox}

\subsection{Program setup}
The simplest \pythia user code comprises three main sections --- initialization, the event loop, and final statistics.  A skeleton of a simple program is as below. Note that the skeleton program should compile but not produce any reasonable output, as no reasonable settings are read in.

\begin{codebox}
\begin {verbatim}
#include "Pythia8/Pythia.h" // access to Pythia objects.
using namespace Pythia8;    // allow simplified notation. 

void main() { 

 // --- Initialization ---
  
  Pythia pythia;     // Define Pythia object.
  Event& event = pythia.event; // quick access to current event. 

  // Read in settings 
  pythia.readString("..."); //  line by line...
  pythia.readFile("cardfile.cmnd");   // or via file. 
  
  // Define histograms, external links, 
  // local variables etc. here. E.g.
  int maxEvents = 1000; // The number of events to run.

  pythia.init();   // Initialize
  
  // --- The event loop ---
  
  for(int iEvent = 0; iEvent < maxEvents; iEvent++){ 
 
    // Generate next event; 
    // Produce the next event, returns true on success.  
    if(!pythia.next()) { 
      // Any error handling goes here.	
    }
	
    // Analyse event; fill histograms etc.
	
  } // End event loop.
  
  // ---  Calculate final statistics --- 
  pythia.stat(); 
  
  // Print histograms etc.
  
  return;
}
\end{verbatim}
\end{codebox}

\subsection{Settings}

The internal \pythia event generation is divided into three steps:
\begin{itemize}
	\item Process level, dealing with the hard process.
	\item Parton level, dealing with showers, MPIs,
        colour reconnection, and beam remnants.
	\item Hadron level, dealing with hadronization and further decays
        of the particles produced.
\end{itemize}
Naturally, there are specific settings to control each of these levels. Aside from this, 
there are several classes of settings to address output during initialization and generation
of each event. In the following, we give an overview 
of how these may be used. However, the reader should consult the \htmlmanual
(also accessible from \texttt{share/Pythia8/htmldoc/\-Welcome.html} distributed with the release) 
for a full listing of all available settings and options. Note that all possible setting keys are indexed, and can be searched via the \texttt{Search Docs} box in the upper-right corner of the page.

It is possible to run \pythia entirely with the default settings. The only minimal user input required is the choice of production process. As a default, the incoming beams are both protons with a centre-of-mass energy of 14~TeV with the parton distribution function set to the NNPDF2.3 QCD+QED LO $\alphas(M_Z) = 0.130$ one~\cite{Ball:2013hta}. Furthermore, initial- and final-state radiation is turned on, using the internal \pyt simple shower. MPIs and hadronization are both on by default, and all unstable hadrons with $c\tau_0 < 1000$~mm are decayed to stable ones. The default tune is the Monash 2013 one~\cite{Skands:2014pea}, see \cref{sec:monash-tune}.

\pythia collects settings performing related functions into groups (\eg overarching parton-level settings are named \setting{PartonLevel:*}).  Input strings for changing settings have the form 
\begin{codebox}
settingGroup:nameOfSetting = value
\end{codebox}
\noindent For example, decays of all resonances can be turned off by setting 
\begin{codebox}
\settingval{ProcessLevel:resonanceDecays}{off} 
\end{codebox}
\noindent \pythia supports four different types of settings:
\begin{itemize}
\item \texttt{flag} is a boolean \texttt{true} or \texttt{false}.  Acceptable input alternatives include \texttt{on/off}, \texttt{yes/no}, and \texttt{1/0}.
\item \texttt{mode} is an integer switch enumerating either available options or a wider range of values.  Acceptable values are integers.
\item \texttt{parm} is a real number parameter.
\item \texttt{word} is a character string. It cannot contain single or double quotation marks, or curly braces, \ie \{ \}.
\end{itemize}
It is further possible to have a vector of each of these types. If necessary,
users can define their own settings that can then be used in their code. 

The user can read in settings in one of two ways: either line-by-line with
\texttt{pythia.readString()} calls inside the user \cpp code, or by providing a
plain-text file that is read at run time.  The latter has the advantage of
not requiring a recompilation every time a change is made. It is triggered by
\begin{codebox}
pythia.readFile("cardfile.cmnd");
\end{codebox}
\noindent inside the code.

All settings have reasonable default values enabled, and can furthermore be
defined with maximal and/or minimal values beyond which they cannot be
changed. These can be studied in the \htmlmanual under the
respective parameter. A parameter can be forced outside the allowed bound by
using the keyword \texttt{force}, for example:
\begin{codebox}
PhaseSpace:pTHatMinDiverge force= 0.1
\end{codebox}
\noindent will force the parameter \setting{PhaseSpace:pTHatMinDiverge}, which usually 
has a minimal value of 0.5~GeV, to 0.1~GeV. \emph{The \texttt{force} keyword should be used 
with extreme caution!} The boundaries are there for a reason, and breaking them can make 
the program unstable or invalidate the physics model.

If nothing else is mentioned explicitly, dimensional parameters have units of GeV for energy, 
momentum, and mass, and mm for length and time, with the speed of light $c = 1$ implicit. 
Internal cross sections are book kept in mb, but communication with other programs may 
require conversion from/to other units. 

\subsubsection{Beams and PDFs}\index{PDFs}

The incoming beams are set by providing the PDG code of the incoming
particles to \setting{Beams:idA} and \setting{Beams:idB} (the default
for both is proton \ie \texttt{2212}).  For example, a \ppbar collision
can be set by changing the value of \setting{Beams:idB} to
\begin{codebox}
Beams:idB = -2212
\end{codebox}
\noindent An $\epem$ collision can be set by \settingval{idA}{11}
and \settingval{idB}{-11}. Currently available beams
include protons (2212), neutrons (2112), pions ($\pm 211, 111$), most other
light hadrons (but not necessarily all combinations of them),
electrons (11), muons (13), photons (22), and several heavy-ion species.
The collision energy can then be set by
\begin{codebox}
\settingval{Beams:eCM}{2000.} 
\end{codebox}
\noindent Units of GeV are implicit, as already mentioned. For heavy-ion collisions, 
this is the energy per nucleon-nucleon collision, as per the usual heavy-ion conventions.

By default, collisions are assumed to be in the CM frame. Other options can be set 
with \setting{Beams:frameType}. Using option \settingval{Beams:frameType}{2}
the beam energies can be set separately and \eg a HERA-like beam configuration
can be obtained with
\begin{codebox}
\settingval{Beams:frameType}{2}\\
\settingval{Beams:idA}{2212}\\
\settingval{Beams:eA}{920.}\\
\settingval{Beams:idB}{-11}\\
\settingval{Beams:eB}{27.5}
\end{codebox}
\noindent Furthermore, the beams do not need to be back-to-back but option
\settingval{Beams:frameType}{3} allows for setting also some transverse momentum for the
beams. A particularly useful setting to automatically set beam information when using
external LHE files (see \cref{subsection:lha} for details) is 
\begin{codebox}
\settingval{Beams:frameType}{4}
\end{codebox}
\noindent It is also possible to specify a simple Gaussian spread of incoming beam momentum
and of the interaction vertex position. These can be set by
\setting{Beams:allowMomentumSpread} and \setting{Beams:\-allowVertexSpread} and their
accompanying parameters in the $x,y$, and $z$ directions for each beam.

\index{PDFs}
The applied proton PDF set can be selected with setting \setting{PDF:pSet} which is also
applied for antiprotons and neutrons via isospin symmetry. By default, this sets PDFs
to be the same for beam $A$ and $B$ but it is also possible to set the PDFs for beam $B$
separately using option \setting{PDF:pSetB}. The internal PDF sets can be selected by
setting an integer value for the above options, \eg the current default is set with
\begin{codebox}
\settingval{PDF:pSet}{13}
\end{codebox}
\noindent To use LHAPDF grids instead, \pyt needs either be linked to the LHAPDF library or
one can use the internal implementation for the LHAPDF grid interpolation, see see \cref{sec:interface:lhapdf}. In the first
case, the set is defined with a string 
\setting{LHAPDF6:set/member}, \eg
\begin{codebox}
\settingval{PDF:pSet}{LHAPDF6:NNPDF23\_lo\_as\_0130\_qed/0}
\end{codebox}
\noindent which would correspond to the current default above. Also LHAPDF version 5 is
supported and enabled with keyword \setting{LHAPDF5:set/member}. The internal
interpolation for the LHAPDF~6 format is enabled with \setting{LHAGrid1:filename} and with
this, the default PDF can be obtained with
\begin{codebox}
\settingval{PDF:pSet}{LHAGrid1:NNPDF23\_lo\_as\_0130\_qed\_0000.dat}
\end{codebox}
\noindent The grid file should be located in the folder
\texttt{share/Pythia8/xmldoc} or an absolute path should be provided. These settings change the PDF used throughout the program,
including hard-process generation, MPIs, and ISR. To keep the underlying event
description intact, one can also change the PDFs only for the hard processes by setting
\settingval{PDF:useHard}{on} and selecting the hard PDFs with \setting{PDF:pHardSet}.
All the above options can be used to select the PDFs for hard processes and one can
also include nuclear modifications for these with \settingval{PDF:useHardNPDFA}{on}
or \settingval{PDF:useHardNPDFB}{on}. Similarly, one can select PDFs for other beam types
including pions, pomerons, photons, and leptons, see the
\htmlmanual and \cref{sec:hadronPDFs} for further details.

\subsubsection{Process selection}
\index{Hard Process!setting}
The minimal initialization information required by \pythia to generate
events is which process(es) are to be run. This is done by turning on the
relevant flags. For example, to generate a $\g\g \to \q\qbar$ hard
process, set
\begin{codebox}
\settingval{HardQCD:gg2qqbar}{on}
\end{codebox}
\noindent A full list of internally defined processes is available
in \cref{sec:hardProcesses}.

It is possible to turn on more than one
process at a time. \pythia will then generate events for each process in
proportion to their cross sections. Some extra switches are also available
for processes that are often grouped together, \eg 
\begin{codebox}
\settingval{HardQCD:all}{on}
\end{codebox}
\index{Phase-space generation!Cuts}\noindent will turn on all QCD $2 \to 2$ quark/gluon production processes. Since these 
processes are divergent in the $\pT \to 0$ limit, it is necessary to introduce a 
lower transverse-momentum cutoff \setting{PhaseSpace:pTHatMin}. Note that such a parton-level 
cut does not directly translate into a cut on jet properties, since intermediate parton showers, 
MPIs, hadronization effects, and jet finders will distort the original simple process. Further details are available in \cref{subsection:hardQCD,subsection:phasespacecuts}.

Several choices of renormalization and factorization scale are
available. For \mbox{$2\rightarrow2$} processes, these can be set via
\setting{SigmaProcess:renormScale2} and
\setting{SigmaProcess:factorScale2}, respectively. The default for the
renormalization scale is the geometric mean of the squared transverse
masses of the two outgoing particles. The default of factorization
scale is set at the smaller of the two squared transverse masses. The possible options
are listed in \cref{subsection:couplingsscales}.

\subsubsection{Soft processes}

The bulk of the total cross section in high-energy hadronic collisions
is not associated with a visible hard process. A reasonably
complete and consistent description of these relevant processes is instead
obtained with 
\begin{codebox}
\settingval{SoftQCD:all}{on}
\end{codebox}
\noindent This includes elastic, single- and double-diffractive, and
non-diffractive processes, which alternatively could be switched on
individually. The inelastic processes, \ie the diffractive and
non-diffractive ones, include a modelling of MPIs, which does include
a tail of high-$\pT$ processes. Thus \setting{HardQCD:all} becomes a subset
of the \setting{SoftQCD:all} total cross section, and one should not mix
\setting{SoftQCD} and \setting{HardQCD} processes. Colour screening ensures
that the hard processes here are damped appropriately at low $\pT$
values, as described in \cref{subsection:mpi}. 

At very low collision energies the perturbative processes are gradually
phased out and only truly soft processes remain. This occurs \eg in
hadronic rescattering, or in the final stages of the evolution of a
cosmic-ray cascade in the atmosphere. To simulate low-energy collisions
directly, use 
\begin{codebox}
LowEnergyQCD:all = on
\end{codebox}
\noindent or related \setting{LowEnergyQCD:*} flags to turn on only a subset
of the available processes. These are assumed to be accurate below 10~GeV.
It is possible to simultaneously turn on both \setting{LowEnergyQCD:*} and
\setting{SoftQCD:*} processes, in which case a mix of the two is used at
intermediate energies.

A number of other processes are available, including numerous non-QCD
processes which may not be applicable for proton beams. See
\cref{sec:SMprocesses} for a complete list of included standard-model
processes, and \cref{sec:BSMprocesses} for a list of BSM processes.

\subsubsection{Parton- and hadron-level settings}

The primary switches for parton showers are
 \begin{codebox}
\settingval{PartonLevel:ISR}{on|off} \\
\settingval{PartonLevel:FSR}{on|off}
\end{codebox}
\noindent\pythia has two other showers available, aside from the ``simple showers''. The choice of shower model can be performed with 
\begin{codebox}
\settingval{PartonShowers:model}{1|2|3}
\end{codebox}
\noindent where the default (\texttt{1}) corresponds to the ``old'' simple shower,  (\texttt{2}) corresponds to \vincia, and  (\texttt{3}) to the \dire shower.

\noindent Finally, the primary switch for hadronization is 
 \begin{codebox}
\settingval{HadronLevel:all}{on|off}
\end{codebox}

\subsubsection{Particle data}
\label{sub:SettingParticleData}\index{Particle data}
All known information regarding particles (mass, charge, decay width, branching fractions, \etc) is stored within the \texttt{ParticleData} class.  Each particle has the following basic properties:
\begin{itemize}
\item \texttt{id} holds the PDG identity number of the particle. 
\item \texttt{name} is a string containing the name of the particle. Particle and antiparticle names
  are stored separately, with \texttt{void} returned when no  antiparticle exists.
\item \texttt{spinType} in the form of an integer equal to $(2s +1)$.
\item \texttt{chargeType} is three times the electric charge.
\item \texttt{colType} is the colour representation (\texttt{0:} uncoloured,  \texttt{(-1)1:} (anti-) triplet, \texttt{2:} octet,  \texttt{(-3)3:} (anti-) sextet).
\item \texttt{m0} is the nominal mass in GeV.
\item \texttt{mWidth} is the Breit--Wigner width in GeV.
\item \texttt{mMin, mMax} are the limits for mass generated by the Breit--Wigner.
\item \texttt{tau0} is the proper lifetime in mm.
\item \texttt{mayDecay} sets whether the particle is allowed to decay.
\item \texttt{isVisible} sets whether the particle is to be considered visible by the detector. 
\end{itemize} 
Other than these, there are a few special properties related to external decays which can be found in the \htmlmanual. Any property of a particle can be changed by setting:
 \begin{codebox}
\texttt{NN:Property = value} 
\end{codebox}
\noindent where \texttt{NN} is the PDG ID of the particle.
 
\index{Particle decays}
The next critical piece of information for a particle is its decay table. The decay table is comprised of decay modes (or decay channels), each of which has the following properties:
\begin{itemize}
\item \texttt{onMode} sets whether this decay channel is open where \texttt{0} is off, \texttt{1} is on, \texttt{2} on for the particle but not for the antiparticle, and \texttt{3} is on for the antiparticle but not for particle.
\item \texttt{bRatio} sets the branching ratio for the channel.
\item \texttt{meMode} sets how this decay is handled, in particular whether internal matrix element reweighting is available to account for mass or angular correlations.  The default is \texttt{0} and corresponds to flat phase space. See \cref{dec:summary} for available matrix-element modes for particles and \cref{prop:width} for available matrix-element modes for resonances.
 \item \texttt{multiplicity} sets the number of daughters, the maximum allowed is eight.
 \item \texttt{product(i)} is an array that holds the PDG IDs of the daughter particles; empty slots are set to zero.
\end{itemize}

Several shortcuts exist to quickly set up the decay table of a particle.  For example, deleting the existing decay table to start anew can be done by using the following.
 \begin{codebox}
NN:oneChannel = onMode bRatio meMode product1 product2 ...  \\
NN:addChannel = onMode bRatio meMode product1 product2 ...
\end{codebox}
\noindent Branching fractions are automatically rescaled such that the sum is one. Certain modes can be turned on or off based on the identity of the products by using the following shortcuts
 \begin{codebox}
NN:offIfAny   = product1 product2 ...\\ 
NN:onIfAny    = product1 product2 ... \\
NN:onPosIfAny = product1 product2 ... \\
NN:onNegIfAny = product1 product2 ...
\end{codebox}
\noindent This turns on the mode if any of the products in the list
matches one in the \texttt{product(i)} array. Note that \texttt{onPos...}
(\texttt{onNeg...}) above means that setting only applies to the
decays of the (anti)particle. Further shortcuts (to select based on
matching all products \etc) can be found in the \htmlmanual.

Adding new particles can be done either by directly calling
\texttt{ParticleData::addParticle} from the program or by using the SLHA interface with a \texttt{QNUMBERS} block
(\cf\cref{sec:slha}).

\subsection{Analysis of generated event \label{sec:eventAnalysis}}

A generated ``event'' is essentially a list of particles --- initial, final, or intermediate --- that are generated sequentially based on probabilistic calculations. A user will mostly be interested in studying kinematic variables constructed from the momenta of initial or final-state particles. The following three classes will be useful in constructing such variables.  The full list of available classes and methods for analysing an event is available in the \htmlmanual.

\subsubsection{The \texttt{Vec4} class}\index{4-vectors}\index{Vec4@\texttt{Vec4}}

The \texttt{Vec4} class is designed to hold the four-momentum (or indeed any other four-vector quantity that may be needed) of the particles in the collision event.  Some useful methods are
\begin{itemize}
\item \texttt{px(), py(), pz(), e()} to access the individual components.
\item \texttt{mCalc()} for calculated mass $\sqrt{E^2 - p_x^2 - p_y^2 -p_z^2}$.
\item \texttt{pT()} and \texttt{pAbs()} for the transverse momentum and the absolute value of the three-momentum, respectively.
\item \texttt{theta(), eta(), phi()} for the polar and azimuthal angles, rapidity and pseudorapidity, respectively.
\item \texttt{rot(double theta, double phi) } to rotate the three-momentum.
\item \texttt{bst(const Vec4\& p)} and \texttt{bstback(const Vec4\& p) } to boost the current vector by $\vec \beta = \pm \frac{\vec p}{E}$.
\end{itemize}

\subsubsection{The \texttt{Particle} class}\index{Particle properties@\texttt{Particle}}

The \texttt{Particle} class forms the fundamental particle unit, multiples of which are assembled in the form of an ``event''.  Each \texttt{Particle} has the following properties:
\begin{itemize}
\item \texttt{id()} for the PDG code.
\item \texttt{status()} for the status of the current particle (initial, final, stable, or intermediate \etc, see the \htmlmanual for the full status codes).  For most users, the only relevant check is, if the number is greater than zero, which denotes a stable, final-state particle. This can also be determined directly by asking \texttt{isFinal()}.
\item \texttt{p()} returns a four-vector whereas \texttt{px(), py(), pz(), e()} can be used directly to access components.
\item \texttt{mother1(), mother2()} refer to the indices of the first and last mother, with several special rules.
   \texttt{motherList()} returns a vector of all the mother indices, circumventing the need to know these rules.
\item \texttt{daughter1(), daughter2()} refer to the indices of the first and the last daughter, with several special rules (all contiguous indices in between are daughters of said particle). \texttt{daughterList()} returns a vector of all the daughter indices, circumventing the need to know these rules
\item \texttt{vProd()} for the production four-vertex.
\item \texttt{tau() } is the lab-frame lifetime in mm/c.
\end{itemize}

\subsubsection{The \texttt{Event} class}
\index{Event record}

Finally, we come to the main result of the program which is held in a class called \texttt{Event}, representing a collision event. It contains a dynamic array (\texttt{vector}) of particles along with helper methods that are useful to extract information from the array. A single \texttt{Pythia} instance contains two \texttt{Event}s, called \texttt{process} and \texttt{event}.  The first of these, \texttt{process}, contains only the hard process whereas the second \texttt{event} contains the full history of the collision event.  The user usually does not need to manually add or remove particles from either of these arrays. The individual particles can be accessed simply by using their index in the event (\eg \texttt{pythia.event[i]}).  All methods corresponding to the particle then can be accessed \eg \texttt{pythia.event[i].phi()} accesses the azimuthal angle $\varphi$. Some useful methods beyond those given for individual particles are:

\begin{itemize}
\item \texttt{detaAbs(int i1, int i2) } and \texttt{dphiAbs(int i1, int i2)} to obtain $\Delta \eta$ and $\Delta \varphi$ between two particles in the event.
\item \texttt{REtaPhi(int i1, int i2) } for the $R$ distance between two particles.
\end{itemize}
\noindent with more given in the \htmlmanual.

\noindent Several useful functions that take an \texttt{Event} as in input are available to the user to construct important quantities, \eg \texttt{SlowJet} is a sequential clustering algorithm that can be used to form jets from final-state particles whereas \texttt{Sphericity} and \texttt{Thrust} classes calculate these inclusive variable.  

\subsection{Program output}

The most basic level of output that can be requested is a listing of the full event (when inside the event loop), which is done simply by

\begin{codebox}
pythia.event.list()
\end{codebox}
\noindent A printout of the statistics, \ie number of tried and accepted events, as well as the 
number of events produced for each process and the resulting cross section can be obtained by using

\begin{codebox}
pythia.stat()
\end{codebox}
\noindent For hard processes with \eg \pT cuts, the cross section must
be calculated by Monte-Carlo integration. This is done automatically as
events are generated. After generation, the total cross section and 
its statistical error can be accessed by calling respectively:
 \begin{codebox}
 pythia.info.sigmaGen() \\
 pythia.info.sigmaErr()
 \end{codebox}

\noindent Pythia also provides rudimentary built-in histogramming via the \texttt{Hist} class. The main methods of interest are
\begin{itemize}
\item \texttt{Hist(string title, int numberOfBins, double xMin, double xMax, bool logX)} the constructor, defining a histogram.
\item \texttt{fill(double value, double weight = 1.0)} to fill the histogram with an optional weight. 
\item \texttt{table(string fileName, bool printOverUnder = false, bool xMidBin = true)} to\\ output the histogram as a table.
\end{itemize}
Piping the histogram object directly to the standard output (\texttt{std::cout <\,< myhist; }) will also give a rudimentary ASCII output of the histogram. There are methods that will also generate \python \textsc{Pyplot} code for cleaner graphical representations.

\subsubsection{Messages, warnings, and errors}
\index{Error messages}

\pythia provides four basic levels of diagnostic output that are available in the \texttt{Info} class. All such generated output is provided in a summary at the end of each run and can be useful as a sanity check or as debug information. The main categories are:
\begin{description}
\item[Abort] means that something went seriously wrong, either in initialization or generation. In the former case, event generation cannot begin. In the latter case, the event is flawed, and should be skipped. In either case the respective method \texttt{Pythia::init()} or \texttt{Pythia::next} will return \texttt{false}, to allow the user to react. There are occasions where an abort may be deliberate, such as when a file of Les-Houches events is read and the end of the file is reached.
\item[Error] typically means that something went wrong during event generation, but the program will backup and try again. In cases where this is not possible, a separate \textbf{Abort} will be issued. A typical run can issue several errors, without it being a problem, unless the program aborts. If encountering unusually many errors, it can be a good idea to check if any run parameters are set to unreasonable values, making a calculation unable to converge. The user can set the maximum number of errors to allow before the entire run is aborted via the \setting{Main:timesAllowErrors} parameter.
\item[Warning] is less severe. Typically the program will try again with a good chance of success. Usually no action needs to be taken by the user.
\item[Message] represents informative outputs that confirm \eg reading of an external file. Verbosity of messages can be set separately for each module that provides this function (\eg \setting{SLHA:verbose} can be set to zero for a silent read.)
\end{description}

\subsection{Advanced settings examples}

The example use cases given  above, are enough for performing simple tasks with \pythia. In most cases, however, when a user wants to apply specialized built-in physics capabilities, the application is more complicated, generally scaling with the complexity of the required tasks. For this purpose, \pythia ships with a large number of examples (in the \texttt{examples/} sub-directory), intended to showcase various applications.

In this section we provide a thorough explanation of two such advanced use
cases, to highlight the versatility of the distributed code.
Settings for matching and merging are presented in \cref{sec:standalone:matchmerge},
while \cref{sec:standalone:variableBeams} discusses options for changing
the beam configuration on an event-by-event basis. 

\subsubsection{Matching and merging settings}\label{sec:standalone:matchmerge}
\index{Matching and merging}

\pythia offers implementations of a large variety of matching and merging schemes. This allows both flexibility, but crucially also cross-checks of the results of combining fixed-order perturbative calculations with the event-generator machinery.

\paragraph{\powheg matching}\index{powheg@\powheg} allows for the
combination of specialized next-to-leading order calculations with
\pythia. To facilitate the matching, \pythia offers vetoed parton
showers via so-called \texttt{PowhegHooks}. This tool is available for
the default showers as well as \vincia\footnote{For \vincia,
  \texttt{PowhegHooks} should be swapped for
  \texttt{PowhegHooksVincia}. All settings listed here retain their
  importance with \vincia. More details can be found in
  \citeref[appendix A]{Hoche:2021mkv}.} and prevents the over counting of emissions. It can be enabled with the setting:
\begin{codebox}
	 \settingval{POWHEG:veto}{0|1}
\end{codebox}
\noindent Since it is not strictly guaranteed that the first shower emission can be considered the hardest emission according to the \powheg criteria, the number of emissions to be subjected to vetoed showering can be adjusted by:
\begin{codebox}
	\settingval{POWHEG:vetoCount}{value}
\end{codebox}
\noindent Furthermore, vetoed showering only needs to be applied to Born-type configurations, which can be tagged by the minimal number of partons in the process:
\begin{codebox}
	\settingval{POWHEG:nFinal}{value}
\end{codebox}
\noindent Vetoed showering relies on comparing the hardness of an emission to an allowed maximal hardness. The definition of ``hardness'' is determined by:
\begin{codebox}
	\settingval{POWHEG:pTdef}{0|1|2}
\end{codebox}
\noindent Values other than $1$ are discouraged. The definition of the ``maximal hardness'' can be adjusted with:
\begin{codebox}
	\settingval{POWHEG:pThard}{0|1|2}
\end{codebox}
\noindent where values other than $0$ only serve testing purposes. Finally, the setting:
\begin{codebox}
	\settingval{POWHEG:pTemt}{0|1|2}
\end{codebox}
\noindent determines for which sets of particles the hardness comparison should be applied, with a value of $0$ strongly recommended. A few further, more advanced, settings are listed in the \htmlmanual.

\paragraph{\mcatnlo matching} \index{NLO matching}\index{MCatNLO@\mcatnlo} employs shower-specific fixed-order calculations, which handle the overlap between shower and fixed-order calculation by explicit subtraction. When interfacing these calculations, it is paramount to guarantee consistency of settings between the fixed-order calculation and the parton shower, for any aspects that might have an impact at the NLO level.
No new settings need be introduced in \pyt. The relevant settings to produce consistent results depend on the shower and the \mcatnlo provider. When using \mcatnlo inputs with \pyt's simple showers, a minimal set of consistent settings is:
\begin{codebox}
\settingval{SpaceShower:pTmaxMatch}{1} \\
\settingval{SpaceShower:pTmaxFudge}{1} \\
\settingval{TimeShower:pTmaxMatch}{1} \\
\settingval{TimeShower:pTmaxFudge}{1} \\
\settingval{SpaceShower:MEcorrections}{off} \\
\settingval{TimeShower:MEcorrections}{off} \\
\settingval{TimeShower:globalRecoil}{on} \\
\settingval{TimeShower:weightGluonToQuark}{1}  \\
\end{codebox}
\noindent Please refer to the \htmlmanual for further details.

\paragraph{CKKW-L merging} \index{CKKW(-L) merging}allows for the
combination of several multi-jet tree-level fixed-order calculations with each other and the wider \pythia environment. For example, calculations of Drell--Yan lepton-pair production at hadron colliders in association with zero, one, two, or more additional jets can be combined. In this context, ``additional jets'' refers to further QCD partons, as well as $\W$ and $\Z$ bosons, in the case of simple showers and \dire. 

The inputs for multi-jet merging need to be regularized to avoid soft/collinear configurations. The regularization cut also acts as the criterion to distinguish between fixed-order and parton-shower phase space regions --- the so-called merging scale.
If the input events are regulated by a $\kT$ cut, the following flag can be used to interpret the merging scale in terms of the $\kT$ definition:
\begin{codebox}
	\settingval{Merging:doKTMerging}{on|off}
\end{codebox}
\noindent For the simple showers, the merging scale definition may also be set in terms of the shower evolution variable $\pT$ by setting:
\begin{codebox}
	\settingval{Merging:doPTLundMerging}{on|off}
\end{codebox}
It must be emphasized that this option is naturally not available within \vincia's merging.
The simple shower also offers further built-in merging scale definitions, and the option to supply a pointer to a user-defined \texttt{MergingHooks} class to implement new merging scale definitions. The value of the merging scale separating fixed-order and parton-shower regions must be specified via the parameter:
\begin{codebox}
	\settingval{Merging:TMS}{value}
\end{codebox}
\noindent The merging further requires the definition of the ``process'' through the string:
\begin{codebox}
	\settingval{Merging:Process}{value}
\end{codebox}
\noindent where \texttt{value} should identify the particles of the lowest-multiplicity process partaking in the merging. The process is used under the assumption that each event will contain exactly the specified particles, and potentially further particles that are considered as additional radiation. Looser process definitions are possible through the use of ``particle containers'', and the ``guess'' option, see the \htmlmanual. 
Finally, the number of additional jets must be set via:
\begin{codebox}
	\settingval{Merging:nJetMax}{nJets}	
\end{codebox}
\noindent Other settings are documented in the \htmlmanual.

\paragraph{Sector merging}\index{Sector merging}\index{CKKW(-L) merging} The \vincia antenna shower in \pythia comes with its own implementation of the \ckkwl merging algorithm, which differs from the one implemented for the simple showers. The main difference is that \vincia's sector showers are maximally bijective, \ie possess a minimal number of possible histories that lead to any given multi-parton configuration. As such, they are specifically designed for merging with high-multiplicity matrix elements for which the complexity grows factorially with the number of possible shower histories, \cf\cref{sec:VinciaMatchMerge}.

Sector merging may be enabled by using \vincia with its sector shower option turned on\footnote{The sector shower flag is listed only for completeness --- sector showers are switched on in \vincia by default.} and switching on merging:
\begin{codebox}
	\settingval{PartonShowers:model}{2} 
	
	\settingval{Vincia:sectorShower}{on} 
	
	\settingval{Merging:doMerging}{on} 
\end{codebox}
\noindent By default, it is then assumed that the merging scale is defined in terms of \vincia's evolution variable, \cf\cref{sec:Vincia}. Other definitions (such as a \kT regularization) may be used via the appropriate settings listed for the simple showers above.

While the merging-scale value and the number of additional jets must be set in exactly the same way as listed for the simple showers above, an important difference pertains to the syntax of the process definition. Different to the \texttt{Merging:Process} setting in the default merging implementation, the whole string must be encased in curly braces when using \vincia:
\begin{codebox}
	\settingval{Merging:Process}{\{ value \}} 
\end{codebox}
\noindent In addition, particles must be specified one at a time and be separated by a white-space character. The initial and final state should be separated by \texttt{>} and exactly two initial-state particles must be specified. It must be emphasized that a process string in the ``default'' syntax cannot be processed by \vincia and will lead to an abort.

More advanced settings can be found in the \htmlmanual.

\paragraph{UMEPS merging}\index{UMEPS merging} extends \ckkwl tree-level merging, by ensuring that inclusive cross sections for $n$ additional jets are not changed by the inclusion of calculations for $m>n$ extra jets. This is achieved by introducing subtractions that act to remove the effect of higher-multiplicity events from lower-multiplicity inclusive cross sections. As an extension to \ckkwl, \umeps shares the settings of the former. Beyond these settings, the different stages of \umeps merging can be invoked by:
\begin{codebox}
	\settingval{Merging:doUMEPSTree}{on|off}
\end{codebox}
\noindent which yield \ckkwl-reweighted tree-level results (up to small differences in Sudakov reweighting), and by:
\begin{codebox}
	\settingval{Merging:doUMEPSSubt}{on|off}
\end{codebox}
\noindent which produce the necessary subtractions. Depending on the example main program, these two stages may directly be mixed internally, so that only the first setting may be necessary.

\paragraph{NLO merging}\index{NL3 merging}\index{UNLOPS merging} extends the leading-order merging machinery of \pythia with (externally generated) next-to-leading order QCD event samples. As an extension of LO machinery, NLO merging inherits many of the settings of LO merging. The result of NLO merging is an inclusive calculation that recovers NLO QCD accuracy for inclusive cross sections with $n\leq n_\mathrm{NLO}$ additional partons, and LO (QCD) accuracy for inclusive cross sections with $n_\mathrm{NLO}<m\leq n_\mathrm{LO}$ jets. The maximal number of jets for which NLO samples are available ($n\leq n_\mathrm{NLO}$) has to be set by using:
\begin{codebox}
	\settingval{Merging:nJetMaxNLO}{value}
\end{codebox}
\noindent \pythia offers two NLO merging schemes as part of its core code: \nlthree and \unlops. Other NLO merging schemes (such as the FxFx scheme) can be embedded with the help of \texttt{UserHooks}. \nlthree merging is a straight-forward extension of \ckkwl, and mixes augmented \ckkwl-reweighted tree-level events with events from NLO samples. The reweighted LO stage is enabled by using the flag:
\begin{codebox}
	\settingval{Merging:doNL3Tree}{on|off}
\end{codebox}
\noindent while the processing of NLO samples requires setting the switch:
\begin{codebox}
	\settingval{Merging:doNL3Loop}{on|off}
\end{codebox}
\noindent Typically, NLO input samples contain not only NLO corrections, but tree-level contributions as well. If this is the case, then explicit removal of tree-level contributions from the NLO sample is necessary to avoid double counting. This subtraction is enabled by using the flag:
\begin{codebox}
	\settingval{Merging:doNL3Subt}{on|off}
\end{codebox}
\noindent Note that this subtraction is not related to any of the \umeps subtractions, but rather a necessity due to the structure of available inputs. \nlthree only supports the use of the \setting{Merging:\-doPTLundMerging} merging scale definition.

\unlops merging is an extension of \umeps that --- like \umeps at leading order --- ensures that NLO inclusive cross sections are exactly retained, with the help of unitarity subtractions. Due to this,
\unlops merging proceeds in four phases. The reweighting of tree-level inputs is enabled by:
\begin{codebox}
	\settingval{Merging:doUNLOPSTree}{on|off}
\end{codebox}
\noindent while the processing of NLO samples is produced when using:
\begin{codebox}
	\settingval{Merging:doUNLOPSLoop}{on|off}
\end{codebox}
\noindent Both of the former stages should then be accompanied by subtractions to ensure the correctness of the inclusive cross section. Subtractive leading-order samples are produced when using
\begin{codebox}
	\settingval{Merging:doUNLOPSSubt}{on|off}
\end{codebox}
\noindent while subtractive NLO events are enabled by:
\begin{codebox}
	\settingval{Merging:doUNLOPSSubtNLO}{on|off}
\end{codebox}
\noindent Depending on the example main program, these two tree-level-dependent stages, as well as the two NLO-dependent stages, may directly be mixed internally, so that only the first two settings may be necessary in practice. \unlops supports the use of the \setting{Merging:doPTLundMerging} merging-scale definition natively. Other merging-scale definitions (embedded by custom \texttt{MergingHooks} classes) can be enabled by setting
\begin{codebox}
	\settingval{Merging:unlopsTMSdefinition}{value}
\end{codebox}
to a non-zero \texttt{value}.

\subsubsection{Variable energies and beam particles}\label{sec:standalone:variableBeams}

By default, the beam configuration is initialized at one specified energy.
In some cases, however, one may need to generate events across a range of
energies. In \pythia, this feature is enabled by setting
\begin{codebox}
\settingval{Beams:allowVariableEnergy}{on}
\end{codebox}
\noindent When this is enabled, the MPI machinery for \texttt{SoftQCD} will be
initialized at a grid of energies ranging from 10~GeV up to the maximum
energy specified by \setting{Beams:eCM}. This way, interpolation can be used
to efficiently find the relevant coefficients at each particular energy.
(The \texttt{LowEnergyQCD} code is intended for energies below 10~GeV
where MPIs are irrelevant, and no specific initialization is needed.)
Events can then be generated using one of the variant
\texttt{Pythia::next} methods below, corresponding to the kinematics setup
specified by \setting{Beams:frameType}. In other cases, it is also necessary
to change the beam particle types on an event-by-event basis. One example
of a relevant use case is hadronic cascades in a medium like the 
Earth's atmosphere or a particle detector. A number of settings must be
explicitly switched on to enable this feature:
\begin{codebox}
\settingval{SoftQCD:all}{on} \\
\settingval{LowEnergyQCD:all}{o}n \\
\settingval{Beams:allowVariableEnergy}{on} \\
\settingval{Beams:allowIDAswitch}{on} 
\end{codebox}
\noindent This will initialize the MPI machinery for a set of
some 20 different common hadrons. To switch beam
configurations, use one or more of the following variants of the \texttt{Pythia::set} methods
\begin{codebox}
pythia.setBeamIDs( idA, idB = 0) \\
pythia.setKinematics( eCM) \\
pythia.setKinematics( eA, eB) \\
pythia.setKinematics( pxA, pyA, pzA, pxB, pyB, pzB) \\
pythia.setKinematics( pAin, pBin)
\end{codebox}
\noindent that match the \texttt{Beams:frameType} set. After calling these methods, all subsequent events called with \texttt{next} will use the updated configuration, unless the \texttt{set} call was unsuccessful. The first method
preserves the kinematics of the previous event, modulo the change
of masses. In this framework, currently
only $\p/\n/\pbar/\nbar$ is supported for \texttt{idB}. An optional parameter
\texttt{procType} can be passed to \texttt{next}, and is used to generate an event of a particular type, such 
as non-diffractive or single-diffractive on a specified side.

For applications such as cascades in a medium, the decision whether a
variable-type interaction should occur or not must be based on the
relevant cross section. To this end, the parameterizations outlined in
\cref{subsection:sigmaother} and
\cref{subsubsection:lowenergyprocesses} can be accessed by using the
\begin{codebox}
pythia.getSigmaTotal( idA, idB, eCMAB, mixLoHi = 0) \\
pythia.getSigmaPartial( idA, idB, eCMAB, procType, mixLoHi = 0)
\end{codebox}
\noindent methods. Here, the default \texttt{mixLoHi = 0} gives a smooth
interpolation between the low-and high-energy descriptions.

Typically, the MPI initialization is the slowest step also in a normal
LHC run setup, and with variable particles and energies it will take
several minutes. It is possible to speed up the initialization process
by saving the MPI parameterizations to disk. This is done using the
\setting{MultipartonInteractions:reuseInit} option, which can take the
following values:
\begin{itemize}
\item \texttt{0} (default): MPIs are reinitialized every time.
\item \texttt{1}: MPIs are reinitialized and the parameterization is saved to disk.
\item \texttt{2}: The MPI parameterization is loaded from disk. If the data file
does not exist, initialization fails.
\item \texttt{3}: The MPI parameterization is loaded if the file exists, otherwise
it is reinitialized and saved to disk.  
\end{itemize}
When using non-zero values, the file name \texttt{MultipartonInteractions:initFile}
to save/load from must be specified. 

\subsection{Advanced usage}
Often, the user might want to use \pyt to simulate physics effects that are not already implemented in the standard release.  We therefore provide several ways of extending \pyt capabilities.  The event generation process can be interrupted at various points (\eg after hard scattering, after first branching in the parton shower, \etc) by using ``user hooks''.  These can be used to reweight (or veto) events and change distributions accordingly.  Additionally, any extra production process or decay of a new particle can be implemented by inheriting from \pyt classes that provide cross section calculation or decay width calculation machinery. We refer to any processes implemented this way as ``semi-internal''.  Finally, when extending capabilities, one may wish to have run-time user-input information in the same way as \pyt settings.  We therefore provide some placeholder settings as well as methods to add custom settings keys that can be used to accompany any new functionality. 

\subsubsection{User-defined settings}
Should the user require additional settings be provided via a card file, some spares have been made available following the same schema as the normal \pyt settings: three each (\texttt{N = 1, 2, 3}) of boolean flags via \setting{Main:spareFlagN}; integer modes via \setting{Main:spareModeN}; floating point parameters via \setting{Main:spareParmN}; and strings via \setting{Main:spareWordN}. These can all be set in the card file and interpreted by the user to suit their needs. To add completely new settings keywords, the user can use corresponding methods in the \texttt{Settings} class, \eg
\begin{codebox}
addFlag(string key, bool default)
\end{codebox}
\noindent to add a boolean, \ie a \texttt{Flag}, and
\begin{codebox}
addParm(string key,\! double\!  default, bool\!  hasMin,\!  bool\!  hasMax,\!  double\!  min,\!  double\!  max)
\end{codebox}
\noindent to add a double-precision parameter. For further fine-grained control or using the comma-separated vector type settings, we advise the user to refer to the methods documented in the \texttt{Settings Scheme} section of the \htmlmanual.

\subsubsection{User hooks}\index{User hooks}
\label{subsection:userhooks}

User hooks are placeholders where the user can interrupt normal \pyt program flow to customize behaviour.  The behaviour of the hook (\ie the position in program flow where it is designed to interrupt) is set by functions of the type \texttt{canVetoX} where \texttt{X} indicates one of the pre-defined locations.  An accompanying \texttt{doVetoX} is then executed during every instance of the \texttt{X}.  A user defines a hook by creating a class inheriting from \texttt{UserHooks}, overriding one or more of the hook methods, and passing an  object of that class to a \texttt{Pythia} instance:
\begin{codebox}
  pythia.setUserHooksPtr(make\_shared<MyUserHooksClass>());
\end{codebox}
\noindent It is also possible to add more than one \texttt{UserHooks} object as follows:
\begin{codebox}
  pythia.addUserHooksPtr(make\_shared<AnotherUserHooksClass>());
\end{codebox}
\noindent Note, however, that this may give rise to ambiguities if
several objects have overridden the same hook function. For the
standard \texttt{doVetoX} functions, \texttt{X} will be
vetoed if any of the objects veto, while for some hook methods for which it is not possible to deduce a reasonable combination. In the latter case, \pyt will issue a
warning during initialization.

\pyt provides user hooks for ten cases: interruption while switching between main-generation levels (\eg process to parton); during parton-level evolutions based on $p_T$ or after a step; vetoes for ISR or FSR emissions; to modify cross section or phase space sampling; after resonance decays; to modify shower scales; to allow colour reconnections; to enhance certain rare splittings (\eg $g \rightarrow b \bar b$); and finally, to modify hadronization. The details of each of these hooks can be found in the \htmlmanual. Using these hooks to modify parton level emissions (\eg to match matrix-element contributions from different orders) is discussed in \cref{sec:using-weights}. Here we discuss a simple case of modifying a resonance decay (\eg to select certain kinematics or decay modes).  The \texttt{Pythia::process} event contains the hard scattering process and decay of resonances produced in the hard scattering.  Defining:
\begin{codebox}
  \begin {verbatim}
bool MyHook::canVetoResonanceDecays() {
  // By default returns false.  
  // Set to true to run following method 
  // after each resonance decay
  return true;
}

bool MyHook::doVetoResonanceDecays(Event& process){
  // Look through the process to check 
  // for desired characteristics.
  // Return false to accept the event.
  // Return true to veto the event.
  return false;
}
\end{verbatim}
\end{codebox}
\noindent this method can be used \eg with an LHE file (an LHE event is always stored in the \texttt{process} before migrating it to the full \texttt{event})  that already has decayed resonances that are decayed again by \pyt.

\subsubsection{Semi-internal processes and resonances}
\index{Resonance decays!Semi-internal}
\index{Decays!Resonances@of Resonances}
\label{sec:semi-internal}
While \pyt provides a large number of models, built-in production
processes and resonances, it is oftentimes necessary to either modify
existing processes or add new ones.  The class structure provided by
\pyt can be easily inherited from to include new processes.  Any
new particles produced can be implemented as new resonances.

For a resonance, there are three levels of methods used when
calculating decay widths in various channels.  The first, \texttt{initConstants()} is run once per resonance and can be used to set
couplings or any other properties that do not depend on kinematics.
The second \texttt{calcPreFac()} has access to the kinematic configuration
(masses of particles and phase-space variables), whereas the third,
\texttt{calcWidth()} has access to all information and usually contains a
case-wise calculation of the decay width in all channels.  When there
is no flavour-dependent factor in the calculation, \texttt{calcPreFac()}
can be used to set the internal variable \texttt{widNow} (inherited from
\texttt{ResonanceWidths}) which serves as the calculated width for a
given channel. The example program \texttt{main22} provides a working
example of a new resonance.

 \begin{codebox}
\begin {verbatim}
class NewResonance : public ResonanceWidths {

public:

  // Constructor.
  NewResonance(int idResIn) {initBasic(idResIn);}

private:

  // Locally stored properties and couplings.
  double coupling1, coupling2;

  // Initialize constants.
  virtual void initConstants();

  // Calculate various common kinematic factors 
  // for the current mass.
  virtual void calcPreFac(bool = false);

  // Calculate width for each channel.
  virtual void calcWidth(bool = false);

};
\end{verbatim}
\end{codebox}

Once the resonance is set up, it can be added to the \pyt particle data table before initializing using
\begin{codebox}
\begin{verbatim}
  ResonanceWidths* newResonance = new NewResonance(pid); 
  // Where pid is the integer PDG id. 
  pythia.setResonancePtr(newResonance);
\end{verbatim}
\end{codebox}
\noindent This will automatically call the relevant width calculation
functions on initialization and calculate the total width of the
particle based on all open channels.  The use may have to set up the decay
table (\ie a list of open channels) using the commands in \cref{sub:SettingParticleData} if the particle is not part of the PDG
standard~\cite{Zyla:2020zbs}.

Once all new particles are set up, production modes can be set up by
inheriting from \pyt's \texttt{SigmaProcess} class and its
derivatives.  For $2\rightarrow 1$, use \texttt{SigmaProcess}, for
$2\rightarrow 2$ use \texttt{Sigma2Process}, and use \texttt{Sigma3Process}
for $2\rightarrow 3$.  All relevant kinematic variables are already
set up and will be filled event by event based on \pyt's phase-space
generator.  The production process should be set up before calling
\texttt{init()} in the main \texttt{Pythia} class. The kind of incoming particles needed for the
production are set by the return value of \texttt{inFlux()}; options are
\texttt{qqbar}, \texttt{qqbarSame} (same flavour $\q\qbar$), \texttt{qg} ($\q\g$
and $\qbar\g$), \texttt{ffbar} (includes quarks $\q \qbar$ and leptons
$\ell \bar \ell$), $\g\g$ (gluons), and a few more.

\begin{codebox}
\begin{verbatim}
class Sigma1qqbar2NewResonance : public SigmaProcess {

public:

  // Constructor.
  Sigma1qqbar2NewResonance() {}

  // Initialize process.
  virtual void initProc();

  // Calculate flavour-independent parts of cross section.
  virtual void sigmaKin();

  // Evaluate sigmaHat(sHat). 
  // Assumed flavour-independent so simple.
  virtual double sigmaHat() {return sigma;}

  // Select flavour, colour and anticolour.
  virtual void setIdColAcol();

  // Info on the subprocess.
  virtual string name() const {return "q qbar -> NewResonance";}
  virtual int    code() const {return 10000;}
  virtual string inFlux() const {return "qqbarSame";}
  virtual int    resonanceA() const {return 1000025;}

  // Set internally shared variables (like couplings) 
  // as protected or private
  ...
}
\end{verbatim}
\end{codebox}

Similar to the new resonance-width calculation, there are three
progressive methods that can be used to optimize running time.
First, \texttt{initProc()} is called once per run and can be used to set
constants or couplings based on input parameters.  Second, \texttt{sigmaKin()} can be used to set up kinematic factors for
unresolved processes that do not rely on flavour information of the
incoming and outgoing states.  Third, \texttt{sigmaHat()} can be used
to calculate the full contribution of the phase-space point, which is
returned as a double-precision floating-point number.  The return
value should be the value of $\frac{d\sigma}{dt}$ for $2 \rightarrow 2$
and $|\mathcal{M}|^2$ for $2\rightarrow 3$ processes, respectively.  If
the user wishes to input the matrix element squared instead of
$\frac{d\sigma}{dt}$, then they should also override \texttt{bool
  convertM2()} to return \texttt{true} (see \texttt{include/Pythia8/SigmaProcess.h} for full class definition and
explanatory comments). Finally, an important step before the process
is usable is to set the incoming and outgoing colours (\ie colour
topology) and flavours where necessary.
\begin{codebox}
\begin{verbatim}
void Sigma1qqbar2NewResonance::setIdColAcol() {

  // Flavours simply to be copied from incoming 
  // quark ids i.e. id1, id2
  setId( id1, id2, idNew);

  // Colour flow topologies. Swap when antiquarks. 
  // Say NewResonance is an octet
  // col1, acol1, col2, acol2, colRes, acolRes
  setColAcol( 1, 0, 0, 2, 1, 2);
  if (id1 < 0) swapColAcol();

}
\end{verbatim}
\end{codebox}

The new process can now be added to the \pyt process array by declaring:
\begin{codebox}
\begin{verbatim}
  SigmaProcess* sigma1Res = new Sigma1qqbar2NewResonance();
  pythia.setSigmaPtr(sigma1Res);
\end{verbatim}
\end{codebox}

\subsubsection{Multithreading}\index{Multithreading}
In most cases, events are generated independently of each other. This means
that in principle, event generation can easily be split across multiple
threads in order to speed up generation. In practice, each \texttt{Pythia}
object contains an internal state, and is therefore not thread-safe.
A straightforward workaround is to create multiple \texttt{Pythia} objects,
each initialized with its own random seed using \setting{Random:seed}
and \setting{Random:setSeed}{on}.

Starting from \pyt~8.307, the \texttt{PythiaParallel} class provides
a framework for doing this. This class is intended to provide a lightweight
solution to easily enabling parallelism for simple studies. 
Objects of this class are constructed and initialized
similarly to normal \texttt{Pythia} objects, but rather than having a
\texttt{next()} method that generates a single event, it provides the
\texttt{run} method, which generates a number of events in parallel.
The way this works is that the \texttt{PythiaParallel} object creates and 
keeps track of a number of \texttt{Pythia} sub-objects. These
sub-objects create events in parallel. Whenever an event is generated, the
\texttt{Pythia} object that generated it is passed to the user so that
the resulting event can be analysed. This process then continues until
a required number of events has been generated, as specified by the
\setting{Main:numberOfEvents} setting. The following snippet gives an
example of how to generate events using this class.

\begin{codebox}
  \begin {verbatim}
  #include "Pythia8/Pythia.h"
  // PythiaParallel.h must be included explicitly.
  #include "Pythia8/PythiaParallel.h"
  using namespace Pythia8;
  
  void main() { 
    // The PythiaParallel object is created 
    // and initialized as normal.
    PythiaParallel pythia;
    pythia.readString("SoftQCD:nonDiffractive = on");
    pythia.readString("Main:numberOfEvents = 10000");
    pythia.init();
    
    // Example: plot charged multiplicity
    Hist nCh("Charged multiplicity", 100, -0.5, 399.5);
    
    // This defines the callback that will analyse events.
    function<void(Pythia& pythiaNow)> callback = 
      [&nCh](Pythia& pythiaNow) {
      int nChNow = 0;
      for (int i = 0; i < pythiaNow.event.size(); ++i)  
        if (pythiaNow.event[i].isFinal() && 
          pythiaNow.event[i].isCharged()) nChNow += 1;
      nCh.fill(nChNow);
    };
    
    // Generate events in parallel, using 
    // the specified callback for analysis.
    pythia.run(callback);

    // Print histogram.
    cout << nCh;
  }
  \end{verbatim}
\end{codebox}

\noindent
In this example, the \texttt{callback} is defined via an anonymous function
(also known as a lambda function). It would also be possible to define it
as a named function, \eg with signature
\begin{codebox} 
void callback(Pythia\& pythiaNow) { ... }
\end{codebox}
\noindent The advantage of using an anonymous function is that it can
directly access local variables such as the \texttt{nCh} histogram, which
is captured by reference according to the \texttt{[\&nCh]} specifier.
It is not necessary to actually save this anonymous function in the
\texttt{callback} local variable. They can instead be passed directly to
\texttt{run}, which would make the structure of the code more similar to
running with \texttt{Pythia::next}. Further examples on using the
\texttt{PythiaParallel} class are included with the \pyt~8.307 distribution.

By default, the framework tries to identify the number of available hardware
threads and use the maximum degree of parallelism. Alternatively, the
number of threads can be fixed using the \setting{Parallelism:numThreads}
setting. This can be useful in order to limit the computational
resources spent on generation, and is mandatory on systems where
the number of threads cannot be detected.

\begin{figure}
	\begin{center}
		\includegraphics[width=1.00\textwidth]{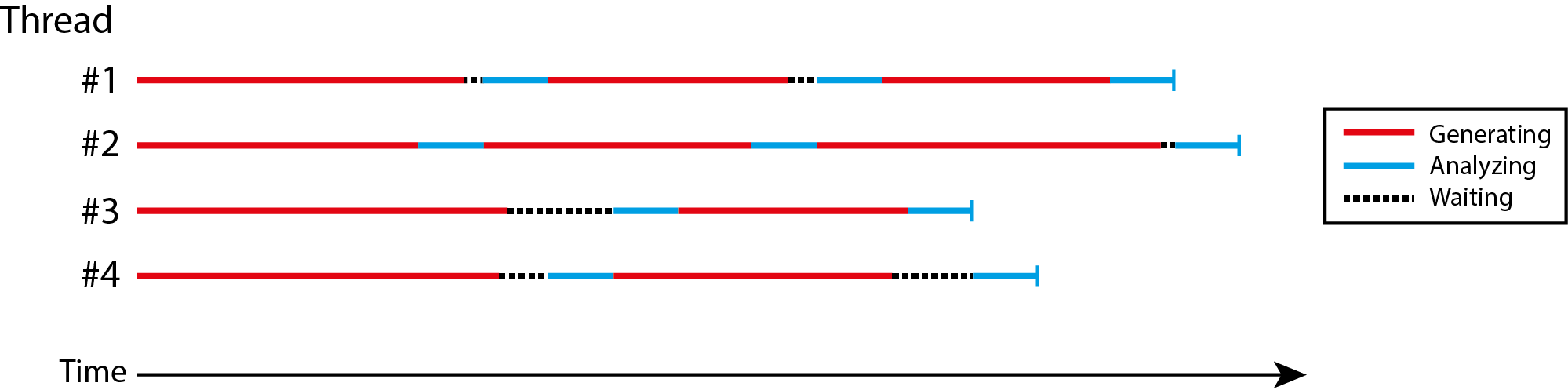} \hfill
	\end{center}
	\caption{\label{fig:parallelism}
  Illustration of how 10 events are generated and processed in parallel
  by four threads. The red lines indicate that a thread is generating an
  event. The blue lines indicate that a thread is analysing the event. 
  Note that two threads are not allowed to analyse at the same time; the
  dashed black lines indicate that a thread is done generating the event,
  and is waiting for another thread to finish its analysis. }
\end{figure}

Although event generation is done in parallel, the analysis is
synchronized by default so that only one event is processed at the same
time. In the example above, this ensures that it is not
possible for two threads to simultaneously write to the
\texttt{nCh} histogram. An illustration of this is shown
in \cref{fig:parallelism}. Usually, the analysis is much faster
than the actual event generation, and this does not have a significant
impact on the run time. However, if the analysis is slow or if
the number of threads is very large, the different threads may spend
a non-negligible amount of time waiting for other threads to finish the
processing. In this case, the run time can be improved by setting
\settingval{Parallelism:processAsync}{on}, which will cause the generated
events to be also processed in parallel. It is then up to the user to ensure
mutually exclusive access to thread-unsafe resources such as histograms.

It is also possible to use external libraries to perform naive parallelization. Several
examples using \openmp are available in the \pyt distribution.

\subsection{Event weight handling}\index{Weights}
\label{sec:using-weights}

By default, \pyt produces unweighted events. This means that every event produced by the generator represents an equal share of the total cross section of the chosen processes. However, some settings and functionalities require the use of weighted events. Weighted events no longer represent equal shares of the total cross section, but are augmented with a corrective event weight that needs to be taken into account when filling histograms of physics observables.

Event weights are useful in different scenarios, which are listed below. One of the key advantages is the use of parameter or setting variations. Rather than regenerating events for different settings and choices of parameters, a vector of corrective event weights can be included in the event generation, reducing the total computation time significantly compared to the generation of separate event samples.

A detailed list of available event weights, related settings and how to access them is available in the \htmlmanual. In the following, we provide an overview of process-specific weights and how they can be accessed. Furthermore, we describe the automated-variation weights and the weight container, which collects all weights in a common structure.

\subsubsection{Overview of process specific weights}
\index{Weights}

\pyt collects the available event weights in a single nominal event weight, which is accessible through the \texttt{Info::weight()} function. In a usual setting, this weight is set to $1$ and thus uninteresting. Several functionalities and settings lead to a modification of this weight though, in which case the weight must be included when filling histograms.

\begin{itemize}
  \item Biased phase-space point selection allows for the reduction of statistical fluctuations for specific kinematic configurations. The corrective weight needs to be included to ensure that the overall distributions are not changed.

  \item If Les-Houches events are used as input, some strategies allow for negative weights, which will be included in the event weight and need to be taken into account. For the strategies $4$ and $-4$, the event weight has units pb, and is converted to mb upon output.

  \item For heavy-ion collisions, \pyt allows a Gaussian sampling of the impact parameter space, leading to weighted events.

  \item In rare cases, the initial over-estimate of the differential cross section might for specific phase-space points lie below the correct differential cross section. In these cases, a weight above $1$ is provided to compensate for this violation.

  \item Enhanced parton-shower emissions (\cf\cref{sec:biasedKernels}) need to be corrected for with a weight to ensure that the distributions remain unchanged when improving the statistical relevance of rare emissions.

  \item Multi-jet merging requires event weights to account for Sudakov factors and the running of coupling parameters. For the leading-order merging schemes CKKW-L and UMEPS, these are by default included in \texttt{Info::weight()}. For the next-to-leading-order multi-jet merging schemes NL3 and UNLOPS, the merging weight needs to be included and is available from \texttt{Info::mergingWeightNLO()}.
\end{itemize}

\subsubsection{Automatic weight variations}
\index{Uncertainties}\index{Weights}

In addition to the nominal weight, additional weights can be provided
to take into account variations of settings and parameters. Filling
histograms with these respective weights allows for an estimation of the
corresponding distributions without rerunning \pyt. Additional
variation weights are available from the parton shower, multi-jet merging and external LHEF
input.  

The parton shower currently allows for renormalization-scale
variations and non-singular term variations in both initial- and
final-state radiation, discussed further in \cref{sec:showerVariations}. Besides, it
allows for the variation of PDF members of LHAPDF~6 families. Details
on the usage of these variations can be found in the \htmlmanual. The
physics background is described in \citeone{Mrenna:2016sih}.  

The multi-jet merging schemes CKKW-L, UMEPS, NL3, and UNLOPS also allow for re\-normal\-iz\-ation-scale variations. Furthermore, variations of the UNLOPS merging scheme itself are available. For details, see \citeone{Gellersen:2020tdj} and the \htmlmanual. The renormalization-scale variations in the merging are automatically combined with corresponding variations from LHEF input and the parton shower.

With the availability of variation weights from different sources, \pyt~8.3 introduces a common structure, the weight container, to make all these weights available to the user. This structure is also used for writing available event weights to \hepmc output. The naming conventions are based on \citeref[p. 162]{Andersen:2014efa}. If multi-jet merging is activated, combined weights for renormalization-scale variations in LHEF input, parton shower, and merging are included. Custom weights from LHEF input or the parton shower are presented with a prefix to emphasize that further processing or combining might be necessary. While \hepmc output automatically contains these variation weights, the user can access them directly through the following methods:
\begin{codebox}
int Info::numberOfWeights() \\
string Info::weightNameByIndex(int i) \\
double Info::weightValueByIndex(int i) \\
vector<string> Info::weigthNameVector() \\
vector<double> Info::weightValueVector() 
\end{codebox}
The first entry of this weight vector, or correspondingly the weight with index $0$ is the nominal weight, including the weights from all the above-mentioned sources.

\subsection{Tuning \pyt}\index{Tuning}
\newcommand{\chisq}{\ensuremath{\chi^2}\xspace}
\newcommand{\MC}{\ensuremath{\text{MC}}\xspace}
\newcommand{\w}{\vec{w}}
\newcommand{\calO}{\mathcal{O}}
\newcommand{\calS}{\mathcal{S}}

By default, \pythia operates using a particular set of run-time
parameters that determine the behaviour of the physics models.
A set of parameters that is chosen based on a comparison of \pythia
predictions
to data is generically named a ``tune''. As the name suggests, the
procedure to obtain a tune is similar to adjusting the pegs on a stringed instrument
to achieve a certain sound.   However, there is no universal agreement
on what constitutes a good tune in contrast to a good sound.
The goal of tuning is to find an ``optimal'' set of physics
parameters, $\p^*$, that  minimizes the difference between the
experimental data and the simulated data from the event generator.
In practice, this difference is defined as follows: 
\index{Uncertainties!Tuning@in Tuning}\begin{equation}
\chisq_{\MC}(\p,\w) = \sum_{\calO\in\calS_\calO}
  \sum_{b\in\calO}w_{\calO_b}\frac{(\MC_b(\p)-{\cal O}_b)^2}{\Delta \MC_b(\p)^2+\Delta\calO_b^2}~, \label{eq:MC}
\end{equation}
where $\calS_\calO$ is the set of observables used in the tune, $b\in\calO$ denotes the bins in a certain observable $\calO$,
and $\w$ is a vector of weights $w_{\calO_b}$ for each bin of each observable.
The $\Delta$s are the uncertainties on the simulated data and the observable.
The weights $w_{\calO_b}\geq0$ reflect  how much an observable contributes
to the tune, \ie if $w_{\calO_b}=0$ for some $\calO_b$, then this
observable bin will not influence the  tuning of $\p$, whereas if
$\w_{\calO_b}=1$ then all data is treated equally. The choice of
$\calS_\calO$ and $w_{\calO_b}$ determines a unique tune, and these choices are driven by both theoretical and experimental
considerations.  The variable in \eqref{eq:MC} is called a
``chi-squared'', but due to the presence of weights there is
no guarantee that it will have the
properties of a proper $\chisq$ distribution.

\subsubsection{Comments on the tuning procedure}

There are several aspects of this problem that make it non-trivial.
One is that the model, which is a mixture of theoretically- and
phenomenologically-grounded sub-models, does not describe all
data equally well.    Related to this is that there is no systematic
method for predicting \emph{a priori} where the model will fail ---
the sub-models are often deeply entwined and are not factorizable,
despite our original intention that they should be.   If one had a
numerical estimate of the uncertainty coming from a certain model (not
just its sensitivity to parameter variations, but an estimate of where
it fails), then that could be included in $\Delta \MC_b$, which
usually includes only the uncertainty arising from a finite number of
simulation runs.   Another issue is how the tune will be applied ---
will it be used for generic simulations with lowest-order matrix
elements or with matched and merged predictions with higher-order
matrix elements?

There is more than one way to attack these problems.   One, as
suggested above, is to set $\w_\calO=0$ for some set of observables or
analyses.   Such a decision is made at the beginning of tuning when
one
decides which data is most relevant, \eg Tevatron data or LHC
data from a lower-energy run, minimum bias or inclusive jet data,
\etc
Data can also be removed from the tune 
when it becomes obvious they do not fall within the envelope of the model
predictions.
In practice, it is sometimes found that even problematic data should be included,
with
reduced significance, to improve the overall quality of the tune.
This can be accomplished by adjusting some of the $\w_\calO$ values to
emphasize or de-emphasize certain datasets.
Obviously,
such \textit{a posteriori} manipulation of data is subject to bias and
abuse.
However, one should remember that tuning is not hypothesis testing ---
we do not allow for the possibility that the model is ruled out.
To illustrate two different, but not exhaustive, approaches, we will
describe the \monash and the ATLAS A14 tuning exercises.

\subsubsection{The default \pythia tuning: \monash}
\index{Tuning!Monash tune}\index{Default tune|see{Monash tune}}\index{Monash tune}
\index{Tuning!Default tune|see{Monash tune}}
\label{sec:monash-tune}

The \monash tune is currently the default one.    It was performed
using data from HEPDATA and the PDG.
It was aimed at non-diffractive, 
high momentum-transfer collisions using the leading-order matrix
elements coded in \pyt.
It started from the hypothesis
that hadronization was independent of the environment, and the
related physics parameters could be best constrained using
$e^{+}e^{-}$ data, particularly from LEP (for most observables) and
SLD (for \b-hadron specific observables).   Any modifications to
hadronization
predictions from the breakup of the proton, for example, would be
handled
by explicit models that modified the initial conditions, but not the
mechanism
of hadronization.   Once the observables were selected,
all with either
$\w_\calO=0$ or $1$,; several inconsistent values of particle yields
were adjusted based on common sense.   Physics parameters related
to final state radiation, hadronization, and particle decays were
selected using \cref{eq:MC} as a guide, but without an explicit
global minimization.    
\index{Uncertainties}An additional \textit{ad hoc} ``theory
uncertainty'' of 5\% was added per bin of each histogram used in the
tune
to prevent overfitting.
These parameters were then frozen as a particular \texttt{eetune}.
The tuning of the remaining parameters, specific to hadronic
collisions, began with a choice of PDF, which is an integral part of
any such tune.   The central tune of the NNPDF2.3 PDF set was selected, as it
was
being used in many other theory calculations at the time.  In
particular, the choice was leading order with a value of $\alpha_s$
closer to that found in the \texttt{eetune}.
The tuning of initial-state parameters, such as those related to
initial-state radiation, beam remnants, and multiparton interactions,
proceeded in a similar fashion using LHC data at the highest energy
available.
Scaling of the multiparton-interaction parameter was obtained by
including Tevatron data.
Again, at no point was a global
optimization
of parameters made based on minimization of a \chisq.

\subsubsection{The ATLAS A14 tune}
\index{Tuning!ATLAS A14 tune}

The ATLAS A14 tune took a different approach.   
First, it took the basic \monash parameters as a starting point, with
the goal of optimizing parameters for LHC physics studies. It relied
heavily on the \professor~\cite{Buckley:2009bj} framework. The observables were selected 
and weighted to emphasize high-\pT radiation and some top-quark observables.
It was designed to be used for BSM physics searches, where precision was 
not the main goal. To that end, it minimized \cref{eq:MC} for ten parameters, 
but in an iterative process to select weights that produced a ``reasonable'' fit.
Using \professor, it also produced eigentunes that could be used as alternative tunes to
study sensitivity to the \pyt parameters. However, because of the
inclusion of weights, and since the fit residuals do not appear to be \chisq distributed,
 an \textit{ad hoc} criterion was used to determine these variations. In the process of
selecting data, many observables were included that are obviously
highly correlated. However, those correlations were not reported
consistently by the experiments. As a result, some of the
observables have a hidden weight.

\subsubsection{Automatic tuning approaches}\index{Tuning!Automated approaches}

Automatic tuning approaches can be helpful to circumvent some of the challenges that manual tuning entails, 
like subjectivity based on expert knowledge of models, parameters, constraints, and data, and challenges due 
to a high amount of data sets and parameters to be taken into account. Automatic tuning aims at simplifying 
the tuning procedure and making it more systematic, which is especially helpful when many parameters are to 
be tuned.

A brute force grid-based tuning approach is usually prohibited due to the high computational cost of generator 
runs, especially if many parameters are to be tuned. To circumvent this problem, one can use iterative 
optimization approaches, which can take time due to the serial running with different parameters, but 
focus well on relevant regions in parameter space. Alternatively, one can attempt to parameterize the 
generator response, and to then optimize based on an interpolation. After an initial generator run, which can 
be trivially parallelized, the actual optimization based on the interpolation is much more straight forward.

As outlined in \citeone{Ilten:2016csi}, an iterated Bayesian optimization approach can be employed for event 
generator tuning. A $\chi^2$ value for different parameter values is obtained, and all information is used 
to find the next set of parameters iteratively. This approach thus goes beyond local gradient-based 
optimization, balancing exploration and exploitation.

The \professor toolkit employs a parameterization approach. After an initial parallelized 
MC event-generator run, the generator response is parameterized using a polynomial function. A $\chi^2$ 
optimization is then performed based on this interpolation. This approach allows for several parameters, 
but becomes prohibitive if the parameter space becomes too large. It is then beneficial to tune in successive 
steps based on model and data knowledge.

There are multiple efforts in improving the \professor tuning approach. The \autotunes 
method~\cite{Bellm:2019owc} employs \professor, and goes beyond by automatically identifying subsets of 
correlated parameters that can be optimized successively. The weights are chosen correspondingly to constrain 
sub-tunes by the most relevant experimental data. The \apprentice method~\cite{Krishnamoorthy:2021nwv} goes 
beyond \professor by allowing for more general interpolations, a larger variety of optimization methods,
and automated setting of weights.

Automatic tuning methods can be very useful, and are helpful when many parameters are to be optimized based 
on a large amount of experimental data.  In combination with expert knowledge about the tuned models and the 
experimental data, pitfalls can be avoided, like too strong constraints due to single well-measured 
distributions or unphysical tuning results.

%% file: using-pythia/interface-external.tex
\section{Interfacing to external programs}
\label{section:intext}

In most realistic use-cases, \pythia is not used stand-alone, but rather as part of a large software stack capable of providing everything
from calculation of Feynman rules from a Lagrangian, to detector simulation and analysis, including interfaces between all those steps. 
Technically, \pythia is a \cpp library, and only the 
users' technical proficiency limits the ways the program can be interfaced to other code, thus a manual section describing external interfaces, 
will by definition be incomplete. For practical purposes, however, \pythia comes with a number of interfaces pre-written, and several more
with an official or unofficial ``blessing'' by the developers. These are interfaces which should in general work, and where the \pythia developers will
at take some responsibility for helping users when setting up. Those interfaces are described here, along with an explanation of how
\pythia is expected to interact with them. The section is sub-divided in four. In \cref{sec:generation-tools} we describe file-based or 
run-time based interfaces to external providers of \emph{input} to \pythia, be it external matrix elements, PDFs, or random numbers. In \cref{sec:output-formats} we describe the most often used output formats, such as \hepmc events or \root ``n-tuples''. In \cref{sec:analysis-tools} we describe run-time interfacing with the analysis tools \rivet and \fastjet, and finally in \cref{sec:computing-environments} the use of \pythia through the
\python interface and on multicore HPC architectures is discussed.

\subsection{Generation tools}
\label{sec:externalGenerators}

\label{sec:generation-tools}
Several file-based or run-time interfaces exist. For file-based interfaces, generation steps must be run in a strict sequence. 
For run-time interfaces, we take a \pyt-centric view, \ie that \pyt controls the overall event generation (unless stated otherwise).

\subsubsection{Les Houches Accord and Les Houches Event File functionality}
\label{subsection:lha}\index{LHA|see{Les Houches Accords}}\index{LHEF|see{Les Houches Accords}}\index{Les Houches Accords!LHA}

The \ac{LHA} format~\cite{Boos:2001cv} allows a
factorized event generation chain and is one of the most long-lived
and successful interface agreements in particle physics. Using the LHA
format, complex perturbative calculations can be factored out from the
rest of the event generation chain, and performed by specialized
tools. The basic idea of LHA is a run-time interface between two
generator codes: the ``fixed-order generator'' stores the collision
setup and cross-section information in memory for the ``event
generator'' to read upon initialization (see
\cref{tab:lha-heprup}). At generation time, the individual
phase-space points used in the fixed-order generator are stored
in memory for the event generator to read and process further, \cf\cref{tab:lha-hepeup} for the format definition. Originally, the
in-memory structures were \fortran common blocks (called
\texttt{HEPRUP} for initialization and \texttt{HEPEUP} for event
information). This original format is still used in modern
applications, \eg the interfaces to \madgraph or \powhegbox
discussed below. An example of another in-memory structure is
discussed in \cref{ss:hdf5}.

\index{Les Houches Accords!LHEF}
Although desirable from a computing perspective, run-time interfaces
require programming language-specific in-memory representations. The Les
Houches event file (LHEF) format~\cite{Alwall:2006yp} is a
text-file-based update and extension of LHA, such that no run-time
interface is necessary, making the results somewhat more portable. Les
Houches Event files provide pre-tabulation and storage of phase-space points, thus
enabling the reuse of computationally expensive results.

The LHEF format defines \texttt{XML}-like ``tag'' structures to store
information. As such, all relevant information in a LHEF file is
enclosed in:
\begin{center}
\verb|<LesHouchesEvents version="|$v$\verb|"> ... </LesHouchesEvents>|.
\end{center}
The \texttt{version} can be $v=1.0$~\cite{Alwall:2006yp} or
$v=3.0$~\cite{Andersen:2014efa}.

The \texttt{HEPRUP} initialization information of the LHA is mirrored by a text block bracketed with \verb|<init> ... </init>|, while the \texttt{HEPEUP} event information is captured in a text block enclosed in an \verb|<event> ... </event>| tag. Auxiliary information pertaining to all events can also be stored in a block bracketed with a \verb|<header> ... </header>| tag. The content of each tag may contain further tags, see \cref{tab:lhef-init,tab:lhef-event} for a list
of all recognized tags. 

A basic example \texttt{<header>} block is  
\begin{codebox}
<header>\\
Some auxiliary information that \\
...\\
is not parsed.\\
</header>
\end{codebox}
Such a header would be compliant with all versions of the format. Additional tags may appear in later versions (v.~3) of the format, as shown in the example below
\begin{codebox}
<header>\\
Some auxiliary information that\\
...\\
is not parsed.\\
<initrwgt>\\
  <weightgroup type="alphasVariation">\\
    <weight id="A"> nominal   alphas </weight>\\
    <weight id="B"> decreased alphas </weight>\\
    <weight id="C"> increased alphas </weight>\\
  </weightgroup>\\
</initrwgt>\\
</header>
\end{codebox}
In this particular example, \pythia will be instructed to expect each event to contain a \texttt{<rwgt>} block that contains three \texttt{<wgt>} entries.

In a slight extension of the accord, \pythia will also parse the parts of the \texttt{<header>} block that are enclosed in \verb|<slha> .. </slha>| as if the block contained an \slha file. See \cref{sec:slha} for a description of \slha files.

The \texttt{<init>} block is a mandatory part of any LHE file. A basic example will contain the two beam-particle identifiers, their two energies in \GeV, two PDF-author-group identifiers, two PDF-set identifiers, and weighting information, followed, in a separate line, by cross section, statistical error, and unit weight information, followed by an integer process label:  
\begin{codebox}
<init>\\
   2212   2212 0.4E+04 0.4E+04 -1 -1  21100  21100 -4   1\\
 0.50109086E+02 0.89185414E-01 0.50109093E+02     1234\\
</init>
\end{codebox}
Nowadays, the most common weighting-strategy information (given by $-4$ in the example) allows for both positive and negative event weights, where the average weight gives the cross section of the generated events.
In later versions of the format, the optional \texttt{generator} tag may also be included:
\begin{codebox}
<init>\\
   2212   2212 0.4E+04 0.4E+04 -1 -1  21100  21100 -4   1\\
 0.50109086E+02 0.89185414E-01 0.50109093E+02     1234\\
<generator name="SomeGen1" version="1.2.3"> some additional comments </generator>\\
<generator name="SomeGen2" version="a.x.3"> some other comments </generator>
</init>
\end{codebox}
This tag mainly serves to convey information, and does not affect the file processing through \pythia.

The initialization information is then complemented with a large list of \texttt{<event>} blocks containing the phase-space points. It should be noted that \pythia supports an arbitrary list of attributes of the \texttt{<event>} tag, and further allows ``custom'' additions enclosed in \texttt{<event>} tags:
\begin{itemize}
\item The identifier \verb|#pdf| at the start of a line means the line contains information on PDFs. For example, the line \\
\verb|#pdf     1   -1  0.11  0.3  100  0.5  0.3|\\
will lead to reading/setting the values: ID(particle extracted from beam ``A'') = $1$, ID(particle extracted from beam ``B'') = $-1$; momentum fraction of particle extracted from beam A $x_\mathrm{A} = 0.11$, momentum fraction of particle extracted from beam B $x_\mathrm{B} = 0.3$; factorization scale $\mu_F = 100$ \GeV; value of the parton distribution for beam A $f_\textnormal{A} (x_\textnormal{A},\mu_F) = 0.5$; and value of the parton distribution for beam B $f_\textnormal{B} (x_\textnormal{B},\mu_F) = 0.3$.
\item \texttt{<event>} tags are allowed to enclose two hard-scattering events, as is \eg needed when interfacing to external double-parton scattering codes.
\item In the latter case, the identifier \verb|#scaleShowers| at the start of a line leads to the two subsequent floating-point values being interpreted as parton-shower starting scales
for the first and second hard scattering enclosed by \verb|<event> ... </event>|, respectively.
\item Omitting the incoming particles in the content of the \texttt{<event>} tag can be permissible when interfacing with \pythia to perform only hadronization of resonance-decay systems.
\item The event attributes \texttt{npLO} and \texttt{npNLO} are parsed, and employed when interfacing to \mg5amc.
\end{itemize}
A simple event compliant with both versions of the standard will contain information about the number $N$ of particles in the event, the process label, the ``scale'', and QED and QCD coupling strengths, followed by $N$ lines containing particle information:
\begin{codebox}
<event>\\
       4    1234  5.0  300.0  7.861651E-03  1.084400E-01\\
       2   -1    0    0  101    0  0.000E+00  0.000E+00  3.016E+02  3.016E+02  0.000E+00 0. 9.\\
      -2   -1    0    0    0  102  0.000E+00  0.000E+00 -2.964E+02  2.964E+02  0.000E+00 0. 9.\\
       6    1    1    2  101    0 -1.358E+02 -1.671E+02  1.128E+02  3.000E+02  1.756E+02 0. 9.\\
      -6    1    1    2    0  102  1.358E+02  1.671E+02 -1.076E+02  2.980E+02  1.756E+02 0. 9.\\
</event>
\end{codebox}
For each particle, its identity, status, pair of mothers, pair of colours, momentum, mass, production vertex, and spin are required information.
In version $3.0$ of the standard, further information may be added to an \texttt{<event>}. A more involved example is:
\begin{codebox}
<event type="undecayed\_born\_level\_ttbar">\\
       4    1234  5.0  300.0  7.861651E-03  1.084400E-01\\
       2   -1    0    0  101    0  0.000E+00  0.000E+00  3.016E+02  3.016E+02  0.000E+00 0. 9.\\
      -2   -1    0    0    0  102  0.000E+00  0.000E+00 -2.964E+02  2.964E+02  0.000E+00 0. 9.\\
       6    1    1    2  101    0 -1.358E+02 -1.671E+02  1.128E+02  3.000E+02  1.756E+02 0. 9.\\
      -6    1    1    2    0  102  1.358E+02  1.671E+02 -1.076E+02  2.980E+02  1.756E+02 0. 9.\\
<rwgt>\\
 <wgt id="A"> 5.0 </wgt>\\
 <wgt id="B"> 4.5 </wgt>\\
 <wgt id="C"> 5.5 </wgt>\\
</rwgt>\\
<weights> 1.0 0.7 1.3 </weights>\\
<scales muf="175.0" mur="175.0" mups="300.0" scale\_3="1.0" scale\_4="1.0">\\
content is not parsed\\
</scales>\\
</event>
\end{codebox}
This event contains three auxiliary event weights in the ``detailed format'', as well as three additional event weights in the ``compressed format''. These different ways to transmit event weights do typically not appear together. The ``detailed format'' has become much more widely used. The example above further contains auxiliary scale information through the \texttt{scales} tag. This feature can be used to \eg transfer multiple shower starting scales to \pythia. Starting scales for individual particles in the event can be set by including a \texttt{scales} attribute ending with \_\texttt{iPos}, where \texttt{iPos} is the position of the particle (in the \texttt{<event>}) in question. This functionality is used for MLM jet matching\index{MLM jet matching} with \madgraph, and for \mcatnlo$\Delta$ matching using \mg5amc. At present, \pythia does not support the use of sets of events enclosed in \texttt{<eventgroup>}. Such events sets were originally proposed in~\citeone{Butterworth:2010ym} to collect events that require correlated post-processing. Since the latter is not possible in \pythia, event files containing \texttt{<eventgroup>} tags will be treated as if the \texttt{<eventgroup>} tag was not present.

Finally, note that \pythia will perform momentum-conservation checks on each input \texttt{<event>}. If inconsistencies (\eg due to rounding errors) are found, then actions will be taken to repair the event. This entails enforcing the correct value of particle rest masses, and ensuring that the incoming momentum matches the outgoing momentum.

\begin{table}[H]
  \caption{\label{tab:lha-heprup} The information defining the LHA initialization interface  (in the \texttt{HEPRUP} common block). The suffix \texttt{UP} can be read as ``user process''. At most 100 user processes are allowed. See~\citeone{Boos:2001cv} for details.}
  
  \begin{center}
    \begin{tabular}{c | p{0.7\textwidth}}
      \toprule
      block name & description \\
      \midrule
\texttt{IDBMUP(2)} & pair of two integer values defining the PDG IDs of the colliding beams\\
\texttt{EBMUP(2)} & pair of two floating-point values listing the energies of the two colliding beams in \GeV\\
\texttt{PDFGUP(2)} & pair of two integer values defining the author group of the PDF fit used as the PDF for the colliding beams\\
\texttt{PDFSUP(2)} &  pair of two integer values defining the PDF set used to extract particles from the colliding hadron beams\\
\texttt{IDWTUP} & signed integer value determining how the event weights should be interpreted\\
\texttt{NPRUP} & integer value defining the number of different user processes\\
\texttt{XSECUP(NPRUP)} & list of \texttt{NPRUP} double values giving the cross sections (in units of pb) of the individual user processes\\
\texttt{XERRUP(NPRUP)} & list of \texttt{NPRUP} double values giving the statistical errors associated with the individual user processes\\
\texttt{XMAXUP(NPRUP)} & list of \texttt{NPRUP} double values giving the maximum weight encountered in generating the cross section of the user process\\
\texttt{LPRUP(NPRUP)} & list of \texttt{NPRUP} integer identifiers for the user processes; the identifiers will also feature in the in-memory representation of the phase-space point\\
\bottomrule
\end{tabular}
\end{center}
\end{table}

\begin{table}[H]
\caption{\label{tab:lha-hepeup} The information defining the \lha event information (in the \texttt{HEPEUP} common block). At most 500 particles are allowed. See~\citeone{Boos:2001cv} for details.}
\index{SPINUP@\texttt{SPINUP}}\begin{center}
\begin{tabular}{c | p{0.7\textwidth}}
      \toprule
      block name & description \\
      \midrule
\texttt{NUP} & number of particle entries in the event\\
\texttt{IDPRUP} & identifier of the user process for this event\\
\texttt{XWGTUP} & event weight\\
\texttt{SCALUP} & scale of the event in \GeV\\
\texttt{AQEDUP} & value of the QED coupling for this event\\
\texttt{AQCDUP} & value of the QCD coupling for this event\\
\texttt{IDUP(NUP)} & list of \texttt{NUP} integer values defining the PDG IDs of the individual particles \\
\texttt{ISTUP(NUP)} & list of \texttt{NUP} integer values defining the status (initial state, final state, or resonance) of the individual particles \\
\texttt{MOTHUP(2,NUP)} & pair of two lists of \texttt{NUP} integer values defining the mothers of the particles\\
\texttt{ICOLUP(2,NUP)} & pair of two lists of \texttt{NUP} integer values defining the $N_c\rightarrow\infty$ colour (anticolour) flow indices of the particles\\
\texttt{PUP(5,NUP)} & five lists of \texttt{NUP} double values giving the lab-frame momentum of the particle $(P_x , P_y , P_z , E, M)$ in \GeV\\
\texttt{VTIMUP(NUP)} & list of \texttt{NUP} double values giving the invariant lifetime $c\tau$ (distance from production to decay) in mm \\
\texttt{SPINUP(NUP)} & cosine of the angle between the spin vector of the particle and the three-momentum of the decaying particle, specified in the lab frame\\
\bottomrule
\end{tabular}
\end{center}
\end{table}

\begin{table}[H]
\caption{\label{tab:lhef-init} Allowed tags in the  \texttt{<header>} and \texttt{<init>} blocks of a Les-Houches event file.}
\begin{center}
  \begin{tabular}{c | p{0.6\textwidth}}
    \toprule
    tag name & description\\
    \midrule
\texttt{<header>} & the tag starting the header block, a completely empty header block is allowed \\
\texttt{<initrwgt>}  & optional tag detailing the auxiliary events in the ``detailed LHEF v3.0 format''; the following two tags have to be enclosed in this tag\\
\texttt{<weightgroup>}  &  optional tag defining a group of event weights in the ``detailed LHEF v3.0 format''; this group will contain several instances of the following tag\\
\texttt{<weight id="}$name$\texttt{">} & optional tag defining a particular auxiliary event weight; \pythia expects each event to contain a \texttt{<wgt>} (see \cref{tab:lhef-event})
with \texttt{id}=$name$ for a \texttt{<weight>} with \texttt{id}=$name$\\
\texttt{<init>}  & the tag starting the cross section information and initialization block\\
\texttt{<generator>} & optional tag to transfer information about the generator and generator version used to produce the event sample\\
\bottomrule
\end{tabular}
\end{center}
\end{table}

\begin{table}[H]
  \caption{\label{tab:lhef-event} Allowed tags in the \texttt{<event>} block of a Les-Houches event file.}
  \centering
\begin{tabular}{c | p{0.7\textwidth}}
  \toprule
  tag name & description\\
  \midrule
\texttt{<event>} & the tag starting the event block; an arbitrary number of attributes is allowed \\
\texttt{<rwgt>} & optional tag enclosing a set of event weights in the ``detailed LHEF v3.0 format'', see next tag \\
\texttt{<wgt id="}$name$\texttt{">} & optional tag transmitting the floating-point value of a unique auxiliary event weight as content; the \texttt{id}=$name$ should mirror one of the \texttt{<weight>} tags of the \texttt{<initrwgt>} block (see \cref{tab:lhef-init}) \\
\texttt{<weights>}  & optional tag containing an array of floating-point values for a set of auxiliary event weights in the ``compressed LHEF v3.0 format''\\
\texttt{<scales}  & optional tag allowing additional scale information stored as attributes of the tag \\
\bottomrule
\end{tabular}
\end{table}

\subsubsection{\slha}
\label{sec:slha}\index{SLHA|see{Les Houches Accords}}\index{Les Houches Accords!SLHA}\index{SUSY}

The SUSY Les Houches accord format!\cite{Skands:2003cj,Allanach:2008qq} was designed as a plain-text
interface between supersymmetric spectrum generators, decay packages,
and event generators.  However, it has
since been generalized to contain information for any new physics
model, \cf\eg \citeone{Alwall:2007mw}.

The current SUSY implementation in \pyt is fully general with support for flavour- and R-parity violation.  The physical mass basis for each class of new particles (squarks, sleptons, charginos, and neutralinos, as well as Higgses) is ordered by mass alone.  We refer the reader to the original SLHA2 documentation~\cite{Allanach:2008qq} for the full list of supersymmetric parameters supported by SLHA2.  Here we give a summary of how new parameters can be passed to \pyt, and the modifications made to extend SLHA2 support to be able to read up to 3-dimensional matrix input.  

\index{QNUMBERS@\texttt{QNUMBERS}}An SLHA file contains a number of
pre-formatted 
``blocks''. The three main blocks most often used for passing
information about new particles are \texttt{QNUMBERS}, \texttt{MASS}, and
\texttt{DECAY}. As an example, we show here how a new spin-1 particle
in a colour-octet representation (``heavy gluon'') and a new fermion
(``heavy quark'') can be defined in SLHA~\cite{Alwall:2007mw}. All
characters following a \texttt{\#} symbol are ignored as a comment,
except the first two words after the particle ID code are assumed to
be the name of the particle and, optionally, its antiparticle.  

\begin{codebox}
BLOCK QNUMBERS 9000021 \# HeavyGluon \\
        1 0  \# 3 times electric charge \\
        2 3  \# number of spin states (2S+1) \\
        3 8  \# colour rep (1:singlet, 3:triplet, 8:octet, 6:sextet) \\
        4 0  \# Particle/Antiparticle distinction (0=own anti) \\
\\
BLOCK QNUMBERS 9000006 \# HeavyQuark HeavyQuarkbar \\
        1 0  \# 3 times electric charge \\
        2 2  \# number of spin states (2S+1) \\
        3 3  \# colour rep (1:singlet, 3:triplet, 8:octet, 6:sextet) \\
        4 1  \# Particle/Antiparticle distinction (0=own anti) \\
\end{codebox}
Note that many of the particle ID codes below 3 million, and several above it,
are already in use in \pyt (\eg for hadrons, SM particles, and the MSSM particle
spectrum). To avoid conflicts, it is strongly advised to only use 
codes above 3 million for new BSM particles, and to check in the particle
data table that the codes are not already in use. See also the PDG list of
standard particle ID codes~\cite[sec.~45]{ParticleDataGroup:2020ssz}.
Finally, note that \pyt 
is only able to handle colour singlets, triplets, octets, and sextets. 

The mass block~\cite{Skands:2003cj} contains the mass of the physical
particles and is simply a list containing the particle ID
code and its mass. 
\begin{codebox}
BLOCK MASS \\
9000021  1000.  \# HeavyGluon \\
9000006   450.  \# HeavyQuark \\
\end{codebox}
Note that some matrix-element generators export their complete list of
particle masses in this block, including also those of SM
particles, which may not agree with \pyt's internal values. This can
wreak havoc in unintended places, \eg by overwriting \pyt's
constituent-quark masses\index{Quark masses!Constituent quark masses}
by far smaller current-quark
masses. Therefore, for particles with ID codes less than one million,
\pyt normally ignores SLHA input for any
particle whose default mass in \pyt is smaller than
\settingval{SLHA:minMassSM}{100}~GeV. This allows SLHA input to modify
top and Higgs-boson properties, but not those of $Z$, $W$, and lighter
particles. 

\index{Decays!Resonances@of Resonances}
\index{Resonance decays!SLHA@SLHA \texttt{DECAY} tables}Separate \texttt{DECAY}
blocks~\cite{Skands:2003cj} can be used to 
specify decay tables for both new and existing particles.
(See further \cref{sec:resonances,sec:hardRes} for
more on \pyt's modelling of resonance production and decays.)
The sum of all branching fractions is normalized to one when read in. If
a certain decay channel 
is needed for determining the total width, but is not desired to be
generated in the context of a given run, this can be done by setting
the branching fraction negative.  Each line containing a branching
ratio should also contain the number of daughter particles, followed by
the ID codes of the daughters. Note that only a single decay table
should be provided for each particle type; \pyt does not accept
separate decay tables for antiparticles. However, if different open decay
modes are required for a particle and its antiparticle, this can be
accomplished by using the \pyt \texttt{ParticleData} settings \setting{NN:onIfPos} and
\setting{NN:onIfNeg} which are allowed to override the initial SLHA
settings if \settingval{SLHA:allowUserOverride}{true}.  

\begin{codebox}
\#     PID            Width \\
DECAY 9000021 0.01 \\
\#     BR       NDA      ID1       ID2 \\
      0.67     2        9000006   -9000006 \\
      0.33     2        6         -6 \\
\end{codebox}

When the SLHA interface is used to modify particle data, the
$m_\mathrm{min}$ and $m_\mathrm{max}$ limits used in \pyt's
Breit--Wigner sampling (see \cref{sec:resonances}) 
default to $m_0 \pm \min(5\Gamma_0,m_0/2)$. The  $m_\mathrm{min}$
value is further required to also be above the sum of on-shell masses
for the lightest decay channel. The default values can be modified by the
user, if so desired.

\index{meMode@\texttt{meMode}}The default Breit--Wigner treatment for
decay tables imported via the 
SLHA interface is the simple \settingval{NN:meMode}{100} one with constant
branching fractions, but this can also be changed if desired. 
The phase-space sampling is isotropic, since the SLHA
tables do not convey any differential information. It is up to 
the user to ensure that  the final behaviour is consistent with what
is desired and/or to apply suitable post-facto reweightings. Plotting the
generator-level resonance and decay-product mass distributions and \eg mass
differences, effective branching fractions, \etc, may be of assistance
to validate the program's behaviour for a given application.

Note, finally, that the default in \pyt is to ignore SLHA input for all SM
particles except top quarks and Higgs bosons; this protects \pyt's
more sophisticated modelling of \eg $\Z$ and $\W$ decays (as well as its
definitions of quarks, hadrons, and leptons), \cf\cref{sec:hardRes}, from being unintentionally overridden by the
simpler SLHA treatment. Similar to the above, this choice can be changed
by the user if desired, though care must be taken not to
corrupt \pyt's hadron or light-quark particle data. 

Finally, we describe how user-defined blocks may be accessed via the
SLHA class~\cite{Desai:2011su}.  All unknown, \ie user-defined blocks
that can be stored in arrays of up to 3 dimensions are read in via the
test SLHA file and saved under the name following the \texttt{BLOCK}
keyword.  Depending on the dimensions of the box, one of these methods
can be used to access relevant information.  This functionality can be
used with \eg the semi-internal processes described in
\cref{sec:semi-internal} to use SLHA files to read complex
parameter information.  Using the \texttt{slhaPtr} object available to
all production processes inheriting from the \texttt{SigmaProcess} class, a block
with \texttt{blockName} can be accessed using one of the following.
\begin{codebox}
\begin{verbatim}
# Single value
bool slhaPtr->getEntry(string blockName, double& value); 

# 1D array
bool slhaPtr->getEntry(string blockName, int index, double& value); 

# 2D array
bool slhaPtr->getEntry(string blockName, int index1,
                        int index2, double& value); 

# 3D array
bool slhaPtr->getEntry(string blockName, int index1, int index2, 
                       int index3, double& value);
\end{verbatim}
\end{codebox}

\subsubsection{\lhahdffive}
\index{LHAHDF5|see{Les Houches Accords}}\index{Les Houches Accords!LHAHDF5}
\label{ss:hdf5}

In addition to plain-text based ASCII \lhef, \pythia now also supports Les-Houches event input via the \hdffive\index{HDF5} data format, which some matrix-element generation frameworks, such as \sherpa~\cite{Bothmann:2019yzt}, support as an alternative to \lhef event output.

The \hdffive format is an open-source binary data format, organized like a database within a single file. It allows for heterogeneous data storage, which is more compressed than ASCII files. Being indexed in an efficient way, it enables the possibility of data slicing, \ie the reading of data subsets instead of the entire data at once. The \hdffive format is thus well suited for storing large numbers of \lha phase-space points in a more efficient way than text-based file formats, allowing for massively parallelized simultaneous access to a single event file~\cite{Hoeche:2019rti}.

The \lhahdffive reader uses the \href{https://bluebrain.github.io/HighFive/index.html}{HighFive} header library to interface \hdffive. Moreover, the \href{https://www.hdfgroup.org/solutions/hdf5/}{\hdffive} library tools must be installed and an MPI compiler, such as that shipped with \href{https://www.mpich.org/}{\mpich}, is needed. To use the \lhahdffive reader with \pythia, an example configuration command is therefore given by:
\begin{codebox}
	./configure {-}{-}with-mpich[=path] {-}{-}with-hdf5[=path] {-}{-}with-highfive[=path]
\end{codebox}

As a relatively new event file format, the \lhahdffive standard is still undergoing active development. \pythia internally uses a three-digit numbering scheme to distinguish different \lhahdffive versions, characterized as follows:
\begin{description}
	\item[\texttt{0.1.0}] The event file contains an index group, in which the indices of the particles in a single event are stored. The indices refer to the particle group. Weight variations are not supported and event weights are stored as a single floating-point number in the event group.
	\item[\texttt{0.2.0}] The event file does not contain an index group. Weight variations are not supported, and event weights are stored as a single floating-point number in the event group.
	\item[\texttt{1.0.0}] The event file does not contain an index group. Weight variations are supported, and event weights are stored in a (possibly one-dimensional) array in the event group.
\end{description}
Currently, not all event files may have their version number stored. Therefore, the version can be specified in the \pyt input file using \eg \settingval{LHAHDF5:version}{0.2.0}.
If a version number is present in the event file that is used, the user input will be ignored and the one in the event file is used instead.

\subsubsection{\lhapdf}
\index{Les Houches Accords!LHAPDF}\index{LHAPDF|see{Les Houches Accords}}\label{sec:interface:lhapdf}
The \lhapdf package is the community standard for providing external
parton distribution functions to event generators. Two versions of
\lhapdf are supported by \pythia, version $5$~\cite{Whalley:2005nh}, a
legacy \fortran version, and version $6$~\cite{Buckley:2014ana}, with
a more performant modern \cpp implementation. The use of \lhapdf~5 is
discouraged and will be fully removed in the future, but is currently
kept to provide PDFs for resolved photons that are not currently
available in \lhapdf~6. Both versions act as interpolators and
extrapolators, for $x$ and $Q^2$ PDF grids provided by fitting
groups. The \lhapdf libraries do not perform DGLAP evolution, and are
restricted in $x$ and $Q^2$ to the grids provided by each PDF set.

Support for \lhapdf can be enabled during \pythia configuration by,
\begin{codebox}
./configure {-}{-}with-lhapdf5[=path] {-}{-}with-lhapdf6[=path]
\end{codebox}
\noindent
where the \texttt{path} can optionally be provided. If the executable
\texttt{lhapdf-config} is available, the \lhapdf path will be
automatically extracted. Plugin libraries are generated along with the \texttt{Pythia} library which are then loaded at run time when \lhapdf sets are requested by the user. With this interface, it is technically possible
to simultaneously use both an \lhapdf~5 and \lhapdf~6 PDF, but this
is strongly discouraged. For all PDFs, proton or otherwise, \lhapdf
sets can be selected via setting the relevant configuration key to the
value \setting{LHAPDF5:set/member} or \setting{LHAPDF6:set/member},
where \texttt{set} is the name of the PDF set to use and
\texttt{member} is the numerical member of that set. If
\texttt{member} is not supplied, the nominal member is assumed. The
example \texttt{main52} demonstrates this syntax, while the example
\texttt{main51} shows how PDF classes can be used independently of a
main \texttt{Pythia} instance.

Every \lhapdf set has a range of validity, given by the minimum and
maximum $x$ and $Q^2$ values of the grids provided. By default,
\pythia freezes these PDF sets at all boundaries for the set, \ie for
$x < x_{\min}$ the PDF value is fixed at $x_{\min}$ and for $Q <
Q_{\min}$ the PDF value is fixed at $Q_{\min}$. It is possible to
enable extrapolation below $x_{\min}$ by setting the
\setting{PDF:extrapolate} flag. This flag applies universally to all
PDF sets, both internal and external. Extrapolation should be enabled
with care, as the extrapolation is PDF set and \lhapdf version
dependent, and in many cases may return nonsensical results. Note
that extrapolation for the remaining boundaries, $x_{\max}$,
$Q_{\min}$, and $Q_{\max}$, is never performed. These values are
always frozen at the limits of validity.

The standardized \texttt{LHAGrid1} format used by \lhapdf~6 allows for
\pythia to use grids from \lhapdf~6 sets without requiring the
\lhapdf~6 library. Simple cubic interpolation is performed in $\ln(x)$
and $\ln(Q^2)$, where all $Q^2$ sub-grids must have the same $x$-value
structure. When less than four $Q^2$ sub-grids are available, linear
interpolation is used instead. All relevant PDF sets can use this
interpolation by setting the relevant PDF configuration key to the
value \setting{LHAGrid1:file}, where \texttt{file} is the full name of
the PDF set file. If \texttt{file} begins with \texttt{/}, then an
absolute file path is used, otherwise the \texttt{file} is assumed to be in the \texttt{share/Pythia8/pdfdata} directory.

\subsubsection{\powheg}\index{powheg@\powheg}
A large number of processes utilizing the \powheg method (positive
weight hardest emission
generator)~\cite{Nason:2004rx,Frixione:2007vw,Alioli:2010xd} are
available via the \powhegbox package~\cite{Jezo:2015aia}. The physics
behind the matching and merging of the hard processes generated by
this package with the \pyt parton shower is detailed in
\cref{section:matchmerge}. Here, technical details on how results
from \powhegbox matrix elements may be technically interfaced with
\pyt are given.

The \powhegbox package uses a common \fortran code structure, which is
then duplicated with process-specific modifications in individual
matrix elements, \eg \texttt{dijets} which produces NLO dijet
events. These individual matrix elements are then compiled to create
executables which when run, take input cards from the user and produce
\lhef output, see \cref{subsection:lha} for details on this
format. This output file can then be directly read into \pyt via the
\setting{Beams:LHEF} setting. Direct \powhegbox input, without
correctly setting up matching, will result in double counting of
emissions. A special \texttt{UserHooks} class, \texttt{PowhegHooks} in
\texttt{Pythia8Plugins}, provides a common interface for appropriately
matching \powhegbox output with the \pyt parton shower. In
\texttt{main31} a full example is given, demonstrating how dijet
events produced from the \texttt{dijets} \powhegbox matrix element
can be correctly passed through \pyt to produce full events.

In some cases, particularly within large experimental frameworks,
users may wish to directly access the \fortran common blocks of a
\powhegbox executable, passing the event by memory to \pyt, rather
than through \lhef output. By default, \powhegbox builds only
executables. However, it is possible to modify the \texttt{Makefile}
via the command,
\begin{codebox}
sed -i "s/F77= gfortran/F77= gfortran -rdynamic -fPIE -fPIC -pie/g" Makefile
\end{codebox}
\noindent so that the executables can also be used as shared libraries. When
modified accordingly, these executables can be linked against \pyt
interface code to produce libraries that can be loaded directly by
\pyt at run time. Run-time loading, rather than dynamic linking, is used
so that multiple \powhegbox processes can be accessed by a single
\texttt{Pythia} instance, without creating symbol collisions between
executables that have common names for global functions and variables.

After appropriately modifying the relevant \powhegbox
\texttt{Makefiles} and compiling executables that can also be used as
shared libraries, the \pyt interface libraries must be created. This
can be configured with \pyt via,
\begin{codebox}
./configure {-}{-}with-powheg-bin=path
\end{codebox}
\noindent where \texttt{path} is the directory containing the \powhegbox
executables. When building \pyt, a plugin library for each \powhegbox in
the provided directory will automatically be created. These plugin
libraries can then be used via the \texttt{PowhegProcs} class provided
in \texttt{Pythia8Plugins} as demonstrated in the example
\texttt{main33}. The program flow is as follows,
\begin{codebox}
Pythia pythia; // Create a Pythia instance. \\
PowhegProcs hvq(\&pythia, "hvq"); // Load the "hvq" plugin library. \\
hvq.readString("configure here"); // Configure the "hvq" plugin \\
hvq.init(); // Initialize the plugins. \\
pythia.init(); // Initialize Pythia.
\end{codebox}
\noindent where the heavy-quark process \texttt{hvq} has been loaded and
configured. It is also possible to include another process,
\begin{codebox}
PowhegProcs dijet(\&pythia, "dijet", "dijetrun");
\end{codebox}
where the additional argument is needed to ensure that the integration
grids from the first process are not overwritten by the second
process.

When using the \texttt{PowhegProcs} method for interfacing with
\powhegbox a \texttt{PowhegHooks} instance is automatically created
and passed to the main \texttt{Pythia} instance. The settings for this
matching hook must be set by the user through either the
\texttt{readString} or \texttt{readFile} methods of the
\texttt{Pythia} instance. In many cases, sensible default values are
set, but some settings are process dependent and must be correctly
configured by the user, \ie \setting{POWHEG:nFinal}.

\subsubsection{\mg5amc}
\label{sec:mg5amc}\index{madgraph@\mg5amc}
\mg5amc~\cite{Alwall:2014hca} is a hard process generator, similar
to \powhegbox, but rather than relying upon individually implemented
processes, it can automatically generate arbitrary processes up to
NLO. There are a number of ways through which \mg5amc can be interfaced
with \pyt.
\begin{enumerate}
\item \madgraph and \amcatnlo themselves can interface with \pyt and pass
  generated hard processes through \pyt to produce full events, all
  within the \mg5amc machinery.
\item \lhef output from \mg5amc can be passed to \pythia, see
  \cref{subsection:lha} for details on reading \lhef
  input.\label{madgraph:lhef}
\item Source code for matrix-element libraries, inheriting from the
  internal \texttt{SigmaProcess} class in \pyt, can be generated by
  \madgraph.
\item The \mg5amc executable can be called from within \pyt via the
  \texttt{LHAup\-Madgraph} class.\label{madgraph:lhaup}
\item Matrix-element plugins for the \dire and \vincia parton showers
  can be generated by \madgraph, compiled, and then loaded at run time.
\end{enumerate}
The latter three methods are covered in more detail below. In all
cases, it is important that appropriate matching and merging,
see \cref{section:matchmerge}, is configured to ensure there is no
double counting between the generated hard process and the remainder of
the event produced by \pythia.

Semi-internal processes can be passed to \pythia via inheriting from
the \texttt{SigmaProcess} class. The primary method of this class is
\texttt{sigmaHat} where the exact definition depends upon the
final-state multiplicity of the process. Phase-space generation can be
handled by \pythia for $2 \to 1$, $2 \to 2$, and $2 \to 3$ processes,
although the $2 \to 3$ phase-space sampler is not particularly
sophisticated. When necessary, users can provide custom external
phase-space samplers. Consequently, while $2 \to n$ processes can be
externally supplied, phase-space generation must also be implemented
by the user for $n > 3$.  A full example is given in the example
\texttt{main22} but the general syntax is,
\begin{codebox}
SigmaProcess* userSigma = new UserSigma();\\
pythia.setSigmaPtr(userSigma);
\end{codebox}
where \texttt{UserSigma} is a user-defined process inheriting from
\texttt{SigmaProcess}.

Semi-internal process source code can be generated from within the
\madgraph \python interface as follows.
\begin{codebox}
import model model\_name\\
generate mg5\_process\_syntax\\
add process mg5\_process\_syntax\\
output pythia8 [path\_to\_pythia]
\end{codebox}
A directory containing the output for the process is placed in the
\pythia source directory specified by \texttt{path\_to\_pythia} and an
example is placed in the \texttt{examples} directory.

It is also possible to call \madgraph from within \pythia via the \texttt{LHAupMadgraph} class provided in \texttt{Pythia8Plugins}.
\begin{codebox}
shared\_ptr<LHAupMadgraph> madgraph =\\
\hspace*{5mm} make\_shared<LHAupMadgraph>(\&pythia, true, "madgraphrun", exe);\\
madgraph->readString("generate mg5\_process\_syntax");\\
pythia.setLHAupPtr(madgraph);
\end{codebox}
This interface generates the relevant \madgraph configuration cards,
and then runs the \madgraph executable, specified by \texttt{exe}, to
produce \lhef output that is then read in by \pythia. An attempt is
made to automatically set up matching and merging, but this process
should always be validated by the user. Random-number sequences are
automatically handled, based on the \pythia random-number
generator. Whenever the \lhef input is exhausted, a new call is made
to the \madgraph executable and a new \lhef output is generated.

Finally, it is possible to use \madgraph to generate matrix-element
plugins for use in the \dire and \vincia parton showers. A number of these
plugins are already provided with the \pythia distribution in the
\texttt{plugins/mg5mes} directory. To enable this plugin support, configure \pythia with
\begin{codebox}
./configure {-}{-}with-mg5mes[=path]
\end{codebox}
\noindent
where the path to the matrix-element plugin source-code directories
can optionally be specified. A plugin library for each directory in
the path will be built, which can then be loaded at run time. Just as
for \powhegbox, run-time loading of the matrix elements allows for
multiple plugins to be used with the same instance of
\texttt{Pythia}. For \dire and \vincia, the plugin library to be used
can be specified with the settings \setting{Dire:MEplugin} and
\setting{Vincia:MEplugin} respectively.

New matrix-element plugin libraries can be generated by using the
\texttt{generate} command in the \texttt{plugins/mg5mes} directory. In its simplest
form, the user just needs to specify the process,
\begin{codebox}
./generate {-}{-}process="mg5\_process\_syntax"
\end{codebox}
\noindent but may also specify a model to use, as well as the output
directory. Advanced usage is also possible where a custom \madgraph
card is passed by the user, or the interactive mode of \madgraph is
enabled. Note that this feature requires the use of \href{https://www.docker.com/}{\docker} to
download and run a container with a custom version of \madgraph.

The most common interface to \mg5amc is through text 
files in \lhef format, \cf\cref{subsection:lha}. For easy interfacing
between \mg5amc and \pythia, some custom additions to the file format are 
employed:
\begin{itemize}
\item The event attributes \texttt{npLO} and \texttt{npNLO}
are used to set the number of particles at lowest order for events with
leading-order and next-to-leading order cross sections, respectively.
For the former, \texttt{npLO} amounts to a simple final-state particle count.
For the latter, \texttt{npNLO} gives the number of final-state particles necessary
to define the scattering at Born level. It is assumed that \texttt{npLO}$\geq 0\rightarrow$ \texttt{npNLO}$< 0$ and
\texttt{npNLO}$\geq 0\rightarrow$ \texttt{npLO}$< 0$, meaning that these attributes also act to signal if an event is
a leading-order or next-to-leading-order contribution.
\item Several mechanisms to set the parton shower starting scales for individual particles exist. These rely on
attributes of the \verb|<scales>| tag defined in the \lhef 3.0 format.
\item For the case of MLM matching, the parton-shower starting scale information is also used
to signal whether a particle should not be considered for the MLM jet matching procedure. Particles that
have been assigned a starting scale $\mu>2E_\mathrm{CM}$ will be considered exempt from the MLM jet matching criterion. 
\end{itemize}

\mg5amc further incorporates provisions for automatic NLO+PS matched
calculations within the \mcatnlo\index{NLO matching} approach. The interface between \amcatnlo and  
\pythia typically relies on phase-space points transmitted via \lhef. However, for
special matching tasks, it is possible to invoke \pythia from within \amcatnlo. This
is the case for the \mcatnlo$-\Delta$ matching prescription. The relevant \fortran
code, wrapping \pythia functionality, is shipped within \mg5amc. \pythia can be set up for use within
\mg5amc by setting the configuration flag \setting{Merging:\-runtimeAMCATNLOInterface}. This then allows \mg5amc direct access to select parts of \pythia's internal merging machinery, to \eg enable the extraction of Sudakov form factors. A more detailed introduction may only be relevant to experts in \mg5amc, and may be found in the \htmlmanual.

\subsubsection{\helaconia}\index{Quarkonium}
While \pythia has a complete collection of expandable quarkonia
processes, see \cref{subsection:onia}, it is sometimes necessary to
generate quarkonia states at higher orders or with additional final-state partons. Previous versions of \madgraph were able to produce
arbitrary tree-level quarkonia processes via
\madonia~\cite{Artoisenet:2007qm}, but the current version of
\madgraph no longer has this ability to generate bound heavy-quark
resonances. However, the standalone \helaconia~\cite{Shao:2015vga}
package is able to provide the same functionality of the \madonia
package, and beyond.

The program flow of \helaconia is very similar to that of \madgraph. A
\python interface is used to generate source code which is then
compiled and run to produce \lhef output. This output can then be
provided to \pythia to produce full events with parton showers,
underlying event, and particle decays. The \helaconia syntax is
modelled after the \madgraph syntax, and consequently, the interface is
similar. Unlike \madgraph, \helaconia is not able to produce
semi-internal matrix elements inheriting from the
\texttt{SigmaProcess} class. Instead, \helaconia can be interfaced
either by directly providing \lhef output to \pythia, or using the
\texttt{LHAupHelaconia} class provided in \texttt{Pythia8Plugins}.

The \texttt{LHAupHelaconia} interface is very similar to that of \texttt{LHAupMadgraph},
\begin{codebox}
shared\_ptr<LHAupHelaconia> helaconia =\\
\hspace*{5mm}make\_shared<LHAupHelaconia>(\&pythia, true, "helaconiarun", exe);\\
helaconia->readString("generate ho\_process\_syntax");\\
pythia.setLHAupPtr(helaconia);
\end{codebox}
where \texttt{ho\_process\_syntax} is the \helaconia equivalent for
the \madgraph process syntax. The \helaconia executable must be
available via the string \texttt{exe}. Every time a \pythia event is
generated, the plugin checks if an event is available from an \lhef
file generated by \helaconia. If not, it will automatically run
another batch of events. Random-number seeds and sampling are
consistently handled in the same way as for \texttt{LHAupMadgraph}.

\subsubsection{\evtgen}\index{Evtgen@\evtgen}
For many experimental collaborations, particularly those specializing
in $\gp{B}$-physics, more detailed hadron-decay models are needed than
those provided by default in \pythia. The \evtgen~\cite{Lange:2001uf}
package specializes in $\gp{B}$-hadron decays, including sophisticated
models, spin correlations, and the ability to implement new models. To
include spin correlations \evtgen does not just decay a single
particle at a time, but instead performs the entire decay tree for
each given initial particle. Consequently, decays from \evtgen cannot
be included in \pythia via the provided \texttt{DecayHandler} class,
called during the decay stage of the hadron level, but must rather be
performed after full event generation. Such an interface for \evtgen
is supplied by the class \texttt{EvtGenDecays} provided in
\texttt{Pythia8Plugins}.

In $\gp{B}$-physics, particularly at hadron colliders, one oftentimes
wishes to produce a large sample of events where each event contains
one or more rare signal decays, \eg $\gp{B_\s^0} \to \gp{\mu^+}
\gp{\mu^-}$. The first step, of course, is to generate an event with
at least one signal particle candidate, while the second step is to
force the signal decay for one of these candidates. The weight for an
event containing one candidate with a forced signal decay is simply
the branching fraction for the signal decay. However, when multiple
candidates are present, the event weight becomes slightly more
complex, requiring non-trivial bookkeeping. Consequently, the
\texttt{EvtGenDecays} class in \pythia provides a generalized
mechanism by which to force signal decays for given particle species,
while still providing an appropriate event weight.

Signal particle candidates, $c_i$, do not all need to be the same
particle species. Here, a particle species differentiates not only
between particle types, \eg $\gp{B_\s^0}$ and $\tau^+$, but also
between particles and antiparticles, \eg $\tau^+$ and
$\tau^-$. Additionally, the signal decay for a candidate, with
branching fraction $\mathcal{B}_\mathrm{sig}(c_i)$, can include
multiple channels. Consequently, arbitrarily complex signal decays can
be forced. As an example, events can be required to contain one or
more of the following decays: $\tau^+ \to \bar{\nu}_\tau \pi^+$,
$\tau^+ \to \bar{\nu}_\tau \pi^0 \pi^+$, $\gp{B_\s^0} \to \mu^+
\mu^-$, and $\tau^- \to \tau_\nu \pi^- \pi^- \pi^- \pi^+ \pi^+$. Here,
assuming equal production of the three particle species (which is
almost certainly not the case), the decay $\tau^+ \to \bar{\nu}_\tau
\pi^0 \pi^+$ of the four signal decays will be the most commonly
forced decay. Following this notation, the event weighting is
performed as follows.

\begin{enumerate}
\item An event is generated and all $n$ signal particle candidates,
  $c_i$, are found. If there are no candidates, $n = 0$, then an event
  weight, $\mathcal{W}_\mathrm{event}$, of $0$ is returned.
\item If $n > 0$ then a candidate $c_i$ is randomly chosen with
  probability
  \begin{equation}
    P(c_i) = \frac{\mathcal{B}_\mathrm{sig}(c_i)}{\sum_{j = 1}^m \Big(1 -
      \mathcal{B}_\mathrm{sig}(c_j)\Big)} ~,
  \end{equation}
  where $\mathcal{B}_\mathrm{sig}(c_i)$ is the signal branching
  fraction for each candidate $c_i$.\label{evtgen:choose}
\item A channel is selected for the chosen candidate $c_i$ from one of
  the signal channels contributing to $\mathcal{B}_\mathrm{sig}(c_i)$.
\item Channels for all remaining candidates are selected, using all
  allowed decay channels, not just the signal channels.
\item The number of candidates with a selected signal channel, $m$, is
  determined. The channel selection for the candidates is then kept
  with probability $1/m$. If the channel selection is rejected, the
  algorithm returns to step~\ref{evtgen:choose} and a new set of
  channels is selected.
\item All candidates are decayed via their selected channel and 
  \begin{equation}
    \mathcal{W}_\mathrm{event} = 1 - \prod_{i = 1}^n \Big(1 -
    \mathcal{B}_\mathrm{sig}(c_i)\Big) ~,
  \end{equation}
  is calculated as the event weight.
\end{enumerate}
An unweighted sample of events can be obtained by randomly selecting
events, each with probability
$\mathcal{W}_\mathrm{event}/\mathcal{W}_\mathrm{max}$. The maximum
possible event weight, $\mathcal{W}_\mathrm{max}$, can be determined
by the maximum weight from a sufficiently large sample of events.

To use \evtgen in \pythia, configure \pythia with
\begin{codebox}
./configure {-}{-}with-evtgen[=path]
\end{codebox}
where \texttt{path} optionally provides the path to the \evtgen
installation. Note that \evtgen itself also links against \pythia, so
in some cases it might be necessary to reconfigure \pythia after
installation of \evtgen. A full example using \evtgen is provided in \texttt{main48}. The general syntax is,
\begin{codebox}
EvtGenDecays evtgen(\&pythia, dec, pdl);\\
pythia.next();\\
evtgen->decay();
\end{codebox}
where \texttt{dec} and \texttt{pdl} provide the paths to the \evtgen
decay and particle data files.

\subsubsection{External random-number generators}

\index{Random numbers!External random-number generators}
When including \pyt in a larger software framework, using a single
random-number generator across all components is oftentimes required
to ensure reproducible results. Consequently, an external random-number-generator pointer may be passed for use by a given \pyt
instance.
\begin{codebox}
\begin{verbatim}
pythia.setRndmEnginePtr(rng)
\end{verbatim}
\end{codebox}
\noindent Here, \texttt{rng} is a pointer to an instance of a user-defined
random number generator derived from the \texttt{RndmEngine}
class. The only method that must be implemented by the user is
\texttt{flat} which should return a uniform distribution between $0$
and $1$. \index{Random numbers!Seed}The example below implements a linear congruential
generator with a configurable seed, modulus, multiplier, and
increment.
\begin{codebox}
\begin{verbatim}
class RandomLCG : public RndmEngine {
public:
  long int seed{1}, m{2147483648}, a{1103515245}, c{12345};

  // The only method that needs to be implemented.
  double flat() {
    seed = (a * seed + c) % m;
    return double(seed)/m;
  }
};
\end{verbatim}
\end{codebox}

\index{Random numbers!RndmEngine@\texttt{RndmEngine}}\index{RndmEngine@\texttt{RndmEngine}}Typically, the \texttt{RndmEngine} class can be used to wrap some
other random number generator. An example of this is the
\texttt{MixMadRndm} class which is a wrapper for an implementation of
the MIXMAX
algorithm~\cite{Savvidy:2014ana}.\index{MIXMAX}\index{Random numbers!MIXMAX}
\begin{codebox}
\begin{verbatim}
#include "Pythia8Plugins/MixMax.h"
MixMaxRndm rng(0, 0, 0, 123);
pythia.setRndmEnginePtr(&rng);
\end{verbatim}
\end{codebox}
\noindent The argument to the generator constructor is four seed
values. While this functionality of providing an external random
number generator is useful, it should be treated with care. Some
pseudo-random-number generators implemented in standard packages are
not sufficient for large scale generation, \eg the CLHEP
implementation of the \index{RANLUX}\index{Random numbers!RANLUX}RANLUX algorithm. Consequently, when possible,
the default random number generator in \pyt, based on the RANMAR
implementation of the Marsaglia-Zaman
algorithm~\cite{Marsaglia:1990ig}, is recommended and sufficient for
most physics purposes, see \cref{sec:rng}.

\subsection{Output formats}
\label{sec:output-formats}

\pyt comes with a set of example main programs, and in most of these
the analysis of the produced event is performed directly in the code
there. It is also possible to output the events to be analyzed by
interfacing to external programs and code. For this purpose \pyt can
communicate its events with different output formats as described in
this subsection.

\subsubsection{\hepmc versions 2 and 3}
\label{sec:int-ex-hepmc}\index{HepMC@\hepmc}

The standard format for communicating fully generated events is called
\hepmc~\cite{Dobbs:2001ck,Buckley:2019xhk} and defines a set of \cpp
classes to describe an event and all particles therein. Internally, the
particles are connected by vertex objects using pointers.

The latest version of the \hepmc code is not yet adopted by all LHC
collaboration and \pyt therefore has support for both version 2 (2.06
and later) and version 3. The interface as such is provided at the header file
level using \texttt{Pythia8Plugins/HepMC2.h} or
\texttt{Pythia8Plugins/HepMC3.h}, and the \pyt code itself does not
have any dependencies on these. This means that the \pyt (shared)
library can be built independently of which version of \hepmc
should be used. However, if one wishes to use the example main programs
that show how to use \hepmc\footnote{The example main programs can be
  found in the \htmlmanual under \texttt{Getting Started $\to$
    Examples by Keyword}, search for \texttt{Hepmc}.} the configuring
of \pyt must be done according to
\begin{codebox}
  ./configure -{-}{-}with-hepmc3=/path/to/hepmc/installation
\end{codebox}
or
\begin{codebox}
  ./configure -{-}{-}with-hepmc2=/path/to/hepmc/installation
\end{codebox}

Besides the particles, other information will also be transferred to
the \hepmc format, such as cross sections, parton density information, and
different weights (see \cref{sec:using-weights}). Note,
however, that not all information in the \texttt{Pythia8::Event} is
preserved in the \hepmc output. Notably, the status codes for particles
in \hepmc are only set to \texttt{1} (final state particle), \texttt{2} (decayed
standard model hadron or \tauon or \muon), \texttt{4} (incoming beam), or a
number in the range \texttt{11}--\texttt{200} (generator dependent status of an
intermediate particle, given by the absolute value of the
corresponding \pyt status code).

\subsubsection{Histograms with the \yoda package}\index{YODA@\yoda}

Even though the built-in histogram package might suffice for the most basic
use cases, such as one-dimensional histograms, most users require more advanced
capabilities. Since \pythia is not a statistics or plotting package, we refer the user
to external programs. For slightly more advanced use cases, we recommend interfacing
to the \yoda\footnote{See \url{https://yoda.hepforge.org/}.} histogram package. If installed,
\pythia can be configured with \texttt{{-}{-}with-yoda=/path/to/yoda}, which allows the user to 
create \texttt{Makefile} recipes with access to \yoda histograms easily. The \yoda package will then be accessible
in \pythia as any other \cpp library can be accessed. Questions regarding the \yoda histogram package should be addressed to the \yoda authors.

\subsubsection{Interfacing with \root}\index{root@\root}

For more advanced analyses, many users prefer the \root~\cite{Brun:1997pa} package. \pythia
provides several possibilities to interface with \root, version 6 or higher. Use cases
can roughly be grouped into three categories:
\begin{enumerate}
	\item Using \root as a histogram package inside \pythia.
	\item Using \pythia to generate \root events or ``n-tuples'', which can be post-processed by \root.
	\item Steering \pythia from inside a \root-based framework.
\end{enumerate}
We will here briefly cover the first two use cases, but refer the user to the \root documentation for
using the \pythia interface in \root, where it is extensively documented.

The simplest use case is of the first category which, from a technical point of view, is not too
different from using any other \cpp library along with \pythia. In the example \texttt{main91}, it is
shown how to declare a \root \texttt{TApplication} environment and \root \texttt{TH1F} histograms, to
be filled by \pythia, and displayed on screen. The crucial part is the \texttt{Makefile} recipe. If
\pythia is configured \texttt{{-}{-}with-root}, convenient variables pointing to the \root libraries and the 
\texttt{root-config} script can be used as shown, to compile a \texttt{main} program with the necessary
linking to \root libraries. The generated histograms can then be saved to a \texttt{.root} file for
later access.

Most users already familiar with \root, would rather store event files generated with \pythia (so-called 
``n-tuples'') on disk, which can then be post-processed with a \root-centric analysis framework, often with
auxiliary packages, provided by a large experiment. In such cases, examples \texttt{main92} and \texttt{main93}
can be of inspiration. The \texttt{main92} example shows how to store full events into a \root \texttt{TTree}.
For most realistic use cases, this is not very practical, as such files will quickly grow large, containing a
significant amount of information which is of little relevance to the user. The \texttt{main93} program provides a more 
streamlined interface. In the header file \texttt{main93.h}, two classes \texttt{RootTrack} and \texttt{RootEvent}
are defined. Those classes define what information about each track (including track-level cuts, \eg desired 
acceptance) as well as each event, should be stored in an output \root-file. If \pythia has been configured with
\root, the \texttt{main93} example can be run with an input \texttt{.cmnd} file with the flag \texttt{Main:writeRoot = on},
and the desired information will be stored. If changes are made to the header file, \texttt{main93} must be
recompiled.
For both \texttt{main92} and \texttt{main93}, the compilation recipe in the \texttt{Makefile} is the most difficult
part to set up, as both require generation of compiled and linked \root dictionary libraries with CINT. A user wishing
to go beyond simple extensions of the given examples are encouraged to study the existing \texttt{Makefile} 
recipes, as well as the \root documentation on \texttt{Linkdef.h}. It is kindly requested that queries about \root
dictionary library generation are directed to the \root authors.

\subsection{Analysis tools}
\label{sec:analysis-tools}

The tools included in \pyt are normally enough for doing simple
analyses of the generated events, but for more complicated analyses, or
if direct comparison with data is wanted, the user needs to interface
to external tools. Here we describe some of these
interfaces.

\subsubsection{\rivet versions 2 and 3}
\label{sec:int-ex-rivet}\index{Rivet@\rivet}

The \rivet package~\cite{Buckley:2010ar,Bierlich:2019rhm} is probably
the most convenient way of comparing event-generator models to
experimental data. The program includes a large collection of
experimental analyses encoded (usually by the experiments themselves)
in \cpp classes that read \hepmc input and produce \yoda files that
can be plotted together with the experimental data points (also
provided by the experiments through HEPDATA~\cite{Whalley:1988wi,Maguire:2017ypu}).

Since \rivet only needs \hepmc input, the only thing needed for \pyt is
to write the events to a \hepmc file (see \cref{sec:int-ex-hepmc}) or a \emph{pipe} (which is recommended to
avoid creating unnecessarily large files), and have \rivet take this as
input. Assuming a main \pyt program, \texttt{mymain-hepmc}, that
simply writes \hepmc to the standard output, the commands to do this
are
\begin{codebox}
  mkfifo hepmc-pipe
  
  ./mymain-hepmc > hepmc-pipe \&

  rivet -a SomeAnalysis hepmc-pipe

  rivet-mkhtml Rivet.yoda
\end{codebox}
\noindent where the last command will produce formatted web pages in the
\texttt{rivet-plots} subdirectory, with the comparisons to data.

In \pyt there is also a more direct way of calling \rivet from within
a main program provided. This uses the header file
\texttt{Pythia8Plugin/Pythia8Rivet.h} and provides simple shortcuts as
shown in some of the provided example main programs.\footnote{The
  example main programs can be found in the on-line manual under
  \texttt{Getting Started $\to$ Examples by Keyword}, search for
  \texttt{Rivet}.} To enable this, \pyt must be configured using
\begin{codebox}
  ./configure {-}{-}with-rivet=/path/to/rivet/installation
\end{codebox}
\noindent together with the corresponding \texttt{-{-}{-}with} for the
version of \hepmc that \rivet was configured with.

\pyt currently supports direct linking with both versions 2 and 3 of
\rivet. The additional features in the later version includes the
possibility of using different weights (see \cref{sec:using-weights}), several heavy-ion specific 
features~\cite{Bierlich:2020wms}, and to 
provide options to the analyses. Support for version 2 of \rivet will
likely be dropped in the future.

\subsubsection{\fastjet}\index{fastjet@\fastjet}
The \texttt{fjcore} code is distributed together with the \pythia code by permission 
from the authors. There is also an interface that inputs \pythia events into 
the full \fastjet library, for access to a wider set of
methods, but then \fastjet must be linked by using 
\begin{codebox}
	./configure {-}{-}with-fastjet=/path/to/fastjet/installation {-}{-}with-fastjetlib=/path/to/ \\
	fastjet/library
\end{codebox}

Among \pythia example main programs, \texttt{main71.cc} shows in the case of \W  plus jet production, how  the \fastjet package can be used for analysis of the final state, and \texttt{main80.cc} performs CKKW-L merging\index{CKKW(-L) merging} 
with a merging scale defined in $\kT$, with \texttt{main80.cmnd} and LHE files as input. Also, \texttt{main72.cc} 
compares QCD jet finding in \texttt{SlowJet} and \fastjet, using the header file \texttt{FastJet3.h} present 
in the directory \texttt{Pythia8Plugins} contributed by Gavin Salam~\cite{Cacciari:2011ma}.

\subsection{Computing environments}
\label{sec:computing-environments}

\pyt has been developed as a \cpp library to write and
compile programs to execute standalone on a generic $*$nix
operating system on a generic computer. However, we address
here the rise in
popularity of \python as a development language and a powerful
tool in machine-learning applications.

\subsubsection{\python interface}\index{python@\python}
To meet the growing requirements of a large user base, \pyt includes a flexible \python interface to most frequently used classes,
and thus allows a user to write a \pyt \texttt{main} program entirely in \python. This provides the user direct
access to the wealth of analysis and visualization tools, available through \python libraries, all at run time. A number of \python examples are provided, each a direct translation of their corresponding \cpp counterpart. The interface
is generated with \binder using the \pybind template library. The specific version of \binder and \pybind needed to generate the interface is provided through a small \docker container.

The default interface is a simplified one, with only the core \pyt functionality available. This interface is a trade off between
usability and remaining light weight. The top level \texttt{Pythia} class is available, as well as all relevant \texttt{Event}, \texttt{ParticleData}, and analysis tool related classes. An important feature of the interface is that it is bi-directional, derived classes
in \python can be passed back to \pyt. This is useful, for example, to create a \texttt{UserHooks} derived class (see \cref{subsection:userhooks}). All user interface classes, typically passed to the main \texttt{Pythia} object via pointers in the standard \cpp code, are available through the simplified interface.

A full \python interface can also be generated by the user. Only \docker is required to enable the generation of a new \python interface to \pyt. The following generates the full interface.
\begin{codebox}
cd plugins/python\\
./generate {-}{-}full 
\end{codebox}
It is also possible to generate a user-defined interface which is tailored to a specific use case via the flag \texttt{{-}{-}user=FILE} instead. Here, \texttt{FILE} is a \binder configuration file specified by the user. Note that whenever changes are made to the \pyt \cpp headers, the \python interface must be generated again, whether simplified, full, or user defined.

Installation of the \python interface requires the \texttt{Python.h} header to be available. The \texttt{python-config}
script can be used to find the relevant paths when configuring \pyt. An example configuration for \pyt with \python3.6 could then be:
\begin{codebox}
./configure {-}{-}with-python-config=python3.6-config
\end{codebox}
This would configure \pyt to be built with the default interface, using \python3.6.
After configuring, the compiled \pyt module is available in the \texttt{lib/} directory under the top level \pyt directory. The 
\python installation must have that directory made available, \eg by setting:
\begin{codebox}
\begin{verbatim}
export PYTHONPATH=$(PWD)/lib/:$PYTHONPATH
\end{verbatim}
\end{codebox}
from the top level \pyt directory. After compiling with \texttt{make}, the \python interface should be available. The following example loads the \pyt \python module and prints the internal documentation which includes the available classes, as well as some of the not-so-obvious features.
\begin{codebox}
{>}{>}{>} import pythia8\\
{>}{>}{>} help(pythia8)
\end{codebox}

One of the main reasons for the \python interface is the fast development of a standalone \texttt{main} program
in \python rather than \cpp, allowing for an environment of external tools, which the user might be more familiar with. As an example
of such a program, consider the short \python script below, which will run \pyt to produce a \texttt{numpy} histogram containing
the distribution of charged hadron multiplicity at mid-pseudorapidity in proton collisions at LHC energies.

\begin{codebox}
\begin{verbatim}
# Wrapper around numpy histogram to allow fill functionality.
import numpy as np
class HistoFiller(object):
    def __init__(self, bins):
        self.bins = bins
        self.hist, edges = np.histogram([], bins=bins, weights=[])
        self.widths = []
        for i in range(len(edges)-1):
            self.widths.append(edges[i+1] - edges[i])

    def fill(self, val, w=1.0):
        hist, edges = np.histogram(val, bins=self.bins, weights=w)
        self.hist+=hist

    def get(self):
        scale = 1./sum(self.hist)
        return [h/w*scale for h,w in zip(self.hist,self.widths)],
               [np.sqrt(h)*scale for h in self.hist]

# Set up Pythia and declare histogram.
import pythia8
pythia = pythia8.Pythia()
pythia.readString("SoftQCD:all = on")
pythia.init()
mult = HistoFiller([3.*x for x in range(20)])

# Event loop. Find particles and fill histogram.
for iEvent in range(1000000):
    if not pythia.next(): continue
    nCharged = 0
    for p in pythia.event:
        if p.isFinal() and p.isHadron() and p.isCharged():
            nCharged += 1
    mult.fill(nCharged)

# Plot the histogram using the matplotlib library.
import matplotlib.pyplot as plt
fig = plt.figure()
ax = fig.add_subplot(111)
y, ye = mult.get()
ax.errorbar(mult.xvals,y,xerr=[w/2. for w in mult.widths],
	yerr=ye, drawstyle='steps-mid',fmt='-',color='black')
ax.set_xlabel(r'$dN_{ch}/d\eta$')
ax.set_ylabel(r'$P(dN_{ch}/d\eta)$')
plt.show()
\end{verbatim}
\end{codebox}

%% file: summary/summary.tex
Our goal in writing this manual was to provide reference material for users
and developers of \pythia.  We provided some basic content
that we considered mandatory, such as defining what \emph{is} an event
generator,  how does our code structure reflect the physics, and what
sorts of numerical methods we use in the program.  This is
covered in \cref{part:intro}.  The core of the manual, provided in \cref{part:physics},
describes in detail the phenomenon that is simulated and our
assumptions and approximations. Parton showers and
hadronic or nuclear physics are covered in more detail because these
have been the arenas of more recent development. Other topics are
covered more liberally in the HEP literature, and we hope to have
provided enough outside references.  What is somewhat new compared
to other \pyt manuals is \cref{part:use}, dedicated to the user of \pythia.
Our aim was not to give the user an easy way to skip the description
of physics, but to facilitate the use of the program in real analyses
and investigations.  This part of the manual is the most pragmatic, but also
the one most susceptible to acronyms, initialisms, and jargon.
It is also the most technical in describing our and others' computer
code.

This manual is a snapshot of an evolving entity.  Within a short
period of our concluding statements, new developments will arise that
are not covered in this manual. We hope this continues, even as we
pass the torch to the next generation of \pyt authors and
contributors.

James D. Bjorken (``BJ'' to his generation) wrote of the ``tyranny''
of Monte Carlo in a short paragraph of a larger editorial on the
future of particle physics in 1992~\cite{Bjorken:1992cu}. He lamented the fact that
Monte-Carlo predictions were taken as the truth, event though most
of the prediction was a black-box. Had he read this manual in 2022, we
hope he would understand that the authors have provided a code
that is more democratic, and allows users to liberally test ideas, but
within well-defined boundaries. As such, there is no \emph{single}
\pyt prediction to compare to data.    

%% file: summary/acknowledgements.tex
\section*{Acknowledgements}

A large number of people should be thanked for their contributions to the \pythia event generator. 

First of all, Bo Andersson and G\"osta Gustafson are the originators of the Lund model, and have strongly influenced the development of both early code versions and also recent model additions. Hans-Uno Bengtsson should furthermore be acknowledged as the originator of the \pyt program.

Some made contributions dating way back in the programs' history, others more recently. While praise for the contributions should go to the contributors, blame for mistakes made in modifications to the original code, or failure to keep it up-to-date, should rest with the core authors.

\subsection*{Former authors}

Former \pyt~8 authors, who are no longer active in the field, are:
Stefan Ask, Jesper Roy Christiansen, Richard Corke, Nadine Fischer, and
Christine O.\ Rasmussen. The merging of \vincia into \pythia brought
with it further significant author contributions from Helen Brooks 
in particular.

\subsection*{Further contributions}

The program has received many smaller and larger contributions and bug reports over time, from users to numerous to mention here. They are mentioned in the online update notes as the bug fixes go in, and are all gratefully acknowledged. 

In particular, contributions from the following should be mentioned: Baptiste Cabouat for developing and implementing the initial-final dipole approach, Silvia Ferreres-Sol\'e for implementing the space-time hadronic production points in string fragmentation, and Tomas Kasemets for implementation of new proton PDFs.

Code contributions from the following collaborators and users are also gratefully acknowledged: O.~Alvestad, S.~Baker, B.~Bellenot, R.~Brun, A.~Buckley, M.~Cacciari, L.~Carloni, S.~Carrazza, R.~Ciesielski, V.~Hirschi, N.~Hod, H.~Hoeth, J.~Huston, M.~Kirsanov, A.~Larkoski, B.~Lloyd, J.~Lopez-Villarejo, O.~Mattelaer, M.~Montull, A.~Morsch, A.~Naumann, S.~Navin, P.~Newman, M.~Ritzmann, J.~Rojo, G.~Salam, K.~Savvidy, G.~Savvidy, A.~Singh, G.~Soyez, M.~Sutton, R.~Thorne, and G.~Watt. We also thank J. Altmann and T. Garnett for correction of typos in this manuscript.

Finally, vigilant code tests of \pyt releases by Mikhail Kirsanov, Dimitri Konstantinov, and Vittorio Zecca are gratefully acknowledged.

\subsection*{Financial support}

The Lund and Monash groups have received financial support from the
EU H2020 Marie Sk\l odowska-Curie Innovative Training Network MCnetITN3,
grant agreement 722104.\\\\
\noindent
The Lund group has also received funding from the European Research
Council (ERC) under the European Union's Horizon 2020 research
and innovation programme, grant agreement No 668679 (MorePheno),
and from the Swedish Research Council, contract number 2016-05996.\\\\
\noindent
The Jyv\"askyl\"a group (IH and MU) has been funded as a part of the CoE in Quark Matter of the Academy of Finland.\\\\
\noindent
CB and LL acknowledge support from the Knut and Alice Wallenberg
foundation, contract number 2017.0036.\\\\
\noindent
SC and LL acknowledge support from the Swedish Research Council,
contract number 2020-04869.\\\\
\noindent
ND acknowledges support from the Science and Engineering Research Board, Government of India under Ramanujan Fellowship SB/S2/RJN-070. \\\\ 
\noindent
SM is supported by the Fermi Research Alliance, LLC under Contract No. DE-AC02-07CH11359 with the U.S. Department of Energy, Office of Science, Office of High Energy Physics.\\\\
\noindent
PS acknowledges support from the Australian Research Council via
Discovery Project DP170100708 --- ``Emergent Phenomena in Quantum
Chromodynamics''.\\\\
\noindent
RV acknowledges support from the European Research Council (ERC) under the European
Union’s Horizon 2020 research and innovation programme (grant agreement No. 788223,
PanScales), and from the Science and Technology Facilities Council (STFC) under the grant
ST/P000274/1.\\\\
\noindent
CTP acknowledges support from the Swiss National Science Foundation (SNF) under contract 200021-197130, the Monash Graduate Scholarship, the Monash International Postgraduate Research Scholarship, and the J.~L.~William Scholarship.\\\\
\noindent
IH acknowledges support from the Academy of Finland, project numbers 308301 and 331545, and from the Carl Zeiss Foundation.\\\\
\noindent
MU acknowledges support from the Academy of Finland, project number 336419.\\\\
\noindent
PI acknowledges support from the United States National Science Foundation (NSF) via grant NSF OAC-2103889.

%% file: physics/hard-proc-table.tex
\section{Full list of internal processes}
\label{sec:hardProcesses}
\subsection{Standard model processes}
\label{sec:SMprocesses}

\begin{table}[th!]\index{QCD processes}
\index{Soft QCD processes}\index{Diffraction}
\index{Elastic scattering}\index{Single diffraction}
\index{Double diffraction}\index{Central diffraction}
\caption{List of internal soft QCD processes, see \cref{subsection:sigmatotal} for details and references.
\label{tab:softProcesses-qcd}}
\begin{tabular}{p{0.15\textwidth}p{0.4\textwidth}p{0.3\textwidth}}
\toprule
  process & internal name & code \\
\midrule
 & \texttt{SoftQCD:all} & \\
$A \, B \rightarrow X $ & \texttt{SoftQCD:nonDiffractive} & \texttt{101} \\
$A \, B \rightarrow A \, B $ & \texttt{SoftQCD:elastic} &  \texttt{102}  \\
$A \, B \rightarrow X \, B $ & \texttt{SoftQCD:singleDiffractiveXB} & \texttt{103}  \\
$A \, B \rightarrow A \, X $ & \texttt{SoftQCD:singleDiffractiveAX} & \texttt{104} \\
$A \, B \rightarrow X_1 \, X_2 $ & \texttt{SoftQCD:doubleDiffractive} & \texttt{105}   \\
$A \, B \rightarrow A \, X \, B $ & \texttt{SoftQCD:centralDiffractive} &\texttt{106}  \\
 & \texttt{SoftQCD:singleDiffractive} & \texttt{104, 103} \\
 & \texttt{SoftQCD:inelastic} & \texttt{101, 103, 104, 105, 106} \\
\bottomrule
\end{tabular}
\end{table}

\begin{table}[thp!]\index{QCD processes}\index{Hard QCD processes}
\caption{List of internal hard QCD processes, see \cref{subsection:hardQCD} for details.
\label{tab:hardProcesses-qcd}}
\begin{tabular}{p{0.15\textwidth}p{0.4\textwidth}p{0.15\textwidth}p{0.15\textwidth}}
\toprule
  process & internal name & code & refs. \\
\midrule
 & \texttt{HardQCD:all} & & \\
$\g \g \rightarrow \g \g $ & \texttt{HardQCD:gg2gg} & \texttt{111} & \cite{Combridge:1977dm, Cutler:1977qm, Bengtsson:1982jr} \\
$\g \g \rightarrow \q \qbar $ & \texttt{HardQCD:gg2qqbar} & \texttt{112} & \cite{Combridge:1977dm, Cutler:1977qm, Bengtsson:1982jr} \\
$\q \g \rightarrow \q \g $ & \texttt{HardQCD:qg2qg} & \texttt{113} & \cite{Combridge:1977dm, Cutler:1977qm, Bengtsson:1982jr}\\
$\q \q' \rightarrow \q \q' $ & \texttt{HardQCD:qq2qq} & \texttt{114} & \cite{Combridge:1977dm, Cutler:1977qm, Bengtsson:1982jr, Eichten:1984eu}\\
$\q \qbar \rightarrow \g \g $ & \texttt{HardQCD:qqbar2gg} & \texttt{115} & \cite{Combridge:1977dm, Cutler:1977qm, Bengtsson:1982jr}\\
$\q \qbar \rightarrow \q' \qbar' $ & \texttt{HardQCD:qqbar2qqbarNew} & \texttt{116} & \cite{Combridge:1977dm, Cutler:1977qm, Bengtsson:1982jr, Eichten:1984eu}\\ 
$\g \g \rightarrow \c \cbar $ & \texttt{HardQCD:gg2ccbar} & \texttt{121} &  \cite{Combridge:1978kx}\\
$\q \qbar \rightarrow \c \cbar $ & \texttt{HardQCD:qqbar2ccbar} & \texttt{122} & \cite{Combridge:1978kx}\\
$\g \g \rightarrow \b \bbar $ & \texttt{HardQCD:gg2bbbar} & \texttt{123} &  \cite{Combridge:1978kx}\\
$\q \qbar \rightarrow \b \bbar $ & \texttt{HardQCD:qqbar2bbbar} & \texttt{124} & \cite{Combridge:1978kx}\\
\midrule
$\g \g \rightarrow \g \g \g $ & \texttt{HardQCD:gg2ggg} & \texttt{131} & \cite{Berends:1981rb}\\
$\q \qbar \rightarrow \g \g \g $ & \texttt{HardQCD:qqbar2ggg} & \texttt{132} &  \cite{Berends:1981rb}\\
$\q \g \rightarrow \q \g \g $ & \texttt{HardQCD:qg2qgg} & \texttt{133} & \cite{Berends:1981rb}\\
$\q \q' \rightarrow \q \q' \g $ & \texttt{HardQCD:qq2qqgDiff} & \texttt{134} &  \cite{Berends:1981rb}\\
$\q \q \rightarrow \q \q \g $ & \texttt{HardQCD:qq2qqgSame} & \texttt{135} &  \cite{Berends:1981rb}\\
$\q \qbar \rightarrow \q' \qbar' \g $ & \texttt{HardQCD:qqbar2qqbargDiff} & \texttt{136} &  \cite{Berends:1981rb}\\
$\q \qbar \rightarrow \q \qbar \g $ & \texttt{HardQCD:qqbar2qqbargSame} & \texttt{137} &  \cite{Berends:1981rb}\\
$\g \g \rightarrow \q \qbar \g $ & \texttt{HardQCD:gg2qqbarg} & \texttt{138} &  \cite{Berends:1981rb}\\
$\q \g \rightarrow \q \q' \qbar' $ & \texttt{HardQCD:qg2qqqbarDiff} & \texttt{139}  & \cite{Berends:1981rb}\\
$\q \g \rightarrow \q \q \qbar $ & \texttt{HardQCD:qg2qqqbarSame} & \texttt{140} &  \cite{Berends:1981rb}\\
\bottomrule
\end{tabular}
\end{table}

\begin{table}[thp!]\index{QCD processes}\index{Low-energy QCD processes}\index{Low energy processes}
  \caption{List of internal low-energy QCD processes,
    see \cref{subsubsection:lowenergyprocesses} for details.
  \label{tab:lowEnergyProcesses-qcd}}
\begin{tabular}{p{0.15\textwidth}p{0.45\textwidth}p{0.1\textwidth}p{0.15\textwidth}}
\toprule
  process & internal name & code & refs. \\
\midrule
   & \texttt{LowEnergyQCD:all} & \\
  $A \, B \rightarrow X $         & \texttt{LowEnergyQCD:nonDiffractive}      & \texttt{151} & \cite{Sjostrand:2020gyg} \\
  $A \, B \rightarrow A \, B $    & \texttt{LowEnergyQCD:elastic}             & \texttt{152} & \cite{Sjostrand:2020gyg} \\
  $A \, B \rightarrow X \, B $    & \texttt{LowEnergyQCD:singleDiffractiveXB} & \texttt{153} & \cite{Sjostrand:2020gyg} \\
  $A \, B \rightarrow A \, X $    & \texttt{LowEnergyQCD:singleDiffractiveAX} & \texttt{154} & \cite{Sjostrand:2020gyg} \\
  $A \, B \rightarrow X_1 \, X_2$ & \texttt{LowEnergyQCD:doubleDiffractive}   & \texttt{155} & \cite{Sjostrand:2020gyg} \\
  $N \, N \rightarrow N^*\, N$    & \texttt{LowEnergyQCD:excitation}          & \texttt{157} & \cite{Sjostrand:2020gyg} \\
  $B \, \bar{B} \rightarrow X $   & \texttt{LowEnergyQCD:annihilation}        & \texttt{158} & \cite{Sjostrand:2020gyg} \\
  $A \, B \rightarrow R$          & \texttt{LowEnergyQCD:resonant}            & \texttt{159} & \cite{Sjostrand:2020gyg} \\
\bottomrule
\end{tabular}
\end{table}

\begin{table}[thp!]\index{Weak bosons}
  \caption{List of internal weak-boson processes, see \cref{subsection:EWprocesses} for details.
  \label{tab:hardProcesses-weak}}
\begin{tabular}{p{0.16\textwidth}p{0.455\textwidth}p{0.07\textwidth}p{0.14\textwidth}}
  \toprule
process & internal name & code & refs. \\
\midrule
 & \texttt{PromptPhoton:all} & &\\
$\q \g \rightarrow \q \gamma $ & \texttt{PromptPhoton:qg2qgamma} & \texttt{201} &  \cite{Halzen:1978et, Bengtsson:1982jr}\\
$\q \qbar \rightarrow \g \gamma $ & \texttt{PromptPhoton:qqbar2ggamma} & \texttt{202} &  \cite{Halzen:1978et, Bengtsson:1982jr}\\
$\g \g \rightarrow \g \gamma $ & \texttt{PromptPhoton:gg2ggamma} & \texttt{203} &  \cite{Costantini:1971cj, Berger:1983yi,Dicus:1987fk}\\
$\q \qbar \rightarrow \gamma \gamma $ & \texttt{PromptPhoton:ffbar2gammagamma} & \texttt{204} &  \cite{Berger:1983yi}\\
$\g \g \rightarrow \gamma \gamma $ & \texttt{PromptPhoton:gg2gammagamma} & \texttt{205} & \cite{Costantini:1971cj, Berger:1983yi,Dicus:1987fk}\\
\hline
 & \texttt{WeakBosonExchange:all} & & \cite{Ingelman:1987kh}\\
$\f \f' \rightarrow \f \f'$ & \texttt{WeakBosonExchange:ff2ff(t:gmZ)} & \texttt{211} & \\
$\f_1 \f_2 \rightarrow \f_3 \f_4 $ & \texttt{WeakBosonExchange:ff2ff(t:W)} & \texttt{212} & \\
\midrule
& \texttt{WeakSingleBoson:all} & & \cite{Eichten:1984eu}\\
$\f \fbar \rightarrow \gamma^*/\Z$ & \texttt{WeakSingleBoson:ffbar2gmZ} & \texttt{221} &\\
$\f \f' \rightarrow \Wpm$ & \texttt{WeakSingleBoson:ffbar2W} & \texttt{222} &\\
\hline
$\f \fbar \rightarrow \gamma^* \rightarrow \f' \fbar'$ & \texttt{WeakSingleBoson:ffbar2ffbar(s:gm)} & \texttt{223} & \cite{Bengtsson:1982jr, Eichten:1984eu}\\
$\f \fbar \rightarrow \gamma^*/\Z \rightarrow \f' \fbar'$ & \texttt{WeakSingleBoson:ffbar2ffbar(s:gmZ)} & \texttt{224} & \cite{Bengtsson:1982jr, Eichten:1984eu}\\
$\f_1 \fbar_2 \rightarrow \Wpm \rightarrow \f_3 \fbar_4$ & \texttt{WeakSingleBoson:ffbar2ffbar(s:W)} & \texttt{225} &  \cite{Bengtsson:1982jr, Eichten:1984eu}\\

	\midrule
& \texttt{WeakDoubleBoson:all} & &\\
$\f \fbar' \rightarrow \gamma^*/\Z \, \gamma^*/\Z $ & \texttt{WeakDoubleBoson:ffbar2gmZgmZ} & \texttt{231} &  \cite{Eichten:1984eu,Gunion:1985mc}\\
$\f \fbar' \rightarrow \Z \, \Wpm $ & \texttt{WeakDoubleBoson:ffbar2ZW} & \texttt{232} &  \cite{Eichten:1984eu, Gunion:1985mc}\\
$\f \fbar \rightarrow \Wp \, \Wm $ & \texttt{WeakDoubleBoson:ffbar2WW} & \texttt{233} &  \cite{Eichten:1984eu, Bardin:1994sc}\\
\hline
 & \texttt{WeakBosonAndParton:all} & &\\
$\q \qbar \rightarrow \gamma^*/\Z \, \g $ & \texttt{WeakBosonAndParton:qqbar2gmZg} & \texttt{241} &  \cite{Eichten:1984eu}\\
$\q \g \rightarrow \gamma^*/\Z \, \q $ & \texttt{WeakBosonAndParton:qg2gmZq} & \texttt{242} & \cite{Eichten:1984eu}\\
$\f \fbar \rightarrow \gamma^*/\Z \, \gamma $ & \texttt{WeakBosonAndParton:ffbar2gmZgm} & \texttt{243} &  \cite{Eichten:1984eu}\\
$\f \gamma \rightarrow \gamma^*/\Z \, \f $ & \texttt{WeakBosonAndParton:fgm2gmZf} & \texttt{244} &  \cite{Gabrielli:1986mc}\\
$\q \qbar \rightarrow \Wpm \, \g $ & \texttt{WeakBosonAndParton:qqbar2Wg} & \texttt{251} &  \cite{Eichten:1984eu}\\
$\q \g \rightarrow \Wpm \, \q $ & \texttt{WeakBosonAndParton:qg2Wq} & \texttt{252} &  \cite{Eichten:1984eu}\\
$\f \fbar \rightarrow \Wpm \, \gamma $ & \texttt{WeakBosonAndParton:ffbar2Wgm} & \texttt{253} &  \cite{Eichten:1984eu, Samuel:1990qd}\\
$\f \gamma \rightarrow \Wpm \, \f $ & \texttt{WeakBosonAndParton:fgm2Wf} & \texttt{254} &  \cite{Gabrielli:1986mc}\\
\bottomrule
\end{tabular}
\end{table}

\begin{table}[thp!]\index{Photon-photon collisions}
  \caption{List of internal photon-collision processes, the second code in parenthesis is used to separate photons from beam A and beam B when both are possible, see \cref{subsection:EWprocesses} for details.
  \label{tab:hardProcesses-photon}}
  \begin{tabular}{p{0.15\textwidth}p{0.4\textwidth}p{0.15\textwidth}p{0.15\textwidth}}
    \toprule
process & internal name & code & refs. \\
\midrule
 & \texttt{PhotonCollision:all} & & \cite{Barklow:1990ah}\\
$\gamma \gamma \rightarrow \q \qbar $ & \texttt{PhotonCollision:gmgm2qqbar} & \texttt{261} &\\
$\gamma \gamma \rightarrow \c \cbar $ & \texttt{PhotonCollision:gmgm2ccbar} & \texttt{262} &\\
$\gamma \gamma \rightarrow \b \bbar $ & \texttt{PhotonCollision:gmgm2bbbar} & \texttt{263} &\\
$\gamma \gamma \rightarrow \eplus \eminus $ & \texttt{PhotonCollision:gmgm2ee} & \texttt{264} &\\
$\gamma \gamma \rightarrow \muplus \muminus $ & \texttt{PhotonCollision:gmgm2mumu} & \texttt{265} &\\
$\gamma \gamma \rightarrow \tauplus \tauminus $ & \texttt{PhotonCollision:gmgm2tautau} & \texttt{266} &\\
\midrule
 & \texttt{PhotonParton:all} &  &\\ 
$\g \gamma \rightarrow \q \qbar $ & \texttt{PhotonParton:ggm2qqbar} & \texttt{271} (\texttt{281}) &  \cite{Duke:1982bj}\\ 
$\g \gamma \rightarrow \c \cbar $ & \texttt{PhotonParton:ggm2ccbar} &  \texttt{272} (\texttt{282}) & \cite{Fontannaz:1981ka}\\
$\g \gamma \rightarrow \b \bbar $ & \texttt{PhotonParton:ggm2bbbar} & \texttt{273} (\texttt{283}) & \cite{Fontannaz:1981ka}\\
$\q \gamma \rightarrow \q \g $ & \texttt{PhotonParton:qgm2qg} & \texttt{274} (\texttt{284}) & \cite{Duke:1982bj}\\
$\q \gamma \rightarrow \q \gamma $ & \texttt{PhotonParton:qgm2qgm} & \texttt{275} (\texttt{285}) & \cite{Duke:1982bj}\\
\bottomrule
\end{tabular}
\end{table}

\begin{table}[thp!]
  \caption{List of internal top-production processes and production of fourth-generation fermions. Expressions are from \citeone{Sjostrand:2006za}.
  \index{Top quark}\index{Single top}\index{Top quark!Single top} \index{Fourth-generation fermions}
  \label{tab:hardProcesses-top}}
\begin{tabular}{p{0.27\textwidth}p{0.58\textwidth}p{0.08\textwidth}}
  \toprule
process & internal name & code \\
\midrule
 & \texttt{Top:all} & \\
$\g \g \rightarrow \t \tbar $ & \texttt{Top:gg2ttbar}  & \texttt{601} \\
$\q \qbar \rightarrow \t \tbar $& \texttt{Top:qqbar2ttbar} & \texttt{602}  \\
$\q \q \rightarrow \t q $ & \texttt{Top:qq2tq(t:W)} & \texttt{603} \\
$\f \fbar \rightarrow \gamma/\Z  \rightarrow \t \tbar $ & \texttt{Top:ffbar2ttbar(s:gmZ)} & \texttt{604} \\
$\f \fbar \rightarrow W^\pm\rightarrow  \t \qbar $  & \texttt{Top:ffbar2tqbar(s:W)} & \texttt{605} \\
$\gamma \gamma \rightarrow \t \tbar $ & \texttt{Top:gmgm2ttbar}  & \texttt{606}\\
$\g \gamma \rightarrow \t \tbar $ & \texttt{Top:ggm2ttbar}  & \texttt{607}\\
\midrule
 & \texttt{FourthBottom:all} & \\
$\g \g \rightarrow \b^\prime \bbar^\prime $ & \texttt{FourthBottom:gg2bPrimebPrimebar} & \\
$\q \qbar \rightarrow  \b^\prime \bbar^\prime $ & \texttt{FourthBottom:qqbar2bPrimebPrimebar} & \texttt{801} \\
$\f \fbar \rightarrow  \b^\prime \q $ \mbox{($t$-channel W)} & \texttt{FourthBottom:qq2bPrimeq(t:W)}  & \texttt{803} \\
$\f \fbar \rightarrow  \b^\prime \bbar^\prime $ \mbox{($s$-channel $\gamma/\Z$)}  & \texttt{FourthBottom:ffbar2bPrimebPrimebar(s:gmZ)} & \texttt{804}\\
$\f \fbar^\prime \rightarrow  \b^\prime \qbar$ \mbox{($s$-channel W)}  & \texttt{FourthBottom:ffbar2bPrimeqbar(s:W)}  & \texttt{805}\\
$\f \fbar^\prime \rightarrow  \b^\prime \tbar$ \mbox{($s$-channel W)} & \texttt{FourthBottom:ffbar2bPrimetbar(s:W)} & \texttt{806} \\
\midrule
 & \texttt{FourthTop:all} & \\
$\g \g \rightarrow \t^\prime \tbar^\prime $  & \texttt{FourthTop:gg2tPrimetPrimebar} & \texttt{821}\\
$\q \qbar \rightarrow  \t^\prime \tbar^\prime $  & \texttt{FourthTop:qqbar2tPrimetPrimebar} & \texttt{822}\\
$\f \fbar \rightarrow  \b^\prime \q $ \mbox{($t$-channel W)} & \texttt{FourthTop:qq2tPrimeq(t:W)} & \texttt{823} \\
$\f \fbar \rightarrow  \t^\prime \tbar^\prime $ \mbox{($s$-channel $\gamma/\Z$)} & \texttt{FourthTop:ffbar2tPrimetPrimebar(s:gmZ)}  & \texttt{824}\\
$\f \fbar^\prime \rightarrow  \t^\prime \qbar$ \mbox{($s$-channel W)}  & \texttt{FourthTop:ffbar2tPrimeqbar(s:W)} & \texttt{825} \\
$\f \fbar^\prime \rightarrow  \t^\prime \bbar^\prime$ \mbox{($s$-channel W)} & \texttt{FourthPair:ffbar2tPrimebPrimebar(s:W)}  & \texttt{841}\\
$\f \fbar^\prime \rightarrow  \tau^\prime \bar\nu^\prime$ \mbox{($s$-channel W)}  & \texttt{FourthPair:ffbar2tauPrimenuPrimebar(s:W)} & \texttt{842}\\
\bottomrule
\end{tabular}
\end{table}

\begin{table}[t!]\index{Higgs bosons}\index{VBF}
\caption{List of internal SM-Higgs production processes. See \cref{subsection:Higgs} for details.
  \label{tab:hardProcesses-higgs}}
\begin{tabular}{p{0.3\textwidth}p{0.5\textwidth}p{0.1\textwidth}}
\toprule
  process & internal name & code \\
\midrule
 &   \texttt{HiggsSM:all} \\
$\f \fbar \rightarrow H_\mathrm{SM}  $  & \texttt{HiggsSM:ffbar2H}& \texttt{901} \\
$\g \g \rightarrow H_\mathrm{SM} $ & \texttt{HiggsSM:gg2H}  & \texttt{902}\\
$\gamma \gamma \rightarrow  H_\mathrm{SM} $  & \texttt{HiggsSM:gmgm2H} & \texttt{903}\\
$\f \fbar \rightarrow H_\mathrm{SM} \Z$  & \texttt{HiggsSM:ffbar2HZ} & \texttt{904} \\
$\f \fbar \rightarrow H_\mathrm{SM} \W$& \texttt{HiggsSM:ffbar2HW} &  \texttt{905} \\
$\f \fbar \rightarrow H_\mathrm{SM} \f \fbar $ (ZBF)  & \texttt{HiggsSM:ff2Hff(t:ZZ)} & \texttt{906}\\
$\f \fbar \rightarrow H_\mathrm{SM} \f \fbar $ (WBF) & \texttt{HiggsSM:ff2Hff(t:WW)} & \texttt{907} \\
$\g \g \rightarrow  H_\mathrm{SM} \t \tbar $  & \texttt{HiggsSM:gg2Httbar} & \texttt{908}\\
$\q \qbar \rightarrow  H_\mathrm{SM}  \t \tbar $  & \texttt{HiggsSM:qqbar2Httbar} & \texttt{909}\\
 &  \\
 \midrule
$\q \g \rightarrow   H_\mathrm{SM} q $ & \texttt{HiggsSM:qg2Hq} & \texttt{911} \\
$\g \g \rightarrow  H_\mathrm{SM} \b \bbar $  & \texttt{HiggsSM:gg2Hbbbar} & \texttt{912}\\
$\q \qbar \rightarrow  H_\mathrm{SM}  \b \bbar $ & \texttt{HiggsSM:qqbar2Hbbbar} & \texttt{913}  \\
$\g \g \rightarrow H_\mathrm{SM}  \g  $  & \texttt{HiggsSM:gg2Hg(l:t)} & \texttt{914}\\
$\q \g \rightarrow H_\mathrm{SM} \q $  & \texttt{HiggsSM:qg2Hq(l:t)}& \texttt{915} \\
$\q \qbar \rightarrow  H_\mathrm{SM} g $ & \texttt{HiggsSM:qqbar2Hg(l:t)}  & \texttt{916} \\
 &  \\
 \bottomrule
\end{tabular}
\end{table}

\clearpage
\subsection{Beyond-the-Standard-Model processes}
\label{sec:BSMprocesses}

\begin{table}[thp!]\index{SUSY}
  \caption{List of internal SUSY particle production
  processes. Expressions from \citerefs{Bozzi:2007me, Fuks:2011dg, Desai:2011su}. Particular
  flavour states can be selected using \texttt{IdA}
  and \texttt{idB}, see \cref{subsection:SUSY} or the \htmlmanual~for details.
  \label{tab:hardProcesses-SUSY}}
\begin{tabular}{p{0.25\textwidth}p{0.6\textwidth}}
  \toprule
process & internal name  \\
\midrule
 &  \texttt{SUSY:all}  \\
$\g \g \rightarrow \tilde g \tilde g $ &  \texttt{SUSY:gg2gluinogluino} \\
$\q \qbar \rightarrow \tilde g \tilde g $ &  \texttt{SUSY:qqbar2gluinogluino} \\
$\q \g \rightarrow \tilde q \tilde g $ &  \texttt{SUSY:qg2squarkgluino} \\
$\g \g \rightarrow \tilde q_i \tilde q_j^* $ &  \texttt{SUSY:gg2squarkantisquark} \\
$\q \qbar \rightarrow \tilde q_i \tilde q_j^* $  &  \texttt{SUSY:qqbar2squarkantisquark} \\
$\q \qbar \rightarrow \tilde q_i \tilde q_j^* $ (No EW) &  \texttt{SUSY:qqbar2squarkantisquark:onlyQCD} \\
$\q \qbar \rightarrow \tilde q_i \tilde q_j^* $  &  \texttt{SUSY:qqbar2squarkantisquark} \\
$\q \q \rightarrow \tilde q_i \tilde q_j $ &  \texttt{SUSY:qq2squarksquark} \\
$\q \q \rightarrow \tilde q_i \tilde q_j $ (No EW) &  \texttt{SUSY:qq2squarksquark}:onlyQCD \\
$\q \qbar \rightarrow \tilde \chi_i^0 \tilde \chi_j^0 $ &  \texttt{SUSY:qqbar2chi0chi0} \\
$\q \qbar \rightarrow \tilde \chi_i^\pm \tilde \chi_j^0 $ &  \texttt{SUSY:qqbar2chi+-chi0} \\
$\q \qbar \rightarrow \tilde \chi_i^\pm \tilde \chi_j^\mp $ &  \texttt{SUSY:qqbar2chi+chi-} \\
 &  \\
$\q \g \rightarrow \tilde q \tilde \chi_i^0 $  &  \texttt{SUSY:qg2chi0squark} \\
$\q \g \rightarrow \tilde q \tilde \chi_i^\pm $ &  \texttt{SUSY:qg2chi+-squark} \\
$\q \qbar \rightarrow \tilde \chi_i^0 \tilde g$ &  \texttt{SUSY:qqbar2chi0gluino} \\
$\q \qbar \rightarrow \tilde \chi_i^\pm \tilde g$ &  \texttt{SUSY:qqbar2chi+-gluino} \\
$\f \fbar \rightarrow \tilde \ell_i \tilde \ell_j^* $ &  \texttt{SUSY:qqbar2sleptonantislepton} \\
$\q_i \q_j \rightarrow \tilde q_k^*$ &  \texttt{SUSY:qq2antisquark} \\
&  \\
\bottomrule
\end{tabular}
\end{table}

\begin{table}[thp!]\index{Higgs bosons}\index{VBF}
\caption{List of internal BSM-Higgs production processes. See \cref{subsection:Higgs} for details. Expressions from \citerefs{Huitu:1996su,Barenboim:1996pt}.
  \label{tab:hardProcesses-BSMhiggs}}
\begin{tabular}{p{0.325\textwidth}p{0.35\textwidth}p{0.32\textwidth}}
  \toprule
process & internal name & code \\
\midrule
 &   \texttt{HiggsBSM:all} & \\
 & (replace \texttt{H1} with \texttt{H2} or \texttt{A3}) \\
 &   \texttt{HiggsBSM:allH1} & \\
$\f \fbar \rightarrow H_1 (H_2, A_3)$ & \texttt{HiggsBSM:ffbar2H1}  & \texttt{1001, 1021, 1041} \\
$\g \g \rightarrow H_1 (H_2, A_3) $ & \texttt{HiggsBSM:gg2H1} & \texttt{1002, 1022, 1042}\\
$\gamma \gamma \rightarrow  H_1 (H_2, A_3) $  & \texttt{HiggsBSM:gmgm2H1} & \texttt{1003, 1023, 1043}\\
$\f \fbar \rightarrow H_1 (H_2, A_3) \Z$  & \texttt{HiggsBSM:ffbar2H1Z} & \texttt{1004, 1024, 1044}\\
$\f \fbar \rightarrow H_1 (H_2, A_3) \W$  & \texttt{HiggsBSM:ffbar2H1W} & \texttt{1005, 1025, 1045}\\
$\f \fbar \rightarrow H_1 (H_2, A_3) \f \fbar $ (ZBF)  & \texttt{HiggsBSM:ff2H1ff(t:ZZ)} & \texttt{1006, 1026, 1046}\\
$\f \fbar \rightarrow H_1 (H_2, A_3) \f \fbar $ (WBF)  & \texttt{HiggsBSM:ff2H1ff(t:WW)} & \texttt{1007, 1027, 1047}\\
$\g \g \rightarrow  H_1 (H_2, A_3) \t \tbar $   & \texttt{HiggsBSM:gg2H1ttbar} & \texttt{1008, 1028, 1048}\\
$\q \qbar \rightarrow  H_1 (H_2, A_3)  \t \tbar $ & \texttt{HiggsBSM:qqbar2H1ttbar} & \texttt{1009, 1029, 1049}  \\
&  \\
& \texttt{HiggsBSM:allH+-} & \\
$\f \fbar \rightarrow H^\pm$ & \texttt{HiggsBSM:ffbar2H+-}  & \texttt{1061}\\
$\b \g \rightarrow H^\pm $  & \texttt{HiggsBSM:bg2H+-t} & \texttt{1062}\\
 &  \\
 & \texttt{HiggsBSM:allHpair} \\
$\f \fbar \rightarrow A_3 H_1  $ & \texttt{HiggsBSM:ffbar2A3H1}  & \texttt{1081}\\
$\f \fbar \rightarrow A_3 H_2 $  & \texttt{HiggsBSM:ffbar2A3H2} & \texttt{1082}\\
$\f \fbar \rightarrow H^\pm H_1  $  & \texttt{HiggsBSM:ffbar2H+-H1} & \texttt{1083}\\
$\f \fbar \rightarrow H^\pm H_2 $  & \texttt{HiggsBSM:ffbar2H+-H2} & \texttt{1084}\\
$\f \fbar \rightarrow H^\pm A_3$  & \texttt{HiggsBSM:ffbar2H+H-} & \texttt{1085}\\
 &  \\
$\q \g \rightarrow   H_1  (H_2, A_3) q $  & \texttt{HiggsBSM:qg2H1q} & \texttt{1011, 1031, 1051}\\
$\g \g \rightarrow  H_1  (H_2, A_3) \b \bbar $ & \texttt{HiggsBSM:gg2H1bbbar}  & \texttt{1012, 1032, 1052} \\
$\q \qbar \rightarrow  H_1   (H_2, A_3) \b \bbar $  & \texttt{HiggsBSM:qqbar2H1bbbar} & \texttt{1013, 1033, 1053}\\
$\g \g \rightarrow H_1   (H_2, A_3) \g $ & \texttt{HiggsBSM:gg2H1g(l:t)} & \texttt{1014, 1034, 1054}\\
$\q \g \rightarrow H_1   (H_2, A_3) \q $  & \texttt{HiggsBSM:qg2H1q(l:t)} & \texttt{1015, 1035, 1055}\\
$\q \qbar \rightarrow  H_1   (H_2, A_3) g $  & \texttt{HiggsBSM:qqbar2H1g(l:t)} & \texttt{1016, 1036, 1056}\\
\bottomrule
\end{tabular}
\end{table}

\begin{table}[thp!]\index{Dark matter}
\caption{List of internal processes for dark matter. See \cref{subsection:DarkMatter} and \citeone{Desai:2018pxq} for details.
  \label{tab:hardProcesses-DM}}
\begin{tabular}{p{0.15\textwidth}p{0.5\textwidth}p{0.15\textwidth}p{0.15\textwidth}}
  \toprule
process & internal name & code  \\
\midrule
$\g \g \rightarrow\chi \bar \chi$ &  \texttt{DM:gg2S2XX}  & \texttt{6011} \\
$\g \g \rightarrow \chi \bar \chi j$ &  \texttt{DM:gg2S2XXj}  & \texttt{6012} \\
$\f \fbar \rightarrow \chi \bar \chi $ &  \texttt{DM:ffbar2Zp2XX}  & \texttt{6001}   \\
$\f \fbar \rightarrow \chi \bar \chi $ &  \texttt{DM:ffbar2Zp2XXj}  & \texttt{6002} \\
$\f \fbar \rightarrow \chi \bar \chi j$ &  \texttt{DM:qg2Zp2XXj}  & \texttt{6003} \\
$\f \fbar \rightarrow Z^\prime H $ &  \texttt{DM:ffbar2ZpH} & \texttt{6004} \\
$\q \qbar \rightarrow \Psi \bar \Psi $ &  \texttt{DM:qqbar2DY} & \texttt{6020} \\
\midrule
\end{tabular}
\end{table}

\begin{table}[thp!]\index{Leptoquarks}\index{Exotica!Leptoquarks}\index{Exotica!New gauge bosons}
\caption{List of internal processes mediated by new gauge bosons or leptoquarks. See \cref{subsection:otherexotica}.
  \label{tab:hardProcesses-newgauge}}
\begin{tabular}{p{0.15\textwidth}p{0.4\textwidth}p{0.15\textwidth}p{0.15\textwidth}}
  \toprule
process & internal name & code & refs. \\
\midrule
$\f \fbar \rightarrow \gamma /\Z/Z^\prime  $  &  \texttt{NewGaugeBoson:ffbar2gmZZprime} & \texttt{3001}  & \cite{Ciobanu:2005pv}\\
$\f \fbar \rightarrow W^\prime $ &  \texttt{NewGaugeBoson:ffbar2Wprime} & \texttt{3021}  & \cite{Ciobanu:2005pv}\\
$\f \fbar \rightarrow R_0 $ &  \texttt{NewGaugeBoson:ffbar2R0}  & \texttt{3041} & \cite{Ciobanu:2005pv}\\
 &  &  & \\
\midrule
 &  \texttt{LeftRightSymmmetry:all} & & \cite{Altarelli:1989ff} \\
$\f \fbar \rightarrow \Z_R $ &  \texttt{LeftRightSymmmetry:ffbar2ZR} & \texttt{3101} & \\
$\f \fbar \rightarrow W^\prime $ &  \texttt{LeftRightSymmmetry:ffbar2WR} & \texttt{3102} & \\
$\ell \bar \ell \rightarrow H_L$ &  \texttt{LeftRightSymmmetry:ll2HL} & \texttt{3121} & \\
$\ell \gamma \rightarrow H_L e$ &  \texttt{LeftRightSymmmetry:lgm2HLe} & \texttt{3122} & \\
$\ell \gamma \rightarrow H_L \mu$ &  \texttt{LeftRightSymmmetry:lgm2HLmu} & \texttt{3123} & \\
$\ell \gamma \rightarrow H_L \tau$ &  \texttt{LeftRightSymmmetry:lgm2HLtau} & \texttt{3124} & \\
$\f \fbar \rightarrow \f \fbar H_L$ &  \texttt{LeftRightSymmmetry:ff2HLff} & \texttt{3125} & \\
$\f \fbar \rightarrow H_L H_L$ &  \texttt{LeftRightSymmmetry:ffbar2HLHL} & \texttt{3126} & \\
$\ell \bar \ell \rightarrow H_R$ &  \texttt{LeftRightSymmmetry:ll2HR} & \texttt{3141} & \\
$\ell \gamma \rightarrow H_R e$ &  \texttt{LeftRightSymmmetry:lgm2HRe} & \texttt{3142} & \\
$\ell \gamma \rightarrow H_R \mu$ &  \texttt{LeftRightSymmmetry:lgm2HRmu} & \texttt{3143} & \\
$\ell \gamma \rightarrow H_R \tau$ &  \texttt{LeftRightSymmmetry:lgm2HRtau} & \texttt{3144} & \\
$\f \fbar \rightarrow \f \fbar H_R$ &  \texttt{LeftRightSymmmetry:ff2HRff} & \texttt{3145} & \\
$\f \fbar \rightarrow H_R H_R$ &  \texttt{LeftRightSymmmetry:ffbar2HRHR} & \texttt{3146} & \\
 &  & \texttt{} & \\
\midrule
 &  \texttt{LeptoQuark:all} & \texttt{} & \\
$\q \ell \rightarrow S$  &  \texttt{LeptoQuark:ql2LQ}  & \texttt{3201} & \cite{Hewett:1987yg} \\
$\q \g \rightarrow \ell S$ &  \texttt{LeptoQuark:qg2LQl}  & \texttt{3202} & \cite{Hewett:1987yg}\\
$\g \g \rightarrow S S^*$ &  \texttt{LeptoQuark:gg2LQLQbar}  & \texttt{3203} & \cite{Hewett:1987yg}\\
$\f \fbar \rightarrow S S^*$ &  \texttt{LeptoQuark:qqbar2LQLQbar}  & \texttt{3204} & \cite{Hewett:1987yg} \\
\bottomrule
\end{tabular}
\end{table}

\begin{table}[thp!]\index{Exotica!Excited fermions}
\caption{List of internal processes for excited fermions. See \cref{subsection:otherexotica}.
  \label{tab:hardProcesses-excited}}
\begin{tabular}{p{0.15\textwidth}p{0.5\textwidth}p{0.1\textwidth}p{0.15\textwidth}}
  \toprule
process & internal name & code & refs. \\
\midrule
 &  \texttt{ExcitedFermion:all} & & \cite{Eichten:1983hw, Baur:1989kv} \\
$ \d \g \rightarrow d^*$ &  \texttt{ExcitedFermion:dg2dStar} & \texttt{4001} &\cite{Eichten:1983hw, Baur:1989kv} \\
$ \u \g \rightarrow u^*$  &  \texttt{ExcitedFermion:ug2uStar}& \texttt{4002} & \cite{Eichten:1983hw, Baur:1989kv}\\
$ \s \g \rightarrow s^*$  &  \texttt{ExcitedFermion:sg2sStar}& \texttt{4003} & \cite{Eichten:1983hw, Baur:1989kv}\\
$ \c \g \rightarrow c^*$   &  \texttt{ExcitedFermion:cg2cStar} & \texttt{4004}& \cite{Eichten:1983hw, Baur:1989kv}\\
$ \b \g \rightarrow b^*$ &  \texttt{ExcitedFermion:bg2bStar} & \texttt{4005}&\cite{Eichten:1983hw, Baur:1989kv} \\
$ e \gamma \rightarrow e^*$ &  \texttt{ExcitedFermion:egm2eStar} & \texttt{4011}& \cite{Eichten:1983hw, Baur:1989kv}\\
$ \mu \gamma \rightarrow \mu^*$ &  \texttt{ExcitedFermion:mugm2muStar} & \texttt{4013}& \cite{Eichten:1983hw, Baur:1989kv}\\
$ \tau \gamma \rightarrow \tau^*$ &  \texttt{ExcitedFermion:taugm2tauStar} & \texttt{4015}& \cite{Eichten:1983hw, Baur:1989kv}\\
 &  & & \\
$ \q \q \rightarrow d^* q$ &  \texttt{ExcitedFermion:qq2dStarq} & \texttt{4021}& \cite{Eichten:1983hw, Baur:1989kv}\\
$ \q \q \rightarrow u^* q$ &  \texttt{ExcitedFermion:qq2uStarq} & \texttt{4022}& \cite{Eichten:1983hw, Baur:1989kv}\\
$ \q \q \rightarrow s^* q$ &  \texttt{ExcitedFermion:qq2sStarq} & \texttt{4023}& \cite{Eichten:1983hw, Baur:1989kv}\\
$ \q \q \rightarrow c^* q$ &  \texttt{ExcitedFermion:qq2cStarq} & \texttt{4024}& \cite{Eichten:1983hw, Baur:1989kv}\\
$ \q \q \rightarrow b^* q$ &  \texttt{ExcitedFermion:qq2bStarq} & \texttt{4025}& \cite{Eichten:1983hw, Baur:1989kv}\\
$ \q \qbar \rightarrow e^* e$ &  \texttt{ExcitedFermion:qqbar2eStare} & \texttt{4031}& \cite{Eichten:1983hw, Baur:1989kv} \\
$ \q \qbar \rightarrow \nu_e^* \nu_e$ &  \texttt{ExcitedFermion:qqbar2nueStarnue} & \texttt{4032}& \cite{Eichten:1983hw, Baur:1989kv}\\
$ \q \qbar \rightarrow \mu^* \mu$ &  \texttt{ExcitedFermion:qqbar2muStarmu} & \texttt{4033}& \cite{Eichten:1983hw, Baur:1989kv}\\
$ \q \qbar \rightarrow \nu_\mu^* \nu_\mu$ &  \texttt{ExcitedFermion:qqbar2numuStarnumu} & \texttt{4034}& \cite{Eichten:1983hw, Baur:1989kv}\\
$ \q \qbar \rightarrow \tau^* \tau$ &  \texttt{ExcitedFermion:qqbar2tauStartau} & \texttt{4035}& \cite{Eichten:1983hw, Baur:1989kv}\\
$ \q \qbar \rightarrow \nu_\tau^* \nu_\tau$ &  \texttt{ExcitedFermion:qqbar2nutauStarnutau} & \texttt{4036}& \cite{Eichten:1983hw, Baur:1989kv}\\
$ \q \qbar \rightarrow e^* e^*$ &  \texttt{ExcitedFermion:qqbar2eStareStar} & \texttt{4051}& \cite{Eichten:1983hw, Baur:1989kv}\\
$ \q \qbar \rightarrow \nu_e^* \nu_e^*$ &  \texttt{ExcitedFermion:qqbar2nueStarnueStar}& \texttt{4052} &\cite{Eichten:1983hw, Baur:1989kv} \\
$ \q \qbar \rightarrow \mu^* \mu^*$ &  \texttt{ExcitedFermion:qqbar2muStarmuStar} & \texttt{4053}& \cite{Eichten:1983hw, Baur:1989kv}\\
$ \q \qbar \rightarrow \nu_\mu^* \nu_\mu^*$ &  \texttt{ExcitedFermion:qqbar2numuStarnumuStar} & \texttt{4054}&\cite{Eichten:1983hw, Baur:1989kv} \\
$ \q \qbar \rightarrow \tau^* \tau^*$  &  \texttt{ExcitedFermion:qqbar2tauStartauStar} & \texttt{4055}& \cite{Eichten:1983hw, Baur:1989kv}\\
$ \q \qbar \rightarrow \nu_\tau^* nu_\tau^*$ &  \texttt{ExcitedFermion:qqbar2nutauStarnutauStar} & \texttt{4056} & \cite{Eichten:1983hw, Baur:1989kv} \\
&  & & \\
\bottomrule
\end{tabular}
\end{table}

\begin{table}[thp!]\index{Extra dimensions!Randall--Sundrum resonances}\index{Exotica!Extra dimensions|see{Extra dimensions}}
\caption{\label{tab:hardProcED-RS}List of internal processes for Randall--Sundrum resonances. See \cref{subsection:otherexotica} and \citerefs{Bijnens:2001gh,Bella:2010sc} for details.} 
\begin{tabular}{p{0.12\textwidth}p{0.63\textwidth}p{0.13\textwidth}}
  \toprule
process & internal name & code  \\
\midrule
 &  \texttt{ExtraDimensionsG*:all} & \texttt{} \\
$\g \g \rightarrow G^*$ &  \texttt{ExtraDimensionsG*:gg2G*} & \texttt{5001} \\
$\f \fbar \rightarrow G^*$ &  \texttt{ExtraDimensionsG*:ffbar2G*} &  \texttt{5002} \\
$\g \g \rightarrow G^* \g$ &  \texttt{ExtraDimensionsG*:gg2G*g} &  \texttt{5003} \\
$\g \q \rightarrow G^* \q$ &  \texttt{ExtraDimensionsG*:qg2G*q} &  \texttt{5004} \\
$\q \qbar \rightarrow G^* g$ &  \texttt{ExtraDimensionsG*:qqbar2G*g} &  \texttt{5005} \\
$\q \qbar \rightarrow G_\mathrm{KK} g$ &  \texttt{ExtraDimensionsG*:qqbar2KKgluon*} &  \texttt{5006} \\
& &  \\
\bottomrule
\end{tabular}
\end{table}

\begin{table}[thp!]\index{Extra dimensions!TeV$^{-1}$-sized}\index{Exotica}
\caption{\label{tab:hardProcED-TEV}List of internal processes for TeV$^{-1}$-sized extra dimensions. See \cref{subsection:otherexotica} for details, expressions from \citeone{Ask:2011zs}.} 
\begin{tabular}{p{0.12\textwidth}p{0.63\textwidth}p{0.13\textwidth}}
\toprule
process & internal name & code  \\
\midrule
 $\f \fbar \rightarrow \d \dbar$ &  \texttt{ExtraDimensionsTEV:ffbar2ddbar} & \texttt{5061} \\
 $\f \fbar \rightarrow \u \ubar$ &  \texttt{ExtraDimensionsTEV:ffbar2uubar} &  \texttt{5062} \\
 $\f \fbar \rightarrow \s \sbar$ &  \texttt{ExtraDimensionsTEV:ffbar2ssbar} &  \texttt{5063} \\
 $\f \fbar \rightarrow \c \cbar$ &  \texttt{ExtraDimensionsTEV:ffbar2ccbar} &  \texttt{5064} \\
 $\f \fbar \rightarrow \b \bbar$ &  \texttt{ExtraDimensionsTEV:ffbar2bbbar} &  \texttt{5065} \\
 $\f \fbar \rightarrow \t \tbar$ &  \texttt{ExtraDimensionsTEV:ffbar2ttbar} &  \texttt{5066} \\
 $\f \fbar \rightarrow e^+ e^-$ &  \texttt{ExtraDimensionsTEV:ffbar2e+e-} &  \texttt{5071} \\
 $\f \fbar \rightarrow \nu_e \bar \nu_e$ &  \texttt{ExtraDimensionsTEV:ffbar2nuenuebar} &  \texttt{5072} \\
 $\f \fbar \rightarrow \mu^+ \mu^-$ &  \texttt{ExtraDimensionsTEV:ffbar2mu+mu-} &  \texttt{5073} \\
 $\f \fbar \rightarrow \nu_\mu \bar \nu_\mu$  &  \texttt{ExtraDimensionsTEV:ffbar2numunumubar} &  \texttt{5074} \\
 $\f \fbar \rightarrow \tau^+ \tau^-$ &  \texttt{ExtraDimensionsTEV:ffbar2tau+tau-} &  \texttt{5076} \\
 $\f \fbar \rightarrow \nu_\tau \bar \nu_\tau$&  \texttt{ExtraDimensionsTEV:ffbar2nutaunutaubar} &  \texttt{5076} \\
\bottomrule
\end{tabular}
\end{table}

\begin{table}[thp!]\index{Extra dimensions!Large extra dimensions}\index{Exotica}
\caption{\label{tab:hardProcED-LED}List of internal processes for large extra dimensions. See \cref{subsection:otherexotica} for details.} 
\begin{tabular}{p{0.12\textwidth}p{0.53\textwidth}p{0.09\textwidth}p{0.14\textwidth}}
\toprule
process & internal name & code & refs. \\
\midrule
 &  \texttt{ExtraDimensionsLED:monojet} &  & \cite{Franceschini:2011wr, Bella:2010vi} \\
 $\g \g \rightarrow G g$ & \texttt{ExtraDimensionsLED:gg2Gg} & \texttt{5021} & \\
 $\g \q \rightarrow G q$ & \texttt{ExtraDimensionsLED:qg2Gq} & \texttt{5022} & \\
 $\q \qbar \rightarrow G g$ & \texttt{ExtraDimensionsLED:qqbar2Gg} & \texttt{5023} & \\
 &  & \texttt{} & \\
 $\f \fbar \rightarrow G \Z$ &  \texttt{ExtraDimensionsLED:ffbar2GZ} & \texttt{5024} & \cite{Franceschini:2011wr}\\
 $\f \fbar \rightarrow G \gamma$ &  \texttt{ExtraDimensionsLED:ffbar2Ggamma} & \texttt{5025} & \cite{Franceschini:2011wr}\\
 $\f \fbar \rightarrow \gamma \gamma $ &  \texttt{ExtraDimensionsLED:ffbar2gammagamma} & \texttt{5026} & \cite{Franceschini:2011wr}\\
 $\g \g \rightarrow \gamma \gamma $ &  \texttt{ExtraDimensionsLED:gg2gammagamma} & \texttt{5027} & \cite{Franceschini:2011wr}\\
 $\f \fbar \rightarrow  \ell \bar \ell$ &  \texttt{ExtraDimensionsLED:ffbar2llbar} & \texttt{5028} & \cite{Franceschini:2011wr}\\
 $\g \g \rightarrow  \ell \bar \ell$ &  \texttt{ExtraDimensionsLED:gg2llbar} & \texttt{5029} & \cite{Franceschini:2011wr}\\
 &  & \texttt{} & \\
 &  \texttt{ExtraDimensionsLED:dijets} & \texttt{} & \cite{Franceschini:2011wr}\\
 $ \g \g \rightarrow \g \g $ & \texttt{ExtraDimensionsLED:gg2DJgg} & \texttt{5030} & \\
 $ \g \g \rightarrow \q \qbar$ & \texttt{ExtraDimensionsLED:gg2DJqqbar} & \texttt{5031} & \\
 $ \q \g \rightarrow \q \g$& \texttt{ExtraDimensionsLED:qg2DJqg} & \texttt{5032} & \\
 $ \q \q \rightarrow \q \q$ & \texttt{ExtraDimensionsLED:qq2DJqq} & \texttt{5033} & \\
 $ \q \qbar \rightarrow \g \g$ & \texttt{ExtraDimensionsLED:qqbar2DJgg} & \texttt{5034} & \\
 $ \q \qbar \rightarrow \q^\prime \qbar^\prime $ & \texttt{ExtraDimensionsLED:qqbar2DJqqbarNew} & \texttt{5035} & \\
\bottomrule
\end{tabular}
\end{table}

\begin{table}[thp!]\index{Extra dimensions!Unparticles}\index{Exotica}
\caption{List of internal processes for unparticles. Expressions from \citerefs{Ask:2008fh, Ask:2009pv}, see \cref{subsection:otherexotica}.
 \label{tab:hardProcesses-unpart}}
\begin{tabular}{p{0.15\textwidth}p{0.55\textwidth}p{0.1\textwidth}}
\toprule
process & internal name & code \\
\midrule
 &  \texttt{ExtraDimensionsUnpart:monojet} & \\
$\g \g \rightarrow U g$ & \texttt{ExtraDimensionsUnpart:gg2Ug} &  \texttt{5045} \\
$\g \q \rightarrow U q$ & \texttt{ExtraDimensionsUnpart:qg2Uq} &  \texttt{5046} \\
$\q \qbar \rightarrow U g$ & \texttt{ExtraDimensionsUnpart:qqbar2Ug} &  \texttt{5047} \\
  &  &  \\
$\f \fbar \rightarrow U \Z$  &  \texttt{ExtraDimensionsUnpart:ffbar2UZ} &  \texttt{5041} \\
 $\f \fbar \rightarrow U \gamma$ &  \texttt{ExtraDimensionsUnpart:ffbar2Ugamma} & \texttt{5042}  \\
 $\f \fbar \rightarrow  U \rightarrow\gamma \gamma $ &  \texttt{ExtraDimensionsUnpart:ffbar2gammagamma} &  \texttt{5043} \\
 $\g \g \rightarrow  U \rightarrow \gamma \gamma $ &  \texttt{ExtraDimensionsUnpart:gg2gammagamma} &  \texttt{5044} \\
$\f \fbar \rightarrow  U \rightarrow \ell \bar \ell$ &  \texttt{ExtraDimensionsUnpart:ffbar2llbar} &  \texttt{5048} \\
$\g \g \rightarrow U \rightarrow  \ell \bar \ell$ &  \texttt{ExtraDimensionsUnpart:gg2llbar} &  \texttt{5049} \\
 \bottomrule
\end{tabular}
\end{table}

\begin{table}[thp!]\index{Hidden valleys}\index{Exotica}
\caption{List of internal hidden valley processes, see \cref{subsection:HiddenValleyProcesses} and \citerefs{Carloni:2010tw, Carloni:2011kk} for details.
  \label{tab:hardProcesses-HV}}
\begin{tabular}{p{0.15\textwidth}p{0.5\textwidth}p{0.2\textwidth}}
\toprule
process & internal name & code \\
\midrule
 &  \texttt{HiddenValley:all} \\
$\g \g \rightarrow \d_v \dbar_v $ &  \texttt{HiddenValley:gg2DvDvbar} & \texttt{4901} \\
$\g \g \rightarrow \u_v \ubar_v $ &  \texttt{HiddenValley:gg2UvUvbar} & \texttt{4902} \\
$\g \g \rightarrow \s_v \sbar_v $ &  \texttt{HiddenValley:gg2SvSvbar} & \texttt{4903} \\
$\g \g \rightarrow \c_v \cbar_v $ &  \texttt{HiddenValley:gg2CvCvbar} & \texttt{4904} \\
$\g \g \rightarrow \b_v \bbar_v $ &  \texttt{HiddenValley:gg2BvBvbar} & \texttt{4905} \\
$\g \g \rightarrow \t_v \tbar_v $ &  \texttt{HiddenValley:gg2TvTvbar} & \texttt{4906} \\
$\q \qbar \rightarrow \d_v \dbar_v $ &  \texttt{HiddenValley:qqbar2DvDvbar} & \texttt{4911} \\
$\q \qbar \rightarrow \u_v \ubar_v $ &  \texttt{HiddenValley:qqbar2UvUvbar} & \texttt{4912} \\
$\q \qbar \rightarrow \s_v \sbar_v $ &  \texttt{HiddenValley:qqbar2SvSvbar} & \texttt{4913} \\
$\q \qbar \rightarrow \c_v \cbar_v $ &  \texttt{HiddenValley:qqbar2CvCvbar} & \texttt{4914} \\
$\q \qbar \rightarrow \b_v \bbar_v $ &  \texttt{HiddenValley:qqbar2BvBvbar} & \texttt{4915} \\
$\q \qbar \rightarrow \t_v \tbar_v $ &  \texttt{HiddenValley:qqbar2TvTvbar} & \texttt{4916} \\
$\f \fbar \rightarrow \d_v \dbar_v $ &  \texttt{HiddenValley:ffbar2DvDvbar} & \texttt{4921} \\
$\f \fbar \rightarrow \u_v \ubar_v $ &  \texttt{HiddenValley:ffbar2UvUvbar} & \texttt{4922} \\
$\f \fbar \rightarrow \s_v \sbar_v $ &  \texttt{HiddenValley:ffbar2SvSvbar} & \texttt{4923} \\
$\f \fbar \rightarrow \c_v \cbar_v $ &  \texttt{HiddenValley:ffbar2CvCvbar} & \texttt{4924} \\
$\f \fbar \rightarrow \b_v \bbar_v $ &  \texttt{HiddenValley:ffbar2BvBvbar} & \texttt{4925} \\
$\f \fbar \rightarrow \t_v \tbar_v $ &  \texttt{HiddenValley:ffbar2TvTvbar} & \texttt{4926} \\
$\f \fbar \rightarrow e_v \bar e_v $ &  \texttt{HiddenValley:ffbar2EvEvbar} & \texttt{4931} \\
$\f \fbar \rightarrow \mu_v \bar \mu_v $ &  \texttt{HiddenValley:ffbar2MUvMUvbar} & \texttt{4932} \\
$\f \fbar \rightarrow \tau_v \bar \tau_v $ &  \texttt{HiddenValley:ffbar2TAUvTAUvbar} & \texttt{4933} \\
$\f \fbar \rightarrow \nu_{\tau v} \bar \nu_{\tau v} $ &  \texttt{HiddenValley:ffbar2nuEvnuEvbar} & \texttt{4934} \\
$\f \fbar \rightarrow \nu_{\tau v} \bar \nu_{\tau v} $ &  \texttt{HiddenValley:ffbar2nuMUvnuMUvbar} & \texttt{4935} \\
$\f \fbar \rightarrow \nu_{\tau v} \bar \nu_{\tau v} $ &  \texttt{HiddenValley:ffbar2nuTAUvnuTAUvbar} & \texttt{4936} \\
$\f \fbar \rightarrow \Z_v$ &  \texttt{HiddenValley:ffbar2Zv} & \texttt{4941} \\
\bottomrule
\end{tabular}
\end{table}

%% file: acronym-list.tex
\noindent{\Large \bf List of acronyms}\\
\noindent Below follows a list of standard acronyms used throughout the text, with a reference to
the page where the acronym is first introduced.\\
\begin{acronym}
	\acro{2HDM}{Two--Higgs Doublet Model}
	\acro{AQM}{Additive Quark Model}
	\acro{BSM}{Beyond Standard Model \acroextra{of Particle Physics}}
	\acro{CKKW-L}{Catani--Krauss--Kuhn--Webber--Lönnblad}
	\acro{CM}{Centre of Mass}
	\acro{CMW}{Catani--Marchesini--Webber}
	\acro{CR}{Colour Reconnection}
	\acro{DGLAP}{Dokshitzer--Gribov--Lipatov--Altarelli--Parisi}
	\acro{DIS}{Deep Inelastic Scattering \acroextra{in the context of \ep collisions}}
	\acro{DL}{Donnachie--Landshoff}
	\acro{DM}{Dark Matter}
	\acro{DPS}{Double Parton Scattering}
	\acro{EPA}{Equivalent Photon Approximation}
	\acro{EW}{Electroweak}
	\acro{FF}{Final--Final}
	\acro{FI}{Final--Initial}
	\acro{FSR}{Final--State Radiation}
	\acro{HEP}{High--Energy Physics}
	\acro{HI}{Heavy Ion}
	\acro{HV}{Hidden Valley}
	\acro{IF}{Initial--Final}
	\acro{II}{Initial--Initial}
	\acro{ISR}{Initial--State Radiation}
	\acro{LC}{Leading Colour}
	\acro{LEP}{Large Electron--Positron Collider}
	\acro{LHC}{Large Hadron Collider}
	\acro{LHA}{Les Houches Accord}
	\acro{LHE}{Les Houches Event \acroextra{\textit{also} LHEF: Les Houches Event File}}
	\acro{LO}{Leading Order}
	\acro{LIPS}{Lorentz Invariant Phase Space}
	\acro{LL}{Leading Logarithmic}
	\acro{MB}{Minimum Bias}
	\acro{MBR}{Minimum Bias Rockefeller}
	\acro{MC}{Monte Carlo}
	\acro{MCMC}{Markov Chain Monte Carlo}
	\acro{MEC}{Matrix Element Corrections}
	\acro{MPI}{Multiple Parton Interactions \acroextra{\textit{or} Multi--Parton Interactions}}
	\acro{MSSM}{Minimal Supersymmetric Simplified Model}
	\acro{NLC}{Next--to--Leading Colour \acroextra{\textit{also} NNLC \etc}}
	\acro{NLL}{Next--to--Leading Logarithmic \acroextra{\textit{also} NNLL \etc}}
	\acro{NLO}{Next--to--Leading Order \acroextra{\textit{also} NNLO, N$^{3}$LO \etc}}
	\acro{$NN$}{Nucleon--Nucleon}
	\acro{NRQCD}{Non--Relativistic Quantum Chromodynamics}
        \acro{PDF}{Parton Distribution Function}
	\acro{PDG}{Particle Data Group}
	\acro{RF}{Resonance--Final}
	\acro{RHIC}{Relativistic Heavy Ion Collider}
	\acro{RPV}{R--Parity Violating}
	\acro{SaS}{Schuler and Sjöstrand\acroextra{ \textit{or} DL/SaS}}
	\acro{SIDIS}{Semi--Inclusive DIS}
	\acro{SLHA}{SUSY Les Houches Accord}
	\acro{SM}{Standard Model \acroextra{(of Particle Physics)}}
	\acro{UMEPS}{Unitarised Matrix Element + Parton Shower}
	\acro{UPCs}{Ultra--Peripheral Collisions}
	\acro{QCD}{Quantum Chromodynamics}
	\acro{QED}{Quantum Electrodynamics}
	\acro{QGP}{Quark--Gluon Plasma}
	\acro{SUSY}{Supersymmetry \acroextra{\textit{or} Supersymmetric}}
	\acro{VBF}{Vector Boson Fusion}
	\acro{VMD}{Vector Meson Dominance}
\end{acronym}